\documentclass[manyauthors,nocleardouble]{cernphprep} 
\usepackage[english,russian]{babel}

\usepackage{colortbl}
\usepackage{tablefootnote}
\usepackage{wrapfig}
\usepackage{amssymb}
\usepackage{amsfonts}
\usepackage{amsmath}
\usepackage{epigraph}
\usepackage{longtable}
\usepackage{eurosym}
\usepackage{lscape}
\usepackage[T2A]{fontenc}
\usepackage[utf8x]{inputenc}
\usepackage{wrapfig}
\usepackage{amssymb}
\usepackage{amsfonts}
\usepackage{authblk}
\usepackage{amsmath}
\usepackage[titletoc]{appendix}
\usepackage{lineno}
\usepackage{color}
\usepackage{xcolor}
\usepackage{multirow}
\usepackage{longtable}
\usepackage{lscape}
\usepackage[T2A]{fontenc}
\usepackage[utf8x]{inputenc}
\usepackage{hyperref}
\usepackage{bigstrut}
\usepackage{isotope}
\usepackage{soul}
\originalTeX

\usepackage{ifpdf}
\ifpdf
  \usepackage[pdftex]{graphicx}
  \usepackage{epstopdf}
\else
  \usepackage{graphicx}
\fi
\DeclareGraphicsExtensions{.png,.pdf,.jpg,.mps,.eps,.tiff}
\graphicspath{{./figs/}}

\usepackage{ifpdf}
\ifpdf
  \usepackage[pdftex]{graphicx}
  \usepackage{epstopdf}
\else
  \usepackage{graphicx}
\fi
\DeclareGraphicsExtensions{.png,.pdf,.jpg,.mps,.eps}
\graphicspath{{./fig/}}
\usepackage{xspace} 
\usepackage{subfig}
\usepackage{comment}
\usepackage{fix2col}
\usepackage[sort&compress,square,comma,numbers,merge]{natbib}
\usepackage{amsmath,amssymb}
\usepackage{amsfonts}
\usepackage{upgreek} 
\usepackage{hepnames}
\usepackage{units}
\usepackage{import}
\usepackage{multirow}
\usepackage{booktabs}
\usepackage{color}
\usepackage{tikz}

\input ./sty/mymathphys.sty
\input ./sty/myunits.sty
\usepackage[textsize=tiny]{todonotes}
\newcommand{\red}[1]{{\color{red}#1}}

\begin{document}
\selectlanguage{english}


\hypersetup{pageanchor=false}   
\begin{titlepage}
\PHdate{\today}

\title{Conceptual design of the Spin Physics Detector}
\begin{center}
Version 1.1\\
\end{center}

\begin{center}
The SPD collaboration\footnote{Contact person: A. Guskov (JINR),  Alexey.Guskov@cern.ch } \\
\vspace{5cm}
\vspace{5cm}


\end{center}
\end{titlepage}

\hypersetup{pageanchor=true}


\tableofcontents

\chapter*{Preface}


According to the astrophysical and cosmological data, the relative contribution of visible baryonic matter, the properties of which are determined by strong and electromagnetic interactions, is about 5\% of  the Universe mass. With respect to two other components, dark matter and dark energy, baryonic matter seems to be a well-studied subject. In fact, despite the great advances in quantum chromodynamics made in describing the interaction of quarks and gluons within the framework of the perturbative approach, the question of why nucleons are exactly like we see them, remains open. Understanding the structure and the fundamental properties of the nucleon directly from the dynamics of its quarks and gluons based on the first principles is one of the main unsolved problems of QCD.

The nucleon behaves like a spinning top with a spin of $\hslash/2$. This spin is responsible for such fundamental properties of Nature as the nucleon magnetic moment, different phases of matter at low temperatures, the properties of neutron stars, and the stability of the known Universe. That is why the study of the spin structure of the nucleon is of particular importance. The naive quark model has successfully  predicted most of the gross properties of hadrons, such as charge, parity, isospin and symmetry properties and their relations. Some of the dynamics of particle interactions can  be qualitatively understood in terms of this model as well. However, it falls short in explaining the spin properties of hadrons in terms of their constituents.
Since the famous ''spin crisis'' that began in 1987, the problem of the nucleon spin structure remains one of the most intriguing puzzles in the contemporary high-energy physics. The central problem, for many years attracting both enormous theoretical and experimental efforts, is the problem of how the spin of the nucleon is built up from spins and orbital momenta of its constituents -- the valence and sea quarks and gluons. A full description can be given in terms of the so-called transverse-momentum dependent parton distribution functions. 

Over the last 25 years, both polarized deep inelastic scattering experiments (CERN, DESY, JLab, SLAC) and high-energy polarized proton-proton collisions (RHIC at BNL)  have been the major providers of information about spin-dependent structure functions of the nucleon. Nevertheless, our knowledge of the internal structure of the nucleon is still limited. This is especially true of the gluon contribution. New facilities for spin physics, such as the Electron-Ion Collider at BNL and the fixed-target experiments at CERN LHC are planned to be built in the near future to get the missing information.

The Spin Physics Detector, a universal facility for studying the nucleon spin structure and other spin-related phenomena with polarized proton and deuteron beams, is proposed to be placed in one of the two interaction points of the NICA collider that is under construction at the Joint Institute for Nuclear Research (Dubna, Russia). At the heart of the project there is huge experience with polarized beams at JINR.
 The main objective of the proposed experiment is the comprehensive study of the unpolarized and polarized gluon content of the nucleon.  
Spin measurements at the Spin Physics Detector at the NICA collider have bright perspectives to make a unique contribution and challenge our understanding of the spin structure of the nucleon. 

In this document the Conceptual Design of the Spin Physics Detector is presented.

\chapter{Executive summary}

\begin{figure}[htbp]
  \begin{center}
    \includegraphics[width=0.8\textwidth]{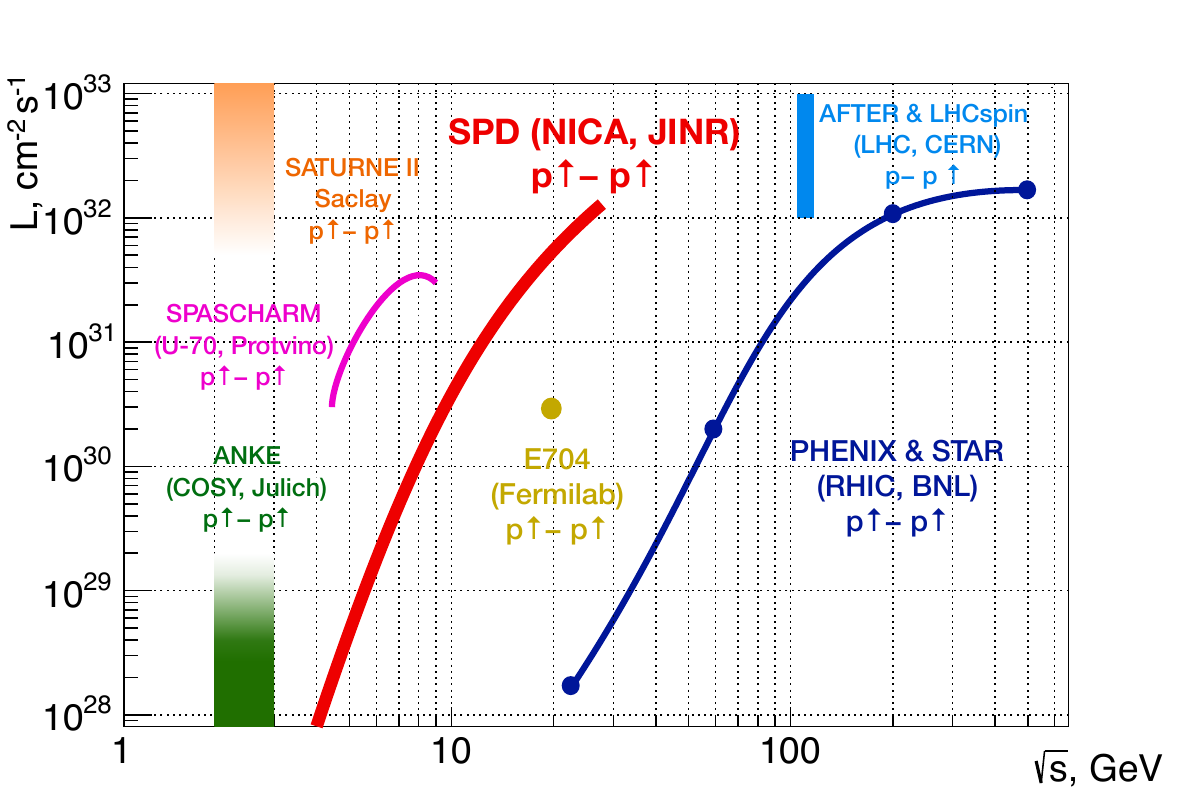}
  \end{center}
  \caption{NICA SPD and the other past, present, and future experiments with polarized protons.}
  \label{S_L}
\end{figure}

The Spin Physics Detector (proto-)collaboration proposes to install a universal detector in the second interaction point of the NICA collider under construction (JINR, Dubna)  to study the spin structure of the proton and deuteron and the other spin-related phenomena with polarized proton and deuteron beams at a collision energy up to 27 GeV and a luminosity up to $10^{32}$ cm$^{-2}$ s$^{-1}$. In the polarized proton-proton collisions, the SPD experiment \cite{SPD_URL} at NICA will cover the kinematic gap between the low-energy measurements at ANKE-COSY and SATURNE and the high-energy measurements at the Relativistic Heavy Ion Collider, as well as the planned fixed-target experiments at the LHC (see Fig.  \ref{S_L}). The possibility for NICA to operate with polarized deuteron beams at such energies is unique. 

 SPD is planned to operate as a universal facility for comprehensive study of the unpolarized and polarized gluon content of the nucleon at large Bjorken-$x$, using different complementary probes such as: charmonia, open charm, and prompt photon production processes. 
 The experiment aims at providing access to the gluon helicity, gluon Sivers, and Boer-Mulders functions in the nucleon, as well as the gluon transversity distribution and tensor PDFs in the deuteron, via the measurement of specific single and double spin asymmetries (see Tab. \ref{gluonPDFs}). The results expected to be obtained by SPD will play an important role in the general understanding of the nucleon gluon content and will serve as a complementary input to the ongoing and planned studies at RHIC, and future measurements at the EIC (BNL) and fixed-target facilities at the LHC (CERN). Simultaneous measurement of the same quantities using different processes at the same experimental setup is of key importance for minimization of possible systematic effects. Other polarized and unpolarized physics is possible, especially at the first stage of NICA operation with reduced luminosity and collision energy of the proton and ion beams.

\begin{table}[htp]
\caption{Gluon TMD PDFs at twist-2. The columns represent gluon polarization, while the rows represent hadron polarization. The PDFs planned to be addressed at SPD are highlighted in red.}
\begin{center}
\begin{tabular}{|c|c|c|c|}
\hline
   &  Unpolarized & Circular & Linear \\
\hline   
Unpolarized & \red{g(x)} &           &   \red{$h_1^{\perp g}(x,k_T)$ }         \\
                     & \red{density} &  & \red{Boer-Mulders function} \\
\hline
Longitudinal &        & \red{$\Delta g(x)$} &       Kotzinian-Mulders       \\
                    &        &  \red{helicity} &     function         \\
\hline                    
Transverse  & \red{$\Delta_N^g(x,k_T)$}    &     Worm-gear       & \red{$\Delta_T g(x)$}   \\
                    &\red{ Sivers function}       &   function          & \red{transversity (deuteron only)},\\
                    &                                         &             & pretzelosity     \\
\hline
\end{tabular}
\end{center}
\label{gluonPDFs}
\end{table}%

 \begin{figure}[!h]
  \begin{center}
    \includegraphics[width=0.8\textwidth]{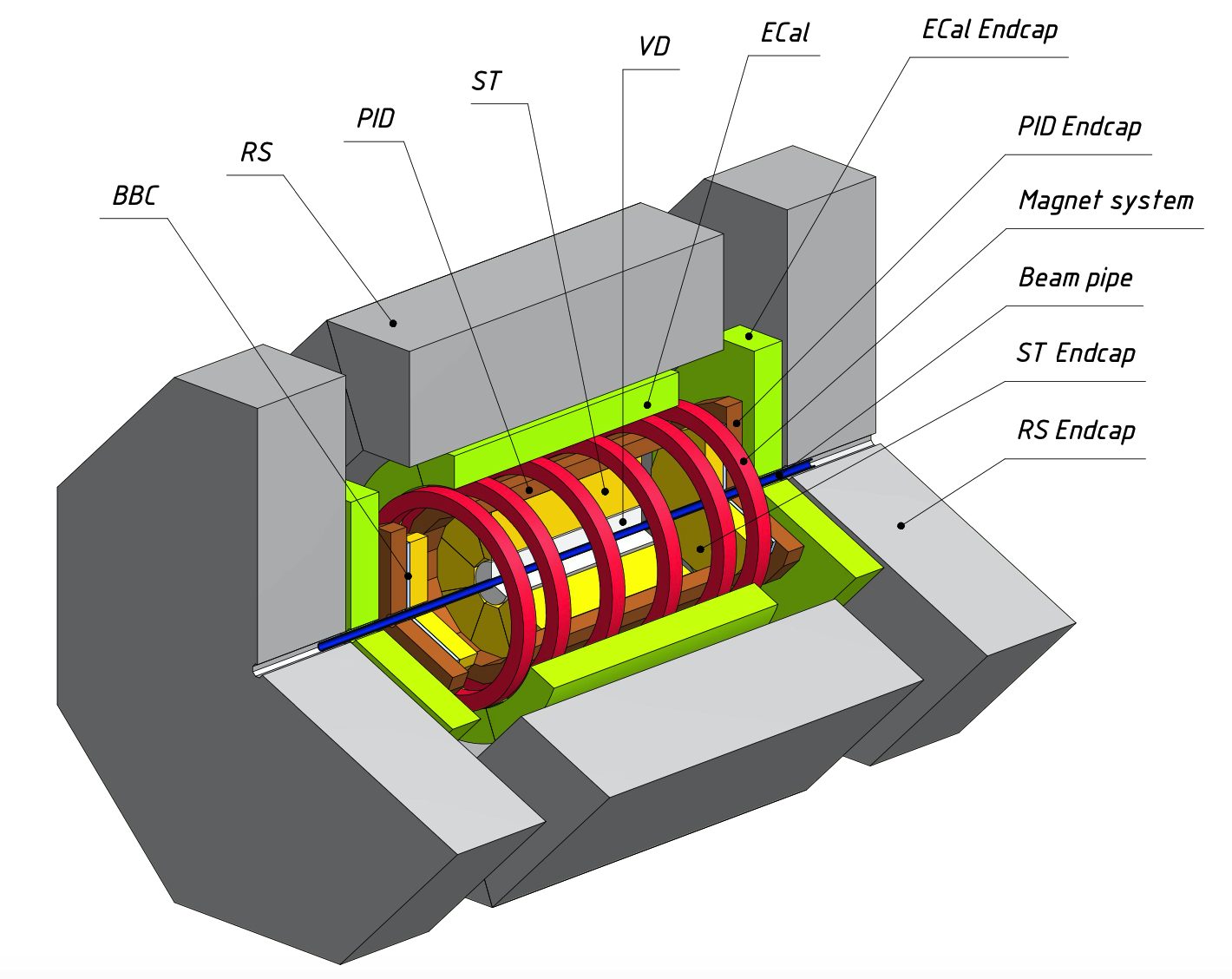}
  \end{center}
  \caption{General layout of the SPD setup.}
  \label{SPD-LA_EX}
\end{figure}
The physics goals dictate the layout of the detector. The SPD experimental setup is being designed as a universal $4\pi$ detector with advanced tracking and particle identification capabilities based on modern technologies. The silicon vertex detector (VD) will provide resolution for the vertex position on the level of below 100 $\mu$m needed for reconstruction of secondary vertices of $D$-meson decays. The straw tube-based tracking system (ST) placed within a solenoidal magnetic field of up to 1 T at the detector axis should provide the transverse momentum resolution $\sigma_{p_T}/p_T\approx 2\%$ for a particle momentum of 1 GeV/$c$. The time-of-flight system (PID) with a time resolution of about 60 ps will provide $3\sigma$ $\pi/K$ and $K/p$ separation of up to about 1.2 GeV/$c$ and 2.2 GeV/$c$, respectively. Possible use of the aerogel-based Cherenkov detector could extend this range. Detection of photons will be provided by the sampling electromagnetic calorimeter (ECal) with the energy resolution $\sim 5\%/\sqrt{E}$. To minimize multiple scattering and photon conversion effects for photons, the detector material will be kept to a minimum throughout the internal part of the detector. The muon (range) system (RS) is planned for muon identification. It can also act as a rough hadron calorimeter. The pair of beam-beam counters (BBC) and zero-degree calorimeters will be responsible for the local polarimetry and luminosity control.  To minimize possible systematic effects, SPD will be equipped with a triggerless DAQ system. A high collision rate (up to 4 MHz) and a few hundred thousand detector channels pose a significant challenge to the DAQ,  online monitoring, offline computing system, and data processing software.

The proposed physics program covers at least 5 years of the SPD running.

 The estimated cost of the Spin Physics Detector at current prices is about 96 M\$.  This value does not cover the R\&D expenses and the construction of the SPD Test zone. Any expenses related to the development and construction of the infrastructure for polarized beams at NICA are also out of this estimation.

\chapter{Physics case \label{chapter2}}
\section{Gluons in proton and deuteron}
Gluons, together with quarks, are the fundamental constituents of the nucleon. 
 They play a key role in generation of its mass and carry about half of its momentum in hard (semi)inclusive processes. 
The spin of the nucleon is also built up from the intrinsic spin of the valence and sea quarks (spin-1/2), gluons (spin-1), and their orbital angular momenta. Notwithstanding the progress achieved during the last decades in the understanding of the quark contribution to the nucleon spin, the gluon sector is much less developed. One of the difficulties is the lack of the direct probes to access gluon content in high-energy processes. 
While the quark contribution to the nucleon spin was determined quite precisely in semi-inclusive deep-inelastic scattering (SIDIS) experiments like EMC, HERMES, and COMPASS, the gluon contribution is still not well-constrained even so it is expected to be significant. 

In recent years, the three-dimensional partonic structure of the nucleon became a subject of a careful study. Precise mapping of three-dimensional structure of the nucleon is crucial for our understanding of Quantum Chromodynamics (QCD). One of the ways to go beyond the usual collinear approximation is to describe nucleon content in the momentum space employing the so-called Transverse-Momentum-Dependent Parton Distribution Functions (TMD PDFs)~\cite{Kotzinian:1994dv,Mulders:1995dh,Boer:1997nt,Goeke:2005hb,Bacchetta:2006tn,Angeles-Martinez:2015sea}.

The most powerful tools to study TMD PDFs are the measurements of the nucleon spin (in)dependent azimuthal asymmetries in SIDIS~\cite{Kotzinian:1994dv,Goeke:2005hb,Bacchetta:2006tn,Bastami:2018xqd} and Drell--Yan processes \cite{Arnold:2008kf,Bastami:2020asv}. Complementary information on TMD fragmentation process, necessary for the interpretation of SIDIS data, is obtained from $e^+e^-$ measurements~\cite{Metz:2016swz}.
Being an actively developing field, TMD physics triggers a lot of experimental and theoretical interest all over the world, stimulating new measurements and developments in TMD extraction techniques oriented on existing and future data from lepton-nucleon, electron-positron and hadron-hadron facilities at BNL, CERN, DESY, FNAL, JLab, and KEK. For recent reviews on experimental and theoretical advances on TMDs see Refs.~\cite{Anselmino:2020vlp,Avakian:2019drf,Perdekamp:2015vwa,Boglione:2015zyc,Aidala:2012mv}.
While a lot of experimental measurements were performed (and are planned) and theoretical understanding was achieved for Leading Order (LO) (twist-2) TMD PDFs such as Sivers, transversity and Boer-Mulders functions of quarks, only few data relevant for the study of gluon TMD PDFs are available~\cite{Adare:2013ekj,Adare:2010bd,Aidala:2018gmp,Aidala:2017pum,Adolph:2017pgv,Szabelski:2016wym}.
 
The simplest model of the deuteron is a weakly-bound state of a proton and a neutron mainly in the S-wave with a small admixture of the D-wave state. This approach is not much helpful in the description of the deuteron structure at large $Q^2$ \footnote{We use $Q^2$ (or $\mu^2$) as a generic notation for the hard scale of a reaction: the invariant mass square of lepton pairs in Drell-Yan processes, $Q^2$, transverse momentum square $p_T^2$ of produced hadron or its mass square $M^2$.}.  
Possible non-nucleonic degrees of freedom in deuteron could play an important role in the understanding of the nuclear modification of PDFs (the EMC effect).
Since the gluon transversity operator requires two-unit helicity-flip it does not exist for spin-1/2 nucleons~\cite{Barone:2001sp}. Therefore, proton and neutron gluon transversity functions can not contribute directly to the gluon transversity of the deuteron. A non-zero deuteron transversity could be an indication of a non-nucleonic component or some other exotic hadronic mechanisms within the deuteron.

Most of the existing experimental results on spin-dependent gluon distributions in nucleon are obtained in the experiments at DESY (HERMES), CERN (COMPASS), and BNL (STAR and PHENIX). Study of polarized gluon content of the proton and nuclei is an important part of future projects in Europe and the United States such as AFTER@LHC and LHCSpin at CERN, and EIC at BNL~\cite{Hadjidakis:2018ifr,Aidala:2019pit,Accardi:2012qut}. Notwithstanding the gluons in nucleon were successfully probed in SIDIS measurements, hadronic collisions have an important advantage since they probe the gluons at the Born-level without involving the EM couplings. 

\subsection{Gluon probes at NICA SPD \label{probes}}

\begin{figure}[!t]
  \begin{minipage}[ht]{0.50\linewidth}
  \center{\includegraphics[width=1\linewidth]{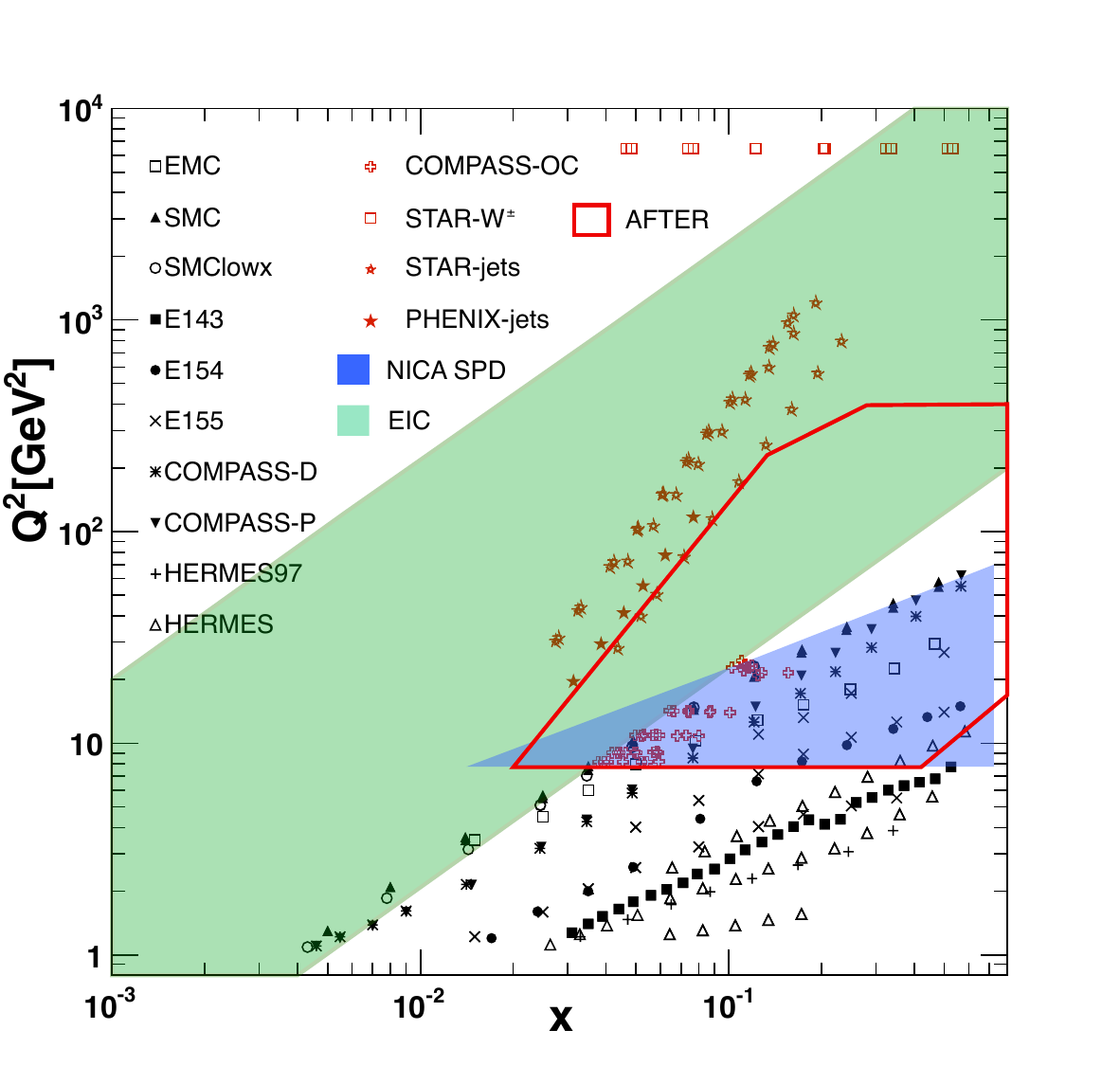}\\ (a)}
  \end{minipage}
  \begin{minipage}[ht]{0.46\linewidth}
 \center{ \includegraphics[width=1\linewidth]{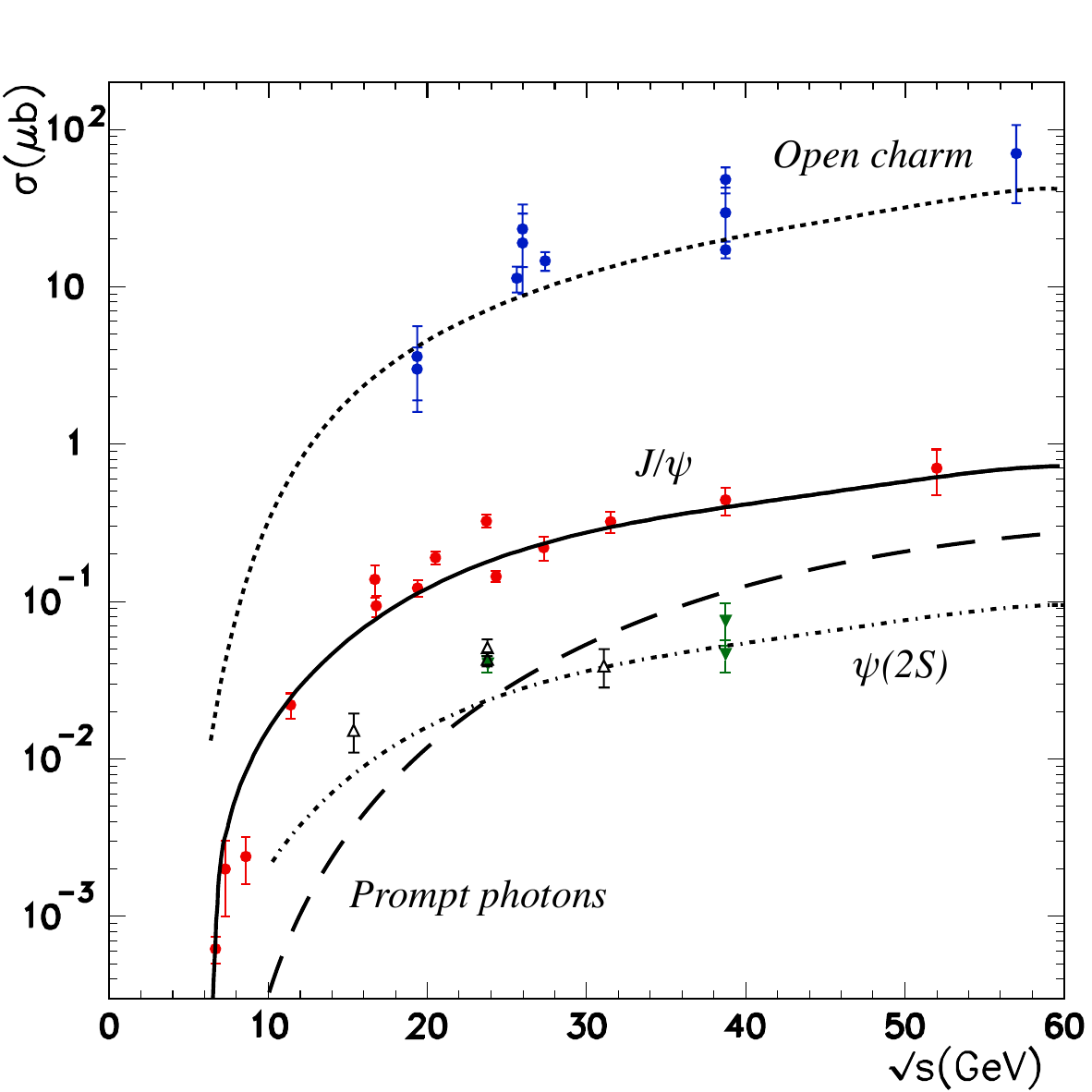}\\ (b)}
  \end{minipage}
    \caption{(a) The kinematic coverage, in the ($x$, $Q^2$) plane, of the hadronic cross-section data for the processes commonly included in global QCD analyses of polarized quark (black) and gluon (red) PDFs~\cite{Nocera:2014gqa}. The kinematic domain expected to be covered by NICA SPD by charmonia, open charm and prompt-photon production is shown in blue. (b) Cross-section for the processes of open charm, $J/\psi$, $\psi(2S)$ and prompt photons ($p_T>3$~GeV) production as a function of center-of-mass energy (based on~\cite{BrennerMariotto:2001sv}).}
  \label{fig:kinematic}  
\end{figure}

The polarized gluon content of proton and deuteron at intermediate and high values of the Bjorken $x$ will be investigated using three complementary probes: inclusive production of charmonia, open charm, and prompt photons. 
Study of these processes is complementary to such proven approaches to access the partonic structure of the nucleon in hadronic collisions as the inclusive production of hadrons with high transverse momentum and the Drell-Yan process. Unfortunately, the latter one is unlikely to be accessible at SPD due to the small cross-section and unfavourable background conditions.
 For effective registration of each aforementioned gluon probes, the SPD setup is planned to be equipped with a range (muon) system, an electromagnetic calorimeter, a time-of-flight system, straw tracker, and a silicon vertex detector. Nearly a $4\pi$ coverage of the setup and a low material budget in the inner part of the setup should provide a large acceptance for the detection of the desired final states.  
In Fig.~\ref{fig:kinematic}(a) the kinematic phase-space in $x$ and $Q^2$ to be accessed by the SPD is compared to the corresponding ranges of previous, present and future experiments. Parameters of the experimental facilities planning to contribute to gluon physics with polarized beams are listed in Tab. \ref{tab:facilities}. 
 Figure \ref {fig:kinematic}(b) illustrates the behavior of the cross-sections for the inclusive production of $J/\psi$, $\psi(2S)$, $D$-mesons and high-$p_T$ prompt photons in $p$-$p$ collisions as a function of $\sqrt{s}$.

\begin{table}[t!]
\caption{Main present and future actors in gluon spin physics.}
\begin{center}
\begin{tabular}{|l|c|c|c|c|c|}
\hline \bigstrut 
 Experimental  & SPD         & RHIC~\cite{RHIC} & EIC~\cite{Accardi:2012qut}& AFTER  & SpinLHC \bigstrut \\  
 facility            &  @NICA ~\cite{Meshkov:2019utc}    &       &         &   @LHC~\cite{Hadjidakis:2018ifr}  &    \cite{Aidala:2019pit}    \bigstrut \\
\hline  
Scientific center & JINR     & BNL               & BNL & CERN & CERN   \bigstrut \\
\hline
Operation mode & collider & collider        & collider & fixed  & fixed \bigstrut \\
                           &             &                   &             & target & target  \bigstrut \\
\hline                           
Colliding particles& $p^{\uparrow}$-$p^{\uparrow}$ & $p^{\uparrow}$-$p^{\uparrow}$  & $e^{\uparrow}$-$p^{\uparrow},d^{\uparrow},^3$He$^{\uparrow}$ & $p$-$p^{\uparrow}$,$d^{\uparrow}$  & $p$-$p^{\uparrow}$ \bigstrut  \\                       
\& polarization      & $d^{\uparrow}$-$d^{\uparrow}$&                                 &                &                 &        \bigstrut \\
     & $p^{\uparrow}$-$d$, $p$-$d^{\uparrow}$&                                 &                &                      &   \bigstrut \\
\hline
Center-of-mass            &  $\leq$27 ($p$-$p$) &  63, 200,      & 20-140 ($ep$)&  115 & 115 \bigstrut \\
energy $\sqrt{s_{NN}}$, GeV&  $\leq$13.5 ($d$-$d$)     &   500      &       &   &  \bigstrut \\
                           &  $\leq$19 ($p$-$d$)     &         &       &   &  \bigstrut \\
\hline
Max. luminosity,                  & $\sim$1 ($p$-$p$)  &  2  & 1000 &up to & 4.7  \bigstrut \\   
10$^{32}$ cm$^{-2}$ s$^{-1}$  & $\sim$0.1 ($d$-$d$) &        &  &   $\sim$10 ($p$-$p$) &  \bigstrut \\
\hline
Physics run            & $>$2025  & running & $>$2030  &  $>$2025 & $>$2025 \bigstrut \\
\hline
\end{tabular}
\end{center}
\label{tab:facilities}
\end{table}%
\begin{figure}[!t]
  \begin{minipage}[ht]{0.325\linewidth}
    \center{\includegraphics[width=1\linewidth]{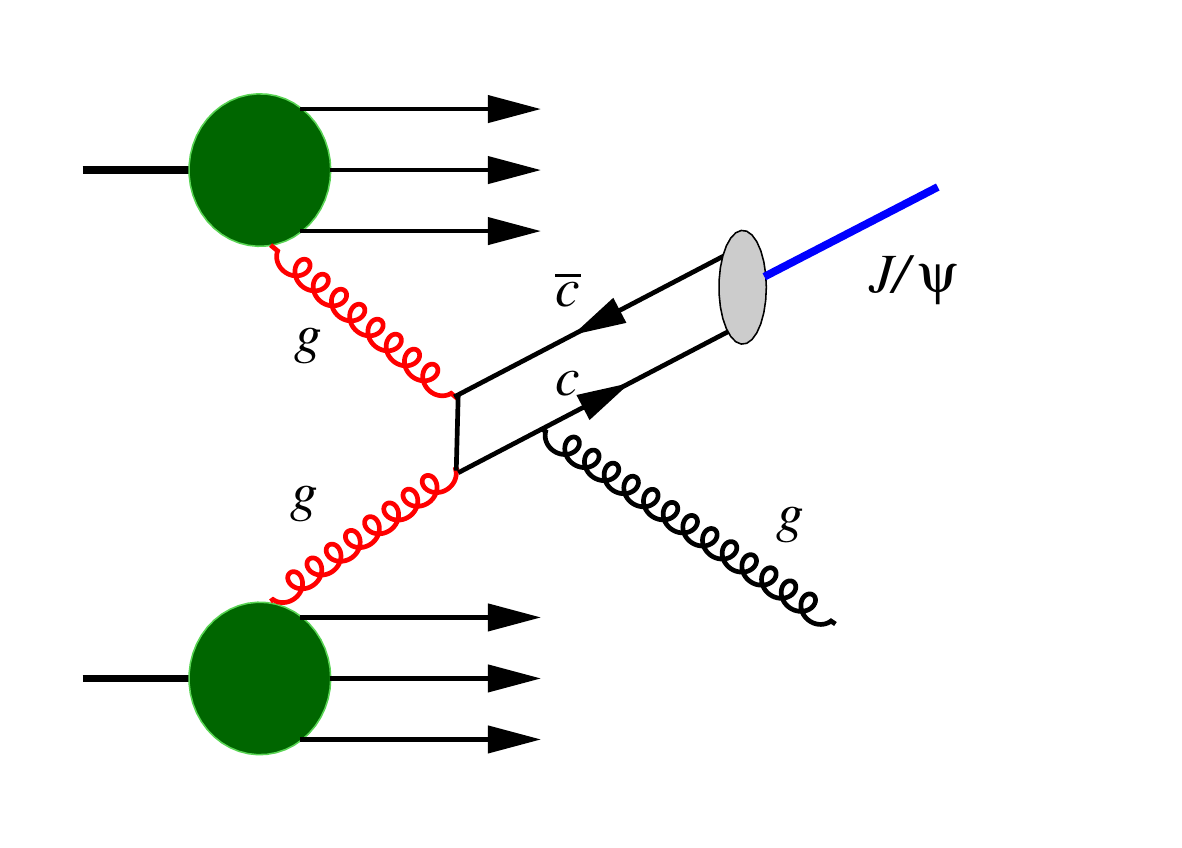} \\ (a)}
  \end{minipage}
  \begin{minipage}[ht]{0.325\linewidth}
    \center{\includegraphics[width=1\linewidth]{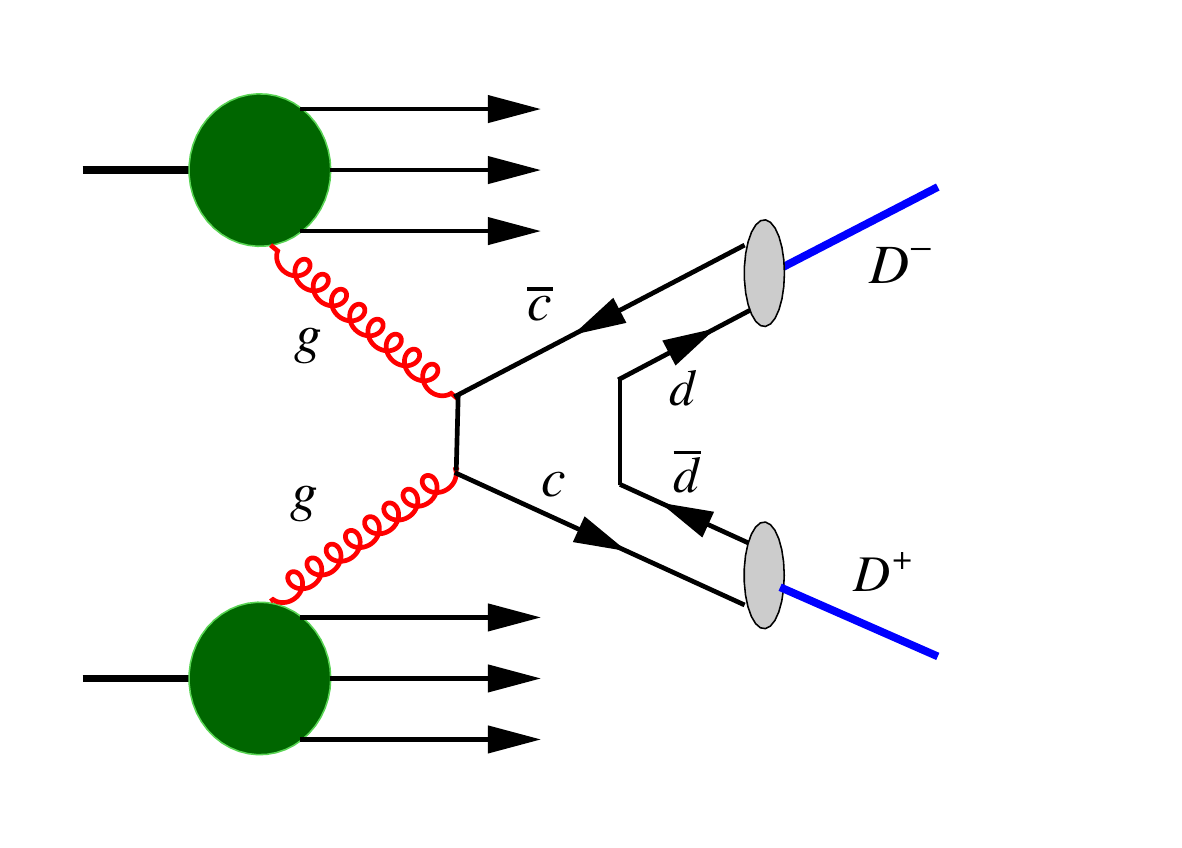} \\ (b)}
  \end{minipage}
  \begin{minipage}[ht]{0.335\linewidth}
    \center{\includegraphics[width=1\linewidth]{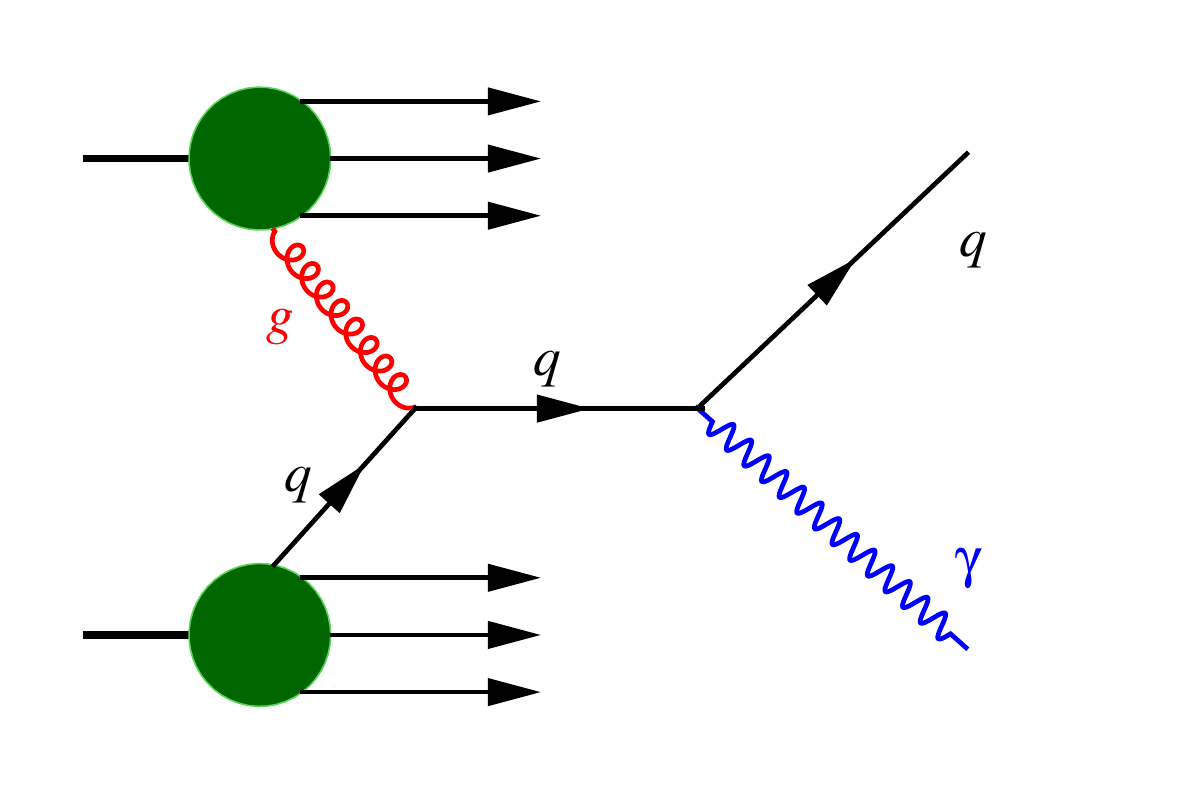} \\ (c)}
  \end{minipage}
  \caption{ Diagrams illustrating three probes to access the gluon content of proton and deuteron in polarized collisions at NICA SPD: production of (a) charmonium, (b) open charm, and (c) prompt photons.}
  \label{fig:diagrams-gg}  
\end{figure}

We do not discuss separately such gluon probe as inclusive production of neutral and charged pions and other light mesons, for which the $qg\to qg$ hard process dominates in a certain kinematic region. They have been successfully used to access the polarized gluon content of the proton at the RHIC experiments and can of course  be used at SPD for this purpose. Registration of these processes does not impose additional specific requirements on the experimental setup and can be performed in parallel with the aforementioned main probes.

\subsubsection{Charmonia production\label{probes_ch}}

 From the experimental point of view, for considered energies, hadronic production of charmonia seems to be particularly suited to access gluon content in hadrons. Production of prompt $J/\psi$-mesons looks most attractive, since large data set of $J/\psi\to\mu^+ \mu^-$ (the branching fraction $BF$ is 0.06 \cite{Zyla:2020zbs}) events is accumulated in beam-dump experiments with proton and pion beams at $\sqrt{s}$ close to 20 GeV. However $J/\psi$-meson is not the cleanest probe of the proton structure, since a significant fraction (about $20\%$ ~\cite{Abt:2008ed}) of $J/\psi$-mesons observed in hadronic collisions is produced indirectly through decays of $\chi_{cJ}$ and $\psi(2S)$ (the so-called feed-down contribution), and modeling of this contribution introduces additional uncertainties in theoretical calculations. Hence, to provide additional constraints to production models, it is important to study the production of $\chi_{cJ}$ and $\psi(2S)$ separately, through their decays $\chi_{cJ}\to \gamma J/\psi$ ($BF=0.014$, $0.343$ and $0.19$ for $J=0,1,2$ \cite{Zyla:2020zbs}) and $\psi(2S)\to J/\psi \pi^+ \pi^-$ ($BF=0.347$ \cite{Zyla:2020zbs}). The latter state is of special interest, because it is essentially free from feed-down contamination from higher charmonium states, due to the proximity of $D^0\overline{D}^0$-threshold. However, the separation of the $\chi_{c0,1,2}$ signals is a challenging experimental task due to the small mass difference between the states and low energy resolution of the electromagnetic calorimeters for soft photons.  

 Besides, from the theoretical point of view the task of accessing gluon distributions using heavy quarkonia is rather challenging. The heavy quark-antiquark pair couples directly to gluons from initial-state hadrons (Fig.~\ref{fig:diagrams-gg}(a)) and its production can be calculated perturbatively, because the hard scale of the process is limited from below by the heavy quark mass, providing direct access to polarized and un-polarized gluon distributions. However, the process of transition of the heavy quark-antiquark pair into a physical bound-state is not well understood at present and can become a source of significant theoretical uncertainties. We review the modern status of the theory of quarkonium production in more detail in Sec.~\ref{quarkonia_th} to explain the latter point. 

 Therefore, quarkonium production can be used to study the structure of hadrons only with a great caution and only if results consistent with other probes will eventually emerge. The studies of hadronic structure and heavy quarkonium production mechanism should become complementary. But for now the most reasonable phenomenological strategy for measurements at SPD concerning quarkonia is to study yields and polarization of different quarkonium states in a wide kinematic range, at various energies, and in polarized as well as unpolarized hadronic collisions, to constrain the theoretical models.  When the theory of production of heavy quarkonia is firmly established -- it will become an invaluable tool to study the details of hadronic structure.

\subsubsection{Open charm production\label{probes_och}}
It is well-known that heavy flavor production offers direct probes of the gluon distributions in hadrons. The basic mechanism responsible for charm pair production in $pp$ collisions is the gluon fusion (GF, see Fig.~\ref{fig:diagrams-gg}(b)). In the framework of pQCD, the GF contributes to the hadron cross-section as ${\cal L}_{gg}\otimes {\hat \sigma}_{c\bar{c}}$, where the gluon luminosity ${\cal L}_{gg}$ is a convolution of the gluon densities in different protons, ${\cal L}_{gg}=g\otimes g$. At leading order in pQCD, ${\cal O}(\alpha_s^2)$, the partonic cross-section ${\hat \sigma}_{c\bar{c}}$ describes the process $gg\rightarrow c\bar{c}$.  

The GF contribution to the charmonia production in $pp$ collisions has the form ${\cal L}_{gg}\otimes {\hat \sigma}_{(c\bar{c})+X}\otimes W_{c\bar{c}}$. At the Born level, the partonic cross-section ${\hat \sigma}_{(c\bar{c})+X}$ is of the order of $\alpha_s^3$ because its basic subprocess is $gg\rightarrow (c\bar{c})+g$. Moreover, the quantity $W_{c\bar{c}}$, describing the probability for the charm pair to form a charmonium, imposes strong restrictions on the phase space of the final state.\footnote{To form a charmonium, the momenta of the produced quark and antiquark should be sufficiently close to each other.} For these two reasons, the $\alpha_s$-suppression and phase space limitation, the cross-sections for charmonia production are almost two orders of magnitude smaller than the corresponding ones for open charm, see Figs.~\ref{fig:kinematic} (b). 


To analyze the kinematics of a $D\overline{D}$ pair, each $D$-meson has to be reconstructed. The decay modes $D^+\to \pi^+K^-\pi^+$ (BF=0.094) and $D^0\to K^-\pi^+$ (BF=0.04) can be used for that. To suppress a combinatorial background SPD plans to use the search for a secondary vertex of a $D$-meson decay that is about 100 $\mu$m far from the interaction point (the $c\tau$ values are 312 and 123 $\mu$m for the charged and neutral $D$-mesons, respectively). The identification of a charged kaon in the final state by the time-of-flight system would also help to do that. The production and decay of $D^*$-mesons could be used as  additional tags for open-charm events. Single-reconstructed $D$-mesons also carry reduced but still essential information about gluon distribution that is especially important in the low-energy region lacking statistics.

\subsubsection{Prompt photon production} \label{probes_pp}
Photons emerging from the hard parton scattering subprocess, the so-called prompt photons, serve as a sensitive tool to access the gluon structure of hadrons and hadron-hadron collisions.
Inclusive direct photon production proceeds without fragmentation, i.e. the photon carries the information directly from the hard scattering process. Hence this process measures a combination of initial $k_T$ effects and hard scattering twist–3 processes. There are two main hard processes for the production of direct photons: gluon Compton scattering, $gq(\bar{q})\rightarrow \gamma q(\bar{q})$ (Fig.~\ref{fig:diagrams-gg}(c)), which dominates, and  quark-antiquark annihilation, $q \bar{q} \rightarrow \gamma g$. Contribution of the latter process to the total cross-section is small. 

\begin{figure}[!t]
  \begin{minipage}[ht]{0.50\linewidth}
  \center{\includegraphics[width=1\linewidth]{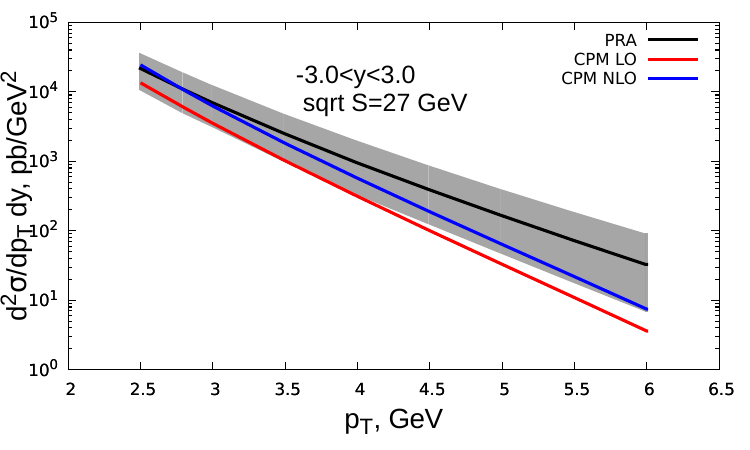}\\ (a)}
  \end{minipage}
  \begin{minipage}[ht]{0.46\linewidth}
 \center{ \includegraphics[width=1\linewidth]{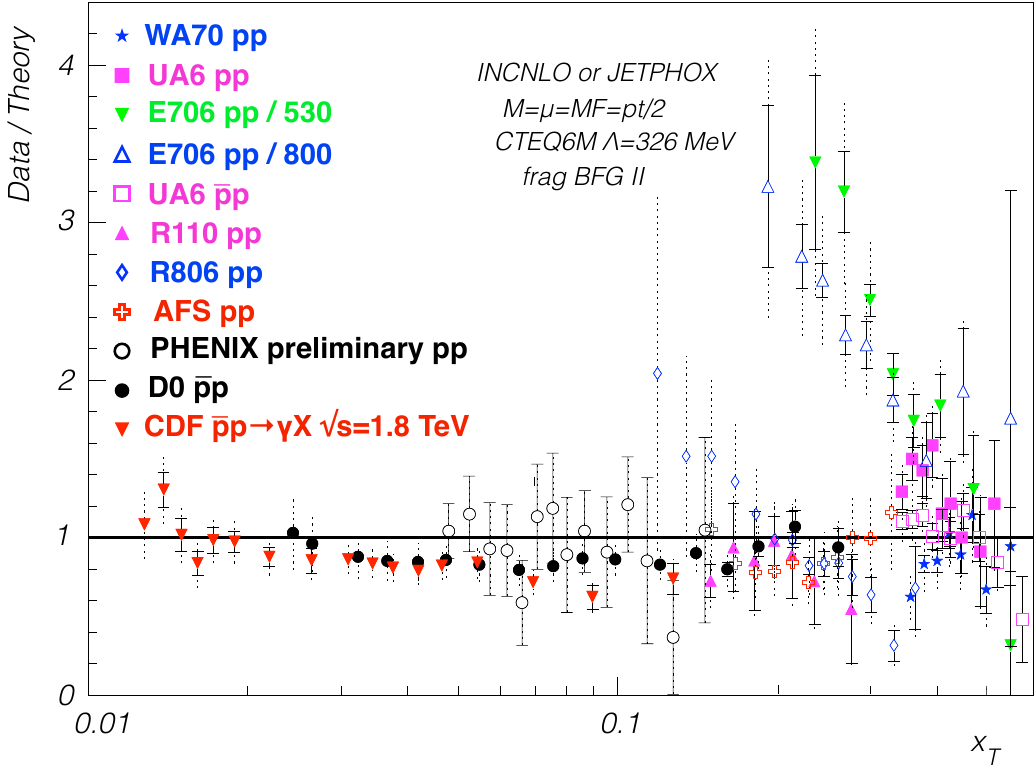}\\ (b)}
  \end{minipage}
    \caption{(a) Prediction for prompt photon transverse momentum spectrum at $\sqrt{s}=27$ GeV obtained in LO (red line) and NLO (blue line) approximations of CPM and LO of PRA (black line). Uncertainty bands for PRA predictions are due to factorization/renormalization scale variation only. (b) Data-to-theory ratio for the fixed-target and collider experiments \cite{Aurenche:2006vj}.}
  \label{fig:NICA-photon-pT}  
\end{figure}

Theoretical predictions for inclusive prompt photon production are shown in Fig. \ref{fig:NICA-photon-pT}(a) as transverse momentum spectrum at the energy $\sqrt{s}=27$ GeV. Calculations are performed in LO and NLO approximations of the Collinear Parton Model (CPM), as well as in the Parton Reggeization Approach (PRA), which is a QCD and QED gauge-invariant version of $k_T$-factorization. They include direct and fragmentation contributions, the latter one is about 15-30 \%. The K-factor between LO and NLO calculations in the CPM slightly depends on $p_{T\gamma}$ and equals about 1.8 \cite{Wong:1998pq}. LO prediction of PRA coincides with the result of NLO CPM calculation at moderate transverse momenta ($p_{T}<4$ GeV) while at higher $p_T$ PRA predicts somewhat harder $p_T$-spectrum. 

In experiments prompt photons are detected alongside with a much larger number of photons from decays of secondary $\pi^0$ and $\eta$ mesons (minimum-bias photons). The main challenge is to subtract these decay contributions to obtain the photons directly emitted from hard collisions. This kind of background is especially important at small transverse momenta of produced photons ($p_T$) and gives the lower limit of the accessible $p_T$ range. Therefore the prompt-photon contribution with $p_T\leq 2-3$ GeV is usually unreachable in the experiment~\cite{Vogelsang:1997cq}. Figure \ref{fig:NICA-photon-pT}(b) \cite{Aurenche:2006vj} presents the comparison of the $p_T$ spectra ($x_T=2 p_T/\sqrt{s}$) measured in a wide kinematic range of $\sqrt{s}$ in different fixed-target and collider experiments and the theoretical NLO calculations performed within the JETPHOX package \cite{Binoth:1999qq}.  While high-energy collider results exhibit rather good agreement with expectations, situation at high-$x_T$ is not pretty good.
The results of the E706 ($\sqrt{s}=31.6$ and $38.8$ GeV) \cite{Apanasevich:1997hm} and R806 ($\sqrt{s}=63$ GeV) \cite{Anassontzis:1982gm} experiments break out the trend and demonstrate some ''slope''. It could be an indication of possible systematic effects that have not been yet fully understood.

 A pair of prompt photons can be produced in hadronic interactions in $q\bar{q}$ annihilation, quark-gluon scattering, and gluon-gluon fusion hard processes (at the leading, next-to-leading, and next-to-next-leading orders, respectively).
The double prompt photon production in nucleon interactions at low energies is not yet well-studied experimentally. The production cross-section for proton-carbon interaction at $\sqrt{s}=19.4$ GeV/$c$ has been measured by the CERN NA3 experiment~\cite{Badier:1985vm}. Based on this result we can expect the cross-section of the double photon production with $p_T>2$~GeV/$c$ for each photon on the level of about 0.5~nb.

\begin{table}[!h]
\caption{Expected cross-section and counts for each of the gluon probes per one effective year of SPD running ($10^7$ s). Detector acceptance and reconstruction efficiency are not taken into account.} 
\begin{center}
\begin{tabular}{|l|c|c|c|c|}
\hline \bigstrut 
                                &  $\sigma$\textsubscript{27\,GeV},  & $\sigma$\textsubscript{13.5\,GeV}, & $N$\textsubscript{27\,GeV}, & $N$\textsubscript{13.5\,GeV}  \bigstrut \\                    
   Probe                   &      nb ($\times$BF)                         &  nb   ($\times$BF)                             &   10$^6$             &     10$^6$       \bigstrut \\
\hline
\hline
Prompt-$\gamma$ ($p_T>3$ GeV/c) & 35          & 2 & 35 &  0.2 \bigstrut \\
\hline
\hline
$J/\psi$                    & 200          & 60 &  &   \bigstrut \\
~~~$\to\mu^+\mu^-$ & 12         & 3.6 & 12 & 0.36  \bigstrut \\
\hline
 $\psi(2S)$    &  25      &  5     &  &   \bigstrut \\
~~~$\to J/\psi \pi^+\pi^- \to \mu^+\mu^- \pi^+\pi^-$            &  0.5      &  0.1     & 0.5 & 0.01  \bigstrut \\
~~~$\to\mu^+\mu^-$                  &  0.2      &  0.04     & 0.2 & 0.004  \bigstrut \\
\hline
$\chi_{c1}$ + $\chi_{c2}$ & 200& & & \bigstrut \\
~~~$\to \gamma J/\psi \to \gamma \mu^+\mu^-$ & 2.4 & & 2.4 & \bigstrut \\
\hline
$\eta_c$ & 400 & & & \bigstrut \\
~~~$\to p\bar p$ & 0.6 & & 0.6 & \bigstrut \\
\hline
\hline
Open charm: $D\overline{D}$ pairs       			     & $14000$& 1300&   & \bigstrut \\
Single $D$-mesons       			     & & &  & \bigstrut \\
~~~$D^{+}\to K^-2\pi^+$ ($D^{-}\to K^+2\pi^-$)           &   520            &   48      & 520 & 4.8  \bigstrut \\
~~~$D^{0} \to K^- \pi^+$ ($\overline{D}^{0} \to K^+ \pi^-$)           &   360             &    33     & 360 & 3.3 \bigstrut \\

\hline
\end{tabular}
\end{center}
\label{tab:crossections}
\end{table}%

Estimations of the expected event rates are evaluated for $p$-$p$ collisions at $\sqrt{s}=27$ and $13.5$~GeV for the projected integrated luminosity 1.0  and 0.1~fb$^{-1}$, respectively that corresponds effectively to one year of data taking (10$^7$ s).  The results are listed in Tab.~\ref{tab:crossections}.

 \begin{figure}[!h]
  \begin{minipage}[ht]{0.50\linewidth}
    \center{\includegraphics[width=1\linewidth]{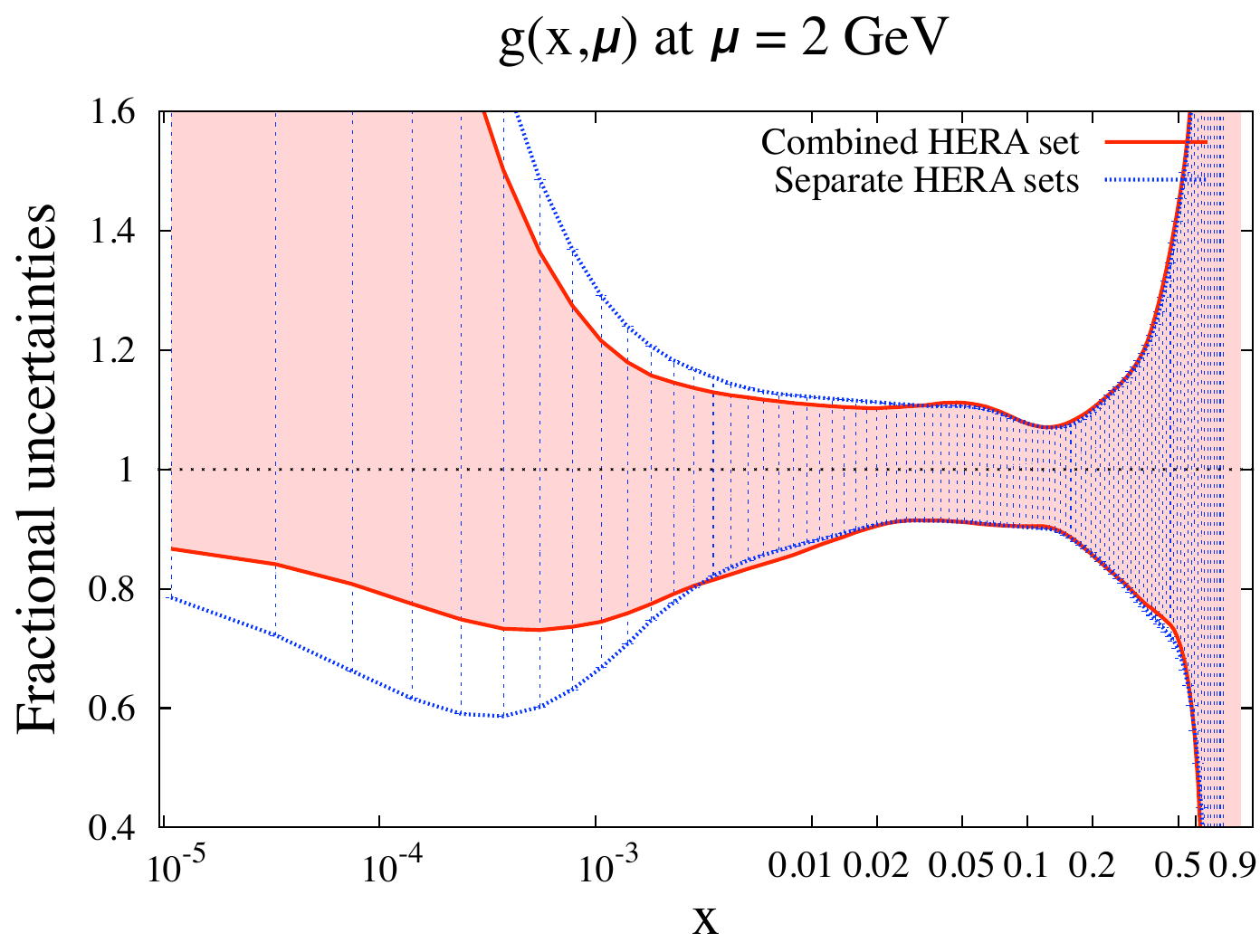} \\ (a)}
  \end{minipage}
  \hfill
  \begin{minipage}[ht]{0.50\linewidth}
  \vspace{+1pt}
    \center{\includegraphics[width=1.\linewidth]{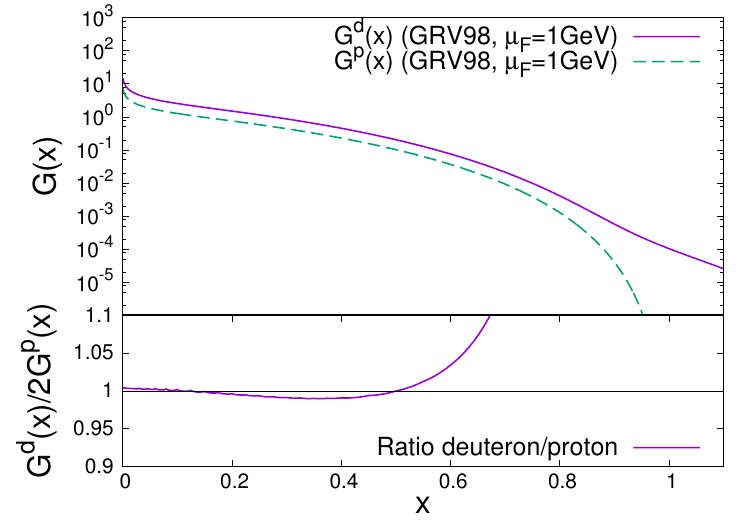} \\ (b)}
  \end{minipage}
  \caption{(a) Uncertainty of unpolarized gluon PDF based on HERA data ($\mu=2$ GeV)~\cite{Lai:2010vv}. (b) Gluon PDF in the deuteron in comparison with the nucleon~\cite{Brodsky:2018zdh}.}
  \label{fig:unpolarized} 
\end{figure}


\subsection{Gluons at large $x$ \label{un_pdf}}

The gluon PDF is one of the less known parton distributions in the proton because the available data constrain weakly the quantity $g(x,Q^2)$, particularly for $x$ greater than 0.5~\cite{Abt:2020mnr,Sirunyan:2017azo}. In the high-$x$ region, the gluon density is usually parameterized as  $g(x, Q^2)\sim (1-x)^L$, and values of $L$ extracted from global fits differ considerably from each other. In particular, obtained results for $L$ vary from 3 to 11 at $Q^2=1.9$ GeV$^2$~\cite{Abdolmaleki:2019tbb}.

To improve the situation with large $x$, one needs precise data on the heavy flavor production at energies not so far from the production threshold. Concerning the open charm production in $pp$ collisions, the corresponding cross-sections are poorly known for $\sqrt{s}<27$~GeV~\cite{Lourenco:2006vw,Accardi:2016ndt}.\footnote{On the contrary, few data exist for the $J/\psi$ production cross-section down to threshold, see Fig.~\ref{fig:kinematic}(b).} 
Presently, the only available measurements for this region were performed by the E769 experiment, which corresponds to three hundred events collected in $pA$ collisions~\cite{Alves:1996rz}. Unfortunately, the E769 results have large uncertainties, which is enough to estimate only the order of magnitude for the $pp\rightarrow c\bar{c}X$ cross-section at $\sqrt{s}\approx 20$~GeV. For this reason, future studies of the open charm production at SPD in $pp$ and $dd$ collisions for $\sqrt{s}\leq 27$~GeV are of special interest. In particular, they will allow to reduce significantly the present uncertainties in the gluon density (and $\alpha_s$) at a GeV scale, especially for high $x$.

Detailed information on the gluon distribution at large $x$ is very important for various phenomenological applications. For instance, it is of current interest to estimate the $b\bar{b}$ pair production cross-section at NICA energies. Such predictions, however, are not presently reliable due to their strong dependence on the exponent $L$ which is poorly known. Another example is the DGLAP evolution of the PDFs. Using precise data on $g(x,Q^2)$ (and $\alpha_s$) at $Q^2\sim m_c^2$ ($m_c$ is the mass of the c-quark) as boundary conditions in the DGLAP equations, one could reduce essentially the uncertainties in evolution of PDFs for higher values of $Q$.

From the theoretical point of view, the threshold behavior of cross-sections is closely related to the so-called infrared renormalon problem. It is well known that radiative corrections to the production cross-sections contain the mass (or threshold) logarithms whose contribution is expected to be sizable near the threshold. These logarithms are usually taken into account within the soft gluon resummation (SGR) formalism~\cite{Sterman:1986aj,Catani:1989ne,Contopanagos:1996nh,Kidonakis:1998nf, Ivanov:2001wm}. Formally resummed cross-sections are, however, ill-defined due to the Landau pole contribution, and few prescriptions have been proposed to avoid the renormalon ambiguities~\cite{Catani:1996yz,Berger:1996ad,Kidonakis:2000ui,Forte:2006mi}. Unfortunately, numerical predictions for heavy quark production cross-sections may significally depend on the choice of resummation prescription. Undoubtedly, anticipated data from SPD on the charm production not so far from the production threshold will provide an excellent test for these prescriptions.

Another interesting problem that could be addressed at NICA SPD is to probe the intrinsic charm (IC) content of the proton~\cite{Brodsky:1980pb,Brodsky:1981se}. The IC contribution to open charm production is expected to be sizable near the threshold because its PDF, $c(x,Q^2)$, is predicted to be harder than the gluonic one. The IC could be also accessed via double $J/\psi$ production.
As a result, the IC density in the proton can be dominant at sufficiently large $x$ independently of its overall normalization~\cite{Ananikyan:2007ef}. To visualize the IC component, one needs to collect enough events like $D\overline{D}$ pair  produced in $pp\rightarrow D\overline{D}$  with a large overall $x_F$ close to 1. Those events are predicted to be very rare within the gluon fusion mechanism and would directly indicate the five-quark component in the proton, $|\,uudc\bar{c}\rangle$.
 
Investigation of the open charm production in $p$-$p$, $p$-$d$ and $d$-$d$ collisions might be one of the key points in the NICA SPD program. The motivation is twofold. On the one hand, production of $D$-mesons in $p$-$p$ collisions is practically unmeasured at NICA energies. On the other hand, these presently unavailable data on open charm production rates are strongly necessary for determination of the gluon density $g(x,\mu)$ at large $x$ where this PDF is practically unknown.

Moreover, anticipated results on the open charm production are very important for many other current issues in particle physics: from infrared renormalon ambiguities in cross-sections to intrinsic charm content of the proton.

\subsection{Tests of TMD factorization with gluon probes \label{un_tmd}}
The description of hard inclusive processes in hadron collisions is based on factorization theorems. The formulation of factorization theorems in terms
of the TMD PDFs of quarks and gluons is the most important step towards
studying the 3D structure of hadrons and the nature of their spins.
The conventional TMD-approach~\cite{Collins:2011zzd} can be applied for study of the processes
with colorless final states with transverse momenta much smaller
than the relevant scale of hadron interactions, $q_T\ll Q$. In recent
years a substantial success was achieved in the quark sector of TMD PDFs
related with their correct theoretical definition and the connection
with experimentally observed cross-sections within the framework of
factorization theorems~\cite{Angeles-Martinez:2015sea}. In the case of unpolarized
hadron collisions, in the leading twist approximation the production cross-section is a function of two independent
TMD PDFs, i.e. distribution
functions of unpolarized quarks $ f_1^q $ and distribution functions
of transversely polarized quarks $h_1^{\bot q}$ (referred to as Boer-Mulders function) in unpolarized nucleons. For description of
cross-sections in collisions of polarized hadrons, the number of TMD
PDFs increases.

However, the situation with gluon TMD PDFs is significantly different.
Until recently, gluon TMD PDFs were used only within the framework
of phenomenological models of the type of the Generalized Parton Model (GPM), in which the factorization formula of the Collinear Parton Model
 is applied if small (non-perturbative-origin) transverse
momenta of gluons from colliding hadrons are available.

The proof of the factorization theorem for processes with gluon TMD
PDFs, as well as the formulation of evolution equations for them,
have been presented relatively recently in~\cite{Echevarria:2015uaa}, where they were
applied to describe the Higgs boson production with small
transverse momenta. However, hard processes in which detailed
information on gluon TMD PDFs can be obtained primarily, include the
processes of production of heavy mesons ($D$, $B$) and heavy
quarkonia ($ J / \psi$, $\Upsilon$, $\eta_c$, $\eta_b$, \ldots). In these
processes, there are two non-perturbative mechanisms to be
factorized: the emission of soft gluons in the initial state and the
formation of a colorless hadron in the final state. Even in the case
of heavy meson production with small transverse momenta when their
spectrum is determined only by a non-perturbative
$q_T-$distribution of initial gluons, for factorization of hard and
soft interactions it is not enough to use the TMD PDFs formalism, the
introduction of new non-perturbative process-dependent hadron
observables, the so-called TMDShFs (TMD shape functions)~\cite{Echevarria:2019ynx,Fleming:2019pzj} is
needed. Moreover, the differential cross-section for the process of
production of the state $ \cal Q $ in a collision of unpolarized
hadrons is written as\footnote{Here $f_1^g(x) \equiv g(x)$.}
\begin{equation}
\frac{d\sigma}{dy d^2 q_T} \sim  f_1^g \otimes f_1^g \otimes S_{\cal
Q} - w_{UU}\otimes h_1^{\bot g} \otimes h_1^{\bot g} \otimes S_{\cal
Q},
\end{equation}
where $S_{\cal Q} $ is the polarization-independent TMDShFs of this
process and $ w_{UU} $ is the universal contribution weight function of
linearly polarized TMD PDFs.

The factorization theorem contains three or more non-perturbative hadronic
quantities at low transverse momenta: gluon TMD PDFs and
TMDShFs. Thus, the phenomenological extraction of gluon TMDs from
quarkonium production processes is still possible, i.e., a robust
factorization theorem can potentially be obtained in any particular
case of heavy meson production. However one also needs to model and
extract the involved TMDShFs.

\subsection{Linearly polarized gluons in unpolarized nucleon \label{un_BM}}

The search for the polarized quarks and gluons in unpolarized hadrons is of special interest in the studies of the spin-orbit couplings of partons and understanding of the proton spin decomposition. The corresponding intrinsic transverse momentum $\vec{k}_{T}$ dependent distributions of the transversely polarized quarks, $h_{1}^{\perp q}(x,\vec{k}_{T}^2)$, and linearly polarized gluons, $h_{1}^{\perp g}(x,\vec{k}_{T}^2)$, in an unpolarized nucleon have been introduced in Refs.~\cite{Boer:1997nt} and~\cite{Mulders:2000sh}. Contrary to its quark version $h_{1}^{\perp q}$ the TMD density $h_{1}^{\perp g}$ is $T$- and chiral-even, and thus can directly be probed in certain experiments.

Azimuthal correlations in heavy quark pair production in unpolarized $ep$ and $pp$ collisions as probes of the density $h_{1}^{\perp g}$ have been considered in Refs.~\cite{Boer:2010zf,Pisano:2013cya}. For the case of DIS, the complete angular structure of the pair  production cross-section has been obtained in terms of seven azimuthal modulations. However, only two of those modulations are really independent; they can be chosen as the $\cos \varphi$ and $\cos 2\varphi$ distributions, where $\varphi$ is the heavy quark (or anti-quark) azimuthal angle~\cite{Efremov:2017ftw, Ivanov:2018tvg}.\footnote{The function $h_{1}^{\perp g}$ can also be determined from measurements of the Callan-Gross ratio in DIS~\cite{Efremov:2018myn}.}

To probe the TMD distributions, the momenta of both heavy quark and anti-quark, $\vec{p}_{Q}$ and $\vec{p}_{\bar{Q}}$, in the process $pp\rightarrow Q\bar{Q}X$ should be measured (reconstructed). For further analysis, the sum and difference of the transverse heavy quark momenta are introduced,
\begin{align} \label{TMD1}
\vec{K}_{\perp}&=\frac{1}{2}\left(\vec{p}_{Q\perp}-\vec{p}_{\bar{Q}\perp}\right), &
\vec{q}_{T}&=\vec{p}_{Q\perp}+\vec{p}_{\bar{Q}\perp},
\end{align}
in the plane orthogonal to the collision axis. The azimuthal angles of $\vec{K}_{\perp}$ and $\vec{q}_{T}$ are denoted as $\phi_{\perp}$ and $\phi_{T}$,  respectively.

The angular structure of the $pp\rightarrow Q\bar{Q}X$ cross-section has the following form:
\begin{equation} \label{TMD2}
{\rm d}\sigma_{pp}\propto A(q_T^2) + B(q_T^2) q_T^2 \cos2(\phi_{\perp}-\phi_{T})+ C(q_T^2) q_T^4 \cos4(\phi_{\perp}-\phi_{T}).
\end{equation}
Assuming factorization for the TMD distributions, the terms $A$, $B$ and $C$ can schematically be written as the following  convolutions~\cite{Pisano:2013cya}:
\begin{eqnarray}
&&A\propto f_1^{q}\otimes f_1^{\bar{q}}\otimes A_q + f_1^{g}\otimes f_1^{g}\otimes A_g+ h_{1}^{\perp g}\otimes h_{1}^{\perp g}\otimes A_g^{\perp}, \nonumber\\
&&B\propto h_{1}^{\perp q}\otimes h_{1}^{\perp \bar{q}}\otimes B_q+ f_{1}^{g}\otimes h_{1}^{\perp g}\otimes B_g, \label{TMD3}\\
&&C\propto h_{1}^{\perp g}\otimes h_{1}^{\perp g}\otimes C_g. \nonumber
\end{eqnarray} 
The order $\alpha_s^2$ predictions for the coefficients $A_i$, $B_i$ and $C_i$ ($i=q,g$) in Eqs.\eqref{TMD3} are presented in Ref.\cite{Pisano:2013cya}. Using these results, one can, in principle, extract the densities $h_{1}^{\perp q}(x,\vec{k}_{T}^2)$ and $h_{1}^{\perp g}(x,\vec{k}_{T}^2)$ from azimuthal distributions of the $D\overline{D}$ pairs produced in $pp$ collisions.

Another processes proposed to probe the linearly polarized gluons in unpolarized proton are:  pseudoscalar $C$-even quarkonia (such as $\eta_c$ and $\chi_c$)~\cite{Boer:2012bt}, di--gamma ($pp\rightarrow \gamma\gamma X$)~\cite{Qiu:2011ai} and $J/\psi$-- pair ($pp\rightarrow J/\psi\, J/\psi\, X$)~\cite{Lansberg:2017dzg} production. These reactions are however strongly suppressed in comparison with $pp\rightarrow D\overline{D}X$.

\subsection{Hadron structure and heavy charmonia production mechanisms\label{quarkonia_th}}

In this section we give a short review of modern status of the theory of heavy quarkonium production with an emphasis on possible applications of heavy quarkonium measurements for studies of the gluon content of hadrons. 

Production of heavy quarkonia proceeds in two stages: first, a heavy quark-antiquark pair is produced at short distances, predominantly via gluon-gluon fusion but also with a non-negligible contribution of $q\bar{q}$ and $qg$-initiated subprocesses. The second stage is hadronization of quark-antiquark pair into a physical quarkonium state, which happens at large distances (low scales) and is accompanied by a complicated rearrangement of color via exchanges of soft gluons between the heavy quark-antiquark pair and other colored partons produced in the collision. Existing approaches, aimed to describe hadronization stage, such as Non-Relativistic QCD factorization (NRQCD-factorization)~\cite{Bodwin:1994jh} and (Improved-) Color-Evaporation Model (CEM)~\cite{Barger:1979js, Barger:1980mg, Gavai:1994in, Ma:2016exq} are currently facing serious phenomenological challenges (see e.g. recent reviews~\cite{Brambilla:2010cs,Lansberg:2019adr}). NRQCD-factorization is challenged by the long-standing ``polarization puzzle''~\cite{Butenschoen:2012px,Butenschoen:2012qr} and violation of Heavy-Quark Spin Symmetry relations between Long-Distance Matrix Elements (LDMEs) of $\eta_c$ and $J/\psi$~\cite{Butenschoen:2014dra}, while CEM usually rather poorly reproduces the detailed shapes of inclusive $p_T$-spectra of charmonia and bottomonia and, unlike NRQCD-factorization~\cite{Sun:2014gca,He:2019qqr}, significantly under-predicts bulk of cross-section for pair hadroproduction of $J/\psi$ even at NLO in $\alpha_s$~\cite{Lansberg:2020rft}. Presently, the study of the heavy-quarkonium production mechanism is an active field of research, with new approaches, such as subleading-power fragmentation~\cite{Kang:2011mg} and Soft-Gluon Factorization~\cite{Ma:2017xno,Li:2019ncs,Chen:2020yeg}, being proposed recently. 
  
   \begin{figure}

 \begin{minipage}[ht]{0.49\linewidth}
    \center{\includegraphics[width=\linewidth]{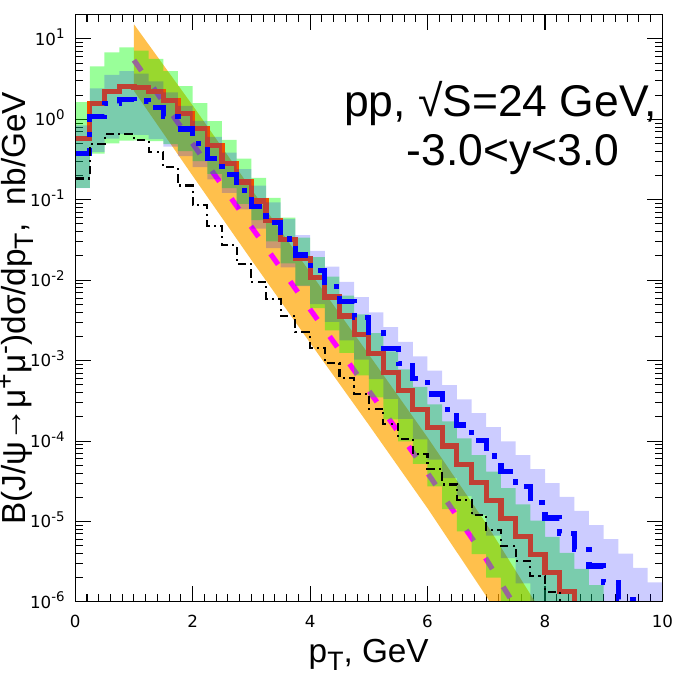} \\ (a)}
\end{minipage} \hfill 
 \begin{minipage}[ht]{0.49\linewidth}
   \vspace{2pt}
    \center{\includegraphics[width=\linewidth]{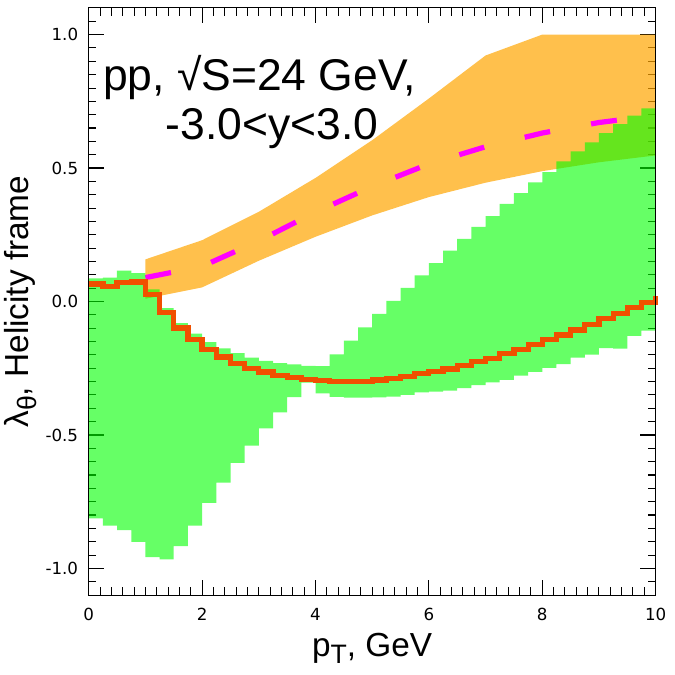} \\ (b)}
\end{minipage}
     \caption{Theoretical predictions for inclusive $J/\psi$ $p_T$-spectrum (a) and $p_T$-dependence of polarization parameter $\lambda_\theta$ (b) in various models: NLO of Collinear Parton Model + NRQCD-factorization (thick dashed line with orange uncertainty band)~\cite{Butenschoen:2010rq,Butenschoen:2011yh}, LO of PRA~\cite{Karpishkov:2017kph} + NRQCD-factorization (thick solid histogram with a green uncertainty band)~\cite{Saleev:2012hi,Karpishkov:2020wwe}, and LO PRA~\cite{Karpishkov:2017kph} + Improved Color Evaporation Model (thick dash-dotted histogram with blue uncertainty band)~\cite{Cheung:2018tvq}. The contribution of $q\bar{q}$-annihilation channel to the central ICEM prediction is depicted by the thin dash-dotted histogram. Uncertainty bands are due to factorization/renormalization scale variation only. }
     \label{fig:NICA-Jpsi-pT}
 \end{figure}
  
 Due to the above-mentioned problems and the multitude of competing theoretical approaches and the models available on the market, our lack of quantitative understanding of the mechanism of hadronization can become a source of significant theoretical uncertainties if quarkonium production is to be used as a tool to study the proton structure. The Fig.~\ref{fig:NICA-Jpsi-pT} provides an insight on this situation at NICA SPD. In this figure, predictions of three models for the $p_T$-spectrum (Fig.~\ref{fig:NICA-Jpsi-pT}(a)) and $p_T$-dependence of the polarization parameter $\lambda_\theta$ (Fig.~\ref{fig:NICA-Jpsi-pT}(b)) are compared. The first one relies on the NLO calculation in Collinear Parton Model (with LO being $O(\alpha_s^3)$, see Fig.~\ref{fig:diagrams-gg}(a)) to describe short-distance part of the cross-section and uses the NRQCD-factorization formalism for the long-distance part, with LDMEs of the latter tuned to charmonium production data in hadronic collisions, DIS and $e^+e^-$-annihilation~\cite{Butenschoen:2010rq,Butenschoen:2011yh,Butenschoen:2012px,Butenschoen:2012qr}. In the second prediction, the short-distance part of the cross-section is calculated in the LO ($O(\alpha_s^2)$ for color-octet and $P$-wave contributions and $O(\alpha_s^3)$ for color-singlet $S$-wave ones) of PRA~\cite{Karpishkov:2017kph}, while  LDMEs in this calculation had been fitted to the charmonium hadroproduction data from RHIC, Tevatron and LHC~\cite{Saleev:2012hi,Karpishkov:2020wwe}. The third prediction is performed in the LO ($O(\alpha_s^2)$) of PRA with the same unintegrated PDFs as for the second one, but interfaced with an improved Color-Evaporation Model (ICEM) of Ref.~\cite{Cheung:2018tvq} for description of hadronization. Non-perturbative parameters of the ICEM had been taken from the Ref.~\cite{Cheung:2018tvq} where they had been fitted to charmonium hadroproduction data at Tevatron and LHC energies. Predictions of all three models for inclusive $J/\psi$ $p_T$-spectrum at NICA SPD appear to be consistent within their uncertainty bands. However, the structure of these predictions is significantly different, with NRQCD-based predictions being dominated by gluon-gluon fusion subprocess, while ICEM prediction containing significant contamination from $q\bar{q}$-annihilation (thin dash-dotted histogram in the Fig.~\ref{fig:NICA-Jpsi-pT}(a)), which reaches up to 50\% at low $p_T<1$ GeV/$c$ and contributes up to 10\% at higher $p_T>3$ GeV/$c$. Also ICEM tends to predict significantly harder $p_T$-spectrum at $p_T>5$ GeV/$c$, than NRQCD-based PRA prediction which was performed with the same unintegrated PDFs. 

 The discussion above shows, that the $J/\psi$ $p_T$-spectrum can be reliably predicted only in a limited range of transverse momenta, approximately from 3 to 6 GeV at $\sqrt{s}=24$ GeV. At higher $p_T$ the shape of the spectrum becomes highly model-dependent and at lower $p_T<M_{J/\psi}$ the TMD-factorization effects (including possible violation of factorization, see~\cite{Echevarria:2019ynx,Fleming:2019pzj}) come into the game and the contribution of $q\bar{q}$-annihilation subprocess becomes uncertain. Nevertheless, predictions and measurements of rapidity or $x_F$-differential cross-sections even in this limited $p_T$-range could help to further constrain the gluon PDF, e.g. to rule-out the extreme values of $L$ in the $x\to 1$ asympthotics of the PDF $\sim (1-x)^L$.   
  
  Predictions of NLO CPM and LO of PRA for polarization parameter $\lambda_\theta$ (see the Fig.~\ref{fig:NICA-Jpsi-pT}(b)) are significantly different, with PRA predicting mostly un-polarized production ($\lambda_\theta\simeq 0$) while CPM predicts transverse polarization ($\lambda_\theta=+1$) at high $p_T$. Disagreement of the predictions for polarization parameters  mostly reflects the difference of LDMEs obtained in two fits and their large uncertainty bands are due to significant uncertainties of LDMEs. Measurements of heavy quarkonium polarization at NICA energies will provide additional constraints on models, however due to well-known problems with description of polarization at high energies~\cite{Butenschoen:2012px,Butenschoen:2012qr} constraints coming from polarization measurements should be interpreted with great care and one should try to disentangle conclusions for gluon PDF from the results related to heavy quarkonium polarization.

 We would like to emphasise here also, that studies of $\eta_c$-production at NICA SPD will be instrumental for better understanding of its production mechanism. If the color singlet mechanism dominated NRQCD prediction turns out to be correct, then $\eta_c$ production becomes a unique instrument to study the gluon content of the proton without introducing additional free-parameters, such as the color octet LDMEs, to the analysis.

\subsection{Non-nucleonic degrees of freedom in deuteron \label{un_non}}
The naive model describes the deuteron as a weakly-bound state of a proton and a neutron mainly in S-state with a small admixture of the D-state. However, such a simplified picture failed to describe the HERMES experimental results on the $b_1$ structure function of the deuteron~\cite{Airapetian:2005cb}. Modern models treat the deuteron as a six-quark state with the wave function 
\begin{equation}
|6q\rangle =  c_1 | NN\rangle + c_2 |\Delta \Delta \rangle + c_3 | CC \rangle,
\end{equation}
 that contains such terms as the nucleon $| NN\rangle$, $\Delta$-resonance $|\Delta\Delta\rangle$  and the so-called hidden color component $| CC \rangle$ in which two color-octet baryons combine to form a color singlet~\cite{Harvey:1988nk}. Such configurations can be generated, for example, if two nucleons exchange a single gluon. The relative contribution of the hidden-color term varies from about 0.1\% to 80\% in different models~\cite{Miller:2013hla}. The components other than $|NN\rangle$ should manifest themselves in the high-$Q^2$ limit. Possible contributions of the Fock states with a valent gluon like $|uuudddg\rangle$ could also be discussed~\cite{Hoyer:1997rh,Brodsky:2018zdh}.

 The unpolarized gluon PDF of the deuteron in the light-front quantization was calculated in the Ref.~\cite{Brodsky:2018zdh} under the approximation where the input nuclear wave function is obtained by solving the nonrelativistic Schrödinger equation with the phenomenological Argonne v18 nuclear potential as an input. Gluon PDFs calculated per nucleon are very similar for the proton one in the range of small and intermediate $x$ values while for $x>0.6$ the difference becomes large due to the Fermi motion (see Fig.~\ref{fig:Deuteron}(a)). A similar work was performed in Ref.~\cite{Mantysaari:2019jhh} for determination of the spatial gluon distribution in the deuteron for low-$x$ that could be tested in the $J/\psi$ production at EIC. Today the gluon content of deuteron and light nuclei becomes the matter of interest for the lattice QCD studies~\cite{Winter:2017bfs}. Apart from the general understanding of the gluon EMC effect, the measurement of the gluon PDF at high-$x$ for deuteron could provide a useful input for high-energy astrophysical calculation~\cite{Brodsky:2018zdh}.
 
 SPD can perform an explicit comparison of the differential inclusive production cross-sections $d\sigma/dx_{F}$ for all three gluon probes: charmonia, open charm, and prompt photons using $p$-$p$ and $d$-$d$ collisions at $\sqrt{s_{NN}}=13.5$~GeV and possibly below. Such results could be treated in terms of the difference of unpolarized gluon PDFs in deuteron and nucleon.

\subsection{Gluon polarization $\Delta g$ with longitudinally polarized beams \label{hel}}

The gluon helicity distribution function $\Delta g(x)$ is a fundamental quantity characterizing the inner structure of the nucleon.
It describes the difference of probabilities to find in the longitudinally polarized nucleon a gluon with the same and opposite spin orientations.
 The integral $\Delta G=\int \Delta g(x) dx$ can be interpreted as the gluon spin contribution to the nucleon spin. After the EMC experiment discovered that only a small part of proton spin is carried by the quarks~\cite{Ashman:1989ig}, the gluon spin was assumed to be another significant contributor. So $\Delta G$ is a key ingredient of the nucleon helicity sum rule
\begin{equation}
\frac{1}{2} = \frac{1}{2}\Delta \Sigma + \Delta G + L_{q}+L_{g},
\end{equation}
where $\Delta \Sigma \approx 0.25$~\cite{Aidala:2012mv} is the net contribution from the quark spin and $L_{q}$, $L_{g}$ represent the contributions of the orbital angular momenta of quarks and gluons, respectively.

\begin{figure}[!t]
  \begin{minipage}[ht]{0.49\linewidth}
    \center{\includegraphics[width=1\linewidth]{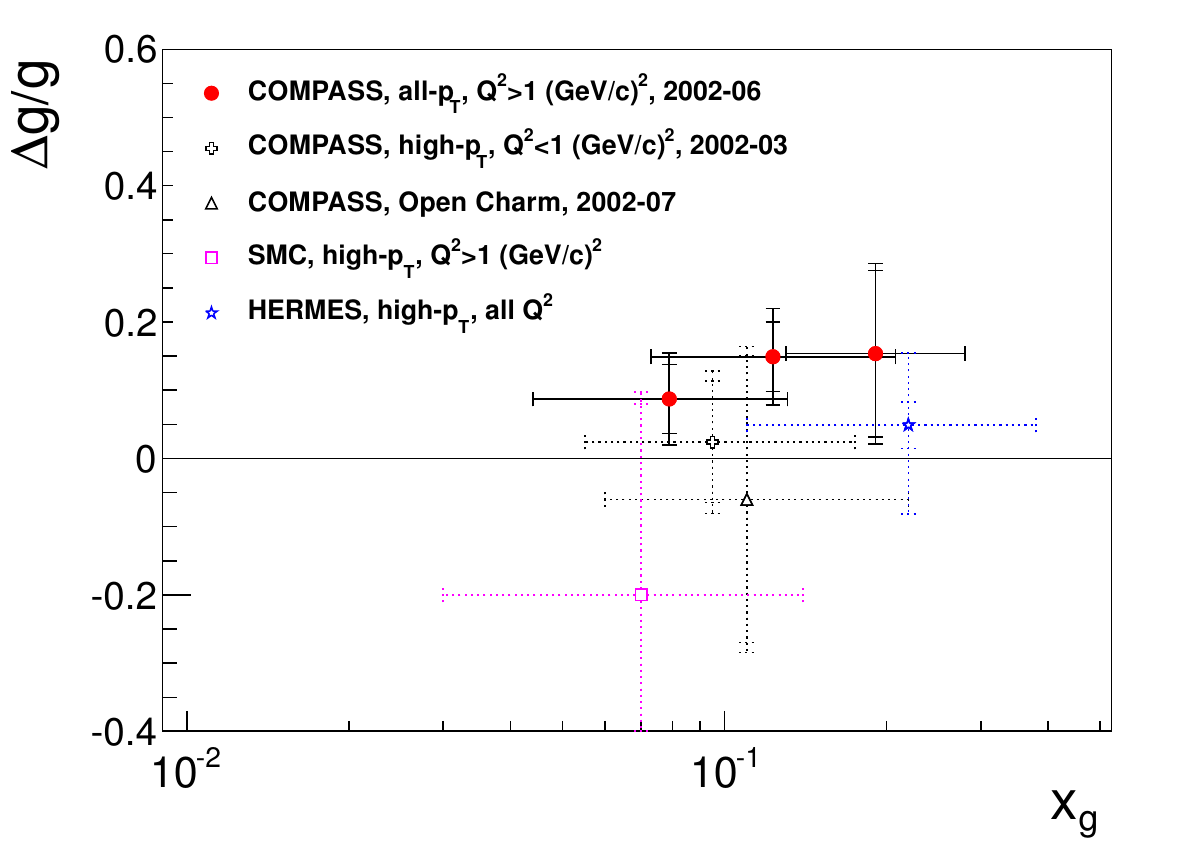} \\ (a)}
  \end{minipage}
  \hfill
  \begin{minipage}[ht]{0.49\linewidth}
    \center{\includegraphics[width=1\linewidth]{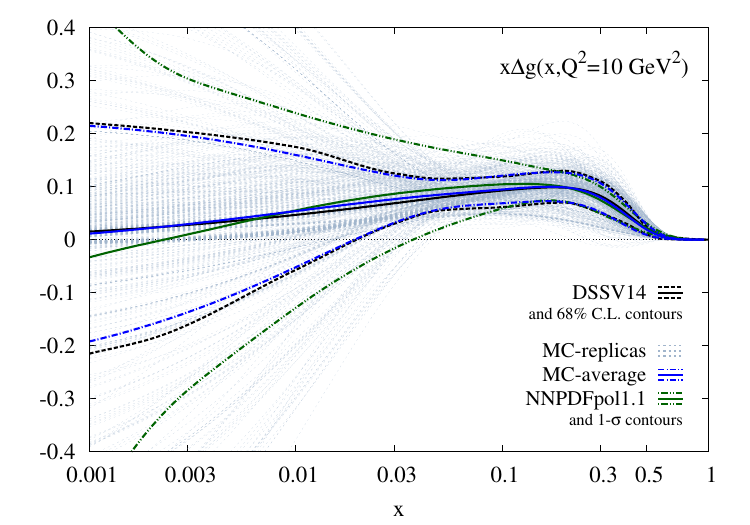} \\ (b)}
  \end{minipage}
  \caption{ (a) SIDIS data on $\Delta g(x)/g(x)$ extracted in LO~\cite{Adolph:2015cvj}. (b) Global fit results for the gluon helicity distribution $\Delta g(x)$~\cite{deFlorian:2019zkl}. }
  \label{fig:dgg_1}  
\end{figure}

The first attempt to measure the gluon polarization in the nucleon was made by the FNAL E581/704 Collaboration using a 200~GeV polarized proton beam and a polarized proton target~\cite{Adams:1994bg}. They measured the longitudinal double-spin asymmetries $A_{LL}$
     for inclusive multi-$\gamma$ and $\pi^0\pi^0$ production to be consistent with zero within their sensitivities. In the following years a set of SIDIS measurements was performed by the HERMES~\cite{Airapetian:2010ac}, SMC~\cite{Adeva:2004dh} and COMPASS~\cite{Ageev:2005pq, Alekseev:2009ad,Adolph:2012vj,Adolph:2012ca,Adolph:2015cvj} experiments. The production of hadron pairs with
high transverse momenta and the production of the open charm where the photon-gluon fusion mechanism dominates were studied.
 It was figured out that with a large uncertainty the value of $\Delta G$ is close to zero. Nevertheless, for gluons carrying a large fraction $x$ of the nucleon momentum, an evidence of a positive polarization has been observed, see Fig.~\ref{fig:dgg_1}(a). 
New input for $\Delta G$ estimation was obtained from the measurement of the $A_{LL}$ asymmetries in the inclusive production of high-$p_T$ neutral ~\cite{Adare:2014hsq, Adare:2008aa, Adare:2008qb,Adam:2018cto} and charged pions \cite{Acharya:2020fgb}, $\eta$-mesons~\cite{Adare:2014hsq}, jets~\cite{Djawotho:2013pga}, di-jets \cite{Adamczyk:2016okk, Adam:2018pns},  heavy flavors~\cite{Adare:2012vv}, and, recently, $J/\psi$-mesons~\cite{Adare:2016cqe} in polarized $p$-$p$ collisions at RHIC. The new data in general are in agreement with SIDIS measurements, which demonstrates the universality of the helicity-dependent parton densities and QCD factorization. 

\begin{figure}[!t]
  \begin{minipage}[ht]{0.58\linewidth}
    \center{\includegraphics[width=1\linewidth]{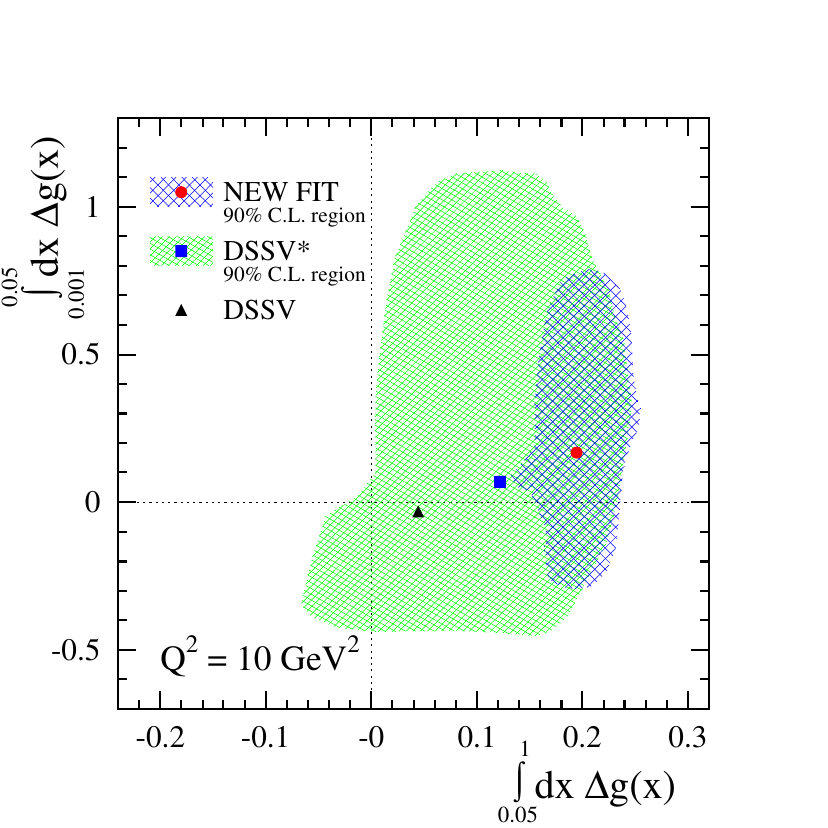} \\ (a)}
  \end{minipage}
  \hfill
  \begin{minipage}[ht]{0.41\linewidth}
    \center{\includegraphics[width=1\linewidth]{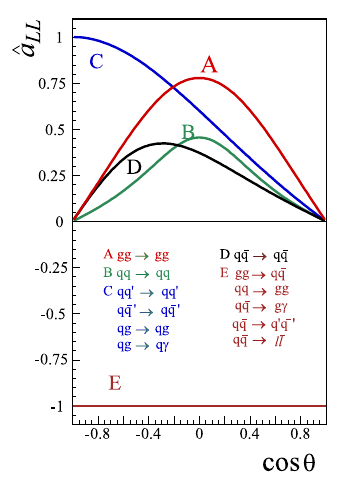} \\ (b)}
  \end{minipage}
  \caption{ (a) Estimates of contributions of low-$x$ and high-$x$ kinematic ranges into $\Delta G$ for the DSSV series of the global fit.  The 90\% C.L. areas are shown~\cite{deFlorian:2014yva}. (b) Partonic longitudinal double-spin asymmetries $A_{LL}$ for different hard processes as a function of center-of-mass scattering angle~\cite{osti_15011678}.}
  \label{fig:dgg_2}  
\end{figure}

At the moment the most recent sets of polarized PDFs extracted in the NLO approximation are LSS15~\cite{Leader:2014uua}, DSSV14~\cite{deFlorian:2014yva,deFlorian:2019zkl}, NNPDF-pol1.1~\cite{Nocera:2014gqa}, and JAM17~\cite{Ethier:2017zbq}. To obtain them, different approaches, parameterizations, and sets of experimental data were used, see Ref.~\cite{Ethier:2020way} for more details. Fit results for $\Delta g(x)$ from  DSSV14 and NNPDF--pol1.1 are presented in Fig. \ref{fig:dgg_1}(b)~\cite{deFlorian:2019zkl}. The RHIC $p$-$p$ data put a strong constraint on the size of $\Delta g(x)$ 
in the range $0.05<x<0.2$ while a constraint on its sign is weaker
as soon as in some processes only $\Delta g$ squared is probed (see details below).
 The small $x$ region remains still largely unconstrained and could be covered in future by measurements at EIC~\cite{Accardi:2012qut}. 
The region of high $x$ is covered at the moment only by SIDIS measurements which still lack a proper NLO description~\cite{deFlorian:2008mr}.
The uncertainty of the contribution to $\Delta g$ from the kinematic range $0.001<x<0.05$ vs. the corresponding contribution from the range $x>0.05$ for the DSSV global fits is shown in Fig.~\ref{fig:dgg_2}(a)~\cite{deFlorian:2014yva}.

\begin{figure}[!h]
  \begin{minipage}[ht]{0.498\linewidth}
    \center{\includegraphics[width=1\linewidth]{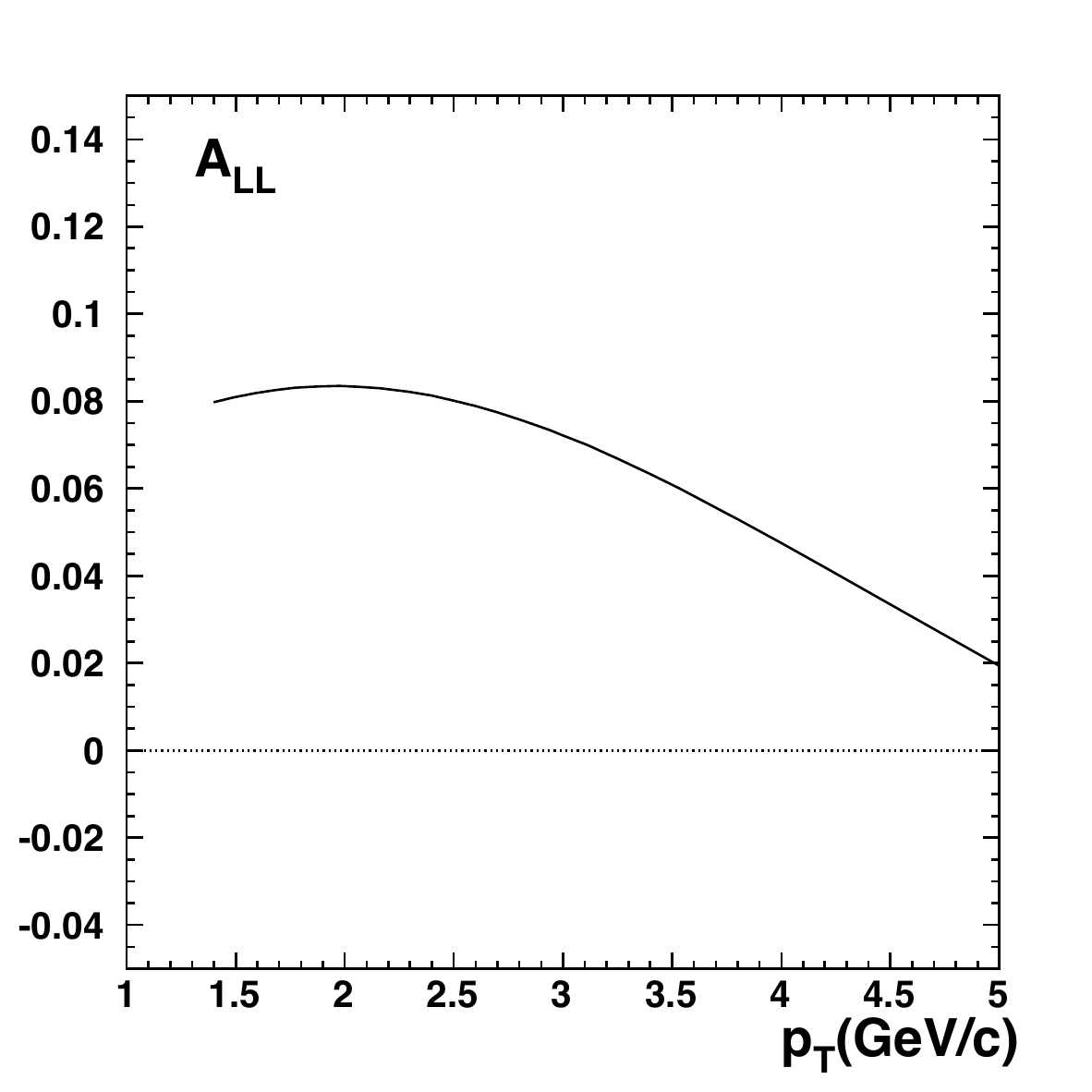} \\ (a)}
  \end{minipage}
  \hfill
  \begin{minipage}[ht]{0.498\linewidth}
    \center{\includegraphics[width=1\linewidth]{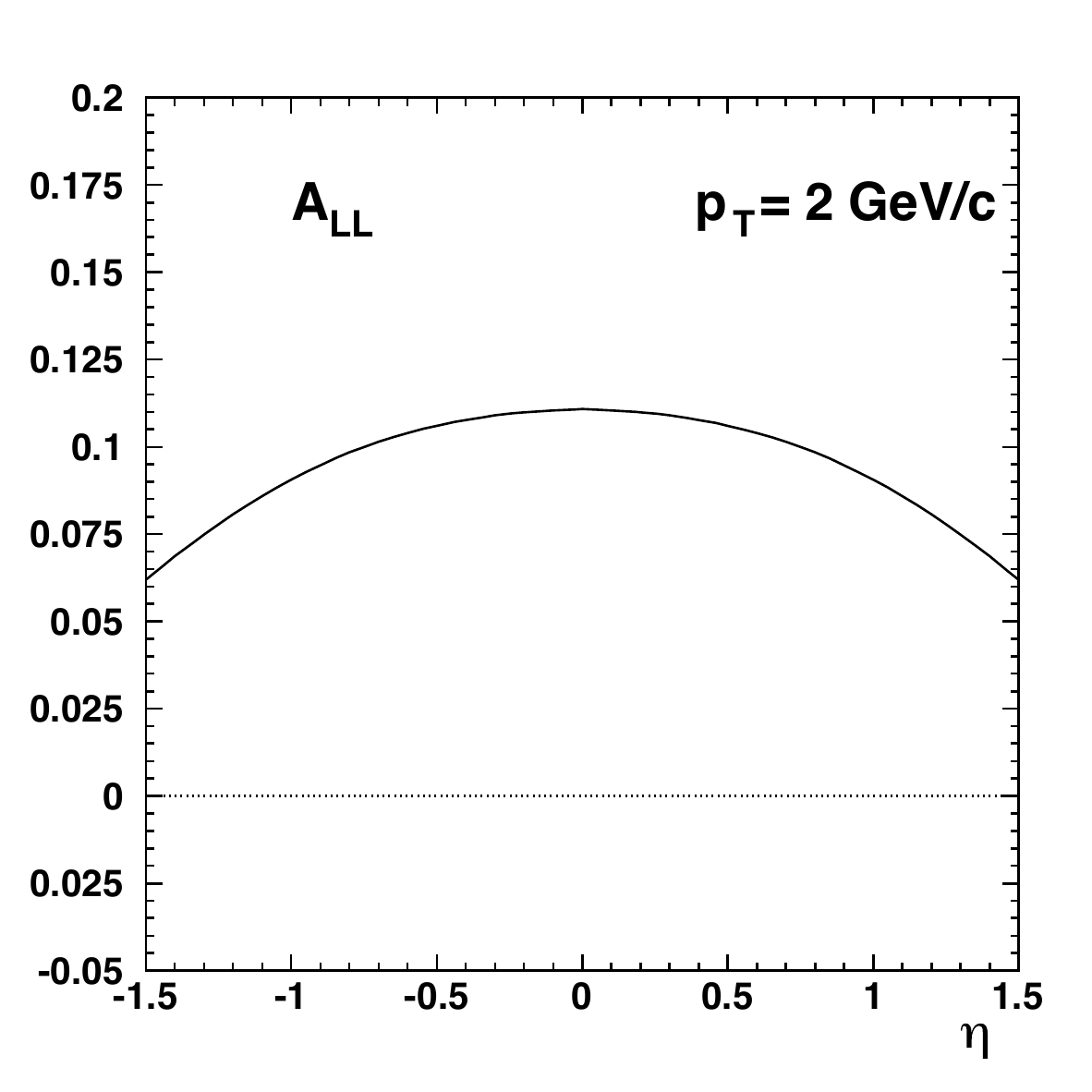} \\ (b)}
  \end{minipage}
  \caption{ Longitudinal double spin asymmetry $A_{LL}$  for inclusive $J/\psi$ production calculated for $p$-$p$ collisions at $\sqrt{s}=39$~GeV in the LO approximation as a function of a) transverse momentum $p_T$ and b) pseudorapidity $\eta$~\cite{Anselmino:1996kg}.}
  \label{fig:ALL_JPSI}  
\end{figure}

\begin{figure}[!h]
  \begin{minipage}[ht]{0.50\linewidth}
    \center{\includegraphics[width=1\linewidth]{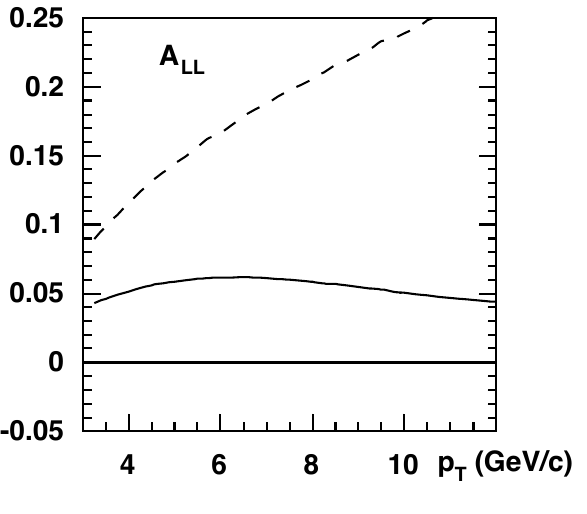} \\ (a)}
  \end{minipage}
  \hfill
  \begin{minipage}[ht]{0.50\linewidth}
    \center{\includegraphics[width=0.82\linewidth]{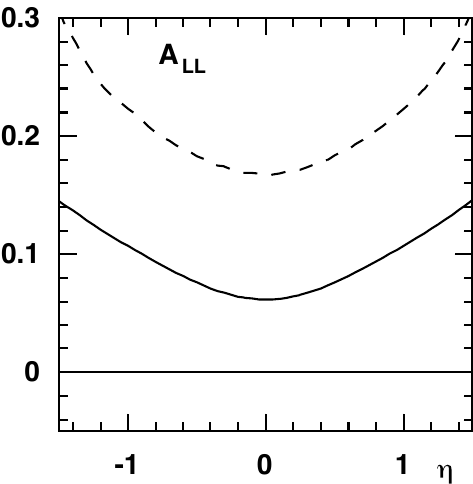} \\ (b)}
  \end{minipage}
  \caption{ Longitudinal double spin asymmetry $A_{LL}$ for inclusive prompt-photon production calculated for $p$-$p$ collisions at $\sqrt{s}=39$~GeV in the LO approximation as a function of a) transverse momentum $p_T$ and b) rapidity $\eta$ ($p_T=6$ GeV/$c$)~\cite{Anselmino:1996kg}.}
  \label{fig:ALL_PP}  
\end{figure}

In case of the longitudinally polarized $p$-$p$ collisions the asymmetry $A_{LL}$ is defined as
\begin{equation}
A_{LL}=\frac{\sigma^{++}-\sigma^{+-}}{\sigma^{++}+\sigma^{+-}},
\label{eq:ALL}
\end{equation}
where $\sigma^{++}$ and $\sigma^{+-}$ denote the cross-sections with the same and opposite proton helicity combinations, respectively. For the prompt
photons produced via the gluon Compton scattering 
\begin{equation}
A^{\gamma}_{LL}\approx \frac{\Delta g(x_1)}{g_(x_1)}\otimes A_{1p}(x_2) \otimes \hat{a}_{LL}^{g q(\bar{q}) \to \gamma q (\bar{q})} + (1 \leftrightarrow 2).
\end{equation}
Here $A_{1p}(x)$ is the asymmetry well-measured in a wide range of $x$ and $\hat{a}_{LL}^{g q(\bar{q}) \to \gamma q (\bar{q})}$ is the asymmetry 
of the corresponding hard process. 
The Fig.~\ref{fig:dgg_2}(b) shows the behavior of $\hat{a}_{LL}$ for different hard processes as a function of the center-of-mass scattering angle.
For charmonia and open charm production via the gluon-gluon fusion process the expression for the corresponding asymmetry reads
\begin{equation}
A^{c\bar{c}}_{LL}\approx \frac{\Delta g(x_1)}{g(x_1)}\otimes \frac{\Delta g(x_2)}{g_(x_2)} \otimes \hat{a}_{LL}^{gg\to c\bar{c}X}.
\end{equation}
This asymmetry on the one hand is more sensitive to the gluon polarization than the corresponding one for the prompt photons due to the quadratic dependence on $\Delta g$. 
 On the other hand the sign of the $\Delta g$ value can not be determined from it. So the measurements with prompt photons and heavy-quark states are complementary. The contribution of $q\bar{q}$ annihilation processes to the above-mentioned asymmetries is negligible despite $\hat{a}_{LL}=-1$ because of the smallness of the sea-quark polarization in the nucleon.
 
 It is important to emphasise that a sizable systematic uncertainty of $A_{LL}$ measurements in the inclusive $J/\psi$ production comes from our limited knowledge of charmonia production mechanisms including the feed-down contribution. Each of them has different partonic asymmetries $\hat{a}_{LL}$~\cite{Leader:2001gr}. 
 For the $\Delta g$ estimation in Ref.~\cite{Adare:2016cqe} the value of $\hat{a}^{J/\psi}_{LL}$ has been forced to $-1$. The SPD setup will have the possibility to reconstruct $\chi_{cJ}$ states via their radiative decays and resolve $J/\psi$ and $\psi(2S)$ signals in a wide kinematic range and disentangle contributions of different production mechanisms.
 The quality of the $\Delta g$ estimation could be significantly improved by measuring $A_{LL}$ separately for each charmonium state.
 
Predictions for the longitudinal double-spin asymmetries $A_{LL}$ in $p$-$p$ collisions can be found in Refs.~\cite{Feng:2018cai} ($J/\psi$) and~\cite{ Vogelsang:2000bx} (prompt photons). They mostly cover the kinematic range of the RHIC experiments. Some estimates for $A_{LL}$ in charmonia~\cite{Anselmino:1996kg} and prompt-photon~\cite{Gordon:1996ft, Gordon:1996rk,Anselmino:1996kg} production at $\sqrt{s}=39$~GeV (see Figs.~\ref{fig:ALL_JPSI} and \ref{fig:ALL_PP}, respectively) have been done in preparation of the unrealized HERA-$\vec{N}$ project.

The authors of the Ref.~\cite{Xu:2004es} proposed to extract information about the gluon helicity $\Delta g$ via the study of the production of high-$p_T$ prompt photons accompanied by $\Sigma^+$ hyperons. To do that, the single longitudinal spin asymmetry $A^{\gamma\Sigma}_L$ and the polarization of the produced $\Sigma^+$ hyperons should be measured. However, further elaboration of this method is needed.

\subsection{Gluon-related TMD and twist-3 effects with transversely polarized beams \label{tmd_t3}}

One of the promising ways to investigate the spin structure of the nucleon is the study of transverse single-spin asymmetries (SSAs) in the inclusive production of different final states in high-energy interactions. The  SSA $A_N$ is defined as
\begin{equation}
A_{N}=\frac{\sigma^{\uparrow}-\sigma^{\downarrow}}{\sigma^{\uparrow}+\sigma^{\downarrow}},
\end{equation}
where $\sigma^{\uparrow}$ and $\sigma^{\downarrow}$ denote the inclusive production cross-sections with opposite transverse polarization of one of the colliding particles. At the moment, more than forty years after the transverse spin phenomena were discovered, a wealth of experimental data indicating non-zero $A_N$ in the lepton-nucleon and nucleon-nucleon interactions was collected. However, our understanding of the SSA phenomenon is not yet conclusive.

Theoretically two dual approaches are used to explain the transverse single-spin asymmetries: the collinear twist-3 formalism and the transverse momentum dependent (TMD) factorization approach. In the first one at large transverse momenta $p_T \gg \Lambda_{QCD}$ of a produced particle, the collinear factorization involving twist-3 contributions for three-parton (Efremov-Teryaev-Qiu-Sterman) correlations~\cite{Efremov:1981sh, Efremov:1984ip, Qiu:1991pp,Efremov:1994dg} is used. Here $\Lambda_{QCD}\approx 200$ MeV  is the QCD scale. An alternative approach assumes the TMD factorization, valid for $p_T\ll Q$, where the SSAs come from the initial-state quark and gluon Sivers functions or the final-state Collins fragmentation functions.
The Sivers function $f_{1T}^{\perp,q(g)}(x,k_{T})$ is a TMD PDF that describes the left-right asymmetry in the distribution of the partons w.r.t. to the plane defined by the nucleon spin and momentum vectors. Originating from the correlation between the spin of the nucleon and the orbital motion of partons, it is an important detail of the three-dimensional picture of the nucleon. 
 This function is responsible for the so-called Sivers effect (for both quarks and gluons) that was first suggested in~\cite{Sivers:1989cc} as an explanation for the large single transverse spin asymmetries $A_N$ in the inclusive production of the nucleon.
More on the theoretical and experimental status of the transverse spin structure of the nucleon can be found in Refs.~\cite{Perdekamp:2015vwa, Boer:2015vso}.
%
The first attempt to access the gluon Sivers function (GSF) studying azimuthal asymmetries in high-$p_T$ hadron pair production in SIDIS of transversely polarised deuterons and protons, was performed by COMPASS~\cite{Adolph:2017pgv}.
Using neural network techniques the contribution originating from Photon–Gluon Fusion (PGF) subprocess has been separated from the leading-order virtual-photon absorption and QCD Compton scattering subprocesses.
The measured combined proton-deuteron PGF-asymmetry was found to be negative and more than two standard deviations below zero, which supports the possible existence of a non-zero Sivers function. In the meantime, COMPASS did not see any signal for the PGF Collins asymmetry, which can analogously be related to the gluon transversity distribution. COMPASS studied GSF also through Sivers asymmetry in the $J/\psi$-production channel~\cite{Szabelski:2016wym}, again obtaining an indication of a negative asymmetry.

\begin{figure}[!h]
  \begin{minipage}[ht]{0.5\linewidth}
    \center{\includegraphics[width=1\linewidth]{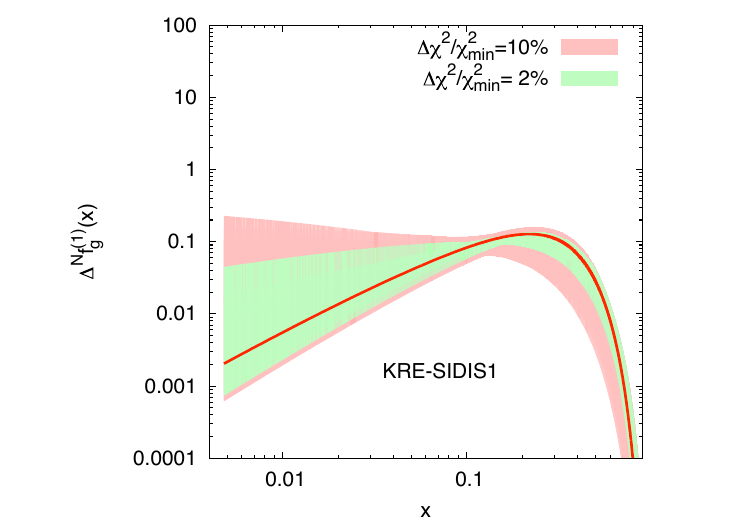} \\ (a)}
  \end{minipage}
  \hfill
  \begin{minipage}[ht]{0.5\linewidth}
    \center{\includegraphics[width=1\linewidth]{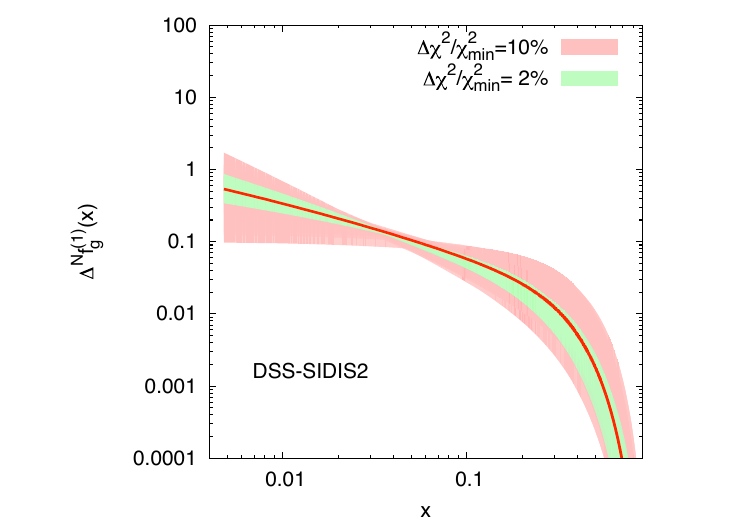} \\ (b)}
  \end{minipage}
  \caption{ The first $k_T$-moment of the gluon Sivers function for SIDIS1~\cite{Anselmino:2005ea} and SIDIS2~\cite{Anselmino:2008sga} extractions of the quark Sivers functions~\cite{DAlesio:2015fwo}.}
  \label{fig:GSF} 
\end{figure}

Recently, in Ref.~\cite{DAlesio:2015fwo} a first estimate of the GSF was obtained using the midrapidity data on the $A_N$ SSA, measured in $\pi^0$ production at RHIC~\cite{Adare:2013ekj}.
The extraction was performed within the GPM framework using GRV98-LO set for the unpolarized PDF and available parameterizations for the quark Sivers functions (SIDIS1 from Ref.~\cite{Anselmino:2005ea} and SIDIS2 from Ref.~\cite{Anselmino:2008sga}). The two parameterizations were obtained using different options for fragmentation functions, namely Kretzer~\cite{Kretzer:2000yf} and DSS07~\cite{deFlorian:2007aj} sets, which give significantly different results for gluons. The latter point has a strong impact on the extracted GSF especially in low-$x$ region.
First $k_T$-moments of the GSF $\Delta^{q(g)}_N (x,k_{T})$ for the SIDIS1 and SIDIS2 sets are shown in Fig. \ref{fig:GSF} (a) and (b), respectively.

The gluon Sivers function is expected to satisfy the positivity bound defined as two time the unpolarized TMD gluon distribution.
However, some theoretical expectations are that the gluon Sivers function at relatively high $x$ is about 1/3 of the quark one~\cite{Boer:2015vso}.

Several inclusive processes were proposed to access the gluon-induced spin effects in transversely polarized $p$-$p$ collisions. Single spin asymmetries for production of charmonia ~\cite{Godbole:2017syo} (RHIC, AFTER), open charm ~\cite{Anselmino:2004nk, Koike:2011mb,Kang:2008ih,Godbole:2016tvq} (RHIC)~\cite{Godbole:2016tvq} (AFTER),  and prompt photons ~\cite{Qiu:1991pp, Hammon:1998gb} (E704), ~\cite{Kanazawa:2012kt} (RHIC) were estimated using both approaches for the experimental conditions of the past, present, and future experiments. 
     
The SSA $A^{J/\psi}_{N}$ in the $J/\psi$ production was measured by PHENIX in the $p$-$p$ and $p$-$A$ collisions at $\sqrt{s_{NN}}=200$ GeV/$c$~\cite{Adare:2010bd, Aidala:2018gmp}. The obtained values for $A^{J/\psi}_{N}$ are consistent with zero for negative and positive $x_F$.
The consistent with zero result for SSA $A^{\gamma}_{N}$ with isolated direct photons in midrapidity region was obtained recently by PHENIX \cite{Acharya:2021fsd}.

\begin{figure}[!h]
  \begin{minipage}[ht]{0.5\linewidth}
    \center{\includegraphics[width=1\linewidth]{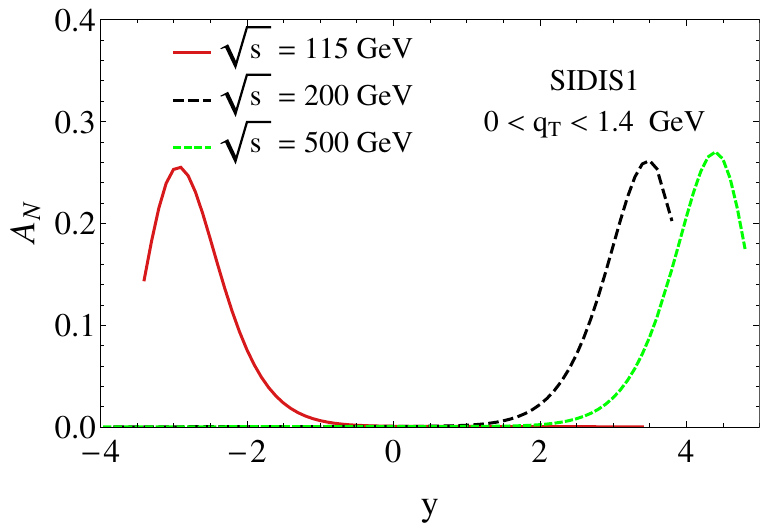} \\ (a)}
  \end{minipage}
  \hfill
  \begin{minipage}[ht]{0.5\linewidth}
    \center{\includegraphics[width=1\linewidth]{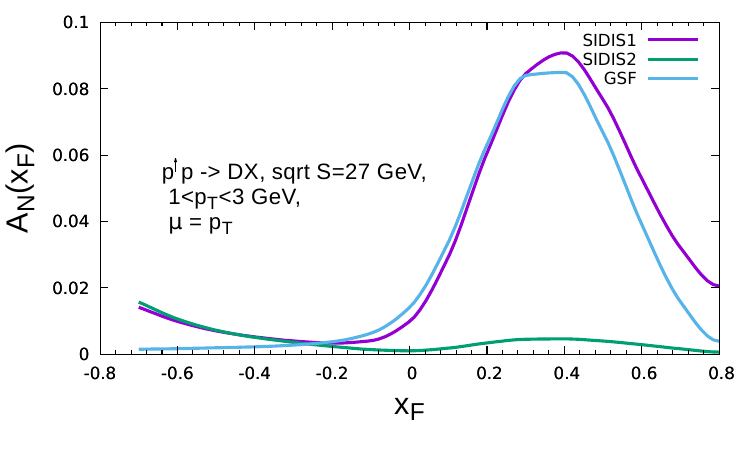} \\ (b)}
  \end{minipage}
  \caption{(a) Predictions for $A^{J/\psi}_{N}$ for $\sqrt{s}=115$ GeV (AFTER), 200~GeV and 500~GeV (RHIC) as a function of rapidity $y$\cite{Godbole:2017syo}. 
  (b)  Sivers effect contribution to the $A_N^D$ asymmetry calculated within the Generalized Parton Model.}
  \label{fig:AN_D} 
\end{figure}

\begin{figure}[!h]
  \begin{minipage}[ht]{0.50\linewidth}
  \center{\includegraphics[width=1\linewidth]{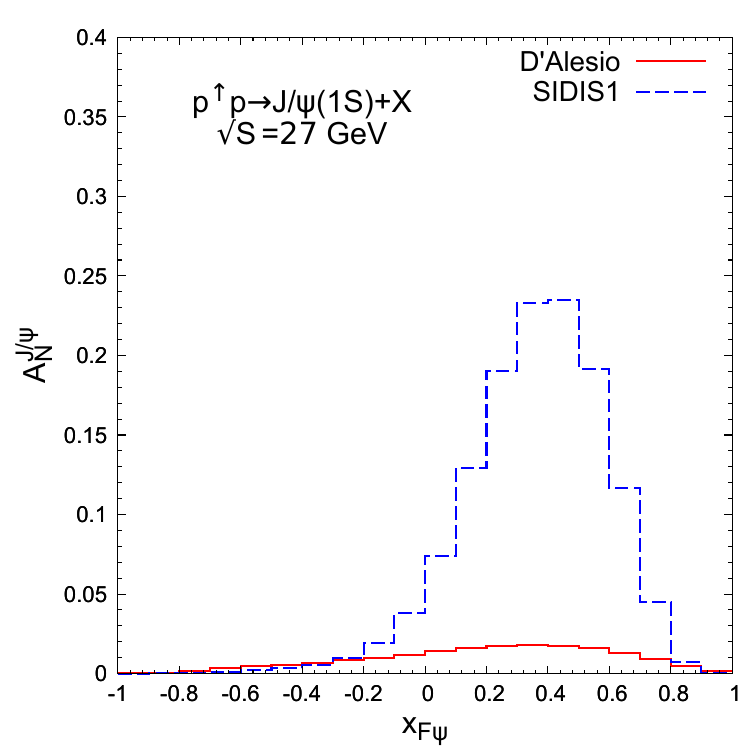} \\ (a)}
  \end{minipage}
  \hfill
  \begin{minipage}[ht]{0.50\linewidth}
 \center{\includegraphics[width=1\linewidth]{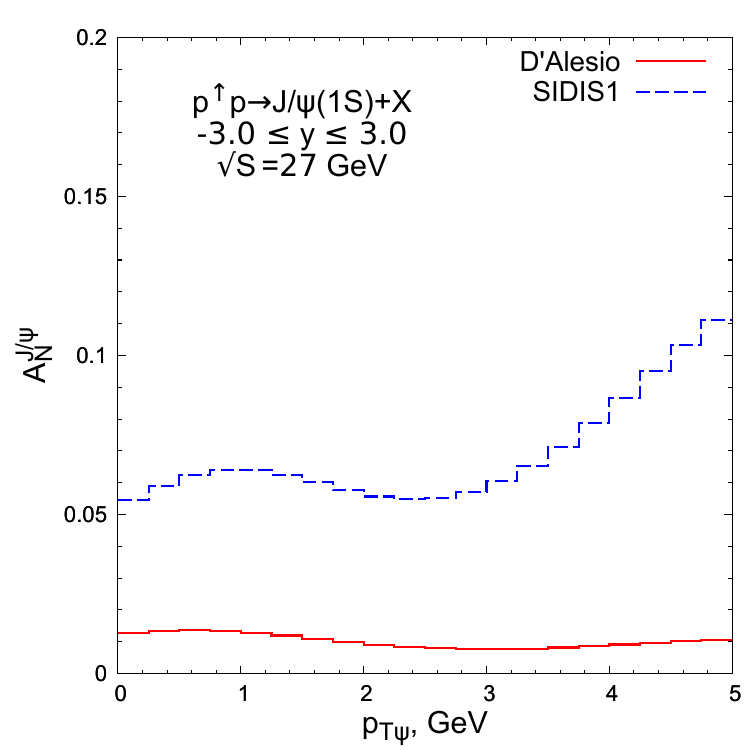} \\ (b)}
  \end{minipage}
  \caption{Predictions for $A_N^{J/\psi}$ as function of $x_F$ (b) and
$p_T$ (b) in $p$-$p$ collisions at the energy $\sqrt{s}=27$ GeV
obtained in GPM + NRQCD approach. 
  }
  \label{fig:AN_Jpsi_VS} 
\end{figure}

Theoretical predictions~\cite{{Godbole:2017syo}} based on the Color Evaporation Model with TMD approach and the gluon Sivers function from Ref.~\cite{Boer:2003tx} for different center-of-mass energies are shown in Fig.~\ref{fig:AN_D}(a) as functions of rapidity $y$.
Since the $J/\psi$ production mechanism is not well understood, the measurement of the $A^{J/\psi}_{N}$ may bring a valuable input to that matter as well.
Predictions for $A^{J/\psi}_{N}$ in proton-proton collisions at NICA energy $\sqrt{s}=27$ GeV, obtained in GPM $+$ NRQCD approach, as
function of $x_F$ and $p_T$ are shown in the Figure (\ref{fig:AN_Jpsi_VS}). For
comparison, results are presented for SIDIS1 \cite{Anselmino:2005ea} and D'Alesio et al.
\cite{DAlesio:2017rzj,DAlesio:2019gnu} parameterizations of proton Sivers
function.
     
 A measurement with open-heavy hadrons (both $D$- and $B$-mesons) was performed at RHIC (PHENIX, $\sqrt{s}=200$ GeV)~\cite{Aidala:2017pum} using high-$p_T$ muons from their semileptonic decays. 
The obtained results are affected by relatively large statistical uncertainties and do not exhibit any significant non-zero asymmetry. Nevertheless, the results do not contradict the predictions of the twist-3 approach from Ref.~\cite{Koike:2011mb}. The Sivers effect contribution to the $A_N^D$ asymmetry calculated within the Generalized Parton Model for $\sqrt{s}=27$ GeV is presented in Fig.~\ref{fig:AN_D}(b).
 

\begin{figure}[!h]
  \begin{minipage}[ht]{0.50\linewidth}
  \center{\includegraphics[width=1\linewidth]{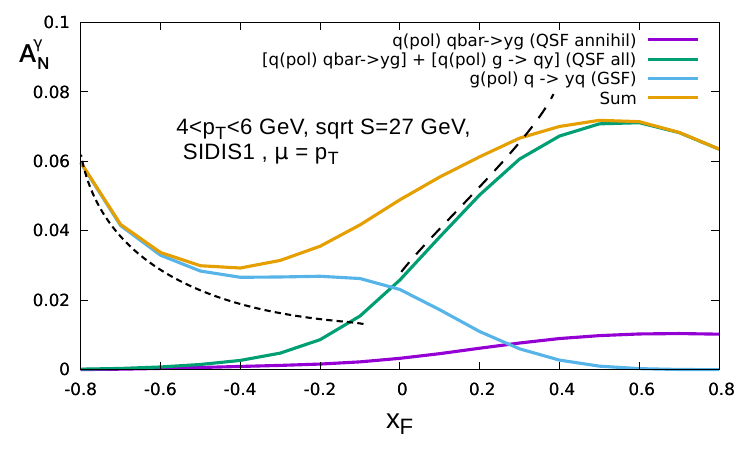} \\ (a)}
  \end{minipage}
  \hfill
  \begin{minipage}[ht]{0.50\linewidth}
 \center{\includegraphics[width=1\linewidth]{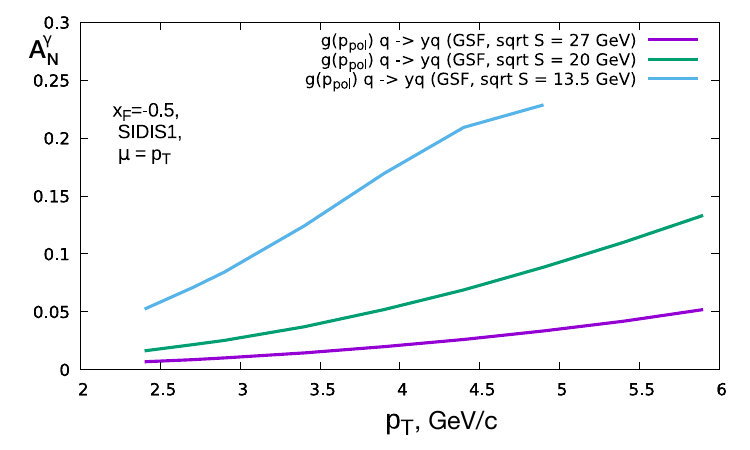} \\ (b)}
  \end{minipage}
  \caption{(a) $x_F$-dependence of the asymmetry $A_N^{\gamma}$ calculated basing on the SIDIS1 Sivers function for $\sqrt{s}=27$ GeV and $4<p_T<6$ GeV/$c$. Gluon and quark contributions are shown separately by color solid lines. Dashed lines illustrate the twist-3 predictions for  $\sqrt{s} = 30$ GeV and $p_T = 4$ GeV/$c$ for negative~\cite{Hammon:1998gb} and positive~\cite{Qiu:1991pp} values of $x_F$. (b) $p_T$-dependence of the $A_N^{\gamma}$ asymmetry for different values of $\sqrt{s}$ at $x_F=-0.5$.}
  \label{fig:AN_PP} 
\end{figure}

The measurement of the $A^{\gamma}_N$ SSA with prompt photons provides a unique opportunity to study the Sivers PDF and twist-3 correlation functions
, since the corresponding hard process does not involve fragmentation in the final state and thus is exempt from the Collins effect.
The first attempt to measure $A^{\gamma}_N$ at $\sqrt{s}=19.4$ GeV was performed at the fixed target experiment E704 at Fermilab in the kinematic range $-0.15<x_F<0.15$ and 2.5 GeV/$c$ $<p_T<$ 3.1 GeV/$c$. The results were consistent with zero within large statistical and systematic uncertainties~\cite{Adams:1995gg}. Figure~\ref{fig:AN_PP}(a) shows the expected $A_N^{\gamma}$ asymmetry as a function of $x_F$ for $\sqrt{s}=27$~GeV based on the SIDIS1 extraction of the gluon Sivers function. Quark and gluon contributions from the gluon Compton scattering, dominating at positive and negative values of $x_F$, respectively, are shown separately. The $q\bar{q}$ annihilation contribution is also presented.  Dashed lines illustrate the twist-3 predictions for  $\sqrt{s} = 30$ GeV and $p_T = 4$ GeV/$c$ for negative~\cite{Hammon:1998gb} and positive~\cite{Qiu:1991pp} values of $x_F$. The $p_T$ dependence of the $A_N^{\gamma}$ asymmetry at $x_F=-0.5$ is shown for different values of $\sqrt{s}$ in Fig.~\ref{fig:AN_PP}(b).
 

     

\subsection{Gluon transversity in deuteron \label{trans}}
\begin{figure}[!h]
  \begin{minipage}[ht]{0.50\linewidth}
    \center{\includegraphics[width=1\linewidth]{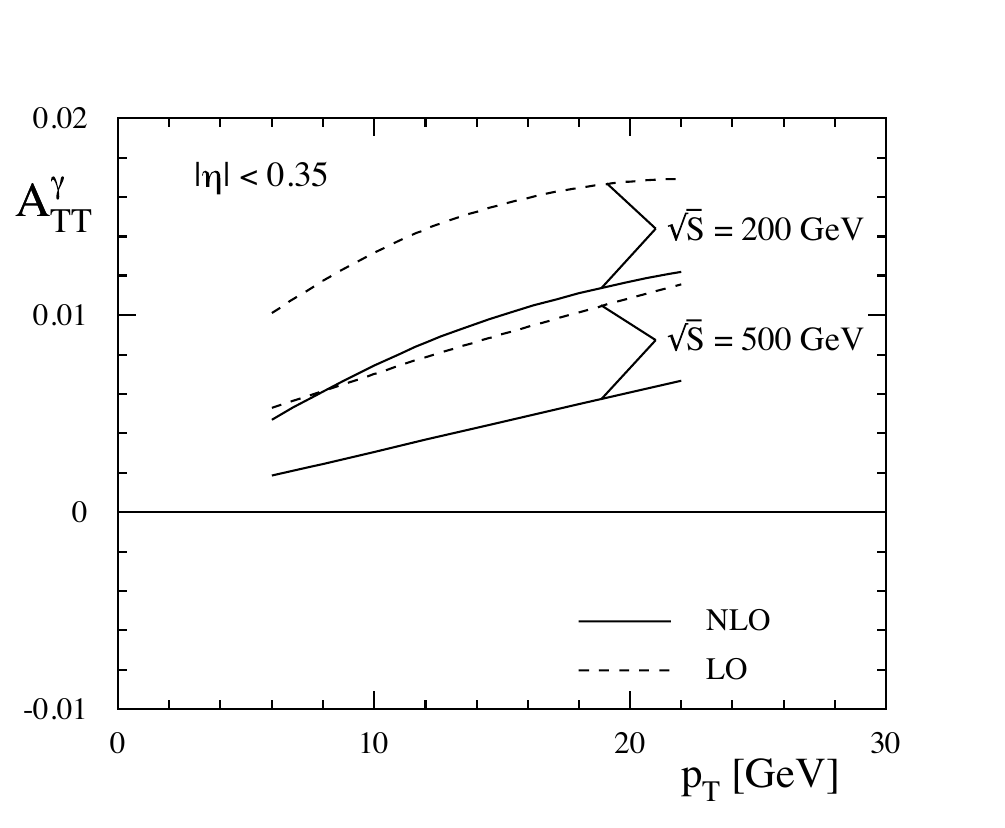}\\ (a)}
  \end{minipage}
  \hfill
  \begin{minipage}[ht]{0.50\linewidth}
  \vspace{-14pt}
   \center{\includegraphics[width=1.\linewidth]{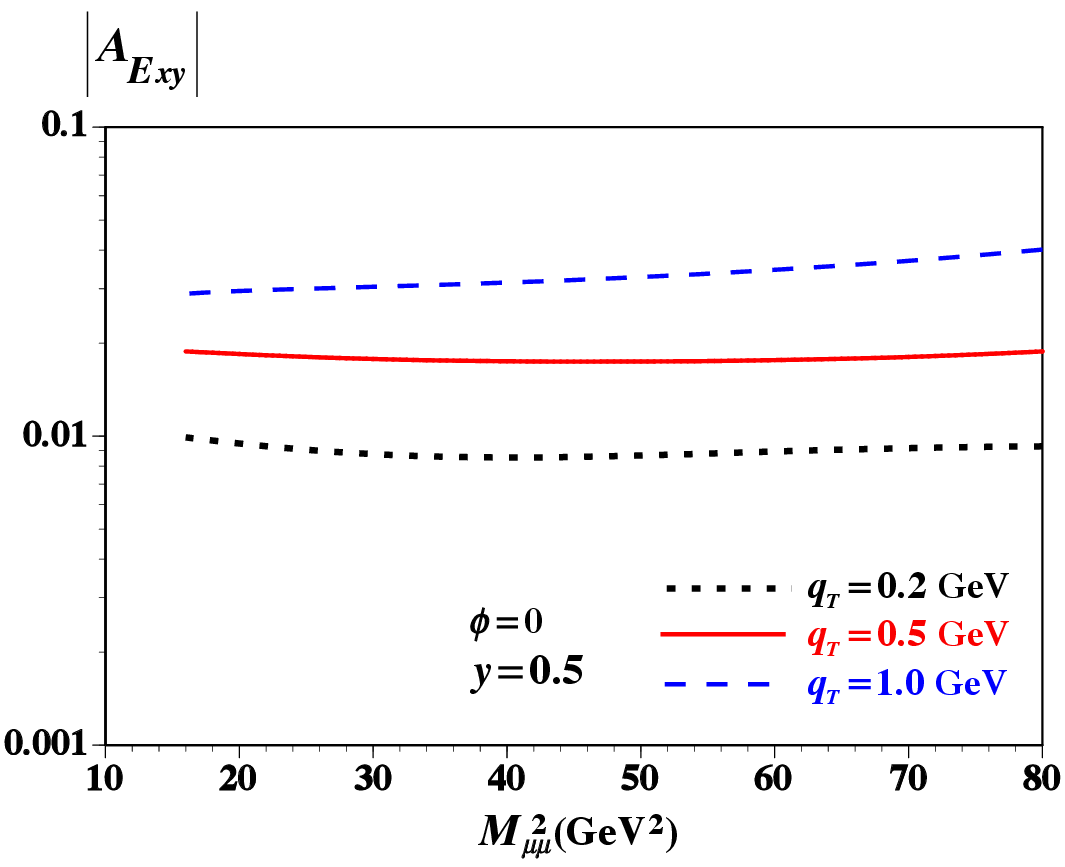} \\ (b)}
  \end{minipage}
  \caption{(a) $A^{\gamma}_{TT}$ asymmetry for the prompt-photon production at 200 and 500 GeV coming from the $q\bar{q}$ annihilation process calculated in LO~\cite{Soffer:2002tf}  and NLO~\cite{Mukherjee:2003pf}. (b) Spin asymmetries $| A_{E_{xy}} |$ for 
the proton-deuteron Drell-Yan process as a function of the dimuon mass squared for the dimuon rapidity $y=0.5$ and different transverse momenta $q_T$ \cite{Kumano:2019igu}. }
  \label{fig:Deuteron} 
\end{figure}

The transversity function $\Delta_T q(x)$ is defined for partons as the difference of probabilities to find in a transversely polarized nucleon a parton with the same and opposite spin orientations. In spite of the definition is similar to the helicity function $\Delta q(x)$, the transversity describes a completely different aspect of the nucleon spin structure. This function is known quite well after a series of SIDIS and Drell-Yan experiments. As soon as the transversity is related with the spin flip, for the spin-1/2 nucleon only a quark contribution ($\Delta s=1$) is possible while $\Delta s=2$ for the spin-1 gluons is forbidden in the twist-2. Nevertheless, a tiny nonzero gluon transversity is allowed due to higher-twist effects and  possible physics beyond the Standard model like electric dipole moment of the neutron~\cite{Kumano:2019igu}. The transverse double spin asymmetry $A_{TT}$ defined for interaction of transversely polarized hadrons by the similar manner as $A_{LL}$ is a way to access the transversity. But due to the absence of a gluonic contribution in the leading order in the case of the nucleon interactions $A_{TT} \ll A_{LL}$. As an example, the asymmetry $A^{\gamma}_{TT}$ for the prompt-photon production at 200 and 500~GeV coming from the $q\bar{q}$ annihilation process calculated in LO~\cite{Soffer:2002tf} and NLO~\cite{Mukherjee:2003pf} is shown in Fig.~\ref{fig:Deuteron}. 

The situation changes \cite{Jaffe:1989xy} for the spin-1 deuteron where a gluon component not embedded into the nucleons is possible. So in the collision of transversely polarized deuterons a nonzero contribution of the gluon transversity $\Delta_T g(x)$ to $A_{TT}$ asymmetries is possible already in the twist-2. At the moment there is no experimental data on the gluon transversity in the deuteron.  So, a measurement of the double transverse spin asymmetries $A_{TT}$ in the gluon-induced processes at polarized $d$-$d$ collisions at NICA SPD could be a way to access the $\Delta_T g(x)$.

 The gluon-induced (NLO) Drell-Yan process $qg\to q\gamma^*\to q\mu^+\mu^-$ was proposed in Ref.~\cite{Kumano:2019igu} as a way to access it in the  collisions of linearly polarized deuterons and unpolarized protons at the SpinQuest experiment at Fermilab ($\sqrt{s_{NN}}=15$ GeV).  Under assumption that $\Delta_T g(x)=\Delta g(x)$, the asymmetry 
 \begin{equation}
A_{E_{xy}}=\frac{d\sigma (E_x)-d\sigma (E_y)}{d\sigma (E_x)+d\sigma (E_y)}
\end{equation}
could reach the level of a few percent as it is shown in Fig. \ref{fig:Deuteron}~(b). At NICA SPD, the $J/\psi$, open charm and prompt-photon production in $p$-$d$ and $d$-$d$ collisions with a linearly polarized deuteron can be used to access the gluon transversity. The asymmetry $A_{E_{xy}}$ in these processes is expected to be of the same order of magnitude. 

\subsection{Deuteron tensor polarization and shear forces \label{TensorD}}

The availability of tensor polarized deuteron beam opens a possibility to study shear forces generated by quarks and gluons~\cite{Teryaev:2019ale}.  
The natural way to get the traceless part of the energy-momentum tensor related to shear is provided just by tensor polarization, as the relevant tensor $S^{\mu \nu}$ is a traceless one by construction. The contribution of the "tensor polarized" parton distribution $C^T$ \cite{Hoodbhoy:1988am,Close:1990zw} (introduced as
an "aligned" one~\cite{Efremov:1981vs}) is constrained by the zero sum rule~\cite{Efremov:1981vs} for its second moment (complementing the Close-Kumano sum rule \cite{Close:1990zw})which may be decomposed into quark and gluon components~\cite{Efremov:1994xf}:

\begin{eqnarray}
\label{sr}
\sum_{i=q, \bar q} \int_0^1 C^T_i (x) x dx = \delta_T (Q^2), \\
\int_0^1 C^T_G (x) x dx = - \delta_T (Q^2).
\end{eqnarray}
As a result, the matrix elements of energy momentum tensors of quarks and gluons look like
\begin{eqnarray}
\sum_{i} \langle P, S|T_{i}^{\mu \nu}|P, S \rangle _{Q^2}=
2 P^{\mu} P^{\nu} (1-\delta(Q^2)) +2 M^2 S^{\mu \nu} \delta_T(Q^2)\\
\langle P, S|T_g^{\mu \nu}|P, S \rangle _{\mu ^2}=
2 P^{\mu} P^{\nu} \delta(Q^2) - 2 M^2 S^{\mu \nu} \delta_T(Q^2),
\end{eqnarray}
where the second terms describe the average (integrated over transverse distance) shear force. Here $M$ is the nucleon mass. 

The zero sum rules (\ref{sr}) were later interpreted~\cite{Teryaev:2009zz} as yet another manifestation of Equivalence Principle (EP), as it was done earlier~\cite{Teryaev:1999su} for Ji sum rules. In turn, the smallness of $\delta_T$, compatible with the existing HERMES data, was suggested~\cite{Teryaev:2009zz} to be the new manifestation of Extended Equivalence Principle (ExEP)~\cite{Teryaev:2003ch,Teryaev:2006fk, Teryaev:2016edw} valid separately for quarks and gluons in non-perturbative QCD due to the confinement and chiral symmetry violation. It was originally suggested for anomalous gravitomagnetic moments~\cite{Teryaev:2003ch, Teryaev:2016edw}.
In particular, it provides the rotation of spin in the terrestrial experiment with the angular velocity 
of Earth rotation. Let us stress, that it may seem trivial if spin is considered just as a vector. However, 
it becames highly non-trivial if the measurement of spin by the device rotating together with Earth is taken into account. This is a particular example of the practical importance of the quantum theory of measurement. Another example may be represented by the Unruh radiation in heavy-ion collisions~\cite{Prokhorov:2019cik}, which implies that the particles production may be also considered as a quantum-mechanical measurement in the non-inertial hadronic medium. 

Recently, ExEP was also discovered for the pressure~\cite{Polyakov:2018exb}.

To check ExEP for shear force one may use future studies of DIS at JLab and of Drell-Yan process with tensor polarized deuterons~\cite{Kumano:2017uel}~\footnote{Complementary probes are provided by vector mesons~\cite{Teryaev:2006fk}.}.

Note that tensor polarized parton distribution may be also measured in {\it any} hard process with the relevant combination of deuteron polarizations, in particular, for large $p_T$ pions production, providing much better statistics. The correspondent quantity can be the P-even Single Spin asymmetry
\begin{equation}
A_T=\frac{d \sigma(+)+ d \sigma(-)- 2 d \sigma(0)}{d \sigma(+)+d \sigma(-) + d \sigma(0)} \sim \frac{\sum_{i=q,\bar{q},g} \int d\hat \sigma_i C^T_i(x)}{\sum_{i=q,\bar{q},g} \int d\hat \sigma_i q_i(x)} ,
\end{equation}
where the differential cross-section with definite polarization of deuteron appears. 

Note that due to 
the tensor polarization tensor being traceless the sum rule for the three mutually orthogonal orientations 
of coordinate frame is valid~\cite{Efremov:1981vs}:
\begin{equation}
\sum_{i} S^i_{zz}=0.
\end{equation}
As a result, the leading twist kinematically dominant "longitudinal" tensor polarization can be obtained by 
accelerating {\it transverse} polarized deuterons which will be accessible at NICA.

\subsection{Summary \label{sum}}
The proposed measurements at SPD will be carried out performing high-luminosity $p$-$p$, $p$-$d$ and $d$-$d$ collisions at the center-of-mass energy up to 27~GeV using longitudinally or transversely polarized proton and vector- or tensor-polarized deuteron beams. The SPD experiment will have a unique possibility to probe gluon content employing simultaneously three gluon-induced processes: the inclusive production of charmonia ($J/\psi$ and higher states), open charm production, and production of the prompt photons.
The kinematic region to be covered by SPD is unique and has never been accessed purposefully in polarized hadronic collisions. The data are expected to provide inputs for gluon physics, mostly in the region around $x\sim 0.1$ and above. 
The expected event rates for all three aforementioned production channels are sizable in the discussed kinematic range and the experimental setup is being designed to increase the registration efficiency for the final states of interest. 

The main proposed measurements to be carried out at SPD are summarized in Table~\ref{tab:summary_g}. More detailed review of the possible polarized gluon physics at NICA SPD could be found at \cite{Arbuzov2020}.

\begin{table}[htp]
\caption{Study of the gluon content in proton and deuteron at SPD.}
\begin{center}
\begin{tabular}{|l|c|c|}
\hline
Physics goal & Observable & Experimental conditions  \bigstrut \\
\hline
Gluon helicity $\Delta g(x)$ & $A_{LL}$ asymmetries & $p_L$-$p_L$, $\sqrt{s}=$27 GeV \bigstrut \\
\hline
Gluon Sivers PDF  $\Delta^{g}_N (x,k_{T})$,      & $A_N$ asymmetries,    & $p_T$-$p$, $\sqrt{s}=$27 GeV \bigstrut \\
Gluon Boer-Mulders PDF $h_1^{\perp g}(x,k_T)$ &   Azimuthal asymmetries        &  $p$-$p$, $\sqrt{s}=$27 GeV \bigstrut \\
TMD-factorization test         &  Diff. cross-sections,               &      $p_T$-$p$,        energy scan                                \bigstrut \\
                                            &      $A_N$ asymmetries          &         \bigstrut \\          
\hline
Unpolarized gluon  &       & $d$-$d$, $p$-$p$, $p$-$d$ \bigstrut \\   
density $g(x)$ in deuteron   &  Differential               &                       $\sqrt{s_{NN}}=$ 13.5 GeV                             \bigstrut \\         
Unpolarized gluon  & cross-sections  & $p$-$p$, \bigstrut \\   
density $g(x)$ in proton   &                 &                       $\sqrt{s}\leq$ 20 GeV                             \bigstrut \\      
\hline  
Gluon transversity $\Delta g_T(x)$ & Double  vector/tensor & $d_{tensor}$-$d_{tensor}$, $\sqrt{s_{NN}}=$ 13.5 GeV \bigstrut \\
                                                        &asymmetries &  \bigstrut \\
''Tensor porlarized'' PDF $C_G^T(x)$ & Single  vector/tensor  & $d_{tensor}$-$d$, $p$-$d_{tensor}$ \bigstrut \\
                                                           &asymmetries & \bigstrut \\
\hline
\end{tabular}
\end{center}
\label{tab:summary_g}
\end{table}%

The study of the gluon content in the proton and deuteron at NICA SPD will serve as an important contribution to our general understanding of the spin structure of hadrons and QCD fundamentals. The expected inputs from the SPD will be highly complementary to the ongoing and planned measurements at RHIC, future facilities such as EIC at BNL and fixed-target LHC projects at CERN. 


\section{Quarks in proton and deuteron}
\subsection{Single-transverse spin asymmetries in the production of light mesons}
The single-transverse spin asymmetries (STSA) in the inclusive production of light mesons are the simplest spin observables in hadronic scattering and also related with the Sivers, Collins, and Boer-Mulders transverse momentum dependent functions discussed in the previous section. But unlike the case of charmonia, open charm and prompt photon production, quarks are the main contributors to the corresponding asymmetries. The first result for $A_N$ was reported by the E704 collaboration for the production of the charged and neutral pions in $p^{\uparrow}$-$p$ and $\bar{p}^{\uparrow}$-$p$ collisions at $\sqrt{s}=19.4$ GeV, which is up to 30\% in the forward direction \cite{Adams:1991rw,Adams:1991cs,Bravar:1996ki,Adams:1997dp}. Similar values were reported by the RHIC experiments for higher energies \cite{Heppelmann:2009cp,Arsene:2008aa,Abelev:2008af,Adler:2005in,Adams:2003fx}.  

Understanding of STSAs in the light mesons production is possible basing on the QCD factorization approach that separates the cross-sections into perturbatively calculable partonic-level cross-section and non-perturbative physics encoded in parton distribution functions (PDFs) and fragmentation functions (FFs). Being included into the  global analysis  together with other available data they are the important source of information about quark TMD functions. The JAM20 \cite{Cammarota:2020qcw} is a recent example of such global QCD analysis where the available data for  $p$-$p$ collisions have been combined with SSA data in SIDIS, Drell-Yan pair production, and  $e^+e^-$ annihilation.

\begin{figure}[h]
\begin{center}
\includegraphics[scale=0.18]{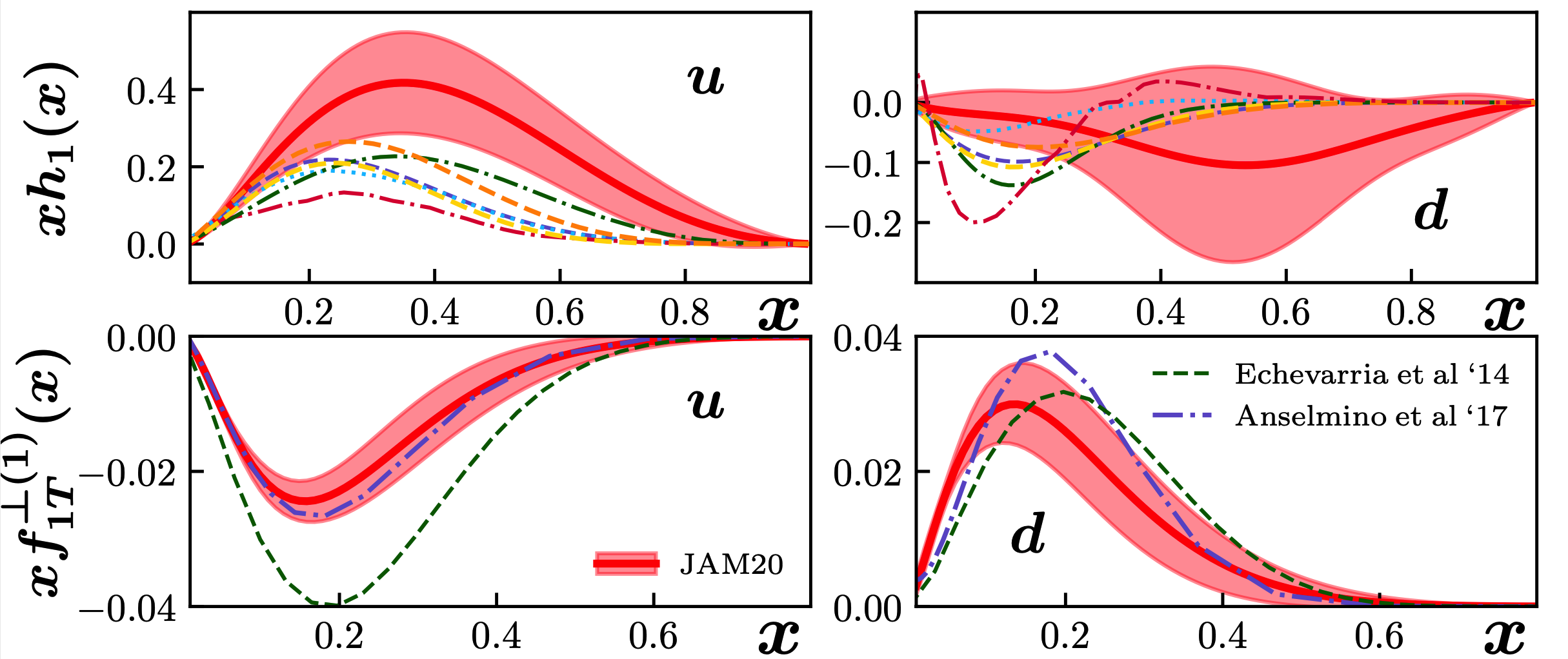}
\caption{The extracted from the global analysis JAM20 (JAM collaboration) \cite{Cammarota:2020qcw} functions $h_1(x)$ and $f_{1T}^{\perp (1)}(x)$ at $Q^2=4$ (GeV/c)$^2$ together with the corresponding results of other groups.}
\label{fig:quarks_fit1}
\end{center}
\end{figure}

\begin{figure}[h]
\begin{center}
\includegraphics[scale=1]{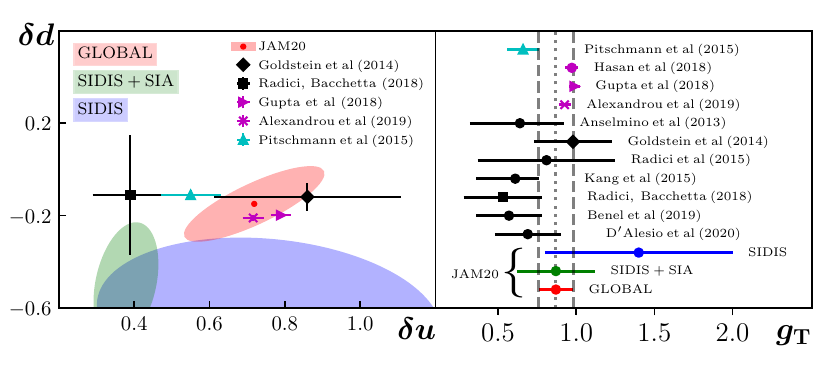}
\caption{The results of the global analysis JAM20 (JAM collaboration) \cite{Cammarota:2020qcw} along with others from phenomenology (black), lattice QCD (purple), and Dyson-Schwinger (cyan)  for the tensor charges $\delta u$, $\delta d$ and $g_T=\delta u - \delta d$ at $Q^2=4$ (GeV/c)$^2$.}
\label{fig:quarks_fit2}
\end{center}
\end{figure}

The first momenta $h_1(x)$ and $f_{1T}^{\perp (1)}(x)$ of the TMD pretzelosity $h_1(x,k_T)$ and Sivers $f_{1T}^{\perp (1)}(x, k_T)$ functions obtained from the global analysis JAM20  together with $1\sigma$ uncertainties at $Q^2=4$ (GeV/c)$^2$ are presented in Fig. \ref{fig:quarks_fit1}. The so-called tensor charges 
\begin{equation}
\delta q = \int_0^1 (h_1^q(x)-h_1^{\bar{q}}(x))dx
\end{equation}
for $u$ and $d$ quarks and $g_T=\delta u - \delta d$ are shown in Fig. \ref{fig:quarks_fit2}. New data on the STSA in the light mesons production in the SPD energy range (and especially high-$p_T$ data) are very much in demand for such kind of global analyses \cite{Prok2020}.

\subsection{Drell-Yan pair production}
Production of Drell-Yan (DY) pairs in polarized hadronic collisions $pp\to \gamma^* \to \mu^+\mu^-$ is a promising way to touch the TMP PDFs of valence quarks and sea antiquarks by measuring the azimuthal asymmetries. A tiny DY cross-section and a huge combinatorial background coming from decays of secondary pions and kaons into muons make this task rather difficult. A typical detector configuration for such kind of studies at $\sqrt{s}\sim 20$ GeV is a fixed-target beam-dump setup where due to the Lorentz boost most of secondary pions and kaons are stopped in a thick absorber before decaing. At the moment only the COMPASS experiment at CERN has presented the results for the three azimuthal asymmetries measured in pion-induced polarized DY \cite{Aghasyan:2017jop, Parsamyan:2019wwd}. The observed glimpse of the sign change  in the Sivers asymmetries is found to be consistent with the fundamental prediction of QCD that the Sivers TMD PDF extracted from the DY has a sign opposite to the one extracted from the
SIDIS data. Unique results for the Sivers functions of $\bar{u}$ and $\bar{d}$ are expected from the SpinQuest experiment at Fermilab \cite{Brown:2014sea, Chen:2019hhx}. 
\begin{figure}[h]
\begin{center}
\includegraphics[scale=0.15]{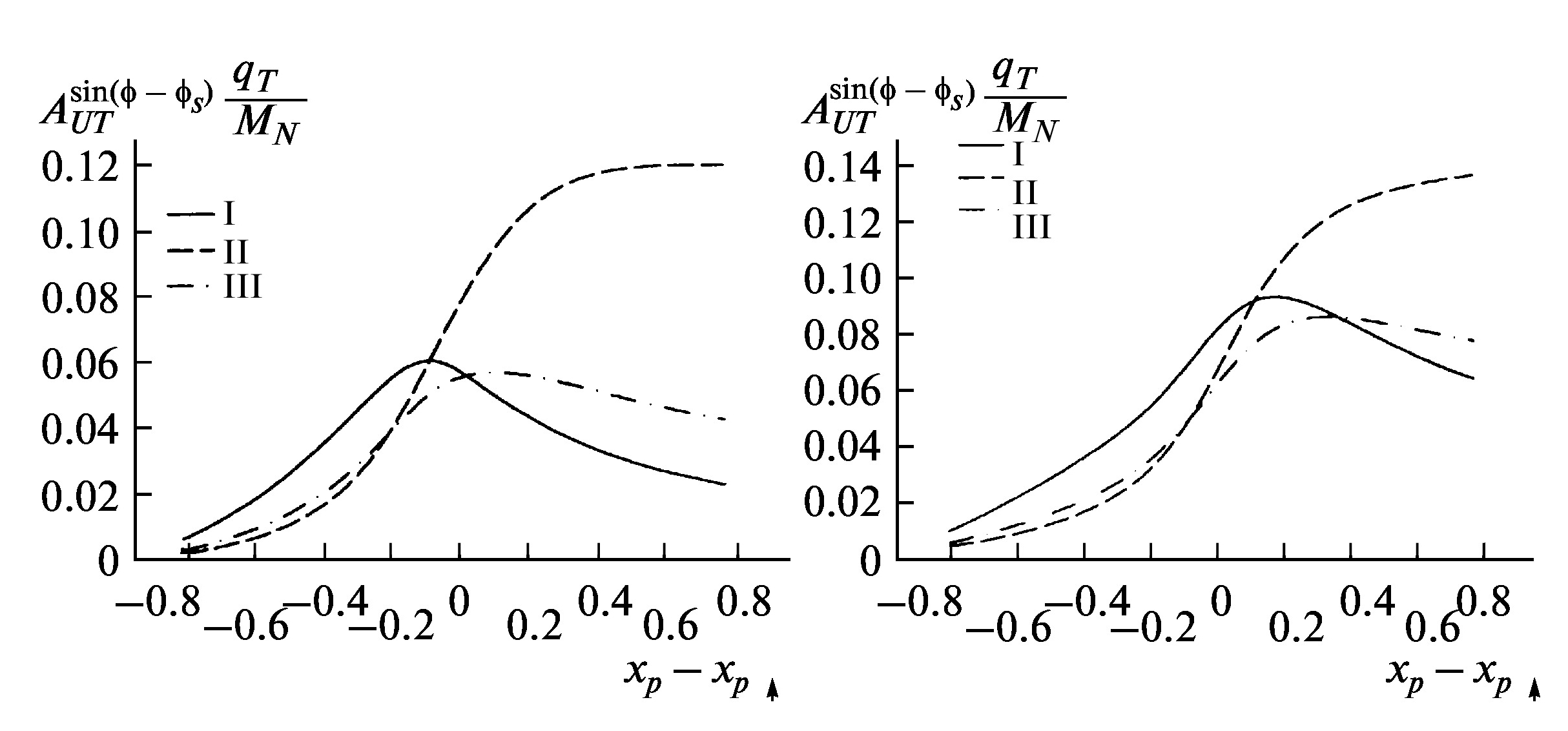}
\caption{Estimated Sivers asymmetries for the NICA conditions $s=20~GeV$, $Q^2=4~GeV^2$ (left) and 
$s=20~GeV$, $Q^2=15~GeV^2$ (right).
Fits for the Sivers functions are taken from \cite{Collins:2005ie}.}
\label{fig:theor-sivers}
\end{center}
\end{figure}

\begin{figure}[h]
\begin{center}
\includegraphics[scale=0.15]{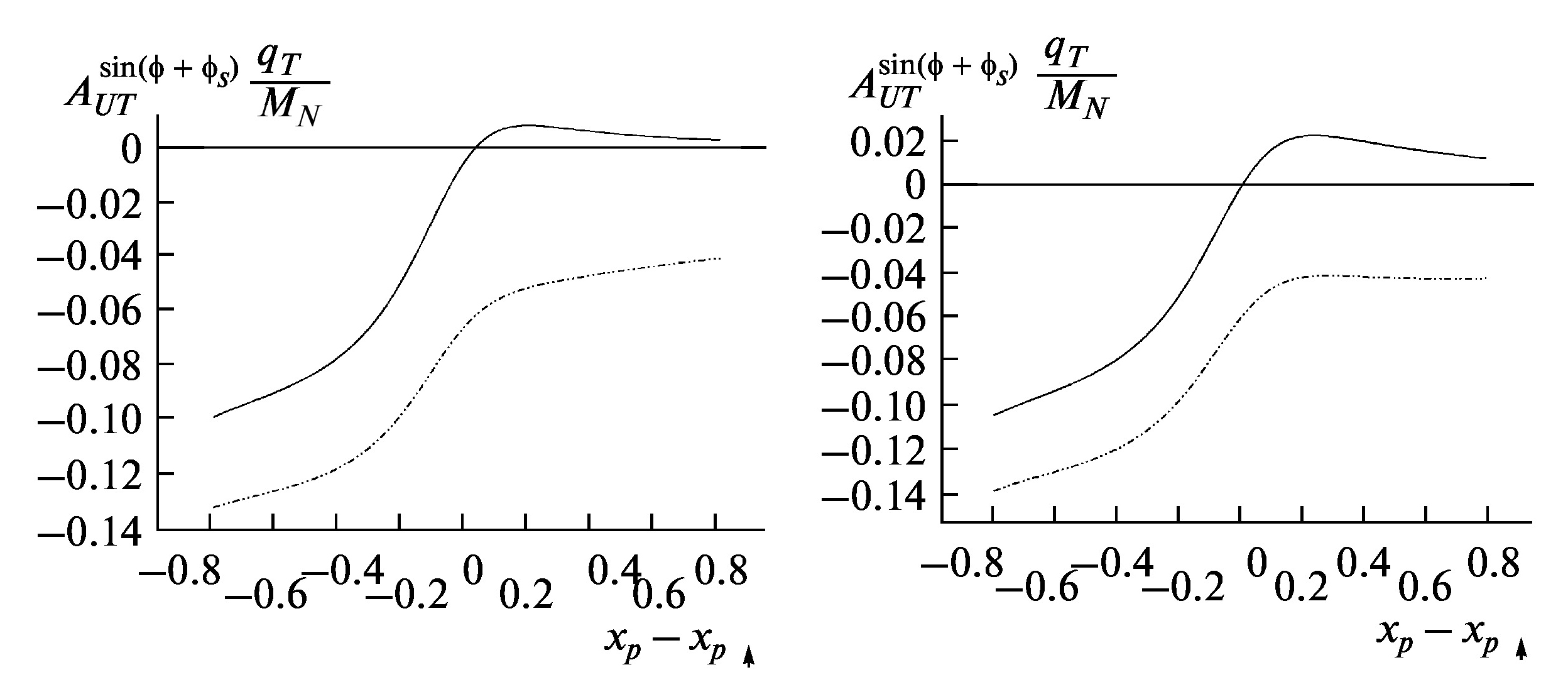}
\caption{Estimated Boer-Mulders asymmetries for the NICA conditions.}
\label{fig:theor-boer}
\end{center}
\end{figure}
Unfortunately, the Spin Physics Detector cannot use the advantage of fixed-target beam-dump setups and the expected background conditions for the Drell-Yan measurements are rather untoward. However, further improvement of the experimental techniques and analysis procedures could give a chance to access polarized DY at SPD. The estimated Sivers and Boer-Mulders asymmetries for the SPD conditions are presented in Fig. \ref{fig:theor-sivers} and \ref{fig:theor-boer}, respectively.

\subsection{Generalized parton distributions}
\begin{figure}[h]
  \begin{minipage}[ht]{0.32\linewidth}
    \center{\includegraphics[width=1\linewidth]{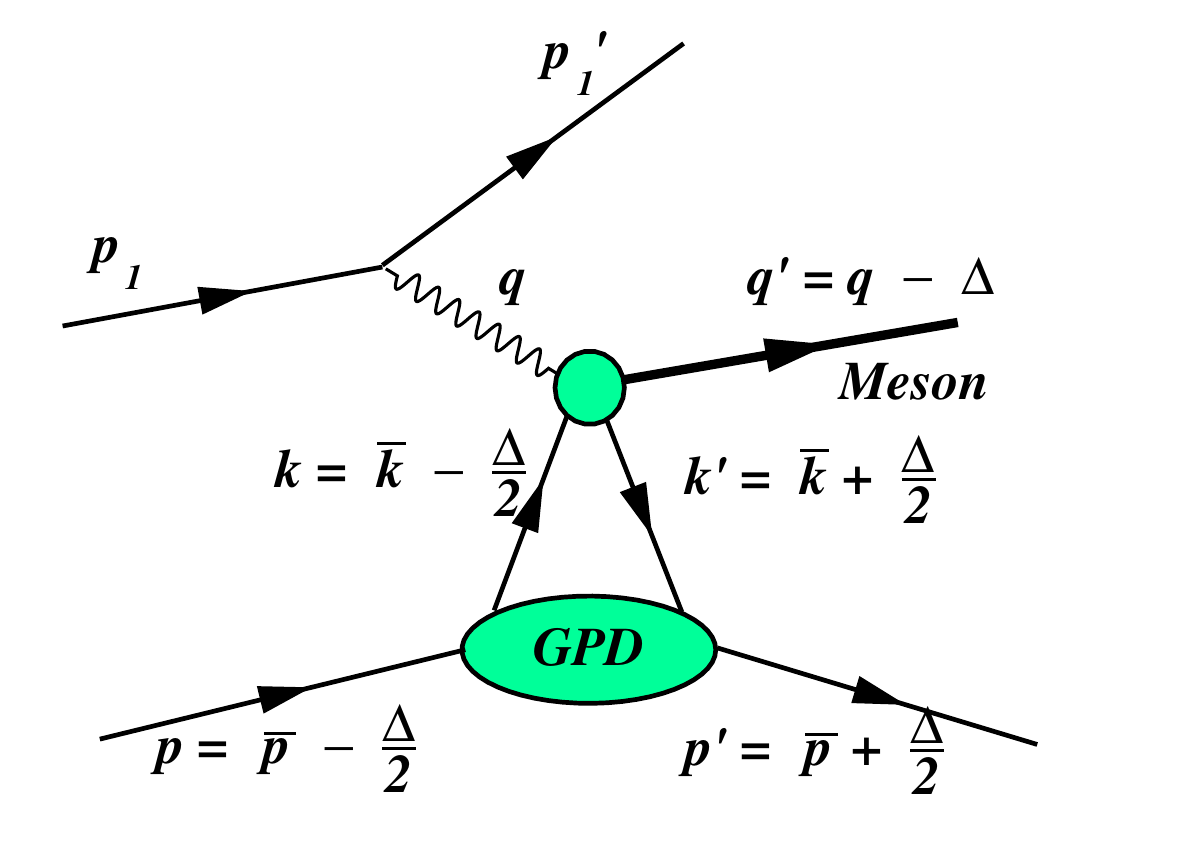} \\ (a)}
  \end{minipage}
  \hfill
  \begin{minipage}[ht]{0.32\linewidth}
    \center{\includegraphics[width=1\linewidth]{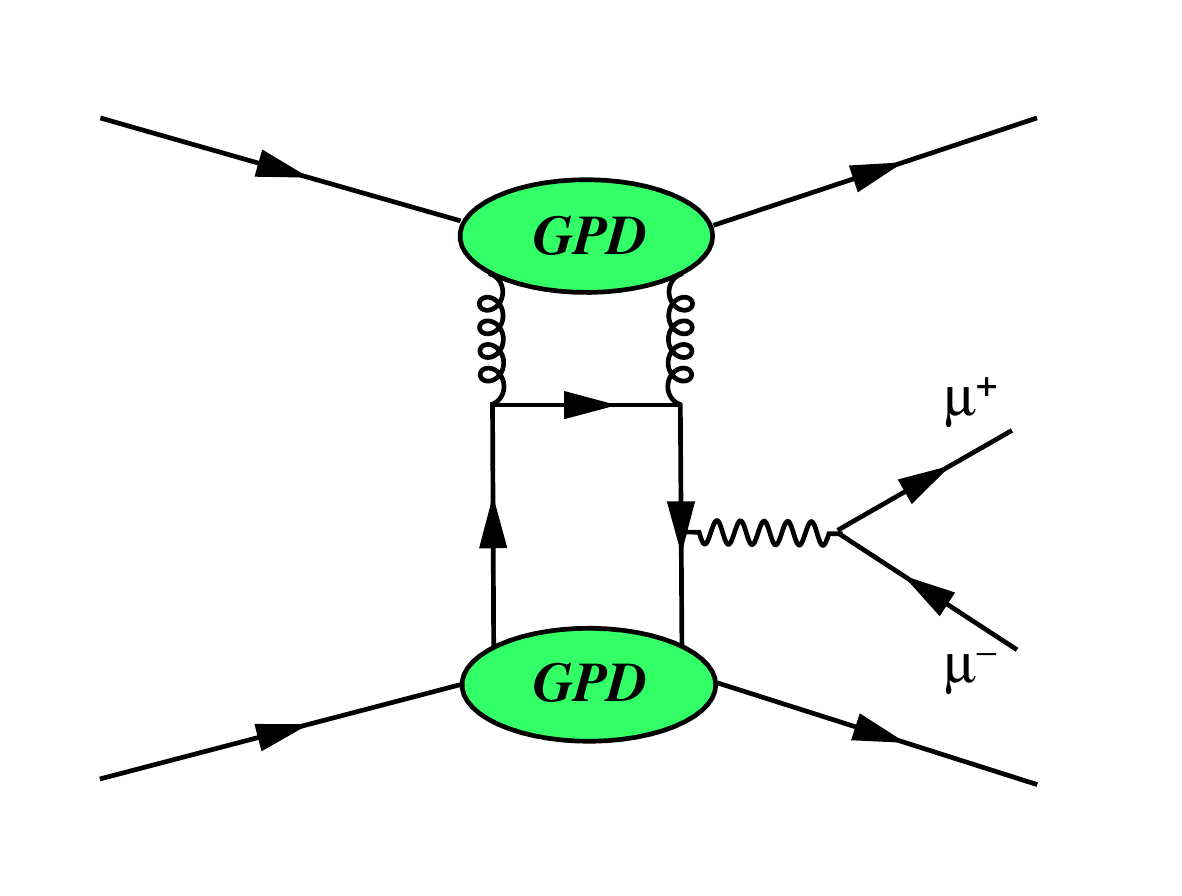} \\ (b)}
  \end{minipage}
  \begin{minipage}[ht]{0.32\linewidth}
    \center{\includegraphics[width=1\linewidth]{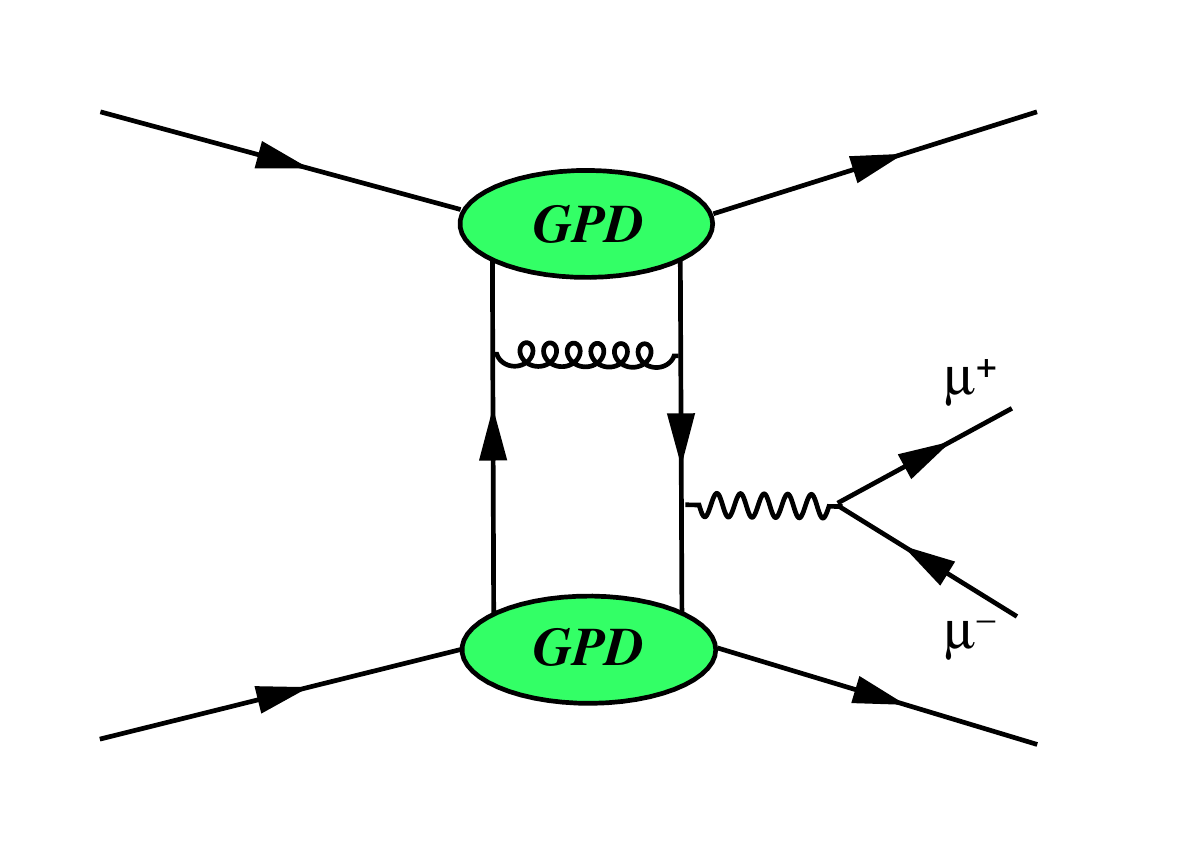} \\ (c)}
  \end{minipage}
  \caption{(a) Vector meson production at NICA via photoproduction mechanism or odderon exchange.
  (b, c) Drell-Yan process with gluon and quark GPDs.
  \label{fig:gpd1}}
\end{figure}

The concept of Generalized Parton Distributions (GPDs) is a complementary to the TMD PDFs approach to describe the three-dimensional structure of hadrons. Study of the deeply virtual meson production (DVMP) is one of the proven ways to access GPDs. This process has been investigated at HERMES \cite{Airapetian:2008aa}, COMPASS \cite{Alexeev:2019qvd,Akhunzyanov:2018nut}, CLAS \cite{Park:2017irz}  using electron and muon beams. An exclusive electromagnetic process $pp\to ppM$ shown in Fig. \ref{fig:gpd1}(a), where the first proton radiates a 
photon with low virtuality that interacts with the other proton and produces a meson, could be used to access the Generalized Parton Distributions at SPD. At the SPD energies, the
meson photoproduction amplitude can be presented in a factorized
form as a convolution of the hard scattering part which can be
calculated perturbatively and the GPDs
\cite{Goloskokov:2006hr,Goloskokov:2007nt}. In the case of vector mesons  production, the odderon exchange (that could be described as an exchange by at least 3 gluons) is also possible and the interference of these two channels is a matter of special interest. Ultraperipheral  $p$-$A$ collisions at SPD, which enhance the photoproduction contribution by several orders of magnitude could also be considered. In addition, ultraperipheral processes could be used to test the most general non-perturbative concept of the Generalized Transverse Momentum dependent Distributions (GTMD). This possibility was explored for high energies in Ref. \cite{Hagiwara:2017fye} but the approach could be extended down to the SPD energies.

The exclusive production of the $J/\psi$ meson can be studied at SPD at energies $W=\sqrt{(q+p)^2} \sim 5 \div 15~\mbox{GeV}$. Here
$q$ and $p$ are the 4-momenta of a virtual photon (odderon) and a proton, respectively.
The large meson mass makes it possible to perform perturbative calculations at sufficiently low $Q^2$, where  
 the photon exchange should dominate.  The corresponding cross-section is estimated to be of about
$\sigma_{J/\psi} \sim 10 ~\mbox{nb}$. The main contribution to the cross-section is coming from the gluon GPDs.




The  exclusive Drell-Yan (exDY) process was proposed for the study of GPDs in $p$-$p$ collisions in Ref. \cite{Teryaev:2005uj}.
The kinematics of this process is defined by convolution of two GPDs. Both quark and gluon 
GPDs contribute to the exDY cross-section via the diagrams shown schematically in Fig. \ref{fig:gpd1}(b, c).

 Investigation of the
cross-section determined by two-GPDs effects is in progress now
\cite{Goloskokov:2020bsx}.  It is shown that the gluon and sea quark  GPDs lead to
the cross-section which does not decrease with the growth of energy. The exDY cross-section $d \sigma/dQ^2$ at the NICA kinematics $\sqrt{s}=24~ \mbox{GeV}$ and $Q^2=5~
\mbox{(GeV/c)}^2$  is estimated as $5~
\mbox{pb/(GeV/c)}^2$ which is much smaller with respect to the inclusive Drell-Yan cross-section. Nevertheless, the exclusivity requirement applied in the analysis of the future SPD dimuon data could increase the signal-to-background ratio. It should be mentioned that $J/\psi$ could also be produced exclusively in a similar way.

\subsection{Polarized fragmentation functions}
The information on the polarization of produced partons can be transmitted to the asymmetries of final state hadrons. This process is controlled by the polarized fragmentation functions. 

JINR played the pioneering role in that activity started with the notion of jet handedness 
\cite{Efremov:1978qy, Efremov:1992pe} which was explored in $Z_0$ decays \cite{Efremov:1994zs}. 
To measure this quantity one should explore the P-odd mixed productz of hadron (usually pions) momenta in the jets and compare the number of these products with different signs indicating the right and left triples:   
 
\begin{equation}
A =\frac{N_L-N_R}{N_L+N_R}.
\end{equation}

In the case of transverse quark polarization the triple may contain its direction as well as that of jet momentum, so that only the pion momentum is varying. This is the case of Collins function \cite{Collins:1992kk} revealed in the specific angular modulations.
The Collins functions are systematically studied in $e^+e^-$ annihilation, SIDIS and hadronic processes.

Another opportunity to form the P-odd combination of momenta is the dihadron and interference fragmentation functions \cite{Jaffe:1997hf}. 
The dihadron fragmentation function was effectively used for transversity measurement \cite{Bacchetta:2011ip}.

The polarized fragmentation functions for both helicity and transversity quark distributions can be used also in the case of $\Lambda$-hyperon production 
\cite{Ma:2001rm}.

The polarized quark and gluon fragmentation functions may be explored in the production to tensor polarized vector mesons \cite{Schafer:1999am}.

The gluon fragmentation to the transverse polarized quarkonium 
\cite{Cho:1994gb} may be also considered as polarized fragmentation function. It is known to produce the dominant transverse polarization of colour-octet states.

The exploration of polarized fragmentation function in hadronic processes should involve the measurement of rapidity dependence of the hard process, which should be essential in separation of various partonic subprocesses. 

\subsubsection{$\Lambda$- polarization}

The $\Lambda$-hyperon is an ideal testing ground for spin studies since it has a rather simple spin structure in the naive quark parton model. Furthermore, its self-analyzing decay makes polarization measurements experimentally feasible. Measurements of the $\Lambda$ polarization in hadronic collisions can provide a clear answer to the question of whether polarized $u$ and $d$ quarks can transfer polarization to the final state $\Lambda$. Polarization of the $\Lambda$ was the subject to close study in the polarized leptoproduction (COMPASS \cite{Alekseev:2009ab,Moretti:2019peg}, HERMES \cite{Airapetian:2007mx}, E665 \cite{Adams:1999px}), and $p$-$p$ collisions (E704 \cite{Bravar:1995fw, Bravar:1997fb}, STAR \cite{Abelev:2009xg}, PHENIX \cite{Han:2007zzc}). 

There is a well known experimental fact that the hyperons produced in the unpolarized $p$-$p$ collisions are polarized transversely to the production plane \cite{Bunce:1976yb}. That is the result of higher-twist effects.
Polarization measurement of  the $\Lambda$-hyperons produced inclusively in $p$-$p$ collisions with one of the protons longitudinally polarized, 
\begin{equation}
p^{\uparrow} p → {\Lambda}^{ \uparrow}X,
\end{equation}
could provide information about the spin-dependent fragmentation function $D_{q,g\to \Lambda}(z)$ that describes fragmentation of quarks and gluons into the $\Lambda$.

The E704 measurement \cite{Bravar:1997fb} of the parameter $D_{NN}$, characterizing the fraction of incident (polarized) proton  polarization  transferre to the inclusively produced $\Lambda$, was performed at $\sqrt{s}=19.4$ GeV.  The unexpectedly large value, up to about 30\%, was found for the large $p_T$. The obtained result can not be treated easily within the existing models.

\section{ Tests of the QCD basics at low energies}
Quantum chromodynamics has a remarkable success in describing the high-energy and large-momentum transfer processes, where partons in hadrons behave, to some extent, as free particles and, therefore, the perturbative QCD approach can be used. Cross-section of a process in QCD is factorized  into two parts: the process-dependent perturbatively-calculable short-distance partonic cross-section (the hard part) and universal long-distance functions, PDFs and FFs (the soft part).
 Nevertheless, a largest fraction of hadronic interactions involves  low-momentum transfer processes in which the effective strong coupling constant is large and the description within a perturbative approach is not adequate. A number of (semi-)phenomenological approaches have been developed through the years to describe strong interaction in the non-perturbative domain starting from the very basic principles. They successfully describe such crucial phenomena, as the nuclear properties and interactions, hadronic spectra,  deconfinement, various polarized and unpolarized effects in hadronic interaction, etc. The transition between the perturbative and non-perturbative QCD is also a subject of special attention.  In spite of a large set of experimental data and huge experience in few-GeV region with fixed-target experiments worldwide, this energy range still attracts both experimentalists and theoreticians.  In addition, the  low-energy physics at SPD is some kind of a bridge to the physics program of MPD, another experiment at NICA \cite{MPD_CDR,MPD_WP,MPD_URL}. In this Section, we discuss certain problems that could be addressed at the initial stage of SPD operation with a reduced collision energy and luminosity, realizing, however, that this list is not  exhaustive.

\subsection{Elastic scattering}
\subsubsection{Large-angle  $p$-$N$ elastic scattering}

The $p$-$p$ and $p$-$n$ elastic scattering at high energy $\sqrt{s}=5\div7$ GeV  and large transferred momentum $-t= 5\div10$ GeV$^2$ is powered by the 
 short-range properties of $N$-$N$ interactions corresponding to a  small separation between nucleons 
 $r_{NN}\sim \hslash/\sqrt{-t}\leq 0.1$ fm.
There are three following  aspects of QCD dynamics in  these processes.
 ({\it i}) First,   the differential cross-section $d\sigma^{pp}/dt({s},\theta_{cm})$ at a fixed angle $\theta_{cm}\sim 90^\circ$ on the whole
   follows the  pQCD  constituent counting rules $d\sigma^{pp}/dt({s},\theta_{cm})\sim s^{-10}$ \cite{Allaby:1968zz,Akerlof:1967zz,Perl:1969pg,Stone:1977jh}. However, a clear  deviation from this  prediction
    in the form of oscillations with an increasing energy is observed in the region
     $s=10\div 40$ GeV$^2$ \cite{Allaby:1968zz,Akerlof:1967zz,Perl:1969pg,Stone:1977jh}. The irregularity in the energy dependence is at the level of $\sim 50 \%$ in the region, where the magnitude of the $p$-$p$ elastic cross-section falls down by 8 orders of magnitude.
 ({\it ii}) Second, anomalous polarization asymmetries were observed in hard $p$-$N$ scattering at
 $p_{lab}=11.75$ GeV/c 
 \cite{Crabb:1978km,Crosbie:1980tb,Court:1986dh}.
The elastic $p$-$p$-cross-section with spins of protons parallel and normal   
    to the scattering plane  is almost four times larger than the cross-section 
    with antiparallel spins.
    The challenge is that in order to generate  such a large
    polarization effect, one  needs to have a large contribution  from  the double spin-flip  helicity amplitude $\phi_2$ or a negligible contribution from the helicity conserving
    $\phi_1$ amplitude.
   However, in pQCD, in contrast,  $\phi_2$ is the most suppressed and $\phi_1$ is the largest \cite{Sargsian:2014bwa}. 
   Predicted within the pQCD  (quark-interchange model), the double spin asymmetry $A_{NN}=1/3$  does  not depend on energy \cite{Brodsky:1979nc},
    \cite{Farrar:1978by}, whereas the
     measured asymmetry demonstrates ''oscillating`` energy dependence. ({\it iii}) The third QCD aspect of hard NN scattering is related to the Color Transparency phenomen (CT), that is, a  reduction of the absorption in the nuclear medium of hard produced hadrons, both mesons and  baryons \cite{Mueller}, \cite{Brodsky}. 
     The initial and final hadrons  participating in hard process in point-like configurations dictated by mechanism of high momentum transfer,
 have  small color dipole  momenta and, therefore, small interaction cross-section with the nuclear medium.
     These expectations resulted in   huge  theoretical and experimental activities in the 90's. While the CT effect is observed for the hard production of the $q\bar q$ systems, the similar effect for $qqq$  is elusive. The data \cite{Mardor:1998zf,Aclander:2004zm} on the reaction $p+A\to pp+X$ on the $^{12}$C  and $^{27}$Al   show  an ''oscillatory'' effect again, i.e.  the transparency increases
     with an increasing momentum up  to $p_{lb}= 9$ GeV/$c$, and  then decreases  below the  Glauber calculation predictions at 14 GeV/c.
      An attempt to connect  all the three above-mentioned aspects together into one approach was undertaken in Ref. \cite{Brodsky:1987xw}  and, recently, in \cite{Sargsian:2014bwa}. However, the recent measurement of the cross-section
       of the reaction $e~ ^{12}C \to e  p X$ at $Q^2=8\div14$ (GeV/$c$)$^2$ 
        \cite{Bhetuwal:2020jes}
         shows   no CT effect and this fact  raises new  questions to the  analysis made  in  \cite{Sargsian:2014bwa}.
         Nevertheless, according to \cite{Brodsky:1987xw},   the observed 
         large variations in  spin correlations of $p$-$p$ elastic scattering are consistent  with formation in the s-channel  of the ''octoquark'' resonances $uuds\bar s uud$  and    $uudc\bar c uud$ near the strangeness and  charm production thresholds, respectively. The  variations with an increasing energy are explained as a result of interference of the pQCD background amplitude  with  non-perturbative resonant amplitudes.
         Furthermore, the model \cite{Brodsky:1987xw} provides a description of the oscillations in the unpolarized differential $p$-$p$ elastic cross-section. One should mention, however, 
         that  another explanation of the oscillation effect in the  $d\sigma^{pp}/dt({s},\theta_{cm})$
         was  suggested  in  Ref. \cite{Ralston:1982pa}.

     A different spin-isospin structure of the   transition matrix elements
     for the  near threshold  $J/\psi$ production in $p$-$n$ and $p$-$p$ collisions \cite{Rekalo:2002ck} means
    that  spin observables in $p$-$n$ elastic scattering
     can give independent information on the considered dynamics.       
      Data on  these observables  are  almost absent in the considered  energy region. 
       A task  to get such  data from the  $p^{\uparrow}d^{\uparrow} \to pnp$  reaction is accessible for NICA SPD.

\subsubsection{Polarized $p$-$d$ elastic scattering within the Glauber model
and $pN$ spin amplitudes}
Nucleon-nucleon elastic scattering contains fundamental information on the dynamics of the $NN$ interaction and  constitutes a basic process in the physics of atomic nuclei and  hadrons. Full information about the 
spin amplitudes of $p$-$p$ and $p$-$n$ elastic scattering
 can be obtained, in principle,  from a complete polarization experiment, which, however, requires to measure dozens of  independent observables at a given  collision energy and constitutes a too complicated experimental task.
A systematic reconstruction of these amplitudes from the scattering data is provided by the SAID data base \cite{Arndt:2007qn} 
and covers laboratory energies up to $3$ GeV ($p_{lab}\approx 3.8$~GeV/c) for $pp$ and $1.2$ GeV ($p_{lab}\approx 1.9$~GeV/c)
for $p$-$n$ scattering. At higher energies there are only non-complete data on $p$-$p$ scattering, whereas the information 
about the $p$-$n$ system is very scarce. In the literature there are several models and corresponding
parametrizations for $p$-$N$ amplitudes. Some of them are obtained 
 in the eikonal approach  for the lab momentum of $6$ GeV/c \cite{Sawamoto:1979cb} and for the LHC energies \cite{Selyugin:2009ic}. 
 Within the Regge 
phenomenology, a parametrization is obtained  
 for $3$-$50$ GeV/c (corresponding to $2.77 < \sqrt{s} < 10$ GeV)\cite{Sibirtsev:2009cz}
and for the values of $s$ above $6$ GeV$^2$ ($p_{lab} \ge 2.2$~GeV/$c$)
in Ref.~\cite{Ford:2012dg}. 
A possible way to check the existing parametrizations is to study the spin effects in proton-deuteron ($p$-$d$) and deuteron-deuteron ($d$-$d$) 
elastic and quasi-elastic scattering. At high energies and small four-momentum transfer $t$, $p$-$d$ scattering can be described 
by the Glauber diffraction theory of multistep scattering, which involves on-shell $p$-$N$ elastic scattering amplitudes as input data. 
Applications of this theory with spin-dependent effects 
included \cite{Platonova:2016xjq} indicate a good agreement with 
the $p$-$d$ scattering data at energies of about $1$~GeV, if the SAID data on $p$-$N$ scattering amplitudes are used as input 
for the calculations \cite{Temerbayev:2015foa, Mchedlishvili:2018uur,Platonova:2019yzf}.
 
 The spin-dependent Glauber theory 
\cite{Platonova:2016xjq,Temerbayev:2015foa} has been recently applied
\cite{Uz-Jh-TemerB}
to calculate the spin observables of $p$-$d$ 
elastic scattering at $3\div50$ GeV/c, utilizing the $pp$ elastic scattering amplitudes $f_{pp}$ 
established and parametrized in Ref.~\cite{Sibirtsev:2009cz} within  
the Regge formalism.
The Regge approach allows one to 
construct $p-n$, $f_{pn}$, (and ${\bar p}-n$, 
$f_{{\bar p}n}$) amplitudes together with $p-p$ amplitudes 
 at small $-t$ and large  $s$. 
This feature allows one to perform a test of a broad  set of $p$-$N$
amplitudes and applicability of the Regge model itself to $p$-$N$ elastic scattering.
However, in view of the scarce experimental information about the spin-dependent $p$-$n$ amplitudes and taking into account 
that the spin-independent parts of the $p$-$p$ and $p$-$n$ amplitudes at high energies are approximately the same, it was  assumed  in 
\cite{Uz-Jh-TemerB}, as  a first approximation, 
 that $f_{pn}=f_{pp}$.   
The unpolarized differential cross-section, vector ($A_y^p$, $A_y^d$)
and tensor ($A_{xx}$, $A_{yy}$) analyzing powers and some 
spin correlation parameters
($C_{x,x}$, $C_{y,y}$, $C_{xx,y}$, $C_{yy,y}$)
\footnote{ We use here  notations of Ref. \cite{Ohlsen:1972zz}}
of $pd$ elastic scattering were calculated 
 at $p_l=4.85$ GeV/$c$ and 45 GeV/$c$ at $0<-t< 2$ GeV$^2$, 
 using $p$-$N$ amplitudes  from  \cite{Sibirtsev:2009cz}. 
 As shown in Ref. \cite{Uz-Jh-TemerB}  
 the available data on $p$-$d$ elastic differential cross-section in the forward hemisphere are well described by this model.
 Most sensitive to the spin-dependent $pN$ amplitudes are the vector analyzing
 powers $A_y$ and the spin correlation parameters $C_{x,x}$ and $C_{y,y}$.
 Thus, even the measurement of the ratio $A_y^d/A_y^p$ at low $t$ gives valuable information on the transverse spin-spin term
 in NN-amplitudes \cite{Uzikov:2019hzx}.
 In contrast, the tensor analyzing powers $A_{xx}$ and $A_{xx}$ are very
 weakly sensitive to those amplitudes  and weakly change with  increasing energy. 
The polarization observables  calculated in \cite{Uz-Jh-TemerB}
can be measured at  NICA SPD
and will provide a  test of the used $p$-$N$ amplitudes. 
The corresponding differential cross-section  is rather  large  in the considered region $p_{lab}=3\div50$ GeV/$c$   and   $|t|=0\div2$ GeV$^2$ being   
 $d\sigma/dt >0.1$ mb/GeV$^2$.   The expected counting rate 
 $N$
 at $p_{lab}=50$ GeV/$c$  (${q_{pp}^{cm}}= 5$ GeV/$c$)  for the luminosity $L= 5\times 10^{30}$ cm$^{-2}$s$^{-1}$ and
 for the solid angle $\Delta\Omega=0.03$ is $N \geq 10^2$ s$^{-1}$. 
 
The $p$-$N$ helicity amplitudes   $\phi_5$ and $\phi_1 +\phi_3$, which can be tested in the above described procedure  
are necessary   in the search for time-reversal 
invariance effects in double-polarized $pd$ scattering \cite{Uzikov:2015aua,Uzikov:2016lsc}.
 The  data
  of the spin-correlation parameters on $p$-$p$  elastic scattering  being analyzed
  in the framework of the eikonal model
  \cite{Selyugin:2009ic}
   will allow one to obtain  the  space structure of the
  spin-dependent hadron forces \cite{Selyugin:2010xv}.

\subsubsection{Elastic  $d$-$d$ scattering.}

The spin observables of  $d$-$d$- elastic scattering  in forward hemisphere   also can be used to  test   spin-dependent
amplitudes of $p$-$N$ elastic scattering  since the Glauber model can be applied to the description of these observables. 
 Unpolarised  differential  cross-section of the $d$-$d$ elastic scattering  in forward hemisphere  measured
  at energies $\sqrt{s}=53\div63$ GeV \cite{Goggi:1978ae}
  was well described by  the modified Glauber theory  including Gribov inelastic  corrections.  At lower energies corresponding to the SPD NICA region, one may expect that inelastic corrections are not important, that can be checked by direct calculation of unpolarised  cross-section and subsequent comparison with
  the data. In this calculations  the above considered  spin dependent amplitudes of the $pd$ elastic scattering 
  \cite{Uz-Jh-TemerB} can be used as input for the  Glauber calculations of the $dd$ scattering.
  
        At large scattering angles  $\theta_{cm}\sim 90^\circ$  
      the  $pd\to pd$ and $dd\to dd $  processes are  sensitive to the short-range (six-quark) structure of the deuteron. Therefore, measurement of
      the differential cross-sections or spin observables
      of these processes at large  $\theta_{cm}$  will be  important to search for  non-nucleonic degrees of freedom of the deuteron. 

\subsubsection{Elastic small-angle  $p$-$p$ scattering and periphery of the nucleon}

         The first evidence of the pion cloud effect in the
diffractive scattering
$|t|\sim 0.1$~GeV$^2 \approx 4m_\pi^2$
has been found in the ISR measurements
\cite{Amaldi:1979kd}.
A theoretical study  of the  effect
was provided by Anselm and Gribov \cite{Anselm:1972ir} and recently 
in Refs.
\cite{Khoze:2014nia} and \cite{Jenkovszky:2017efs}).
The  oscillation effect observed at ISR was studied later on in  Protvino 
and one more oscillation in the differential cross-section was found there 
at $|t|\sim 0.5$~GeV$^2$ located at a higher $t$,
it might be related to somewhat heavier mesons around the proton.
     The oscillation effect of the $pp$ scattering amplitude
          at a small momentum transfer was found also in 
         the analysis of the recent high-precision experimental data
         of the TOTEM collaboration at $\sqrt{s} = 13 $ TeV
         \cite{Selyugin:2019ybv}. 
         The effect is         related to the behavior of the
          hadron potential at large distances \cite{Selyugin:2020jnt}.
    The SPD experiment \cite{Ljvov2020} can provide new precise data on small-angle
elastic $p$-$p$-scattering to explore this phenomenon.
For this purpose,  measurements in the kinematic region of
$|t| \sim 0.1\div0.8$~GeV$^2$ will be performed. They will simultaniousely detect  in coincidence
elastically scattered protons at angles $\theta \sim 3\div10^\circ$
 with an  accuracy of $t$ determination higher than
$\sim 0.02$~GeV$^2$.
 
 \subsection{Single-spin asymmetries at low energies}
A systematic study of such single-spin phenomena as the transverse single-spin polarization of hadrons ($A_N$) and the polarization of hyperons ($P_N$) in $p$-$p$, $d$-$d$, $C$-$C$, and $Ca$-$Ca$
collisions is proposed. A systematic study means a detailed study of the dependence of the observed $A_N$ and $P_N$ for dozens of reactions on variables, such as collision energy ($\sqrt{s}$), Feynman variable ($x_{\rm{F}}$), transverse momentum ($p_T$), the atomic weights of the colliding particles ($A_1$ and 
$A_2$), the multiplicity of charged particles ($N_{ch}$) in the event, and the centrality of collisions. The study of a large number of reactions will reveal the dependence of $A_N$ and $P_N$ on the quantum numbers (spin, isospin, flavor, etc.) of the hadrons participating in the reaction.  A systematic study also implies a global analysis of all available single-spin data within a
certain model in order to identify general behavior  and the mechanism of the origin of polarization phenomena. 

One of such models is the chromomagnetic quark polarization (CPQ) model \cite{Abramov:2008zz}. The CPQ model assumes the presence of an inhomogeneous circular transverse chromomagnetic field $\rm\bf{B}^{\it{a}}$ in the interaction region of colliding hadrons. The interaction of the chromomagnetic moments of test quarks, which later form the observed hadron, with the field $\rm\bf{B}^{\it{a}}$ leads, as a result of the Stern-Gerlach effect, to the appearance of spin effects (with non-zero $A_N$ and $P_N$). The spin precession of test quarks leads to the phenomenon of oscillations  $A_{N}(x_{\rm{F}})$   and $P_{N}(x_{\rm{F}})$   depending on the Feynman variable $x_{\rm{F}}$, and the frequency of these oscillations depends on the number of spectator quarks, color charges of quarks and antiquarks, and the direction of their motion in the c.m. of reactions. The frequency of these oscillations is a linear function of the number of quarks and antiquarks - spectators interacting in pairs with each of the test quarks, taking into account the color state of the pair. The highest oscillation frequencies are expected in the case of antibaryon production in baryon collisions and in ion collisions. 

The CPQ model also predicts for a number of reactions such a phenomenon as the resonance dependence of $A_N$ and $P_N$ on energy ($\sqrt{s}$), which occurs if the sign of the color charge of the test quark and spectators is opposite. The most interesting reaction in this respect is the production of anti-lambda in various initial states of the beams of the NICA collider, for which the resonance energy is close to 7 GeV in the center-of-mass system.

The threshold dependence of $A_N$ on the hadron production angle in the center-of-mass system is also predicted. An example of the manifestation of the threshold dependence $A_N$ is the reaction  
$p^{\uparrow} p(A) \rightarrow \pi^{-} X$, for which the threshold angle is 74$^{o}$, since the test quark is the $d$-quark, which is heavier than the $u$-quark.

An important advantage of hyperons is the ability to measure  $A_N$ and $P_N$ 
through the angular distribution of their deck products, which allows the comparison of these observables with the model predictions.


The rate of pion production in $p$-$p$ collisions varies from $3\times10^{7}$ s$^{-1}$ at 23 GeV  to $2\times10^{5}$ s$^{-1}$ at 7 GeV. In C + C and Ca + Ca collisions, it will be three orders of magnitude lower.
 The rate of production of hyperons is two orders of magnitude lower than that of pions. Antihyperons are produced 5 to 10 times less frequently than hyperons. 

\subsection{Exclusive hard processes with deuterons}

Questions involved in the studies of the short-range nuclear structure and the understanding of the microscopic   nucleon structure and dynamics of large-momentum transfer processes are delicately intertwined: 
the understanding of hard dynamics of two-body processes is also necessary for precision 
 studies of the short-range nuclear structure. Exclusive large $t$ reactions like a $dp\to ppn$ process can address many of these questions. The advantages of such a reaction are a good knowledge of the non-relativistic deuteron wave function and the ability to choose both kinematics sensitive to the  dynamics of elastic $N$-$N$ scattering and the kinematics sensitive to the short-range deuteron structure.The collider kinematics presents a number of advantages, as all particles in the considered reactions have large momenta and 
 hence can be easily detected.

\subsubsection{Probing dynamics of $N$-$N$ interaction}   

 The simplest kinematics is production  of two approximately back-to-back  nucleons  
with large transverse momenta and a spectator nucleon with 
a longitudinal momentum 
  $p \sim p_{d}/2$ 
 and a transverse momentum  of
$\ge$ 200 MeV/c \cite{Frankfurt:1994nw, Frankfurt:1996uz}. 
 
 In the impulse approximation this process corresponds to elastic scattering of the  projectile proton off the quasi-free nucleon of the target. In this kinematics   soft rescatterings of the initial and final nucleons, which accompany the hard $pp (pn)$ reaction, are large. The eikonal approximation, which accounts for relativistic kinematics as dictated by the  Feynman diagrams, reveals the important role played by the initial and final state interactions in the angular and momentum dependences of the differential cross-section in the well-defined kinematics. 
 The condition for the applicability of the generalized eikonal approximation \cite{Frankfurt:1996xx} is  that 
 the c.m. scattering angle and invariant mass of the two-nucleon system are large enough, so that $-t, -u \ge \mbox{2 GeV}^2$. 
 
 It was suggested in \cite{Mueller, Brodsky} that nucleons in the elementary  reaction  interact in small-size configurations with a small cross-section -- the  color transparency phenomenon. This effect is suppressed by the space-time evolution of nucleon wave packets \cite{Farrar:1988me,Frankfurt:1988nt}. However  the effect of evolution  is small for the deuteron, where typical distances between nucleons in the rescattering amplitude  are  $\le$ 1.5 fm. Hence the discussed process allows one  to measure the wave packet size of a nucleon practically right in the interaction point.

It was pointed out  that the hard dynamics in $p$-$p$ and $p$-$n$
elastic scattering may be rather different \cite{Granados:2009jh}.
Hence it would be instructive to compare the channels  where $pp$ and $pn$ 
are produced with a large $p_T$.

Experiments with polarized beams would greatly add to this program: due to a better  separation of kinematic domains where impulse approximation, double and triple scattering dominate, the studies of $p^{\uparrow}d^{\uparrow} \to  pNN$ processes will allow one to investigate the spin structure of $p$-$p$ and $p$-$n$ elastic  scattering  at large $t$ (the latter is practically not known). Also, it would be possible to find out whether   the  
strong difference between the cross-sections of  elastic scattering of 
protons with parallel and antiparallel spins \cite{Crabb:1978km}   involves collisions of protons  in configurations with  sizes depending  on the spin orientation.  

Moreover, it would be possible to study the effects of coherence  in the channels, where the exchange by gluons in the t-channel is not possible, for example, $pd\to \Delta NN$. In particular, it would be possible to test the effect of chiral transparency suggested in \cite{Frankfurt:1996ai} -- suppression of the pion field in the nucleons experiencing large $-t$ scattering.

\subsubsection{ Probing microscopic deuteron structure}

It is established now that  the dominant source of the short range correlations (SRC) in nuclei are proton-neutron correlations with the same quantum numbers as the deuteron and with a high-momentum tail similar to that in the deuteron (see review in \cite{Frankfurt:2008zv,Hen:2016kwk}). Hence the deuteron serves as a kind of the hydrogen atom of the SRC physics. Only after it has been tested experimentally that the approximations currently used for the description of the $p$-$d$ reaction work well, it will be possible to perform high-precision studies of the SRC in heavier nuclei.

 It was demonstrated in  Ref.\cite{Frankfurt:1994nw,Frankfurt:1996uz}  that 
    under specific kinematical conditions (in particular, low transverse momenta of slow nucleons in the deuteron rest frame)  the effect of the initial and final state interactions can be accounted for by rescaling the cross-section calculated within the plane wave impulse approximation.  
 In this kinematics it would be possible to check the universality of the wave function -- in particular, its independence on the momentum transfer in the elementary reaction. Such factorization is expected to break down at sufficiently large $-t$ and  $-u$, where scattering involves interaction of nucleons in the small-size configurations (the color transparency mode) since the small-size configurations are suppressed in bound nucleons with suppression growing with the nucleon off-shellness
\cite{Frankfurt:1988nt}.

Studies of the non-nucleonic configurations in   the deuteron as well as relativistic effects
in the scattering off a polarized deuteron are a separate matter of interest.
In particular, it would be possible to search  for non-nucleonic degrees of freedom, like 6-quark, two  $\Delta$ isobars via a production reaction $p d\to \Delta^{++} p \Delta^- $
with 
 $\Delta^{++} $ and proton back to back and $\Delta^-$ being slow in the deuteron rest frame.  
 
\label{uz-ladyg} 
\subsection{Scaling behavior of exclusive reactions with the lightest nuclei and spin observables}

 The structure of the lightest nuclei at short distances $r_{NN}<0.5$ fm or high relative momenta
 ($q> \hslash /r_{NN}\sim$ 0.4 GeV/$c$ constitutes a fundamental problem in nuclear physics. One of 
  the most important questions is  related to the search for the onset of a transition region
  from the meson-baryon to the quark-gluon picture on nuclei.  A definite signature for transition to the
  valence quark region is given the constituent counting rules (CCR) \cite{Matveev:1973ra,Brodsky:1973kr}.  According the dimensional scaling, the differential cross-section
 of a binary reaction  at a sufficiently high incident energy can be parametrized as
 $d\sigma/dt\sim s^{-(n-2)} f(t/s)$,
 where $n$ is the sum of constituent quarks in all participants, $s$ and $t$ are the Mandelstam variables. 
 Many hard processes with free hadrons are consistent with the CCR at energies of several GeV.
 The CCR properties of the reactions with the lightest nuclei were  observed
 in photodisintegration of the deuteron $\gamma d\to pn$ at $E_\gamma =1\div4$ GeV and
 $^3\mbox{He}$ nucleus $\gamma~^3\mbox{He}\to ppn$, $\gamma~ ^3\mbox{He} \to dp$. Earlier data on the reaction
 $dd\to tp$, $dd\to ~^3\mbox{He} ~n$  \cite{Bizard:1980aj} and $pd\to pd$, as was shown in Ref.\cite{Uzikov:2005jd},
 also follow the CCR  behavior $s^{-22}$ and $s^{-16} $, respectively, at surprisingly low energies of 0.5 GeV.  Recently, the 
  CCR behavior  of the reaction $pd\to pd$ was observed in
  \cite{Terekhin:2018gst,Terekhin:2019bsc} at
   higher energies. On the other hand, the reaction with pion production
   $pp\to d\pi^+$ does not follow the CCR rule demonstrating the differential
   cross-section $\sim s^{-9}$ instead of $s^{-12}$. One possible way to explain this is  a partial restoration of chiral symmetry 
   at  high enough excitation energy \cite{Uzikov:2016myj}.
   However, a systematic study  of these properties of the  reactions with the lightest nuclei are absent. Therefore, it is important to know whether the reaction 
   $pn\to d\rho^0$  follows the CCR  behavior and at what 
    minimal energy there is  a CCR onset. Assuming the model of the vector meson dominance and taking into account the observed   CCR behavior of the $\gamma d\to pn$ reaction, one may expect
    the $\sim s^{-12}$ dependence of the cross-section of the  reaction  $pn\to d\rho^0$.
   Furthermore, a possible  relation between  the CCR behavior of the unpolarized cross-section and the spin observables of the same reaction  are practically not known.  The  NICA SPD  facility  provides a good opportunity  for this study, using polarized beams  in $p$-$p$, $d$-$d$,  and $p$-$d$ collisions.

The tensor $A_{yy}$ and vector $A_{y}$ analyzing power in $d$-$p$- elastic scattering
obtained at $60^\circ$, $70^\circ$, $80^\circ$ and $90^\circ$ in the center-of-mass system versus 
transverse momentum $p_T$ \cite{Kurilkin:2012wk,Kurilkin:2011zz} demonstrates negative and positive asymptotics, respectively. Note that the negative sign of $A_{yy}$ is observed also in the deuteron inclusive breakup 
at a large $p_T$ \cite{Ladygin:2005wq},\cite{ayy}. It would be interesting to extend the range of the measurements
to a larger $p_T$, where the manifestation of non-nucleonic degrees of 
freedom is expected. New precise measurements with small statistical and systematic uncertainties
at energies higher than  $\sqrt{s}\ge$3.3~GeV and at different scattering angles  are required to make a conclusion about the validity of the CCR \cite{Matveev:1973ra,Brodsky:1973kr} in $dp$  elastic scattering. We also propose to measure different vector and tensor analyzing powers in $d$-$p$  elastic scattering at the SPD energies. 

The measurements of $d$-$p$ elastic scattering can be performed either with polarized deuterons and unpolarized protons, or with unpolarized deuterons and polarized deuterons. The $d$-$p$ elastic
scattering events can be selected, using cuts on the azimuthal and polar scattering  angles correlations. 
The vector $A_y$ and tensor $A_{yy}$ and $A_{xx}$ analyzing powers 
will be measured simultaneously in the case of vertically polarized deuteron beams. 
The precision on the tensor   $\Delta A_{yy}\sim$0.09 and $ \Delta A_{xx}\sim$0.09 and on the 
vector $\Delta A_y\sim$0.03 analyzing powers can be achieved  for the scattering angle 
$\sim$90$^\circ\pm$5$^\circ$  
at $\sqrt{s}\sim$4.5~GeV ($p_T\sim$1.7~GeV/$c$) for 30 days of the beam time at the luminosity 
$L\approx$10$^{29}cm^{-2} s^{-1}$. We expect $\sim$70\% of the beam polarization from the 
ideal values of polarization for different spin modes.
The counting rate has been estimated using the
$d$-$p$ elastic scattering cross-section
parameterization from Ref.\cite{Terekhin:2019bsc}.
The spin correlations can be obtained in quasi-free $d$-$p$ elastic scattering using $d$-$d$ collisions.

\subsection{Vector mesons and the open charm near the threshold}
 The study of charm production (hidden and open) and backward vector meson production at SPD will take full advantage of the possibility of using polarized $p$, $d$ beams (as well as heavier ions) in a kinematical region where the data on cross-sections are scarce and polarization effects are mostly unmeasured.  In general, threshold meson production in $N$-$N$ collisions gives a deeper insight in the reaction mechanisms, as it is shown by the experimental programs at different proton accelerators, such as SATURNE and COSY.



\subsubsection{Charm production \label{Section:charm} }

The  production mechanisms for  charmonium and $D$ $(D^*)$ mesons in nucleon-nucleon collisions is poorly  understood. Charm quarks do not preexist  in the nucleon as valence quarks: how they are formed and how they hadronize is an open question.  To interpret the production and the propagation of charm in heavy-ion collisions as a  probe of quark-gluon plasma (QGP), it is necessary to have a solid theoretical background based on the understanding of elementary processes. 

Experimental data and theoretical studies of $J/\psi$ production  in different processes and of its decays exist: for a review, see \cite{Vogt:1999cu}, and for the most recent data collection -- \cite{Maltoni:2006yp}. In the threshold region, the final particles are produced in the $S$-state and the spin structure of the matrix element is largely simplified.  The effective proton size, which is responsible for charm creation, has to be  small, $r_c\simeq 1/m_c \simeq 0.13$ fm, where $m_c$ is the $c$-quark mass, selecting small impact parameters  \cite{Brodsky:2000zc}. The $S$-wave picture can, therefore, be applied for $q\le m_c$, where $q$ is the norm of the $J/\psi-$ three-momentum in the reaction center of mass (CMS).  The momenta of the produced particles are small, but the mechanisms for the production of charmed quarks must involve large scales. In Ref. \cite{Rekalo:2002ck}, the near-threshold $J/\psi-$ production in nucleon-nucleon collisions was analyzed in the framework of the general model independent formalism, which can be applied to any reaction $N~N\to N~N~V^0$, where $V^0=\omega$, $\phi$, or $J/\psi$.  Such reactions show large isotopic effects: a large difference for $pp$- and $pn$-collisions, which is due to the different spin structure of the corresponding matrix elements at the threshold: $\sigma(np\to np J/\psi) / \sigma(pp\to pp J/\psi)=5.$

In Ref. \cite{Rekalo:2002ck} an estimation for the $J/\psi$ production was suggested from the comparison of  the cross-sections for the $\phi$ and $J/\psi $ production in $pp$ collisions. 
For the same value of the energy excess, $Q=\sqrt{s}-2m-m_V$, taking into account the different phase space volumes, coupling constants  for the decay $V\to \pi\rho$,  monopole-like phenomenological form factor for the vertex $\pi^*\rho^*V$, with virtual $\pi$ and $\rho$,
one finds the following simple parameterization for the cross-section, holding in the near-threshold region only:
\begin{equation}
\sigma [nb]=0.097 (Q [\mbox{GeV}])^2.
\label{eq:JPSI}
\end{equation}
In Ref.  \cite {Craigie:1978bp}, a parameterization of the exponential  form 
\begin{equation}
\sigma [nb]= a e^{-bM_{J/\psi}/\sqrt(s)}
\label{eq:Craigie}
\end{equation}
was suggested. The values $a$= 1000 [nb], and $b$=16.7 GeV reproduce the experimental data above the threshold rather well.

\begin{figure}[!b]
  \begin{minipage}[ht]{0.325\linewidth}
    \center{\includegraphics[width=1\linewidth]{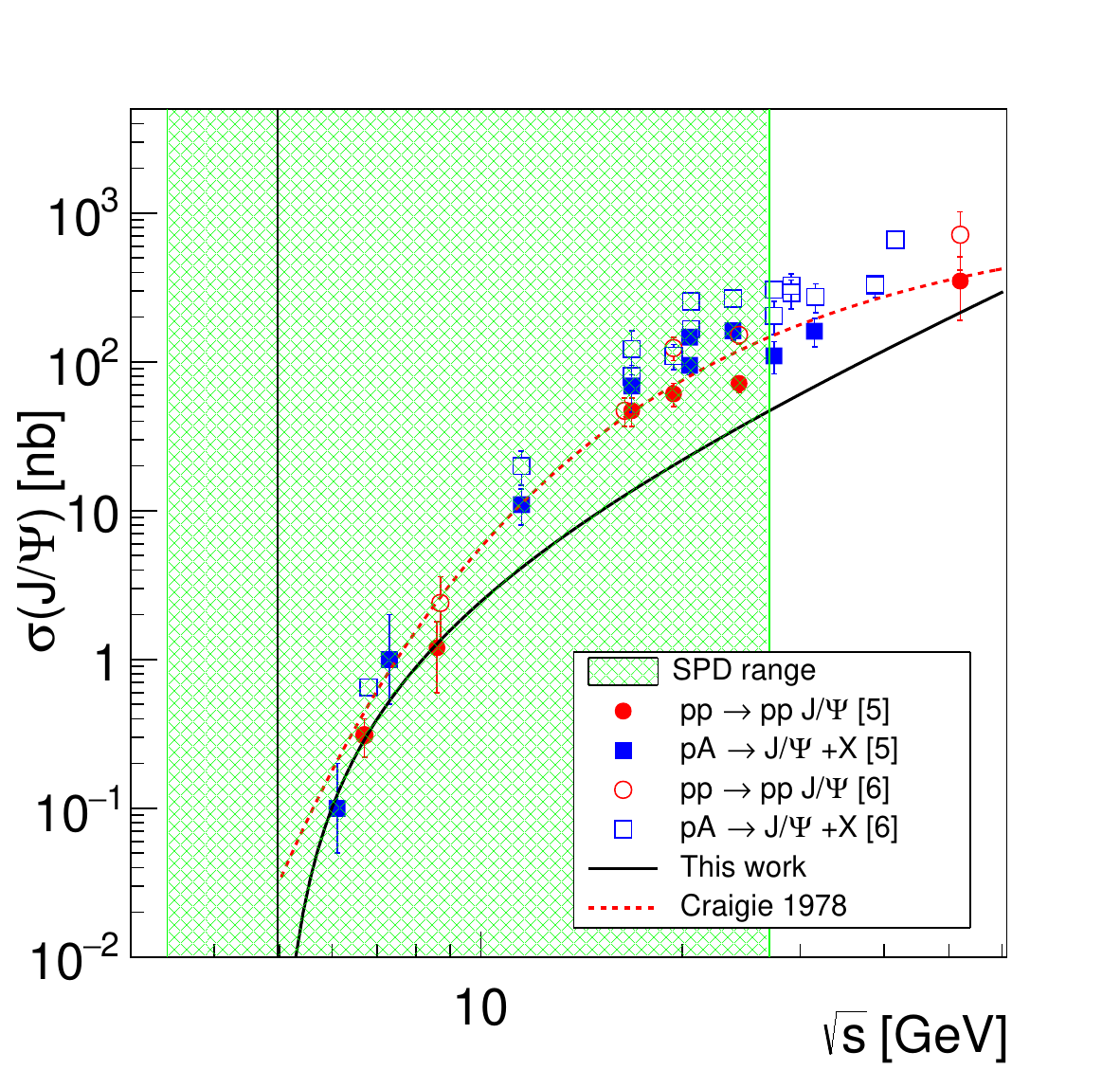} \\ (a)}
  \end{minipage}
  \begin{minipage}[ht]{0.325\linewidth}
    \center{\includegraphics[width=1\linewidth]{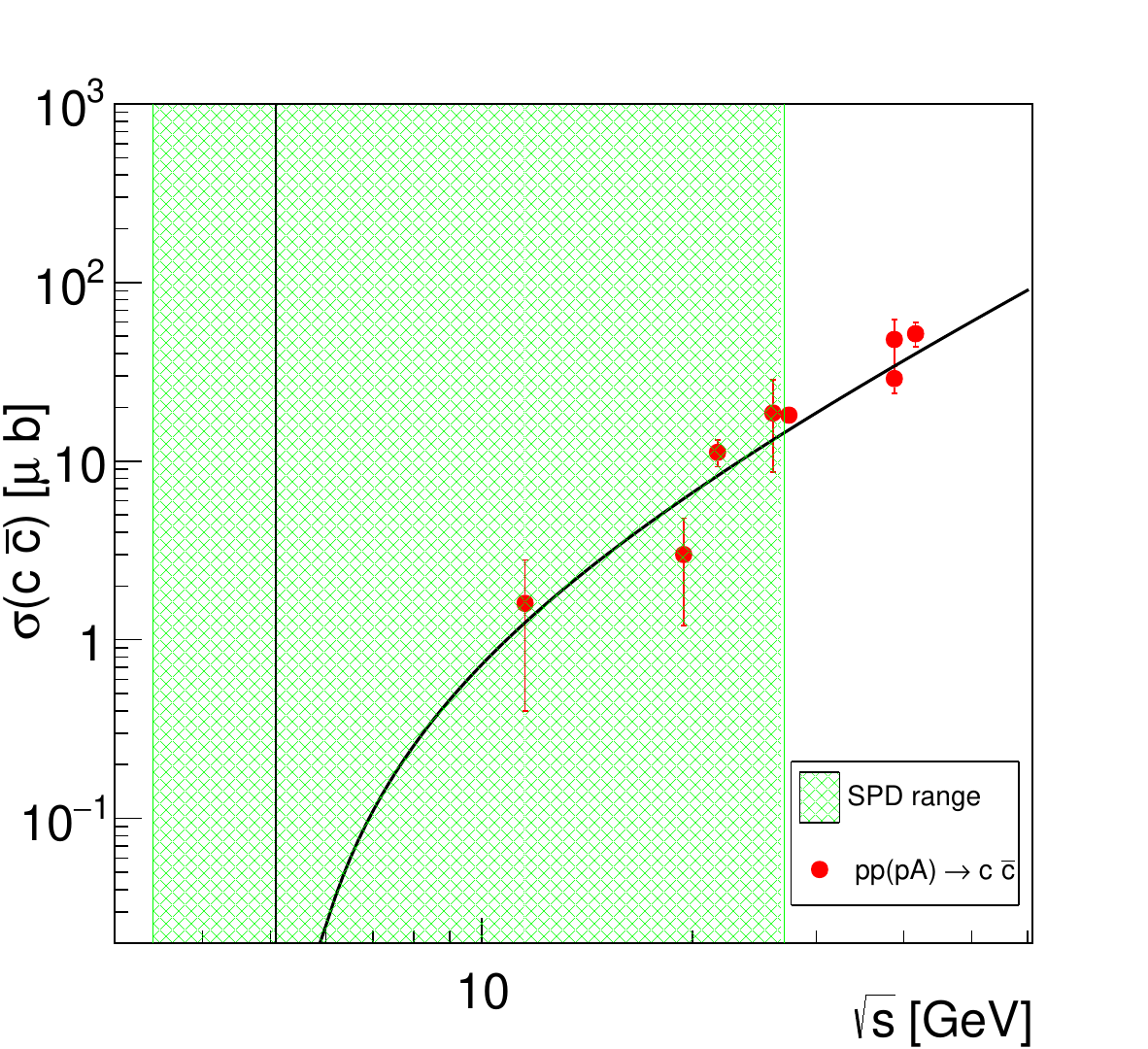} \\ (b)}
  \end{minipage}
  \begin{minipage}[ht]{0.325\linewidth}
    \center{\includegraphics[width=1\linewidth]{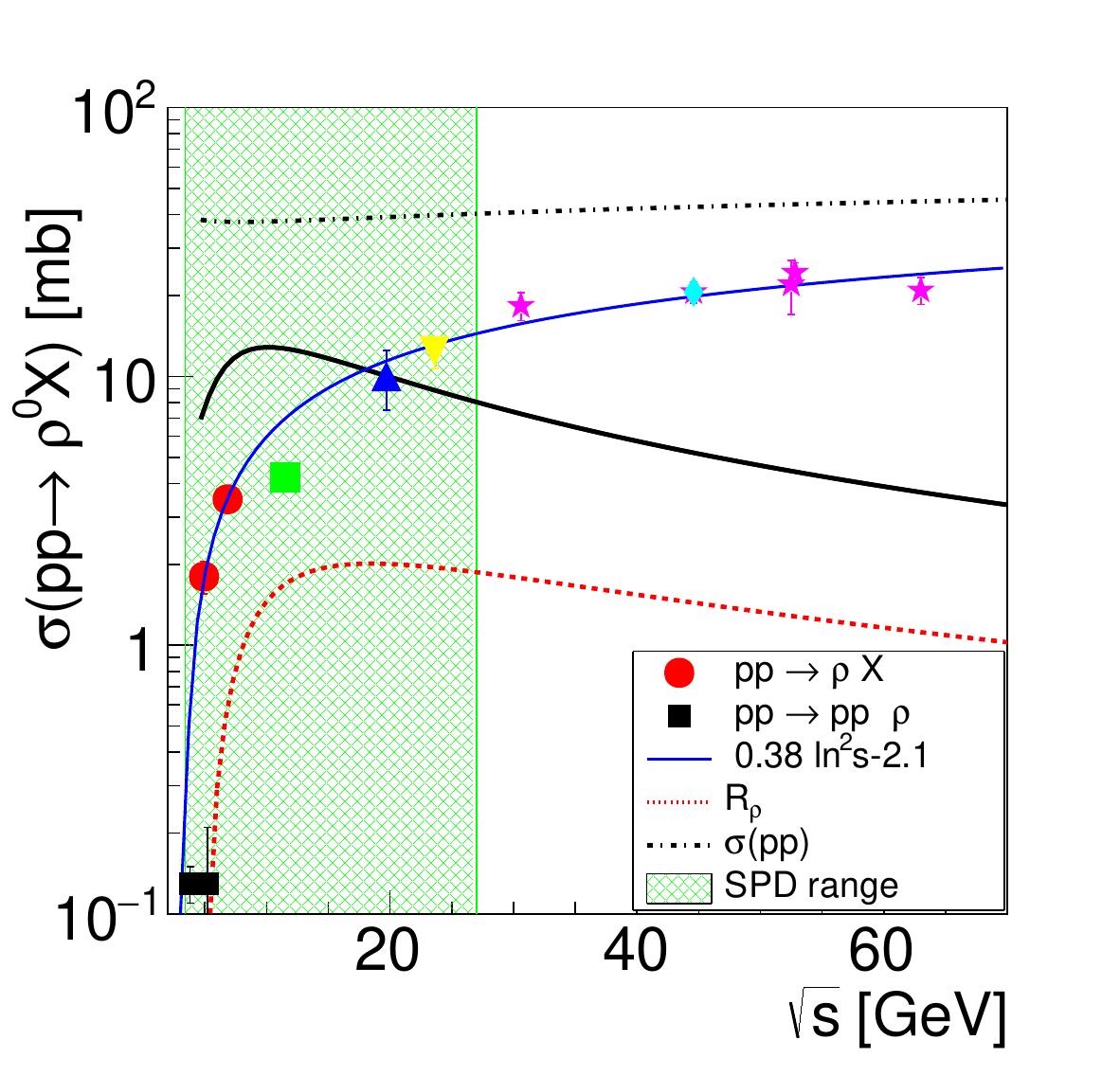} \\ (c)}
  \end{minipage}
  \caption{ (a) Experimental data on $J\psi$ production in $p$-$p$ (red circles) and $p$-$A$ (blue squares) reactions,  from the recollections in Refs. \cite{Vogt:1999cu} (filled symbols) and \cite{Maltoni:2006yp} (open symbols). The solid line is the calculation from Ref. \protect\cite{Rekalo:2002ck}.
  (b) Total charm production in $p$-$p$ and $p$-$A$ collisions. The data are from  Ref. \protect\cite{Louren_o_2006}. The line is a threshold parametrization (see text). (c) Cross-section for $\rho$-meson production in $p$-$p$ collisions: inclusive (different symbols and colors from different experiments) and exclusive data from $pp\to pp\rho$ (black squares). The calculation from Ref.  \cite{Kuraev:2013izz} shown as a black line. The red dashed line is the renormalization factor. The black dash-dotted line is the total $pp$ cross-section. The first red point is the inclusive measurement from Ref. \cite{Rustamov:2010zz}. The blue line is the  parametrization from Ref. \protect\cite{Albrow:1978uu}. The green filled region represents the SPD range. 
  }
  \label{fig:vector}  
\end{figure}

In Fig. \ref{fig:vector}(a), the data for $p p \to J/\psi   p p$ (red circles) and $p A \to J/\psi   X$ (blue squares) are plotted from the recollections in Refs. \cite{Vogt:1999cu} (filled symbols) and \cite{Maltoni:2006yp} (open symbols). Different symbols differentiate $J/\psi$ production in $pp$ or (extrapolated from) $pA$ collisions. The data, mostly collected at CERN, are reconstructed from the measurement using models and/or assumptions, and the compiled total cross-section for $J/\psi$ production may differ up to a factor of two. For example, the original reference for the measurement from Protvino at $\sqrt{s}=11.5$ GeV \cite{ANTIPOV1976309} gives $\sigma (pp\to (J/\psi\to \mu^+\mu^-)X) =9.5\pm 2.5$ nb, whereas the same experimental point is referenced as  $\sigma =11\pm 3$ nb, in Ref. \cite{Vogt:1999cu} and $\sigma =20\pm 5.2$ nb, in Ref. \cite{Maltoni:2006yp}.  The cross-section from Ref. \cite{Rekalo:2002ck} is also plotted in Fig. \ref{fig:vector}(a) (solid line). 

Taking the value of luminosity $L = 10^{30}$ cm$^{-2}$ s$^{-1}$, one expects 3 counts/hour  for such a process with a cross-section of the order of 1 nb. This number is not corrected for the detector efficiency and reconstruction with identification, for example, in a missing mass. The reconstruction of $J/\psi$ through its decay into a lepton pair, that is, the preferred mode, requires two additional orders of magnitude, as the branching ratio is $(\simeq 5.9\pm  0.5)10^{-2}$. 

In Ref.  \cite{Rekalo:2002ck}, it was shown  that  only one polarization observable, the  $J/\psi$-polarization, is identical for  $p$-$p$ and $p$-$n$ collisions: the  $J/\psi$ meson is transversely polarized, even in collisions of unpolarized nucleons.

Open charm production, $N~N\to N~\bar{D}~\Lambda_C(\Sigma_C)$,  gives information on scattering lengths, effective radius, hadronic form factors, and coupling constants and is also related to the dynamics of charm creation in $N$-$N$, $N$-$A$, $A$-$A^*$ collisions. The spin and isospin structure of the matrix element for the reactions $N~N\to \Lambda_C(\Sigma_C) ~\bar D ~N$  was derived for open charm in Ref. \cite{Rekalo:2002wg}. A detailed estimation of cross-sections and the expressions of the polarization observables can be found there. The existing information and estimations indicate that the near-threshold cross-section can be of the order of microbarns. The threshold cross-section normalized at the lowest existing value is plotted in Fig. \ref{fig:vector}(b), where the insert highlights the threshold region. 

The  charm production near threshold cross-section follows the behavior: 
\begin{equation}
\sigma [\mu b]=0.03 (Q [\mbox{GeV}])^2.
\label{eq:open}
\end{equation}
It is plotted in Fig. \ref {fig:vector}(b) over a collection of data from Ref. \cite{Louren_o_2006} reanalyzed from several experiments on charm production in $p$-$p$ and $p$-$A$ collisions at different facilities. 
We stress that these are difficult measurements, with low counting rates and huge backgrounds, but even setting upper limits will be important, as no data at all are present in the threshold region.

The understanding of charm production (open or hidden) should unify different steps: parton-level hard process with production of $c\overline{c}$ pairs, after hadronization of
$c\overline{c}$ into $J/\psi$ or into charmed hadrons (mesons and baryons), including the final state interaction of the produced charmed hadrons with other particles. The relatively large transferred momenta involved in most processes of $J/\psi$ production in hadron-hadron collisions allow one to treat the first step in the framework of perturbative QCD. But the applicability of QCD is not so straightforward for the description of the $c$-quark hadronization. In this respect, precise data collected in the SPD energy range will bring important information, especially if covering a wide range above the threshold. 
\subsubsection{Backward meson production}

Larger counting rates  are expected for light meson productions, since cross-sections are of the order of mb. The $\rho^0$ meson production in elementary collisions and on nuclei has been discussed, 
for example, in Ref. \cite{Sibirtsev:1997ac} and references therein. The $\rho^0$ inclusive cross-section has been measured at different accelerators since the 70's, mostly at CERN \cite{Blobel:1973wr}, and more recently by the HADES collaboration \cite{Rustamov:2010zz}. In Ref. \cite{Albrow:1978uu}, the following parametrization was suggested 
\begin{equation}
\sigma (pp\to\rho^0 X) =  (0.38\pm 0.02)\ln^2 s -(2.1\pm 0.4).
\label{eq:Albrow}
\end{equation}
In Ref. \cite{Kuraev:2013izz}, a specific kinematics,  the backward light meson production in $p$-$p$ or $p$-$A$ collisions, was discussed in similarity to the ''quasi-real electron method'', where a hard photon is produced on the collision of electrons on any target \cite{Baier:1973ms}.  Two important characteristics have been proved for the electron case: (i) the collinear emission probability has a logarithmic enhancement; (ii) the cross-section can be factorized in a term related to the probability of the meson emission  with a given energy at a given angle from the beam particle, and a term related to the interaction of the beam remnant after emission on the target. 

In hadron case, the cross-section can be written as:
\begin{eqnarray}
d\sigma^{pT\to h_+X}(s,x)&=&\sigma^{nT\to X}(\bar{x}s)d W_{h_+}(x), \nonumber \\
d\sigma^{pT\to h_0X}(s,x)&=&\sigma^{pT\to X}(\bar{x}s)d W_{h_0}(x),
\end{eqnarray}
where $h$ is a hadron, $x$ $(\bar{x}=1-x)$ is the energy fraction carried by the meson (the beam remnant).  $d W_{\rho}(x)$ can be inferred using the QED result corrected by a renormalization factor in order to account for the emission of $n$ real soft neutral pions escaping the detection.

The prediction of the model for backward  $\rho$-meson production in $p$ $p$ collisions is shown in Fig. \ref{fig:vector}(c) as a thick solid black  line. The red dashed line is the renormalization factor, integrated over $x$. The total $p$-$p$ cross-section is the black dash-dotted line. The blue line is the  parameterization of the inclusive $\rho$ cross-section  from Ref. \protect\cite{Albrow:1978uu}).  The available data are also shown as different symbols and colors for inclusive measurements and as black squares for exclusive $\rho$ production. Backward production can be of the order of several mb, and, therefore accessible at SPD even with the initial lower luminosity. Collecting precise, systematic data should help to refine the models and is of great interest for the collision on heavy targets as well. Backward kinematics could constitute an original  contribution to the field, offering an alternative possibility to produce neutron beams.

\subsection{Central nucleon-nucleon collisions}

The main experimental basis for clarification of the non-perturbative QCD (NPQCD) baryon structure is the baryon spectroscopy and the short-range nucleon-nucleon interaction. The more the nucleons are overlapped during the collision, the higher sensitivity of the latter is to the NPQCD structure. The maximum sensitivity can be reached in conditions of overlapping of the quark core of nucleons and sufficiently long time of this overlapping. Unfortunately, these conditions are practically not met in the available nucleon-nucleon experimental data: at relatively low energies the effective momentum transfers are not suffi\-ciently high, and at high energies the contents of colliding nucleons diverge too quickly.  This circumstance explains why the region of the \emph{NN} collisions at distances smaller than the radius of the nucleon core still remains unexplored. Access to this area is possible through the central collisions (CC) of the nucleons at adequate energies. The collisions are usually named central if the corresponding impact parameter \emph{R} is small, $R < r_{core} \approx 0.4$ fm.

Overlapping of the nucleon cores can be achieved at the center-of-mass energies $\sqrt{s}_{min} = U_{rep}(0) + 2m_N$, where $U_{rep}(0)$ is the repulsive potential of the \emph{NN} interaction at a zero distance,  $U_{rep}(0) \approx 1$\,GeV. Then the minimal energy of interest is $\sqrt{s}_{min}  \approx 2.9$ GeV. At the energies less than 7.5\,GeV (corresponding to  the chiral symmetry breaking momentum $\Lambda_{\chi SB} \approx 1.2$\,GeV/c \cite{Shuryak:1981fz, Manohar:1983md}) the resulting intermediate state is an excited (6q)* system  of six chiral consti\-tuent quarks interacting via goldstone boson, gluon exchange and con\-finement potential. This interaction is supposed to be much more intensive than in the perturbative quark-gluon system, and provides therefore relatively long lifetime  of the system, sufficient for manifestation of the NPQCD structure. Under certain conditions, it can even produce quasi-bound states, resonance dibaryons. It should be stressed that the (6q)* system  under consideration is characterized by very high baryon and energy densities, since two baryons and the whole center-of-mass energy is concentrated in a small volume of about $4/3 \pi (r_{core})^3$ size.

Decay of the (6q)* system leads to reconstruction of hadronic states in the form
 \begin{equation}
 \label{Kom1}
 p  p \to (6q)^* \to N ~ N  ~ Mesons,
 \end{equation}
where \emph{Mesons}
denotes the system of light mesons, predominantly pions.

Peripheral $N$-$N$ collisions proceed mainly via production of excited baryons \emph{N*} in the intermediate state
 \begin{equation}
 \label{Kom2}
 p  p \to \{(N  N^*)~ \mbox{or}~ (N^*  N^*)\} \to N ~ N ~ Mesons
 \end{equation}
and have, in general, the final state similar to that in the central collisions (\ref{Kom1}). Therefore, in order to distinguish the central collision process (\ref{Kom1}) from the peripheral (\ref{Kom2}), one needs special centrality criteria. According to  \cite{Komarov:2017qnd,Komarov:2020qra}, there are two such criteria:
A) using of the reaction
\begin{equation}
\label{Kom3}
  N  N \to d(90^{\circ}) ~ Mesons,
\end{equation}
where $d(90^{\circ}$) is a deuteron emitted at the angle close to $90^{\circ}$ 
\footnote{or reaction $N  N \to \{pp\}_{S_0}(90^{\circ}) ~ Mesons$, where $\{pp\}_{S_0}$ is a proton pair in the $^1S_0$ state};
B) smallness of the interaction region size $r_{int} < r_{core}$,
where $r_{int} =1/(-Q^2)^{1/2}$ with $ Q = P_1 - D/2$. Here $ P_1$ is the four-momentum of one of the initial nucleons and $D$ is the four-momentum of the final joined nucleon pair.

Evaluation of feasibility of the experiments with the above centrality criteria shows \cite{Komarov:2020qra} that at the expected luminosity \cite{Meshkov:2019utc} the event rate at SPD could be significant. 

The following goals can be aimed at, in particular, in the  experiments with central collisions:
\begin{itemize}
\item  study of known and search for new dibaryon resonances in the region of $\sqrt{s} \approx 2.5\div7.5$\,GeV;
\item  search for the predicted dominance of the $\sigma$-meson production
\cite{Faessler:2005qq};
\item  search for the expected effects caused by the chiral symmetry partial restoration (drop of mass and width of mesons) ~\cite{Hatsuda:1984jm,Blaschke:2006ss};
\item  study of the energy dependence of the reaction (\ref{Kom3}) cross-section, which is sensitive to the strength of the confinement forces and the value of the chiral symmetry breaking momentum;
\item  first measurement of the analyzing power of the reaction (\ref{Kom3}) for transverse and longitudinal beam polarization.
\end{itemize}

It is worth to  mention that experiments of this kind have never been carried out systematically. There exists a possibility of observing  new unexpected effects that can induce new approaches in solving the fundamental problems of the non-perturbative QCD.

\subsection{Onset of  deconfinement in $p$-$p$ and $d$-$d$  central collisions}

\begin{figure}[!t]
  \begin{minipage}[ht]{0.5\linewidth}
    \center{\includegraphics[width=1\linewidth]{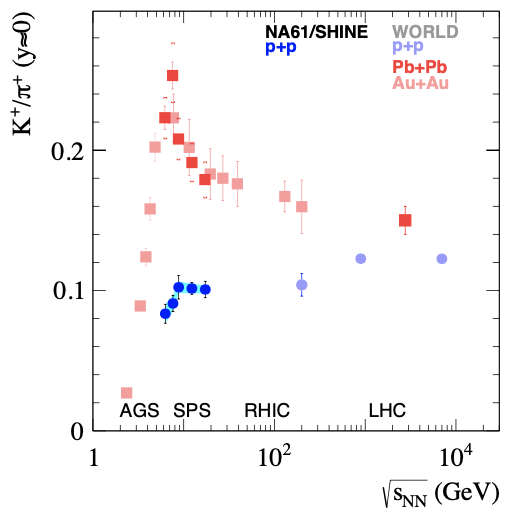} \\ (a)}
  \end{minipage}
  \begin{minipage}[ht]{0.5\linewidth}
    \center{\includegraphics[width=1\linewidth]{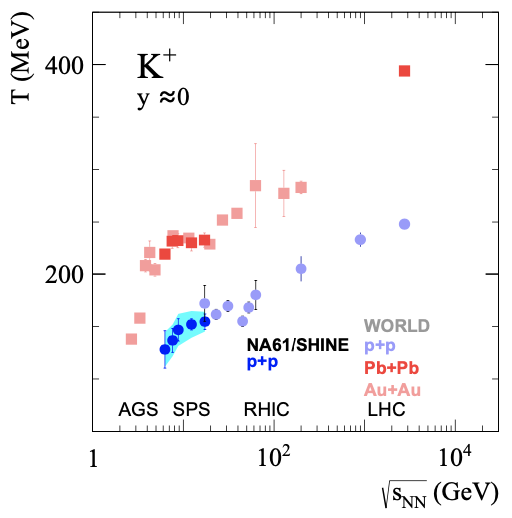} \\ (b)}
  \end{minipage}
  \caption{ Energy dependence of the $K^+/\pi^+$ ratio  (a) and the inverse slope parameter $T$ of transverse mass spectra (b) at mid-rapidity.
 	The NA61/SHINE results for inelastic $p$-$p$ interactions are shown together with the world data \cite{Aduszkiewicz:2019zsv}.
  }
  \label{fig_QGP_K_over_pi_NA61}  
\end{figure}

A study of the phase diagram of strongly interacting matter by varying interaction energy at central collisions
of heavy ions is a primary goal of the NICA MPD experiment \cite{MPD_CDR}.
A structure in the energy dependence of several observables in the range of $\sqrt{s_{NN}}\approx$~7$\div$12\,GeV 
was predicted for the transition to a deconfined phase.
However, recently NA61/SHINE found  intriguing similarities in $p$-$p$ interactions, where
no deconfinement transition is expected \cite{Aduszkiewicz:2019zsv}.
It can be interesting to study this effect in $p$-$p$ and $d$-$d$ interactions in the first phase of SPD.
These measurements can serve as an important crosscheck for the results of NA61/SHINE and, potentially, results of MPD.

The energy dependence of the $K^+/\pi^+$ ratio and inverse-slope parameters  of transverse-mass spectra (the so-called effective temperature $T$) of kaons at mid-rapidity are shown in Fig.\,\ref{fig_QGP_K_over_pi_NA61}. 
The results for heavy ion 
collisions are plotted for comparison.
The  $K^+/\pi^+$ ratio in heavy ion collisions shows the so called horn structure. 
Following a fast rise the ratio has maximum at around 8\,GeV and 
then settles to a plateau value at higher energies.
Meanwhile, the collision energy dependence of the $T$ parameter  shows the so-called step structure
at about the same value of $\sqrt{s_{NN}}$.

The $K^+$ yield is proportional to the overall strangeness production and pions can be associated with the total entropy produced in the reaction. Thus, the $K^+/\pi^+$ production ratio can be a good measure of strangeness-to-entropy ratio,
which is different in the confined phase (hadrons) and the QGP (quarks, anti-quarks, and gluons). 
The $K^+$ is a proper observable for this measurement, because the anti-hyperon yield is small and the main carriers of anti-strange quarks are $K^+$ and $K^0$ with $\langle K^+ \rangle \approx \langle K^0\rangle$ due to approximate isospin symmetry. 
Thus, the $K^+$ (or $K^0$) yield counts about half of all $ s\bar{s}$ pairs produced in the collisions and contained in the reaction products.
In contrast, the fraction of strange quarks carried by $K^-$ (or $\bar{K}^0$) and hyperons is comparable, 
which makes the structure in the $K^-/\pi^-$ distribution less pronounced.

 
A hydrodynamic expansion model of the strongly interacting matter is used for description of  $A$-$A$ collisions.
This model is replaced in $p$-$p$ collisions by excitation of resonances or strong fields between colour charges of quarks and diquarks (strings) \cite{Andersson:1983ia} which makes them qualitatively different.
However, the  similarity of the transition energy in the central heavy ion collisions and the break energy in $p$-$p$ interactions observed by NA61/SHINE provokes the question of whether there is a common physics origin of the two effects.
This makes the precision measurement of the kaon-to-pion ratio in $p$-$p$ and $d$-$d$ interactions  
an interesting topic for the first phase of SPD, when the beam polarization is not  available yet.

\subsection{Study of the lightest neutral hypernuclei with strangeness $-1$ and $-2$}
It is no doubt that the question of stability of the $A=4$ double-$\Lambda$ hypernuclei would be crucial for our understanding of the role played by hyperons in nuclear matter~\cite{Garcilazo:2012qv,Gal:2013bw,Garcilazo:2013zwa}. While it is still controversial for model calculations of such a four-body problem in the mode of weak
binding~
\cite{Filikhin:2002wp,Nemura:2002hv,Garcilazo:2012qv},
\cite{Garcilazo:2014lva},
\cite{Hiyama:2014cua,Garcilazo:2014lva}, 
we suggest that certain general properties arising from the weakly binding systems involving the 2-body and 3-body bound-state energies provide  guidance for a possible stability of 
$\isotope[4][\Lambda\Lambda]{n}$
~\cite{Richard:2014pwa,Richard:2014txa}. Meanwhile, we propose a sensitive reaction process for the search for 
$\isotope[4][\Lambda\Lambda]{n}$
in deuteron-deuteron scatterings, 
i.e. 
\begin{equation}
dd\to K^+ ~ K^+ ~\isotope[4][\Lambda\Lambda]{n},
\end{equation}
which is accessible at NICA. After all, it would rely on the experimental study to decide the dedicated dynamics for such an exotic system.

The quantum numbers of the ground state $\isotope[4][\Lambda\Lambda]{n}$ will favor $J^P=0^+$, where the neutron pair and $\Lambda$ pair have spin 0, and namely, their spins are anti-parallel, respectively. At the same time, the total isospin is $I=1$. Thus, the total wavefunction of the ground state is anti-symmetric under the interchange of the two nucleons or the two $\Lambda$.

The most ideal reaction for producing $\isotope[4][\Lambda\Lambda]{n}$ should be $dd\to K^+ ~ K^+ +T$, which is an extremely clean process, since the background processes involving the $K^+K^-$ productions become irrelevant. Here $T$ is a doubly-strange tetra-baryon system. It makes the measurement of the missing mass spectrum recoiling against the $K^+K^+$ pairs sensitive to the existence of any pole structure in the $nn\Lambda\Lambda$ system. 

In Ref.~\cite{Richard:2014pwa}, we have shown that the energy region above $E_{cm}\simeq 5.2$ GeV is favored to produce $\isotope[4][\Lambda\Lambda]{n}$ with the total cross-section of about 2.5 nb. Here, based on the same analysis, we make a rough estimate of its production rate at the kinematics of SPD.

 We estimate that the total cross-section at $E_{cm}=6.7$ GeV, where the luminosity is $L=10^{30}\mbox{cm}^{-2}\mbox{s}^{-1}$,  will drop by about one order of magnitude compared with the peak value of about 2 nb. Thus, the events expected in effective one-year runtime ($10^7$ s) are
\begin{equation}
  N=\sigma_{total}\times L\times t= 0.2\mbox{~nb}\times 10^{30}\mbox{~cm}^{-2}\mbox{~s}^{-1}\times 10^7 \ \mbox{~s}\simeq 2000 \ ,
\end{equation}
which is sufficient for establishing the state. 
Even though the detection efficiency will reduce the events, there will be tens to hundreds of events to  count.

Apart from the process of $d d\to K^+  K^+ ~(n,n,\Lambda,\Lambda)$, it is also interesting to look at the proton-proton collisions, $p p\to K^+ K^+ \Lambda \Lambda$, where the missing mass spectrum of $K^+K^+$ also provides a clean and direct way to search for the di-baryon $\Lambda\Lambda$, or to study the $\Lambda\Lambda$ interactions. For the proton-deuteron collisions, the double $K^+$ channel is $pd\to K^+ K^+ n \Lambda\Lambda$. The recoiled part of  the double $K^+$ is $n\Lambda\Lambda$, which literally can produce the exotic $H$ di-baryon. A direct measurement of such a system would provide rich information about both $\Lambda\Lambda$ and $n\Lambda$ 
interactions~\cite{Rijken:2010zzb,Polinder:2007mp},
\cite{Haidenbauer:2013oca,Haidenbauer:2015zqb}, \cite{Acharya:2019yvb,Lonardoni:2013rm}.
Nevertheless, notice that the final states have access to the $nK^+$ invariant mass spectrum. The exclusive measurement of this process can also tell whether the light pentaquark state $\Theta^+(1540)$ exists or not.


\label{sec:multiquark_intro}
\subsection{Multiquark correlations and exotic state production} 
Multiquark correlations in the collisions of particles and nuclei play an important role in the understanding of QCD. Multiquark correlation phenomena may be divided into three classes. 
The first one can be related to parton distribution functions (PDFs) of the colliding hadrons and nuclei. 
In the leading twist approximation, in the nuclear PDFs there is a contribution at large $x > 1$, which is related with objects known as fluctons~\cite{Efremov:1982tn}, or few-nucleon short-range correlations~\cite{Frankfurt:1988nt}. Beyond the leading twist, two- or three-quark correlations in parton distributions of hadron and nuclei are related with higher twist contributions.
The second one is related to parton subprocesses. Namely, when multiparton scattering occurs, that is, e.g.,
when two partons  from each colliding object  scatter off each other simultaneously.
The third class can be related to production of exotic multiquark resonance states, e.g., pentaquark and tetraquark states. 
Below one can find a brief possible outline for the studies at SPD, which can shed light on all the three classes of multiquark phenomena mentioned above. 

\label{sec:multiquark_fluctons}
\subsubsection{Multiquark correlations: fluctons and diquarks}
Nuclear fluctons consist of nucleons compressed in distances
comparable with nucleon size, so that the flucton with five or six nucleons could be considered as a cold dense baryon matter since the effective nuclear density~\cite{Andrianov:2016qgy}  would be as high as that in the core of neutron stars~\cite{Frankfurt:2008zv}. Fluctons are directly connected with cumulative hadron production in the nuclear fragmentation region~\cite{Baldin:1971zga,Baldin:1974sh}. 
The flucton  approach~\cite{Efremov:1987mx}, which is based on hard QCD-factorization and EMC-ratio constraints, predicts 
an extra nuclear quark sea, which has a rather hard momentum distribution:   the extra nuclear sea $x$-slope is equal to the $x$-slope of the valence quarks. It leads to "superscaling" for cumulative hadron production at $x > 1$ in the nuclear fragmentation region: the $x$-slope of all cumulative hadron distributions including the "sea" ones~\cite{Efremov:1987mx,Kim:2018saa} are the same. The superscaling phenomenon  was experimentally confirmed by the ITEP group~\cite{Boyarinov:1988ee,Boyarinov:1991up}.
In high-$p_T$ cumulative processes at the central region, other contributions  should be added to the contribution of the nuclear PDFs at $x > 1$, such as the contributions from the PDFs of the other colliding object and possible intranuclear rescattering effects~\cite{Efremov:1985cu,Braun:1994bf}. Therefore, beyond the nuclear fragmentation region, one should observe deviations from superscaling for cumulative production.  
Another aspect of multiquark correlations is two-quark correlations (diquark states) in baryons \cite{Anselmino:1992vg}. This is an important source of high-$p_T$ baryon production~\cite{Breakstone:1985ef,Kim:1987ec,Kim:1988gg}. Being a higher-twist the diquark contribution can describe  
the strong scaling violation for baryon production in hard processes at SPD energies~\cite{Kim:1987ec,Kim:1988gg,Kim:2020a}.

\label{sec:multiquark_scattering}
\subsubsection{ Multiparton scattering}
Measuring a few-particle correlation at SPD, one can study multiparton scattering processes~\cite{Kim:2020a}, which are related to 2D- and 3D- PDFs. It is also significant for production of multiquarks systems~\cite{Efremov:1987eh,Kim:2020a}.    

\label{sec:multiquark_states}
\subsubsection{ Multiquark exotic state production}
Multiparton scattering~\cite{Efremov:1987eh} provides a unique opportunity to study production of various multiquark states, such as, e.g., in Refs.~\cite{Azimov:2015ypa,Barabanov:2014lka,Barabanov:2016jjv}  at the SPD energies.
For multiquark systems with possible diquark structure~\cite {Jaffe:2003sg,West:2020rlk} it can be an issue of particular interest ~\cite{Efremov:1987eh,Kim:2020a}. 

  Near the thresholds of heavy quarks production, where relative velocities 
   of final particles vanish,  formation of new type of resonances, such as $J/\psi$-$N$, is expected \cite{Brodsky:1989jd,Brodsky:1995ds,Tsushima:2011kh,Brodsky:2016tew}. This question became especially interesting after 
   pentaquark observations at LHCb 
    \cite{Aaij:2015tga,Aaij:2019vzc}  in the decay $\Lambda_b^0\to J/\psi p K^-$. 
   
   The enhancement effect was observed at the thresholds of the reactions $e^+e^- \to p\bar p$ and  $e^+e^- \to \Lambda \bar \Lambda$
     \cite{Baldini:2007qg} and  also in the decay $J/\psi\to p\bar p\gamma$ \cite{BaldiniFerroli:2011zz}.  
  Furthermore, the double  spin correlation
   $A_{NN}$ measured in large angle ($\sim 90^\circ$) $p$-$p$ elastic scattering
    \cite{Court:1986dh}
    demonstrates  an enhancement near the strange  ($\sqrt{s}=2.5$ GeV) and charm ($\sqrt{s}=5$ GeV) thresholds, respectively, in the two-baryon system.  According to \cite{Brodsky:1987xw},   the observed strong spin correlations are consistent  with formation  of ``octoquark''resonances $uuds\bar s uud$  and   $uudc\bar c uud$ in the $s$-channel.   SPD has a possibility  of searching for  such states.
   
To summarize this section, SPD, together with the study of the inclusive 
particle production and few-particle correlations at different kinematic regions in $p$-$p$ collisions, as well as in the cumulative processes with light nuclei, has a unique opportunity to test various aspects of multiquark correlations: from the cold dense baryon matter to the exotic multiquark resonance production.

\subsection{Yield of antiprotons in hadronic collisions for astrophysical dark matter search}
Dark matter (DM) is a long-standing mystery in cosmology. It makes up more than 26\% of the Universe \cite{Ade2014}, yet we still do not know its identity. Evidence of DM is mostly gravitational, \emph{e.g.} the rotation curves of spiral galaxies and the mass discrepancy in galaxy clusters \cite{Majumdar2014}. The most favored candidate for DM is the Weakly Interacting Massive Particles \cite{Buchmueller2017}. Different search approaches are employed to search for DM, each with its own underlying paradigm. The main approaches are collider searches, direct detection, and indirect detection. The latter includes astrophysical searches that seek to detect potential anomalous signatures that are hypothetically produced via pair annihilation and decay of DM particles, in the cosmic ray (CR) spectrum \cite{Buchmueller2017}. Naturally, these experiments track different signal components. However, the chance of detection is thought to be higher for rare antimatter components, such as antiprotons. Recently, the AMS-02 experiment \cite{Aguilar2013} has measured a cosmic antiproton flux with unprecedented precision over a wide energy range (from 1 to 450 GeV) \cite{Aguilar2016}. However, we still cannot confirm or rule out an antiproton signature in these measurements due to several sources of uncertainties \cite{Giesen2015}.

Secondary antiprotons are produced in collisions of primary CRs with the interstellar medium (ISM). To be able to detect any anomalous signal, we first need to subtract the flux of antiprotons produced by these CR-ISM collisions. Even though there are several sources of uncertainty standing in the way of pinpointing what this ordinary flux is---such as propagation parameters, solar modulation, and primary spectra slopes---the most significant uncertainty, which ranges from 20\% to 50\% according to the energy, comes from antiproton-production cross-sections \cite{Giesen2015}. Almost 70\% of the secondary antiproton yield is produced in $p$-$p$ collisions. However, the existing datasets for this production channel are incredibly scarce, and mostly date back to before 1980. Moreover, all old datasets did not account for the hyperon decay or isospin effect \cite{Mauro2014}. As for other production channels, data are almost non-existent. Thus, if we were to catch up with the accuracy of AMS-02 measurements, we would have to perform a new precision measurements of antiproton-production cross-sections in $p$-$p$ collisions, as well as other production channels (\emph{e.g.} $p$-$d$, $p$-$ ^{\,3}He$, $p$-$^{\,4}He$, and $^{4}He$-$^{\,4}He$). It is also hoped to study the contribution of different production mechanisms, such as hyperon (namely, $\bar{\Lambda}$ and $\bar{\Sigma}^{-}$) and neutron decays. The kinematic range that needs to be covered to achieve that has already been outlined \cite{Donato2017}.

Preliminary MC studies \cite{Guskov2019,Alexakhin2020} show that at the SPD energy range, the production rate is $>10^{5}\text{ s}^{-1}$, which would minimize the statistical uncertainty. In addition, the $4\pi$ angular acceptance will allow SPD to access a wider kinematic range, in terms of transverse momenta, in comparison with fixed-target experiments operating at the same energy level. With a precise TOF (time-of-flight) system ($\sigma_{TOF}\sim70$ ps), $K^{-}/\bar{p}$ separation can be achieved with a high purity of up to $\sim3.5$ GeV/$c$. SPD can also contribute to the measurement of hyperon-decay contribution via reconstruction of secondary vertices \cite{Alexakhin2020}. To summarize, the SPD detector can make a sizable contribution to the search for physics beyond the standard model in terms of the astrophysical search for DM.
\chapter{Polarized beams \label{chap:beams}}
\section{Available species and types of collisions}
The basic specification to the available polarization states and combinations is the following:
\begin{itemize}
\item protons (possibly helium-3 ions at the second stage):  vector polarization, longitudinal and transverse direction in respect to the particle velocity;

\item deuterons: vector and tensor polarization, vertical direction of polarization, longitudinal polarization up to about a 4 GeV/$c$ momentum, and longitudinal polarization at discrete momentum of about 13 GeV/$c$;
\item possibility of colliding any available polarized particles: $p$-$p$ with a luminosity up to $10^{32}$ cm$^{-2}$ s$^{-1}$ and  $d$-$d$, $p$-$d$, $p$-$^3He$, $d$-$^3He$ with a luminosity of $10^{30}$ cm$^{-2}$ s$^{-1}$ at the collision energy equivalent to $p$-$p$ collisions; 
\item possibility of asymmetric collisions should be considered as an option for the future development of the facility;
\item for essential reduction of a systematic uncertainty, a spin-flipping system is implemented, which allows one to rotate the polarization direction with a step of 90$^{\circ}$  simultaneously in all bunches within a minimal time period.
\end{itemize}
Feasibility of the technical implementation of  the above-mentioned conditions is presented in Refs. \cite{Kovalenko2014, Filatov:2020ygb,Kovalenko:2018tyu,Filatov:2020xsy}.
%
\begin{figure}[!h]
  \begin{minipage}[ht]{1\linewidth}
    \center{\includegraphics[width=0.8\linewidth]{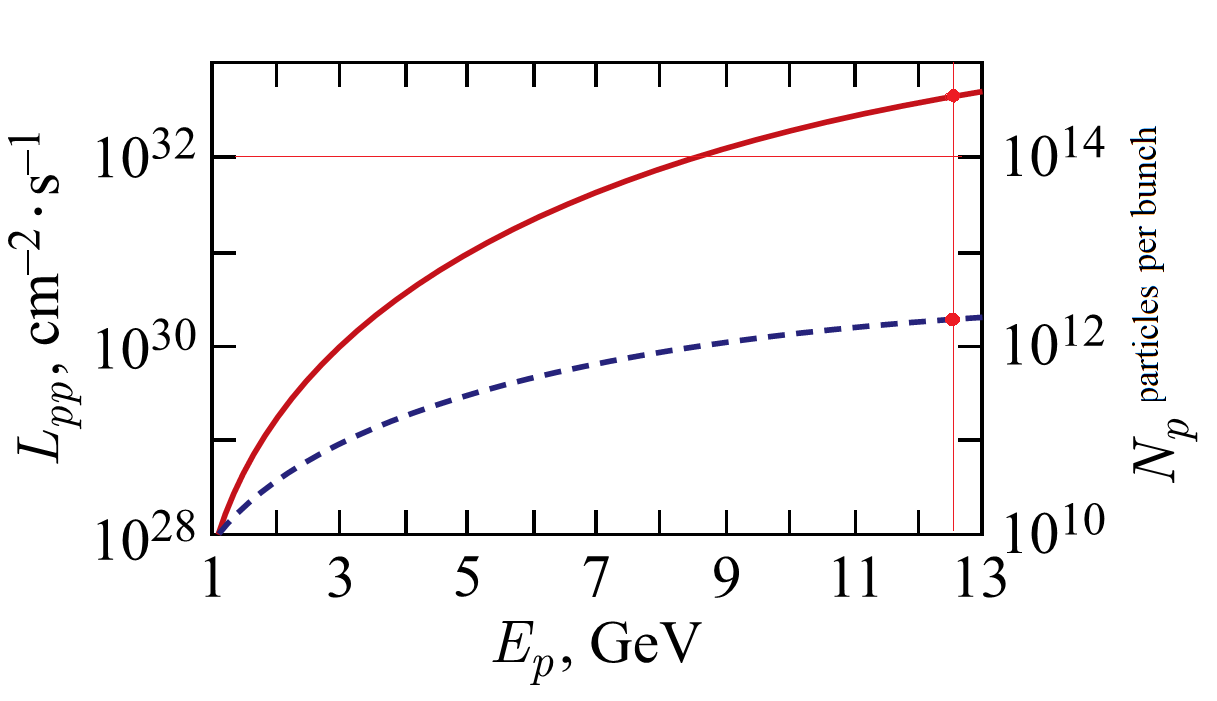}}
  \end{minipage}
  \caption{ Normalized dependence of the luminosity $L_{pp}$ (the red curve and the left scale) and the beam intensity $N_p$ (the blue curve and the right scale) on the proton kinetic energy in the $p$-$p$-collision \cite{Meshkov:2019utc}. }
  \label{fig:Luminosity_1}  
\end{figure}

\begin{table}[!bp]
\caption{Main parameters of the NICA operation cycle.}
\begin{center}
\begin{tabular}{|c|c c|}
\hline
Parameter & \multicolumn{2}{c|}{Injected beam energy} \\
                  &    2 GeV & 7.5 GeV \\         
\hline                  
Nuclotron dipole field ramp up/down, T/s & \multicolumn{2}{c|}{0.6/1} \\
Dipole magnetic field, T &	0.42 &	1.22 \\
Number of accelerated particles per pulse	 &  \multicolumn{2}{c|}{$7\times10^{10}$}\\ 
Number of cycles to store $2~\times~ 10^{13}$ particles &	 \multicolumn{2}{c|}{2 $\times$ 285 (per both rings)} \\
Storage time, s	& 923.5 &	2291 \\
Other time factors, s& 	 \multicolumn{2}{c|}{ $\sim$50} \\
Luminosity lifetime (30\% polarization degradation), s & \multicolumn{2}{c|}{	5400} \\
Beam deceleration to the new injection portion, s &	$\sim$25	& $\sim$12 \\
Total cycle duration, s &	$\sim$ 6500	& $\sim$ 7850 \\
Cycle time efficiency, \% &	83.1 &	69.8 \\
Double polarized collision efficiency FOM, \% &  \multicolumn{2}{c|}{ $\sim$ 0.5} \\
Luminosity averaged over cycle,  $10^{32}~ cm^{-2} s^{-1}$	& 2.9 &	2.45 \\
\hline
\end{tabular}
\end{center}
\label{Acc_cycle}
\end{table}%

\section{Beam structure, intensity, and luminosity}
The beam structure of polarized proton and deuteron beams at the first stage will correspond to the one optimized for the NICA heavy-ion mode. Some of the important for SPD operation parameters in the case of a bunched beam are the following: number of bunches -- 22, bunch length $\sigma = 60$ cm, collider orbit length -- 503 m, bunch velocity  $v\approx c = 3\times 10^8$ m/s,  revolution time $\tau= 1.67\times10^{-6}$ s, bunch revolution frequency $f \approx 0.6$ MHz, time gap between bunches $\Delta\tau = 76.0\times 10^{-9}$ s, crossing angle at the SPD interaction point -- $0^{\circ}$. 
The dependence of the $p$-$p$-collision luminosity on the energy and number of protons is presented in Fig. \ref{fig:Luminosity_1} \cite{Meshkov:2019utc}.

As it is clear from the calculations, the luminosity level of $1\times10^{30}$  cm$^{-2}$s$^{-1}$  is reached at a bunch intensity of $10^{11}$ polarized protons, whereas to obtain the level of $1\times10^{32}$  cm$^{-2}$s$^{-1}$ a multi-bunch storage mode should be used \cite{Kovalenko_poster}. 
The average luminosity in this case is reduced due to  the cycle time efficiency. Basic factors for that are  presented in the Table  \ref{Acc_cycle}.

The effectiveness (Figure of Merit, FOM) of a collider experiment with two polarized beams depends not only on the collision luminosity $L$, but also on the polarization degrees $P_1$ and $P_2$ of the colliding beams \cite{Bazilevsky:2014pfa}: 
\begin{equation}
FOM = LP_1^2 P_2^2.
\end{equation}
The statistical uncertainty of a measurement is proportional to $1/\sqrt{FOM}$, so the FOM describes sensitivity of the experiment or resolution. The value of the FOM depends on a specific scheme applied for suppression of depolarizing effects in each element of the accelerator complex such as the ion source, beam transportation lines, the Nuclotron and collider rings. It also depends on the beam store procedure, needed to provide the specified luminosity and other conditions of the experiment as well.  The strong dependence of  the FOM on the polarization (the fourth power) requires a  high level of attention to maintain the polarization degree at all stages of the operation.

\section{Polarization control and monitoring}
\subsection{Transportation of polarized ions in the accelerator complex}
Polarized  protons and deuterons from the SPI are first accelerated in the linac LU-20M and afterwards are injected and accelerated in the Nuclotron to the specified energy and extracted to the collider via the long transfer line. 
The main tasks at this stage are the following:  i) preservation of the beam polarization during acceleration in the Nuclotron (and in the collider as well) and ii) polarization control in the collider mode. Moreover, it is necessary to adjust the polarization direction in the transfer line and the other points of the collider orbit.
\begin{figure}[!b]
  \begin{minipage}[ht]{1\linewidth}
    \center{\includegraphics[width=0.8\linewidth]{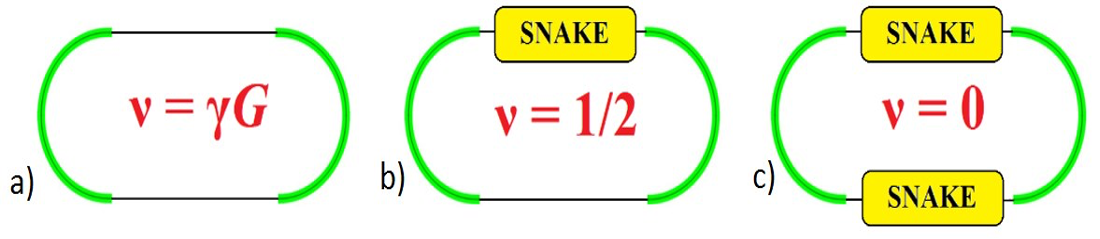}}
  \end{minipage}
  \caption{ Possible operating modes with polarized beams at the NICA collider: a) without snakes, b) with one snake, c) with two snakes. }
  \label{fig:32}  
\end{figure}

\subsection{Spin dynamic modes in the NICA collider}
From the spin dynamics point of view, the NICA collider can operate in two modes, namely: in a Distinct Spin mode (DS-mode) and in the Spin Transparency mode (ST-mode) \cite{Filatov:2020xsy,  ST-NICA}.
In the DS-mode periodic motion of the spin along the particle orbit is the only possible way, i.e. -- stationary magnetic structure selects the only one stable direction of the polarization vector in any point of the particle orbit and the fractional part of the spin tune is not equal to zero.  In the ST-mode the direction of the spin vector is reproduced in any point at every turn, i.e. the magnetic structure of the accelerator ( or the storage ring) is transparent for the spin -- the fractional part of the spin tune is equal to zero \cite{newKondratenko2020}.

The main difference between the DS- and ST-mode can be observed while manipulating the spin direction during physics data taking. In the ST-mode, the spin motion is very sensitive to magnetic field changes, because particles move in the vicinity of the integer resonance.  In the case use of the additional ''weak'' magnetic field, the rotating spin at small angles  $\phi \ll1$ provides the needed polarization direction at any specified point of the collider \cite{newFilatov2013351,newKovalenko20152031}. It is possible to use a pair of solenoids with a field integral of 1 $T\times m$, which would introduce negligible distortions to the closed particle orbit in order to produce the necessary variation of the spin angle in the NICA collider over the momentum range up to 13.5 GeV/$c$. 
In the case of the DS-mode similar procedure will require spin rotators based on a strong field, rotating  the spins at the angles of $\phi \sim 1$. Thus, in  case of changing the polarization direction from the longitudinal to the transverse one, it would be necessary to apply the transverse field with the total integral of 20$\div$30 $T\times m$, which would result in strong distortions of the closed particle orbit. The amplitude of the distortions can reach tens of centimeters at low energies. Thus, efficient polarization control of ions, especially deuterons, by means of quasi-stationary weak fields is possible only if the ST-mode is used.

The configuration of the NICA collider with two solenoid snakes inserted into its straights allows one to operate with polarized proton and deuteron beams in the both DS- and ST-modes (see Fig. \ref{fig:32}) \cite{newDerbenev20211,newButenko20141162}.

The stable polarization direction in the NICA collider with snakes switched off  is vertical at any point of the orbit in the NICA collider, whereas the spin tune is proportional to the particle energy: $\nu=\gamma G$, where  $G$ is the anomalous part of the gyromagnetic ratio and $\gamma$ is the relativistic Lorentz factor.  The collider operates in the DS-mode practically over the total energy range, because $\gamma \neq k/G$, where $k$ is an integer. At integer spin resonances $\gamma G = k$ the collider operates in the ST-mode. After switching one of the two snakes on, the collider operates in the DS-mode with the spin tune equal to half, whereas if two snakes are switched on, it operates in the ST-mode with the spin tune equal to zero. 

\subsection{Specifications to the polarized beams in the collider}
Different experiments are planned with polarized proton, deuteron, and helium-3 (in the future) beams to identify and study various  observables for multiple physics tasks: Drell-Yan, $J/\psi$, high-$p_T$ hadrons, exotic states, etc. The polarization control system should satisfy the following  main conditions: 
\begin{itemize}
\item obtain both longitudinal and transverse polarization in the MPD and SPD detectors with a polarization degree not less than 70\% and a polarization lifetime of not less than the beam lifetime;
\item provide collision luminosity of $\sim10^{30}\div10^{32}$ cm$^{-2}$s$^{-1}$ over the particle momentum range from 2 to 13.5 GeV/$c$;
\item provide the particle energy scan with a step of 1$\div$2 GeV within the energy range of 7$\div$27 GeV and 0.3 MeV at lower energies. The deuteron beam at $\sim$5.6 GeV/u is a special case;
\item provide a possibility of operation in the asymmetric particle momentum mode;
\item make simultaneous spin-flips for all bunches in the case  of Spin Flipping experiments (SF-system).
\end{itemize}

\subsection{Polarization control in the NICA collider without snakes in the ST-mode}
 The ST-mode is enabled at the discrete energy points corresponding to the integer spin resonances: $\gamma = k/G$. For protons, the number of points corresponding to the ST-mode is 25 starting from the minimal energy $E_{kin}^{min}=108$ MeV with a step of $\Delta E=523$ MeV. For deuterons, there is only one point $E_{kin}=5.63$ GeV/u corresponding to the momentum 13 GeV/c.
\begin{figure}[!h]
  \begin{minipage}[ht]{1\linewidth}
    \center{\includegraphics[width=0.8\linewidth]{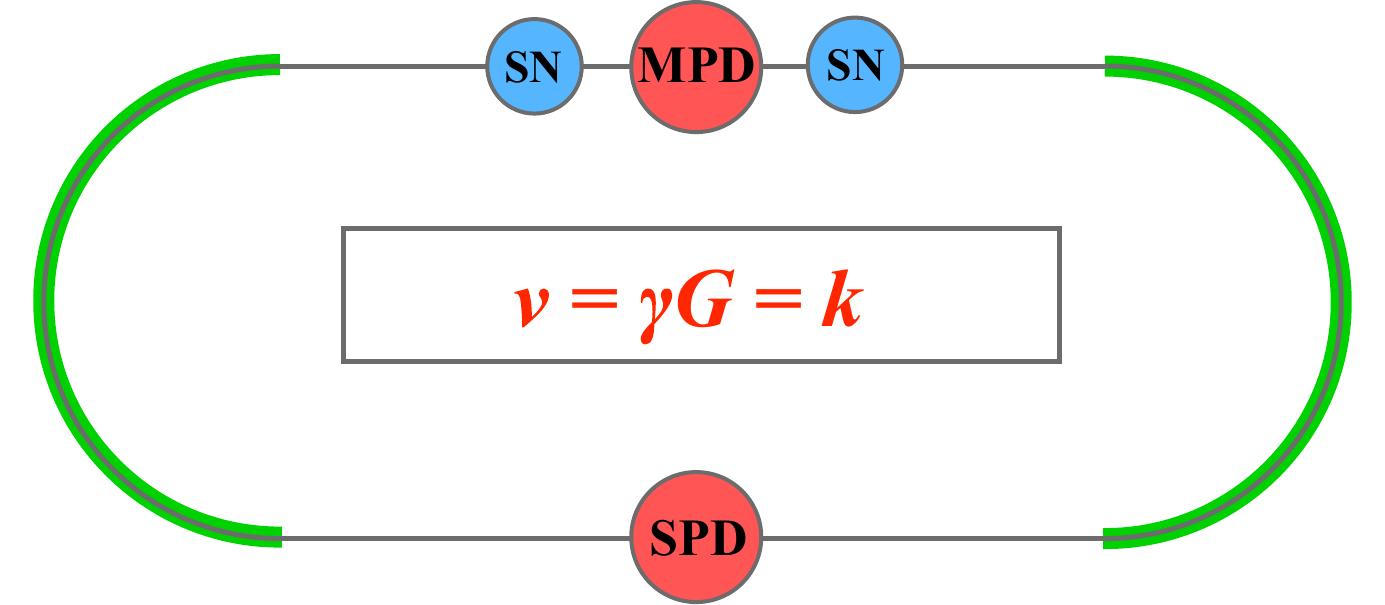} }
  \end{minipage}
  \caption{General scheme of the polarization control at integer spin resonance points.}
  \label{fig:Luminosity_2}  
\end{figure}
\begin{figure}[!h]
    \center{\includegraphics[width=1\linewidth]{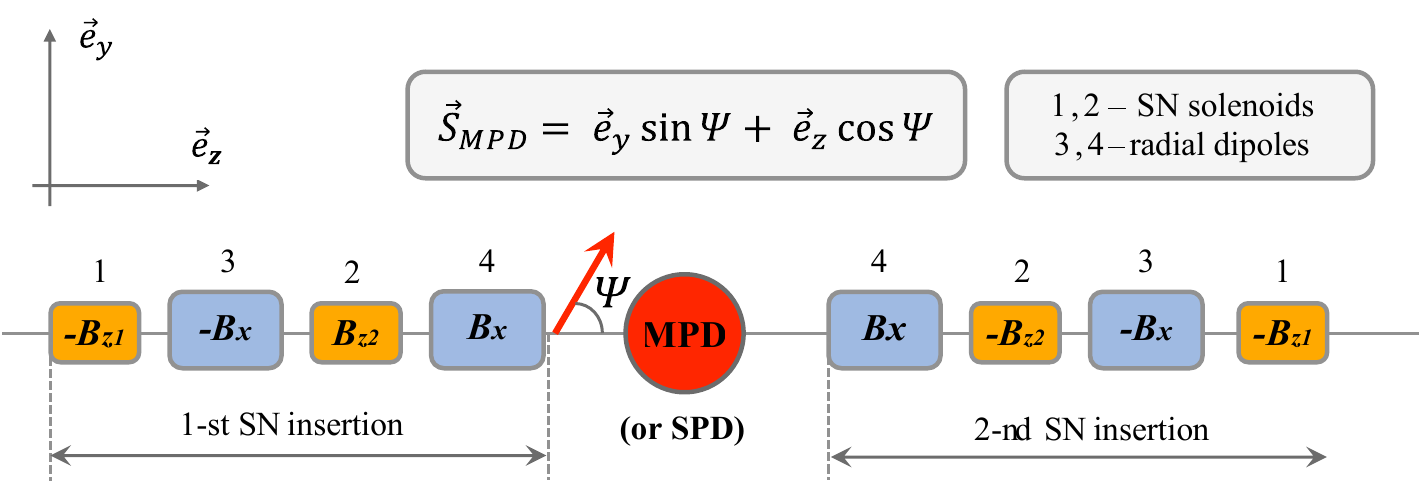}}
     \caption{ Detailed scheme of the spin navigator insertion in the collider in the ST-mode.
       \label{fig:beam2}}  
\end{figure}
A possible scheme of the ion polarization control in the collider at the integer spin resonances is presented in Fig. \ref{fig:Luminosity_2}. Two spin navigator insertions (SN) placed near MPD are used to stabilize the needed polarization direction at any point of the collider ring, including the collision points, at injection, etc. A detailed scheme of the SN’s is presented in Fig. \ref{fig:beam2}. The SN solenoids (marked as 1 and 2) generating longitudinal magnetic fields $B_{z1}$ and $B_{z2}$ are placed among the collider structural magnets (marked as 3 and 4) that generate radial fields $\pm B_x$, deflecting the beams to the collision plane of MPD \cite{ST-NICA}.

\begin{figure}[!h]
    \center{\includegraphics[width=1\linewidth]{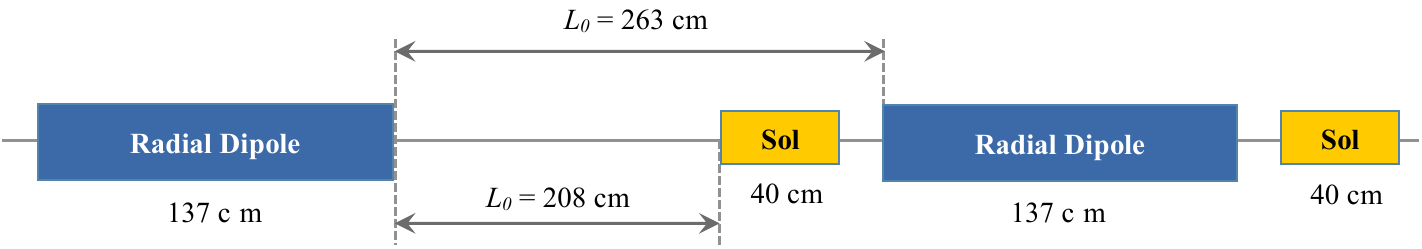}}
     \caption{ Placement of the SN solenoids in the horizontal plane.
       \label{fig:beam3} } 
\end{figure}

The scheme makes the ion polarization control possible in the vertical plane ($yz$) in MPD or SPD ($\Psi$ is the angle between the polarization and the particle velocity vectors). A field integral of 0.6 $T\times m$ of one of the navigator solenoids provides stable operation of the scheme in the entire energy range. The spin tune for protons is  then $\nu = 0.01$. If we limit the field maximum to 1.5 T, the magnetic length of the SN solenoid unit will be 40 cm. The real relative scale of the SN solenoid (40 cm long), the radial dipoles, and the distances between them are shown in Fig. \ref{fig:beam3}.

\begin{figure}[!h]
    \center{\includegraphics[width=1\linewidth]{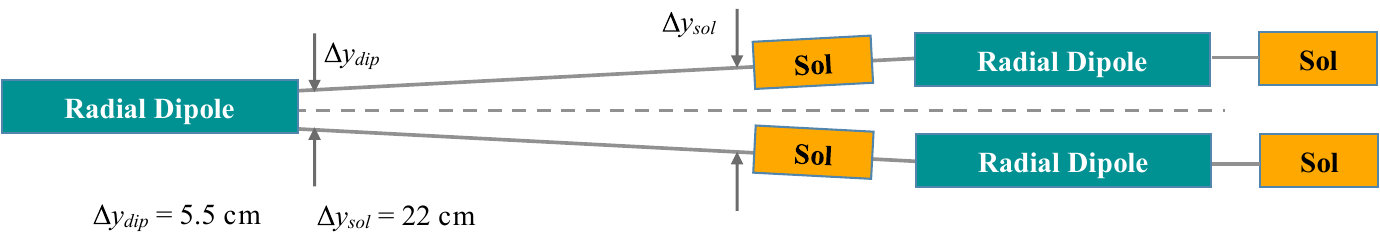}}
     \caption{ Placement of the SN solenoids in the vertical plane together with the radial dipoles.
       \label{fig:beam4}  }
\end{figure}

The scheme of the SN solenoids installation in the vertical plane together with the collider lattice elements is presented in Fig. \ref{fig:beam4}. The beam convergence angle in vertical plane, which is defined by the dipoles with magnetic fields transverse to  the beam axis is $a_x=0.04$ rad. The distance between the collider rings in the vertical plane is 32 cm. The distances  between the closed particle orbits in the vertical plane are $\Delta y_{dip}=L_x a \approx 5.5$ cm and $\Delta y_{sol}= \Delta y_{dip}+2L_1 a\approx 22$  cm at the output of the common radial dipole and at the exits of the  control solenoids, respectively.

\subsection{Ion polarization control in the ST-mode by means of two snakes}
\begin{figure}[!h]
    \center{\includegraphics[width=1\linewidth]{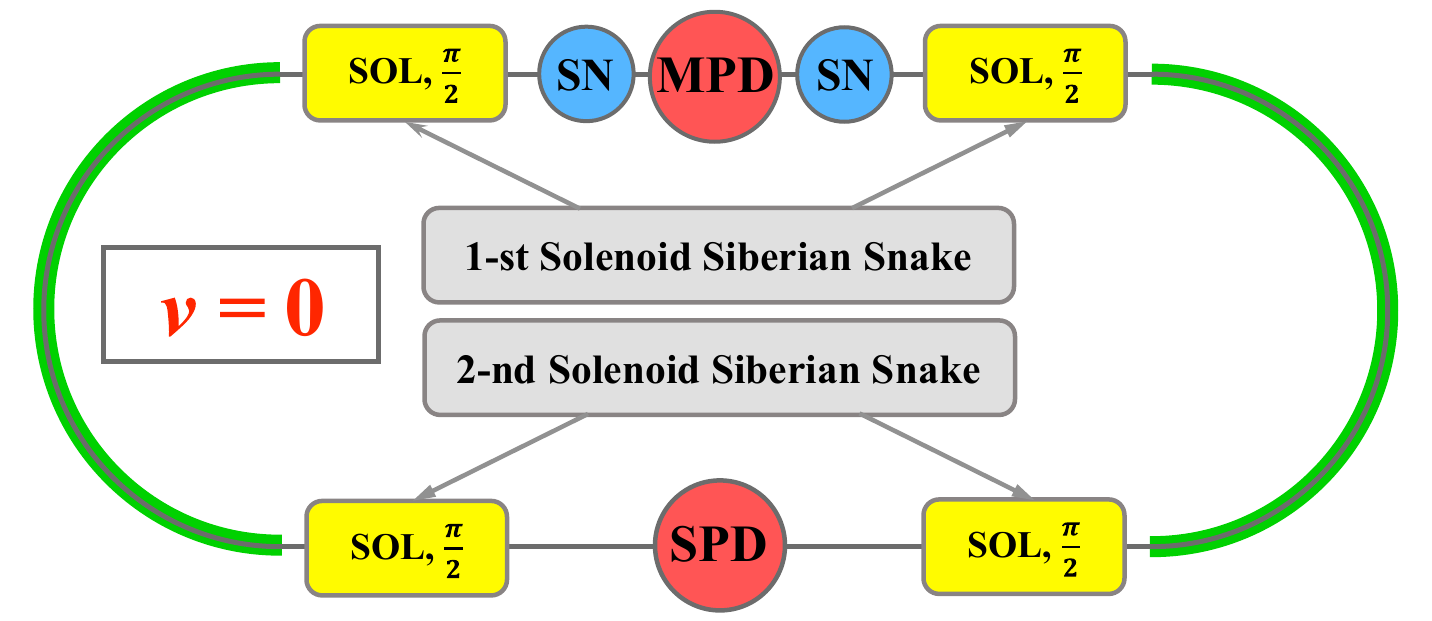}}
     \caption{ Implementation of the ST-mode in the NICA collider.
       \label{fig:beam5}  }
\end{figure}

Two solenoidal snakes installed symmetrically relative to both MPD and SPD setups will provide implementation of the ST-mode in the NICA collider (Fig. \ref{fig:beam5}). 

\begin{figure}[!b]
    \center{\includegraphics[width=1\linewidth]{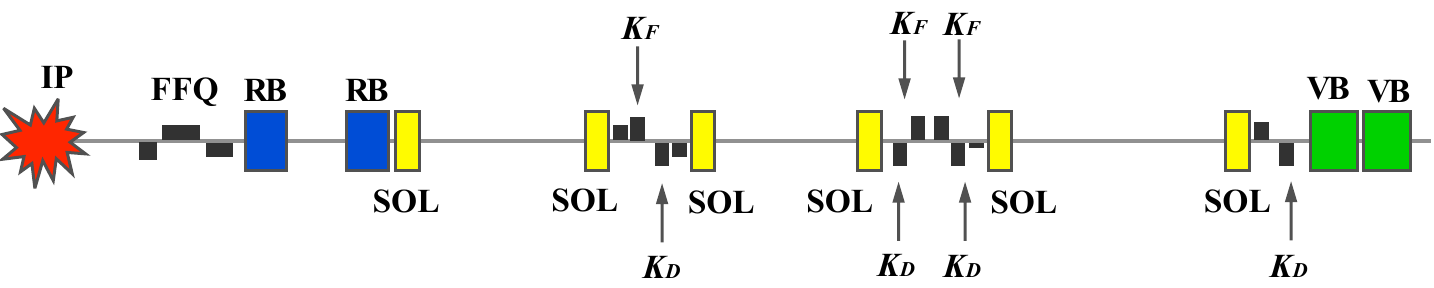}}
     \caption{ Distributed snake (one half) based on short 6 T SC solenoids.
       \label{fig:beam6}  }
\end{figure}

The configuration makes it possible to turn the spin in the vertical plane ($yz$) of MPD or SPD detector, whereas in the collider magnet arcs the polarization vector is moves in the median plane ($xz$) \cite{doi:10.1142/S201019451660096X}.
           The ST scheme with two snakes provides a zero spin tune at any point of the particle energy. It is very important for optimization of the effective NICA operation at the highest possible luminosity of $p$-$p$ collisions, due to a necessity of storing particles at an energy level that provides proper conditions for electron cooling of the stored beam. Only in this case it is possible  to form particle bunches with a high number of particles and a high degree of polarization at low energy (about 1 GeV) with further acceleration up to the experiment energy. 
           The total integral of the longitudinal solenoidal field should reach $4\times25$ T$\times$m per ring at a proton momentum of 13.5 GeV/$c$ and $4\times80$ T$\times$m for deuterons, respectively.  A distributed system consisting of short solenoids is possible, i.e. in the case of 6 T solenoids the total length of 4.2 m is sufficient to form a half-length snake.
        It is possible to adopt the collider lattice structure optimized for a heavy ion beam for the case of the ST-mode at the protons mode over the total energy range. Weak control solenoids barely distort the orbital motion in the collider, whereas strong solenoids of the snakes lead to a strong betatron tunes coupling. As the longitudinal field of the snakes changes proportionally to the particle momentum, the collider magnetic optics stays adequate to the stable motion of a polarized particle during the beam acceleration phase. 
         Matching the solenoids with the collider structure is provided by means of a proper choice of the work point by means of structural KF (focusing) and KD (defocusing) quadrupole lenses. 
        A possible scheme of the distributed snake (one half) based on short 6 T superconducting  (SC) solenoids is shown in Fig. \ref{fig:beam6}. The elements are the following: SOL-SC solenoid, FFQ -- final focus triplet of the collider, VB -- structural dipole magnets; RB -- bending dipoles with a transverse field making for converging bunches in the interaction point \cite{newKovalenko2016012022}.

\subsection{Stability of spin motion}
In the absence of a spin navigator, polarization is strongly influenced by the fields of lattice imperfections  and fields associated with betatron and synchrotron oscillations, i.e. with orbital beam emittances. The magnitude of this effect on polarization is determined by the strength of the zero spin tune resonance. To stabilize the polarization during the acceleration process or when controlling the polarization direction in the ST-mode it will be necessary to maintain the spin tune caused by the spin navigator much higher than the strength of the zero spin tune resonance.  The calculations have shown that the level of $10^{-2}$ for protons and $10^{-4}$ for deuterons would be sufficient \cite{1742-6596-678-1-012023}. These  values impose limitations on the minimum field integral in each of the weak control solenoids – 0.6 T$\times$m.

\subsection{Spin flipping system}

The spin flipping (SF) system makes it possible to carry out spin physics experiments at a much higher level of accuracy \cite{Derbenev}. Being equipped with such a system, the SPD setup will have following  advantages:
\begin{itemize}
\item no need for reversing the polarization direction at the polarized ion source;
\item  no need for bunch-to-bunch luminosity measurements and a bunch monitoring system;
\item the possibility of comparing collisions of bunches of any particle spin directions (vertical-longitudinal, vertical-radial, radial-longitudinal, etc.).
\end{itemize}
The SF-system based on quasi-stationary fields is naturally implemented in the ST collider mode. The pair of SN solenoids provides simultaneous influence on the polarization direction and the spin tune. Thus, the possibility of the spin tune stabilization during the spin flipping  appears, preventing both zero spin tune and higher-order spin resonance crossing. The polarization degree will preserve an exponential accuracy, provided the field of the SN  solenoids changes slowly. A typical flipping time is approximately estimated as 1 ms and 10 ms for the proton and deuteron, respectively.
Enabling the SF-system in the DS-mode will require inserting an RF-module of a several MHz range and the field total integral of 1~T$\times$m, which is a rather challenging technical task.

\subsection{Online control of the polarization in the collider}
A unique possibility of online polarization control becomes available when the collider operates in the ST-mode \cite{Filatov:2020ygb}. Since the field ramp in  the SN solenoids ($t_{change}\sim 0.2$ s) is much larger than the spin precession  period around the induced spin field ($t_{rev}\sim 10^{-4}$ s), any manipulations with the spin direction at the spin tune will  occur adiabatically and the polarization degree during the experiment time will be maintained constant with an exponent accuracy. The direction of the polarization vector will be a function of the SN solenoids field and  can be defined by means of the SN  field measurements. 
Thus, the ST-mode makes it possible to carry out experiments at the NICA collider at a new level of accuracy.


\subsection{Polarized beams dynamics in the Nuclotron}
\begin{figure}[!h]
  \begin{minipage}[ht]{0.49\linewidth}
    \center{\includegraphics[width=1\linewidth]{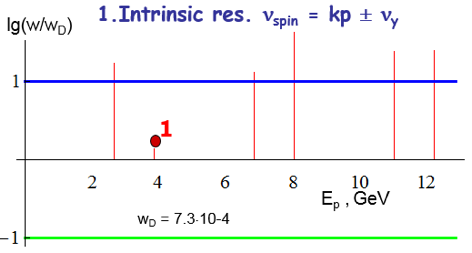} \\ (a)}
  \end{minipage}
  \hfill
  \begin{minipage}[ht]{0.49\linewidth}
    \center{\includegraphics[width=1\linewidth]{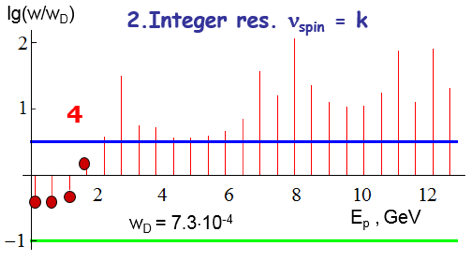} \\ (b)}
  \end{minipage}
  \begin{minipage}[ht]{0.49\linewidth}
    \center{\includegraphics[width=1\linewidth]{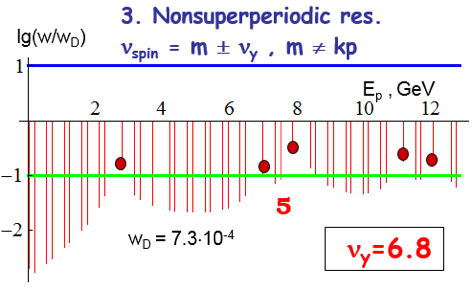} \\ (c)}
  \end{minipage}
  \hfill
  \begin{minipage}[ht]{0.49\linewidth}
    \center{\includegraphics[width=1\linewidth]{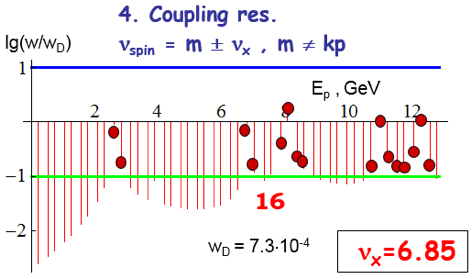} \\ (d)}
  \end{minipage}
  \caption{Linear spin resonances in the Nuclotron in the polarized proton mode.}
  \label{fig:beam7}  
\end{figure}

As it was already mentioned, the stable polarization direction in the Nuclotron is vertical, and the spin tune is proportional to the beam energy: $\nu=\gamma G$, which definitely leads to the crossing of spin resonances during particle acceleration and, as a consequence, to resonance depolarization of the beam.  There is no problem with deuterons:  the only one integer spin resonance at $E_{kin}=5.6$ GeV exists in the NICA energy range. The spin direction at this point can be controlled also by means of a weak solenoid  inserted into the accelerator lattice. The number of different spin resonances in the proton mode is much larger. Logarithmic graphs of linear spin resonance strengths scaled to the specific strength corresponding to complete depolarization of the beam are presented in Fig. \ref{fig:beam7} \cite{newIssinsky1995169,newGolubeva2002,newVokal200948}
. The proton energy range $E_p$ corresponds to the one available at the Nuclotron. Each graph is divided into three areas that correspond to the intermediate crossing (between the horizontal lines), fast crossing (below the green line) and adiabatic crossing (above the blue line). The lines of the fast and adiabatic crossing correspond to 1\% loosing of the polarization degree.

The parameters taken for calculation of the spin resonance strengths were the following:  magnetic field ramp -- 1 T/s; beam emittance (horizontal and vertical) at the injection energy -- 45$\pi$ mm$\times$mrad; quadrupole misalignment errors -- 0.1 mm; errors of angular alignment of the structural dipole and quadrupole magnets -- 0.01 rad; and the relative error of the quadrupole gradients -- 0.001.
The resonances marked with red circles are dangerous and lead to the beam depolarization after their crossing. 

To keep the polarization of the proton beam at a proper level, a partial Siberian snake based on a solenoid will be used. Two options have been considered: i) using a weak 5\% snake with a field integral of 0.65 T$\times$m, which can save the proton beam polarization up to 3.4 GeV/$c$ and  ii) using a 25\% snake ($\sim$ 12 T$\times$m)  \cite{Filatov2014,Filatov,FilatovTUPVA112}. The first one is efficient if the collider operates in the ST-mode with two snakes and injection of the beam is provided at a low energy (around 1 GeV), whereas the  strong enough snake used in the second option could save the polarization over the total energy range in the Nuclotron and is suitable for the operation at integer resonances. The choice of energy points is limited to the points of integer resonances.

\subsection{Operation modes of the NICA collider with polarized beams}
\begin{table}[h]
\caption{Polarization in the SPD and MPD interaction points in the DS- and ST-modes.}
\begin{center}
\begin{tabular}{|c|c|c|c|}
\hline
Collider          & Spin  & Spin   & Polarization direction at   \\ 
configuration &  tune  & mode & SPD and MPD            \\   
\hline
Without snakes & $\gamma G\neq k$ & DS & Vertical at SPD and MPD \\
Without snakes &$\gamma G = k$      & ST  & Any direction at SPD or MPD\\
With one snake (SPD) & 1/2                & DS & Longitudinal at MPD \\
With one snake (MPD) & 1/2                & DS & Longitudinal at SPD \\       
With two snakes           &  0                 & ST  & Any direction at SPD or MPD \\
\hline
\end{tabular}
\end{center}
\label{tab:regimes2}
\end{table}%

\begin{table}[h]
\caption{Features of the operation with polarized protons and deuterons in the NICA collider.}
\begin{center}
\begin{tabular}{|c|c|c|c|c|}
\hline
Collider   & Spin  & SF- & Online  & Energy \\
configuration & mode & system & polarimetry & energy\\
\hline
Without snakes & DS & No & No & No\\
Without snakes & ST &  Yes & Yes & No\\
With one snake & DS & No & No & Yes\\
With two snakes& ST & Yes & Yes & Yes \\
\hline
\end{tabular}
\end{center}
\label{tab:regimes3}
\end{table}%

The NICA collider with two solenoidal snakes will make the following configurations possible (see Table \ref{tab:regimes2}) \cite{ST-NICA}.

If the snakes installed in SPD and MPD sections are switched off, the DS-mode with vertical polarization at any point of the collider orbit is activated. Certain narrow energy gaps where the ST-mode at integer resonances exist, give possibility of having any direction of the polarization in both detectors. After switching one of the snakes on, the collider will operate in the DS-mode with the spin tune 1/2. The snake transforms the spin motion  completely, providing a stable longitudinal direction of the polarization in 
the collider orbit section opposite the snake.

If two dynamic solenoid snakes are switched on, the ST-mode is activated. The spin tune does not depend on the particle energy and is equal to zero, which gives a possibility to obtain any  direction of polarization at any point of the collider orbit. The features of the collider operation in the polarized modes are shown in Table \ref{tab:regimes3}.

It is very important to implement the possibility of polarized beam acceleration in the NICA collider without loosing the polarization degree. The problem of reaching the highest possible luminosity of polarized proton collisions is connected with the particle multi-bunch storage in the collider and electron cooling of the stored beam during the process. The optimal proton beam kinetic energy at the beam injection into the collider is about 1 GeV \cite{Kovalenko:2018tyu, 1742-6596-938-1-012018}.

\subsection{Conclusion}
The proposed scheme of the ion polarization control in the NICA collider is adapted easily to the collider magnetic optics at any modes of the polarization control. Significant advantages could be gained by using the spin transparency mode. A polarization degree of about 70\% is provided at the collision points. The polarization lifetime is expected to last for several hours that is comparable with the beam lifetime. We have not mentioned particular measurement and monitoring systems that should be designed at the stage of the technical project preparation, and namely: precise measurement of the luminosity, 
 absolute polarimeter based on a gas jet, targeting stations, etc.

\chapter{Detector layout \label{ch:DL}}
\section{General design}

The physics tasks presented in Chapter \ref{chapter2}  impose general requirements on the concept of the Spin Physics Detector.  
Unlike the case of high-energy collisions where the collision energy $\sqrt{s}$ is a few orders of magnitude higher than a typical hard scale $Q$
of the studied reactions, at the SPD energies for all the probes planned to be used to access the gluon content of the colliding particles $Q\sim M_{J\psi} \sim 2M_{D}\sim p_{T\gamma~min}$ is just a few times less that $\sqrt{s}/2$.
Therefore, one should expect quite a uniform  distribution of all signal particles (muons from the $J/\psi$ decay, prompt photons, products of $D$-mesons decay, etc.) over the kinematic range. In other words, there is no preferable range in rapidity, which could be specified for each probe for the optimal overall performance. Together with  relatively small cross-sections of the discussed probes, this fact leads one to a requirement of $\sim4\pi$ coverage of the SPD setup.

The Spin Physics Detector must have sufficient tracking capabilities and a magnetic system for spectrometric purposes  for the majority of the addressed physics tasks. It has to be equipped with a  muon system thick enough for effective separation of muons and hadrons to make it possible to deal with the decay $J/\psi\to \mu^+\mu^-$. A precision vertex detector is needed for the recovering of the secondary vertices from the decays of $D^{\pm/0}$ mesons and other short-lived particles. An electromagnetic calorimeter ensures capability to detect signal and background photons. A low material budget and general transparency of the setup should also provide favorable conditions for the photon physics. Hadron identification capability  is needed for any physics task with protons and/or kaons in the final state, in particular,  to enforce a signal-to-background ratio for $D$-mesons selection, and also to improve tracking at low momenta.  Since tiny effects are intended to be investigated, a triggerless DAQ system is planned in order to minimize possible systematic uncertainties of the measurements.

Strict limitations to the SPD detector layout arise from the external conditions, such as the maximal possible load to the floor of the SPD experimental hall (1500 tons together with the lodgement and the detector moving system).  Together with the requirement to have the overall thickness of the muon system not less than 4 nuclear interaction lengths ($\Lambda_I$), this limits the outer size of the SPD detector and the size of the inner part of the detector. The location of the collider infrastructure, in particular, focusing elements, also defines the size of the SPD setup along the beam axis. 
More details could be found in Chapter \ref{chap:beams}.

\begin{figure}[!h]
\begin{center}
  \begin{minipage}[ht]{0.56\linewidth}
    \center{\includegraphics[width=1\linewidth]{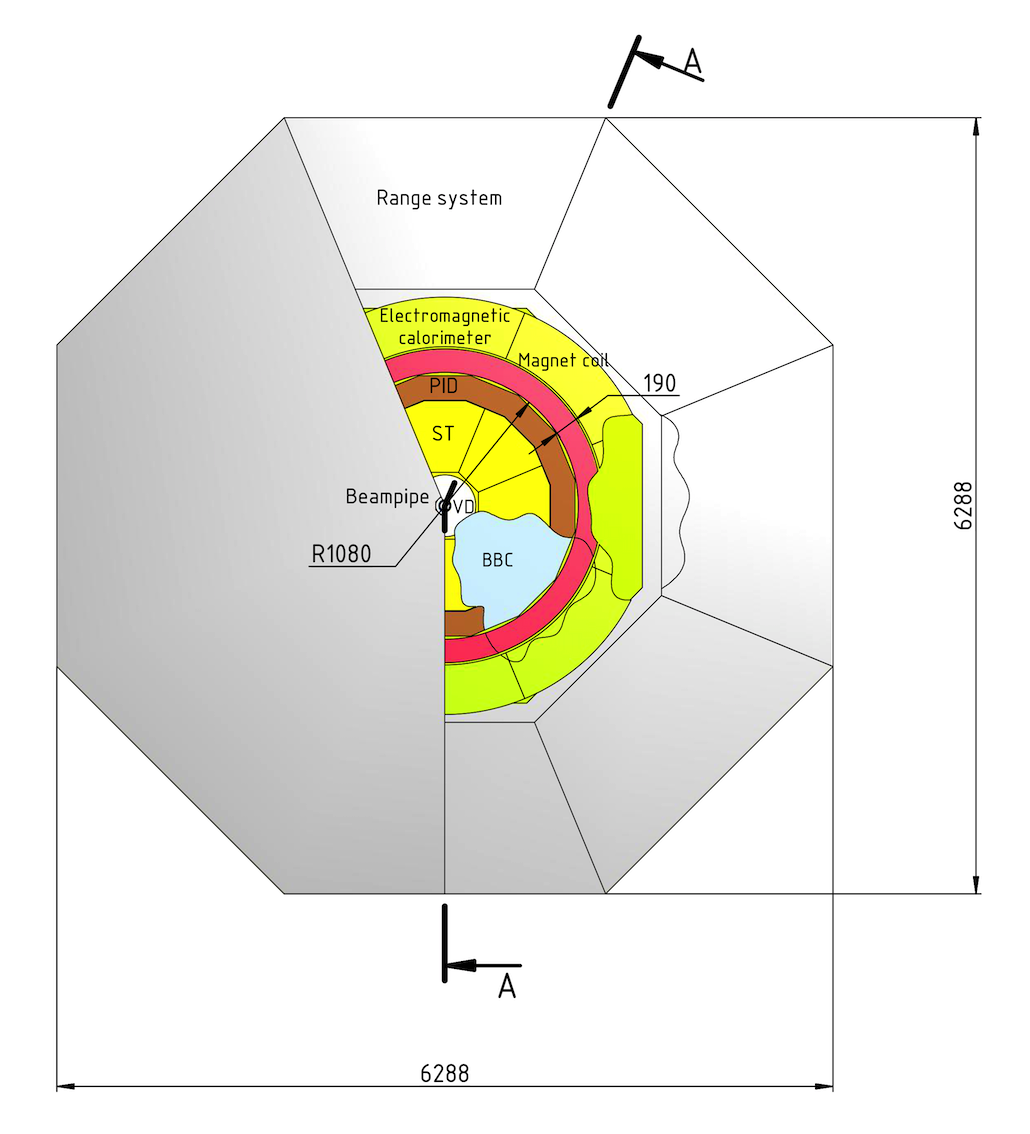} }
  \end{minipage}
  \hfill
  \begin{minipage}[ht]{0.6\linewidth}
    \center{\includegraphics[width=1\linewidth]{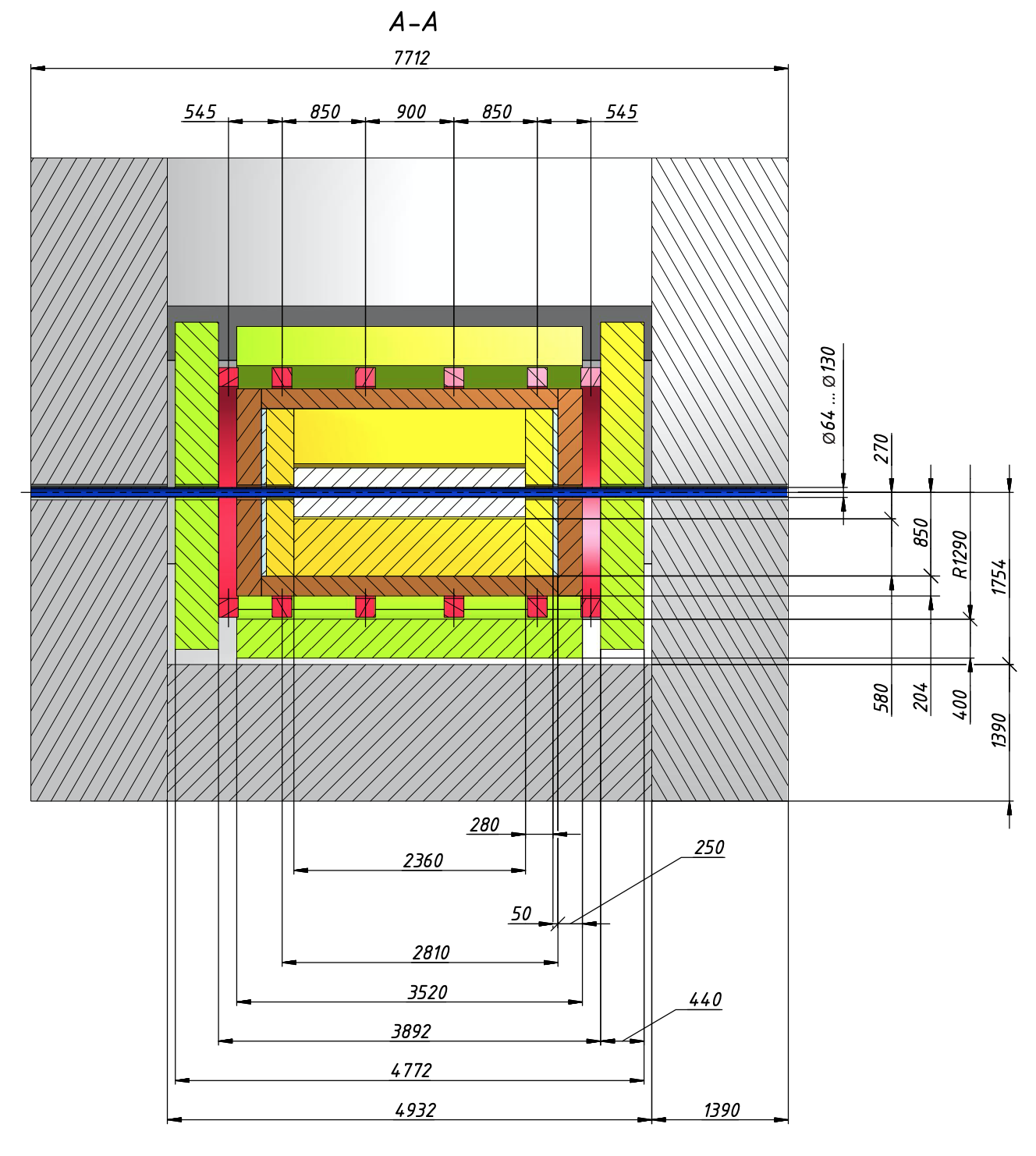}}
  \end{minipage}
  \caption{General layout of the Spin Physics Detector. All dimensions are in millimeters.}
  \label{fig:spd1}  
\end{center}
\end{figure}

The general layout of the SPD is shown schematically in Fig. \ref{fig:spd1}. The detailed description of each subsystem could be found below. Table \ref{tab:main_det} brings together the elements of the SPD physics program and the requirements to the experimental setup.  We would like to  emphasize here that the most part of the SPD gluon program can not be performed at MPD \cite{MPD_CDR}, another detector at the NICA collider, optimized for operation  at a high multiplicity of charged tracks and relatively low luminosity.


\begin{table}[htp]
\caption{Required setup configuration for each  point of the SPD physics program. (+++) - absolutely needed, (++) - extremely useful,
(+) - useful, (-) - not needed.}
\begin{center}
\begin{tabular}{|l|c|c|c|c|c|c|c|c|}
\hline

Program & Vertex     & Straw   & PID      & Electromagnetic & Beam-beam &  Range  \\
               & detector  & tracker & system & calorimeter         & counter         &   system        \\
\hline
\hline
Gluon content with:&            &           &               &                       &                     &           \\
~~~charmonia &    +   &    ++       &   +           &    ++               & +                  &    +++     \\   
~~~open charm&   +++ &  ++       & ++          &    +                  & +                  &    ++     \\
~~~prompt photons& +& +         &  -            &   +++               & +                  &    -      \\
\hline
SSA for $\pi$ and $K$& + & ++  &   +++      &    ++                & +                  &    -     \\
Light vector meson production & +& ++ &  -         &    +                  & +                  &  - \\
Elastic scattering                    &  +  &  ++ & -        &     -           &  +++             &   -  \\
$\bar{p}$ production       &      +     & ++   &  +++   &  ++         &   +                &  -   \\
Physics with light ions    &    ++     & +++ &  +       &  ++         &   ++                &  +   \\
\hline                 

\end{tabular}
\end{center}
\label{tab:main_det}
\end{table}%

\section{Magnetic system}

The SPD Magnetic System (MS) should satisfy the following criteria:
\begin{itemize}
\item  minimization of the material inside the detector inner part;   
\item  a magnetic field integral of (1$\div$2) T$\cdot$ m along the particle tracks, whereas the peak value of the field should be limited to ~ 0.8 T over the straw tracker volume;
\item  minimization of the total weight, the cross-section of the current coil (coils), and the overall amount of the MS material, i.e. the MS should have perfect mechanics.
\end{itemize}

Several options of MS’s were considered:
\begin{enumerate}
\item  Solenoid -- a uniform multi-turn coil placed between the ECal and the muon range (RS) systems;
\item  Toroidal MS (inside ECal): 3$\times$8 = 24 coils forming a toroidal distribution of the field in the detector volume only, 
\item  Hybrid system consisting of a combination of  toroidal coils in the barrel and solenoidal ones in the front/rear parts. Both  room and cryogenic temperatures were considered;
\item  System of 4 separate coils inside the ECal: a) all coils are connected in series, and b) right and left hand pairs are connected opposite to each other;
\item  Hybrid  system consisting of a combination of 8 toroidal coils in the barrel and  2 pairs of separate solenoidal coils in the front/rear parts. Both  room and cryogenic temperatures were considered;
\item System of 6 separate coils placed between the ECal and the RS system of the reduced diameter.
\end{enumerate}

    
   Thus, more than 10 different options of the 3D magnetic field configurations were analyzed. The calculated field maps were  used for the Monte Carlo simulation of the SPD performance (see Chapter \ref{MC}). Conceptual analysis of the considered MS systems was performed also. Certain data were reported at the European Conference on Applied Superconductivity EUCAS2019 \cite{kovalenko_perep}.  The general conclusions are briefly summarized below.
\begin{enumerate}
\item 	The most well-known system is a classical solenoid. Experience in design and construction of superconducting solenoids has been collected by numerous groups worldwide, including the NICA MPD. The MPD solenoid manufacturing is completed and the assembling has been started in the experimental hall. The main disadvantage for SPD of a solenoid similar to the MPD MS would be a lot of material in front of the ECal as well as its  high cost. Moreover, the fixed geometry of the field gives no universality of the SPD experimental program.
\item 	The toroidal MS was considered in its “warm” and “cold” options. The warm one was rejected due to the material budget: the necessary ampere-turns led to an unacceptable cross-section of the coils and hence the  amount of copper. The problem is solved partially in the case of superconducting coils. Nevertheless, the complexity of design of the coil system is very high in any case. The most important negative effect can occur due to concentration of the coil material closer to the bunch crossing area.
\item 	The MS consisting of separated coils is absolutely transparent  for the particles passing through the inner volume of the detector and contains “target” material for the secondary particles in the limiting volume at the ECal inner part. The amount of material depends directly on the necessary ampere-turns of the coil  and the achievable current density. The last point gives evidence in favor of the superconducting approach. The magnetic field radial and axial distribution is not so uniform in comparison with a solenoid, especially in the area close to the coils.  Nevertheless, the accuracy of modern 3D calculation codes for non-linear magnetic fields and precise magnetic measurements can guarantee the necessary accuracy of real field mapping.  Optimization of the coil cross-section is also very important. 
\item 	The hybrid MS consisting of a  toroidal system in the barrel part of the SPD and two pairs of separate coil was considered as a compromise, and namely: minimization of the magnetic field  near the polarized particles interaction zone and a solenoidal-type distribution in the front and rear parts of the detector. 
\end{enumerate}
 
\begin{figure}[!t]
  \begin{center}
    \includegraphics[width=0.8\textwidth]{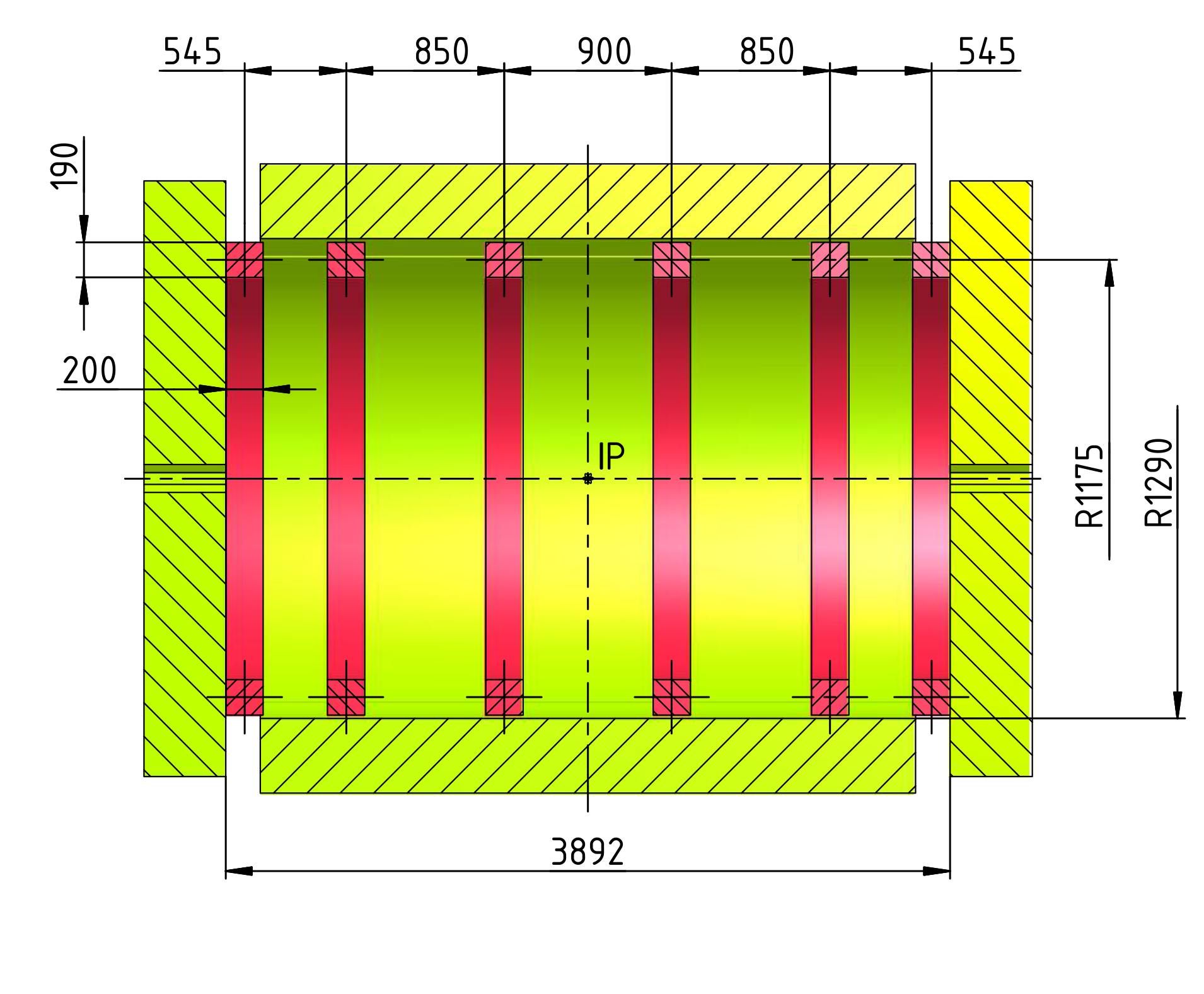}
  \end{center}
  \caption{Geometrical model of the 6-coil magnetic system. All dimensions are in millimeters.}
  \label{pic:mag2}
\end{figure}

 \begin{figure}[!h]
  \begin{minipage}[ht]{0.37\linewidth}
    \center{\includegraphics[width=1\linewidth]{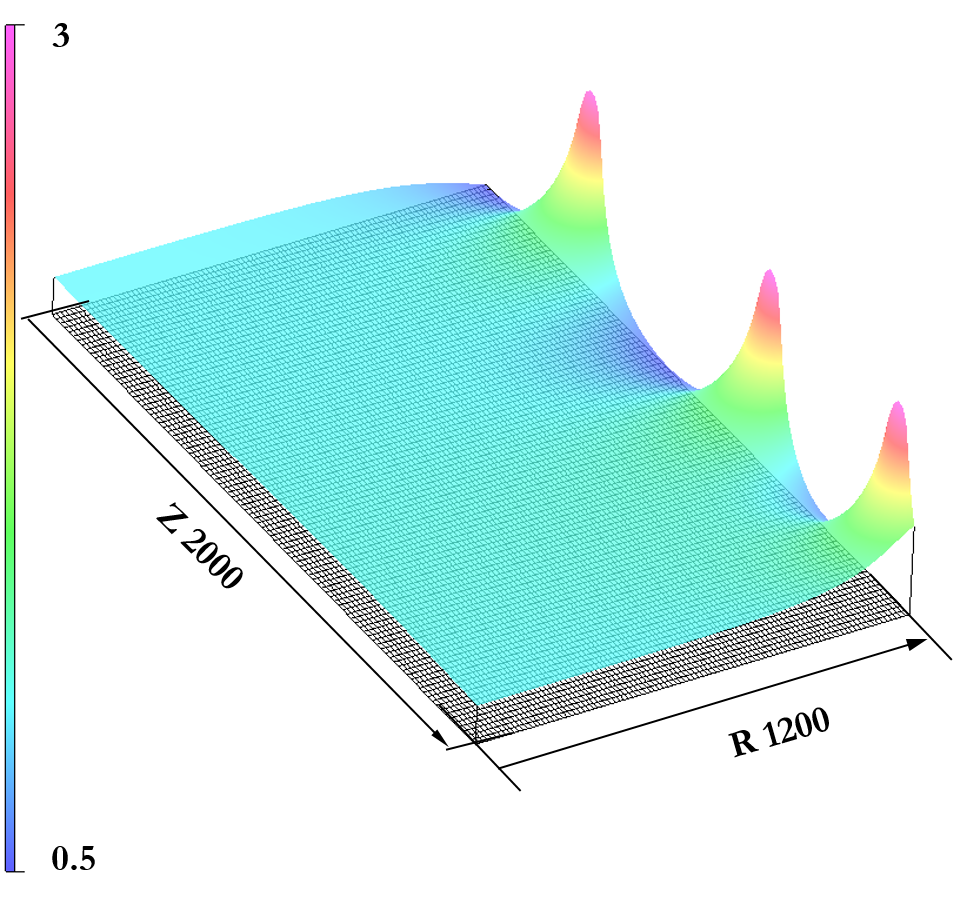} \\ (a)}
  \end{minipage}
  \begin{minipage}[ht]{0.63\linewidth}
    \center{\includegraphics[width=1\linewidth]{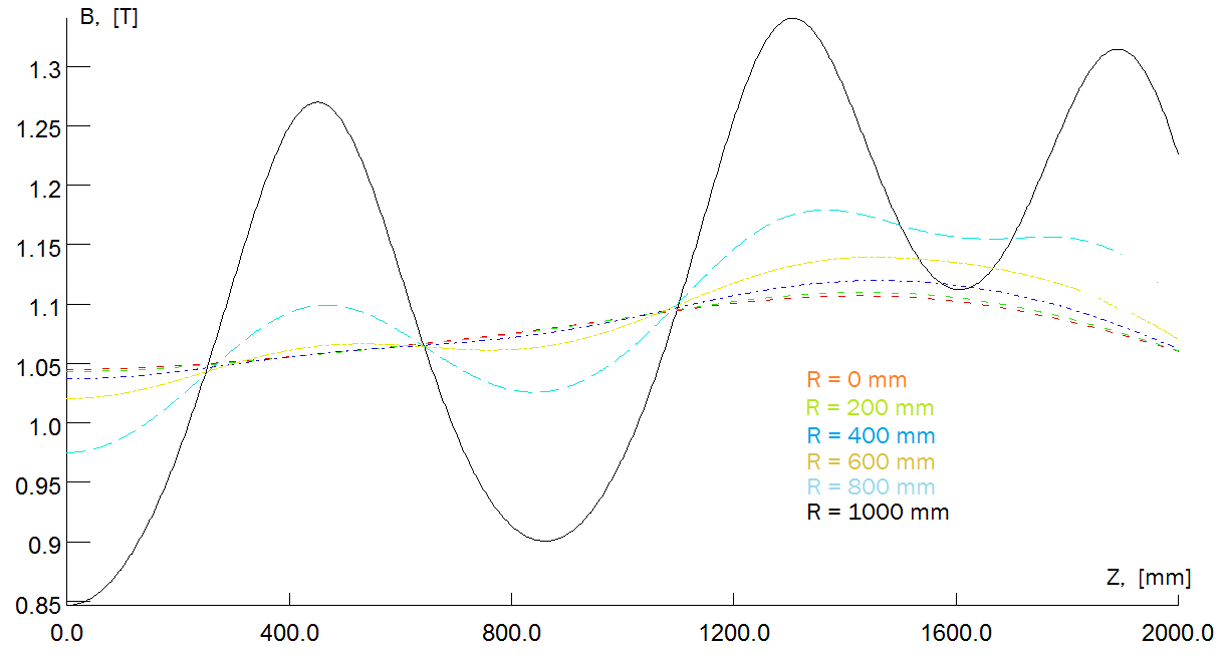} \\ (b)}
  \end{minipage}
  \caption{Field calculation results: (a) $B_z$ as function of $r$, $z$; (b). $B_z$ as function of $z$  at different $r$. All dimensions are in millimeters.}
  \label{pic:mag3}  
\end{figure}

\begin{figure}[!h]
  \begin{minipage}[ht]{0.55\linewidth}
    \center{\includegraphics[width=1\linewidth]{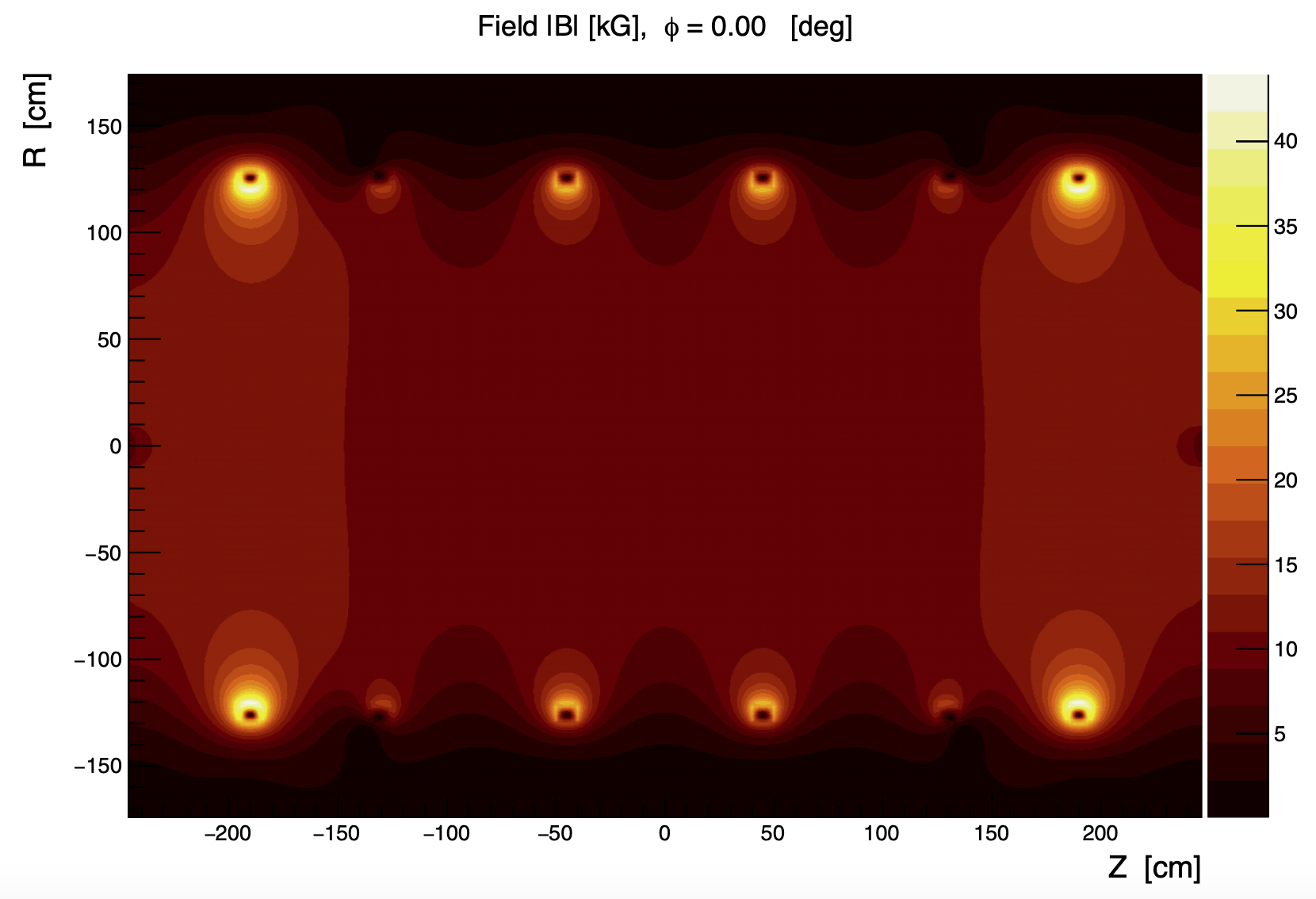} \\ (a)}
  \end{minipage}
  \begin{minipage}[ht]{0.4\linewidth}
    \center{\includegraphics[width=1\linewidth]{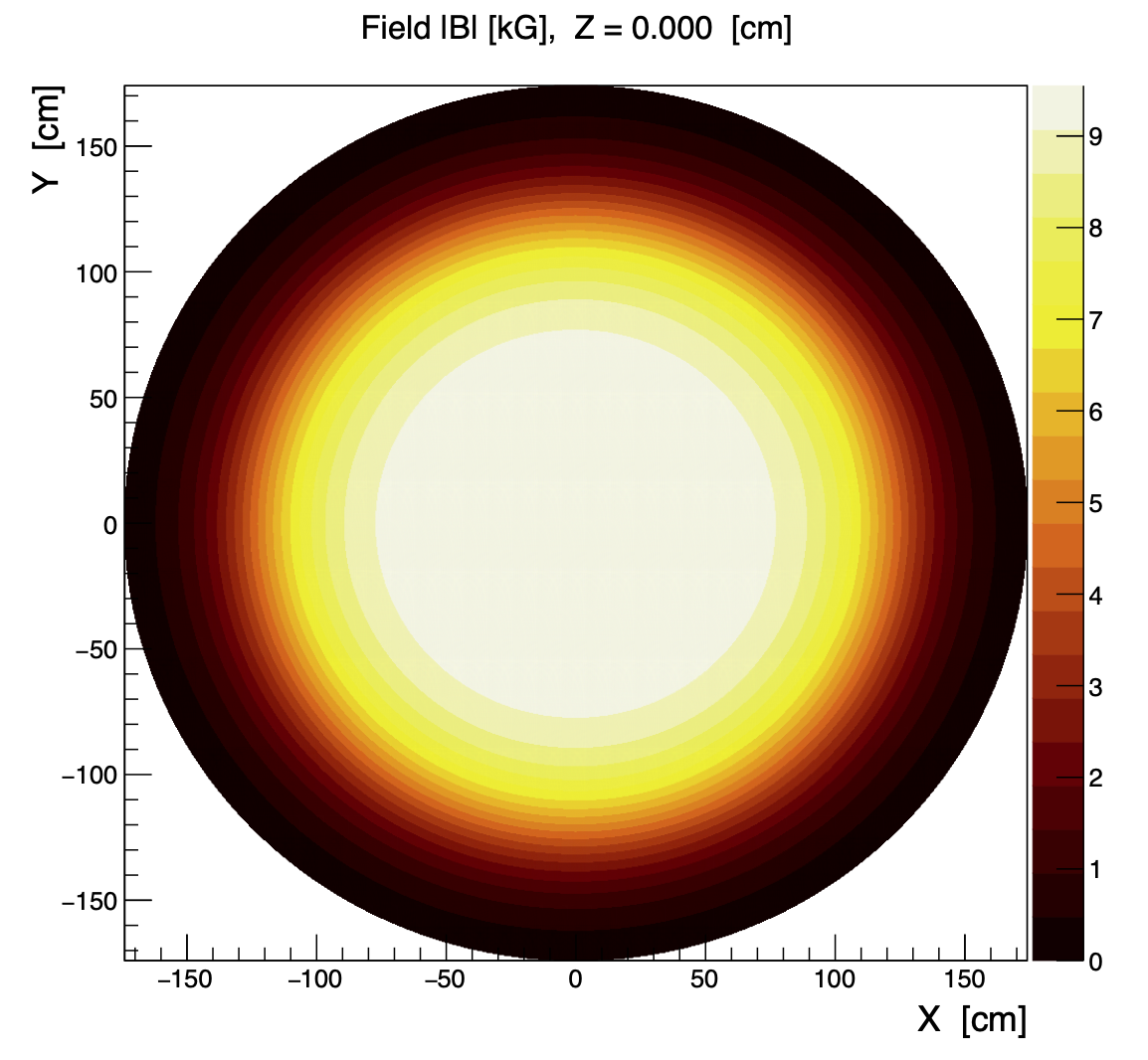} \\ (b)}
  \end{minipage}
  \caption{Magnetic field in the inner part of the SPD setup (a) $z$-$r$ section; (b) $x$-$y$ section at $z=0$. }
  \label{pic:mag5}  
\end{figure}

More advanced analysis of the detector and the collider system has shown that partial compensation of the magnetic field at the axis will give not so many advantages.  However, it would be  more beneficial, somehow, if there was no the spin control system in the collider lattice. The NICA collider will be equipped with such a system. The elements aimed at the particle spin control at the NICA collider were proposed and are under technical design now. The general description of the spin control system is presented in section 3.   Thus, the condition of a “zero” magnetic field along the beam axis is not a critical issue in our case.
   An updated choice of the SPD MS was made in favor of a separate 6-coil design. The geometrical model of the coil system is presented in Fig. \ref{pic:mag2}, and  the field calculation data in Figs. \ref{pic:mag3} and \ref{pic:mag5}.  

As it is clear from Fig. \ref{pic:mag3}, the longitudinal variation of an axial magnetic field is varied from about 5\% at the beam axis to about 12\% at a radial distance of 2 cm from the coil inner turns. The number of 12\% can be further improved by the coils system optimization.    
        We consider the technology of  superconducting coils manufacturing based on a hollow high-current cable similar to the one used for the Nuclotron magnet or the one used in the ITER systems. The manufacturing technology of a hollow cable made of NbTi/Cu composite wires cooled at 4.5 K with a forced helium flow is well developed at the Laboratory. The magnets of the NICA booster and collider are being manufactured  
 at the Laboratory magnet facility (VBLHEP JINR). The coil containing 80 turns will provide 800 kA $\times$ turns and generate the necessary magnetic field in the detector volume. Some of the SPD 6-coil MS are presented in Table \ref{tab:mag1} in comparison with the other detectors. 

\begin{table}[!htp]
\caption{Comparison of the SPD (NICA) and CMS (LHC) magnetic systems.}
\begin{center}
\begin{tabular}{|l|c|c|}
\hline
Parameter	 & SPD/NICA & CMS/LHC \\
\hline
\hline
Size    (diam./length), m/m &	2.9/6.0 & 6.5/12.7 \\
Magnetic system		& 6 coils&	solenoid \\
Peak magnetic field, T&	1.0 (axis)	&4.5\\
Coil average  diam., m & $\sim$2.5	 & $\sim$ 6.5\\
Field volume,	m$^3$ & $\sim$45	& $\sim$414 \\
Stored energy,	MJ	& $\sim$40 &	$\sim$2800 \\
Coil turns &		80 &	2112 \\
Operating current, kA	& 8.75	& 20 \\
Total inductance, H &	 $\sim$0.2 &	12.6 \\
\hline
\end{tabular}
\end{center}
\label{tab:mag1}
\end{table}%

\section{Beam pipe }
A beam pipe separates the detector and high vacuum of the accelerator. It must be mechanically sturdy on the one hand and thin enough in terms of the number of radiation lengths to minimize multiple scattering and radiation effects, on the other hand.
The diameter of the beam pipe is a compromise between the radial size of the beams and the requirement to put coordinate detectors as close to the interaction point as possible for better reconstruction of the primary and secondary vertices. A beryllium beam pipe with outer diameter of about 6 cm with 0.5mm wall thickness is proposed to be used. 

\begin{figure}[h]
  \begin{center}
    \includegraphics[width=1.\textwidth]{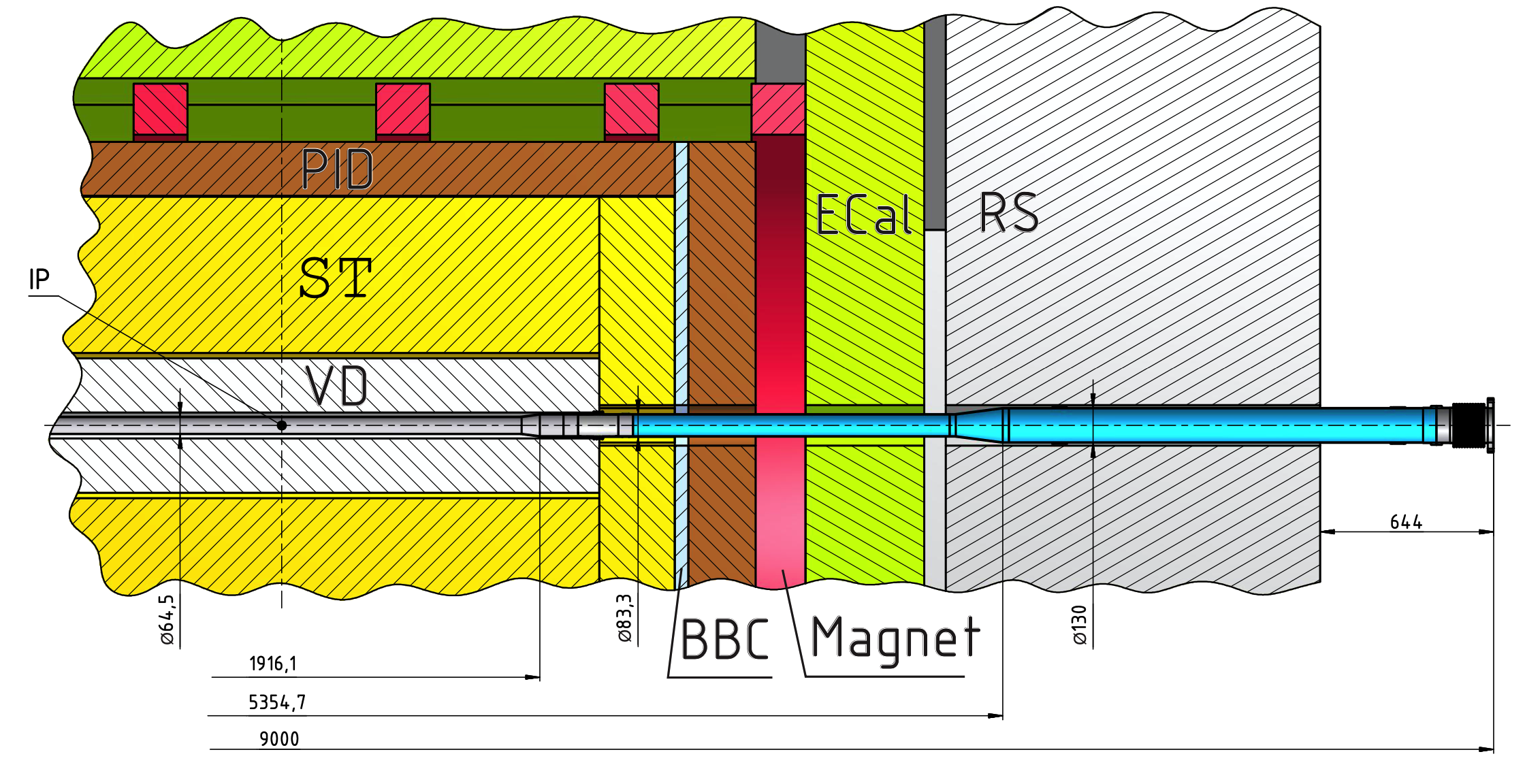}
  \end{center}
  \caption{SPD beam pipe inside the setup. All dimensions are in millimeters.}
  \label{pic:beam_pipe}
\end{figure}

The schematic view of the beam pipe and its positioning inside the SPD is shown in Fig. \ref{pic:beam_pipe}. 

At the first stage of the SPD running a cheap stainless  steel beam pipe could be used. 
\section{Vertex detector \label{sec:vd}}

\begin{figure}[!t]
    \center{\includegraphics[width=0.8\linewidth]{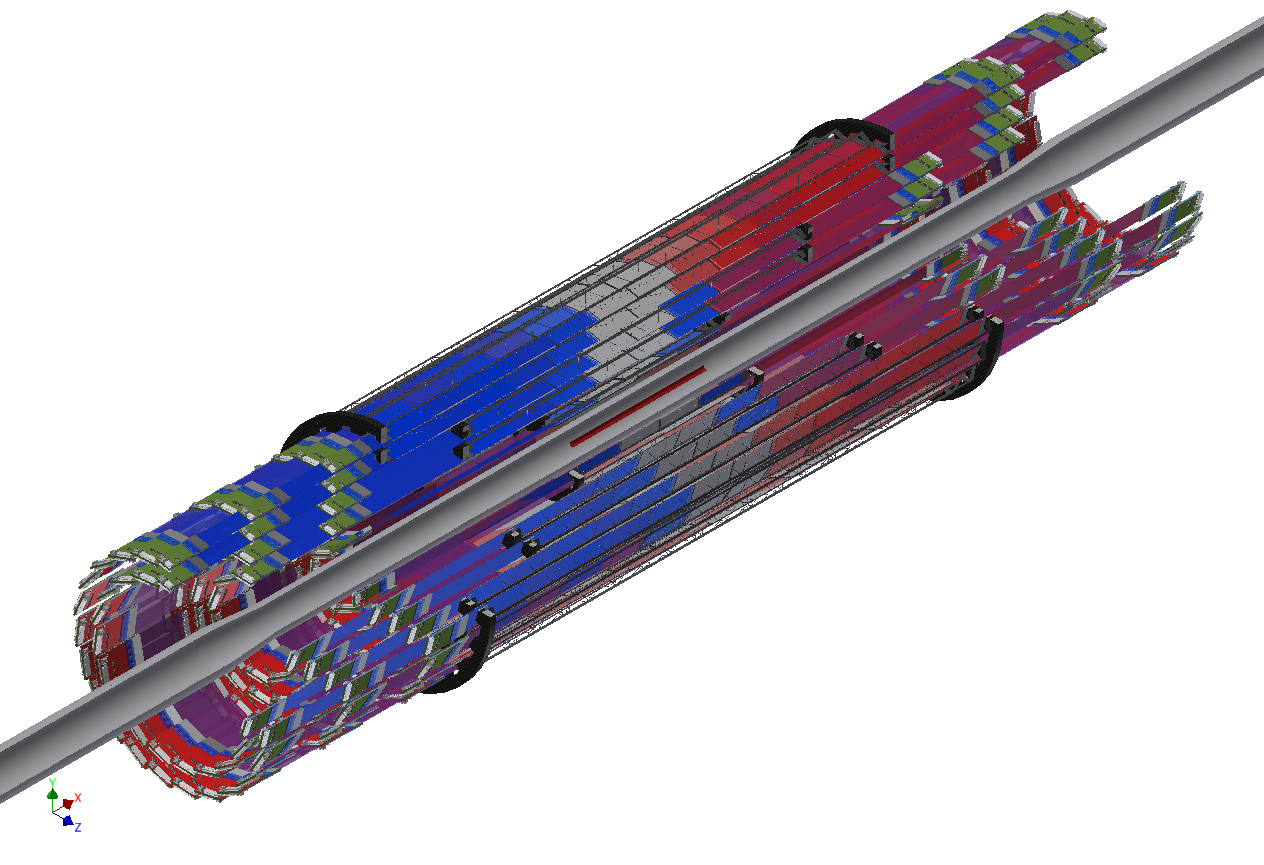}}
  \caption{General layout of the SPD Vertex Detector.}
  \label{fig:vd_0}  
\end{figure}

\subsection{General overview}
\begin{figure}[!h]
  \begin{minipage}[ht]{1.\linewidth}
    \center{\includegraphics[width=1\linewidth]{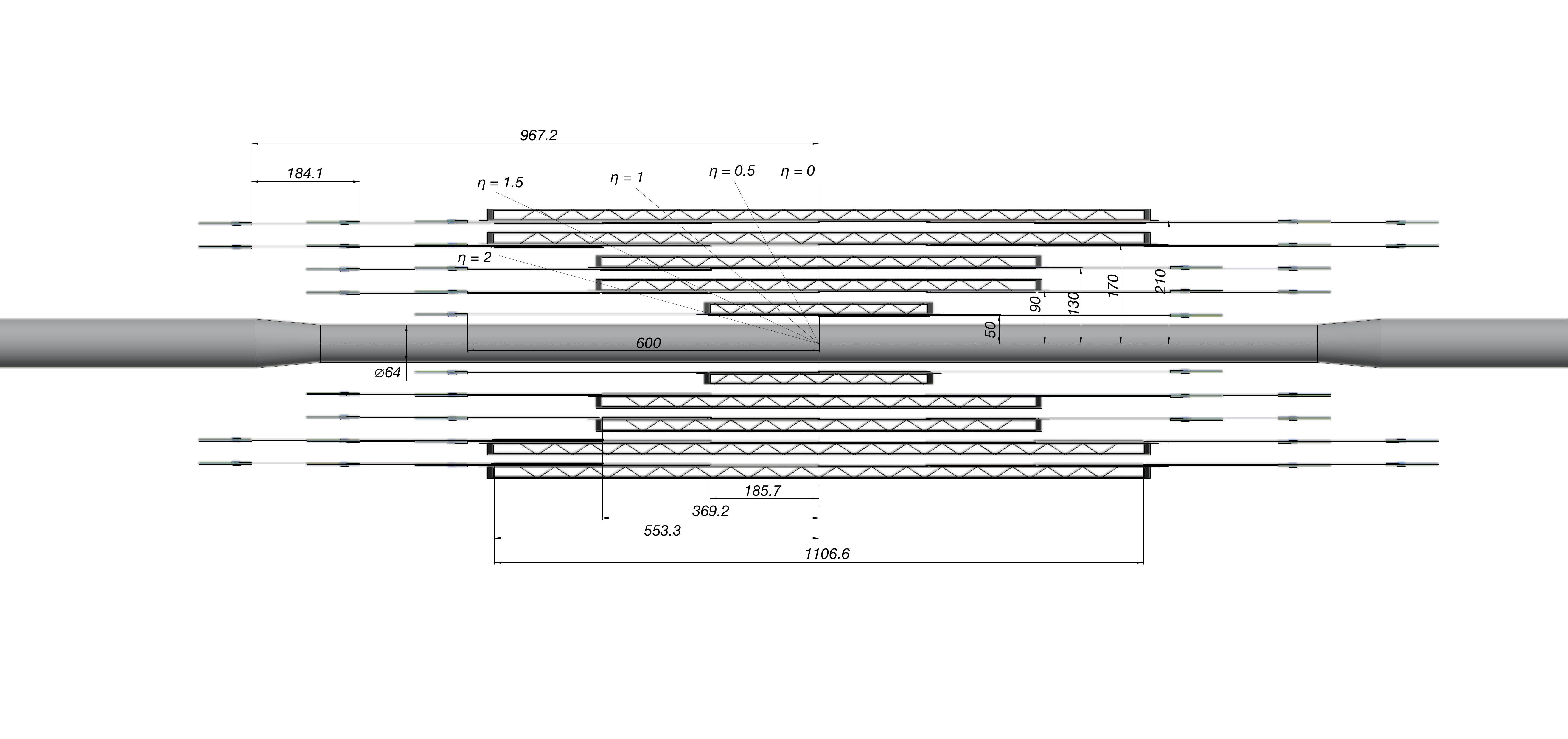} \\ (a)\\}
  \end{minipage}
  \hfill
  \begin{minipage}[ht]{0.99\linewidth}
    \center{\includegraphics[width=1\linewidth]{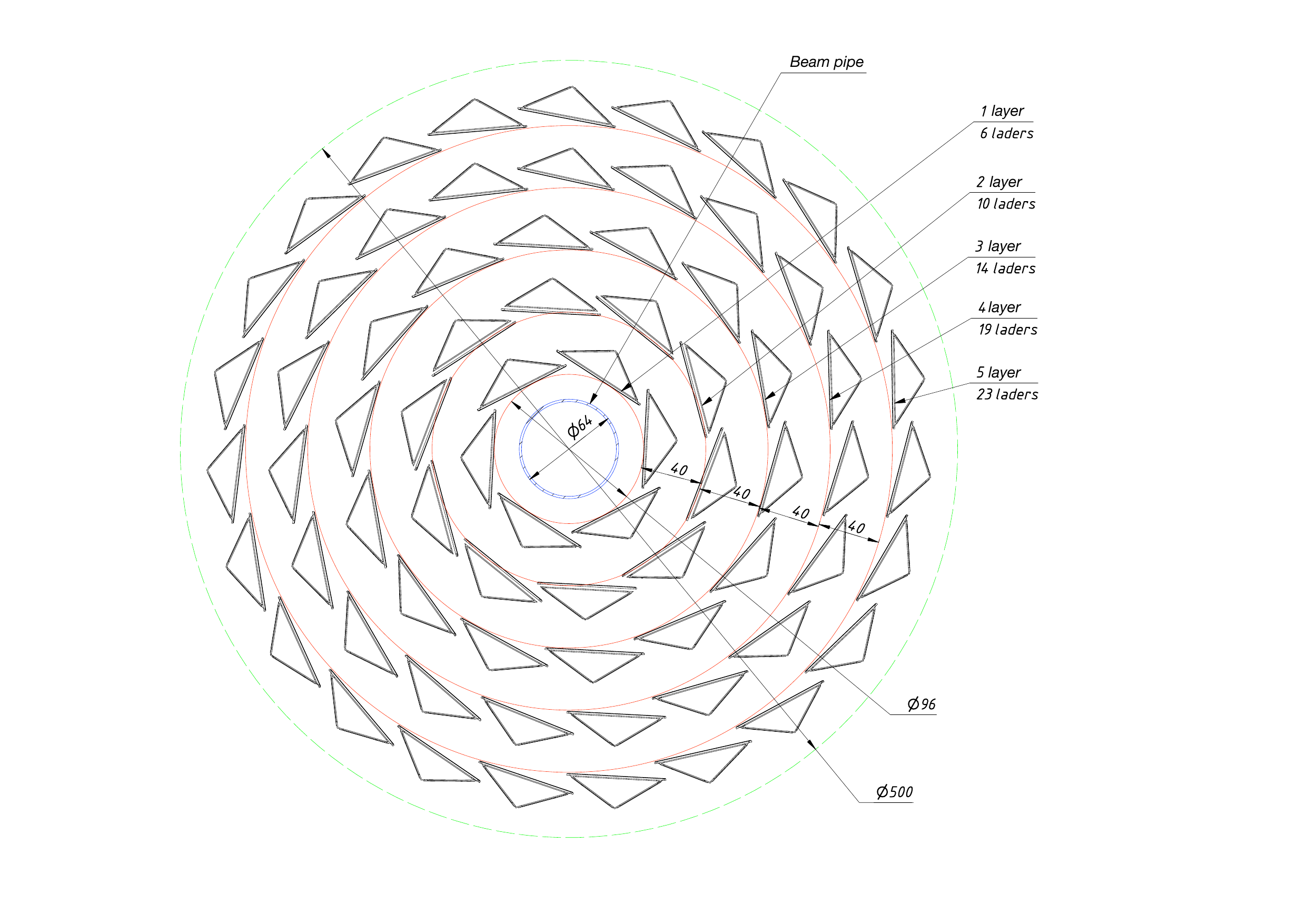} \\ (b)}
  \end{minipage}
  \caption{ Longitudinal (a) and transversal (b) cross-sections of the barrel part of the Vertex Detector. All dimensions are in millimeters.
}
  \label{fig:vd_1}  
\end{figure}

The SPD Vertex Detector (VD) is a silicon-based  part of the spectrometer responsible for  precise determination of the primary interaction point and  
measurement of  secondary vertices from the decays of short-lived particles (first of all, $D$-mesons).
The Vertex Detector is divided into the barrel and two end-cap parts (Fig. \ref{fig:vd_0}). Two different versions of the VD design are discussed: i) five layers based on double side silicon detectors (DSSDs) and ii) three inner layers based on Monolithic Active Pixel Sensors (MAPS) and two outer layers based on DSSDs. 
The VD Barrel consists of five layers based on double side silicon detectors (approximately 4.2 m$^2$). The end-cap regions consist of five disks each (approximately 0.4 m$^2$).  
The VD Barrel covers a radius from 96 to 500 mm (Fig. \ref{fig:vd_1}). All five cylindrical layers are set with rectangular two-coordinate silicon strip detectors and give information on the coordinates of the tracks ($r$,$\phi$,$z$) (which makes it possible  to measure a point in each layer).
The end-cup regions detect particles in the radial region between 96 mm and 500 mm. Each of the five disks is set with a DSSD with concentric ($r$) strips and radial ($\phi$) strips.
The VD has a length of about 1.1 m and covers the region of pseudo-rapidity up to $|\eta|<2.0$. Each DSSD has a 300-$\mu$m thickness and a strip pitch in the range from 95 $\mu$ m to 281.5 $\mu$m. The DSSDs are assembled into detector modules by two detectors per module, forming  18-cm long  strips. 
The detectors and the front-end electronics boards (FEE-PCB) are connected via low-mass polyimide microcables and assembled on the extra-light  carbon fiber mechanical supports with a cooling system in the similar way  as it was done for the ALICE outer  barrel (see Fig. \ref{fig:vd_4}). The relevant numbers for the barrel part of the VD for the DSSD configuration are presented in Tab. \ref{VD_pars}.

From  the general conditions of the SPD setup the VD performance requirements are i) geometry close to $4\pi$; ii) track reconstruction efficiency for muons greater than 99\% at $p\leq 13$ GeV/$c$ (for $0\leq |\eta|\leq 2.5$); iii) low material budget;  iv) coordinates resolutions for vertexing: $\sigma_{r,\phi}< 50~\mu$m, $\sigma_{z} < 100~\mu$m.
The lifetime of the Vertex Detector is required to be not less than 10 years of NICA running.

\begin{table}[!t]
\caption{Relevant numbers for the barrel VD (DSSD configuration).}
\begin{center}
\begin{tabular}{|l|c|c|c|c|c|c|}
\hline
Parameter & Layer 1 &  Layer 2 &  Layer 3 &  Layer 4 &  Layer 5 & Total \\
\hline
\hline
N$_{DSSD}$/module & 2 & 2 & 2 & 2 & 2 & \\ 
N$_{modules}$/ladder & 2 & 4 & 4 & 6 & 6 & \\ 
N$_{ladders}$/layer & 6 & 10 & 14 & 19 & 23 & 72 \\ 
N$_{DSSD}$/layer & 24 & 80 & 112 & 228 & 276  &  720\\ 
N$_{chip}$/module & 10 & 10 & 10 & 10 & 10 & \\ 
N$_{chip}$/layer & 120 & 400 & 560 & 1140 & 1380 & 3600 \\ 
N$_{channel}$/layer & 15360 & 51200 & 71680 & 145920 & 176640 & 460800 \\ 
\hline
\end{tabular}
\end{center}
\label{VD_pars}
\end{table}%

\begin{figure}[!h]
    \center{\includegraphics[width=0.8\linewidth]{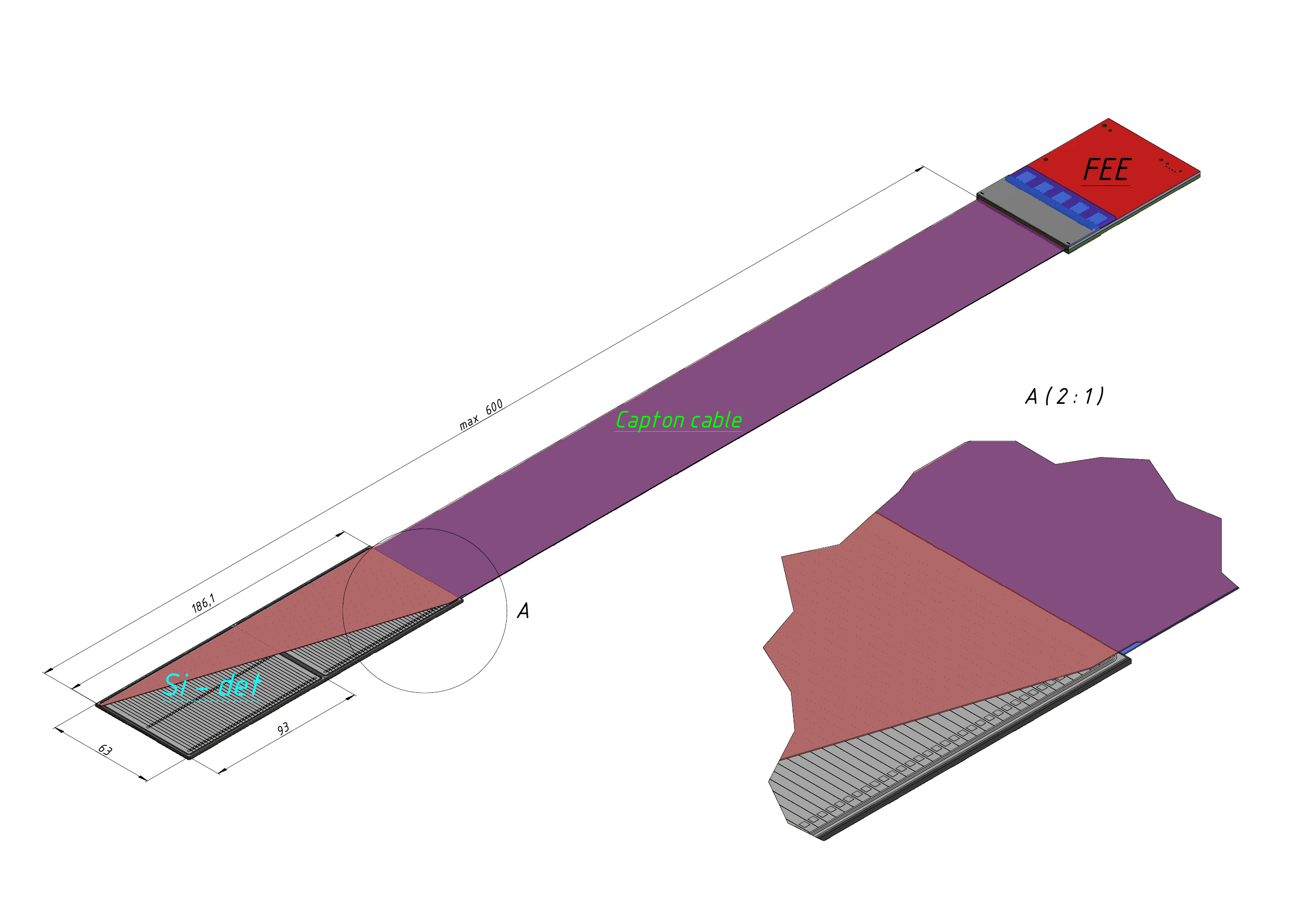}}
  \caption{Concept of the barrel DSSD module. All dimensions are in millimeters.}
  \label{fig:vd_2}  
\end{figure}
\begin{figure}[!bh]
    \center{\includegraphics[width=0.8\linewidth]{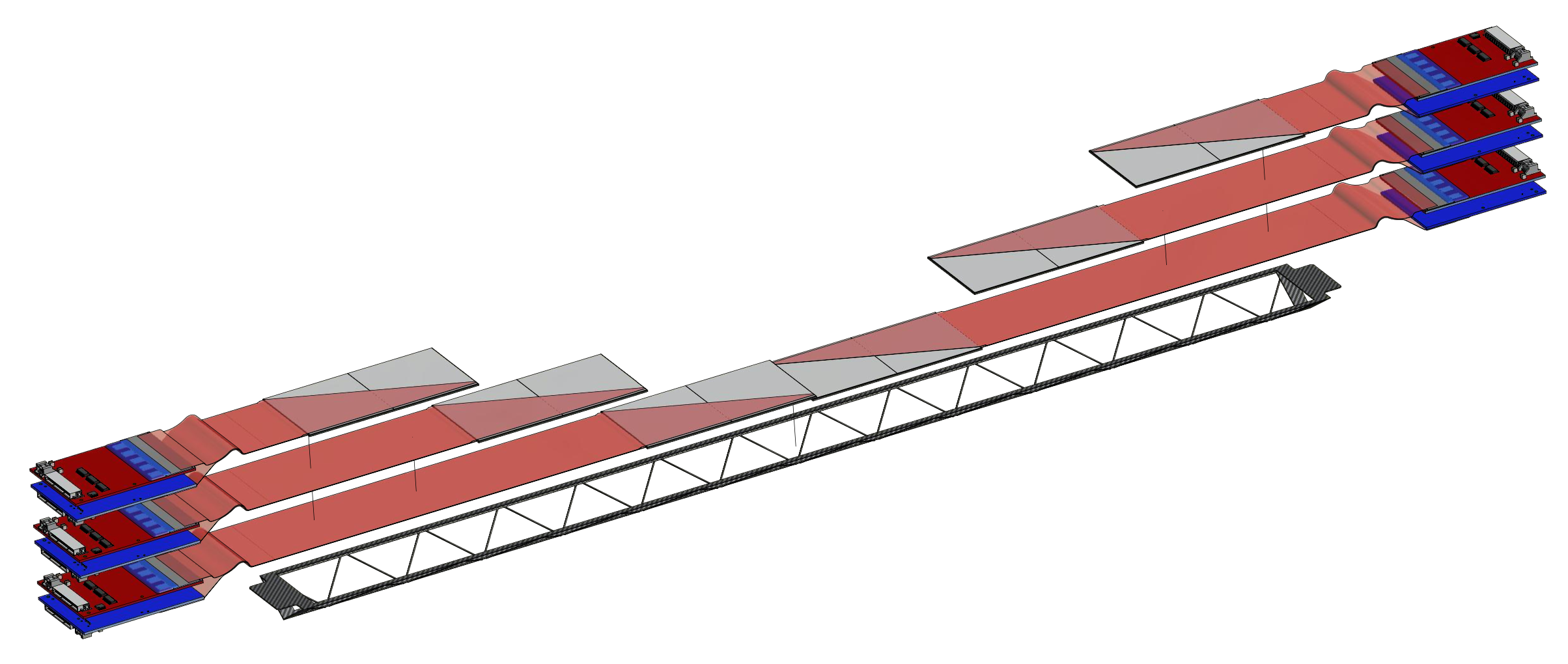}}
  \caption{Conceptual layout  of the barrel ladder.}
  \label{fig:vd_3}  
\end{figure}

\subsection{Double-sided silicon detectors}
The concept of the barrel DSSD module is shown in Fig. \ref{fig:vd_2}. The module consists of two silicon detectors wire bonded strip to strip for the $p+$ side (to reduce the number of readout channels), glued to the plastic frame and connected with two front-end electronic boards via a low-mass polyamide cable. 

The Silicon Detector is made using a planar double-side technology based on the n-type conductivity 6-inch float-zone Silicon wafers (produced by ZNTC, Zelenograd, Russia). Its size is 63x93 mm$^2$ and its thickness is 300 $\mu$m thickness. The pitch for the $p+$ side is 95 $\mu$m and for the $n+$ side 281.5 $\mu$m. The number of strips is 640 and  320 for the $n+$ and $p+$ side, respectively. The stereo angle between the strips is 90 degrees. The excepted spatial resolution for such a detector topology is $pitch_{p(n)+}/\sqrt{12}=$ 27.4 (81.26) $\mu$m for $r-\phi$ and $r-z$ projections, respectively. As mentioned before the barrel DSSD module contains two DSSDs ($p+$ strips wire bonded strip to strip) and has 640 strips at each side. 

To bring the front-end electronics out of the tracker volume,  two thin polyimide cables with aluminum traces (for each side of the module) will be used. The cable consists of several layers: signal, perforated or solid dielectric (polyimide), and a shielding layer. Cable pins were designed for the tape-automated  bonding with the detector and the pitch adapter sides. The maximum cable length is 60 cm, and the total thickness of all cable layers is less than 0.15\% of  $X_0$.  

Since the DSSDs have a  DC topology, it is necessary to supply bias voltage to the detector and electrically decouple the DC current from the ASICs electronics inputs. For this purpose, an integrated RC circuit (sapphire plates with Si-epitaxial layer Silicon On Insulator (SOI)) Pitch Adapter (PA) will be used for each side of the module (produced by ZNTC, Zelenograd) designed with different topologies for each side. 
 After the pitch adapter the detector signal goes to ASIC. Table \ref{Tab:VD} shows a possible ASIC readout solution. The optimal choice should be done after the ongoing R\&D.

\begin{table}[!t]
\caption{Possible ASIC readout solution for the Vertex Detector.}
\begin{center}
\begin{tabular}{|l|c|c|c|c|}
\hline
ASIC & APV25	& VATAGP7.3	& n-XYTER &	TIGER \\
\hline
\hline
Number of channels & 128 & 128 &128 & 64 (128) \\
Dynamic range         &   -40fC -- 40fC &	-30fC -- 30fC &	Input current 10 nA,   & 1--50fC \\
                                 &                         &                           &  polarity $+$ and $-$ & \\
Gain	                         &25mV/fC	          &20$\mu$A/fC	      &                                  &	10.35mV/fc  \\
Noise	                 &246 e$^{-}$+36 e$^-$/pF	&  70e$^-$+12 e$^-$/pF   &	900 e$^-$  at 30pF &	2000 e$^{-}$  at 100pF \\                        
Peaking time             &	50ns      &	50ns/500ns	         & 30ns/280ns &	60ns/170ns \\
Power consumption  &	1.15mW/ch. &	2.18mW/ch.	&10mW/ch. &	12mW/ch. \\
ADC                        &	No &	No	& 16fC, 5 bit &	10-bit \\
                                &             &         &                  &    Wilkinson ADC  \\          
TDC                        &	No &	No	& &	10-bit  \\
                                &             &         &                  &    Wilkinson ADC  \\    
\hline
\end{tabular}
\end{center}
\label{Tab:VD}
\end{table}%

\subsection{Mechanical layout}

\begin{figure}[!t]
    \center{\includegraphics[width=0.75\linewidth]{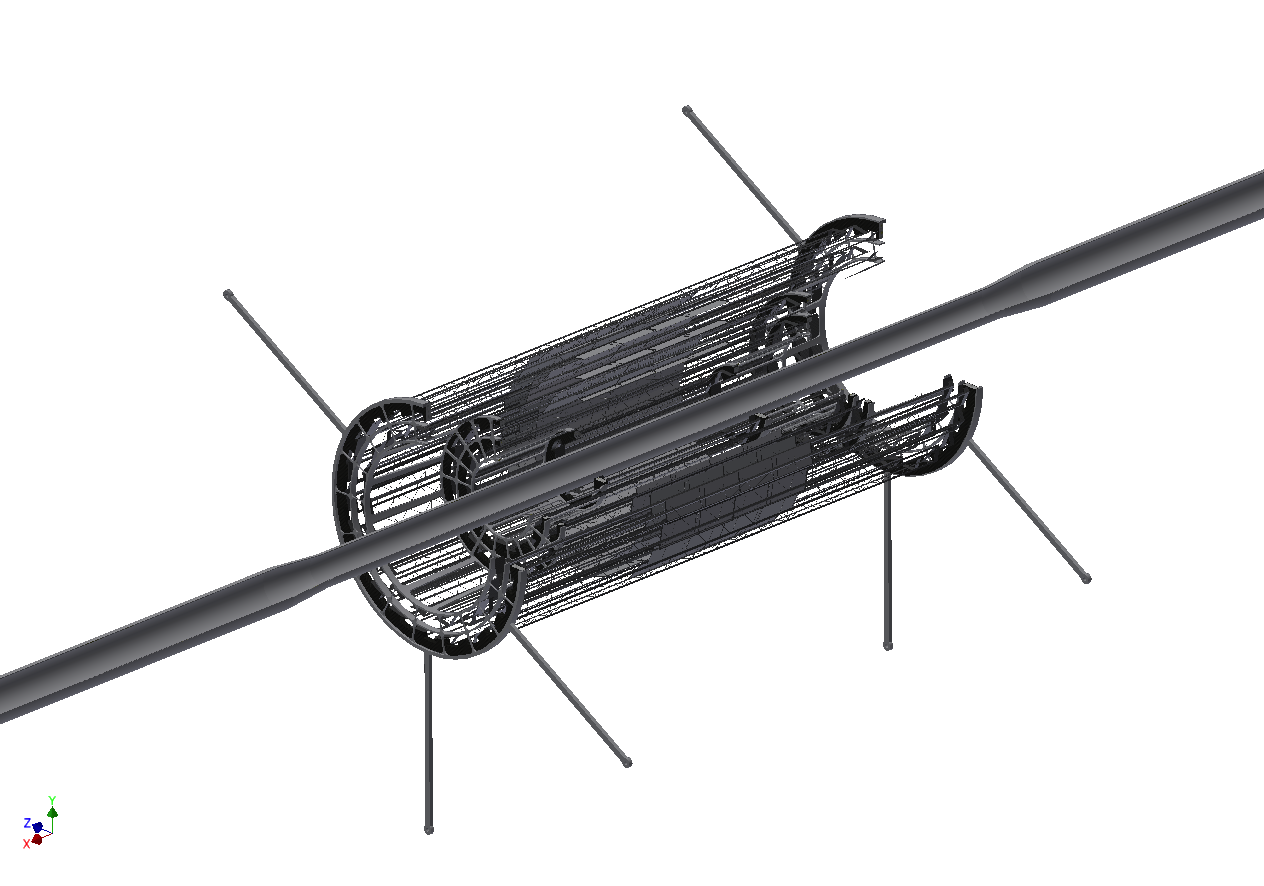}}
  \caption{Schematic exploded view and cross section of the Outer Barrel module similar to ALICE  outer barrel  design.}
  \label{fig:vd_4}  
\end{figure}

The concept of the barrel DSSD ladder is shown in Fig. \ref{fig:vd_3}. The silicon modules are laying on a carbon fiber support from center to edge. The detectors are connected with the FFE via thin low-mass cables. The front-end electronics is located at the edges of the ladder and is placed in the conical caves as shown in Fig. \ref{fig:vd_4} to provide a connection to the voltage supply, DAQ,  and the cooling ASIC chips subsystems.

\subsection{MAPS option}
\begin{figure}[!h]
    \center{\includegraphics[width=0.75\linewidth]{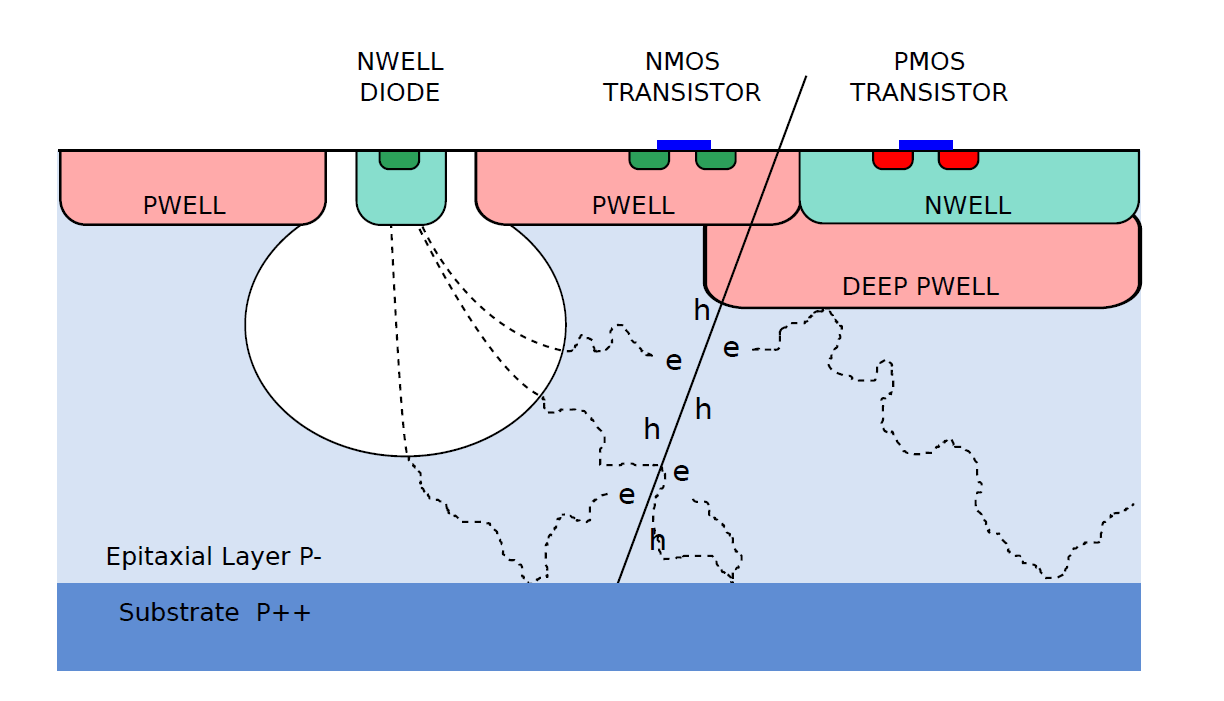}}
  \caption{Scheme of a MAPS pixel \cite{Abelevetal:2014dna}. }
  \label{fig:vd_5}  
\end{figure}

\begin{figure}[!h]
    \center{\includegraphics[width=0.75\linewidth]{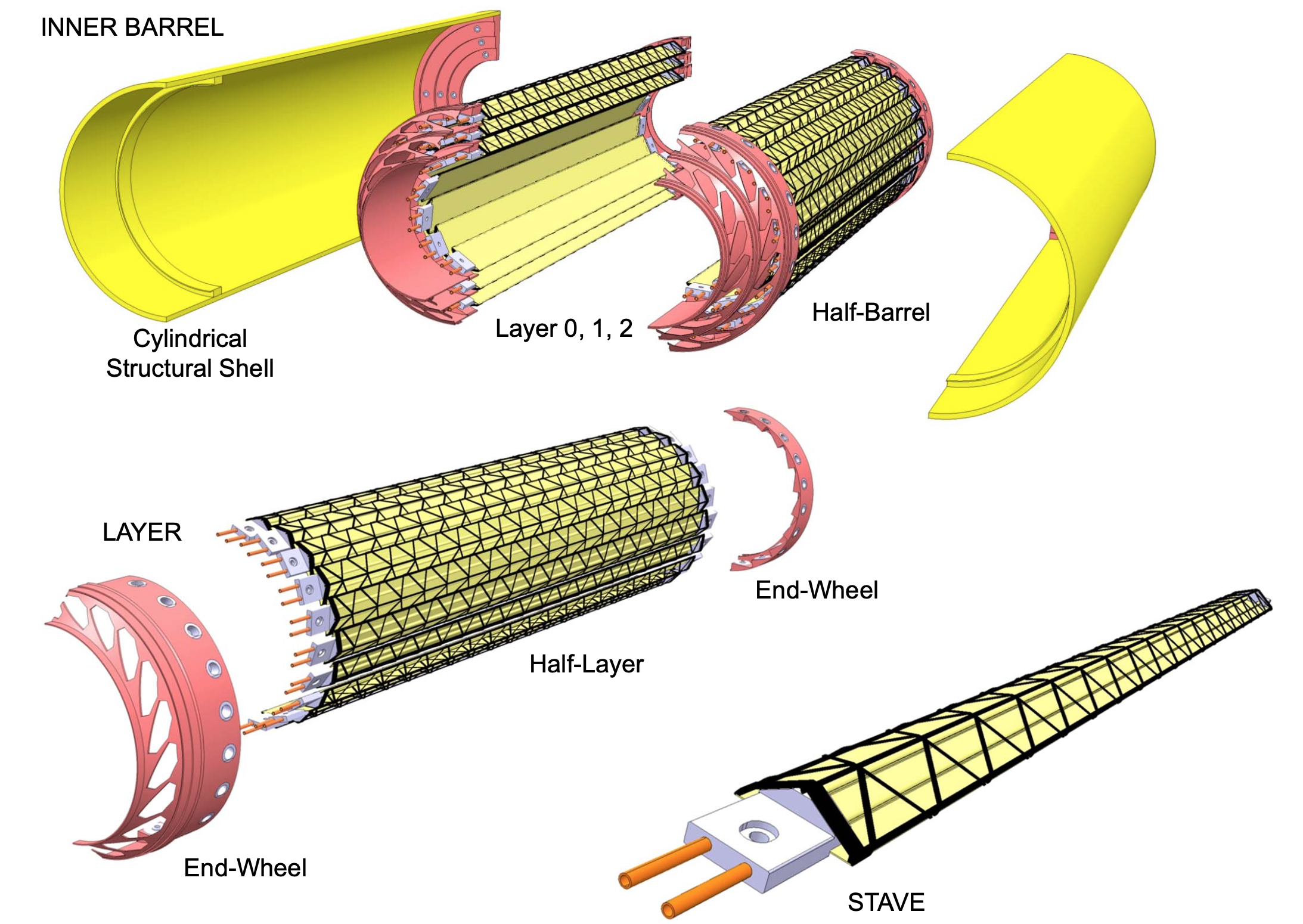}}
  \caption{ ALICE ITS2 Inner Barrel design \cite{Abelevetal:2014dna}.}
  \label{fig:vd_6}  
\end{figure}

	To improve the spatial resolution of the vertex reconstruction the two inner layers could be replaced by the MAPS-ALPIDE detectors, designed and produced for the ALICE experiment basing on the TowerJazz technology \cite{Abelevetal:2014dna}. Each chip has the pixel size of 29 $\mu m\times 27\mu m$. The scheme of a MAPS pixel is shown in Fig. \ref{fig:vd_5}. 
	One of the options of the  MAPS sensors  layout could be based on  the current ITS2 Inner Barrel design of ALICE \cite{Abelevetal:2014dna}, that has  achieved the record minimal level of material budget and provides  precise determination of  $D$-meson decay vertices, (see Fig. \ref{fig:vd_6}). 
	The impact of such replacement to the vertex reconstruction and, in turn, to the proposed physics with $D$-mesons is discussed in Chapter \ref{MC}. 
	The more advanced option like ALICE ITS3 \cite{Musa:2703140} could be also considered for the inner layers of SPD. 
 
\subsection{Cost estimate}
Preliminary cost estimation for the DSSD and DSSD+MAPS configurations of the Vertex Detector is presented in Tab. \ref{tab:costVD}. 
The total cost of the detecting elements in both cases is about 9.5 M\$. As for the FFE, very rough estimation for both variants gives 6.5 and 7 M\$, respectively. 

\begin{table}[htp]
\caption{Cost estimation for the VD configurations (DSSD / DSSD+MAPS).}
\begin{center}
\begin{tabular}{|c|c|c|c|c|c|c|}
\hline 
Layer  & Number of & Number of  & Barrel              & End-cup              & Cost            & Cost \\
           & sensors     & ASICs         & area, m$^2$    & area, m$^2$       & barrel, M\$  & end-cup, M\$ \\
\hline 
\hline
1         & 36 / 440    & 180 / 0       &  0.22 / 0.20     &  0.38 / 0.38     &    0.40 / 0.36  &  0.69 / 0.69\\
2         & 80 / 840    & 400 / 0       &  0.47 / 0.38     &                        &      0.86 / 0.69 &  \\
3         & 112 / 1736    & 560 / 0       &  0.66 / 0.78     &       &    1.20 / 1.42   &   \\
4         & 228 / 228    & 1140 / 1140       &  1.34 / 1.34     &      &    2.43 / 2.43   &  \\
5         & 368 / 368    & 1840 / 1840       &  2.16 / 2.16     &     &    3.92 / 3.92   &   \\
\hline
Total    & 824 / 3612 & 4120 / 2980      & 4.85 /  4.86     & 0.38 / 0.38  & 8.81 / 8.82 & 0.69 / 0.69 \\
\hline
\end{tabular}
\end{center}
\label{tab:costVD}
\end{table}%

\section{Straw tracker}

 The purpose of  Straw Tracker (ST)  is to reconstruct  tracks of  primary and secondary particles with high efficiency, 
  to measure their momenta with high precision based on a track curvature in a magnetic field, and participate in particle identification via energy deposition ($dE/dx$) measurement. 
 A spatial resolution of ST is expected to be about 150\,\textmu{}m and the drift time is about 120 ns for tubes of 1-cm diameter.
 The detector is planned to be built of low-mass straw tubes similar to the ones used in many modern experiments such us NA62 \cite{Sergi:2012twa}, COMET \cite{Nishiguchi:2017gei}, SHiP \cite{Anelli:2015pba}, Mu2e \cite{Lee:2016sdb}, COMPASS \cite{Bychkov:2002xw,Platzer:2005jp}, and NA64 \cite{Volkov:2019qhb}. The technology is quite well established and a detailed  R\&D is not needed. The concept of the SPD ST is similar to the ATLAS TRT \cite{Abat:2008zzb,Abat:2008zz,Abat:2008zza} and PANDA \cite{Gianotti:2013ipa} straw trackers.

\subsection{Straw technology}

The straw manufacturing process in general follows the procedure developed for NA62 \cite{Glonti:2018foc}.
The  tubes are manufactured of PET foil 36\,$\mu$m thick, coated on one side with two thin metal layers 
(0.05 $\mu$m of Cu and 0.02 $\mu$m of Au) in order to provide electrical conductivity of the cathode and to improve the gas (Ar+CO$_2$ mixtute) impermeability. 
NA62 has demonstrated that these straws can be operated in vacuum \cite{Sergi:2012twa}. 
A leak rate of only about 7 mbar/min for the whole detector (7168 straws) was measured \cite{Glonti:2018foc}. 

A few straws with a diameter of 10\,mm were used for dedicated mechanical tests. 
They were cut in 20 segments of about 25 cm long and tested under overpressure until the breaking point. 
The other straws were cut to 5.3 m and the cut ends were preserved for further analysis.
The breaking pressure was found to be 9~bar on average and no one sample broke under 8.5 bar. 
The quality control procedure was the same as for NA62 straws. 
During the ultrasonic welding process the seam quality was verified by a digital microscope (recorded to file for each straw). 
Furthermore, the seam quality was checked by an operator in real time. 
All 50 tubes produced so far have good seams. 

\begin{figure}[ht]
    \begin{minipage}{0.39\linewidth}
    \center{\includegraphics[width=1\linewidth]{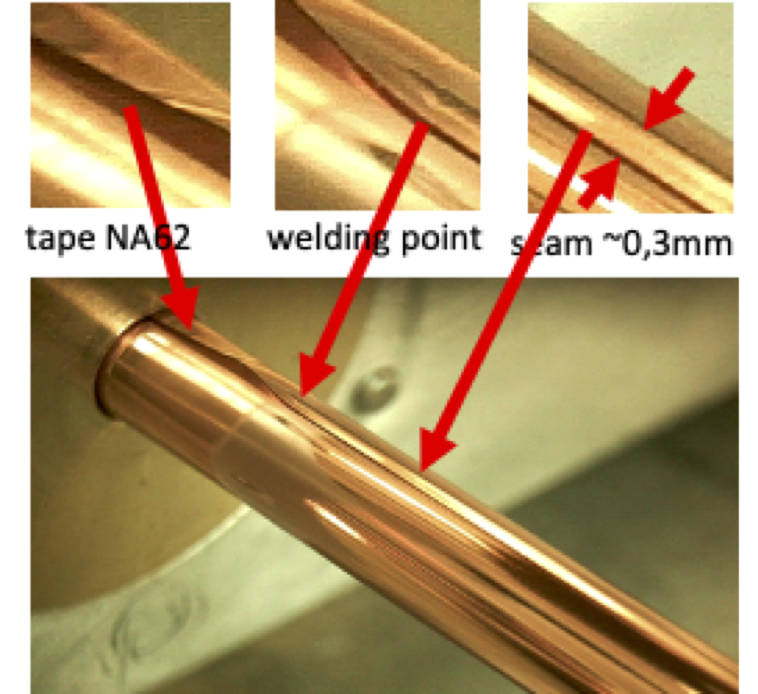} \\ (a)}
  \end{minipage}
  \hfill
  \begin{minipage}{0.59\linewidth}
    \center{\includegraphics[width=1\linewidth]{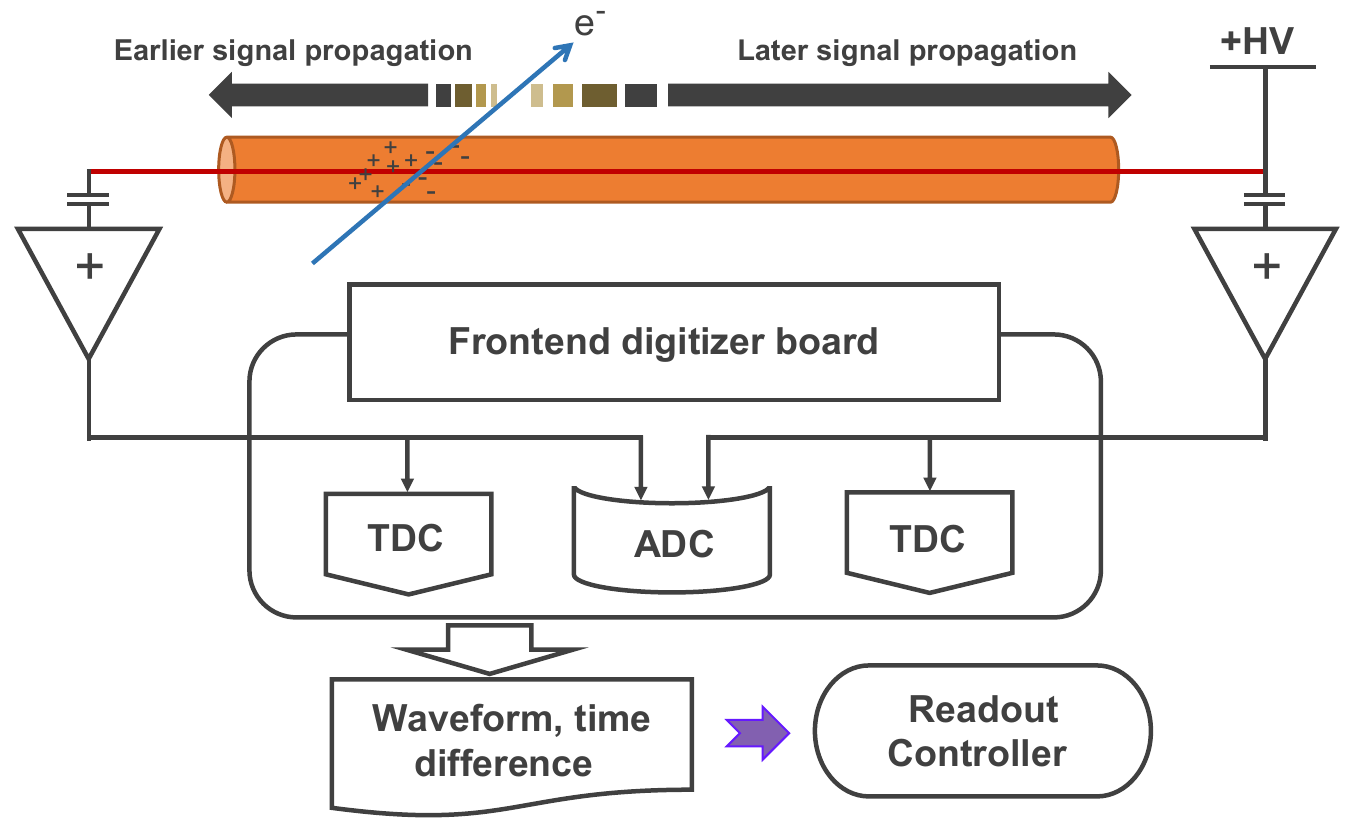} \\ (b)}
  \end{minipage}
\caption{Individual straw tube of the SPD ST (a).
	Schematic representation of a 2-side straw readout (b).}
\label{fig:straw1}
\end{figure}

Several measurements and tests are performed post-fabrication. 
The seam width and straw inner diameter are measured by an optical method. 
The cathode electrical DC resistance is measured. 
The elongation and breaking force are both measured on the test samples (cut straw ends). 
The straws undergo a long-term overpressure test with temporary end-plugs glued into both ends of each straw. 
An overpressure test to $\Delta P\approx2$ bar is performed for a period of about 1 hour. 
Subsequently, the straw is subjected to a long term overpressure test at $\Delta P\approx2$ bar for a period of at least 30 days. 
Gas leak estimation is obtained by measuring the loss of pressure over time. 
The local straw deformation is measured under an applied weight of 300 g, 
and the pressure is derived from the calibrated relation between loss of pressure and deformation. 
Design of an individual straw tube is shown in Fig.\,\ref{fig:straw1}\,(a).

\subsection{General layout of the ST}

The mechanical construction of the SPD Straw Tracker is based on engineering solutions which were already efficiently applied in ATLAS and PANDA experiments. The ST consists of three parts: a barrel part and two end-caps. 
The barrel  has the external radius of 850 mm and the internal hole with radius of 270~mm.  It is subdivided azimuthally into  8 modules, 
each with 30 double layers of straw tubes. 
Each tube has a diameter of 10\,mm.
Four modules are fixed together by a carbon fiber frame,
thus forming a pair of independent semi-cylinders.  
This design provides a possibility to assemble and disassemble the ST in the presence of the beam pipe. 
 
Each module is enclosed  in a 400-$\mu$m carbon fiber capsule. The capsule provides  the positioning of individual straw tubes with 50 $\mu$m accuracy. One side and two ends of the capsule have 5\,mm holes where straw end-plugs will be fixed. 
FE electronic boards which will be connected to these plugs will additionally serve as capsule covers, 
thereby isolating the inner volume from the external environment.
Long straws oriented along the beam line will be read out from both ends in order to obtain an additional coordinate along the straw axis via mid-time calculation as shown in Fig.\,\ref{fig:straw1}\,(b). 
While short straws oriented perpendicular to the beam line will have electronics only on one side.
Each capsule contains about 1500 tubes with parallel and 6000 tubes with perpendicular to the beam orientations.
The total number of electronic channels per capsule is  9000. Thus, the total number of channels in the barrel part of  ST is 72 000.

\begin{figure}[!t]
\centerline{\includegraphics[width=1.\linewidth]{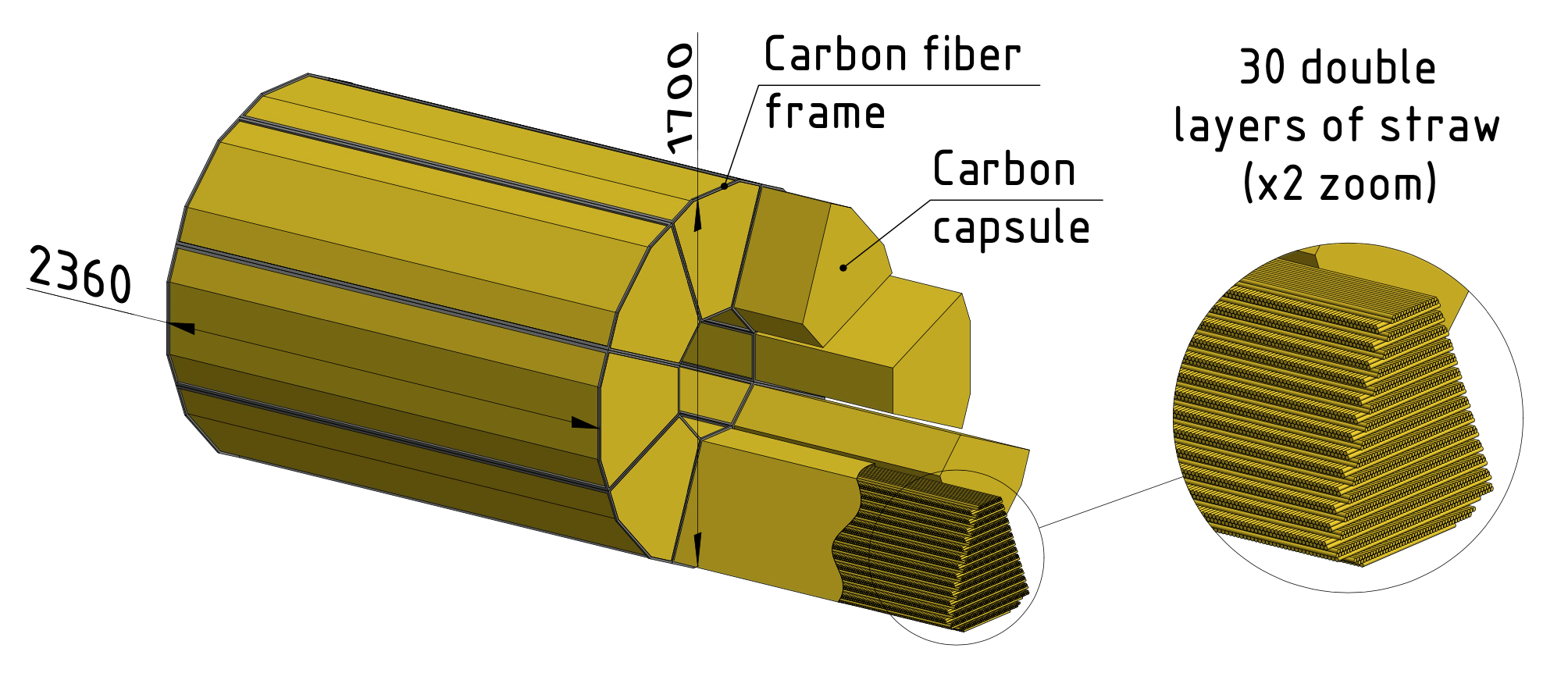}}
\caption{Layout of the barrel part of  ST which shows 8 modules with 30 double-layers of straw in each capsule.
	Straws of adjacent double layers oriented perpendicular to each other. All dimensions are in millimeters.}
\label{fig:Straw2}
\end{figure}

The rigidity of the structure is assured by the low overpressure of gas inside the tubes and their fixation inside the capsule volume. 
The anode wire positioning accuracy is achieved by the wire fixation in the carbon fiber covers. 
The capsule also serves for thermostabilization of the gas mixture inside detector volume and for protection of the straw surface from humidity. 
The layout of the barrel part of the Straw Tracker is shown in Figure\,\ref{fig:Straw2}. 

\begin{figure}[!h]
\centerline{\includegraphics[width=1.\linewidth]{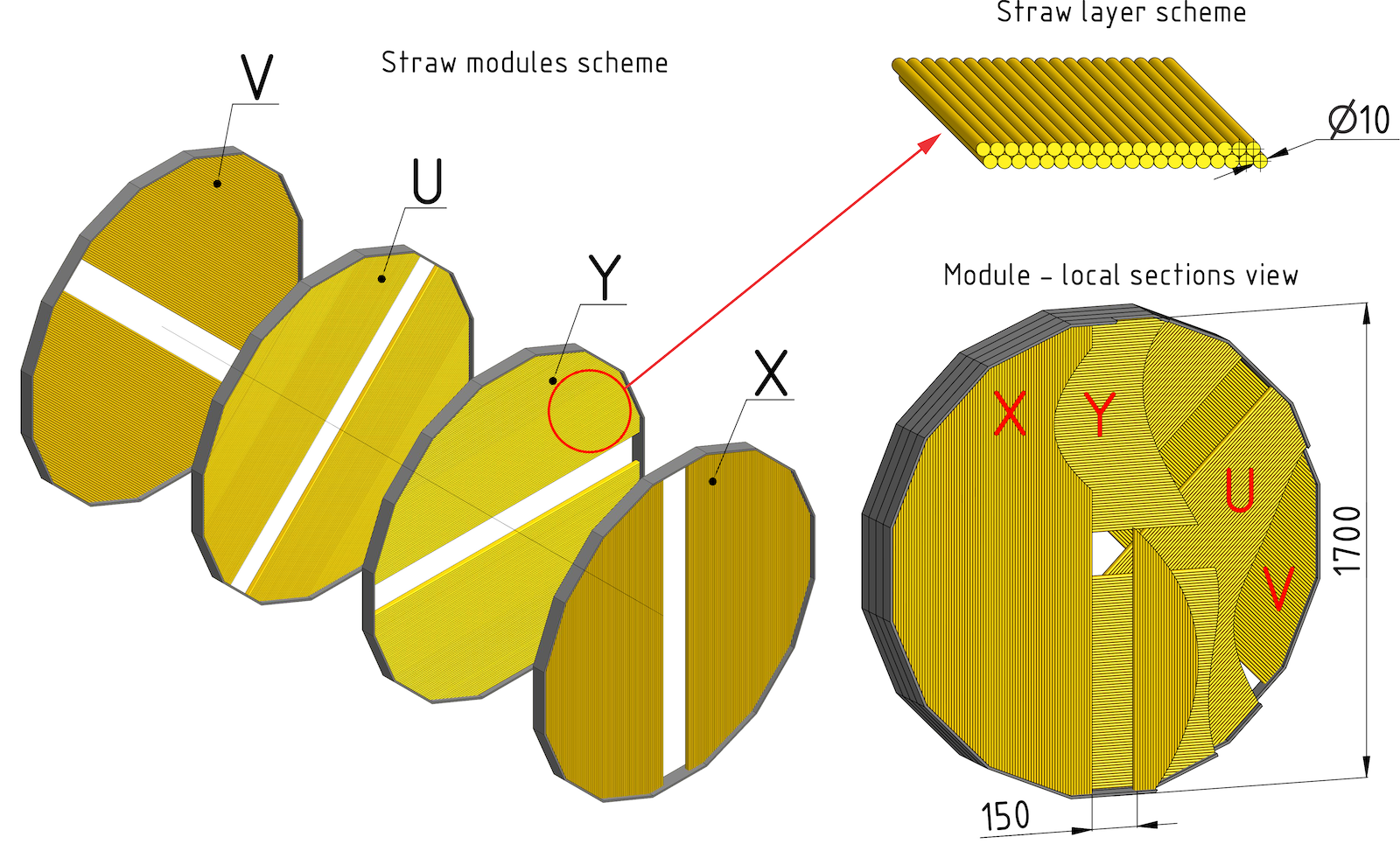}}
\caption{Layout of one end-cap part of the ST.
It includes 3 modules along the beam axis where every module consists of 4 identical hexadecimal cameras (X, Y, U and V). All dimensions are in millimeters.}
\label{fig:Straw3}
\end{figure}

Each end-cap part of  ST comprises 3 modules along the beam axis.  
Each module consists of 4 identical hexadecimal cameras (for measurement of the 4 coordinates: X, Y, U and V). 
By construction, the cameras are divided into halves. Each module has a technological hole $\O=160$\,mm for the beam pipe.   
The mechanical support is provided by  carbon fiber frames.  The total amount of electronic channels  in both end-caps is 7200.
The layout of one straw end-cap  is shown in Fig. \ref{fig:Straw3}.

\subsection{ Front-end electronics}

 The Straw Tracker is designed for precision measurements which require excellent spatial, angular and timing
resolutions to meet physics goals. 
Moreover, an amount of charge collected by anode will be used to identify particles.
In view of this, requirements for the straw readout electronics are the following:
\begin{itemize}
\item to measure time and energy deposit ($dE/dx$);
\item time resolution not worse than 1\,ns for the drift measurements
and 0.1\,ns for the case of the 2 side readout to determine the coordinate along the wire;
\item low threshold to identify a charge from primary electron-ion clusters;
\item dynamic range of about 1000;
\item low power consumption to reduce heating.
\end{itemize}

Two options of electronics are under consideration now. The first one is the front-end electronics designed for the
NA64 experiment \cite{Volkov:2019qhb}. It is a 32-channel amplifier-discriminator board based on AST-1-1 chip, developed by
the Institute of Nuclear Problems of the Belarusian State University. The amplifier sensitivity is K=100~mV/$\mu$A 
(20 mV/fC). The discriminator threshold is adjustable in the range (2$\div$20)~fC. The delay of LVDS output
signal is 6 ns. The amplifier has an ion tail compensation (BLR). The LVDS output signals are sent to the
64-channel time-to-digital converters (TDC).

The second option is to take the front-end electronics designed for DUNE experiment \cite{Acciarri:2016ooe}, 
which is based on 64 channel VMM3 a custom Application Specific Integrated Circuit (ASIC), developed by BNL for the LHC experiments at CERN.  
A low power consumption and a low per-channel cost (about 0.9 \$/ch) of the chip are valuable futures for a compact multichannel detector readout. Each channel has ADC and TDC circuits. Fast serial outputs are used for readout.
 Each of the 64 ASIC channels is highly configurable and combines a preamplifier shaping circuit with an ADC to allow independent digitization of triggered input signals. These digitized signals can be output with four different data readout options, which provides flexibility to accommodate different detector requirements and data rates.
Each input channel has an individual preamplifier and dedicated digitizing logic. Each channel can be configured to accommodate a variety of input signal sizes, polarity and capacitance. The preamplifier shaping circuit can be configured to use one of four different peaking times (25, 50, 100, and 200 ns) and eight gain settings (0.5, 1, 3, 4.5, 6, 9, 12, 16 mV/fC). A channel-specific discriminator triggers on input signals above a configurable threshold to initiate digitization of the amplified pulse with a 10-bit Analogue to Digital Converter (ADC). Discriminator thresholds are adjusted by a global 10-bit Digital to Analogue Converter (DAC) with additional channel-specific 5-bit trimming DACs. These features allow the VMM3 to satisfy the SPD TR requirement of measuring the collected charge and signal time in each channel.
An equivalent noise charge of better than
$1000e^-$ can be achieved with input capacitance less than 100~pF. 
On the basis of this performance, it is reasonable to expect that VMM3
can meet the SPD TR requirement of low charge threshold for the straw tube gain greater than
$G=10^4$ and the input capacitance less than 100~pF.

It was shown that the time resolution better than 1 ns was obtained for 6 pF input capacitance and input charge greater than 1 fC.
A much better time resolution can be achieved  for higher input signal amplitudes. 
This suggests that VMM3 can satisfy the SPD TR requirements for the time resolution 
with sufficiently high gain and appropriate input capacitance. 
The channel thresholds are individually adjusted by a global 10-bit DAC and an individual
channel 5-bit trimming DAC. This suggests that the required low charge threshold can be achieved.

\subsection{Cost estimate}
Presently, we estimate the cost of the SPD ST development and construction at 2.4 M\$.

\section{Electromagnetic calorimeter}

The calorimeter should meet the criteria imposed by the physical goals of the SPD experiment of different nature and importance. The most important criteria arise from the physical requirements to the accuracies of measurement of energies, trajectories, and timings of photons and electrons. Technological possibilities of modern experimental physics should be taken into account when choosing the calorimeter setup. Price factors should also be considered to ensure the feasibility of the project. High multiplicity of secondary particles leads to a requirement of high segmentation and dense absorber medium with a small Molière radius. It is needed in order to have sufficient spacial resolution and a possibility to separate overlapping showers.  The transverse size of the calorimeter cell should be of the order of the Molière radius. A reliable reconstruction of photons and neutral pions is possible only for small shower overlaps. Occupancy should not exceed 5\%, so that it is possible to determine photon reconstruction efficiency with high precision. 

The SPD experiment imposes the following requirements on the calorimeter characteristics:
\begin{enumerate}
 \item reconstruction of photons and electrons in the energy range from 50 MeV to 10 GeV;
\item  energy resolution for the above-mentioned particles: $ \sim 5\%/\sqrt{E} [\textrm{GeV}] $; 
\item  good separation of two-particle showers; 
\item  operation in the magnetic field;
\item  long-term stability: 2$\div$3\% in a six month period of data taking.
\end{enumerate}

The energy range requirement follows from the kinematic range of secondary particles, which are produced in a collision of protons with energy up to 27 GeV and emitted into $4\pi$ solid angle. Good energy resolution is required for identification and quantitative measurement of single photon and neutral pion energies. Good two-particle separation is needed to separate photon showers from the $\pi^0$ decay in order to suppress background events in measurements with prompt photons. Long-term stability is necessary for polarization measurements featuring $\pi^0$ reconstruction in the calorimeter, especially in the end-caps. Calorimeter instability may result in false asymmetry values. While it is essential to meet the physics requirements imposed on the calorimeter design, one should also take into account the cost estimate and the technical feasibility when choosing its granularity, as the larger number of cells leads to larger costs of the manufacturing technology and readout electronics.

\subsection{Overview of the SPD calorimeter}

The SPD electromagnetic calorimeter is placed between the Range System  and the magnet coils, as shown in Figs.\ref{SPD-LA_EX} and \ref{fig:spd1}. 
It consists of a barrel and two end-caps, covering a $4\pi$ solid angle. The outer dimensions of the calorimeter are determined by the inner size of the muon system. The thickness of the calorimeter is determined by the required thickness of the active part and the size of the readout block consisting of photodiode and amplifier boards, as well as by the size of the flexible part of the fibers. 

For efficient absorption of electrons and photons with energies up to 10 GeV, the calorimeter thickness, which is defined by the number of sampling layers, should be at least 18$\div$20 X$_0$ in terms of radiation lengths X$_0$.  For the sampling structure of  a 1.5-mm scintillator and 0.5-mm lead, 200 layers are required for a thickness of 18.6 X$_0$, which sets the length of the active part to 400 mm. The period of the structure is set to 2 mm in order to avoid optical contact between the lead and the scintillator, and because of the connection technique involving special ''Lego'' spikes. The flexible parts of the fibers take up 8 cm. The transverse size of the calorimeter cell should be of the order of the effective Molière radius of the calorimeter medium, which is, in its turn, defined by the scintillator-to-lead sampling ratio. The selected structure has a Molière radius of 2.4 cm. The separation efficiency of two photons with energies from 200 MeV to 500 MeV depends on the cell size and reaches a plateau at a cell size of 40 mm, as was determined in the MC simulation. Therefore, we have selected 40 mm cell granularity for both barrel and end-caps. Cells in the barrel part of the calorimeter have trapezoidal shape in the azimuthal direction to minimize the gaps between the modules. The vertex angle of the trapezoid equals  1.58$^{\circ}$.

\begin{figure}[!t]
    \begin{minipage}{1\linewidth}
    \center{\includegraphics[width=0.7\linewidth]{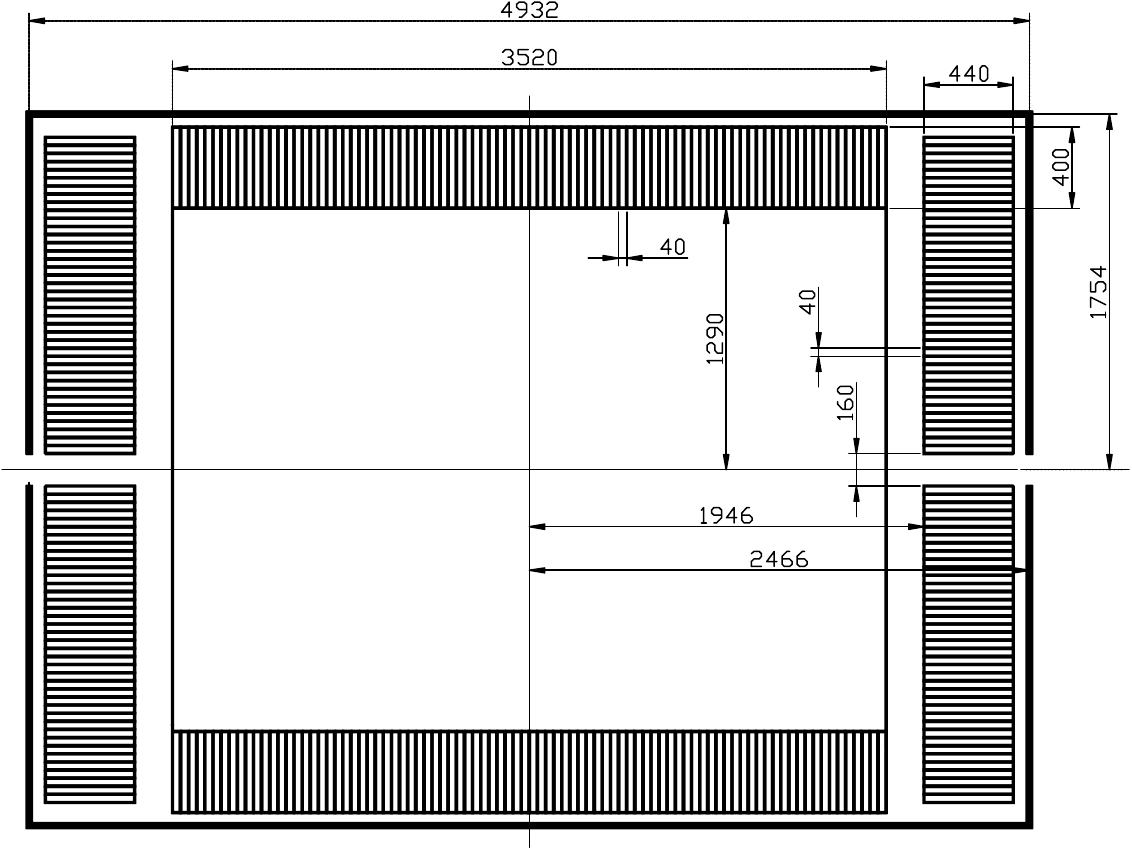} \\ (a)}
  \end{minipage}
  \hfill
  \begin{minipage}{1\linewidth}
    \center{\includegraphics[width=0.7\linewidth]{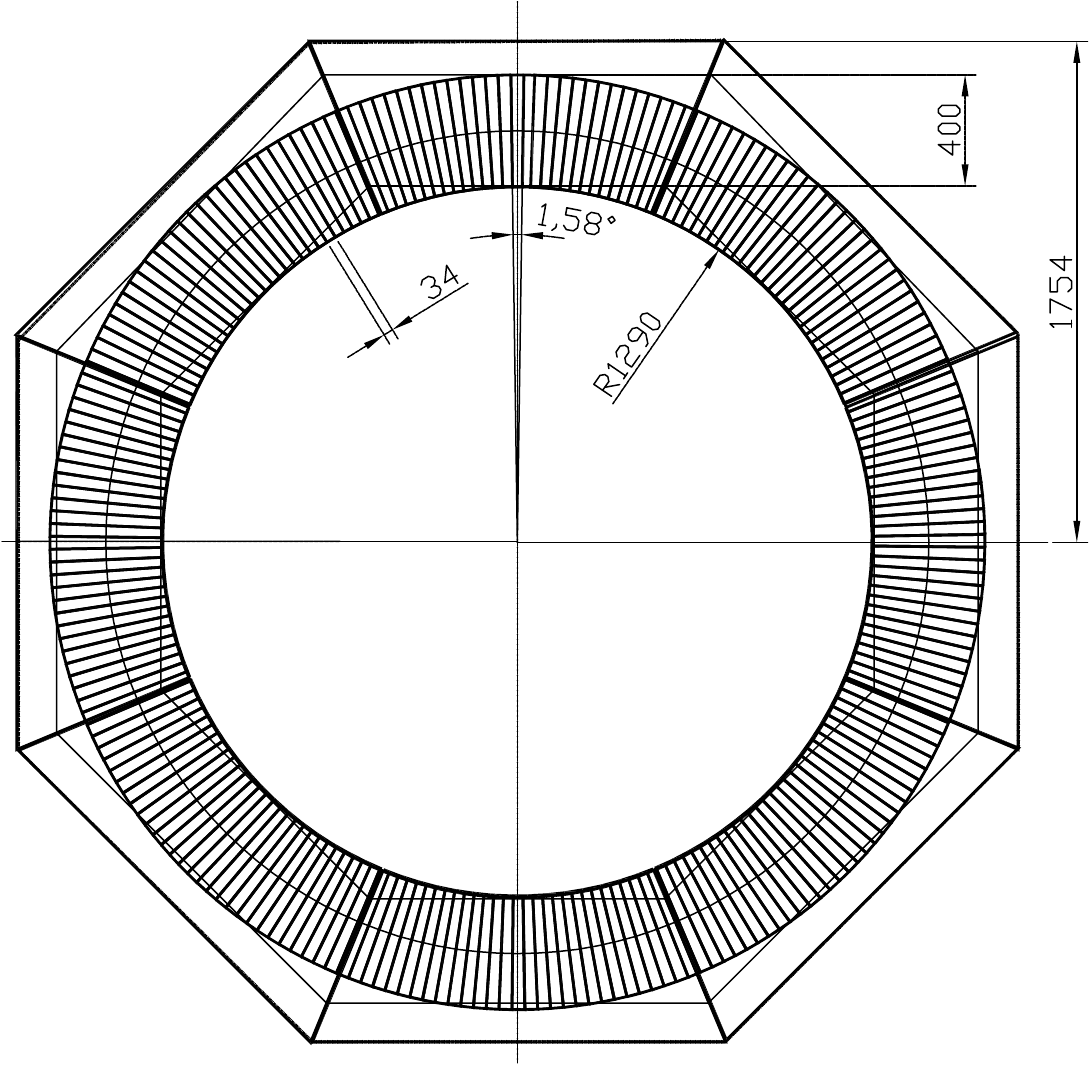} \\ (b)}
  \end{minipage}
  \caption{(a) Barrel and end-cap parts of the calorimeter. The holes of size $160\times160$ $\textrm{mm}^2$ for the beam pipe can be seen in the centers of the end-caps. (b) Schematic drawing of a cross-section of the barrel part of the calorimeter. It is sectioned into 224 azimuthal sectors (8 sections, 28 cells per section) with vertex a angle of 1.58$^{\circ}$. All dimensions are in millimeters. }
  \label{ECAL_G_ecal_full_barrel_endcap}  
\end{figure}

A schematic drawing of the calorimeter, which is limited in size by the muon system, is shown in Fig.\ref{ECAL_G_ecal_full_barrel_endcap}(a). The limits of the calorimeter zone are shown as a thick line. Holes of the size $160\times160$ $\textrm{mm}^2$ in the centers of the end-caps for the beam pipe are shown.

The inner size of the barrel part is limited by radius of the magnetic coils, whereas the outer size is limited by the dimensions of the muon range system. The thickness of the active part is 400 mm, which corresponds to 18.6 X$_0$.  This corresponds to 200 layers of the scintillator and lead of 1.5 mm and 0.5 mm width.

The barrel part of the calorimeter has 19712 cells of trapezoidal shape in the azimuthal direction with a vertex angle of 1.58$^{\circ}$, and a front face size of 34 mm, and rectangular shape in the direction along the beam axis with a size of 40 mm (Fig.\ref{ECAL_G_ecal_full_barrel_endcap}(b)). The total weight of the barrel part is 40 tons.

\begin{figure}[!t]
\centerline{\includegraphics[width=0.7\linewidth]{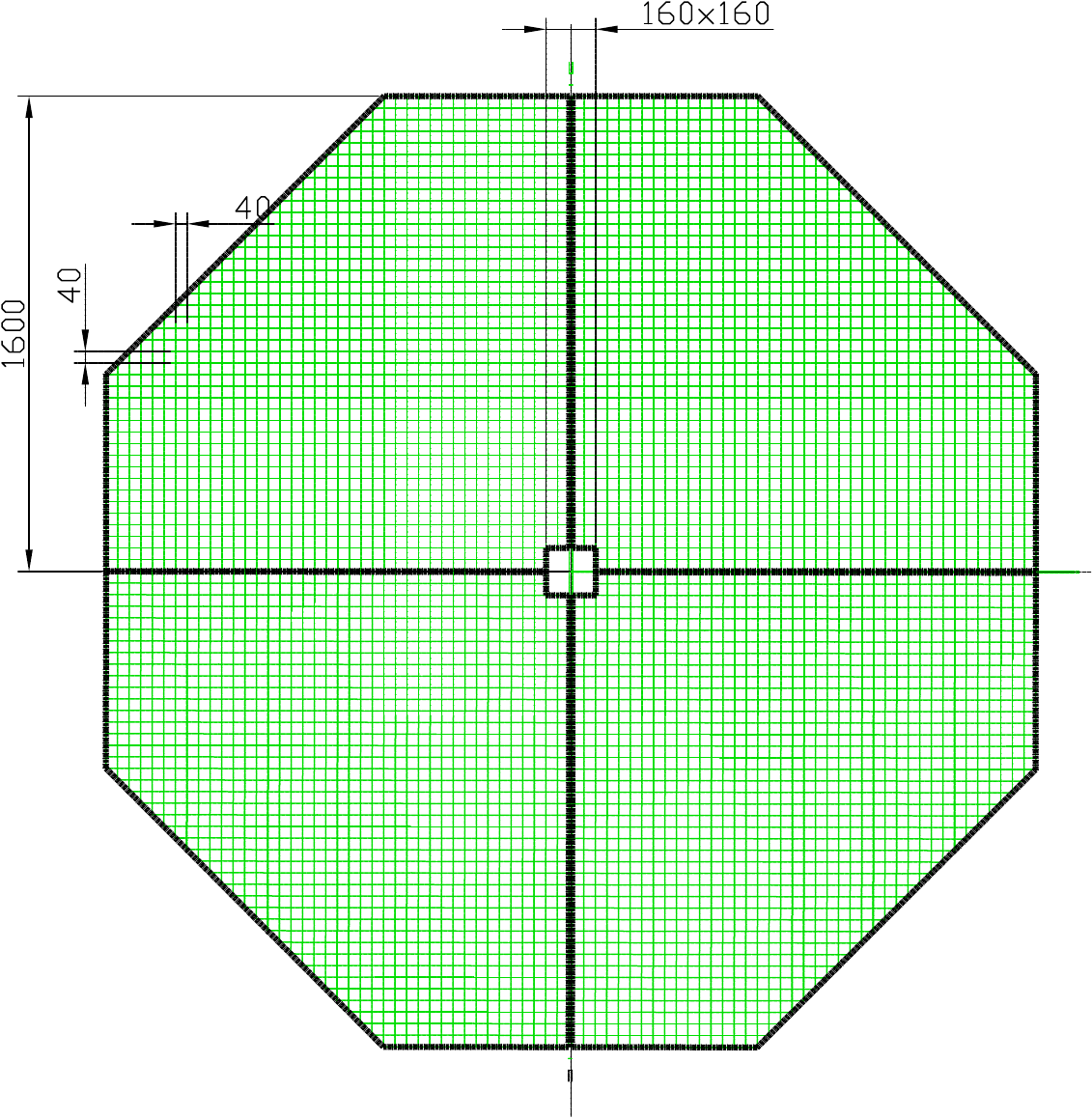}}
\caption{The end-cap part of the calorimeter consists of 4 sectors, 1308 cells each.
In total, there are 5232 cells in one end-cap, and 10464 cells in both end-caps. All dimensions are in millimeters.}
\label{ECAL_G_ecal_full_endcap}
\end{figure}

Each end-cap (one is shown in Fig.\ref{ECAL_G_ecal_full_endcap}) consists of 4 sectors of 1308 cells per sector. The cell cross-section is $40\times40$ $\textrm{mm}^2$.  There is a hole for the beam pipe in the center of each end-cap. The hole has a size of $160\times160$ $\textrm{mm}^2$, which is equivalent to 16 cells. Each end-cap has 5232 cells. The thickness of the active part of an end-cap cell is 440 mm (Fig.\ref{ECAL_G_ecal_full_barrel_endcap}), which corresponds to 20.4 X$_0$.
The weight of one end-cap is 14 tons.  The total weight of two end-caps is, therefore, 28 tons. In total, there are 10464 cells in both end-caps, each with dimensions of $40\times40\times440$ mm$^3$.

The total weight of the calorimeter is 68 = 40 + 28 tons composed of the barrel and two end-caps.  The total number of cells of size about $40\times40$ $\textrm{mm}^2$ is 30176 = 19712 + 10464 for the barrel and the end-caps, respectively.

A possibility to use another type of modules in the central part of the end-cups in order to improve energy and spatial resolution of the calorimeter for energetic photons is also under discussion.

\subsection{Design of the calorimeter module prototype}

The initial version of the module, which was made for testing purposes, consisted of alternating layers of polystyrene scintillator and lead with a thickness of 1.5 mm and 0.3 mm, respectively. The selected number of layers is 220, setting the number of radiation lengths to 12.6$X_0$.  The lead plates are intended to absorb the particle energy and develop an electromagnetic shower, whereas the scintillator plates produce an amount of light proportional to the energy of particles.  The properties of the absorber and the scintillator define the Molière radius, which is equal to 3.5 cm for the selected structure. The energy resolution for 1 GeV photons is assumed to depend on the calorimeter sampling fraction and is expected to be 4.15\%. The test results of the present work are given for this particular design of the module.

The scintillator plates are made of polystyrene beads with an added luminophore admixture of 1.5\% p-Terphenyl and 0.05\% POPOP ($\textrm{C}_{24}\textrm{H}_{16}\textrm{N}_2\textrm{O}_2$) \cite{ECAL_G_scint_IHEP}. It has scintillation time of about 2.5 ns and light output of 60\% of anthracene, which are good results.  The radiation hardness of the scintillator is sufficient for radiation doses up to about 10 Mrad, which is important for operating the calorimeter in the radiation field of secondary particles in the vicinity of the interaction point.

The luminophore admixtures re-emit the energy of excitations in polystyrene in the form of visible light.  The first admixture (p-Terphenyl) emits light with a wavelength of maximum emission at 340 nm. This light is absorbed by the second admixture (POPOP) and is re-emitted into a spectrum with a wavelength of maximum emission of 420 nm, which is seen as a light blue glow.

The light from the scintillator plates is gathered using wavelength shifting fibers (WLS) \cite{ECAL_G_wls}. Fibers of type Y-11(200) manufactured by KURARAY are used.  The fibers absorb the light from the POPOP and re-emit it into a spectrum with a wavelength of maximum emission of 490 nm.  Thirty-six WLS fibers go along each cell, gather in one bundle and transmit light to one multi-pixel 6$\times$6 $\textrm{mm}^2$ photodiode (multi-pixel photon counter, or MPPC). In this prototype, counters of types S13160-6025,  S13160-6050, S14160-6050, and FC-6035 \cite{ECAL_G_sipm} are used. 

The size of the cell for cosmic ray testing with the purpose of estimating signals from the MIP was chosen to be $55\times55$ $\textrm{mm}^2$. It consists of 220 layers of the scintillator and lead with widths of 1.5 mm and 0.3 mm respectively. 

The module consists of 4 cells with a cross-section of 55$\times$55 $\textrm{mm}^2$ combined into one tower with a cross-section of 110x110 $\textrm{mm}^2$ and a length of 440 mm. Nine modules of the calorimeter, each consisting of 4 cells, were manufactured for testing at the experimental test benches in VBLHEP and outside. Four of them were tested on cosmic rays. The test results are shown in Section \ref{ECAL_G_sect_cosmic_ray_test_results}.

\begin{figure}
\centerline{\includegraphics[width=0.7\linewidth]{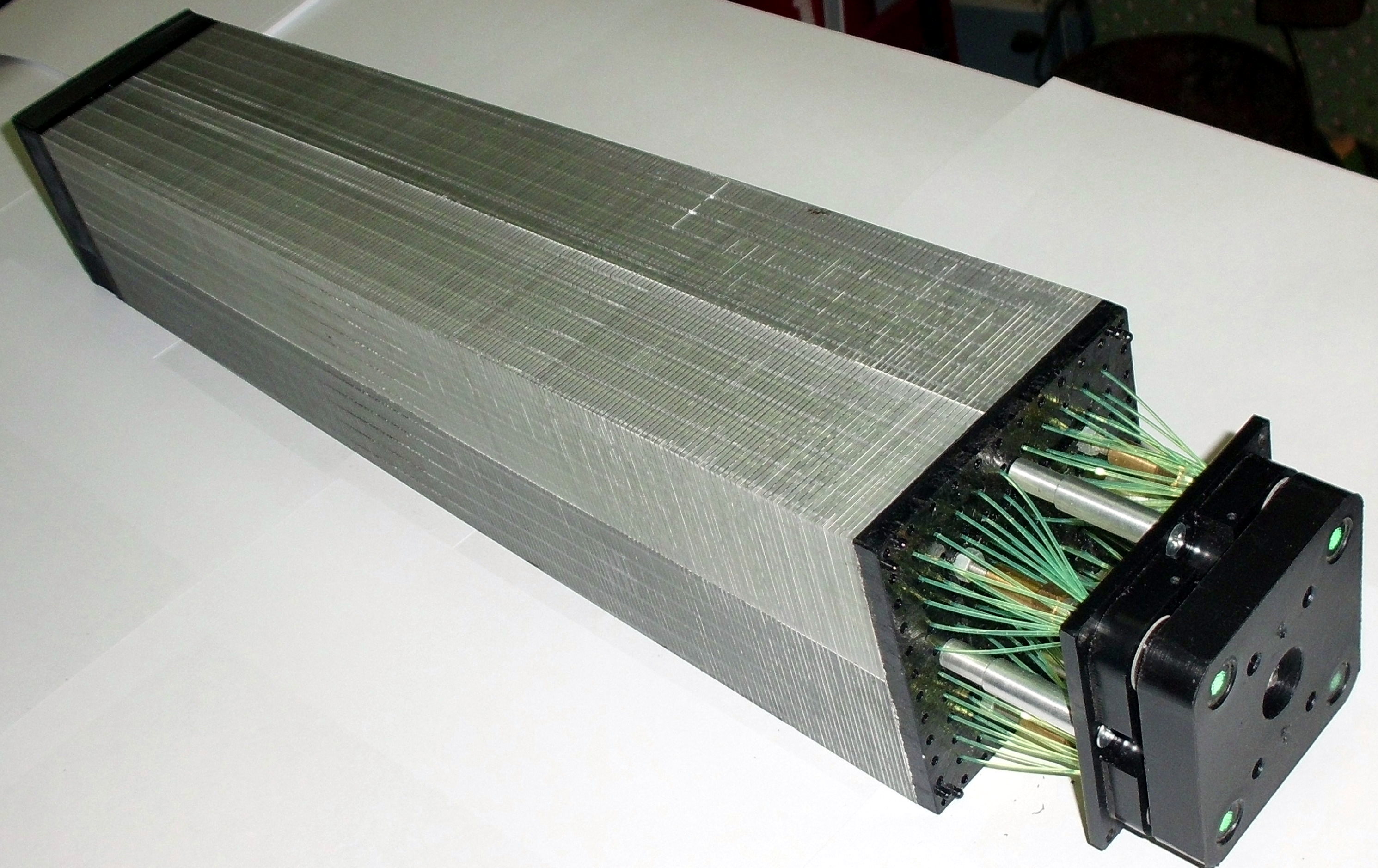}}
\caption{Photo of a single module consisting of 4 cells with 220 layers of the scintillator and the absorber with a thickness of 1.5 mm and 0.3 mm, respectively. Four bundles of fibers for guiding the light to the multi-pixel photon counters (MPPC) can be seen.}
\label{_G_module_photo}
\end{figure}

In the photo (Fig.\ref{_G_module_photo}), a module of trapezoidal shape is shown, which is obtained after milling a rectangular parallelepiped at a ~2 degrees angle.
The cell size of 40$\times$40 $\textrm{mm}^2$ at the front face and 55$\times$55 $\textrm{mm}^2$ at the back face allows one to implement projective geometry (if necessary) in the SPD electromagnetic calorimeter.

\subsection{Multi-pixel photodiodes}

All of the MPPC's, that are used in this prototype have the same size of 6$\times$6 $\textrm{mm}^2$, but have different dynamic and time characteristics. The S13160-6025 series has the best response speed, low capacitance and a large number of pixels, but the largest temperature coefficient of $K_T \sim 0.054$ V/\textdegree C.  The temperature coefficient shows a linear dependence of the breakdown voltage on the temperature and leads to a change in signal amplification. To achieve calorimeter stability of about 2\%, one needs to ensure the stability of the surrounding environment, or use the breakdown voltage compensation scheme $U_{OP}=U_{BR}+\Delta U-K_T\times\Delta T$,  where $U_{OP}$ and $U_{BR}$ are the operation and the breakdown voltages, respectively, $\Delta U$ is a  voltage bias and $\Delta T$ is a deviation of the current temperature from the nominal one, e.g. 20 \textdegree C.

The S14160-6050 series has high photo detector efficiency, but fewer pixels, which is worse in terms of the dynamic range. This series has a small temperature coefficient. An optimal solution would be to manufacture a similar photodiode series, but with a smaller pixel size of 15 $\div$ 20 $\mu$m, which would make them more suitable for a calorimeter.

\subsection{MPPC readout  and the high-voltage control}

\begin{figure}[!h]
  \begin{minipage}[ht]{0.46\linewidth}
  \center{\includegraphics[width=1\linewidth]{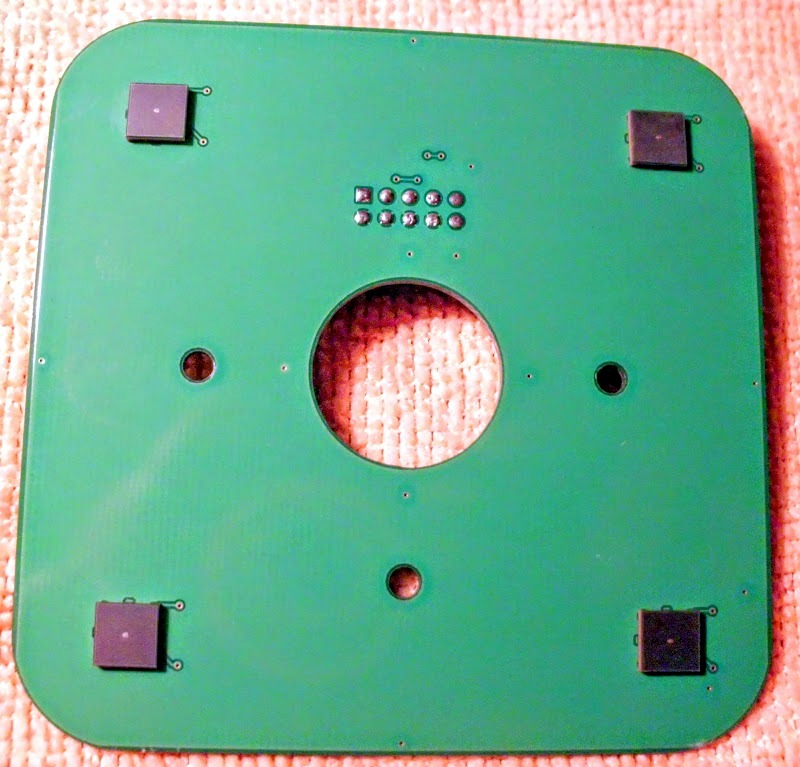}\\ (a)}
  \end{minipage}
  \begin{minipage}[ht]{0.5\linewidth}
 \center{ \includegraphics[width=1\linewidth]{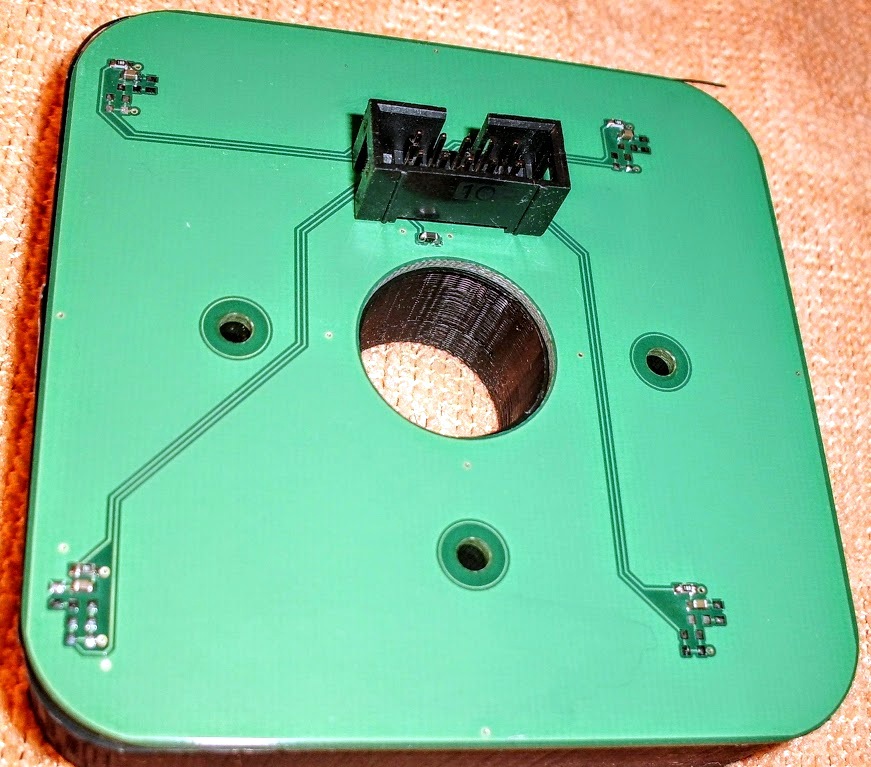}\\ (b)}
  \end{minipage}
    \caption{Printed circuit board with 4 MPPC diodes: front (a)  and back (b) sides. }  
    \label{ECAL_G_circuit_board_MPPC}%
\end{figure}

 Four MPPC's are surface mounted on a circuit board, as shown in Fig.\ref{ECAL_G_circuit_board_MPPC}. A thermistor is also installed to measure the photodiode temperature. The circuit board is connected to a module in such a way that the photodiodes are placed at the positions of fiber bundles. There is no optical contact between the photodiode and the WLS, instead there is an air gap of about 0.1 mm. Optical grease is not used in order to avoid instability in the conditions of light guiding. A light insulating basket made of black plastic is installed on top of the circuit board.
 
\begin{figure}[!h]
  \begin{minipage}[ht]{0.46\linewidth}
  \center{\includegraphics[width=1\linewidth]{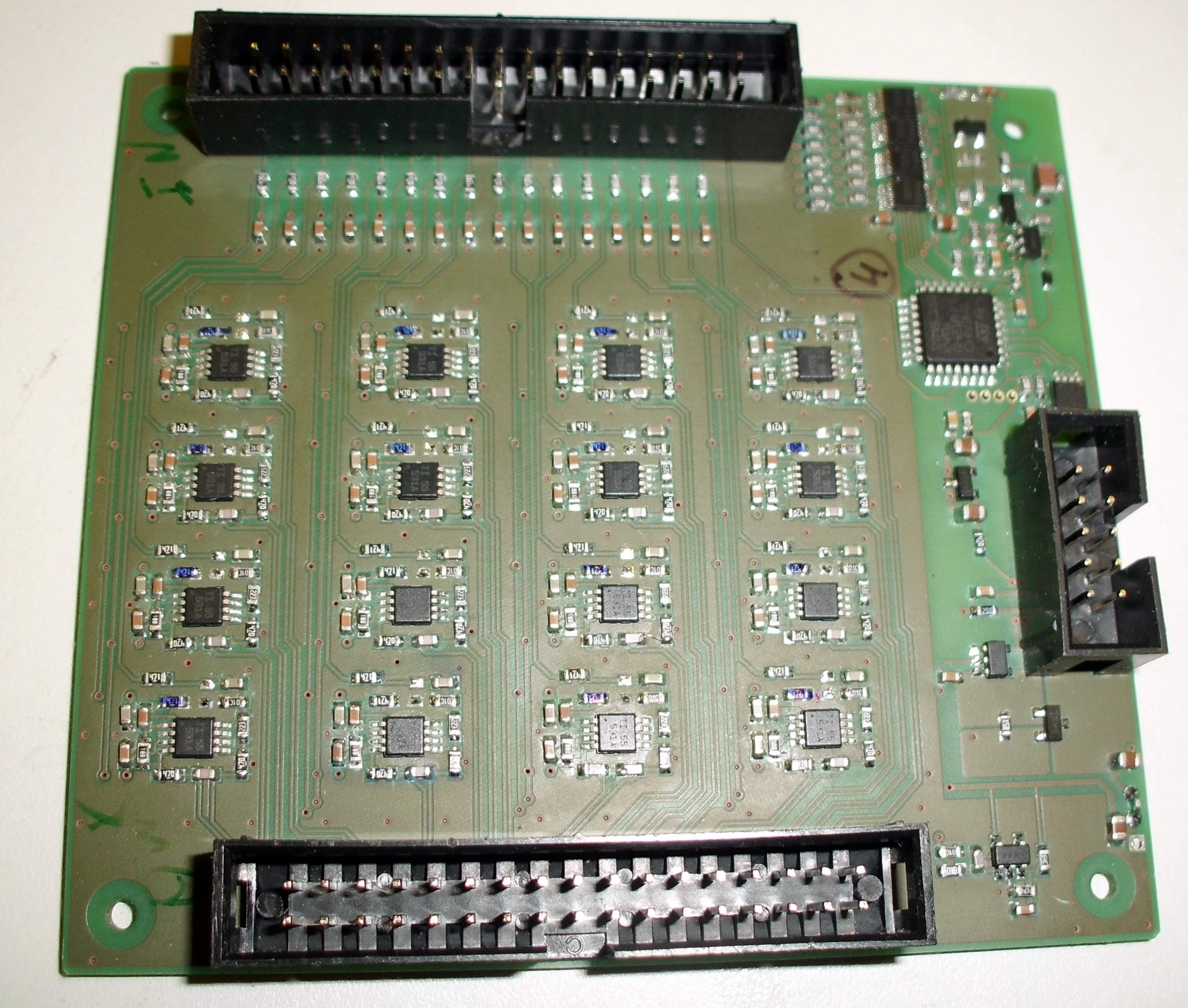}\\ (a)}
  \end{minipage}
  \begin{minipage}[ht]{0.5\linewidth}
 \center{ \includegraphics[width=1\linewidth]{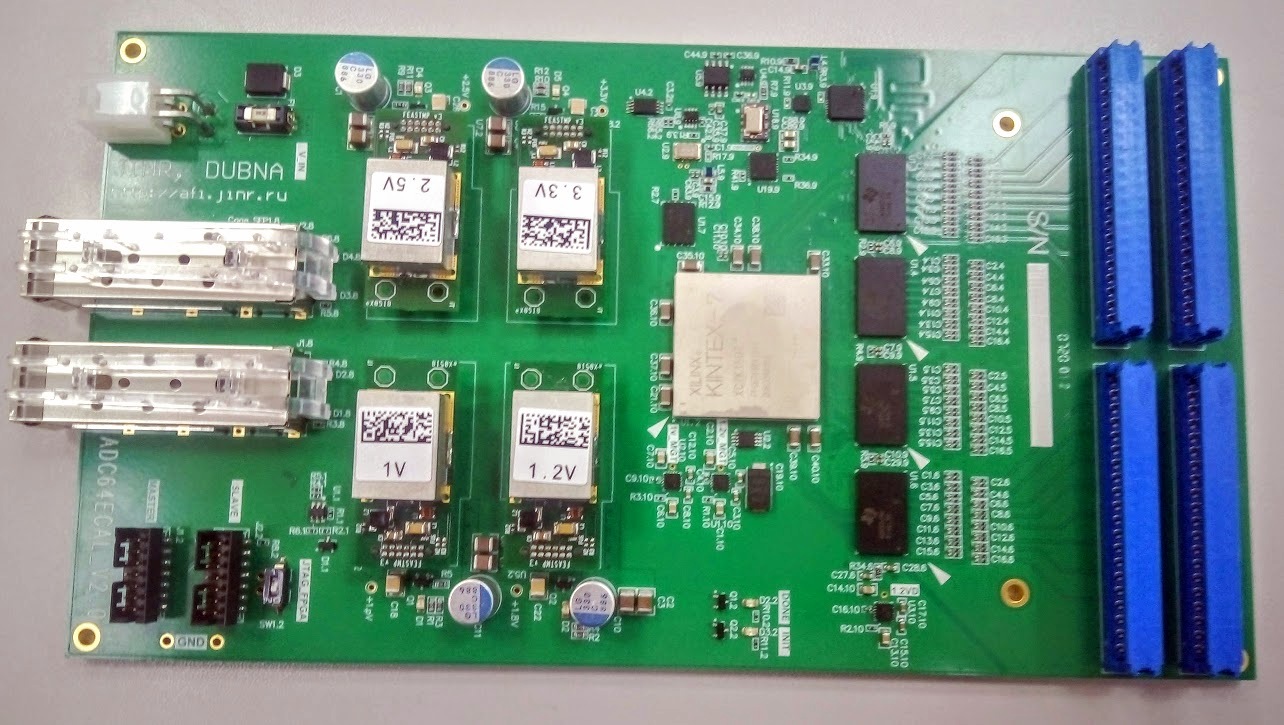}\\ (b)}
  \end{minipage}
    \caption{(a) 16-channel amplifier board is used to control the MPPC High Voltage and transmit signals to the ADC-64. The power consumption is about 16 mW per 16 channels. (b) 64-channel ADC-64- specifically designed for calorimeter operation in the magnetic field.
The power consumption is about 120 W per one board (64 channels).}  
\label{ECAL_G_ampl_board}
\end{figure}

The MPPC are connected to the amplifier board (Fig.\ref{ECAL_G_ampl_board}) using a 1-meter 17-pair flat twisted-pair cable. Five pairs of wires transmit signals to the amplifiers \cite{ECAL_G_HV}. Two wires are used to send base voltage of $\sim 40$ V and connect the thermistor. Channel voltages are transmitted via signal wires as a small bias from 0 to 5 V. This way, the bias voltage can be precisely set in a small range, but with 10-bit precision (i.e. about 5 mV).

The voltage control is implemented at a software level, taking into account the temperature from the thermistor installed on the circuit board.  This allows for operation without special equipment for temperature stabilization. Signal stability of the order of 0.1$\div$0.2\% is achieved during measurement over an extended period of time.

\subsection{Readout electronics}

The readout electronics consists of an analog-to-digital converter ADC-64 \cite{ECAL_G_ADC64} (Fig.\ref{ECAL_G_ampl_board}(a)). The ADC receives  continuous-time samples of the input signal with a fixed frequency and provides full digital representation of signals in time.  Samples are received at a 64-MHz frequency, which corresponds to the time period of 15.625 ns. Each sample is measured with a 12-bit precision. At present, there is an ADC-64-ECal modification, which improves the precision up to 14-bit and significantly extends the range of the measured amplitudes. The new ADC modification also allows for operation in strong magnetic fields, which is necessary for experiments at the NICA accelerator complex. 

An Ethernet connector for data transfer can be seen in Fig.\ref{ECAL_G_ampl_board}~(b), together with  a coaxial input for readout synchronization, which serves as a trigger. The ADC can also operate in streamer mode due to dedicated firmware. Using the White Rabbit technology provides sub-nanosecond synchronization accuracy.

\subsection{Cosmic ray test results}\label{ECAL_G_sect_cosmic_ray_test_results}

\begin{figure}[!t]
\centerline{\includegraphics[width=0.6\linewidth]{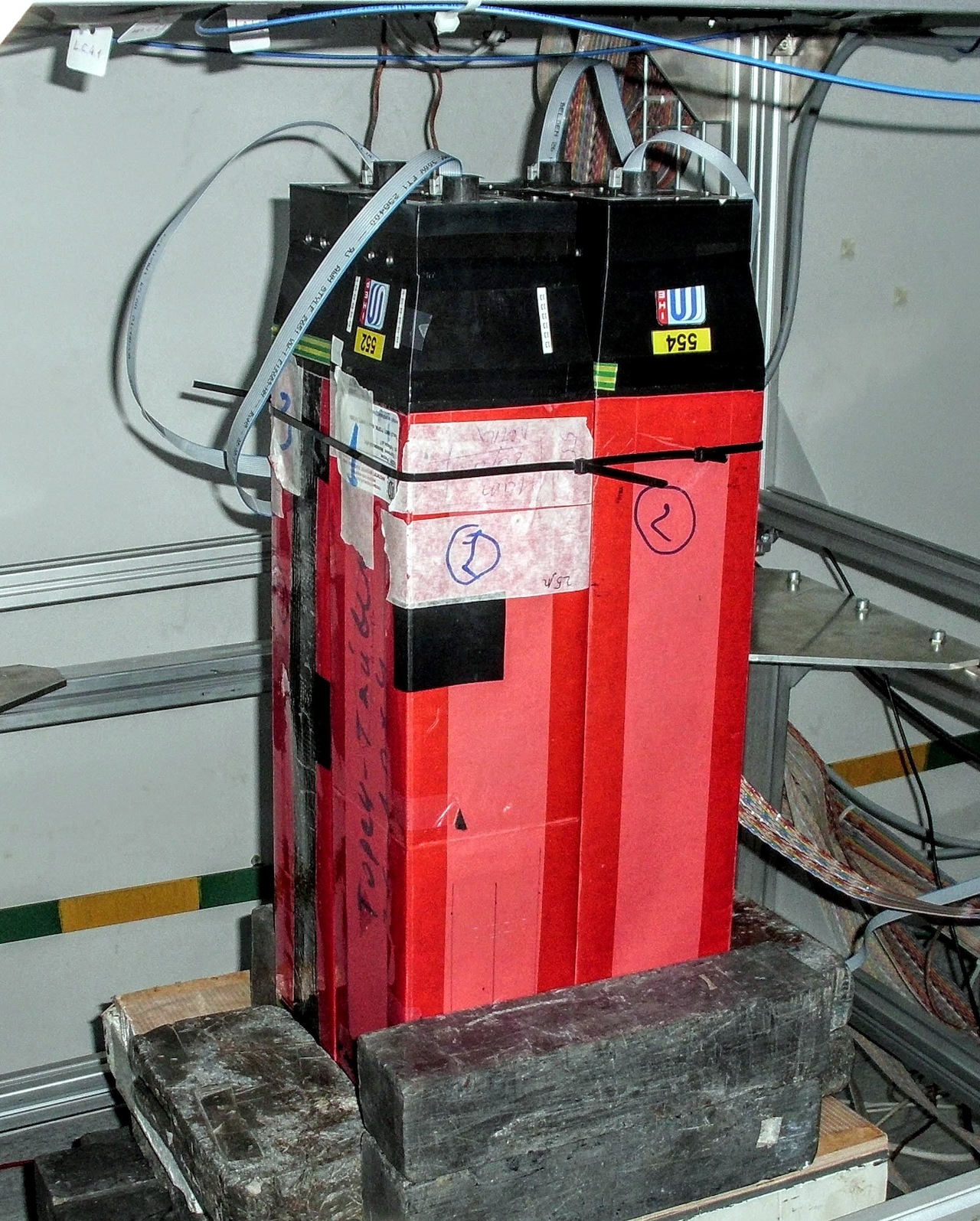}}
\caption{Photo of the calorimeter test setup consisting of 4 modules of the size $11\times11$ cm$^2$, with the total cross-section of $22\times22$ cm$^2$. }
\label{ECAL_G_ecal_setup_cosmic}
\end{figure}

For testing on cosmic rays, a small setup of 4 (Fig.\ref{ECAL_G_ecal_setup_cosmic}) modules (each $11\times11$ $\textrm{cm}^2$), of the total cross-section of $22\times22$ $\textrm{cm}^2$, was used.  The cells, each 55$\times$55 $\textrm{mm}^2$, are assembled in a $4\times4$ setup. The modules are placed vertically, while the direction of the registered cosmic rays is determined by trigger counters. The counters are multi-pixel photodiodes of type FC6035 and size 6$\times$6 $\textrm{mm}^2$. All the photodiodes are included in a coincidence trigger for the ADC. The trigger includes the signal from the generator, which starts the LEDs for control, calibration of calorimeter cells using the estimates of the light yield, and long-term stability control. Data acquisition is conducted at the ADC using the software provided by the developer. During a data taking period of 5-6 days, statistics of the order of million triggers was obtained.

The setup allows one to measure energy depositions and trajectories of cosmic ray particles. 
Relativistic muons with energy above 250 MeV pierce through the calorimeter and form a peak in the deposited energy distribution.  In order to select straight tracks of particles, which pass vertically through one module, only those events are selected, where the number of hits is equal to 1.

Signals obtained on cosmic muons are used for amplitudes alignment and calorimeter energy calibration. Only events with exactly one cell hit are selected. The bordering cells have more events with smaller amplitudes due to angled tracks. We perform calorimeter calibration using only vertical tracks. Each maximum value in terms of ADC units is mapped to the corresponding energy deposition.  The energy scale is determined from the Monte-Carlo simulation as the scale factor between the energy deposition of an electron with a 1-GeV energy, and a relativistic muon with an energy above 1 GeV, in scintillator plates for the given structure.  From this proportion we estimate the MIP signal in this calorimeter to be 240 MeV. This value divided by the position of the muon peak maximum is used as a calibration coefficient for each cell.  This calibration procedure involving the MIP energy deposition is not absolute or conclusive. Primarily, it aligns the amplification coefficients in each cell to ensure equal response of each cell.  The measured electron or photon energy can be further revised by reconstructing neutral pions or calibrating the calorimeter using electron or photon beams of the given energy.

The electromagnetic calorimeter measures electron or photon energy by summing up signals from all 16 cells. Each cell can only contain a fraction of energy deposited by the particle in the calorimeter (if the particle is not a relativistic muon, or a MIP). If the calorimeter is calibrated with a precision of several percent, the total energy weakly depends on the particle angle and the resolution increases only by 1.4\%.

\begin{figure}[!h]
  \begin{minipage}[ht]{0.50\linewidth}
  \center{\includegraphics[width=1\linewidth]{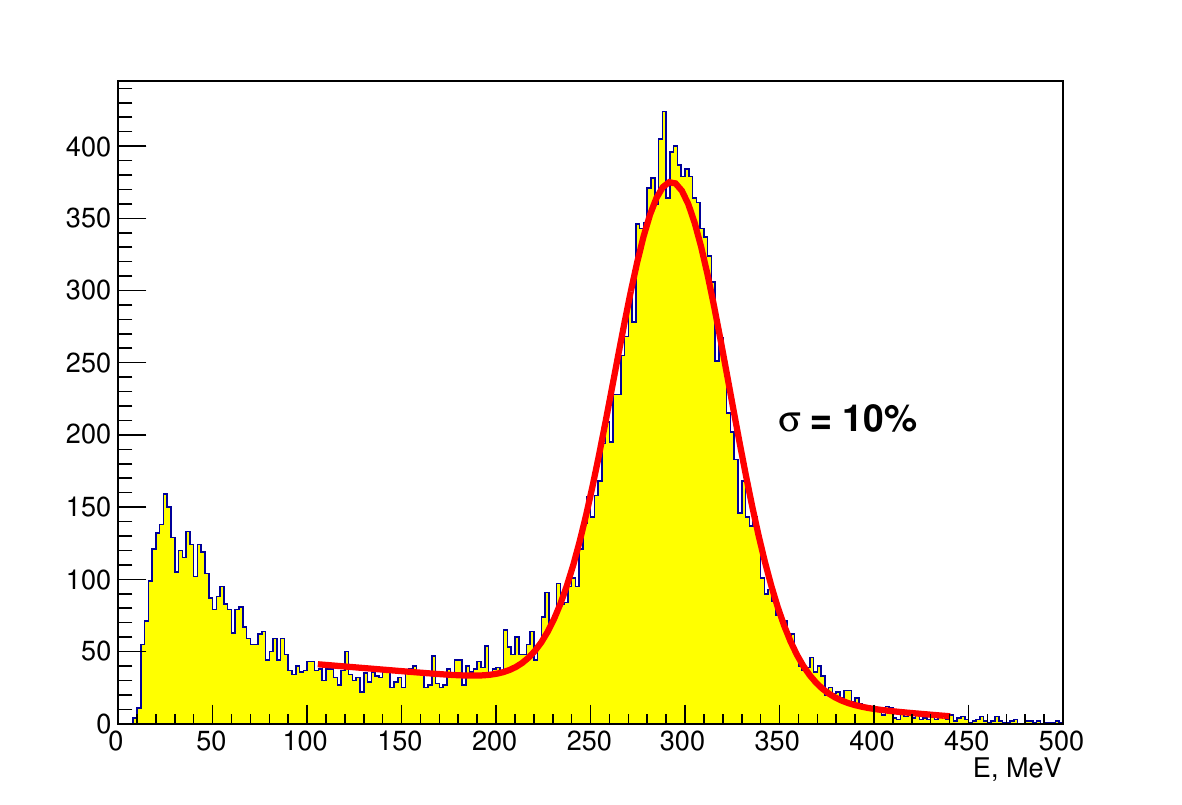}\\ (a)}
  \end{minipage}
  \begin{minipage}[ht]{0.5\linewidth}
 \center{ \includegraphics[width=1\linewidth]{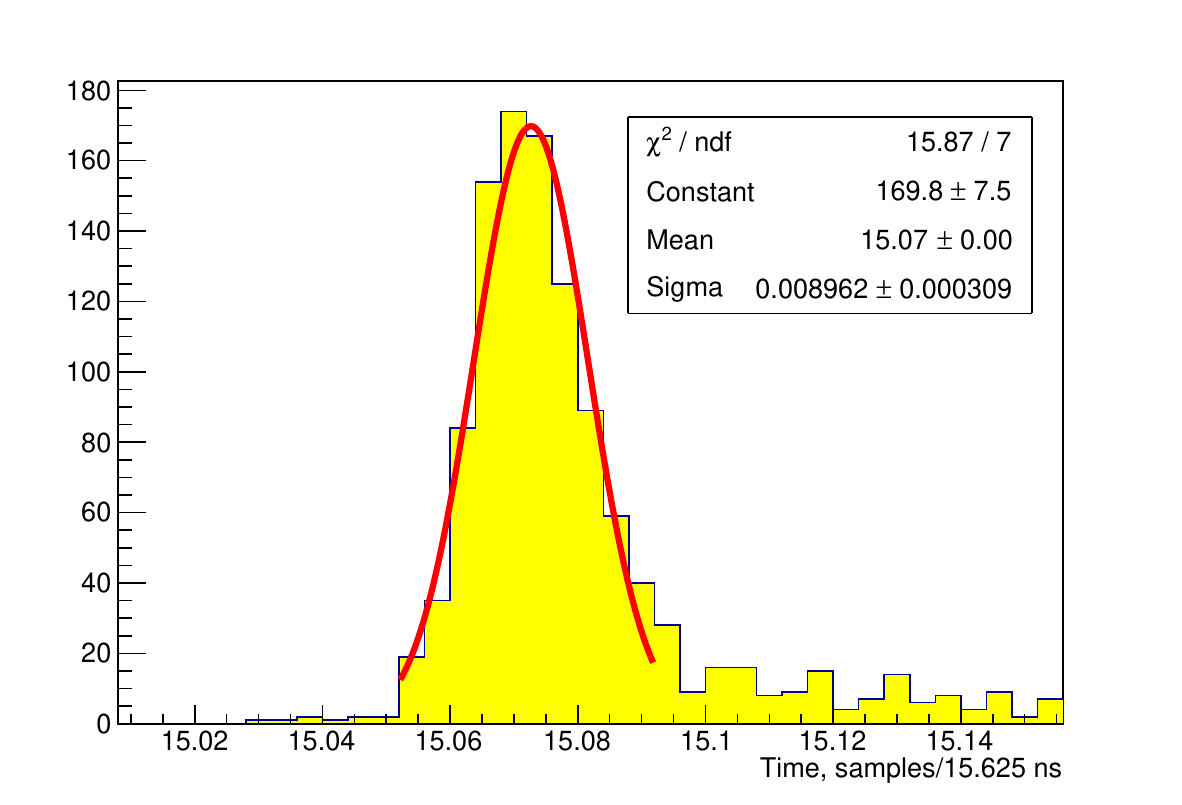}\\ (b)}
  \end{minipage}
    \caption{ (a) Total energy deposition in the calorimeter for the MIP obtained by summing up signals from 16 cells while selecting 1-hit events.  (b) Time resolution for calorimeter cell \#11 is equal to 175 ps.}  
       \label{ECAL_G_total_edep_time_cells} 
\end{figure}

The energy resolution of the calorimeter for vertical cosmic ray particles is 9.6\% (Fig.\ref{ECAL_G_total_edep_time_cells}~(a)).  This number corresponds to energy deposition of 240 MeV. Assuming the resolution depends on the energy as $E^{-1/2}$, the energy resolution at 1 GeV is estimated  to be 5\%. 

Time resolution  for calorimeters  of such types  is about 175 ps for the MIP (Fig.\ref{ECAL_G_total_edep_time_cells}~(b)) and can be improved for high-energy electrons. This can be applied to identify particles in the energy range of 50$\div$1550 MeV. 

\subsection{Dependence of the calorimeter response on the number of photoelectrons}

Cosmic ray testing allows one to obtain dependences of energy and time resolution on the number of photoelectrons ($NPE$) produced during  the MIP passing through a cell.
For each channel, time is calculated as zero intersection of the waveform.  This method, Constant Fraction Discriminator, is used for determining a time value on a constant fraction of the pulse leading edge. The energy and time resolution of the calorimeter depend on the $NPE$ as $1/\sqrt{NPE}$.

Different conditions of light guiding were used in this 4-module calorimeter. These conditions included forming reflective surfaces on the edges of  the WLS, or using the fibers as U-shaped loops. 
Differences in light guiding conditions lead to large variations in the $NPE$ in the range between 1000 and 3000 photoelectrons per MIP for different cells. In terms of the amount of deposited energy, this corresponds to 4000$\div$12000 $NPE$/GeV.

\begin{figure}[!t]
  \begin{minipage}[ht]{0.50\linewidth}
  \center{\includegraphics[width=1\linewidth]{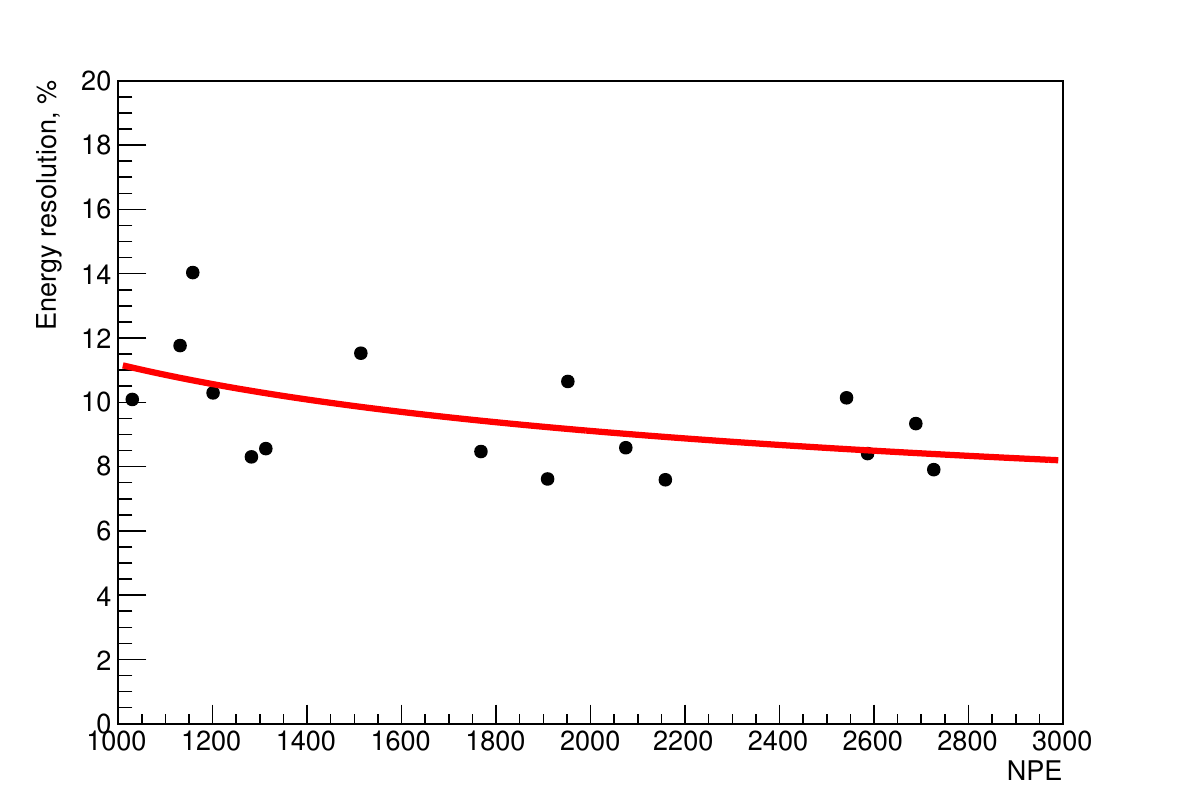}\\ (a)}
  \end{minipage}
  \begin{minipage}[ht]{0.5\linewidth}
 \center{ \includegraphics[width=1\linewidth]{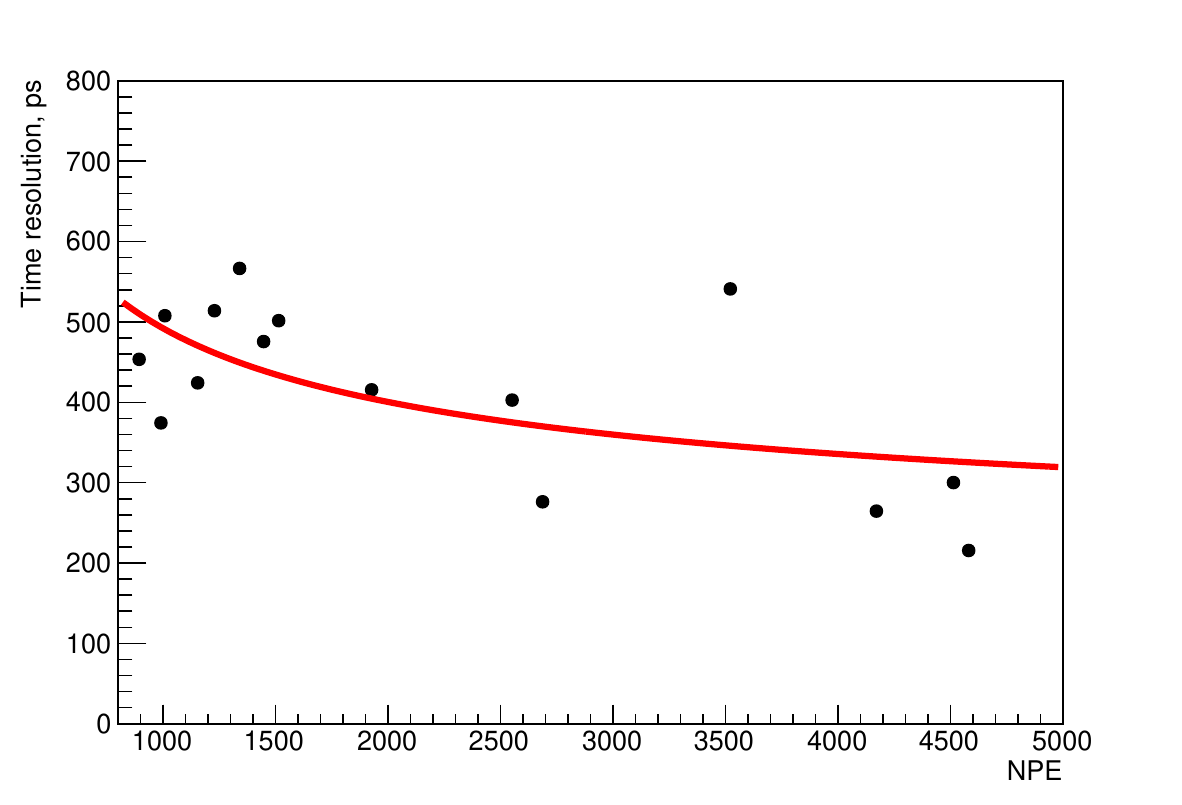}\\ (b)}
  \end{minipage}
    \caption{Dependence of energy (a) and time (b) resolution for different calorimeter cells on the number of photoelectrons ($NPE$).}  
  \label{ECAL_G_npe_dependences}
\end{figure}

Information on the number of photoelectrons for each cell allows one to obtain dependence of energy and time resolution on the $NPE$.  The presented dependences of energy and time resolution are displayed in Fig.\ref{ECAL_G_npe_dependences} and show that the limit for large values of the $NPE$ is $6.2\%$ and $197$ ps for energy and time resolutions, respectively.

\subsection{Long-term stability}
Temperature dependence of calorimeter stability was investigated using daily temperature variations in the range of 18-22 \textdegree C. During the measurement of signals from cosmic ray particles over 5 days, signals from LEDs of a 1-Hz frequency were also measured. 
Photodiode temperature is constantly monitored using a high-voltage system. The voltage bias on photodiodes is corrected during a temperature measurement using a linear dependence: $U_{out}=U_{bias}-k\times(20-T)$.

\begin{figure}[!t]
\centerline{\includegraphics[width=0.7\linewidth]{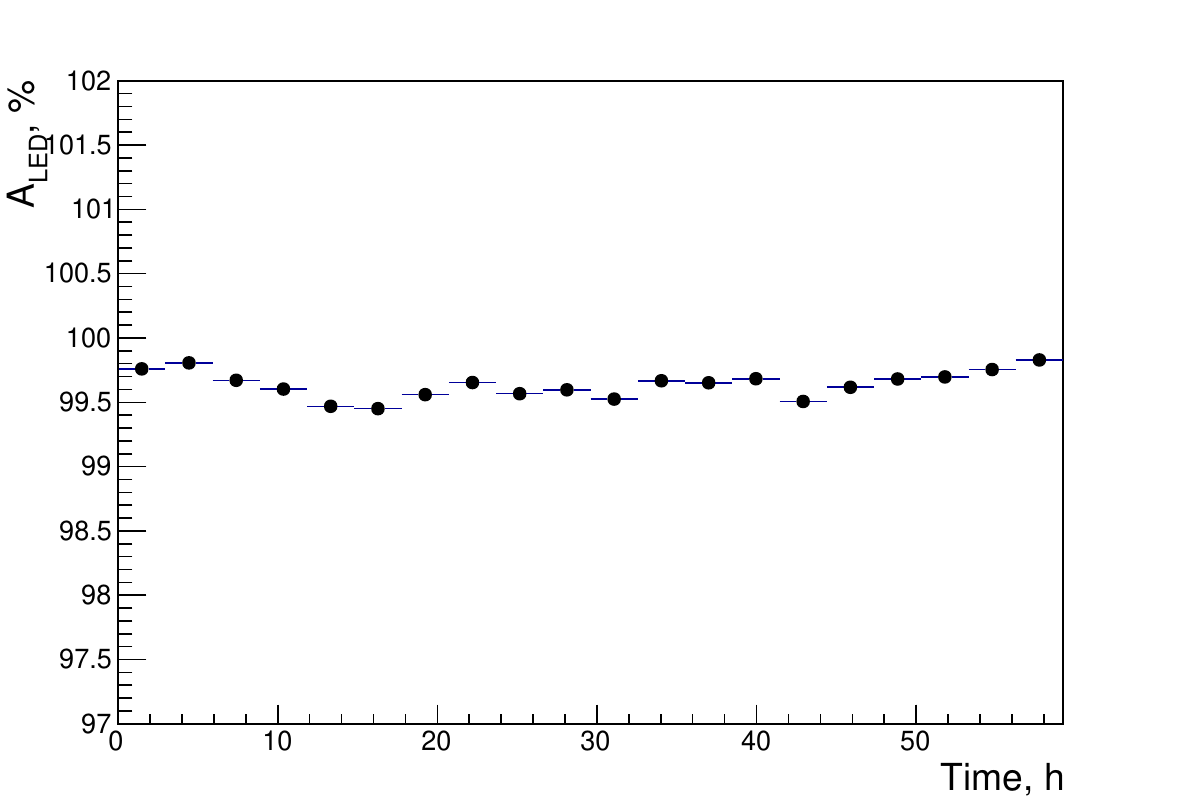}}
\caption{Dependence of the sum (average value) of signals from the calorimeter (in \% with respect to the first 5 minutes of the measurement period) on the time of measurement (in hours) with temperature-dependent voltage compensation.}
\label{ECAL_G_long_term}
\end{figure}

Daily temperature variations during the measurement were about 5\textdegree C. The temperature coefficient $k=0.034$ V/\textdegree C is used for temperature compensation of the operating voltage. After compensation, variations in the signal amplitude are constrained within $\pm$0.4\%. The plots are normalized to the start of the measurement. The first 300 signals (5 min) are used to normalize the full measurement period. The calorimeter can operate with a stability of $\sim$0.8\%  if the temperature compensation of the operating voltage is maintained, as it can be seen from these results and is shown in Fig. \ref{ECAL_G_long_term}. 

\subsection{Cost estimate}

The cost of the calorimeter is proportional to the number of channels. Mechanical assembly of a calorimeter cell from the scintillator and the lead plates costs 300\$ per channel. Another expensive element is the wavelength shifting fibers. For a $40\times40$ $\textrm{mm}^2$ cell,  9 fibers of the total length of 54 m are used. Assuming an average price of 5\$/m, the price per channel amounts to 270\$. The cost of photodiodes depends on the quantity. For purchases of tens of thousands of units, their price is about 30\$ per unit.
The electronics also contributes significantly to the total cost, especially the ADC with a price of 80\$ per channel. The cost of the supply and voltage control systems is 20\$ per channel. The total cost of a calorimeter cell is about 700\$. Thus, the total cost of a 30176-cell calorimeter is 21.1 M\$.

\begin{table}
\caption{\label{ECAL_G_table_ecal_cost}Contributions of separate elements to the cost of the ECal.}
\begin{tabular}{ |c|c|c|c|c|c|c| } 
 \hline
  & ECal cell  & WLS  & ADC & HV  & MPPC  & Total \\ 
 \hline
\centering Cost per 1 cell, \$ & 300 & 270 & 80 & 20 & 30 & 700  \\ 
 \hline
 Cost for 30.176 cells, M\$] & 9.1 & 8.1 & 2.4 & 0.6 & 0.9 & 21.1  \\ 
 \hline
\end{tabular}
\end{table}


\section{Range (muon) system}
\subsection{General description}
The Range System of the SPD detector serves for the following purposes:
 (i) identification of muons in presence of a remarkable hadronic background and (ii) estimation of hadronic energy (coarse hadron calorimetry).
It is important to stress that the system is the only device in the SPD setup, which can identify neutrons (by combining its signals with the electromagnetic calorimeter and the inner trackers). Muon identification (PID) is performed via muonic pattern recognition and further matching of the track segments to the tracks inside the magnets. The precise muon momentum definition is performed by the inner trackers in the magnetic field. The Mini Drift Tubes \cite{Abazov2010, Abazov20101} are used in the Range System as tracking detectors providing two-coordinate readout (wires and strips running perpendicularly). Such readout is mostly needed for the events with high track multiplicity and also for the reconstruction of the neutron space angle.   

As for the design and construction of the present system, we assume to capitalize on the experience gained by the JINR group in the development of the PANDA (FAIR, Darmstadt) Muon System \cite{MUON_TDR}. These two systems (PANDA and SPD), dealing with muons of comparable momentum ranges and solving the same PID tasks, should look very similar in their design and instrumentation.

\subsection{System layout}
\begin{figure}[!h]
  \begin{minipage}[ht]{0.49\linewidth}
   \center{\includegraphics[width=1.0\linewidth]{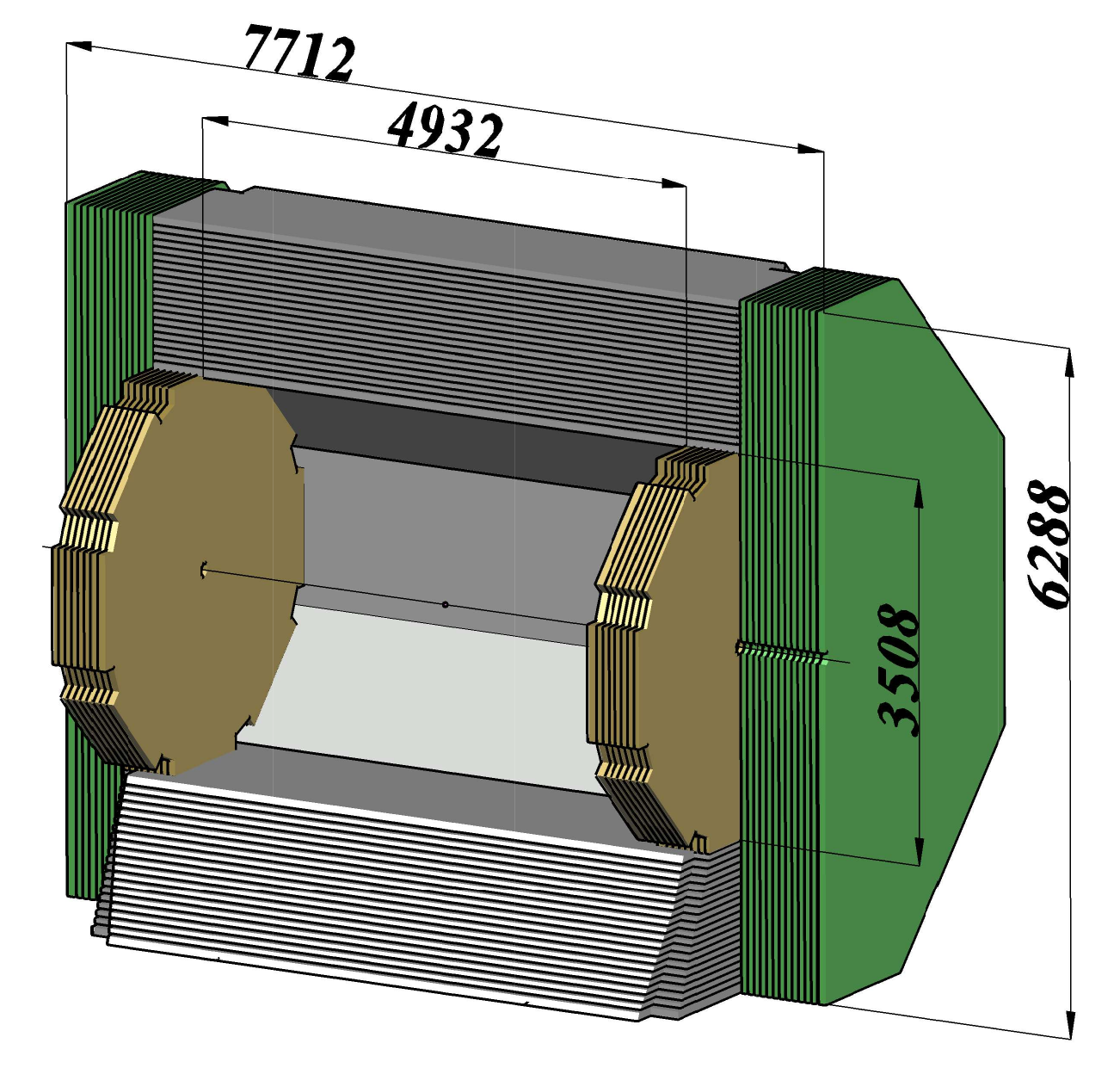} \\ (a)}
  \end{minipage}
  \hfill
  \begin{minipage}[ht]{0.49\linewidth}
    \center{\includegraphics[width=1.0\linewidth]{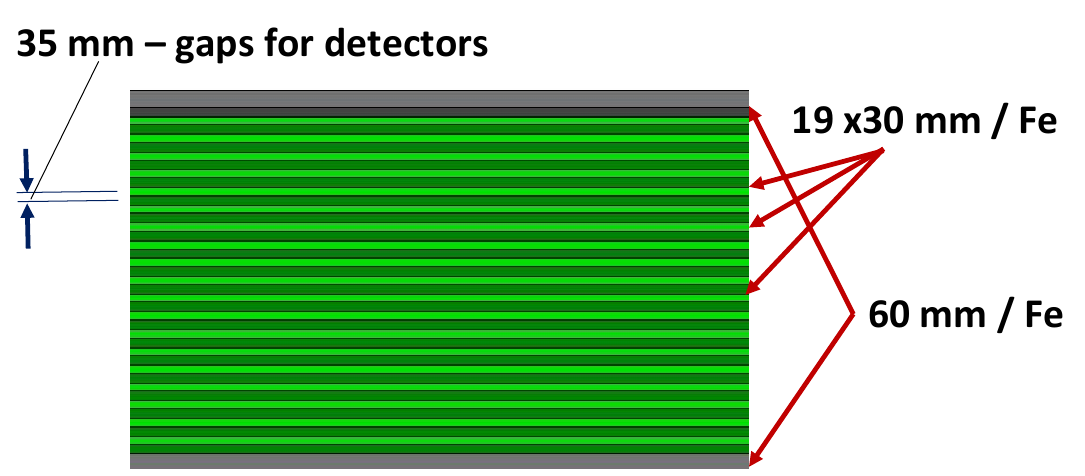} \\ (b)}    
  \end{minipage}
  \caption{3D view (half cut) of the Range (muon) system: (a) Barrel is shown in grey, End Cap Disks $-$ in green, and End Cap Plugs $-$ in yellow; (b) absorber structure.}
  \label{fig:RS_layout} 
\end{figure}

The Range System serves as an absorber for hadrons and a ‘filter’ for muons. It also forms the magnet yoke. It consists of a barrel and two end-caps. Each end-cap, in its turn, consists of an end-cap disk and a plug. The schematic 3D view of the system and its main  sizes are shown in 
Fig.\ref{fig:RS_layout} (a). The absorber structure is shown in Fig.\ref{fig:RS_layout} (b). The outer 60-mm Fe layers are used for bolting the modules together. The interlayer gaps of 35 mm are taken for reliable mounting of the detecting layers comprising the MDTs proper, the strip boards and the front-end electronic boards on them. The 30-mm thickness of the main absorber plates is selected as comparable with muon straggling in steel, thus giving the best possible muon-to-pion separation, and also providing rather good sampling for hadron calorimetry. 

The Barrel consists of eight modules,  and each end-cap disk consists of two halves divided vertically. Such subdivision of the system (14 pieces in total) is chosen to optimize its further assembly and to satisfy the constructional requirements of the SPD experimental hall (cranes capability and  floor load). The total weight of the system is about 810 tons, including 30 tons of detectors. The total number of MDT detectors is about 8000 units. The MDTs are deployed in the following way: along the beam direction in the Barrel, and perpendicular to the beam (horizontally) in the end-caps. 

 

The absorption thicknesses of the barrel and end-caps are selected to be equal  to 4  nuclear interaction lengths ($\lambda_I$) each. It provides uniform muon filtering in all directions. Together with the thickness of the electromagnetic calorimeter ($\sim$ 0.5 $\lambda_I$) the total thickness of the SPD setup is about 4.5 $\lambda_I$.

\subsection{Mini drift tubes detector}
The Mini Drift Tubes (MDT) detector was initially developed and produced at JINR for the Muon System of the D0 experiment at FNAL \cite{Abazov:2005uk}. Later on, an MDT-based muon system was also produced  for the COMPASS experiment at CERN \cite{Abbon:2007pq}. Developed two-coordinate readout modification of the MDT with open cathode geometry and external pickup electrodes was proposed to and accepted by the PANDA collaboration at FAIR for the muon system of their experimental setup. This new version of the MDT is proposed for the  SPD project, as it has  all the necessary features -- radiation hardness, coordinate resolution and accuracy, time resolution, robustness, as well as advanced level of already conducted R\&D within the PANDA project. 

\begin{figure}[!h]
    \center{\includegraphics[width=1.0\linewidth]{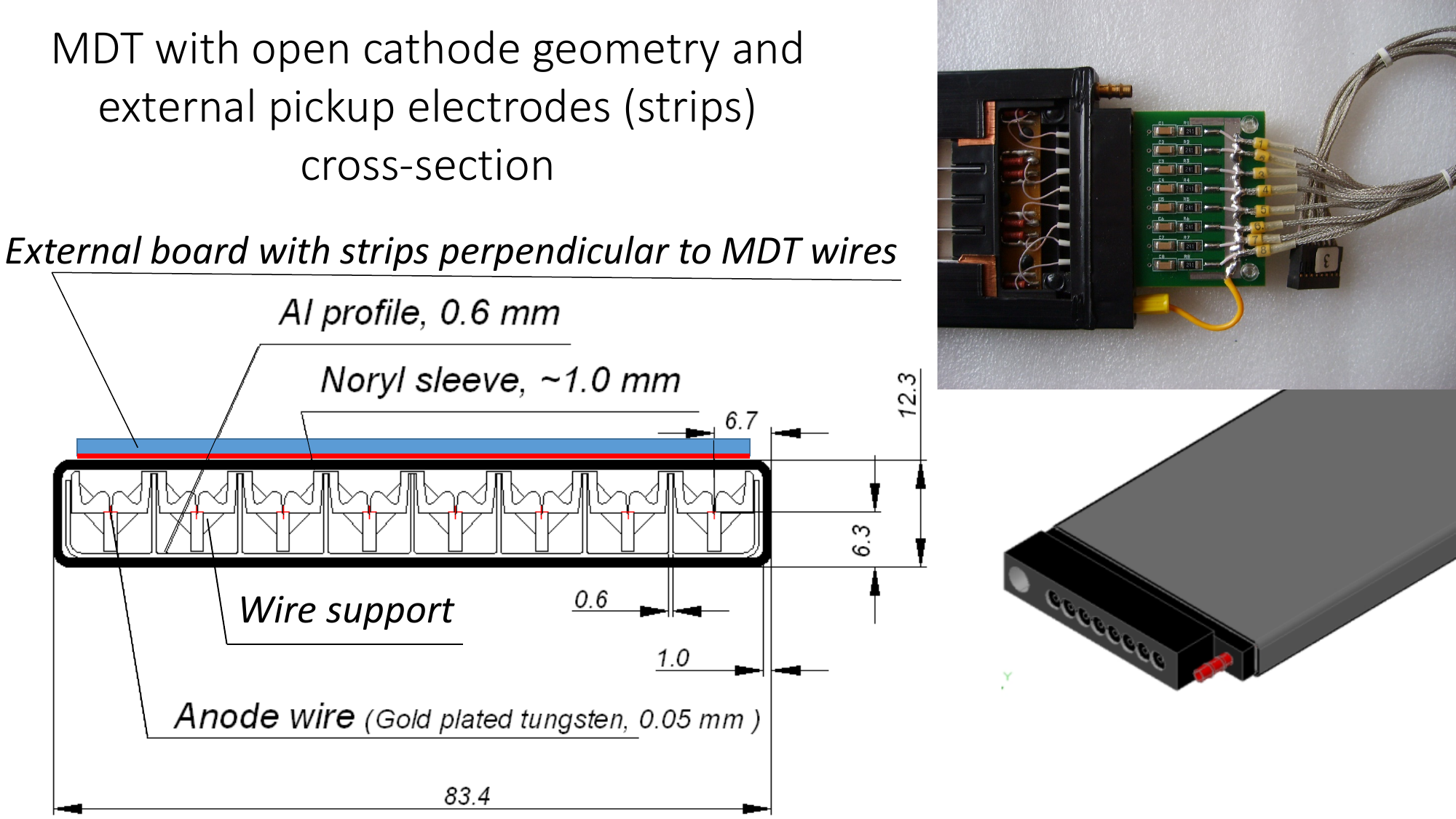}}
     \caption{ Mini Drift Tube with open cathode geometry cross-section (left) and layout (right).}
       \label{fig:MDT}  
\end{figure}

\begin{figure}[!h]
    \center{\includegraphics[width=0.7\linewidth]{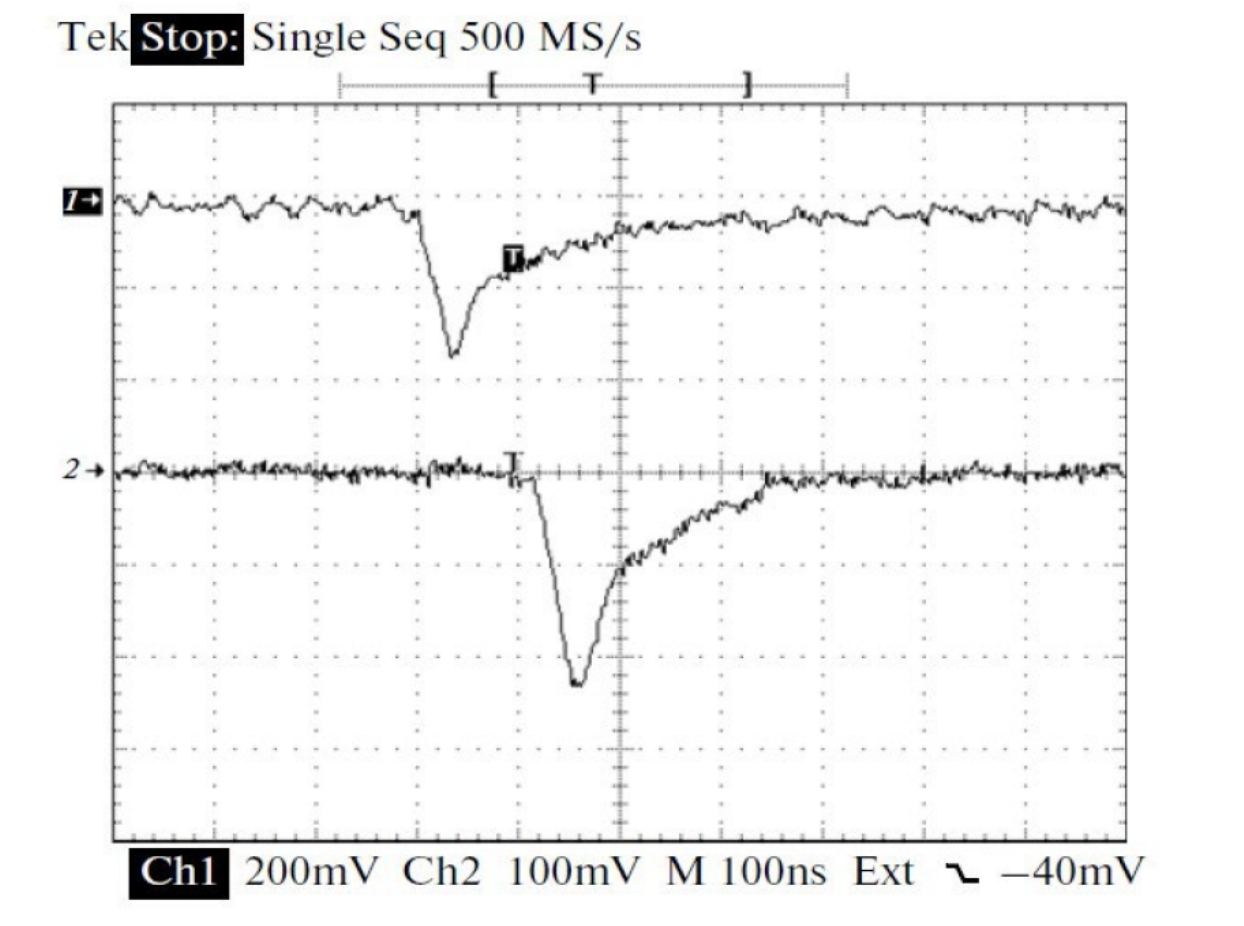}}
     \caption{Oscillograms of single signals: from the anode wire (1) and the strip (2, inverted); the conversion factors are 60 and 480 mV/$\mu$A, respectively.}
       \label{fig:Shape}  
\end{figure}

The cross-section and  layout of the MDT with open cathode geometry are  shown in Fig.\ref{fig:MDT}. 
The detector consists of a metallic cathode (aluminum extruded comb-like 8-cell profile), anode wires with plastic supports, and a Noryl envelope for gas tightness. The comb-like profile of the cathode provides each wire with an opening left uncovered to induce wire signals on the external electrodes (strips) perpendicular to the wires. The strips are applied to obtain the second coordinate readout. The shape of the induced signal repeats the initial one, having the opposite polarity, but the amplitude is about 15$\%$ of the wire signal (see Fig.\ref{fig:Shape}). Thus, the strip signal readout requires higher signal amplification and proper electromagnetic shielding.

\begin{figure}[!h]
    \center{\includegraphics[width=1.0\linewidth]{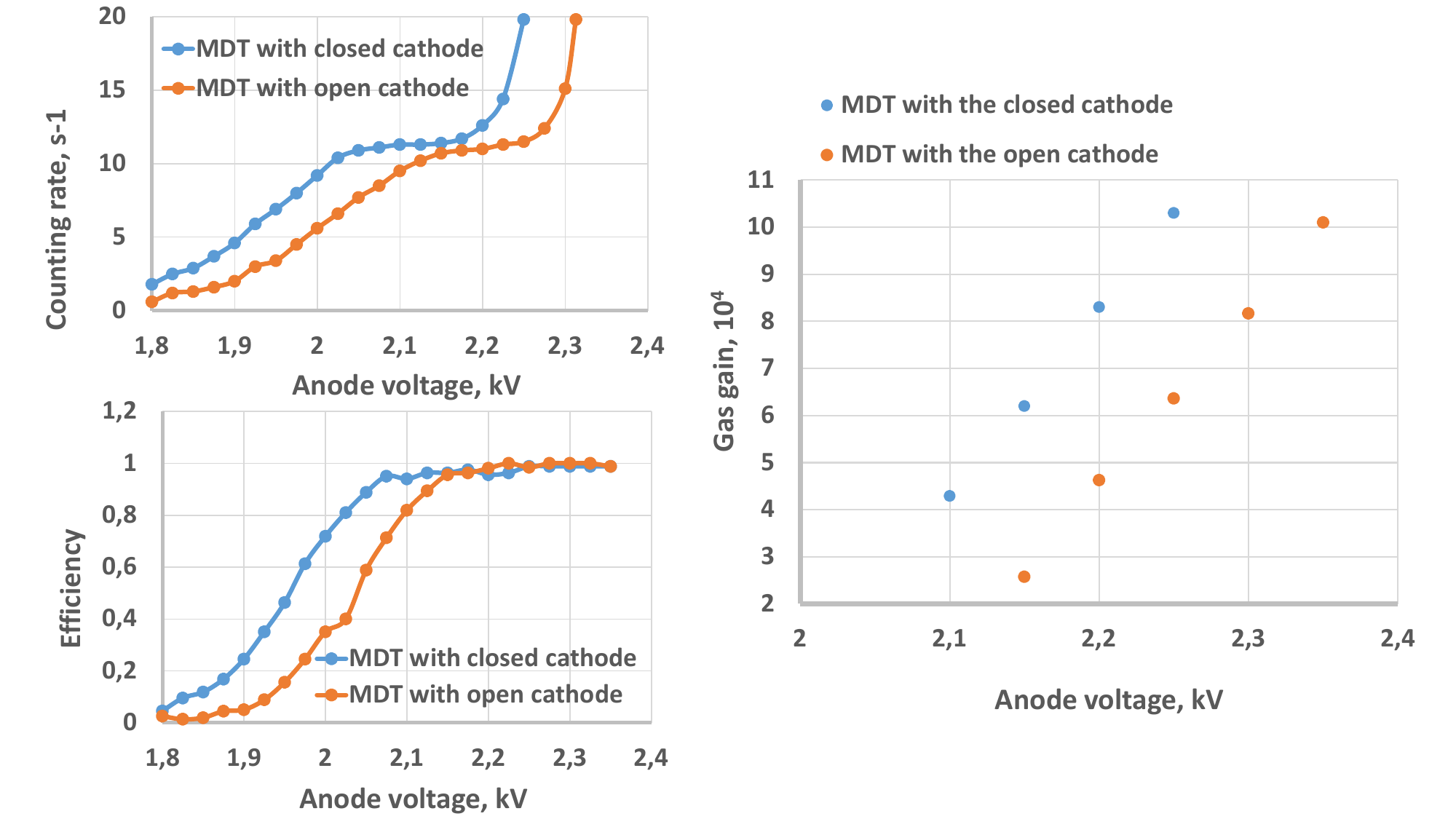}}
     \caption{Comparative plots of the counting rate, efficiency, and gas gain versus the supply voltage for the MDT with closed and open cathode geometry.}
       \label{fig:Rate}  
\end{figure}

Application of an open cathode leads to the loss of the electric field symmetry in each of the 8 detector cells, resulting in lower gas gain for the applied voltage comparing to the standard MDT (cathode openings closed with stainless steel lid). The conducted R$\&$D proved that the MDT with open cathode geometry easily achieves the parameters of the one with a closed cathode at higher voltages. The comparative plots of the counting rate, efficiency, and gas gain for both detector types (see Fig.\ref{fig:Rate}) show that the MDT with open cathode geometry repeats the standard MDT performance at a high voltage shift of +100 V. The drift time and the amplitude spectra of both detector variants also match, if we set this voltage shift between their operating points.

\begin{figure}[!h]
    \center{\includegraphics[width=1.0\linewidth]{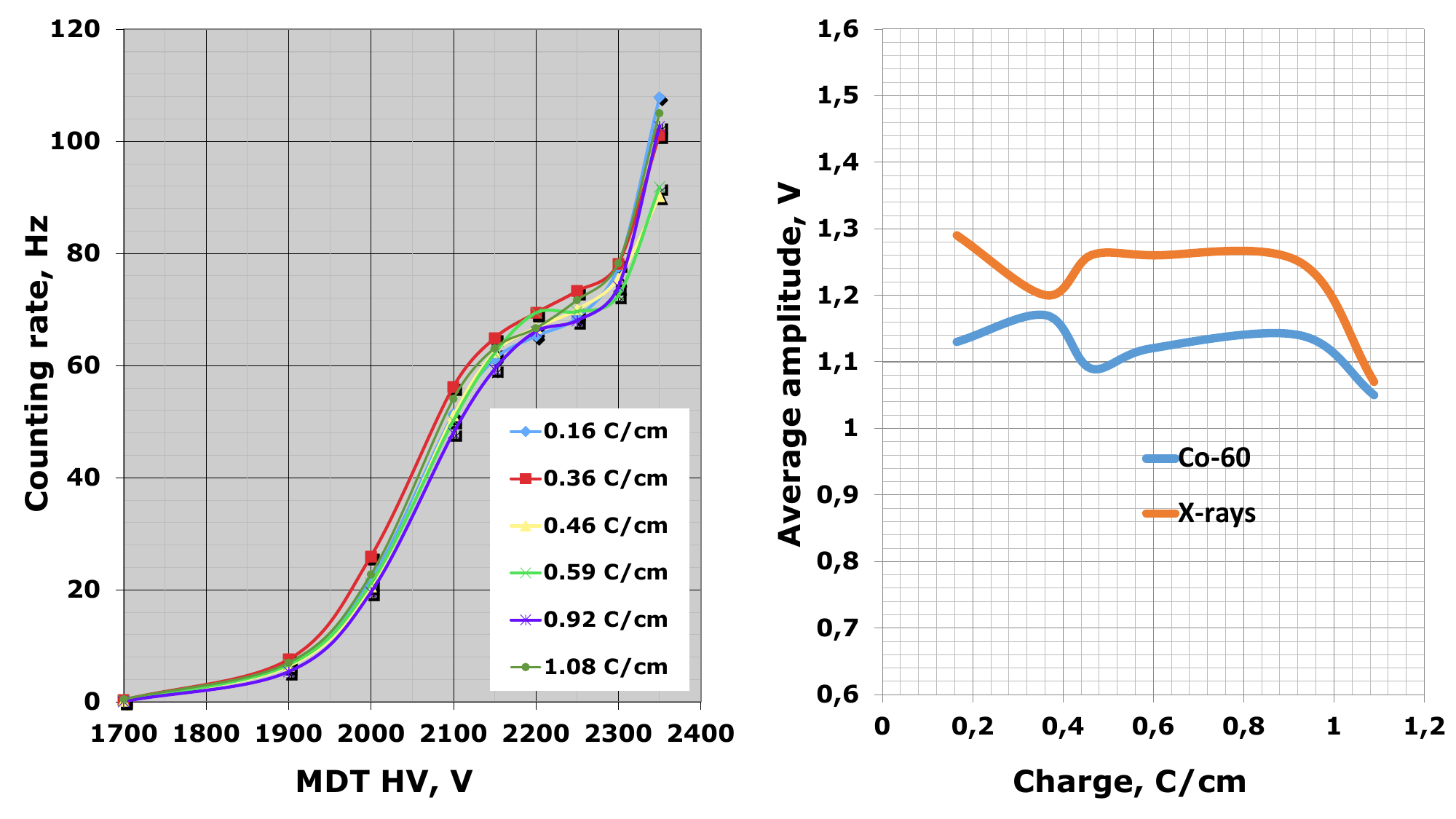}}
     \caption{Counting rate curves for different accumulated charges (0.16$\div$1.08 C/cm) (left); average wire signal amplitudes vs accumulated charge for Co-60 and X-ray sources (right).}
       \label{fig:Curves}  
\end{figure}

According to the results of the MDT (open cathode geometry) ageing tests, accumulation of  a 1 C/cm total charge does not produce any significant effect on the detector performance. To monitor the ageing effects, measurements of the counting rate curves (Co-60 source) together with oscilloscopic observations of the MDT average signals (256 events) for Co-60 and X-rays were made twice a week over the whole period of intense irradiation (see Fig.\ref{fig:Curves}). Later on,  this measurement (with X-rays) was conducted up to 3.5 C/cm of irradiation without any visible degradation of the MDT performance. It should ensure stable MDTs performance for the lifetime of the SPD project.

All R$\&$D studies were made with a gas mixture of 70$\%$ Ar + 30$\%$ CO$_2$ at atmospheric pressure, the one to be used in the proposed SPD Muon System. It is inflammable, radiation hard and fast enough (150-200 ns drift time).
The wire pitch in the present design equals 1 cm, and a 3-cm strip width is selected for the second coordinate. These spatial parameters provide the Range System with coordinate accuracy well enough for identification of muons and give the system the features of a digital hadron calorimeter. 


\subsection{Front-end electronics}
We plan to use the analog front-end electronics (with probable minor modifications) developed for the D0/FNAL and COMPASS/CERN experiments and also accepted by PANDA/FAIR. It is based on two ASIC chips: 8-channel amplifier Ampl-8.3 \cite{Alekseev:2001xn} and 8-channel comparator/discriminator Disc-8.3 \cite{ALEXEEV1999157}.

The HVS/A-8 card serves for two purposes -- as  MDT high voltage distributor and a signal amplifier designed to be the first stage of the Barrel and the End Cap Plugs wire signal readout. It is followed by Disc-8.3 based discriminating electronics (design in process) to fulfill the readout.

The ADB-32 card (initially designed for D0/FNAL)  \cite{Alekseev:2001cg} is used for the End Cap Disks wire readout. It amplifies and discriminates the MDT signal, shaping it to the LVDS standard for further treatment by the digital front-end electronics. 

\begin{figure}[!h]
  \begin{minipage}[ht]{0.49\linewidth}
    \center{\includegraphics[width=1\linewidth]{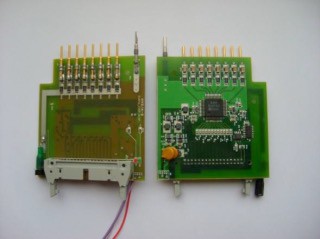} \\ (a)}
  \end{minipage}
  \hfill
  \begin{minipage}[ht]{0.49\linewidth}
    \center{\includegraphics[width=1\linewidth]{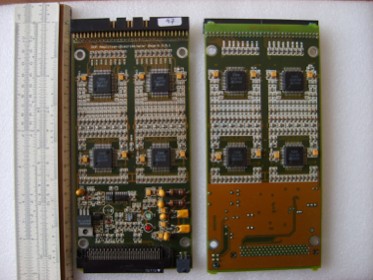} \\ (b)}
  \end{minipage}
  \begin{minipage}[ht]{0.49\linewidth}
    \center{\includegraphics[width=1\linewidth]{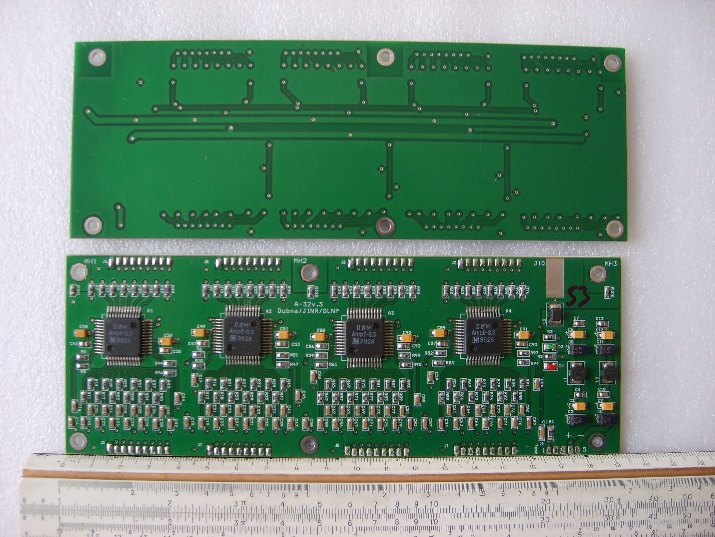} \\ (c)}
  \end{minipage}
  \begin{minipage}[ht]{0.49\linewidth}
    \center{\includegraphics[width=1\linewidth]{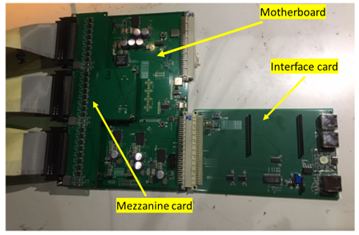} \\ (d)}
  \end{minipage}
  \caption{Front-end analog electronics cards: HVS/A-8 (a), ADB-32 (b) and A-32 (c). (d) Digital 192-channel MFDM module.} 
  \label{fig:RS_FE}  
\end{figure}

An A-32 preamplifier card is used to start the strip signal readout in the whole system. It should be terminated (similarly to wire readout) by Disc-8.3-based discriminating electronics. In case of the End Cap Disks an ADB-32 card will be used for this purpose. The view of the basic FEE cards is shown in Fig.\ref{fig:RS_FE}.

Totally, the Range System has 106000 readout channels (65000 of which are wires and 41000 strips). 

After having been shaped to the LVDS standard, the signals from the analog electronics go to the digital front-end electronics for further treatment.


The digital electronics being created for the Muon System is based on the use of FPGA chips. The prototype of the digital 192-channel MFDM module (see Fig. \ref{fig:RS_FE}~(d)) (Muon FPGA Digital Module) that we have developed includes a XC7A200T chip of the Xilinx Artix 7 family. This unit is functionally, mechanically,  in data format and DAQ interface, compatible with the previously developed   MWDB (Muon Wall Digital Board) unit \cite{Alexeev2005} made on the basis of TDC F1 (ASIC) and  successfully  used for data  readout from the Muon System of the COMPASS experiment (CERN). This approach allows both types of units to be used in the same readout system, thus making it possible for the  new MFDBs cards to be tested under actual operating conditions.

\begin{figure}[!h]
    \center{\includegraphics[width=1\linewidth]{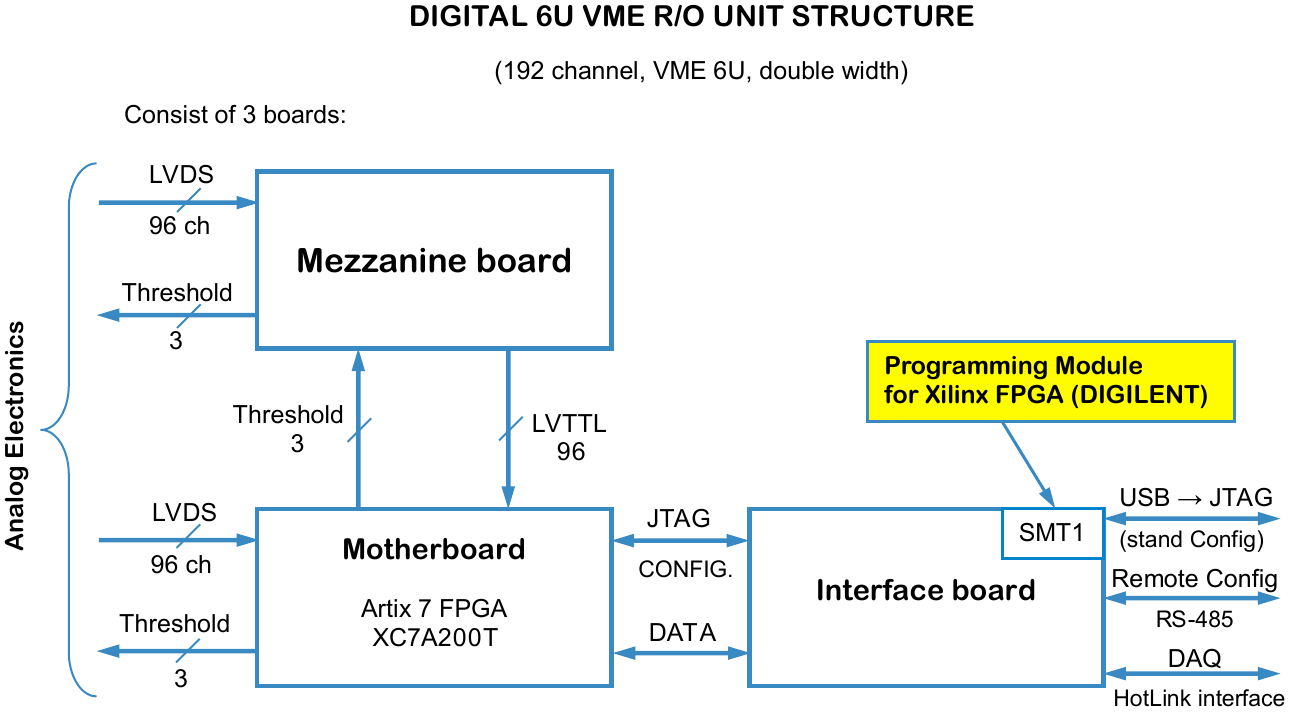}}
\caption{Block-diagram of the MFDM module.}
  \label{fig:MFDM}  
\end{figure}

The unit includes three electronic boards (Fig.\ref{fig:MFDM}): motherboard, mezzanine card, and interface card.
The motherboard accepts 96 LVDS signals from the analog electronics through 3 high-density connectors, converts them to LVTTL levels and writes to the FPGA, and also communicates with two other boards. The mezzanine card also accepts 96 LVDS signals through 3 connectors, converts them to LVTTL levels and transmits through the 120-pin board-to-board connector to the motherboard. The interface card is designed to connect the MFDM module with the DAQ via the HotLink interface (RJ45 connector), to download the firmware to the FPGA from a local computer, as well as to download the firmware via the RS-485 interface (RJ45 connector) from a remote computer.

Tests performed at CERN with the Muon System  prototype on cosmic muons gave encouraging results.
The further tests will be conducted with a prototype of the SPD range system ($\sim$ 1200 channels of wire and strip readout) at the Nuclotron test beam area.

\begin{figure}[!h]
    \center{\includegraphics[width=1.0\linewidth]{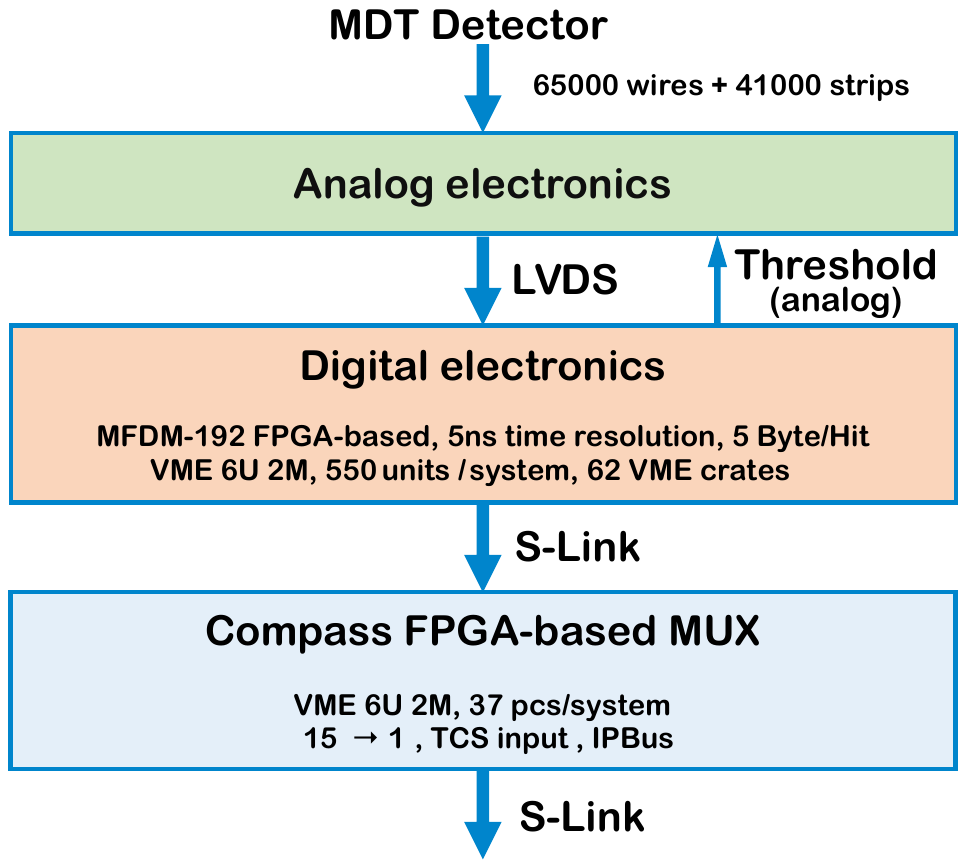}}
     \caption{Data flow diagram – from the RS to the DAQ. }
       \label{fig:DATA_FLOW}  
\end{figure}

In the future, after the final tuning of the unit, we are planning to replace the HotLink interface in the MFDM module with the S-Link interface for direct connection of the Muon System digital electronics to the FPGA-based  SPD DAQ. A general view of the data flow structure for the Muon System is shown in Fig.\ref{fig:DATA_FLOW}.

\subsection{Performance figures }
\begin{figure}[!h]
    \center{\includegraphics[width=0.7\linewidth]{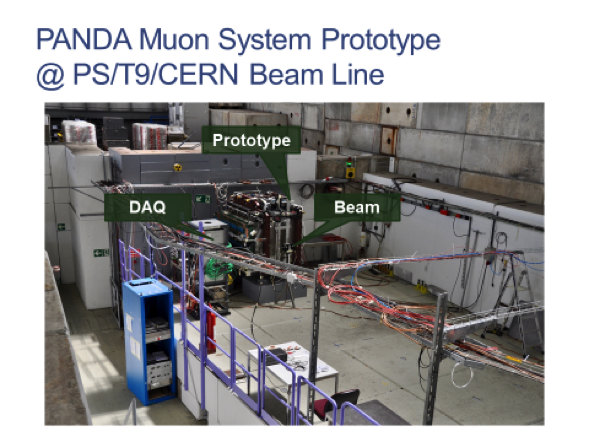}}
     \caption{ Range System prototype (10 ton, 4000 readout channels) at CERN.
       \label{fig:RS_proto}}  
\end{figure}

The evaluation of the main parameters of the proposed Range System is being performed with big prototype installed at CERN within the PANDA program. The prototype (see Fig.\ref{fig:RS_proto}) has a total weight of about 10 tons (steel absorber and detectors with electronics) and comprises 250 MDT detectors with 4000 readout channels (2000 for the wires and 2000 for the strips, 1 cm wide). It has both samplings (3 cm and 6 cm) present in the system (Barrel and End Caps), thus providing an opportunity for direct calibration of the response to muons, pions, protons, and neutrons. 

\begin{figure}[!hbt]
  \begin{minipage}[ht]{0.32\linewidth}
 \center{\includegraphics[width=1\linewidth]{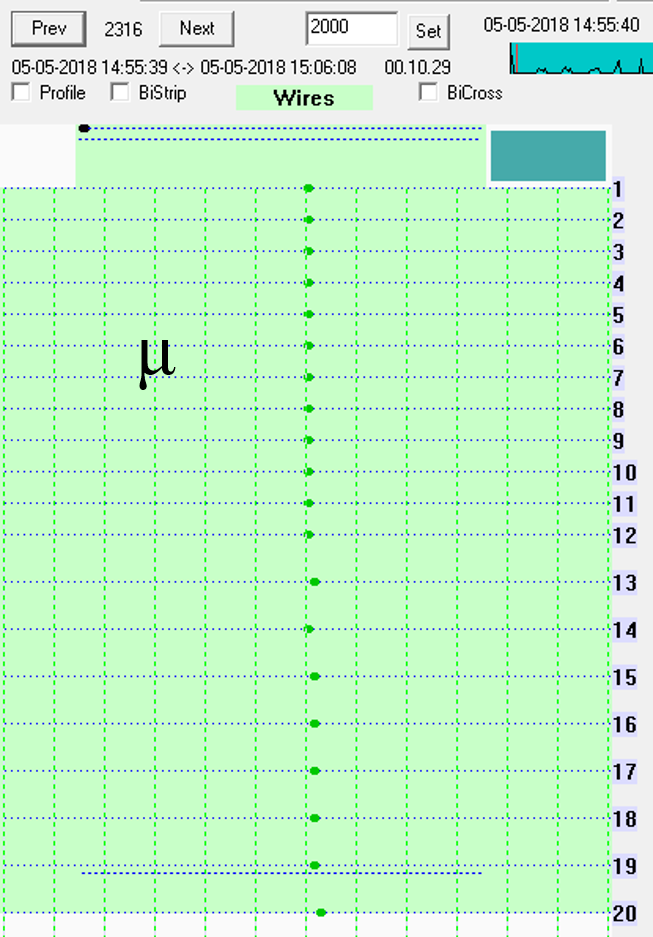} \\ (a)}
 \end{minipage}  
 \hfill
  \begin{minipage}[ht]{0.32\linewidth}
    \center{\includegraphics[width=1\linewidth]{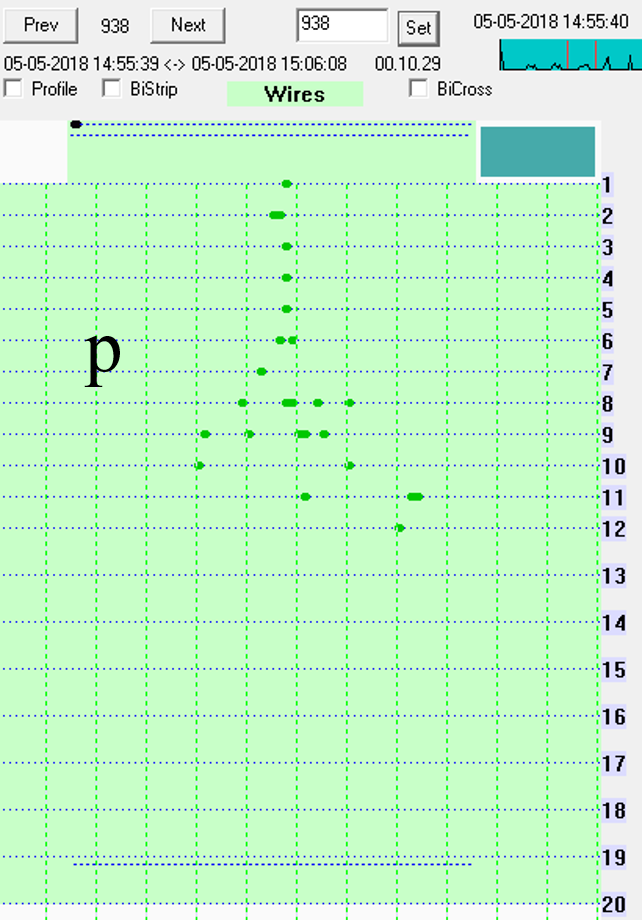} \\ (b)}
  \end{minipage}
   \hfill
  \begin{minipage}[h]{0.32\linewidth}
    \center{\includegraphics[width=1\linewidth]{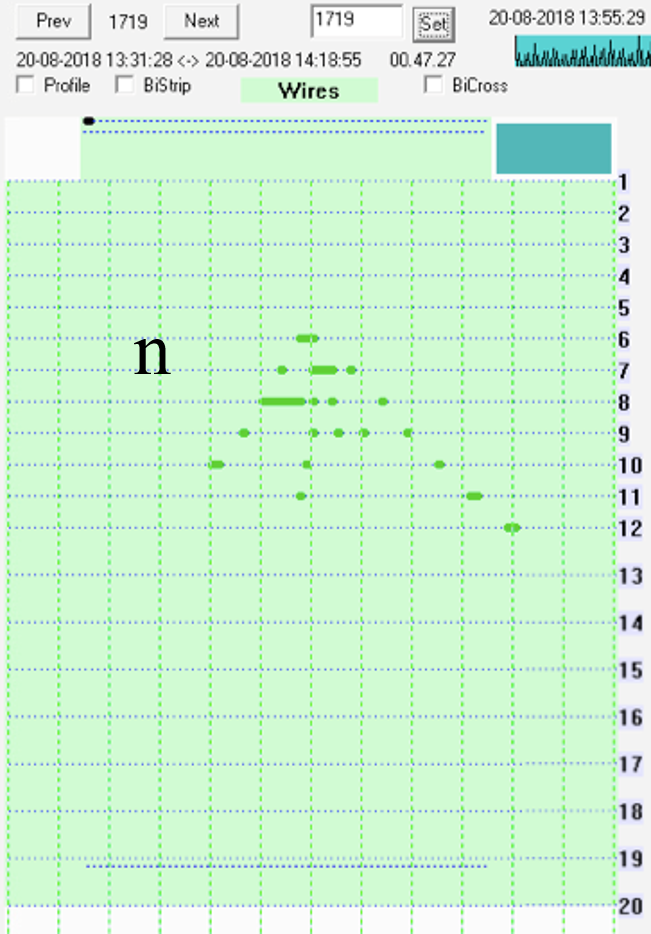} \\ (c)}
  \end{minipage} 
  \caption{Demonstration of PID abilities: patterns for - (a) muon, (b) proton and (c) neutron. }
  \label{fig:RS_PID}  
\end{figure}

Fig.\ref{fig:RS_PID} gives the examples of the prototype response to different particles. The patterns demonstrate excellent PID abilities of the Range System. The data were taken during the May and August runs of 2018 at the T9/PS/CERN test beam. The beam particles hit the prototype from the top of the picture. The beam momentum for all the particles is 5.0 GeV/c. Neutrons were generated by a proton beam on a carbon target placed in the very vicinity of the first detecting layer. The points on the pictures represent hit wires, thus giving the impression of a typical device response with an accuracy  $\sim$ 1 cm.

\subsection{Cost estimate}

The preliminary cost estimate for the Range System is presented in Tab. \ref{RS:cost}.

\begin{table}[htp]
\caption{Cost estimate for the RS.}
\begin{center}
\begin{tabular}{|l|c|}
\hline
 & Cost, M\$ \\
\hline 
\hline
MDT detectors with strip boards  & 2.35   \\
Analog front-end electronics (including cabling) & 2.78 \\
Digital front-end electronics (including VME crates and racks) & 3.05 \\
Yoke/absorber (without supporting structure and movement system) & 6.00 \\
\hline
Total & 14.18 \\
\hline
\end{tabular}
\end{center}
\label{RS:cost}
\end{table}%


\section{Particle identification system }

This Section describes detectors providing identification of particle types such as $\pi$, $K$, $p$. 
The particle identification system of SPD will include 
a time-of-flight detector (TOF) and a Cherenkov threshold detector with aerogel used as a radiator.
Their data will be used in conjunction with the energy loss data registered by the straw tubes.
The latter will serve to identify mainly soft particles which cannot travel more than a meter in the radial direction
because of either decays or interactions. 
Whereas particles with longer trajectories will be able to reach and therefore can be identified using  the TOF and aerogel counters.

The barrel part of the PID system  will be located inside the magnet coils which will limit the radial flight distance for particles to 108\,cm.
The detectors will occupy a maximum radial thickness of 20~cm  between the coils and the Straw Tracker (see Fig.\,\ref{fig:spd1}).
Spatial constraint for the end-caps is  weaker.
In this section, two technologies for  the TOF and one for the Cherenkov aerogel counter are described.
The choice for the baseline option will be made after comparing the respective performances,  costs and  
availability of group which will be able to build the detector.

\subsection{Time-of-flight system}

\begin{figure}[!h]
\centering
\includegraphics[width=\linewidth]{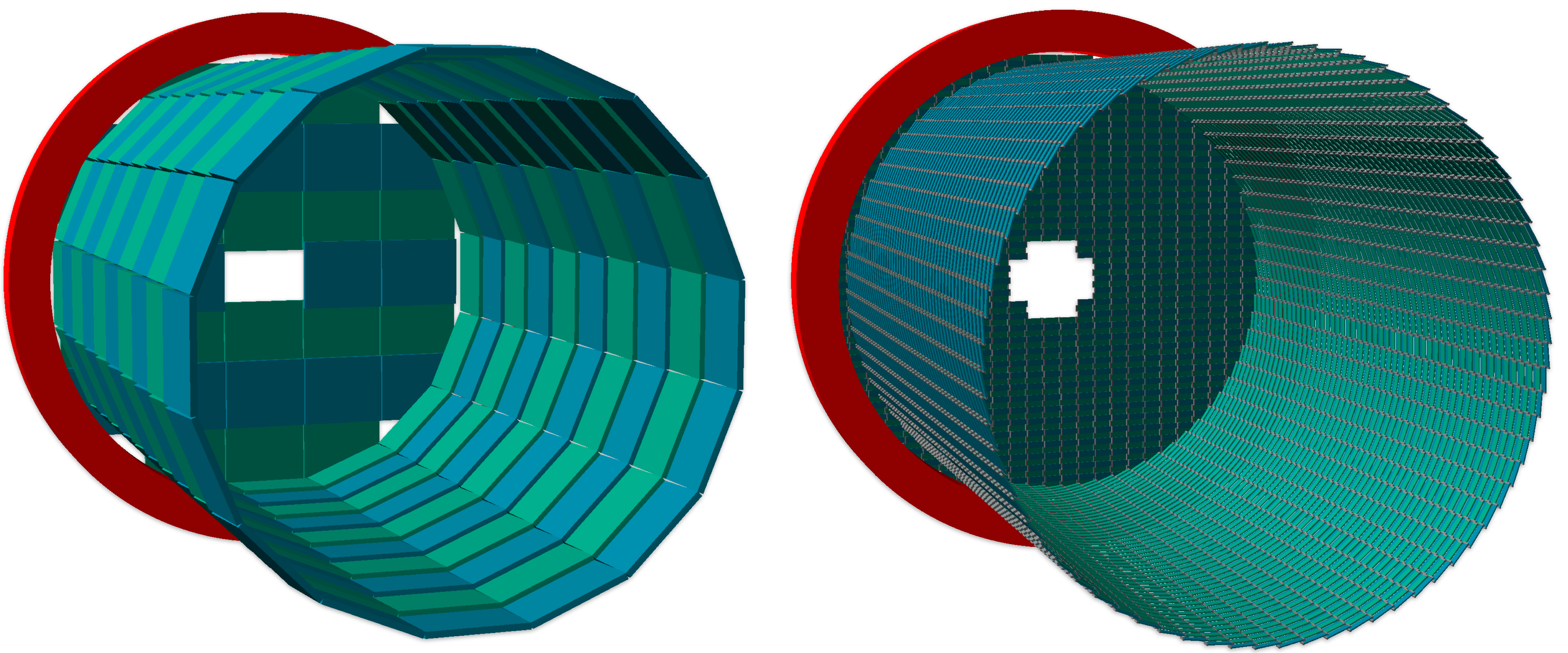}
\caption{Two technologies are being considered  for the time-of-flight system of SPD: 
	 the multigap Timing Resistive Plate Chamber, mRPC (left)
	and the plastic scintillator option (right). Barrel and one of two end-cap parts are shown for both options.
	One of six magnet coils limiting the volume of TOF is shown in red.}
\label{fig:tof_2opts}
\end{figure}

The purpose of the time-of-flight (TOF) system is to discriminate between charged particles
of different masses in the momentum range up to a few GeV/$c$.
A short distance of 108\,cm between the collision point and the TOF dictates the requirement for the time resolution of the TOF to be better than 70\,ps.
In view of  the uncertainty related to the length of the interaction region (about 60\,cm), 
the time $t_0$ which can be assigned to the collision vertex will only be on the order of 1\,ns,
so it is useless for identification purposes. 
Therefore the particle identification (together with the $t_0$ determination) can only be done  for multi-track events, where several particles 
emerged from a same primary vertex
hit active elements of the TOF.  A certain mass hypothesis will have to be applied in this procedure.
For details of this analysis see Section\,\ref{sec:PID} of Chapter \ref{MC}.
In addition to the particle identification,  the detector will also provide a start time to the straw drift tubes.

The TOF system will consists of a barrel and two end-cap parts with an overall active area of 27.1~m$^2$.
The charge particle rate that detector will have to withstand is 0.1~kHz/cm$^2$ for the barrel.
The rate increases rapidly when moving closer to the beam axis. Thus, for  the TOF elements located 
in the end-caps 20~cm off the beam axis, the rate of about 1~kHz/cm$^2$ is expected (see Fig.\,\ref{fig:flux} for details).
Two alternative technologies are being considered for the detector: 
a multigap Timing Resistive Plate Chamber (mRPC)
and a plastic scintillator with Silicon Photomultiplier (SiPM) reading. Both are shown in Fig.\,\ref{fig:tof_2opts}.
These two technologies, in general, can provide about the same efficiency and time resolution, require similar readout 
electronics and have about the same cost per channel. 
Main features as well as pros and cons of both options are listed below:
\begin{itemize}

\item {\it The multigap Timing Resistive Plate Chamber.}
It is a stack of resistive glass plates with high voltage applied to  external surfaces. 
Pickup electrodes are located inside the chamber separating two stacks, 6 gas each.
A fast signal induced on the pickup electrodes by an electron avalanche is further 
transported to FEE located nearby. 
The schematic cross-section of the mRPC is shown in Fig.\,\ref{fig:tof1}.
All aspects of this technology are well tested and deployed in a number of projects
like MPD \cite{TOF_MPD} and BM@N which is described later. 
Overall dimensions of one mRPC is 400$\times$330$\times$25~mm$^3$ which
corresponds to a PCB with 24 readout strips, each 10~mm wide and 400~mm long.
The detector is composed of 220 chambers: 
160  and 30 chambers for the barrel and each end-cap, respectively.
Adjacent mRPCs will be positioned in such a way as to create an overlap of 1 strip at the edge of the active area.
This will ensure the inter-calibration of the mRPCs using tracks crossing both chambers. 
A rectangular shape of each chamber, which is quite large and can not be modified,
creates a certain inconvenience for covering the end-cap parts of detector.
Contributions of all parts of the TOF to the radiation length is about $0.14 \, X_0$ in average \cite{TOF_MPD}.

\item {\it The plastic scintillator option.} 
The basic element of the detector is a plastic scintillator tile with dimensions of 90\,mm$\times$30\,mm×5\,mm.
Scintillation photons produced by an ionizing particle are read out on two ends of tile by an array of 4 SiPMs soldered to custom pre-amplifier PCBs.
A schematic view to the plastic scintillator tile of the TOF/PANDA is presented in Fig.\,\ref{fig:tof_plastic}.
The TOF/SPD detector will be composed of 10.1k staggered  tiles: 
7.3k  and 1.4k tiles for the barrel and each end-cap, respectively.
Heaving much smaller size of a single element of detector, it can easier adopt the cylindrical shape
of magnetic coils and the beam hole.
Furthermore, the system is much lighter in weight than the one of the mRPC and 
the number of radiation lengths for particle crossing the detector is smaller by the factor of 5.
The system manufactured with this technology is easier to maintain than the one of the mRPC
since it does not require  gas circulation, neither high voltage is needed (no trips due to sparks).
This kind of setup, however, was not used  at JINR experiments before
and will require a detailed R\&D. The resistance of scintillator to the radiation level has also to be studied.

\end{itemize}
Dimensions and numbers of the TOF elements for the two options are given in Tab.\,\ref{tab:tof_definition}.
The choice of the technology for the baseline option will be made after more detailed studies of the actual particle rates, comparing the respective performances, the calibration strategy  and the costs.

\begin{table}
\begin{center}
\begin{tabular}{|c|c|c|}
\cline{2-3}
 \multicolumn{1}{c|}{  } & MPRC & Plastic scintillator \\
 \hline
 Active area: & \multicolumn{2}{c|}{  } \\
 barrel + $2 \times$end-cap & \multicolumn{2}{c|}{ 19.8 m$^2$ + 2 $\times$ 3.7 m$^2$ = 27.1 m$^2$ } \\
\hline
 Area of readout element: & strip & tile \\
 pitch $\times$ length &  1.25 cm $\times$ 40 cm = 50 cm$^2$ & 2.9 cm $\times$ 9 cm = 26.1 cm$^2$  \\
\hline
 Size of chamber or tile: & chamber (24 strips) & tile \\
  W $\times$  L $\times$  H & 33 cm  $\times$ 40 cm $\times$ 2.5 cm &  3 cm  $\times$ 9 cm $\times$ 0.5 cm \\
\hline
Number of chambers or tiles: & chambers & tiles \\
 barrel + $2 \times$end-cap &  160 + 2 $\times$ 30  = 220  & 7.3k + 2 $\times$ 1.4k  = 10.1k  \\
\hline
Number of DAQ channels &  10.6k & 20.2k \\
\hline
\end{tabular}
\caption{Dimensions and numbers of the  TOF elements  for two technologies: mRPC and plastic scintillator.}
\label{tab:tof_definition}
\end{center}
\end{table}

\subsubsection{ToF-system based on the mRPC}

The required time resolution for SPD is better than 70 ps while the efficiency of particle registration at high rate (few kHz/cm$^2$) should be above 98\%. Based the experience of building similar systems in such experiments as ALICE \cite{Akindinov:2004fr}, HARP \cite{Ammosov:2009zza}, STAR \cite{STARTOF1}, PHENIX \cite{PHENIXTOF1} and BM@N \cite{BMN1}, a glass multigap  Resistive Plate Chamber could be used as base time detector. For example, the ToF-700 wall in the BM@N experiment, placed at a distance of 8 m  from the target, provides one the $\pi$/$K$ separation up to 3 GeV/$c$ and $p$/$K$ separation up to 5 GeV/$c$ under the assumption that the time resolution of the start timing detector is $<$40 ps.

\begin{figure}[ht]
\centering
\includegraphics[width=0.8\linewidth]{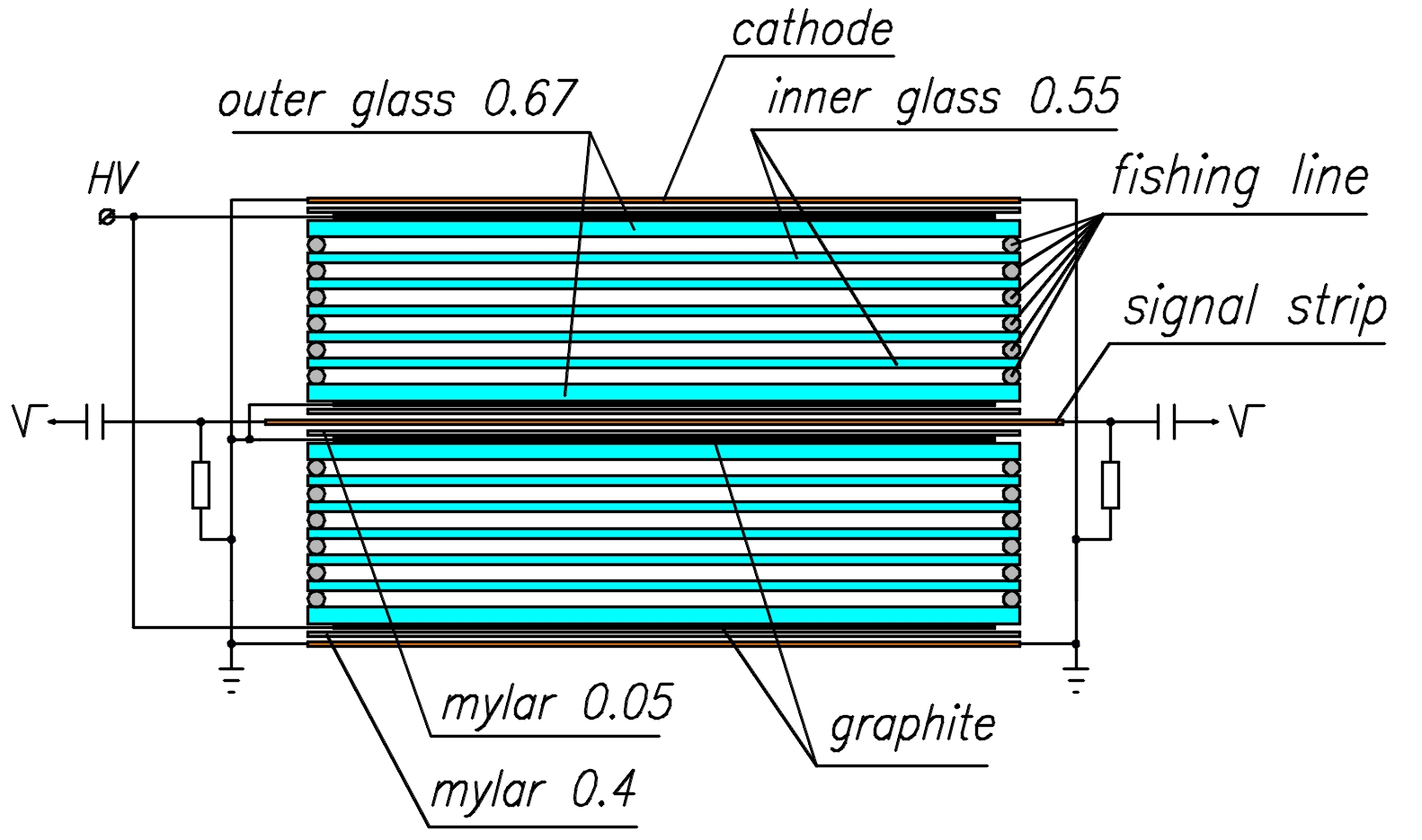}
\caption{Schematic view of the twelve-gap mRPC \cite{Kuzmin:2018nye}.}
\label{fig:tof1}
\end{figure}

The design of the BM@N ToF-700 wall is based on experimental results obtained during multiple tests of various modifications of the glass mRPC exposed in charge particles beam.
The count rate for a standard glass mRPC is limited to several hundreds Hz/cm$^2$ due to the use of conventional float glass plates with a bulk resistivity in the range $10^{12}-10^{13}$ $\Omega \cdot$ cm. Therefore, the extension of the counting rate capabilities of the mRPC has become an important issue.

One of the ways to increase the mRPCs performance at high rates is to use the low resistivity glass (less than $10^{10}- 10^{11}$  $\Omega \cdot$ cm) \cite{Akindinov:2006hs, Ammosov:2007zza, Haddad:2012fx} or ceramics \cite{Akindinov:2017} as the electrode materials. For instance,  time resolutions below 90 ps and efficiencies larger than 90\% were obtained for particle fluxes up to 25 kHz/cm$^2$ for the 10-gap mRPC \cite{Wang:2010bg}. An alternative method is to reduce the glass stack resistance by minimizing  the used electrode
thickness and  increasing a temperature of the glass \cite{GonzalezDiaz:2005xg, Alici:2007zz}. It was shown that such method can provide high time resolution at continuous rate up to ~20 kHz/cm$^2$ \cite{Gapienko:2013nna}.

Schematic cross-section of the mRPC used at BM@N is shown in Fig. \ref{fig:tof1}. It consists of two identical 6-gap stacks with anode strip readout plate in between. The size of mRPC is 473$\times$279$\times$17 mm$^3$ with the working area of 351$\times$160 mm$^2$. Each mRPC has 3210$\times$160 mm$^2$ readout strips with 1 mm gaps between them. 

Each stack is formed by seven glass plates with the $2\times10^{12}$ $\Omega \cdot$cm bulk resistivity. The 0.22-mm gap between the glasses  is fixed by spacers -- usual fishing-lines, which ran directly through the RPC working area. The graphite conductive coating with surface resistivity of $\sim$1 M$\Omega$ is painted to the outer surfaces of the external glass plates of each stack to distribute both the high voltage and its separate ground  to form a uniform electrical field  in the stack sensitive area. 
The anode readout strips plate is a one-sided printed PCB with the thickness of 100 mm. The thickness of the copper layer is 35 $\mu$m. Signals are taken from the both ends of the anode strips. The entire mRPC assembly is put into a gas-tight box. The bottom of box is made of a double-sideed PCB (motherboard) with a thickness of 2.5 mm, the side frame of the box is made of aluminum profile, the top of box is closed by the 1.5-mm aluminum cover. 
The paper \cite{Ammosov:2010zz} presents the performance of the 12-gap mRPC in the range of the counting rate from 0.45 kHz/cm$^2$ up to 10 kHz/cm$^2$ obtained using the secondary muon beam from the U70 accelerator at Protvino. The measurements at different rates were performed in the  temperature range 25-45 $^o$C with the step of 5 $^o$C. The time resolution is reached up to 50-60 ps with good and stable efficiency under the temperature of 40-45 $^o$C. 

Since the particle flux in the SPD experiment is expected to be up to 1 kHz/cm$^2$, a similar approach could be used to build the mRPC-based high-speed TOF with time resolution of about 50\,ps.

\subsubsection{TOF system based on plastic scintillator}

\begin{figure}[t]
\centering
\includegraphics[width=\linewidth]{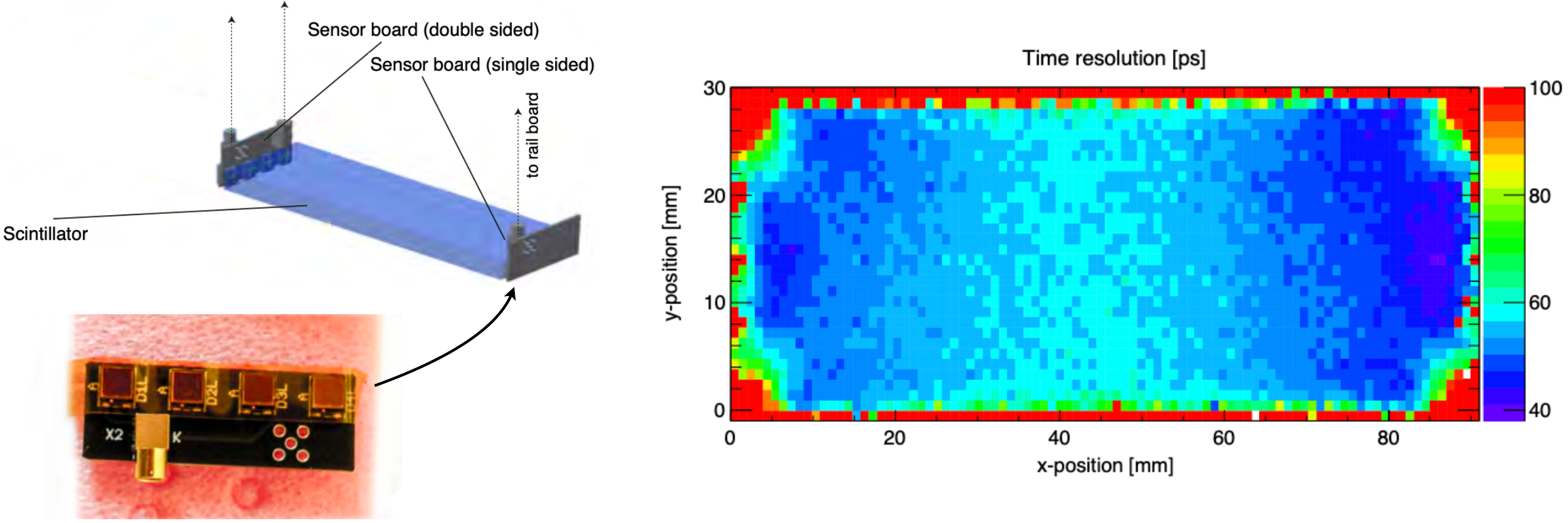}
\caption{Left: schematic view to the plastic scintillator tile and a photo of the 4\,SiPM board of PANDA \cite{PANDA_TOF_CDR}. Right: time resolution obtained from a position scan of a $90\times30\times5$ mm$^3$ EJ-232 scintillator tile read out by Hamamatsu SiPMs attached to opposite sides, 4 SiPMs in series per side \cite{PANDA_TOF_CDR}. }
\label{fig:tof_plastic}
\end{figure}

This option was inspired by the TOF system of the MEGII \cite{MEGII} and PANDA \cite{PANDA_TOF_CDR} experiments.
The surface of the TOF is segmented into many small scintillator tiles 
made  of a fast scintillating organic material. 
The optical readout is performed by Silicon Photomultipliers (SiPM) attached to the ends of every tile. 
A typical number of optical photons produced by a minimum-ionising particle crossing a 1\,cm
plastic scintillator is $\sim$$10^4$. A resulting number of detected photons depends on a signal propagation and photosensor efficiencies. 
Nowadays, large-area SiPMs have appeared on the market at relatively moderate cost and offer several advantages over PMTs: magnetic field tolerance, a much smaller volume and footprint allowing a compact design for bars without light-guides, low operation voltage and high photon detection efficiency (PDE). 
Thus, they ideally meet the requirements dictated by a thousand tile system of the TOF/SPD.

The choice of the scintillator material is primarily driven by the requirement for a short emission time.
Organic scintillators based on a plastic matrix of polyvinyl-toluene, such as
EJ-228 (BC-420)  or EJ-232 (BC-422), have an attenuation length of about 10\,cm,   rise time of 0.5\,ns
and wavelength of maximum emission in UV region of 391\,nm.
They are commonly used for applications as the one discussed in this Section.
Note that, contrarily to the mRPC, the time resolution of a plastic scintillator detector degrades exponentially 
with increase of distance between the interaction point and photosensor.
It is especially crucial for UV photons.
Therefore the choice of scintillator is all the time a compromise between the attenuation length (visible light) and fast emission (UV region). 

In the case of the TOF/PANDA, a scintillator tile with dimensions of $90 \times 30 \times 5$~mm$^3$ 
was read out by 4 Hamamatsu SiPMs coupled to opposite sides 
as shown in Fig.\,\ref{fig:tof_plastic} (left). 
Each SiPM has its sensitive area of $3 \times 3$~mm$^2$, thus the array of four can detect about 
a quoter of photons reaching the edge of tile.
This configuration was chosen as a baseline for estimates for the TOF system of SPD.
The time resolution obtained from a position scan is shown in Fig.\,\ref{fig:tof_plastic} (right).
One can see that the resolution varies from 50\,ps in the near-to-SiPM region to 60\,ps in the center of tile.
The resolution of 100\,ps around the end of tile is, presumably, due to tracks only partially crossing
the volume of scintillator.

Summarizing the advantages of the plastic scintillator option versus the mRPC,
one can say that
the assembling is faster, easier and does not require clean enviroment;
it is easier to maintain the detector (no gas flow, only LV);
the detector can be squeezed within 5\,cm radially, thus leaving space for the aerogel option which is described in the next section;
the circular cross-section of the barrel can be approximated exactly matching the magnet coils from the outside;
the radiation length is only 2\% of $X_0$.
The weak points of the plastic option are: 
an exponentially drop of the resolution vs  distance, which will require larger number of SiPMs for the case of longer tile;
a smaller surface of a single tile, 26\,cm$^2$ vs 50\,cm$^2$ of a mRPC's strip, 
doubles the number of read channels; the organic material is sensitive to radiation.
Although this option is promising it will require a detailed study before being accepted.

\begin{figure}[!ht]
  \begin{minipage}[ht]{0.49\linewidth}
    \center{\includegraphics[width=1\linewidth]{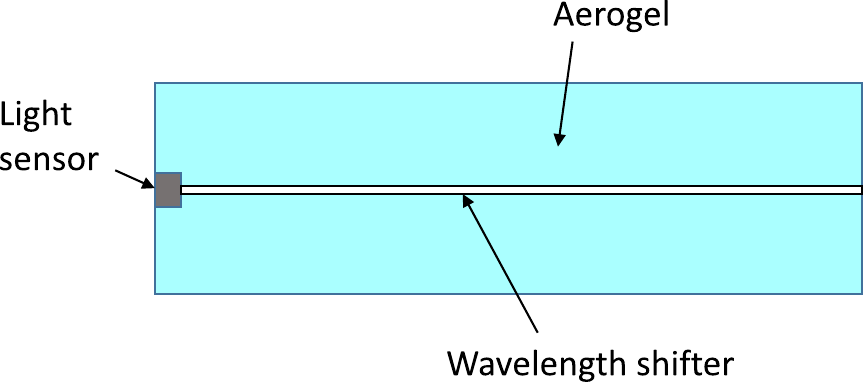} \\ (a)}
  \end{minipage}
  \hfill
  \begin{minipage}[ht]{0.49\linewidth}
    \center{\includegraphics[width=1\linewidth]{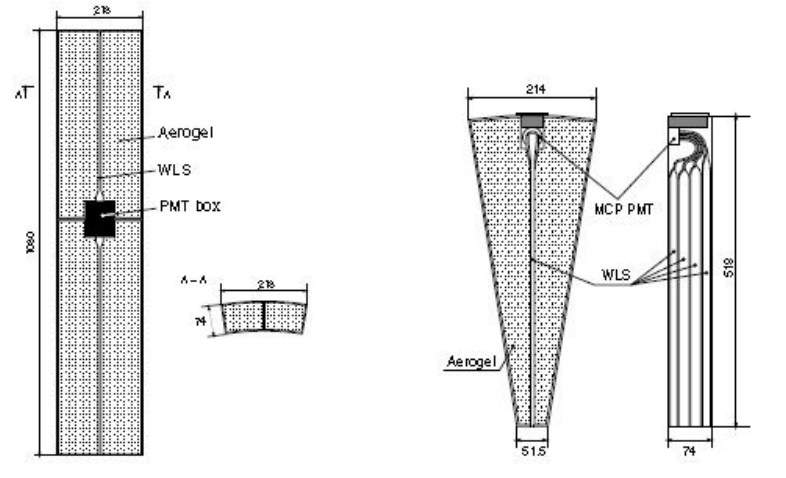} \\ (b)}
  \end{minipage} 
  \vspace{0.2cm}
  
  \begin{minipage}[ht]{0.49\linewidth}
    \center{\includegraphics[width=1\linewidth]{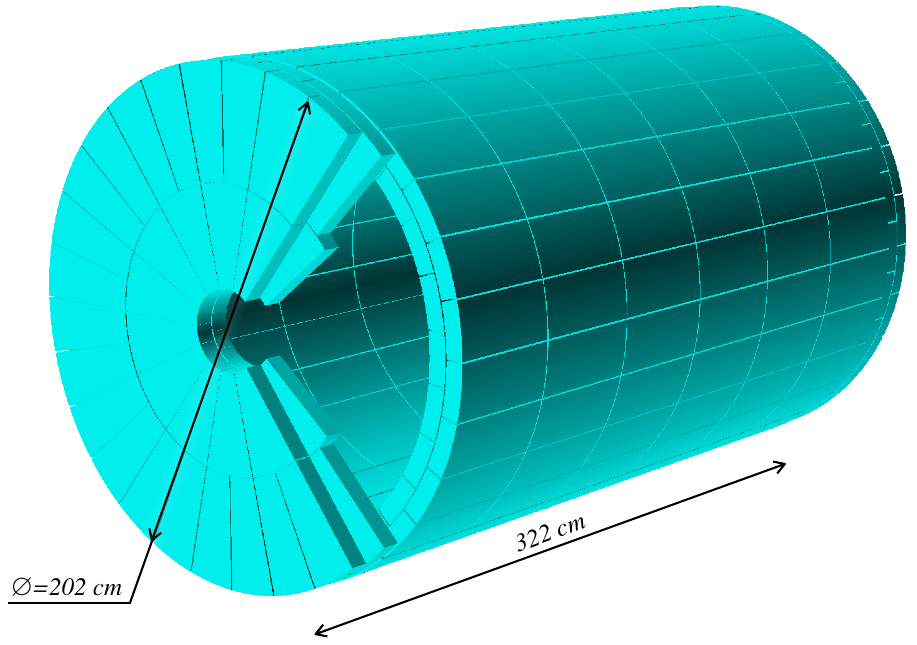} \\ (c)}
  \end{minipage}
  \hfill
  \begin{minipage}[ht]{0.49\linewidth}
    \center{\includegraphics[width=1\linewidth]{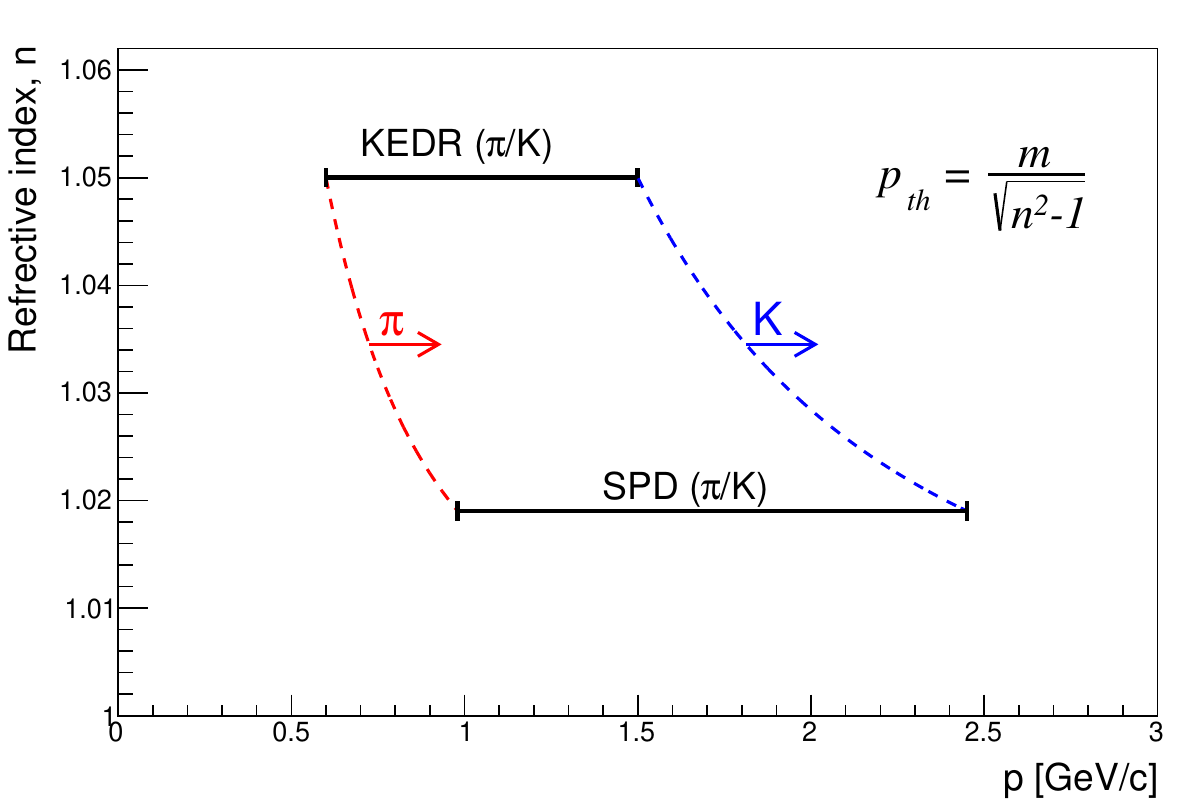} \\ (d)}
  \end{minipage}
  \caption{(a) Principle of ASHIPH operation. 
  	(b) KEDR aerogel counters: two barrel counters in a single housing (left), end-cap counter (right) \cite{KEDR-JINR}.
	(c) Possible arrangement of barrel and end-cap counters in SPD.
	(d) Momentum intervals of KEDR and SPD where the $\pi$/K separation is possible.
	}
  \label{ashiph}  
\end{figure}

\subsection{Aerogel counters}

Another option for the system of particle identification is aerogel Cherenkov counters.
Aerogel is a synthetic porous ultralight material which has found an application,
in particular, as a radiator in Cherenkov counters. Aerogel may have a refractive index
in the range between 1.0006 and 1.2, the exact value of the refractive index being specified
at the production stage. In fact, aerogel fills the gap in the refractive index values
between gases and liquids. This feature of aerogel allows one to use it in Cherenkov counters
for particle identification in conditions when other Cherenkov radiators are not applicable,
for instance, for $\pi$/K separation at the momenta from few hundred MeV/$c$ to about 3\,GeV/$c$.
The selection of the refractive index value defines the region of momenta where the separation is
effective.

There exists a good experience of using threshold aerogel Cherenkov counters, in particular,
in experiments KEDR (BINP, Novosibirsk) \cite{KEDR-JINR}, \cite{KEDR-NIM}, BELLE (KEKB, Tsukuba) \cite{Abashian:2000cg}.
In the BELLE experiment a threshold aerogel Cherenkov counter with refractive index from 1.010 to 1.030
provided $\pi$/K separation in the momentum region up to 3.5 GeV/$c$. In the KEDR detector the aerogel counters
with refractive index 1.05 provided $\pi$/K separation in the range from 0.6 to 1.5\,GeV/$c$.

Aerogel has a short scattering length of light, 12-40\,mm depending on wavelength. Therefore,
directivity of Cherenkov light cannot be used because directivity disappears soon after emission. For this reason
diffusive reflectors are used at the walls. No scintillation has been observed in aerogel. Aerogel samples
suffered from hygroscopicity for a long time, but in the 1990s the technology of hydrofobic aerogel has been developed.

In KEDR, the aerogel counters received the name ASHIPH (Aerogel, SHIfter, PHotomultiplier). In Fig.~\ref{ashiph}\,(a)
a principle scheme of the counter is shown. Cherenkov light from aerogel is captured by a wavelength shifter (WLS).
PMMA light guide doped with BBQ dye is used as wavelength shifter, cross-section of WLS is 3$\times$17\,mm$^2$.
WLS absorbs Cherenkov photons at short wavelengths where Cherenkov radiation is more intensive and re-emit photons at large 
wavelength bringing them to a light sensor. In Fig.~\ref{ashiph}\,(b) the counters of KEDR are shown. The microchannel plate photomultipliers served as light sensors in KEDR, but for later developments the APD were used (BELLE-II), also SiPM are proposed for aerogel detectors in PANDA (GSI) and FARICH (for Super Charm-Tau Factory in Novosibirsk).

If a particle crosses WLS, it produces a much higher signal than the particle traversing aerogel. 
To avoid misidentification, two-layer structure in KEDR is used with shifted layers with respect
to WLS position, so that a particle cannot cross WLS in both layers. The thickness of one layer 
is 74\,mm, a total amount of material in both layers is 0.24$X_0$.

For relativistic cosmic muons that cross both counter layers of KEDR the average number of photoelectrons was $9.3 \pm 0.4$, 
and the detection efficiency $99.3 \pm 0.1$\% at the threshold that equals to 2 photoelectrons. 
For under-Cherenkov-threshold muons ($200 < p_{\mu} < 300$ MeV/$c$) the efficiency was $3 \pm 1$\%.

In SPD aerogel counters {\em a la} ASHIPH can be inserted between the Straw tracker and ECAL. 
Possible arrangement of barrel and end-cap counters in SPD is shown in Fig.~\ref{ashiph}\,(c).
The barrel of SPD is a factor of 2 longer and 50\% wider as compared to the one of KEDR.
Thus the amount of aerogel will have to be three times larger in volume: 1\,m$^3$ and 3.2\,m$^3$ for KEDR and SPD, respectively.
The number of counters will also increase by about the same value.

The value of refractive index can be selected for the momentum region where $\pi$/K separation is the most important. 
As will be shown in Fig.\,\ref{fig:TOF_m}, this separation up to 1\,GeV/c can be provided by the TOF.
Aerogel will be employed for momenta above this value which corresponds to the refractive index equal to 1.02
as shown in Fig.~\ref{ashiph}\,(d). One can see from the figure that kaons begin generating a signal in the counters 
having momenta above 2.5\,GeV/$c$, which defines the region of detector applicability.

The detector is thin (in terms of $X_0$), fast and rather simple in operation. We estimate its cost to be of about 5 M\$.


\section{Beam-beam counters}

\begin{figure}[!h]
  \begin{minipage}[ht]{.9\linewidth}
    \center{\includegraphics[width=1\linewidth]{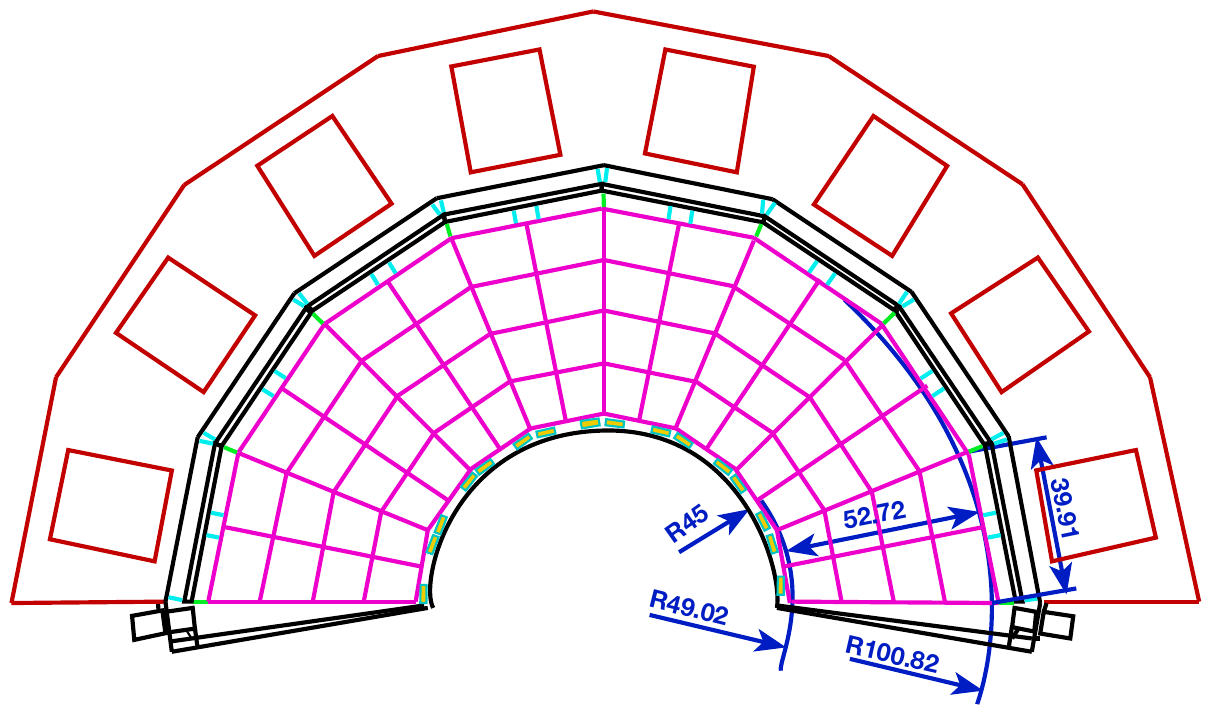}}
  \end{minipage}
  \caption{Beam-Beam Counter: azimuthal and polar angle segmentation \cite{Antropov:1998th}. All dimensions are in millimeters.}
\label{baldin_fig0}
\end{figure}

Two Beam-Beam Counters (BBCs) are planned to be located just in front of the PID system in the end-cups of the SPD setup. 
The detector should consist of two parts: the inner and the outer one, 
which are based on different technologies. 
The inner part of the BBC will use fast segmented MicroChannel Plate (MCP) detectors and should operate inside the beam pipe,
while the BBC outer part will be produced from fast plastic scintillator tiles.
The inner part covers the acceptance 30$\div$60 mrad and should be separated into 
4 layers consisting of 32 azimuthal sectors. The outer part covering the polar angles between 60 and 500 mrad will be divided into
5 or 6 concentric layers with 16 azimuthal sectors in the each of them.
The final granularity is the matter of further optimization for the entire energy range of collisions at SPD. The concept of the BBC is shown in Fig.\ref{baldin_fig0}. The magenta part represents the MCP detector while the internal layer of the outer part is shown in red.

The main goals of the Beam-Beam Counters are: i) the local polarimetry at SPD basing on  the measurements of the azimuthal asymmetries in the inclusive production of charged particles in the collisions of transversely polarized proton beams, ii) the monitoring of beam collisions and iii) participation in the precise determination of the collision time $t_0$ for events in which other detectors can not be used for that (for instance in case of elastic scattering).

Another important goal of the BBCs is fast preselection of different types of events for monitoring purposes. 
The Monte Carlo simulation shows that in the $p$-$p$ collisions at $\sqrt{s}$=27~GeV   
at least one BBC should have a signal in 79\% of events (51\% of events has a signal in the both BBCs).
However, for the hard processes,   in 97\% of events  the only one  
BBC will be hit, while   hits  for both counters could be expected in 68\% of cases.
Therefore, the requirement of the BBC signals allows one to preselect hard processes. 

\subsection{Inner part of the BBC: MCP}

\begin{figure}[!h]
  \begin{minipage}[ht]{0.49\linewidth}
    \center{\includegraphics[width=1\linewidth]{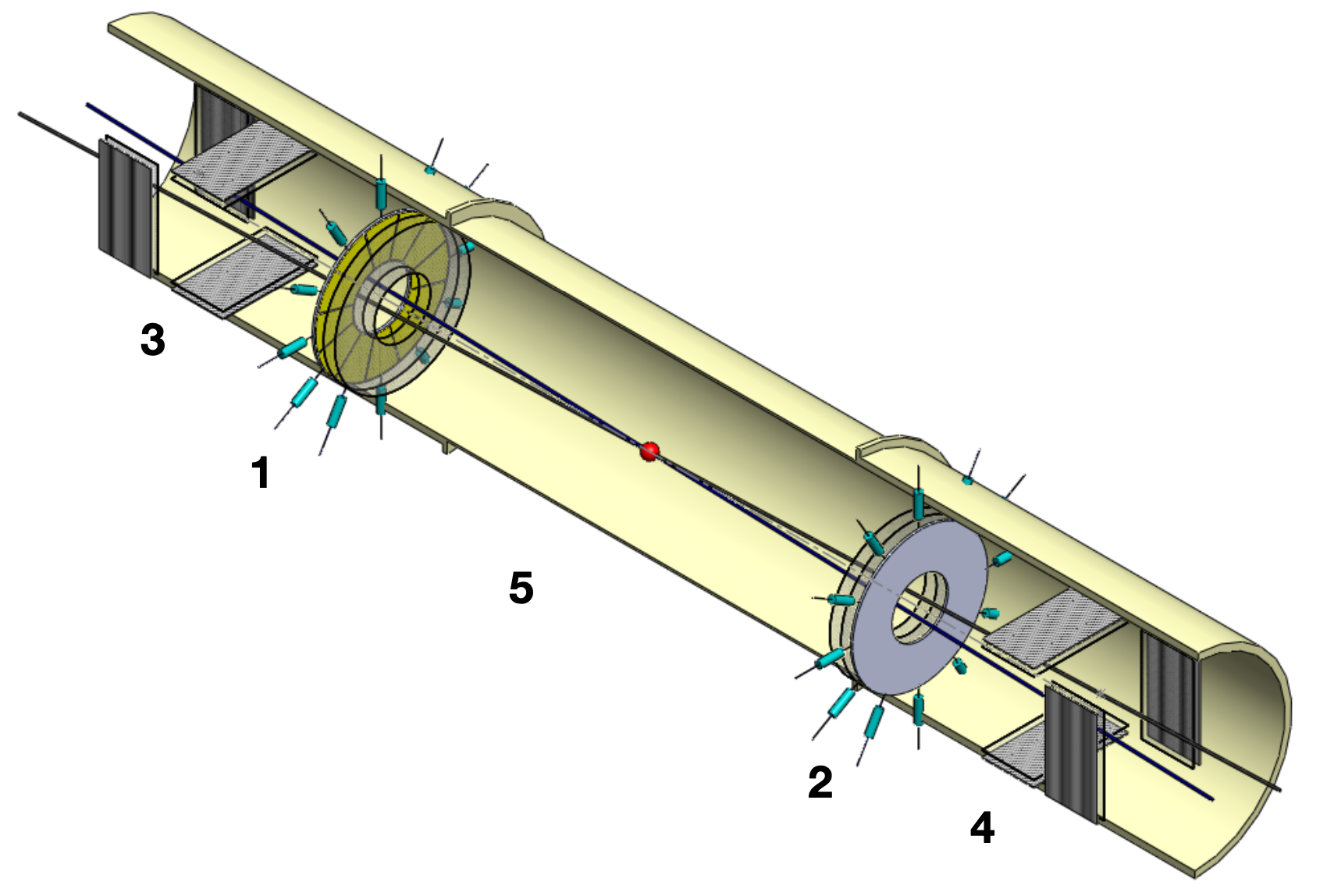} \\ (a)}
  \end{minipage}
  \hfill
  \begin{minipage}[ht]{0.49\linewidth}
    \center{\includegraphics[width=1\linewidth]{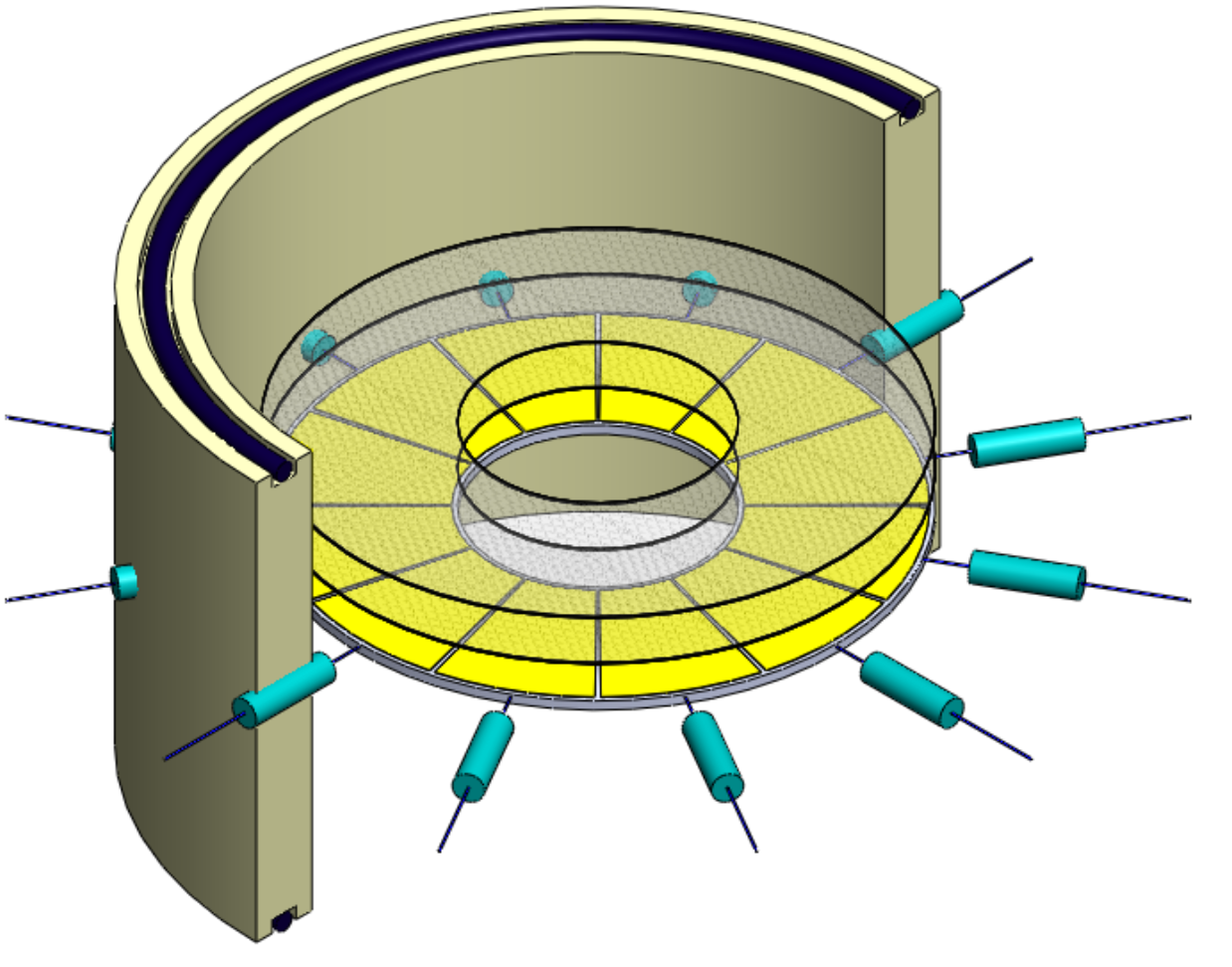} \\ (b)}
  \end{minipage}
  \caption{ (a) Possible layout: the Fast Beam-Beam Collision (FBBC  \cite{Baldin:2019rla}) monitor composed of MCP discs (1 and 2) in combination with the Beam Position devices (BPM) (3 and 4) are situated symmetrically to the Interaction Point (5) inside the vacuum beam pipe of the NICA collider.
(b) Compact module of the Fast Beam-Beam Collision Monitor (FBBC) based on circular MCPs. Sector cathode readout pads and two MCP setups are embedded into a separate flange with a hermetic 50-$\Omega$ signal and HV feedthroughs (the latter are not shown).}
\label{baldin_fig1}
\end{figure}

The concept of the MCP-based Fast Beam-Beam Collision (FBBC) monitor is described in details in \cite{Baldin:2019rla}.
Two ring beam-beam collision detectors (FBBC-left and FBBC-right) could  be located inside the vacuum beam line 
together with two 2D position-sensitive beam imaging detectors (BPMs) (see Fig. \ref{baldin_fig1}).
The FBBC uses the concept of the isochronous multi-pad fast readout and the precise timing determination of short ($\sim$1~ns) MCP signals. Individual fast readout of pads is being also  considered.  New MCPs with the improved characteristics, such as small diameter (8$\mu$m) channels, low resistivity (100$\div$500 M$\Omega$), high gain ($\sim$10$^7$), short fast rise-time ($\sim$0.8ns) signals, could be used.

\begin{figure}[!h]
\begin{center}
  \begin{minipage}[ht]{0.49\linewidth}
    \center{\includegraphics[width=1\linewidth]{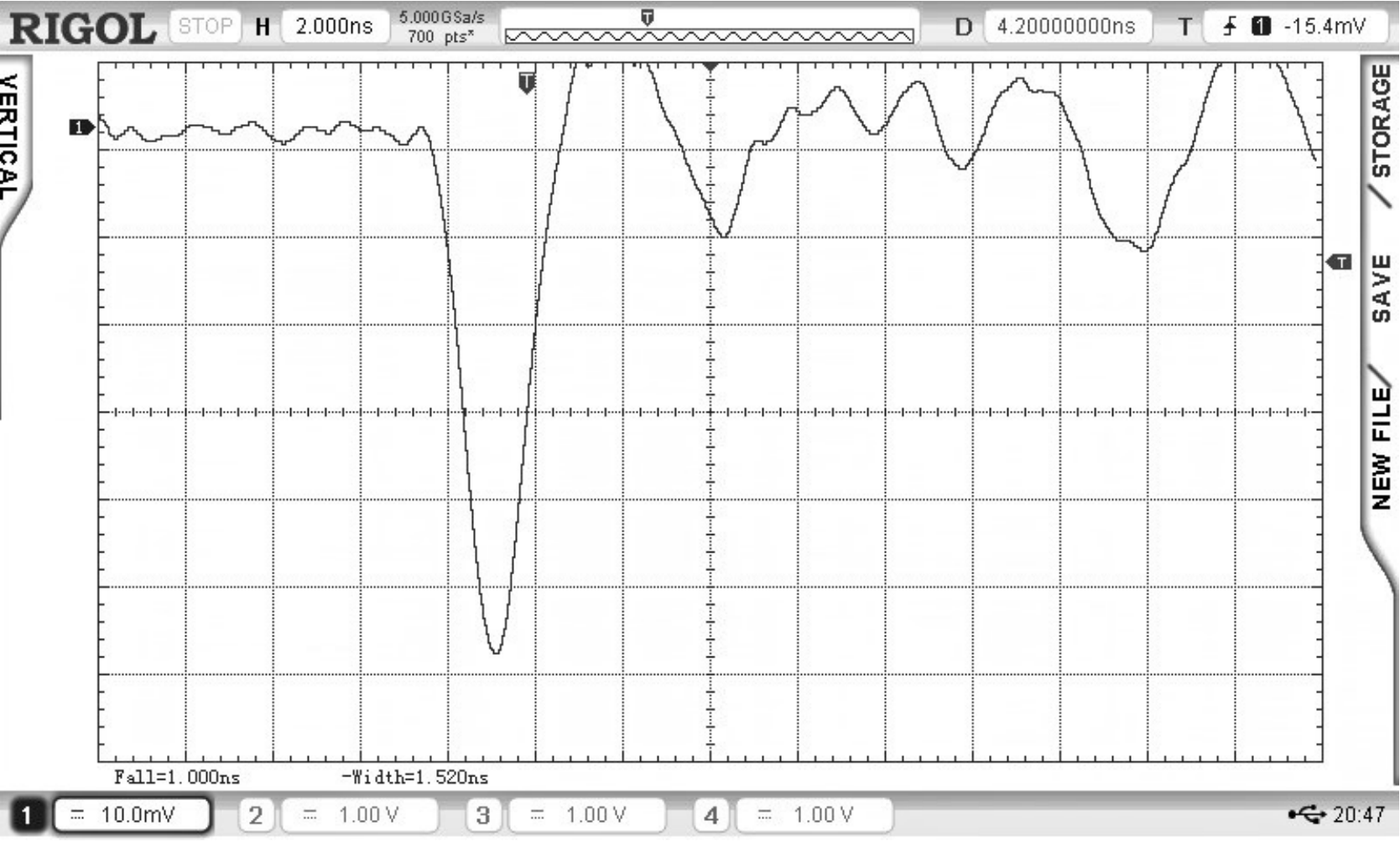}}
  \end{minipage}
  \end{center}
  \caption{ A typical MCP signal the detector  prototype. }
\label{baldin_fig2}
\end{figure}

Fig.\ref{baldin_fig2} shows a typical MCP signal from the detector  prototype.
The colliding beams pass through the central opening of the MCP, and the outer edges of the MCP capture secondary particles flying a definite distance from the interaction point. The signals are recorded at the sectorized cathode and their arrival times are digitized along with the multiplicity information. 
The main feature of the new MCP is a high secondary emission coefficient and fast fronts of the output signal. The new MCPs have a fast leading edge and high gain.

The feasibility of the event-by-event monitoring of the beam-beam interactions at NICA is confirmed both by the previous developments of the UHF-UHV technology and by the beam tests (at JINR and CERN) \cite{Baldin:2019rla,Antropov:1998th, Bondila:2002sy} of the prototype detectors and electronics, as well as by the in-lab tests of new 8-channel MCPs with the improved characteristics.


\subsection{Outer part of the BBC: scintillation tiles} 

The scintillation part of the BBC will consist of tiles viewed by the SiPMs. The measurement of the signal amplitude is required for time-walk correction to improve the time resolution. 
\begin{figure}[!t]
  \begin{minipage}[ht]{0.49\linewidth}
    \center{\includegraphics[width=1\linewidth]{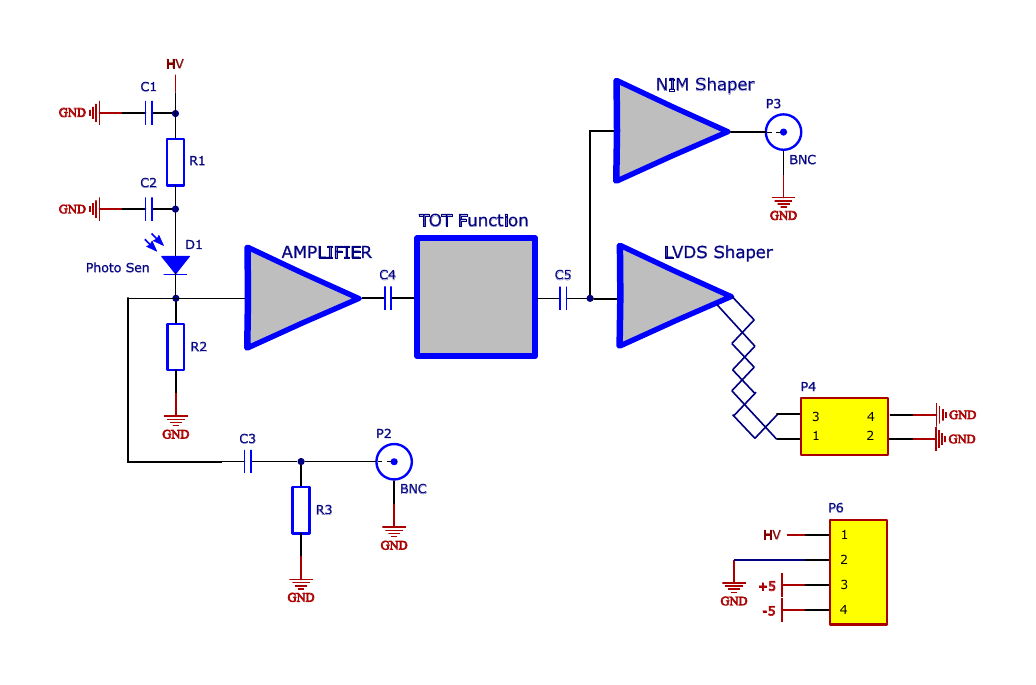} \\ (a)}
  \end{minipage}
  \hfill
  \begin{minipage}[ht]{0.49\linewidth}
    \center{\includegraphics[width=1\linewidth]{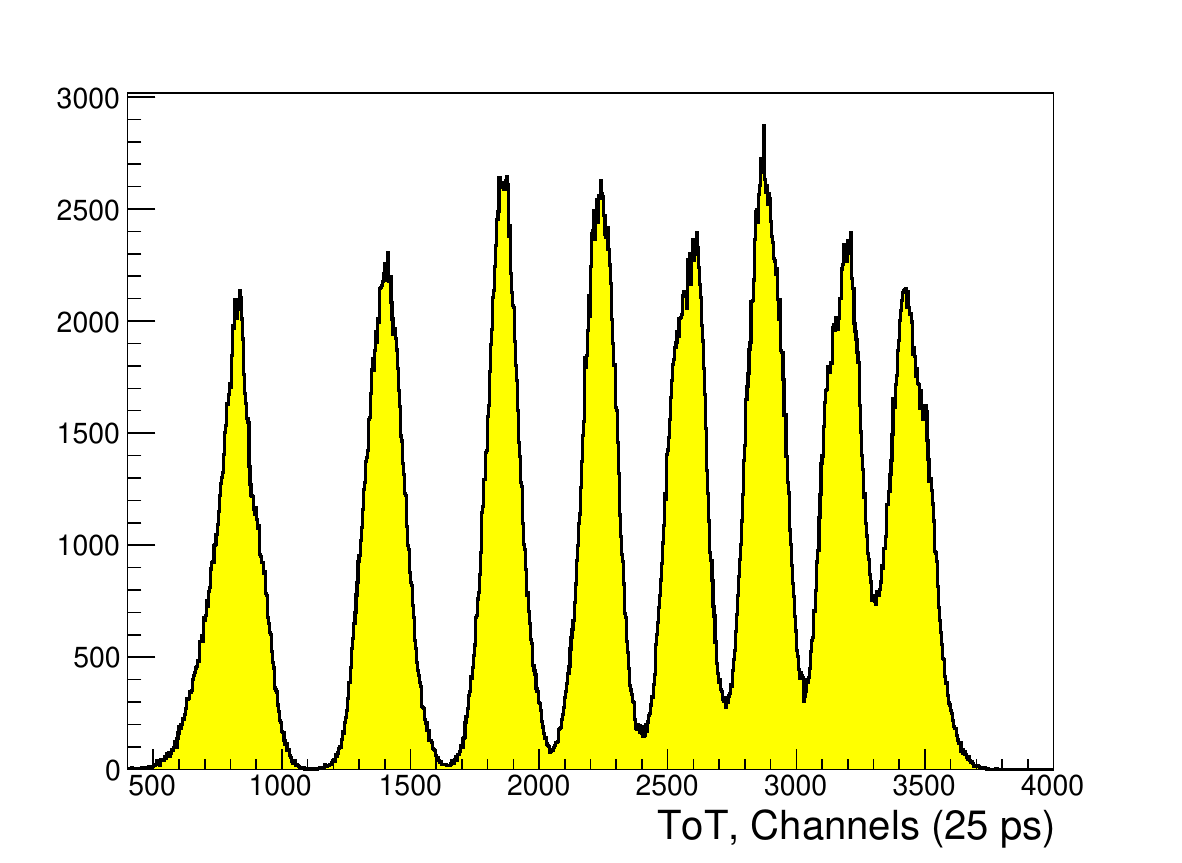} \\ (b)}
  \end{minipage}
  \caption{ (a) Schematic view of the front-end electronics with a ToT function,
 (b)  Distribution of the ToT for LED signal. }
\label{ladygin_fig1}
\end{figure}

With a single-channel prototype of the detector we will be able to measure the amplitude using the developed FEE based on the Time-over-Threshold (ToT) technique. This technique is a well-known method that allows us to measure the energy deposited in the material by reconstructing the given property of the output current pulse – the total charge collected, the pulse amplitude, etc. The ToT method converts the signal pulse height into a digital value in the early stage of the FEE, which greatly simplifies the system in comparison to analog detectors with serial readout through ADCs. The measurement of the ToT is composed of two measurements of time for the signal going above (leading) and returning below (trailing) the given threshold. The first version of the prototype includes a power supply and electronics 
(Fig.\ref{ladygin_fig1}(a)) made on a separate PCB. This PCB is used for each cell of the SiPM. The power supply for the SiPM 
provides a voltage of up to 65 V with an individual channel adjustment within 0-10 V, manual tuning, and a built-in voltmeter for the voltage monitoring
It is possible to connect eight cells simultaneously. The amplifiers used for that do not change the leading edge of the signal. This allows us to get a time stamp of the event. Afterwards, the signal is integrated and transmitted to the comparator.
\begin{figure}[!t]
  \begin{minipage}[ht]{0.49\linewidth}
    \center{\includegraphics[width=1\linewidth]{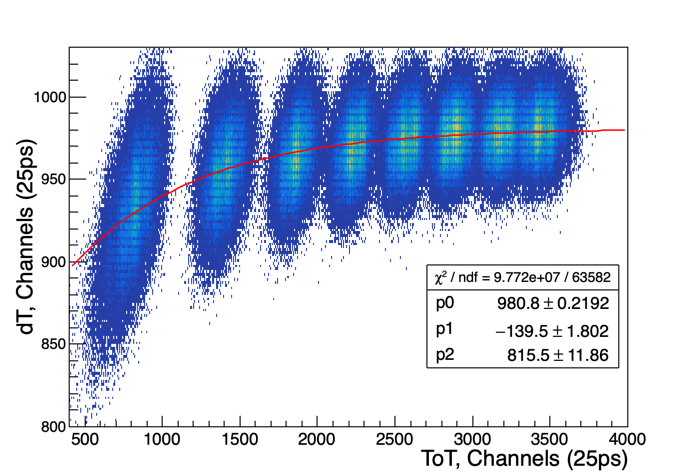} \\ (a)}
  \end{minipage}
  \hfill
  \begin{minipage}[ht]{0.49\linewidth}
    \center{\includegraphics[width=1\linewidth]{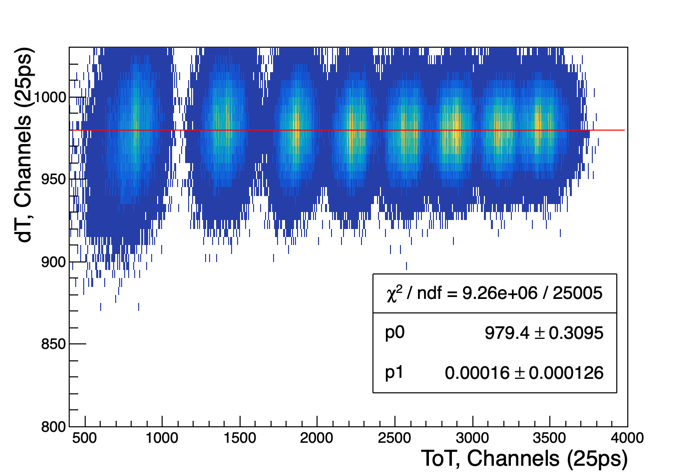} \\ (b)}
  \end{minipage}
  \caption{ (a) $dT$ ($T_{SiPM1} - T_{SiPM2}$) correlation on the ToT.  
 (b) Result after the time-walk correction for the 
$dT$ ($T_{SiPM1} - T_{SiPM2}$) correlation on the ToT. }
\label{ladygin_fig2}
\end{figure}

\begin{figure}[!t]
  \begin{minipage}[ht]{0.49\linewidth}
    \center{\includegraphics[width=1\linewidth]{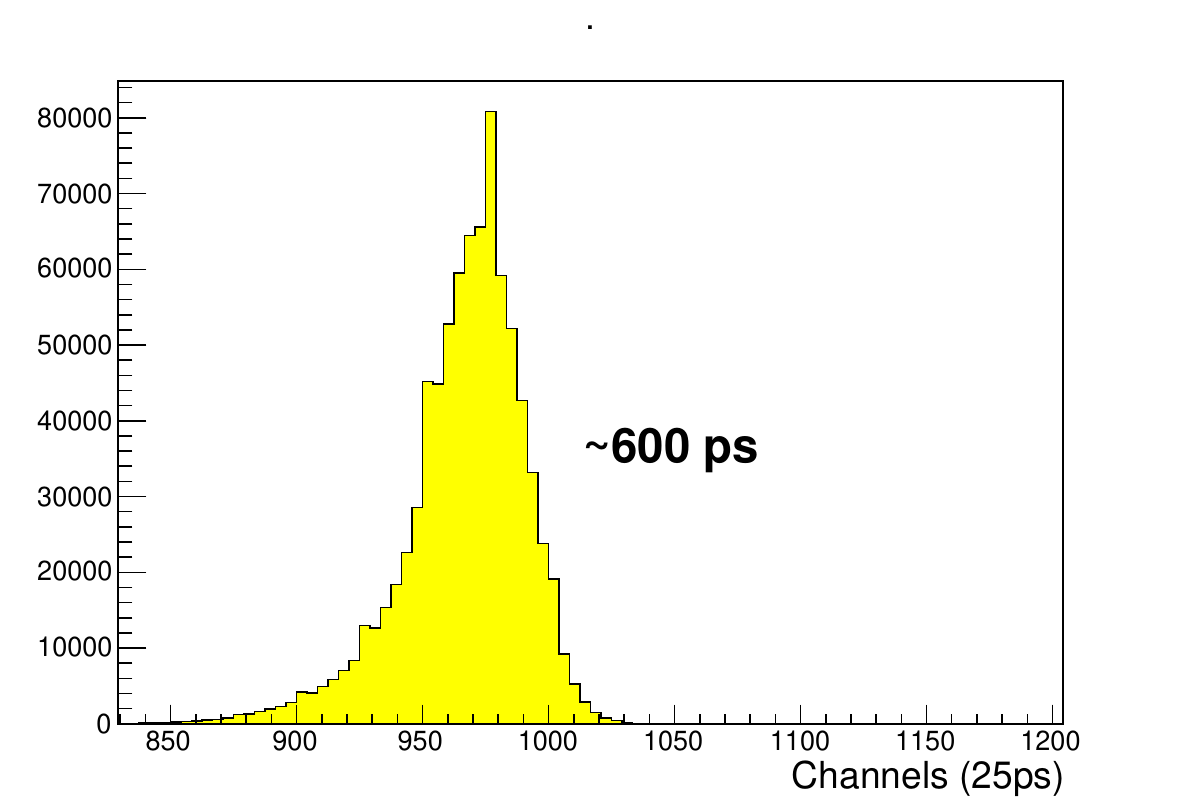} \\ (a)}
  \end{minipage}
  \hfill
  \begin{minipage}[ht]{0.49\linewidth}
    \center{\includegraphics[width=1\linewidth]{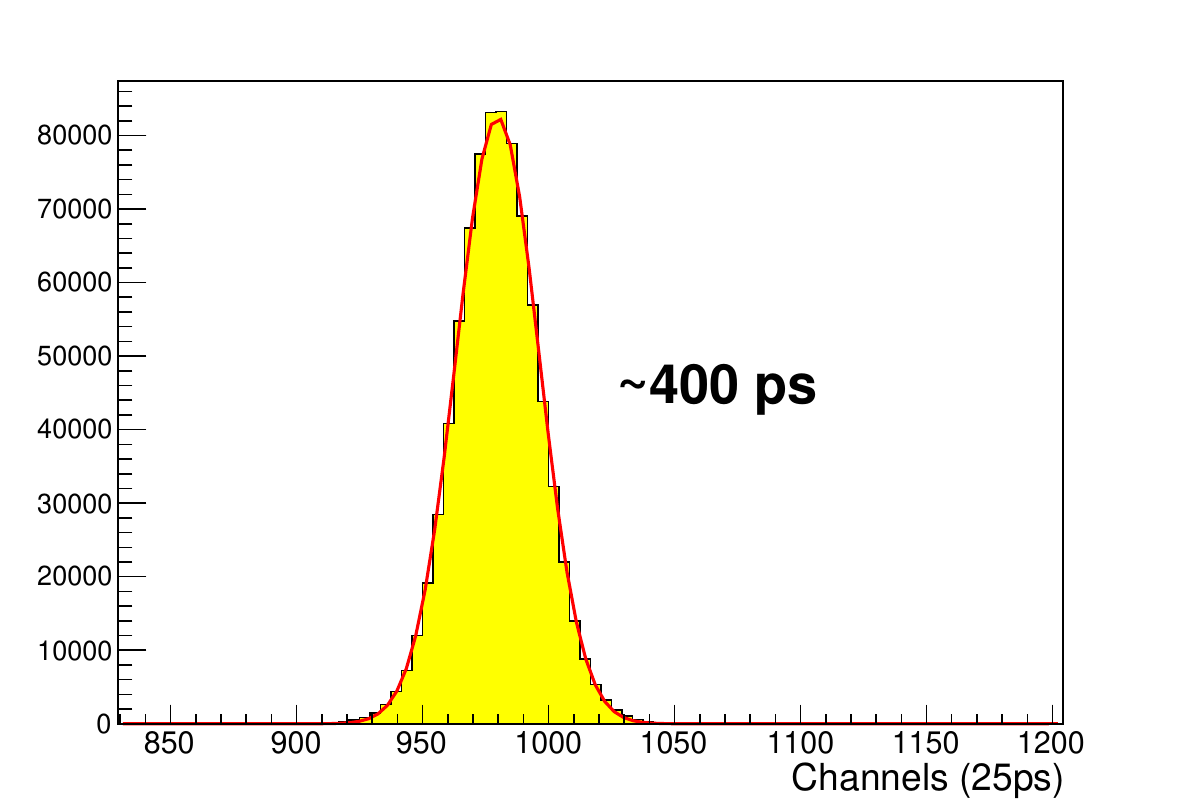} \\ (b)}
  \end{minipage}
\caption {(a) $dT$ ($T_{SiPM1} -T_{SiPM2}$).  
 (b) Result after the time-walk correction for the 
$dT$ ($T_{SiPM1} - T_{SiPM2}$).}
\label{ladygin_fig3}
\end{figure}

The response of the Hamamatsu S12572-010P SiPM \cite{hamamatsu_tot} with the FEE to the LED has been studied. The electrical signal from a LEMO output of the LED was used as a trigger.   The illumination was performed  by uniform light 
in a light-isolated box. 
In addition to the ToT information (Fig.\ref{ladygin_fig1}(b)), the time stamp of the event for each SiPM cell was investigated. The distribution (Fig.\ref{ladygin_fig2}(a)) shows the correlation of these values and that the signal in the region of small amplitudes comes later in time. This is due to signal latency (the so-called time-walking effect). This delay occurs due to the difference between the time when a photon or a charged particle passes through the detecting element and the time when the electronics registers this signal. This leads to deterioration in the time resolution.  After performing the correction (see Fig.\ref{ladygin_fig2}(b)), the time-walking effect has been removed \cite{FEEtot1}.

The time resolution was defined as the RMS and was approximately 600 ps. Taking into account the non-Gaussian waveform (Fig.\ref{ladygin_fig3}(a)) and the fact that the time resolution is not the maximum allowed for this type of the detector, the time-walk correction has been applied.  The most important result of the correction was a time resolution of approximately 400~ps (Fig.\ref{ladygin_fig3}(b)), which is 1.5 times better than the resolution before the correction.

The first version of the prototype using developed front-end electronics based on the Time-over-Threshold method was tested. After the time-walk correction, the time resolution improved up to 400 ps. Taking into account the SiPM suboptimal for precise time measurements, the result is promising.
Further development of the FEE with a ToT function allows using standard TDCs for timing
measurements.

\section{Zero degree calorimeter}

\begin{figure}[!h]
\centerline{\includegraphics[width=0.9\linewidth]{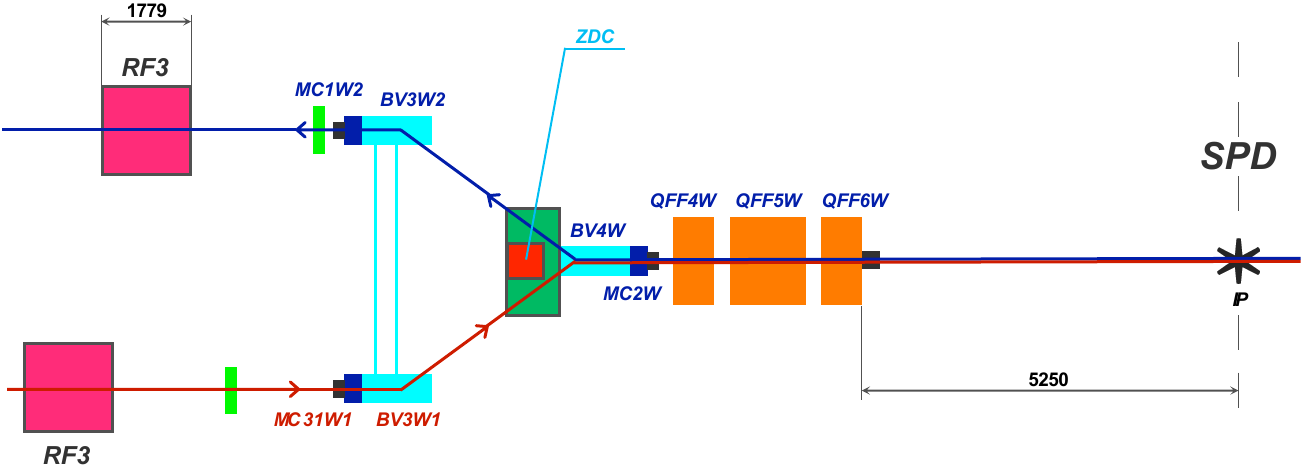}}
\caption{ZDC placement at the collider. All dimensions are in millimeters.}
\label{fig:ZDC0}
\end{figure}

A zero degree calorimeter (ZDC) will be installed in the beam separation areas on both sides of the SPD interaction point as it is shown in Fig. \ref{fig:ZDC0}, where all charged particles originating from the interaction region  are swept out by the strong magnetic field. Main goals of the ZDC are:
\begin{itemize}
\item luminosity measurement;
\item local polarimetry with forward neutrons (see Chapter \ref{ch:pol});
\item spectator neutron tagging;
\item time tagging of the events for event selection.
\end{itemize}

The ZDC is a standard system used for the purposes mentioned above in many collider experiments. The strong magnetic field before a ZDC efficiently removes all charged particles allowing clean measurement of neutrals, so the device can work till very high luminosities. Scattering at zero angle is insensitive to transverse polarization and provides offset-free luminosity measurement, very useful for luminosity crosschecks. The device could provides an additional time stamp for an event. Standard usage of ZDC includes local polarimetry with forward neutrons \cite{Adare:2012vw}. The main purpose is the verification of longitudinal polarization settings. One can expect $1\div 2$ \% asymmetry for very fast neutrons at the position of the detector. Precise tagging of very forward spectator neutrons provides a range of opportunities in study of diffractive processes.
To accomplish these tasks, the following performance parameters should be met:
\begin{itemize}
\item	time resolution $150\div200$ ps;
\item	energy resolution for neutrons $50\div60\%/\sqrt{E} \oplus 8\div10\%$;
\item	neutron entry point spatial resolution 10 mm.
\end{itemize}

 \begin{figure}[!h]
\centerline{\includegraphics[width=0.7\linewidth]{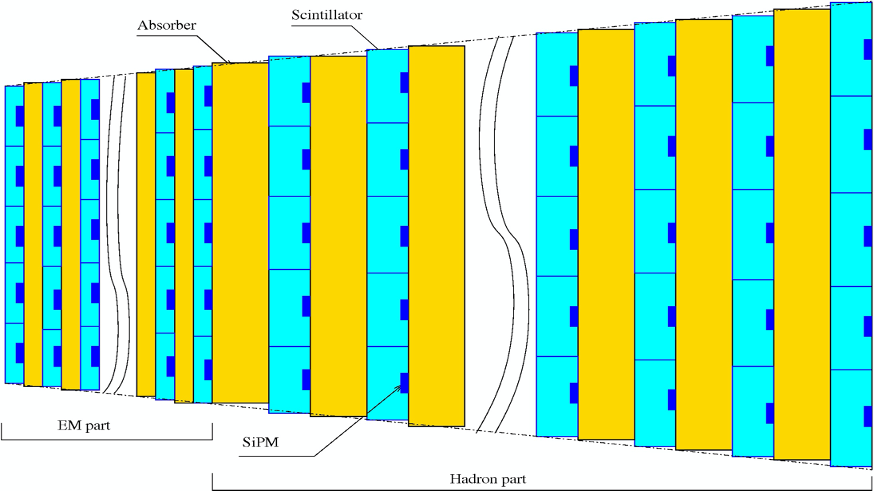}}
\caption{Schematic view of the ZDC.}
\label{fig:ZDC1}
\end{figure}
The design of the Zero Degree Calorimeter was optimized on the basis of the Monte Carlo simulation to obtain the necessary energy, spatial and time resolutions. 
We propose to use a fine segmented calorimeter based on plastic scintillator active tiles with direct SiPM readout and tungsten absorber plates similar to the calorimeter proposed for the CALICE \cite{collaboration:2010hb}. A schematic view of the calorimeter is shown in Fig. \ref{fig:ZDC1}. In order to achieve good energy resolution for photons the first 10 layers will have smaller thickness compared to the rest of the calorimeter (see Tab. \ref{ZDC_pars}). This allows to get reasonable energy resolution for photons, while keeping reasonable number of readout channels. Each scintillator layer has 25 tiles arranged in 5$\times$5 grid with tile size growing from 17$\times$17 mm$^2$ for the first layer to 28$\times$28 mm$^2$ at the last layer. Each tile is covered by a chemically produced thin white reflective layer with a small transparent window for optical readout, done by HAMAMATSU SiPMs S13360-3050PE directly coupled to the tiles. Output signals are digitized by 500 MHz flash ADC 16 channel boards. A fast output for SPD trigger system is also produced. The ZDC will be placed inside the beam separation magnet and its size is limited to 88$\times$88 mm$^2$ at front side and 140$\times$140 mm$^2$ at rear side. The length is limited to 650 mm.

\begin{table}[htp]
\caption{The parameters of the  ZDC layers.}
\begin{center}
\begin{tabular}{|l|c|c|}
\hline
Parameter & Electromagnetic & Hadron \\
                 &  part                    & part       \\
\hline
\hline
Number of layers & 10 & 26 \\
Scintillator thickness, mm & 5 & 10 \\
Absorber thickness, mm & 5 & 10 \\
Total absorber thickness, mm & 45 &  260 \\
Part thickness, $X_0$ & 13 & 75 \\ 
Part thickness, $\lambda_I$ & 0.5 & 2.9 \\
Number of channels  & 250 & 650 \\
\hline
\end{tabular}
\end{center}
\label{ZDC_pars}
\end{table}%

 \begin{figure}[!h]
\centerline{\includegraphics[width=0.9\linewidth]{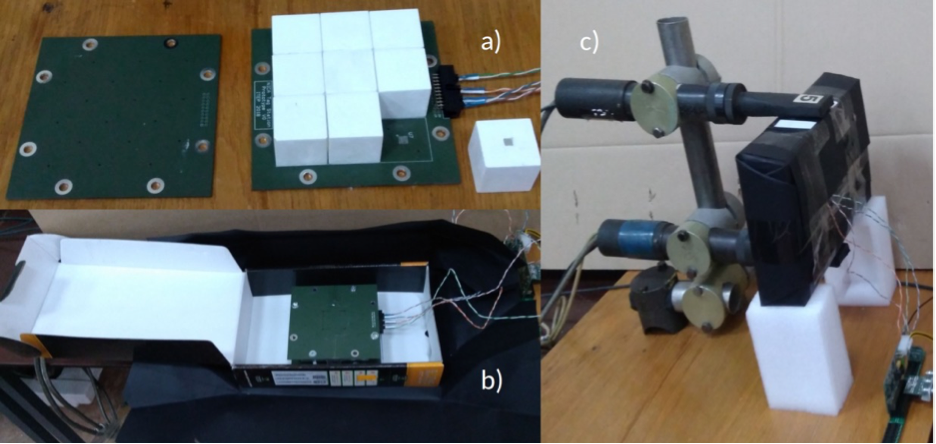}}
\caption{One layer prototype: a) 9 cubes of 30$\times$30$\times$30 mm$^3$ plastic, SiPM board and support board; b) the prototype assembly in a box before wrapping in black paper; c) the box in place for cosmic muon tests.}
\label{fig:ZDC2}
\end{figure}

The expected  energy resolution for neutrons is about $50\%/\sqrt{E} \oplus 30\%$ while the energy resolution for photons is about $20\%/\sqrt{E} \oplus 9\%$. 
Spatial resolution  for photons is estimated to be below 3 mm for 1 GeV and about 1.8 mm for 12 GeV which corresponds to 0.3 and 0.18 mrad for the $L\sim10$ m distance from the interaction point. For neutrons, the space resolution is 10$\div$13 mm within energy range 1$\div$12 GeV.  Longitudinal energy distributions for photons and neutrons are very different and can be used for neutron/photon separation. 

For experimental estimates of the time resolution an assemblage of 9 plastic cubes laid on a printed circuit board with mounted SiPMs and fixed with a support board (see Fig. \ref{fig:ZDC2}) was tested with cosmic muons. Each cube was of 30$\times$30$\times$30 mm$^3$ in size and is chemically covered with a thin light reflecting layer. A numerical simulation has shown the mean number of cells hit in an event is more than 20 for both photons and neutrons. It means the total thickness of scintillator plates of the order of 15 cm (5 times more than for tested prototype). The test results obtained for the prototype wre extrapolated to the full ZDC. Thus the expected time resolution has to be about 150 ps. 

The proposed design of the ZDC calorimeter satisfies most of its physics goals in the limited space inside the magnet. The exception is the energy resolution for neutrons, which is reached by larger calorimeter only. Nevertheless even this goal could be achieved by more elaborated analysis taking advantage from the calorimeter fine granularity.


\chapter{Local polarimetry \label{ch:pol}}
The main goal of the local polarimetry at SPD is  the permanent monitoring of the beam polarization during data 
taking to reduce the systematic uncertainty coming from the beam polarization variation.  
Another task is beam polarization monitoring independent on the major polarimeters (CNI and the absolute one), 
as well as possible usage of this tool to tune
the beam polarization axis.  Since the SPD energy range is relatively new for spin physics, there is a lack of precise polarization data allowing one  to find the explicit solution for the local polarimetry.     

\section{Asymmetry in inclusive production of charged particles} 
 
One of the tools to control the proton beam polarization is the measurement of the azimuthal asymmetry in inclusive production of charged particles in collisions of transverse polarized proton beams.
Such a method is well adopted  at the STAR detector. Two Beam-Beam Counters (BBCs) are  used for this purpose. Each BBC consists of
two zones corresponding to different rapidity range. The inner and outer zones cover 
3.3$<|\eta|<$ 5.0  and 2.1$<|\eta|<$3.3, respectively. The BBCs detect
all the charged particles produced in the forward direction within their acceptance. 

\begin{figure}[h]
  \begin{minipage}[ht]{0.49\linewidth}
    \center{\includegraphics[width=1\linewidth]{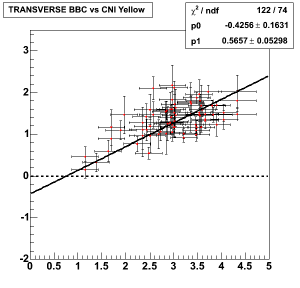} \\ (a)}
  \end{minipage}
  \hfill
  \begin{minipage}[ht]{0.49\linewidth}
    \center{\includegraphics[width=1\linewidth]{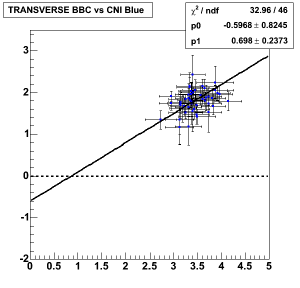} \\ (b)}
  \end{minipage}
  \caption{Correlation of the beam asymmetries measured by the RHIC  $pC$ CNI polarimeter 
\cite{Alekseev:2003cn,Trentalange} and left 
(a) and right (b) STAR BBCs (in arbitrary units). }
  \label{p_ladygin_fig2}  
\end{figure}

The correlation of the  beam asymmetries measured by the RHIC  $pC$ CNI polarimeter 
\cite{Alekseev:2003cn,Trentalange} and the STAR BBCs is demonstrated in Fig.\ref{p_ladygin_fig2}. One can see  that
the measurements by BBCs are sensitive to  the transverse polarization of the colliding beams.  The value of the effective analyzing power $A_N$ for inclusive production of charged particles at $\sqrt{s}$= 200~GeV is about $(6\div 7)\times$10$^{-3}$. At NICA energies it will have, in principle, the same magnitude, or even a larger one due to a larger analyzing power for the $p$-$p$ elastic scattering. Therefore, the BBCs can be used for the local polarimetry at SPD. The design of the SPD BBCs  is described in the previous section. 

 \begin{figure}[h]
  \begin{minipage}[ht]{0.42\linewidth}
    \center{\includegraphics[width=1.\linewidth]{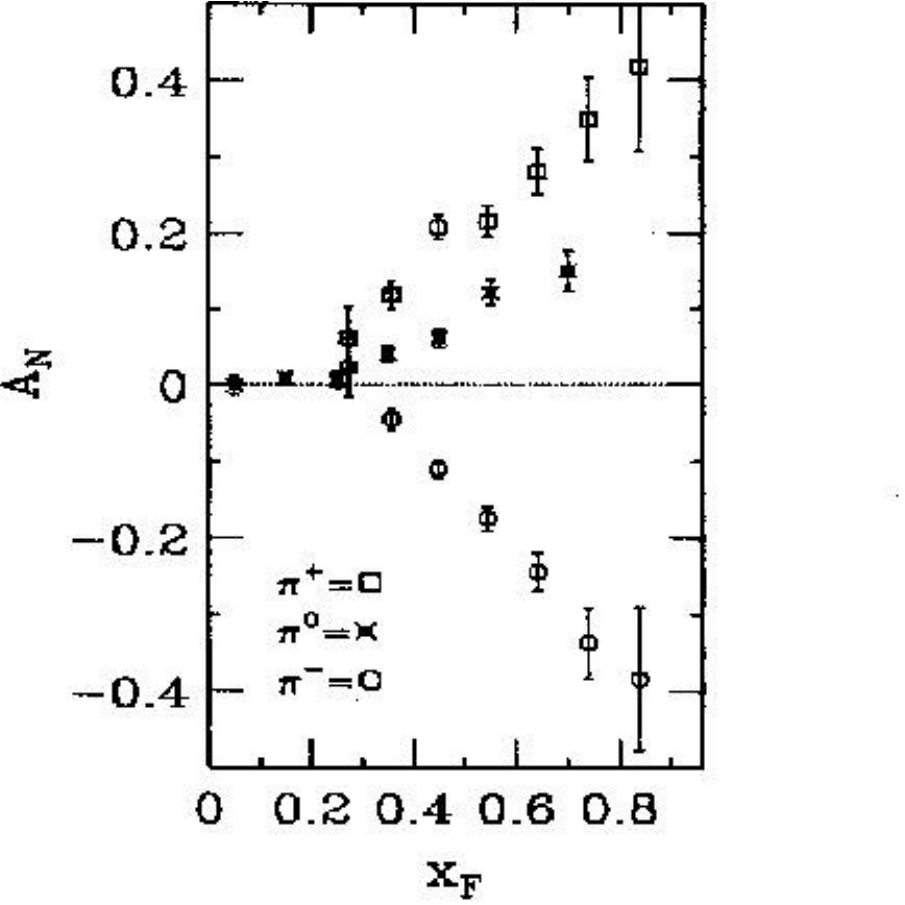} \\ (a)}
  \end{minipage}
  \hfill
  \begin{minipage}[ht]{0.56\linewidth}
    \center{\includegraphics[width=0.85\linewidth]{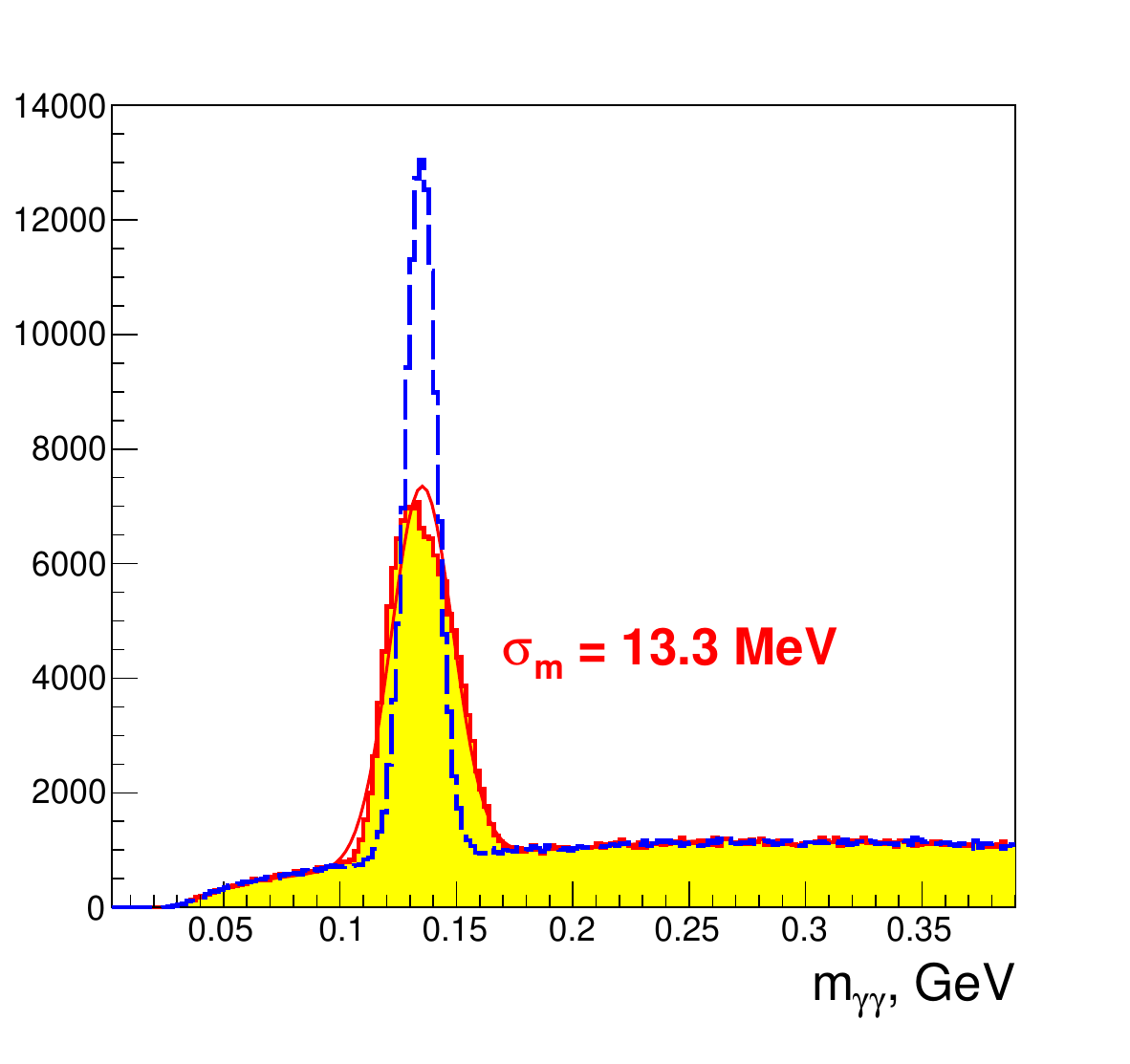} \\ (b)}
  \end{minipage}
  \caption{(a)
  Single transverse spin asymmetry $A_N$ for inclusive pion production in $p$-$p$ collisions at 200 GeV \cite{Adams:1991rw,Adams:1991cs}.. (b) The $\pi^0$ reconstruction in the SPD ECal end-cup with (blue) and without (red) vertex position information.}
  \label{p_ladygin_fig4}  
\end{figure}

\section{Inclusive  $\pi^{0}$ production \label{po0polar}}

One of the reactions to measure and to monitor the vertical component of the 
polarized  proton  beam is the inclusive $pp\to \pi^{\pm,0}X$ reaction. 
Fig.\ref{p_ladygin_fig4}(a) demonstrates the  single transverse spin asymmetries  $A_N$ obtained in the $p$-$p$ collision for 
$\pi^+$, $\pi^0$ and $\pi^-$  inclusive production at 200 GeV ($\sqrt{s}\sim$20~GeV)\cite{Adams:1991rw,Adams:1991cs}.
The data  demonstrate large values of the single transverse spin asymmetries with their signs following the polarization of the 
valence quarks in the pions. 
This regime occurs already at 22 GeV \cite{Allgower:2002qi}  corresponding to 
$\sqrt{s_{NN}}\sim$7 GeV for the collider option.  
Therefore, the inclusive neutral pion production can be used for the polarimetry over the
full energy range of the SPD experiment.

The value of the single transverse spin asymmetry in the $pp\to \pi^{0}X$ reaction
is almost twice smaller than for the charged pions production. However, the $\pi^0$ selection can be done more easily, since it does not require track reconstruction. 

For online 
local polarimetry one can use the parts of the ECal end-cups 
placed around the beam pipe.  Fast $\pi^0$ reconstruction algorithms 
will not include the information on the vertex position along the beam axis, therefore, the width of the $\pi^0$ peak will increase.
The Monte-Carlo results obtained for $\sqrt{s_{NN}}\sim$27 GeV and
 presented in Fig.\ref{p_ladygin_fig4}(b) demonstrate such enlargement.
However, one can see that the selection of $\pi^0$ is good enough for the local polarimetry
purposes. The effective analyzing power $\langle A_N\rangle$ for the kinematic range of produced $\pi^0$ $p_T>0.5$ GeV/$c$ and $x_F>0.5$ is about 0.1. The rate of $\pi^0$ decays reconstructed in the end-caps of the calorimeter provides statistical accuracy of the beam polarization estimation at a few-percent level after 10 minutes of data taking at 10 GeV$<\sqrt{s}\leq 27$ GeV. The corresponding accuracy of the spin direction reconstruction is about a few degrees.

\section{Single transverse spin asymmetry for very forward neutron production}

\begin{figure}[h]
  \begin{minipage}[ht]{0.4\linewidth}
    \center{\includegraphics[width=1\linewidth]{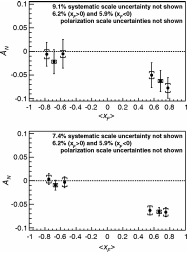} \\ (a)}
  \end{minipage}
  \hfill
  \begin{minipage}[ht]{0.6\linewidth}
    \center{\includegraphics[width=1\linewidth]{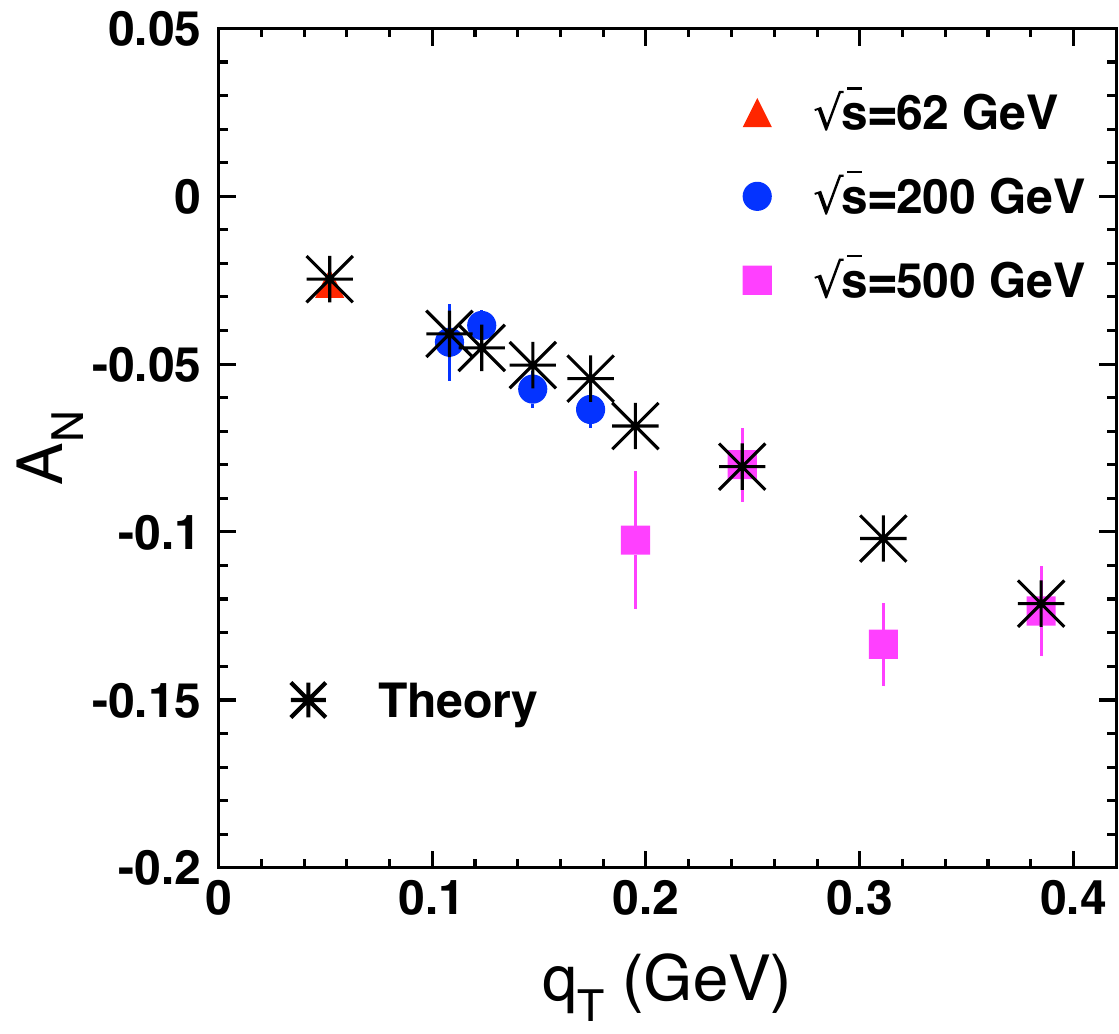} \\ (b)}
  \end{minipage}
  \caption{ (a) $x_F$ dependence of  $A_N$ for neutron production in the (upper) ZDC trigger sample and for the (lower) ZDC$\otimes$BBC trigger sample. 
(b) Single transverse spin asymmetry $A_N$ in the reaction $pp\to nX$ measured at $\sqrt{s}$= 62, 200, 500 ~GeV at PHENIX as a function of the transferred momentum $q_T$. The asterisks show the result of the theoretical calculations \cite{Kopeliovich:2011bx}.}
  \label{p_ladygin_fig3}  
\end{figure}

The energy dependence of the single transverse spin asymmetry, $A_N$,  for neutron production at very forward angles was measured in the PHENIX experiment at RHIC for the polarized $p$-$p$ collisions at $\sqrt{s}$=200  GeV \cite{Adare:2012vw}. The neutrons were observed in the  forward detectors covering an angular range of up to 2.2 mrad.   The observed forward neutron asymmetries are large, reaching $A_N$=−0.08$\pm$0.02 for $x_F$=0.8; the measured backward asymmetries, for negative $x_F$, are consistent with zero.  The results of  the 
$x_F$ dependence of $A_N$ for neutron production in the (upper) ZDC trigger sample and for the (lower) ZDC$\otimes$BBC trigger sample are shown  in Fig.\ref{p_ladygin_fig3}(a). 
The error bars show statistical uncertainties, and the brackets show the $p_T$-correlated systematic uncertainties. The data were obtained for 2 types of triggers: 
the first one is the ZDC trigger for neutron inclusive measurements, requiring an energy deposit in the ZDC to be greater than 5 GeV. The other one was a ZDC$\otimes$BBC
trigger, a coincidence trigger of the ZDC trigger with the BBC hits  defined as one or more charged particles in both of the BBC detectors.

The observed large asymmetry for forward neutron production was discussed within the pion exchange framework, with interference between the spin-flip amplitude due to the pion exchange and the non-flip amplitudes from all Reggeon exchanges. 
The numerical results of the parameter-free calculation of $A_N$ are in excellent agreement with the PHENIX data (see Fig.\ref{p_ladygin_fig3}(b)). One can see that 
$A_N$ is increasing almost linearly as a function of the neutron transverse momentum $q_T$. One can expect the 
$A_N$  value of $\sim-0.02$ at $\sqrt{s}$=27~GeV. Therefore, the   $pp\to nX$  
reaction with neutron emission at very forward angles can be used at SPD at
least at high energy.   

Very forward neutrons are detected by two zero-degree calorimeters (ZDCs) 
\cite{Adler:2000bd} placed in
the gaps between the ion tubes of the colliding beams on the left and right from the center of the detector.  Two ZDCs will be also placed at SPD.   
 These ZDCs can be considered as an additional tool
for the local polarimetry for pp-collisions at the highest
NICA energy.

\chapter{Detector control system}
      The SPD detector control system (DCS) is designed to control the basic operating modes of the detector parts and the detector as a whole, and to  continuously monitor  slowly changing parameters of the detector, engineering means which provide the detector operation, and the environment. The DCS is synchronized with the basic operating modes of the NICA accelerator complex by means of a synchronization subsystem shared between the DCS and the SPD DAQ. The DCS provides parameterization of the managed object (i.e. the SPD detector), implements algorithms for normalization, parameters measurement and control based on these parameters, and generates the necessary sets of abstractions and options for presenting these abstractions to the operator in an intuitive manner.
     Critical values of the parameters going beyond the predefined limits in predetermined situations cause emergency events and initiate procedures for handling such events, including the procedure for an automatical detector shutdown in order to prevent its damage. Parameter values are archived in a database for long-term monitoring of the detector operation and identify possible failures in the operation of the equipment and emergency situations. The configurations of the detector parameters saved in the database make it possible to start the detector promptly and use it with various preset parameters and in various operating modes in accordance with the requirements of a particular physics experiment.
The DCS allows autonomous operation of each detector subsystem at the stage of the initial start-up as well as its  periodic maintenance, calibration sessions, and planned upgrades.
     The number of parameters in the system is expected to be significant, therefore, it is assumed that the system should be extendable and flexibly configurable. Architectural and software solutions based on the event-driven model \cite{Michelson} and client-server and producer-consumer \cite{Etschberger} interaction models should be preferred for communication, when building the general DCS and the control systems of each part of the detector. Centralized systems operating in the master-slave polling mode should be avoided.

\section{DCS concept}

    Most of high-energy physics detectors include parts consisting of similar systems built from devices, sensors, and actuators with similar or identical functionality. This determines parameterization of the entire detector as a managed object. Such systems include:
\begin{enumerate}
\item high voltage (HV) power supply system for powering gas detectors and light (photon) sensors (PMT and SiPM);
\item low voltage (LV)  power supplies for powering magnets, digital and analog electronics;
\item  cryogenic systems;
\item gas supply and mixing systems;
\item vacuum systems;
\item  front-end electronic LV powering control and temperature monitoring;
\item different cooling and temperature control systems;
\item DAQ system;
\item accelerator interface and synchronization;
\item general external electricity and water cooling stations, etc.
\end{enumerate}
          
\begin{figure}[ht]
\begin{center}
\includegraphics[width=0.8\linewidth]{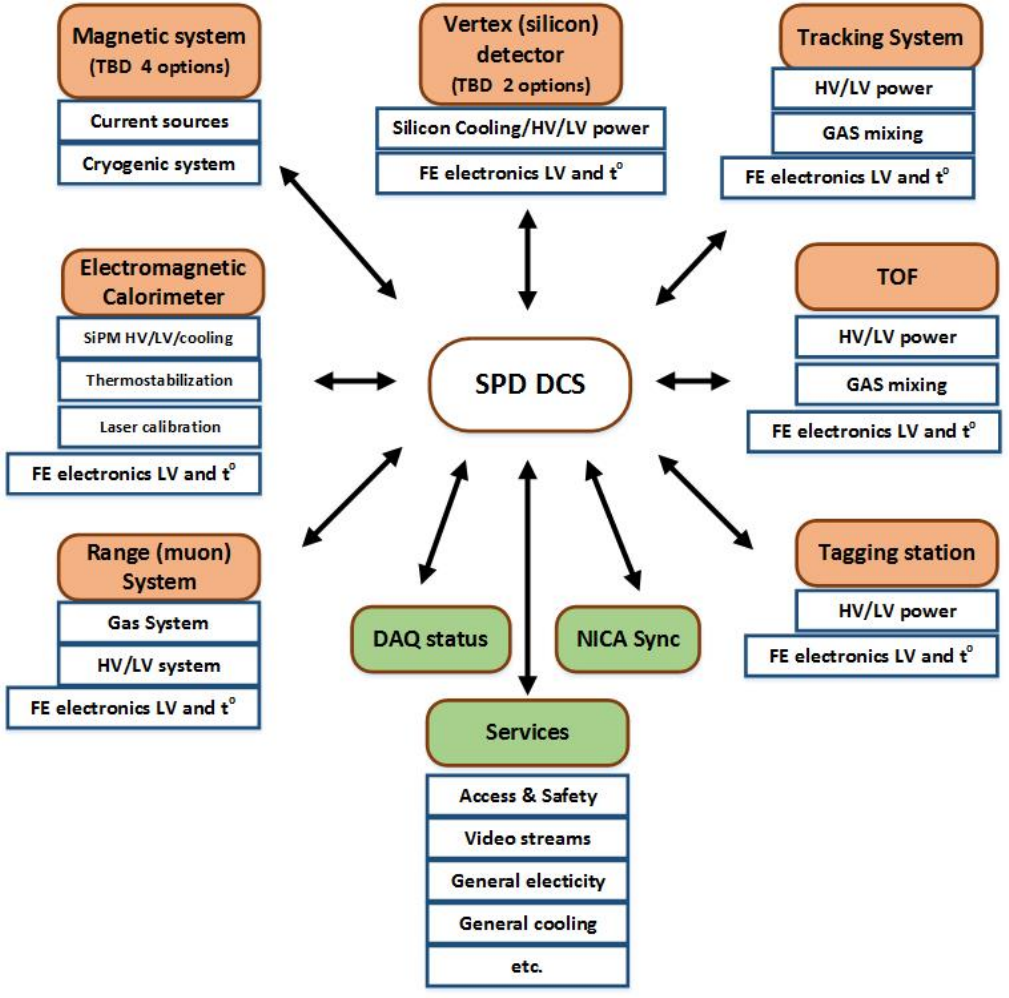}
\caption{SPD detector control system layout.}
\label{fig:dcs1}
\end{center}
\end{figure}

          The SPD detector is no exception and includes almost all of these systems spread among different parts of the detector, as shown in the layout diagram,  Fig. \ref{fig:dcs1}. Each part of the detector refers to one or more subsystems. The composition of the systems will be refined, as the individual parts of the detector are developed.

All the systems can be similarly parameterized and shown to the operator in an intuitive presentation in order to simplify the operator's decision-making algorithm. However, the physical implementation at the hardware level of these elements may vary significantly in different parts of the SPD, because:
\begin{itemize}
\item the parts inherit the experience of their developers gained in previous experiments;
\item hardware and software components are selected based on their cost and availability;
\item parts of the detector are manufactured at different times.
\end{itemize}

Nevertheless, in order to optimize financial and human resources costs for the creation of the entire detector and the DCS, in particular, it is necessary to recommend the developers of the detector parts to strive for standardization of the used hardware and embedded software. This will significantly reduce the efforts put into developing, deploying, and operating the detector and will result in significant cost savings. To achieve these goals, at the stage of prototyping the detector systems it is advisable to work out not only the detector itself, the front-end electronics and the DAQ, but also the slow control systems. This work can be carried out in the beam test zone (BTZ), for which the BTZ slow control system must be made as similar to the final DCS version as possible.

\section{DCS architecture }
\begin{figure}[!hb]
\begin{center}
\includegraphics[width=0.8\linewidth]{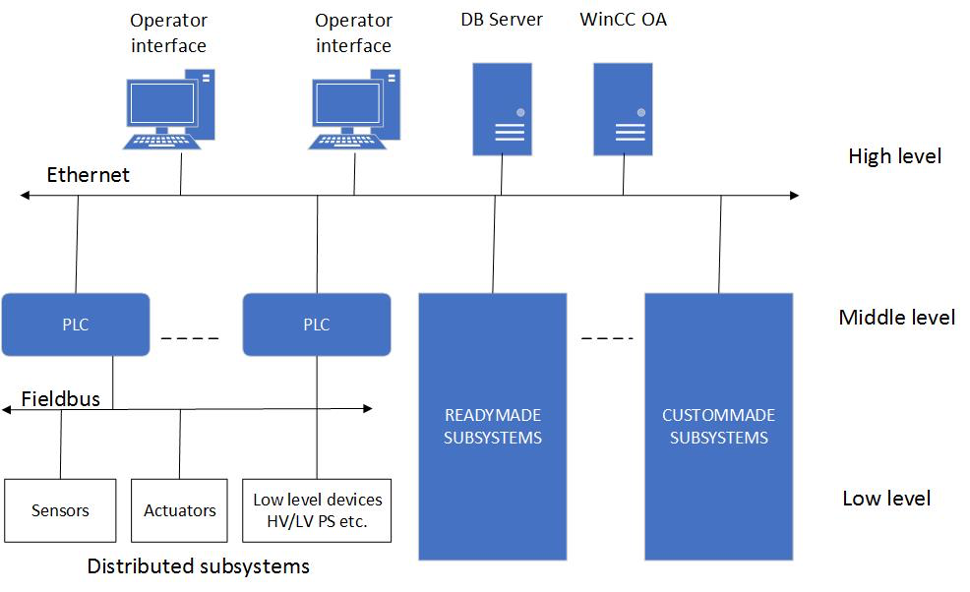}
\caption{SPD detector control system architecture.}
\label{fig:dcs2}
\end{center}
\end{figure}

The detector control system is divided into three logical levels (Fig. \ref{fig:dcs2}). 
The lower level includes measurement channels built into the Front End Electronics (FEE) and Data Acquisition (DAQ) of the detector parts, various stand-alone sensors, I/O devices, and low and high voltage power supplies. The middle level is represented by programmable logic controllers and integrated ready-made and custom-made subsystems (vacuum posts, gas consoles, multichannel ready-made power subsystems, etc.). The interfaces to the FEE and DAQ that provide data for the detector control system are also on this level. The upper level is designed to provide a human-machine interface for operators, implement a database of detector parameters and configurations, communicate with the external world (accelerator, engineering support systems, access system, etc.), and implement macro-control algorithms common for the entire detector. All these levels are connected in a hierarchical network using fieldbuses between the first and the second level, for example, a CAN-bus with a CANopen protocol. An Ethernet LAN is used between the middle and the upper levels. At the top level, special software, such as SCADA (Supervisory Control And Data Acquisition), is used, which provides control, collection and storage of data in real time. It is proposed to use the WinCC OA system common at  CERN, as a SCADA system. We understand that for smooth and reliable communication with the control system of the Nuclotron, a gateway to the Tango Controls \cite{tango,Gorbachev:2016yrd} system should be developed and deployed.

\section{SCADA for the DCS }
WinCC OA (ex PVSS-II) \cite{WinCC,Boterenbrood:530680} is a commercial SCADA system. It is a software component constructor that allows one to use both preinstalled prototypes and templates, and software modules and system components developed in C.  This system is actively used in many experiments at CERN and has support and safety certificates in the Russian Federation. The following properties make WinCC OA an attractive solution to be used in the DCS SPD:

\begin{itemize}
\item object-oriented approach built into the system ensures an efficient development process and the ability to flexibly expand the system;
\item	capability to create distributed systems - up to 2048 WinCC OA servers;
\item	scalability from a simple single-user system to a distributed redundant network system with $>10$ million tags (physical and synthetic parameters);
\item	platform independent system is available for Windows and Linux;
\item	event-driven system;
\item   hot standby and 2x2 redundancy (DRSystem), the required level of availability and reliability;
\item	wide range of drivers and options for communication OPC, OPC UA, S7, Modbus, IEC 60870-5-101/104,DNP3, XML, JSON, SOAP, ...;
\item	support by major manufacturers of electronic devices for building automation systems in high energy physics.
\end{itemize}

\begin{figure}[!h]
\begin{center}
\includegraphics[width=0.8\linewidth]{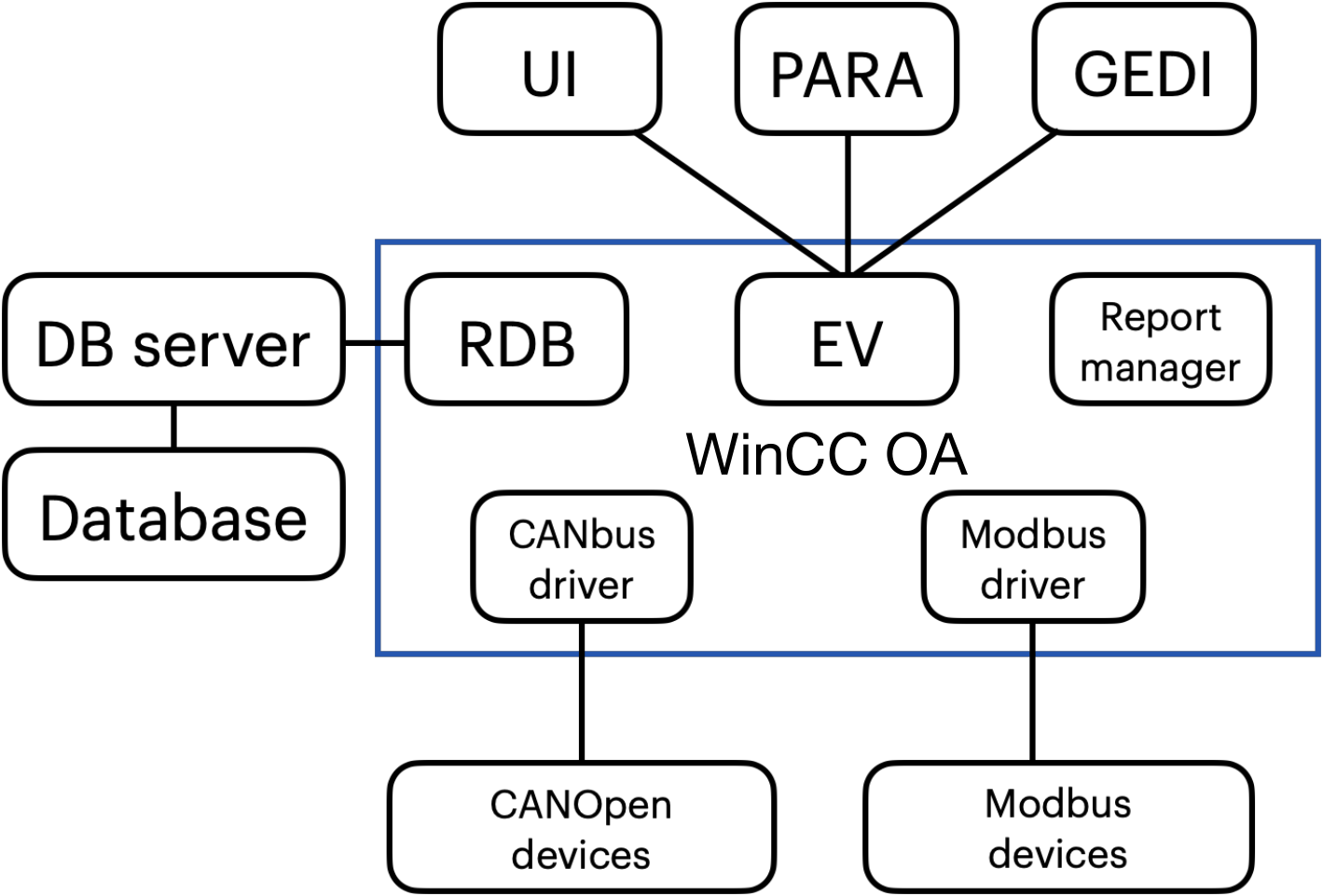}
\caption{SCADA structural scheme of the WinCC OA software.}
\label{fig:dcs3}
\end{center}
\end{figure}

Each functional unit of the system that is software implemented as a separate process is called manager. A set of managers forms a system. Data exchange and communications between managers are done via TCP. The data is exchanged by means of passing events. The system allows parallelizing processes (managers) by running them on different computers with different OS. The system is scalable and balances the load on the control computers. The required managers start only if necessary and multiple instances may run simultaneously. Managers can be distributed across multiple computers/servers. The WinCC OA block diagram is shown in Figure \ref{fig:dcs3}.
The main process is the Event Manager, it contains and manages the process image (current values of all process variables), receives and qualifies data (central message manager), distributes data across other managers, acts as a data server for others, manages users authorization, manages the generation and status of alarm messages.
The Database Manager receives data from the Event Manager and handles it according to its own algorithm. The historical database can use either a proprietary database (HDB) or an Oracle DBMS (the Oracle Real Application Clusters configuration is also supported). Parallel archiving in Oracle and HDB databases is possible. It is also possible to record user-defined data and log system events and messages in an external relational database (MS SQL, MySQL, Oracle, etc).

The WinCC OA Report Manager supports different ways of generating reports: 
\begin{itemize}
\item	in Microsoft Excel format;
\item	in the {\it xml} format with the ability to display in any external tool for working with reports (Eclipse BIRT, Crystal Reports, SYMATIC Information Server etc.), SOAP (Simple Object Access Protocol) protocol is also supported. 
\end{itemize}

Project development for the WinCC OA system is based on an object-oriented approach. In the WinCC OA data model, objects are represented as data points that characterize the image of a specific physical device or process. For each data point (called tag) element, properties and actions, such as signal processing (smoothing, setting limits, etc.), communication with external systems, archiving, generation of alarm messages (alarms), etc. can be defined in accordance to it.  Typing and inheritance are supported, therefore arbitrary hierarchical data structures can be created. Similarly, the principles of inheritance and reusability are implemented for graphical objects. The WinCC OA IDE includes the PARA configuration editor and the GEDI graphical editor of the User Interface Manager (UI) (includes a data model editor, mass configuration tools, administration tools, an interface to version control systems, a debugger, etc.).
Changes to data structures and graphics are applied without restarting the project. Writing custom scripts can be done using CONTROL++ (a programming language, the syntax of which is similar to C/C++). Such scripts can be both event handlers associated with the elements of the graphical interface, and data processing procedures.
The system includes a standard graphical objects library; it can be extended by developing user objects or using the Qt Toolkit widgets. It is also possible to use the JavaScript libraries available on the market or included JavaScript scripts. Thanks to the open API (C++ / C\# API), it is possible to create managers, drivers, widgets, and CONTROL++ extensions. A new set of tools is available for the concept of High Speed Programming implementation, which supports documentation build-up from the source code, unit testing and autocompletion of program structures.

	It is also planned to provide data exchange between the WinCC OA and Tango Controls, which is used as the upper level of the Nuclotron control system. This can be implemented using standard OPC technologies with a client-server architecture, or it can be implemented using SQL tools as a common database for  both SCADA systems used for the accelerator and the detector. The final choice of a suitable solution will be made at the stage of system implementation.

\chapter{Data acquisition system}
\section{Introduction}
The data acquisition system of the SPD should provide continuous data taking, including data readout from the front-end electronics, data consistency check, event building and writing events to a storage. The system should have no dead time or minimal dead time. {\em These features will be implemented with the DAQ operating in a free-running (triggerless) mode.}	

Other important tasks of the DAQ are:
\begin{itemize}
\item initialization of hardware;
\item control and monitoring of the data taking process: control of the status of all hardware devices including front-end electronics, status of software, quality of collected data;
\item monitoring of the parameters characterizing the detector performance (accumulation of time, amplitude and hit distribution histograms, detector rates, etc.);
\item logging of information and errors;
\item distribution of data over computing nodes for further online analysis;
\item etc.
\end{itemize}

The data acquisition system of SPD should withstand the data flux from $p$-$p$, $p$-$d$ or $d$-$d$ interactions at the extreme conditions of high luminosity. At the highest NICA energy and luminosity, $\sqrt{s} = 27$~GeV and $L=10^{32}\;\mathrm{cm}^{-2}\mathrm{s}^{-1}$, the interaction rate within the SPD aperture will be 3~MHz, and the average multiplicity of about 15. This drastically differs from the conditions of another NICA experiment, MPD, where the collision rate of heavy ions is orders of magnitude less, but the multiplicity is much higher.

The structure of DAQ will be similar to recently modernized DAQ of the COMPASS experiment at CERN \cite{Steffen:2016eoe,BODLAK2016976,daq:kon,daq:compass1,daq:compass2,daq:compass3,daq:compass4}. The COMPASS DAQ extensively uses logical programmable integrated circuits FPGA at different levels of the system. This allows one to handle large data streams with minimal latency and provides very good flexibility. Unlike the COMPASS experiment, which uses the beam of the CERN SPS with a spill time structure, the SPD DAQ will deal with a continuous beam.

The DAQ of SPD will operate in a free-running mode, when the readout is not controlled by a trigger system, but occurs with a fixed frequency. It requires all front-end electronics running in a self-triggered mode, and the readout happens synchronously with a common clock distributed by the precise timing system.
All the data received between the acts of readout are accumulated in the memories implemented in the front-end electronics modules and are stored there until the next readout. The readout frequency value will be chosen depending on the detector rates and memory depths available in the front-end cards. The width of the time slice between the successive readouts should be much larger than the response time of the sub-detectors in order to minimize the probability of separating an event into two slices.

Digitization of data and zero suppression occur in the front-end electronics. It is expected that the so-called ”feature extraction algorithm” will be implemented in the front-end electronics of the vertex detector and the calorimeter. This algorithm, which is under development in several collaborations (in particular, PANDA \cite{Kavatsyuk:2010aqa}, COMPASS \cite{daq:mun1,daq:mun2,daq:mun3}), allows transferring only the extracted time and amplitude, instead of many samples of the digitizer, thus greatly decreasing the amount of data to be transferred.

For now the expected data flux in the hardest conditions of the experiment (maximum energy and luminosity) has been estimated without detailed simulations, but using the current knowledge of the sub-detector structure, particle multiplicity per event, hit  multiplicity in different detectors, expectations about the front-end electronics parameters and, where relevant, results of the beam tests at other experiments (MPD, PANDA). The total number of channels to be read out is about 700 thousand, with the major part coming from the vertex detector ($\sim 460$ thousand for the VD strip option). The full numbers are given in Table~\ref{tab:daq-numb}. Preliminary estimation for the data flow is about 20 GB/s including some margin of safety.

\begin{table}
\caption{Summary of detectors outputs to DAQ. Information type: T means time, A -- amplitude (or charge).}
\label{tab:daq-numb}
\renewcommand{\arraystretch}{1.2}
\begin{center}
\begin{tabular}{|l|l|l|l|l|}
    \hline
    \rule{0pt}{14pt}
Sub-detector    & Information & Number & Channels& Number  \\
            & type   & of channels & per FE card& of outputs\\ \hline
Vertex detector 5 DSSD & T + A & 460800 & 640 & 720 \\ \hline
Vertex detector 3 MAPS+ 2 DSSD & T + (T + A) &
(3024 + 596)$^*$
&
& 864 + 596 \\ \hline
Straw tracker   & T + A  & 
79200  & 64  & 1238 \\ \hline
Calorimeter     & T + A & 30176  & 64  & 472 \\ \hline
PID-ToF         & T     & 20200 & 32 & 632 \\ \hline
PID-Aerogel     & T + A &  320  & 32 & 10 \\ \hline
BBC (inner+outer)  & T + (T + A) & 
256 + 192 &  32     &  8 + 6 \\ \hline
Range system    & T     & 106000 & 192  & 553 \\ \hline
ZDC             & T + A & 250 + 650  & 16 & 57 \\ \hline
Total (max)     &       & 698044$^{**}$ & &   4436 \\ \hline
\end{tabular}
\end{center}
\vspace{-1ex}
$^*$ -- number of sensors, see Section \ref{sec:vd} of Chapter \ref{ch:DL}\\
$^{**}$ -- for DSSD version
\end{table}



\section{DAQ structure}
\begin{figure}[p]
\includegraphics[width=\columnwidth]{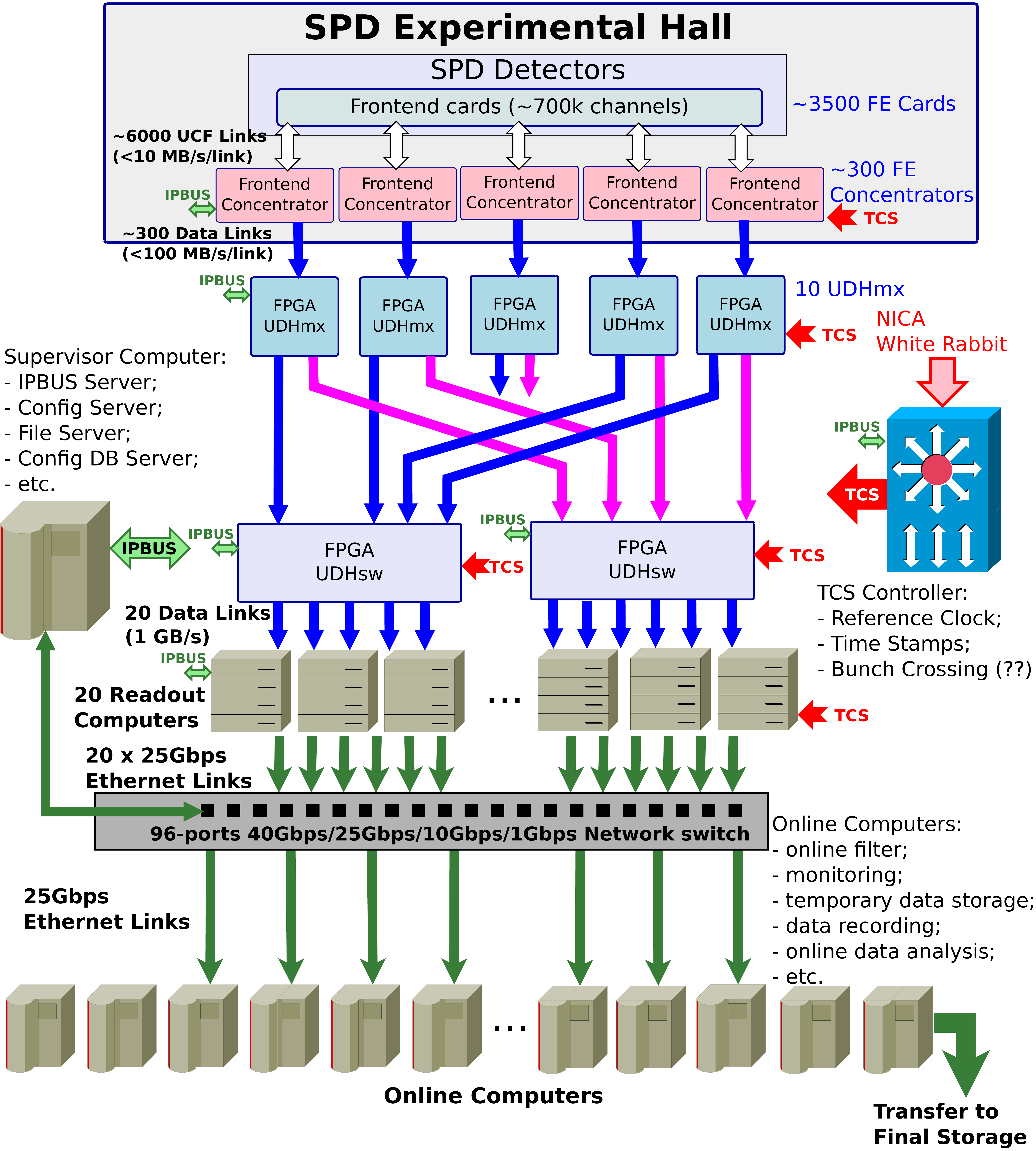}	
\caption{General structure of DAQ-SPD.}
\label{fig:daq-sch}
\end{figure}

The scheme of the DAQ is presented in Fig.~\ref{fig:daq-sch}. The data from the front-end electronics cards come to the detector interface cards (FE concentrators). Now the existing electronics card with 12 input is considering as FE concentrators \cite{daq:kon}. The Data-Handler multiplexers (UDHmx) are configured on the base of FPGA. The multiplexer has 48 high speed input and up to 8 output interfaces. They verify the consistency of data and store them until receiving the readout signal.

The two Data-Handler Switches (UDHSw) function as a $10\times10$ switch and perform event building with a maximum throughput rate of 10 GBytes/second. The UDHSw's perform the final level of event building and distribute the assembled events to 20 readout computers. The Data-Handler Switches and Data-Handler multiplexers are implemented on the same electronics cards by means of different firmware. Each readout computer is equipped with a dedicated PCIe buffer card for data collection. These cards are built on a FPGA chip and are commercially available. The current version of the card used in COMPASS has a bandwidth close to 1~GB/s \cite{daq:compass3}. Finally, the continuous sequence of slices is formed below the Network Switch in each of on-line computers to be used for on-line filtering and event monitoring.

The slow control software accesses the front-end electronics via the FE concentrators using the UDP-based IPBus protocol \cite{Larrea:2015wra}. The interface cards retransmit control and clock signals provided by the time distribution system to the corresponding front-end electronics, and convert the detector information from the detector specific interfaces to a common high-speed serial interface running over optical fiber. It is foreseen to use UCF \cite{Gaisbauer:2016vhl} as a standard high speed link protocol within the DAQ.

The White Rabbit system \cite{Serrano:1215571, WR} is planned to be used at NICA for time synchronization. It provides synchronization for large distributed systems with a time-stamping of 125~MHz, sub-nanosecond accuracy and ${\sim}$10~ps precision. Signals from the White Rabbit system  will be used as an input for the Time Control System (TCS)
\cite{daq:TCS}
which will distribute clock signals through the whole electronic system.

\section{Data format}

\begin{figure}[p]
	\includegraphics[width=\columnwidth]{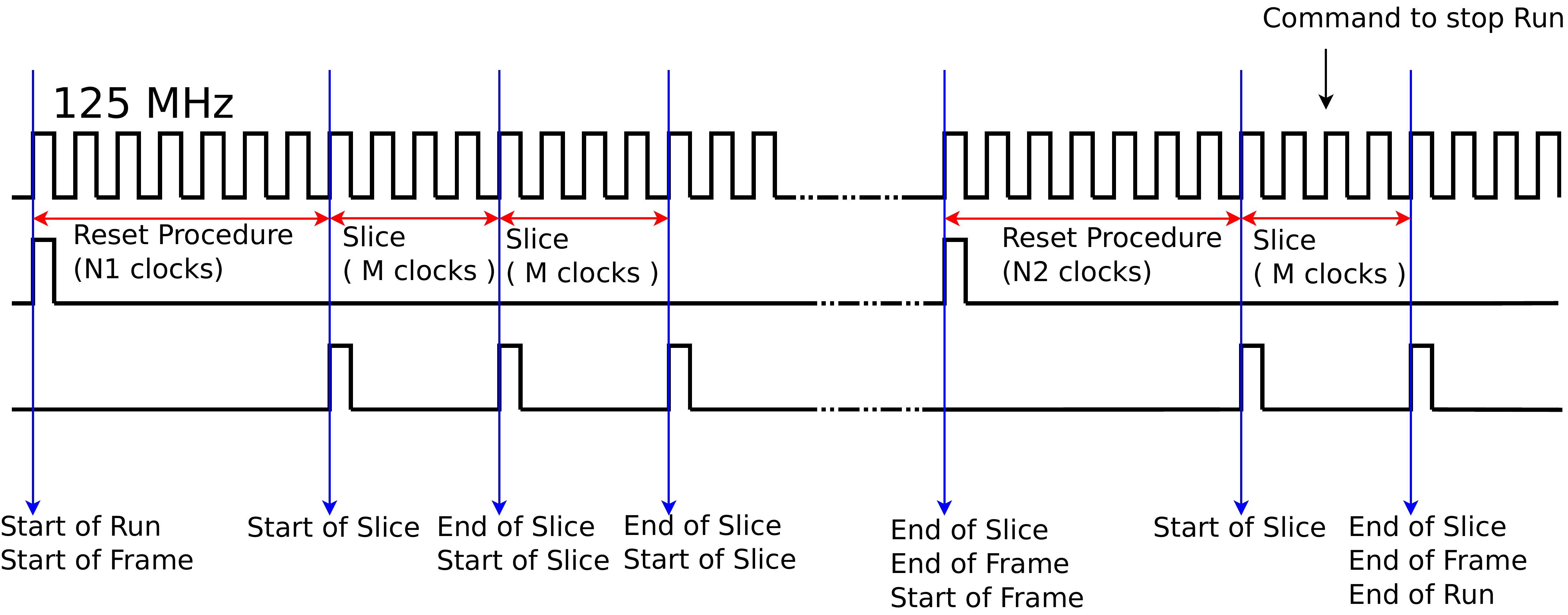}	
	\caption{Time diagram of a sequence of clocks, Slices and Frames within the Run.}
	\label{fig:daq-tcs}
\end{figure}

The time structure of the expected data flow during a run is shown in Fig.~\ref{fig:daq-tcs}. All processes are synchronized with a 125-MHz clock coming from the White Rabbit system. A Run is started after the reset procedure which includes all initialization processes. Afterwards, the continuous date flow is divided into a sequence of time slices. The proposed time slice duration can be selected in the range from 1~\textmu{}s to 8.3~ms and will be chosen according to the data flux and capacity of the whole chain of data collection. Longer slices are preferable, because the longer the slice, the less the probability an event falling into two adjacent slices. The slices have a continuous numbering within a frame, a wider time interval, which can extend from 65~ms to 549.7~s. The slice numbering is reset in every frame by the Start of the frame signal.

\begin{figure}
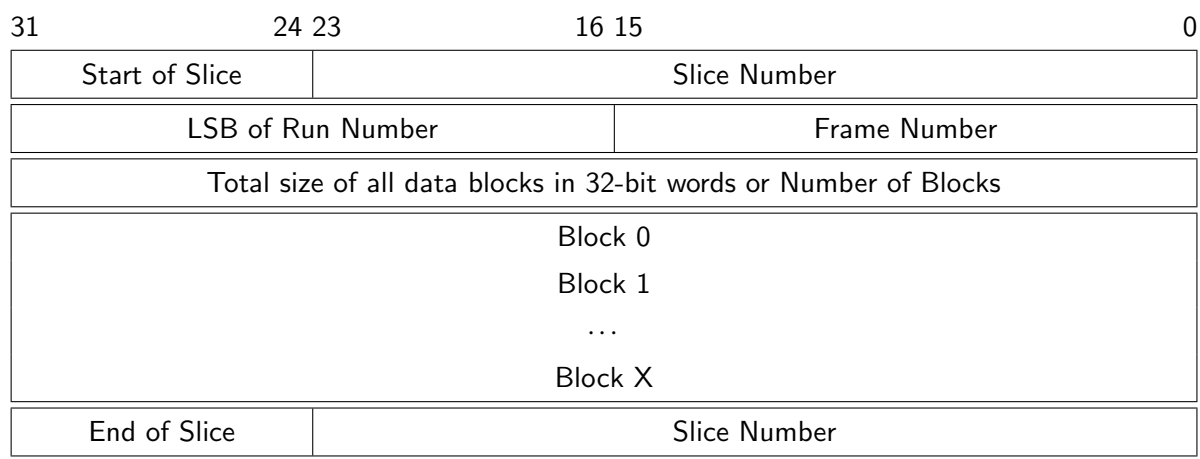

\begin{center}
\sf
\renewcommand{\arraystretch}{1.3}
\begin{tabular}{@{}*{2}{p{0.24\columnwidth}@{~}}@{}p{0.48\columnwidth}@{}}
31\hfill24&23\hfill16&15\hfill0 \\ \hline
\multicolumn{1}{|c}{Start of Slice} & 
\multicolumn{2}{|c|}{Slice Number} \\ \hline \hline
\multicolumn{2}{|c}{LSB of Run Number} & 
\multicolumn{1}{|c|}{Frame Number} \\ \hline \hline
\multicolumn{3}{|c|}{Total size of all data blocks in 32-bit words or Number of Blocks} \\ \hline \hline
\multicolumn{3}{|c|}{Block 0} \\
\multicolumn{3}{|c|}{Block 1} \\
\multicolumn{3}{|c|}{$\cdots$} \\
\multicolumn{3}{|c|}{Block X} \\ \hline \hline
\multicolumn{1}{|c}{End of Slice} & 
\multicolumn{2}{|c|}{Slice Number} \\ \hline 
\end{tabular}
\end{center}
	\caption{Data Format: Slice structure.}
	\label{fig:daq-slice}
\end{figure}

\begin{figure}
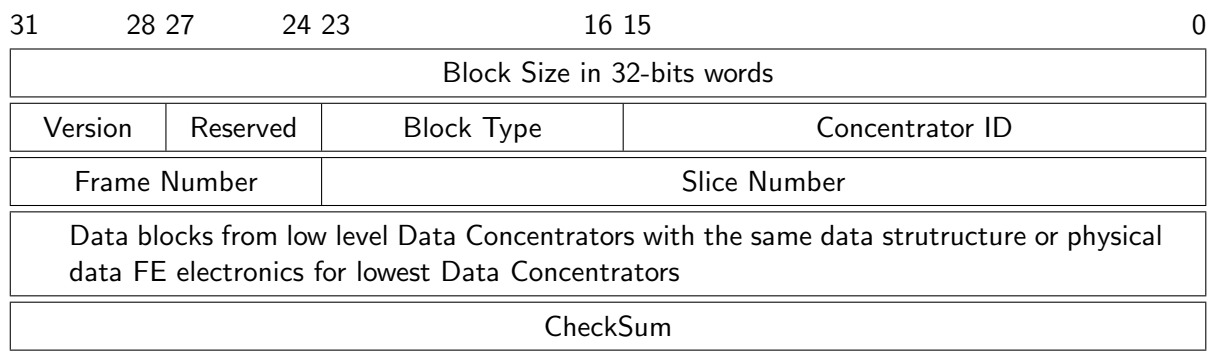

\begin{center}
\sf
\renewcommand{\arraystretch}{1.3}
\begin{tabular}{@{}*{2}{p{0.12\columnwidth}@{~}}@{}p{0.24\columnwidth}@{~}p{0.48\columnwidth}@{}}
31\hfill28&27\hfill24&23\hfill16&15\hfill0 \\ \hline
\multicolumn{4}{|c|}{Block Size in 32-bits words} \\ \hline \hline
\multicolumn{1}{|c}{Version} & 
\multicolumn{1}{|@{}c@{}}{Reserved} & 
\multicolumn{1}{|c}{Block Type} & 
\multicolumn{1}{|c|}{Concentrator ID} \\ \hline \hline
\multicolumn{2}{|c}{Frame Number} & 
\multicolumn{2}{|c|}{Slice Number} \\ \hline \hline
\multicolumn{4}{|@{\qquad}p{0.9\columnwidth}|}
{Data blocks from low level Data Concentrators with the same data strutructure or physical data FE electronics for lowest Data Concentrators} 
\\ \hline \hline
\multicolumn{4}{|c|}{CheckSum} \\ \hline 
\end{tabular}
\end{center}
	\caption{Data Format: structure of Data Blocks of High Level.}
	\label{fig:daq-mux}
\end{figure}

\begin{figure}
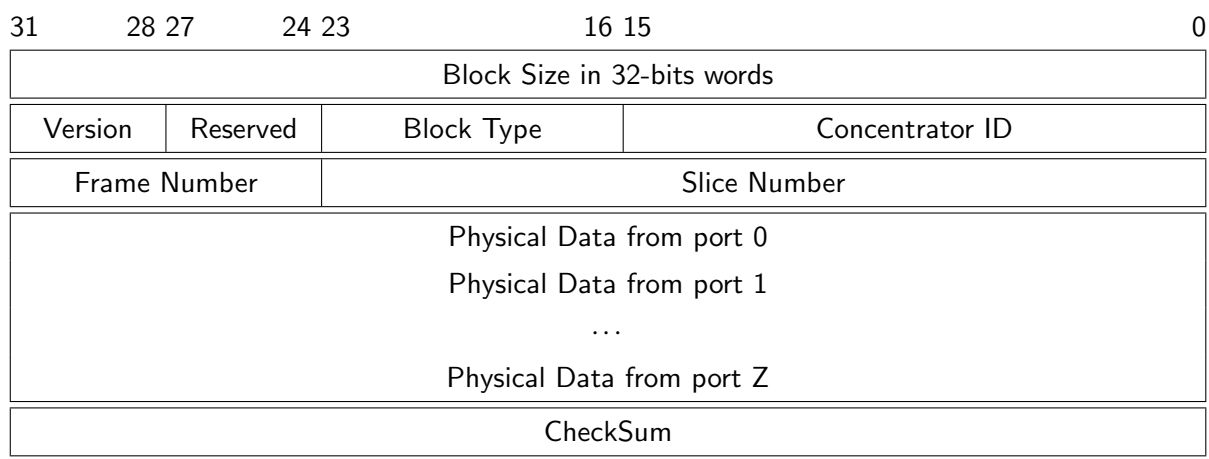

\begin{center}
\sf
\renewcommand{\arraystretch}{1.3}
\begin{tabular}{@{}*{2}{p{0.12\columnwidth}@{~}}@{}p{0.24\columnwidth}@{~}p{0.48\columnwidth}@{}}
31\hfill28&27\hfill24&23\hfill16&15\hfill0 \\ \hline
\multicolumn{4}{|c|}{Block Size in 32-bits words} \\ \hline \hline
\multicolumn{1}{|c}{Version} & 
\multicolumn{1}{|@{}c@{}}{Reserved} & 
\multicolumn{1}{|c}{Block Type} & 
\multicolumn{1}{|c|}{Concentrator ID} \\ \hline \hline
\multicolumn{2}{|c}{Frame Number} & 
\multicolumn{2}{|c|}{Slice Number} \\ \hline \hline
\multicolumn{4}{|c|}{Physical Data from port 0} \\
\multicolumn{4}{|c|}{Physical Data from port 1} \\
\multicolumn{4}{|c|}{$\cdots$} \\
\multicolumn{4}{|c|}{Physical Data from port Z} \\ \hline \hline
\multicolumn{4}{|c|}{CheckSum} \\ \hline 
\end{tabular}
\end{center}
	\caption{Data Format: structure of Low Level Data from FE Concentrators.}
	\label{fig:daq-FE}
\end{figure}

The proposed formats of the collected data are shown in Figs.~\ref{fig:daq-run}--\ref{fig:daq-FE}. The data are formatted at all stages of transfer from the Front-End Concentrators to the Data-Handler Switches. The required headers and checksums are added at all stages.

In Fig.~\ref{fig:daq-run}, the structure of a Run is shown. The Run consists of a sequence of Frames (Fig.~\ref{fig:daq-frame}) numbered from 0 to N, where N, the maximum number of frames in the Run, is assigned by the TCS controller. The Frame consists of a sequence of slices numbered from 0 to K, the maximum number of slices in the Frame, which is also assigned by the TCS controller.

Fig.\ref{fig:daq-slice} shows the structure of the Slice. The Slice contains a sequence of Data Blocks from the Data Concentrators (Fig.~\ref{fig:daq-mux}). Finally, the lowest unit in the Data Format chain is the Data Block of FE Concentrators (Fig.~\ref{fig:daq-FE}) which contains Physical Data from several ports the amount of which depends on the FE card type.

The proposed format provides a unique connection of the physical information to the detectors geometry and the event time.

\begin{figure}
\begin{center}
\sf
\renewcommand{\arraystretch}{1.3}
\centering
\begin{tabular}{*{2}{@{}p{0.24\columnwidth}@{~}}@{}p{0.48\columnwidth}@{}}
31\hfill24&23\hfill16&15\hfill0 \\ \hline
\multicolumn{1}{|c}{Start of Run} & 
\multicolumn{2}{|c|}{Run Number} \\ \hline \hline
\multicolumn{3}{|c|}{Start of Run Time in seconds since DATE} \\ \hline \hline
\multicolumn{3}{|c|}{Frame 0} \\
\multicolumn{3}{|c|}{Frame 1} \\
\multicolumn{3}{|c|}{$\cdots$} \\
\multicolumn{3}{|c|}{Frame K} \\ \hline \hline
\multicolumn{1}{|c}{End of Run} & 
\multicolumn{1}{|@{}c@{}}{LSB of Run Number} &
\multicolumn{1}{|c|}{Run Number} \\ \hline \hline
\multicolumn{1}{|c}{Start of Run} & 
\multicolumn{2}{|c|}{Run Number} \\ \hline \hline
\multicolumn{3}{|c|}{End of Run Time in seconds since DATE} \\ \hline 
\end{tabular}
\end{center}
	\caption{Data Format: Run structure.}
	\label{fig:daq-run}
\end{figure}

\begin{figure}
\begin{center}
\sf
\renewcommand{\arraystretch}{1.3}
\centering
\begin{tabular}{*{2}{@{}p{0.24\columnwidth}@{~}}@{}p{0.48\columnwidth}@{}}
31\hfill24&23\hfill16&15\hfill0 \\ \hline
\multicolumn{1}{|c}{Start of Frame} & 
\multicolumn{1}{|@{}c@{}}{LSB of Run Number} &
\multicolumn{1}{|c|}{Frame Number} \\ \hline \hline
\multicolumn{3}{|c|}{Start of Frame Time in seconds since DATE} \\ \hline \hline
\multicolumn{3}{|c|}{Slice 0} \\
\multicolumn{3}{|c|}{Slice 1} \\
\multicolumn{3}{|c|}{$\cdots$} \\
\multicolumn{3}{|c|}{Slice K} \\ \hline \hline
\multicolumn{1}{|c}{End of Frame} & 
\multicolumn{1}{|@{}c@{}}{LSB of Run Number} &
\multicolumn{1}{|c|}{Frame Number} \\ \hline \hline
\multicolumn{3}{|c|}{End of Frame Time in seconds since DATE} \\ \hline 
\end{tabular}
\end{center}
	\caption{Data Format: Frame structure.}
	\label{fig:daq-frame}
\end{figure}

\section{Cost estimate}
The numbers of modules and their preliminary cost estimation are summarized in Table~\ref{tab:daq-cost}.

\begin{table}
\caption{Cost estimation of the DAQ}
\centering
\label{tab:daq-cost}
\renewcommand{\arraystretch}{1.2}
\begin{tabular}{|p{0.33\columnwidth}||c|c|c|}
\hline
                       & Number &  Cost &   Total \\
                       &        & (k\$) &  (k\$)   \\ \hline
FE concentrators       &  380   &  3.5  &    1330 \\
UDHmx modules          &  12    &  17.5 &    210 \\
UDHsw modules          &   3    &  17.5 &    52     \\
Case for UDHmx modules &   6    &  2.3  &    14     \\
Time Distribution      &   1    &  35   &    35     \\
Online Computers       &   20   &  12   &    240 \\
VME crates             &   45   &  6 &    270     \\ \hline
Consumables            &        &       &    100 \\ \hline
Contingencies          &        & & $\approx$350 (15\%) \\ \hline
Total                  &        &         &    2600 \\ \hline
\end{tabular}
\end{table}
\chapter{Computing and offline software}

\section{SPD computing model}


Expected event rate of the SPD experiment is about 3 MHz ($pp$ collisions at $\sqrt{s}$ = 27 GeV and $10^{32}$ cm$^{-2}$s$^{-1}$ design luminosity). This is equivalent to the raw data rate of 20 GB/s or 200 PB/year, assuming the detector duty cycle is 0.3, while the signal-to-background ratio is expected to be in order of $10^{-5}$. Taking into account the bunch crossing rate of 12.5 MHz, one may conclude that pile-up probability will be sufficiently high. 

\begin{figure}[htbp]
  \begin{center}
    \includegraphics[width=0.8\textwidth]{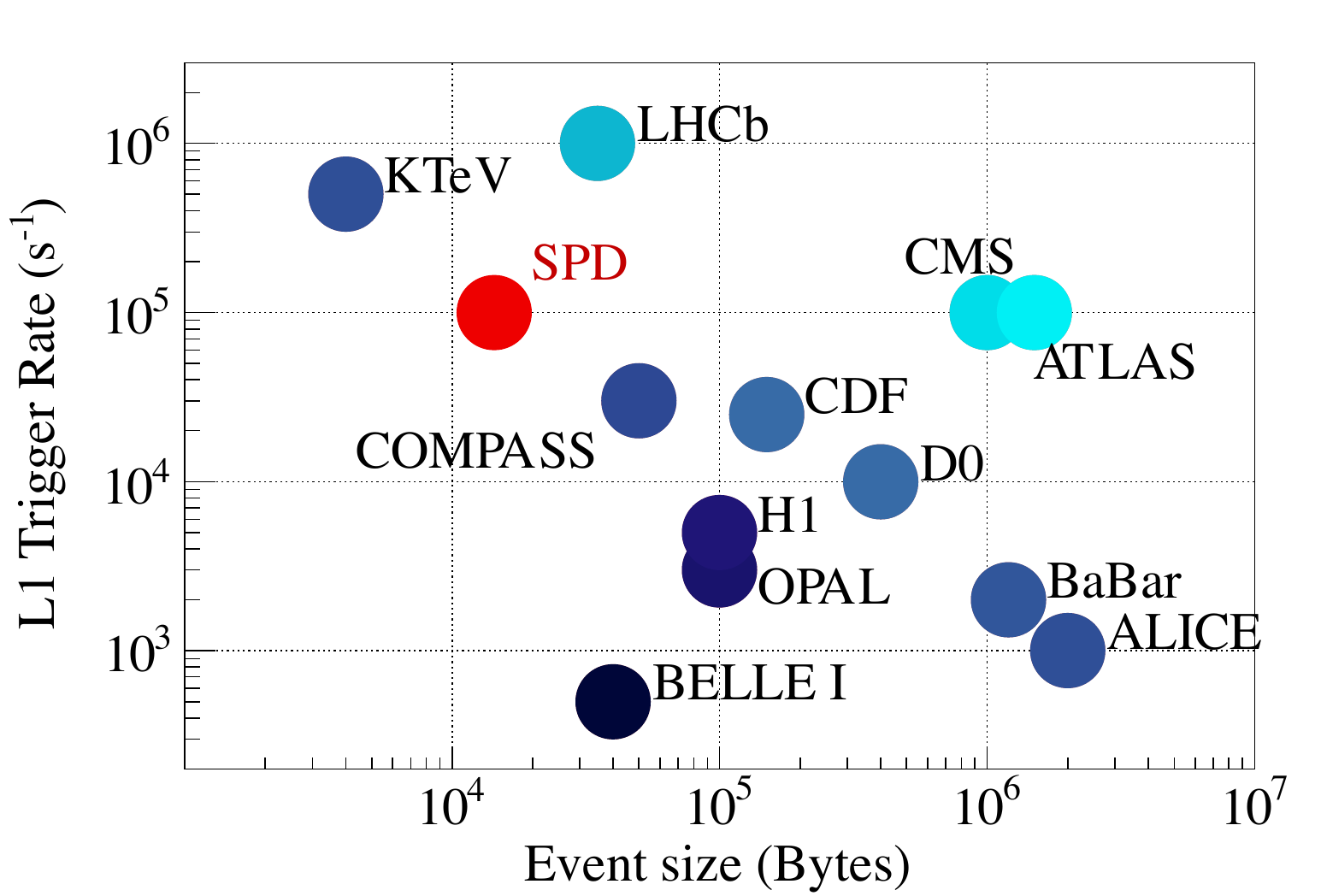}
  \end{center}
  \caption{Expected event size and event rate of the SPD setup after the online filter compared with other experiments \cite{Abbon:2014aex}.}
  \label{fig:rates}
\end{figure}

The key challenge of the SPD Computing Model is the fact, that no simple selection of physics events is possible at the hardware level, because the trigger decision would depend on measurement of momentum and vertex position, which requires tracking. Moreover, the free-running DAQ provides a continuous data stream, which requires a sophisticated unscrambling prior building individual events. That is the reason why any reliable hardware-based trigger system turns out to be over-complicated and the computing system will have to cope with the full amount of data supplied by the DAQ system. This makes a medium-scale setup of SPD a large scale data factory \ref{fig:rates}.

The continuous data reduction is a key point in the SPD computing. While simple operations like noise removal can be done yet by DAQ, it is an online filter that is aimed at fast partial reconstruction of events and data selection, thus being a kind of a software trigger. The goal of the online filter is to decrease the data rate at least by a factor of 50 so that the annual upgrowth of data including the simulated samples stays within 10 PB. Then, data are transferred to the Tier-1 facility, where full reconstruction takes place and the data is stored permanently. Two reconstruction cycles are foreseen. The first cycle includes reconstruction of some fraction of each run necessary to study the detector performance and derive calibration constants, followed by the second cycle of reconstruction of full data sample for physics analysis. The data analysis and Monte-Carlo simulation will likely run at the remote computing centers (Tier-2s). Given the large data volume, a thorough optimization of the event model and performance of reconstruction and simulation algorithms are necessary.

Taking into account recent advances in the computing hardware and software, the investment in the research and development necessary to deploy software to acquire, manage, process, and analyze the data recorded is required along with the physics program elaboration and the detector design. While the core elements of the SPD computing system and offline software now exist as prototypes, the system as a whole with capabilities such as described above is in the conceptual design stage and information will be added to SPD planning documents as it is developed.

\section{Online filter}

The SPD online filter facility will be a high-throughput system which will include heterogeneous computing platforms similar to many high performance computing clusters. The computing nodes will be equipped with hardware acceleration. The software framework will provide the necessary abstraction so that common code can deliver the selected functionality on different platforms.

The main goal of the online filter is a fast reconstruction of the SPD events and suppression of the background ones at least by a factor of 50. This requires fast tracking and fast clustering in the electromagnetic calorimeter, followed by reconstruction of event from a sequence of time slices and an event selection (software trigger). Several consecutive time slices shall be considered, tracker data unpacked and given for a fast tracking. The result of the fast track reconstruction is the number of tracks, an estimate of their momentum and an estimate of primary vertex (to distinguish between tracks belonging to different collisions). Using this outcome, the online filter should combine information from the time slices into events and add a trigger mark. The events shall be separated in several data streams using the trigger mark and an individual prescale factor for each stream is applied.

One of the most important aspects of this chain is the recognition of particle tracks. Traditional tracking algorithms, such as the combinatorial Kalman filter, are inherently sequential, which makes them rather slow and hard to parallelized on modern high-performance architectures (graphics processors). As a result, they do not scale well with the expected increase in the detector occupancy during the SPD data taking. This is especially important for the online event filter, which should be able to cope with the extremely high data rates and to fulfill the significant data reduction based on partial event reconstruction `on the fly'. The parallel resources like multicore CPU and GPU farms will likely be used as a computing platform, which requires the algorithms, capable of the effective parallelization, to be developed, as well as the overall cluster simulation and optimization.
 
Machine learning algorithms are well suited for multi-track recognition problems because of their ability to reveal effective representations of multidimensional data through learning and to model complex dynamics through computationally regular transformations, that scale linearly with the size of input data and are easily distributed across computing nodes. Moreover, these algorithms are based on the linear algebra operations and can be parallelized well using standard ML packages. This approach was already applied successfully to recognize tracks in the BM@N experiment at JINR and in the  BESIII experiment in IHEP CAS in China~\cite{Baranov:2018zcw, Ososkov:2020}. In the course of the project an algorithm, based on recurrent neural networks of deep learning, will be developed to search for and reconstruct tracks of elementary particles in SPD data from the silicon vertex detector and the straw tube-based main tracker. The same approach will be applied to the clustering in the SPD electromagnetic calorimeter, and fast $\pi^0$ reconstruction. The caution is necessary, though, to avoid possible bias due to an inadequacy of the training data to the real ones, including possible machine background and the detector noise. A dedicated workflow that includes continuous learning and re-learning of neuron network, deployment of new versions of network and the continuous monitoring of the performance of the neural networks used in the online filter is necessary and needs to be elaborated.  

Besides the high-level event filtering and corresponding data reduction, the online filter will provide input for the run monitoring by the shift team and the data quality assessment, as well as local polarimetry.

\section{Computing system}
    The projected rate and amount of data produced by SPD prescribe to use high throughput computing solutions for the processing of collected data. It is the experience of a decade of the LHC computing that already developed a set of technologies mature enough for the building of distributed high-throughput computing systems for HEP.

\subsection{Computing model}




The 'online' part of computing systems for the SPD experiment, namely the online filter described above, is an integral part of experimental facilities, connected with the 'offline' part using a high throughput backbone network. The entry point to 'offline' facilities is a high capacity storage system, connected with 'online facility' through a multilink high-speed network. Data from high capacity storage at the Laboratory of Information Technologies will be copied to the tape-based mass storage system for long term storage. At the same time, data from high capacity storage will be processed on different computing facilities as in JINR as in other collaborative institutions.

The hierarchy of offline processing facilities can be introduced:

\begin{itemize}
\item Tier 1 level facilities should provide high capacity long term storage which will have enough capacity to store a full copy of primary data and a significant amount of important derived data;

\item Tier 2 level facility should provide (transient) storage with capacity that will be enough for storing of data associated with a period of data taking;

\item optional Tier 3 level are opportunistic resources, that can be used to cope with a pile-up of processing during some period of time or for special analysis.
\end{itemize}
  
Offline data processing resources are heterogeneous as on hardware architecture level so by technologies and at JINR site it includes batch processing computing farms, high performance (supercomputer) facilities, and cloud resources. A set of middleware services will be required to have unified access to different resources.

\begin{figure}[htbp]
  \begin{center}
    \includegraphics[width=0.8\textwidth]{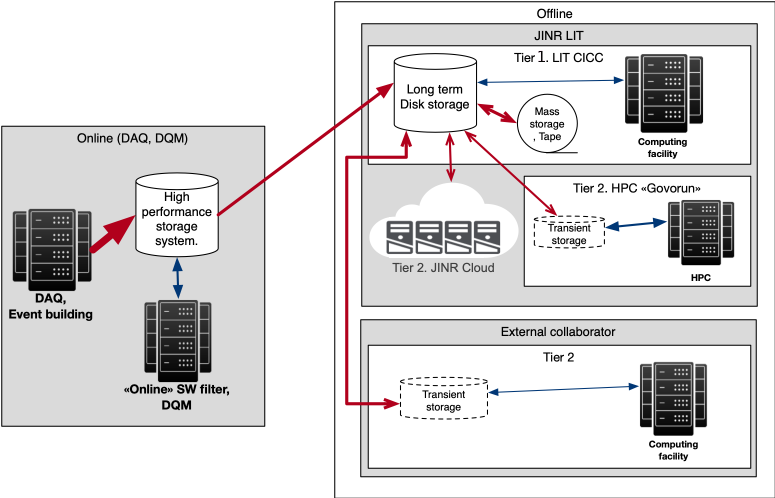}
  \end{center}
  \caption{Scheme of the SPD computing system}
  \label{comp_system}
\end{figure}

\subsection{Computing services}
Computing systems for NICA at JINR are naturally distributed. Experimental facilities and main data processing facilities placed across two JINR sites and, inter alia, managed by different teams. That causes some heterogeneity not only on hardware systems but also on the level of basic software: different OSs, different batch systems etc.

Taking into account the distributed nature and heterogeneity of the existing infrastructure, and expected data volumes, the experimental data processing system must be based on a set of low-level services that have proven their reliability and performance.

It is necessary to develop a high-level orchestrating system that will manage the low-level services. The main task of that system will be to provide efficient, highly automated multi-step data processing following the experimental data processing chain.

\begin{figure}[htbp]
  \begin{center}
    \includegraphics[width=0.8\textwidth]{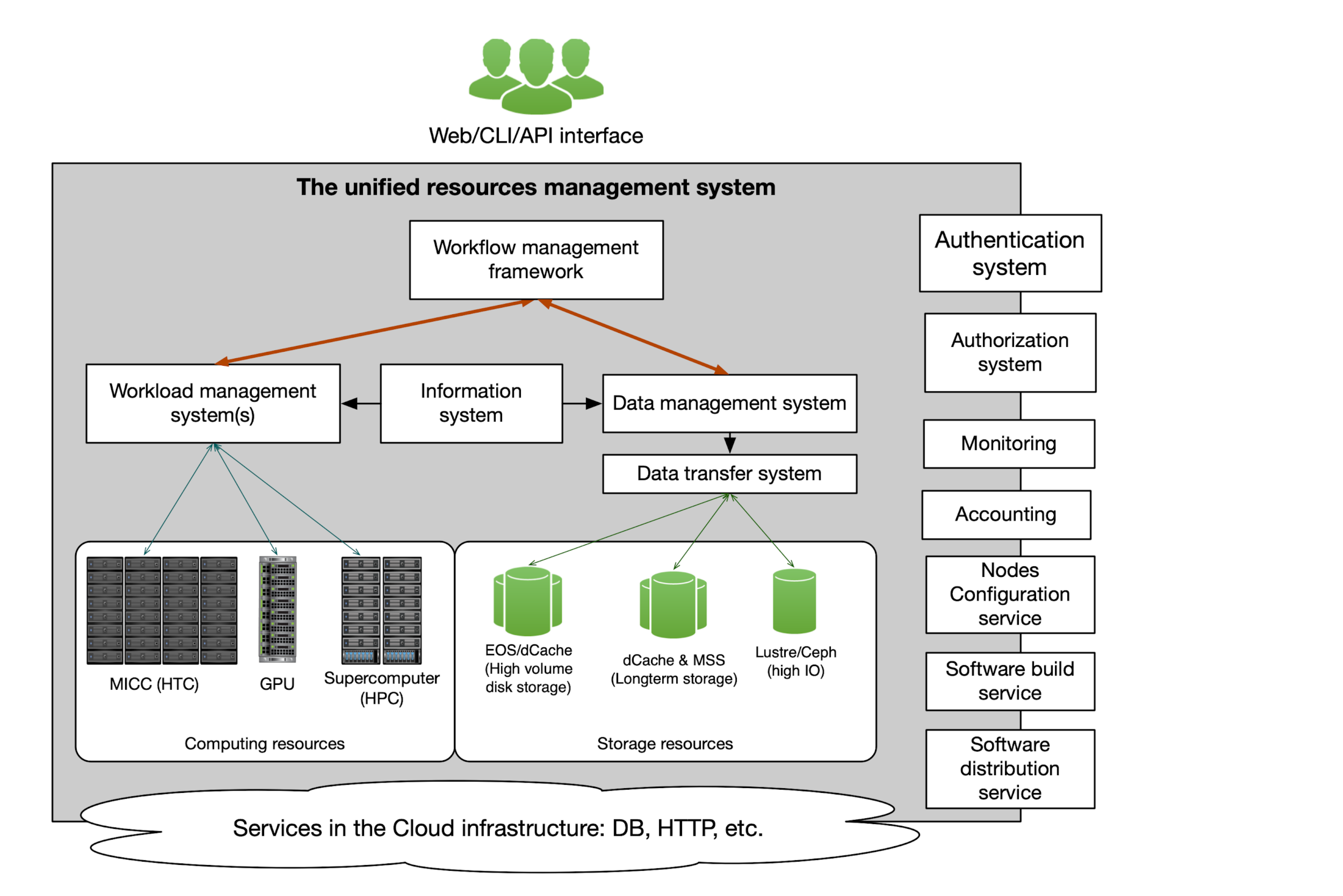}
  \end{center}
  \caption{Distributed SPD computing services}
  \label{comp_services}
\end{figure}
The Unified Resource Management System is a IT ecosystem composed from the set of subsystem and services which should:

\begin{itemize}
\item unify of access to the data and compute resources in a heterogeneous distributed environment;
\item automate most of the operations related to massive data processing;
\item avoid duplication of basic functionality, through sharing of systems across different users (if it possible);
\item as a result - reduce operational cost, increase the efficiency of usage of resources;
\item transparent accounting of usage of resources.
\end{itemize}

Many distributed computing tools have already been developed for the LHC experiments and can be re-used in SPD. For the task management one can use PANDA~\cite{Megino:2015evs} or DIRAC~\cite{Stagni:2017mmz} frameworks. For the distributed data management RUCIO~\cite{rucio} package has been developed. For the massive data transfer FTS~\cite{Frohner:2010zz} can be used. Evaluation of these tools for the SPD experiment and their implementation within the SPD Unified Resource Management System is planned in scope of the TDR preparation.

\section{Offline software}

Offline software is a toolkit for event reconstruction, Monte-Carlo simulation and data analysis. Linux is chosen as a base operating system.

Currently, the offline software of the SPD experiment -- SpdRoot -- is derived from the FairRoot software~\cite{AlTurany:2012gc} and it is capable of Monte Carlo simulation, event reconstruction, and data analysis and visualization. The SPD detector description is flexible and based on the ROOT geometry package. Proton-proton collisions are simulated using a multipurpose generator Pythia8~\cite{Sjostrand:2014zea}. Deuteron-deuteron collisions are simulated using a modern implementation of the FRITIOF model~\cite{Andersson:1986gw, NilssonAlmqvist:1986rx}, while UrQMD~\cite{Bass:1998ca, Bleicher:1999xi} generator is used to simulate nucleus-nucleus interactions. Transportation of secondary particles through the material of the SPD setup and the simulation of detector response is provided by Geant4 toolkit~\cite{Agostinelli:2002hh, Allison:2006ve, Allison:2016lfl}. Track reconstruction uses GenFit toolkit~\cite{Rauch:2014wta} and KFparticle package~\cite{etde_22400191} is used to reconstruct primary and secondary vertices. The central database is going to be established to keep and distribute run information, slow control data and calibration constants.

Recent developments in computing hardware resulted in the rapid increase of potential processing capacity from increases in the core count of CPUs and wide CPU registers. Alternative processing architectures have become more commonplace. These range from the many-core architecture based on x86\_64 compatible cores to numerous alternatives such as other CPU architectures (ARM, PowerPC) and special co-processors/accelerators: (GPUs, FPGA, etc). For GPUs, for instance, the processing model is very different, allowing a much greater fraction of the die to be dedicated to arithmetic calculations, but at a price in programming difficulty and memory handling for the developer that tends to be specific to each processor generation. Further developments may even see the use of FPGAs for more general-purpose tasks. 

The effective use of these computing resources may provide a significant improvement in offline data processing. However, the offline software should be capable to do it by taking advantage of concurrent programming techniques, such as vectorization and thread-based programming. Currently, the SPD software framework, SpdRoot, cannot use these techniques effectively. The studies of the concurrent-capable software frameworks (e.g. ALFA~\cite{Al-Turany:2015ied}, Key4Hep~\cite{key4hep}) are needed to provide input for the proper choice of the offline software for Day-1 of the SPD detector operation, as well as a dedicated R\&D effort to find proper solutions for the development of efficient cross-platform code.

A git-based infrastructure for the SPD software development is already established at JINR~\cite{git}.

\section {Resource estimate}

For the online filter we assume the CPU consumption of 1000 SPD events/core/second. This requires 3000 cores simultaneously for the fast tracking. Taking into account additional expenditures to the event unscrambling and data packing and including a real efficiency of CPU which will be lower than 100\%, one derives the CPU resources for the online filter as 6000 CPU cores. This number sets the upper limit and the required computing power may decrease substantially if an efficient way to use GPU cores is implemented for the event filtration. As for the data storage, a high performance disk buffer of 2 PB capable to keep data of about one day of data taking is needed.  

For the offline computing, the data storage is determined by the data rate after the online filter, or 4 PB/year of raw data. Besides that, we may expect the comparable amount of simulated data and estimate the long term storage as 10 PB/year, assuming two cycles of data processing and possible optimization of the data format and data objects to be stored permanently. We assume that a half of the annual data sample ($\sim$5 PB) is kept on disk storage, and the rest is stored on tape. The CPU power necessary to process the amount of data like this and to run Monte-Carlo simulation is estimated as many as 30000 CPU cores. The summary of computing resources is given in Table.~\ref{cost_tab}. The cost estimate is conservative and will be defined more exactly in the TDR, when detailed hardware solutions and their actual price in the market will be considered.

\begin{table}
  \caption{Required SPD computing resources}
  \label{cost_tab}
  \begin{center}
  \begin{tabular}{lccc}
    \hline
                  & CPU [cores] & Disk [PB] & Tape [PB] \\
    \hline
    Online filter & 6000 & 2 & none \\
    Offline computing  & 30000 & 5 & 9 per year \\
    \hline
    Cost estimate [kUSD] & 4000 & 8000 & 4500 per year \\
  \end{tabular}
  \end{center}
\end{table}

The burden of the SPD computing system operation is a subject of sharing between the computing centers of the participating institutes.

\chapter{Physics performance \label{MC}}
\section{General performance of the SPD setup}
\subsection{Minimum bias events}
\begin{figure}[!b]
    \center{\includegraphics[width=0.8\linewidth]{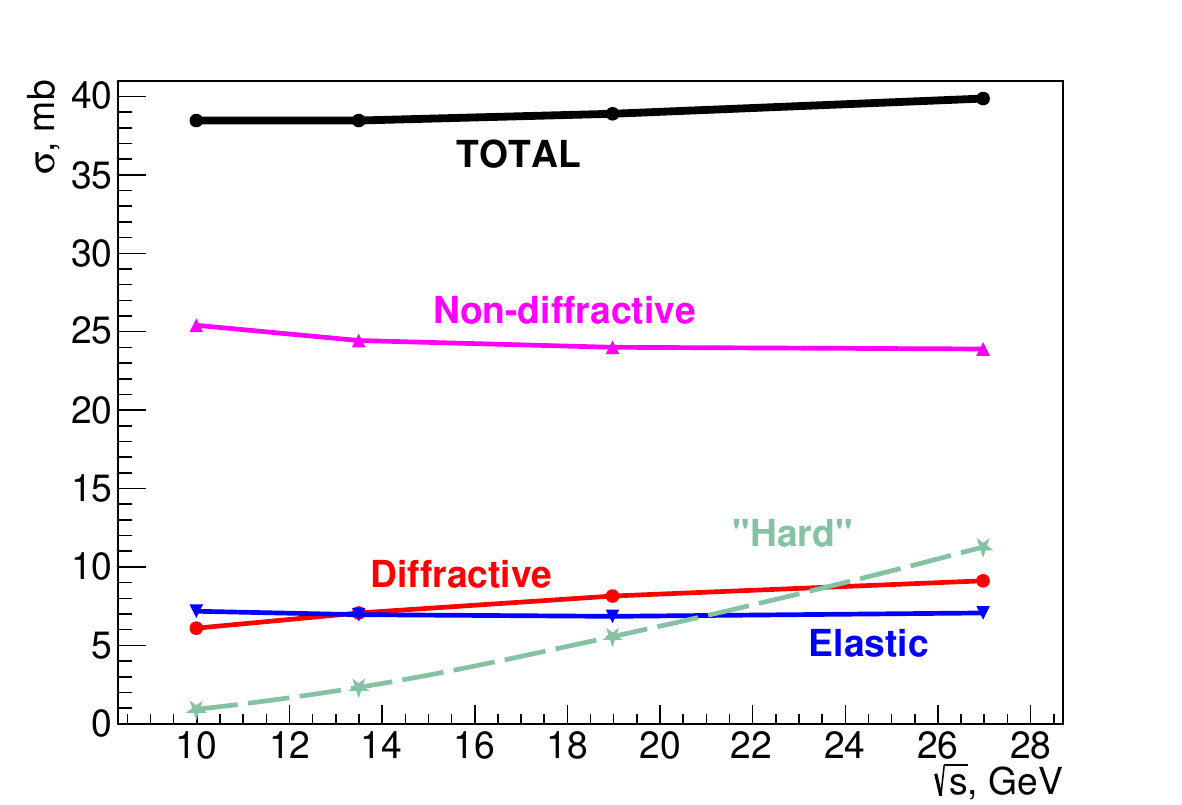}}
  \caption{Main contributions to the total cross-section of $p$-$p$ interaction as a function of $\sqrt{s}$. The ''hard'' cross-section is a part of the non-diffractive one.}
  \label{fig:tot_cross_section}  
\end{figure}

\begin{figure}[!t]
    \begin{minipage}{0.49\linewidth}
    \center{\includegraphics[width=1\linewidth]{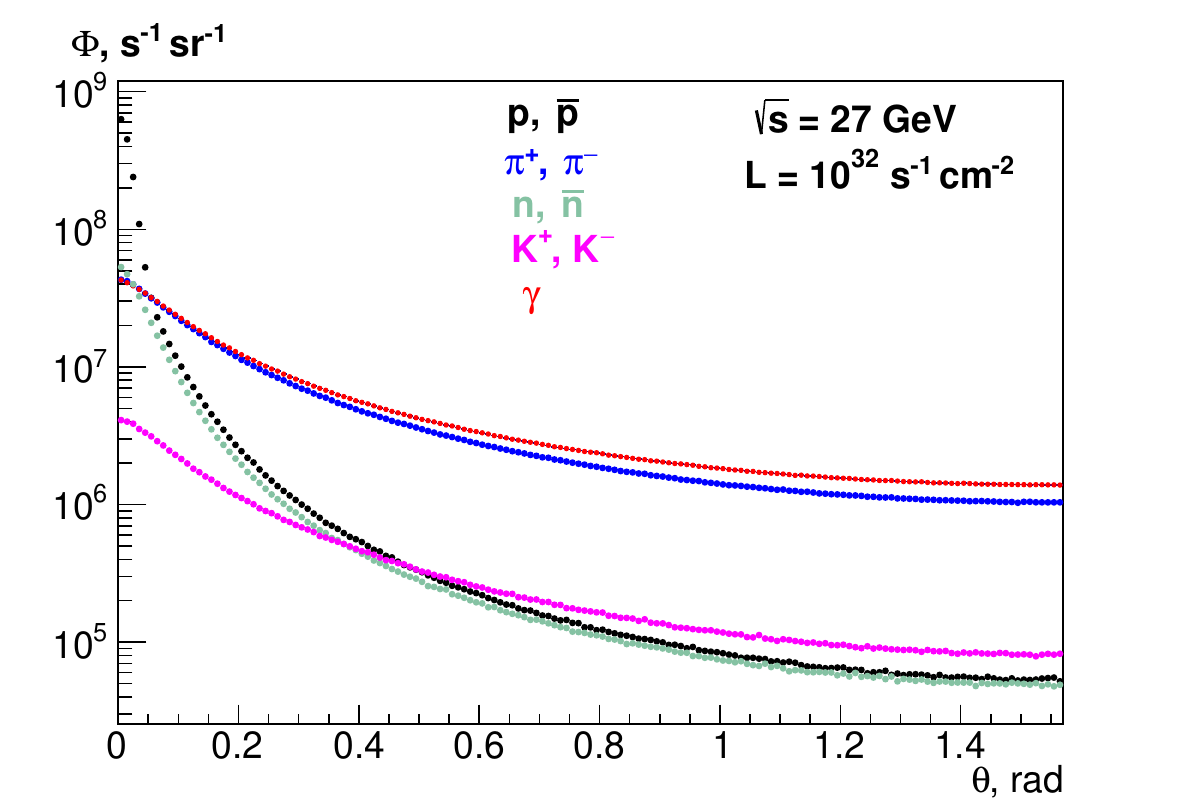} \\ (a)}
  \end{minipage}
  \hfill
  \begin{minipage}{0.49\linewidth}
    \center{\includegraphics[width=1\linewidth]{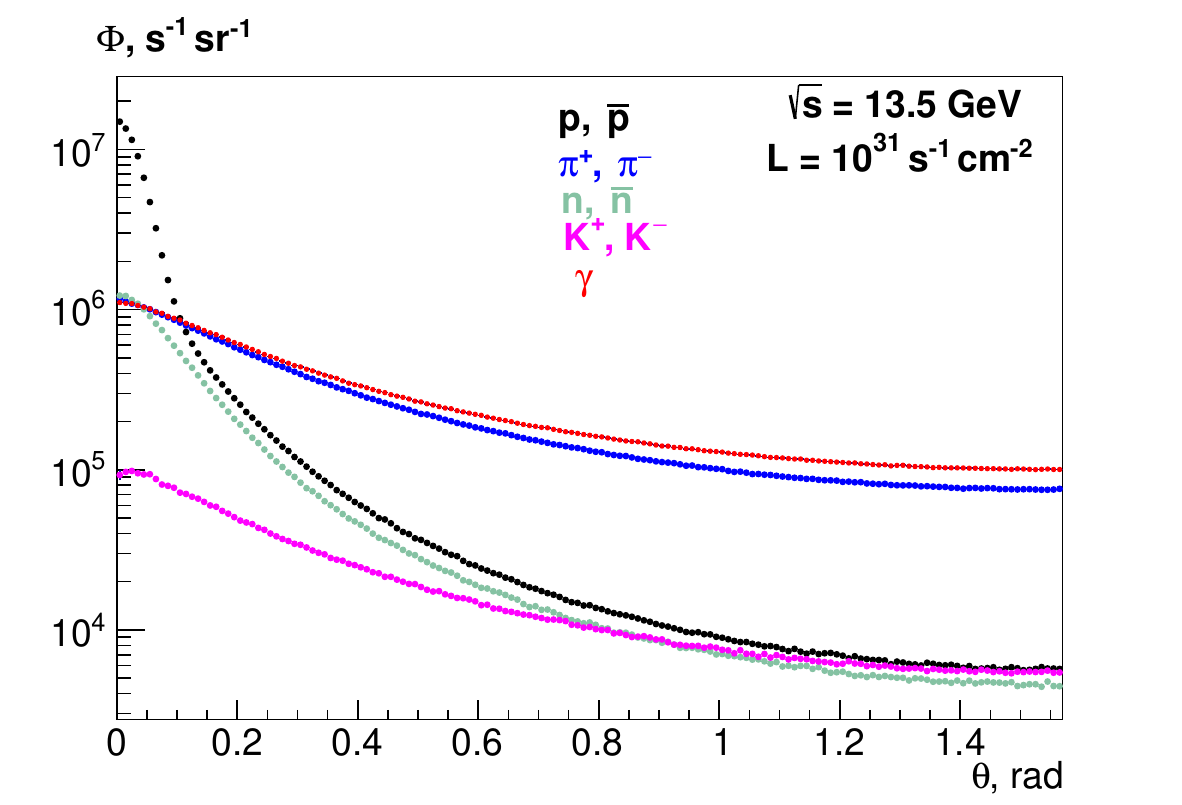} \\ (b)}
  \end{minipage}
  \caption{Fluxes of $p+\bar{p}$, $\pi^{\pm}$,$K^{\pm}$, $n+\bar{n}$ and $\gamma$ as a function of the polar angle $\theta$ for (a) $\sqrt{s}=27$ GeV and (b) 13.5 GeV.}
  \label{fig:angular}  
\end{figure}
\begin{figure}[!b]
  \begin{minipage}{0.49\linewidth}
    \center{\includegraphics[width=1\linewidth]{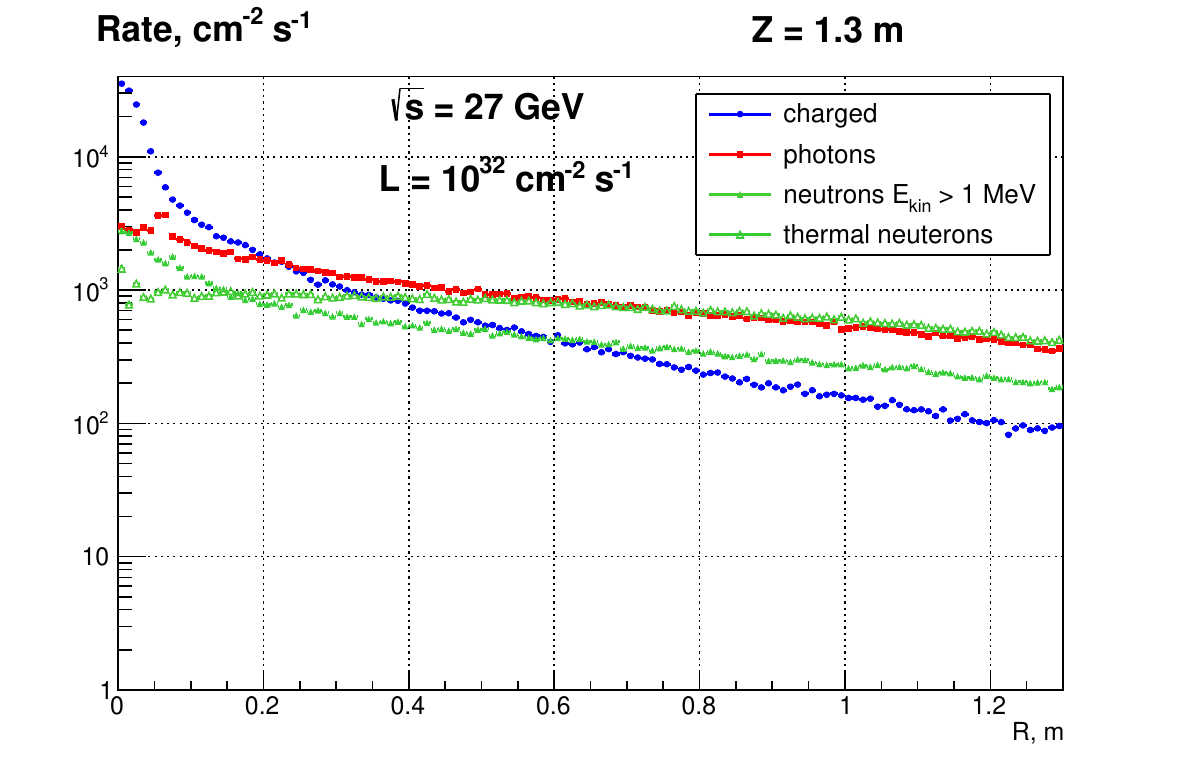} \\ (a)}
  \end{minipage}
  \hfill
  \begin{minipage}{0.49\linewidth}
    \center{\includegraphics[width=1\linewidth]{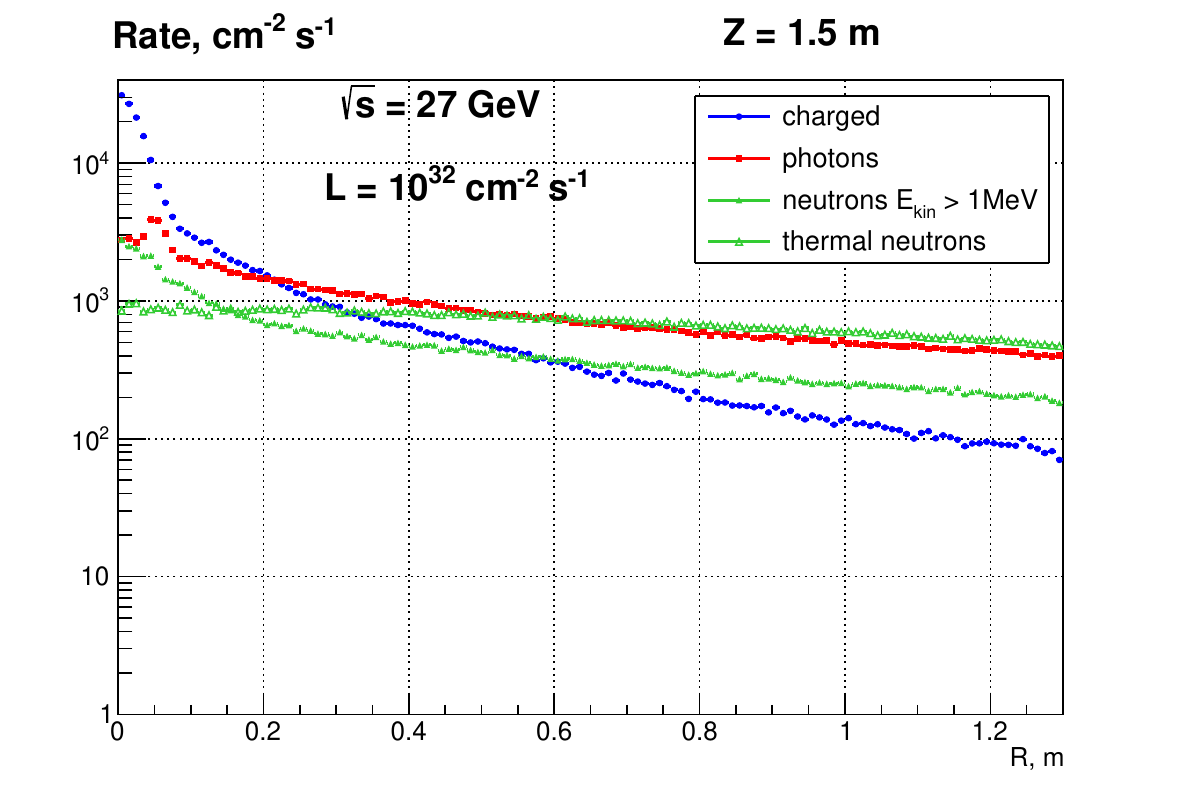} \\ (b)}
  \end{minipage}
    \begin{minipage}{0.49\linewidth}
    \center{\includegraphics[width=1\linewidth]{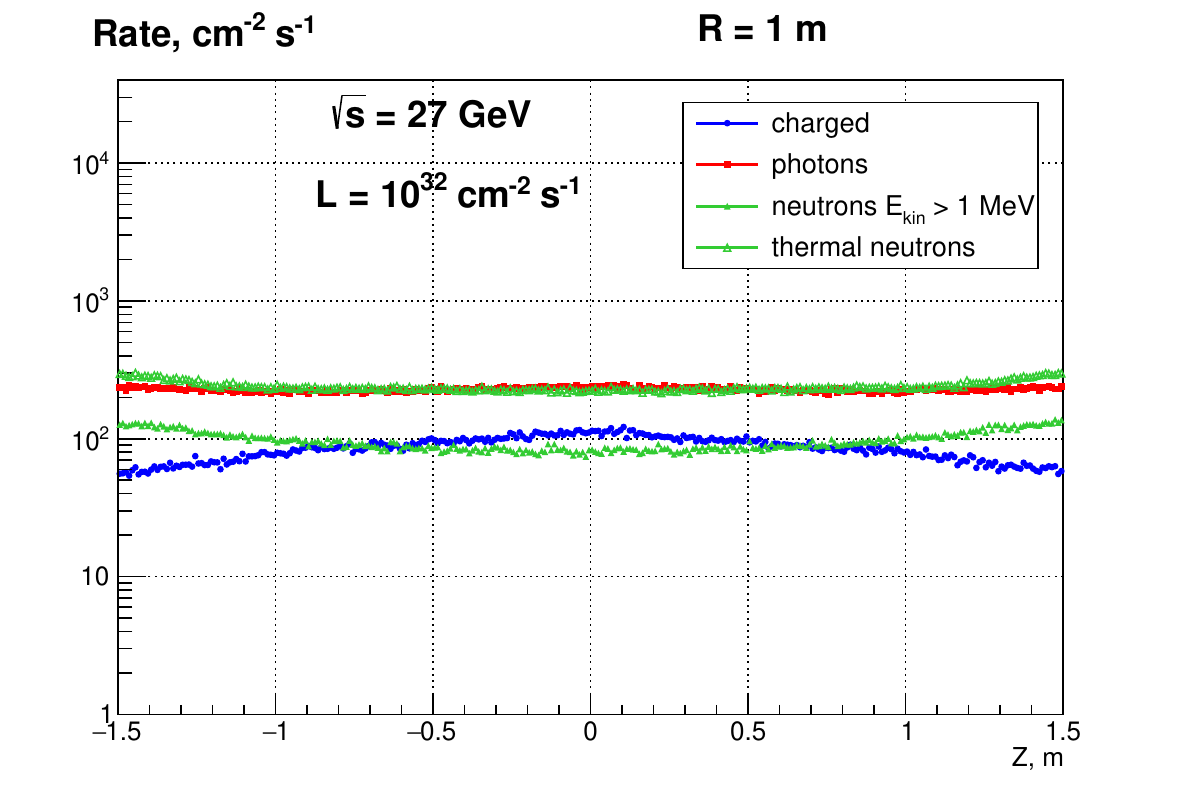} \\ (c)}
  \end{minipage}
  \hfill
  \begin{minipage}{0.49\linewidth}
   \center{\includegraphics[width=1\linewidth]{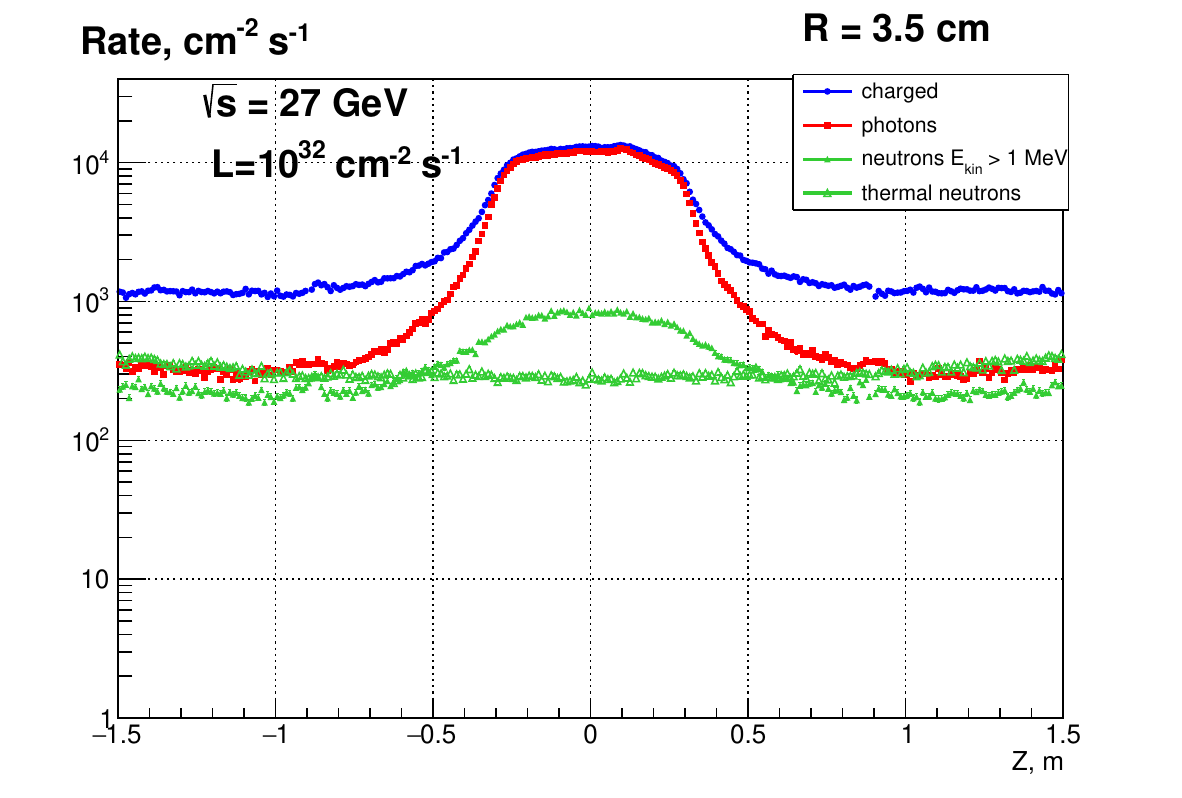} \\ (d)}
  \end{minipage}
  \caption{Flux of charged particles, photons, and neutrons in the radial direction at (a) Z=1.3 m, (b) Z=1.5 m, (c) R=1 m, and (d) R=3.5 cm.}
  \label{fig:flux}  
\end{figure}

The total cross-section of the $p$-$p$ collisions in the full energy range of SPD operation is a constant and equals about 40 mb. The main contributions to that cross-section, i. e. the elastic scattering, the diffractive and non-diffractive processes, are shown in Fig. \ref{fig:tot_cross_section}. The cross-section of ''hard'' processes, the QCD processes with partonic transverse momentum $\hat{p}_T>1$ GeV/$c$, is also shown as a part of  the non-diffractive cross-section. The beam particle collisions in the interaction point are the source of numerous secondary charged and neutral particles in the SPD setup that fully defines our experimental conditions (the load of the detector, radiation environment, etc.). The fluxes of different kinds of the charged and neutral particles produced in the interaction point as a function of the polar angle are shown in Fig. \ref{fig:angular}(a) and (b) for $\sqrt{s}=27$ and 13.5 GeV, respectively.
Table \ref{tab:MB_mult} shows the total cross-section of the $p$-$p$ collisions and the multiplicity of charged and neutral  particles for different collision energies $\sqrt{s}$. 

The secondary interactions in the material of the setup, the multiple scattering, the decays of unstable particles, and the influence of the magnetic field modify  the  radiation environment inside the SPD setup significantly. All these factors are taken into account in Figure \ref{fig:flux} that illustrates the fluxes of the charged particles, photons, and neutrons at different points of the SPD setup for $\sqrt{s}=27$ GeV and $L=10^{32}$  cm$^{-2}$ s$^{-1}$: Z=1.3 m (a),  Z=1.5 m (b), R=1 m (c), and R=3.5 cm (d).

\begin{table}[htp]
\caption{Total cross-section and the average multiplicity of the charged and neutral particles produced in the $p$-$p$ collisions as a function of $\sqrt{s}$.}
\begin{center}
\begin{tabular}{|c|c|c|c|}
\hline
$\sqrt{s}$, GeV & $\sigma_{tot}$, mb & Charged & Neutral ($\gamma$)  \\
             &                            &multiplicity & multiplicity  \\
\hline
13        &   38.4                      & 5.9          & 4.6 (3.8) \\
20        &   38.9                      & 7.2          & 6.0 (5.0)  \\
26         &  39.7                      & 7.8          & 6.5 (5.5)   \\
\hline
\end{tabular}
\end{center}
\label{tab:MB_mult}
\end{table}

\subsection{Tracking}

Traditionally, the track reconstruction procedure is divided into two separate tasks: track finding (or pattern recognition) and  track fitting. Since the track multiplicity in the $p$-$p$ collisions is low enough (see Tab. \ref{tab:MB_mult}) the occupancy of the coordinate detectors is not really a problem. Thus, we hope to have the efficiency of the track finding to be not less than 90\% in the most of our acceptance and we are not paying too much attention to the pattern recognition algorithms now. However, high multiplicity will limit the SPD performance in the case of NICA operation with heavy-ion beams.

\begin{figure}[!b]
    \begin{minipage}{0.49\linewidth}
    \center{\includegraphics[width=1\linewidth]{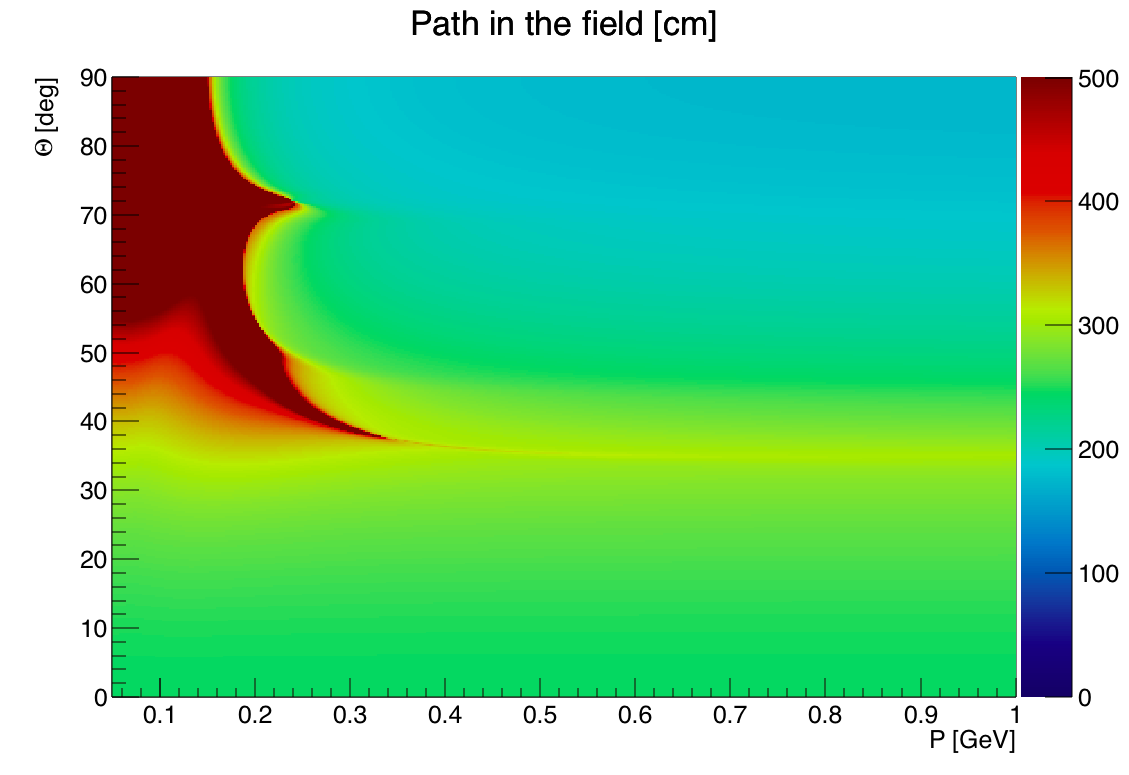} \\ (a)}
  \end{minipage}
  \hfill
  \begin{minipage}{0.49\linewidth}
    \center{\includegraphics[width=1\linewidth]{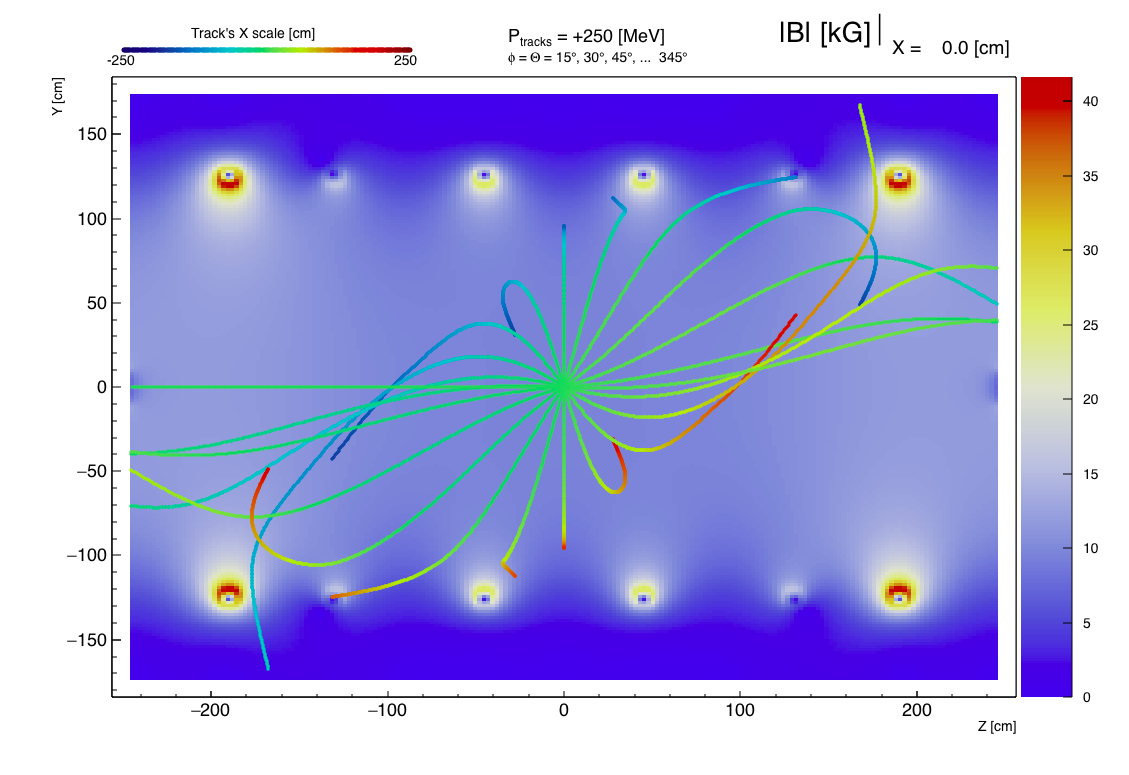} \\ (b)}
  \end{minipage}
  \caption{(a) Track length in the SPD tracking system for charged particles of different momenta emitted from the IP at different polar angles. (b) Trajectories for muons with momentum of 0.25 GeV/$c$ emitted from the IP at different polar and azimuthal angles $\theta=\phi$.  }
  \label{fig:track0}  
\end{figure}

The SPD magnetic system provides the conditions for measurement of the charged particle momenta. The track length in the SPD tracking system for charged particles of different momenta emitted from the interaction point at different polar angles as well as the trajectories for muons with momentum of 0.25 GeV/$c$ emitted at different polar and azimuthal angles $\theta=\phi$ are presented in Fig.  \ref{fig:track0}~(a) and (b), respectivly.

\begin{figure}[!h]
    \begin{minipage}{0.49\linewidth}
    \center{\includegraphics[width=1\linewidth]{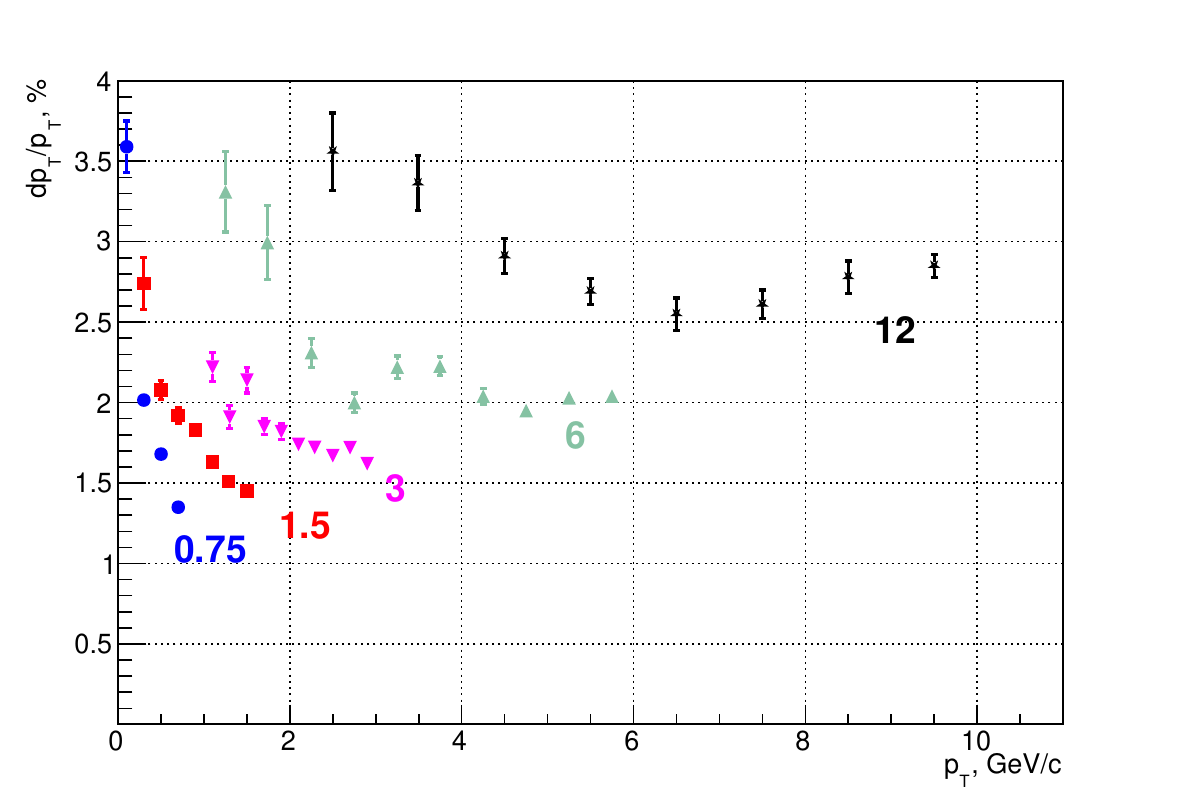} \\ (a)}
  \end{minipage}
  \hfill
  \begin{minipage}{0.49\linewidth}
    \center{\includegraphics[width=1\linewidth]{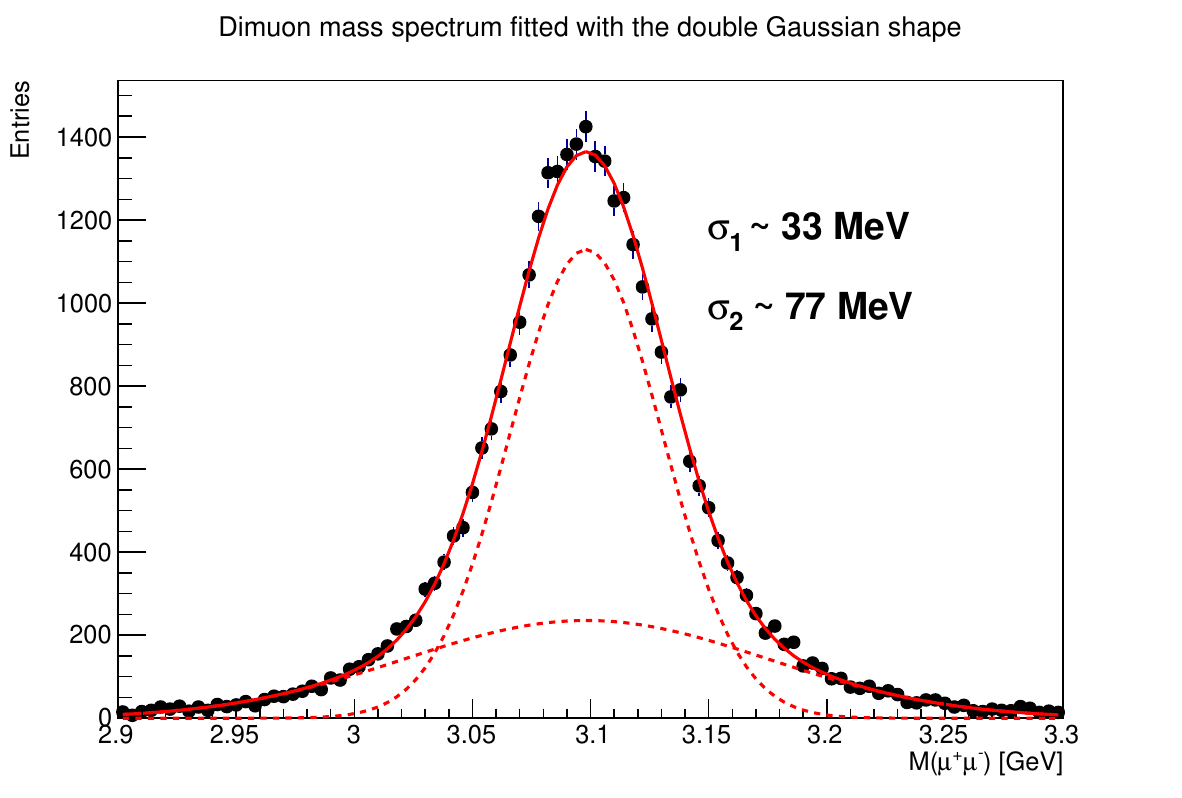} \\ (b)}
  \end{minipage}
  \caption{(a) Expected resolution for the transverse momentum $\sigma_{p_T}/p_T$ of muons with momentum 0.75, 1.5, 3, 6, and 12 GeV/$c$. (b) $J/\psi$ peak from the dimuon decay.}
  \label{fig:track1}  
\end{figure}

The track fitting procedure uses measured hits in the tracking detectors (or simulated points for Monte Carlo events) as an input, and calculates the most probable track parameters at any given point along the track, together with the corresponding covariance matrix. The fitting procedure also takes into consideration such effects related to the particle interaction in the material as multiple scattering, energy losses, and magnitude and configuration of the magnetic field. For the track fitting at SPD the well-known Kalman filter \cite{Kalman60} implemented within the GenFit2 package \cite{Rauch:2014wta} is used. The GenFit2 extrapolates tracks using the standard Runge-Kutta-Nyström method \cite{abramowitz_stegun} modified by Bugge and Myrheim to carry along the Jacobian matrix \cite{Myrheim:1979ng, Bugge:1980iv}.

The expected transverse momentum resolution $\sigma_{p_T}/p_T$ for muons with different momenta for the maximal magnetic field 1.0 T at the beam axis is shown in Fig. \ref{fig:track1}(a). The corresponding resolution for muons emitted at the polar angle $\theta =  90^{\circ}$ could be expressed as
\begin{equation}
\sigma_{p}/p\big|_{\theta=90^{\circ}} = 1.3\% + 0.1\% \times p + 0.003\% \times p^2.
 \end{equation}

The width of the $J/\psi$ peak shown in Fig. \ref{fig:track1}(b) is a good indicator of the tracking performance. The SPD tracking system demonstrates the width at the level of 40 MeV. It is 1.5 times better than at the fixed-target COMPASS experiment with the open setup ($\sim$ 60 MeV \cite{Adolph:2014hba}) and much better than in the fixed-target beam dump experiments like NA3 (80$\div$120 MeV \cite{Badier:1983dg}), COMPASS ($\sim$200 MeV \cite{Aghasyan:2017jop}), SeaQuest ($\sim$150 MeV \cite{SeaQuest}), which worked successfully on the study of the partonic structure of the nucleon at the discussed energy range. 

\subsection{Vertex reconstruction}
\begin{figure}[!h]
    \begin{minipage}{0.49\linewidth}
    \center{\includegraphics[width=1\linewidth]{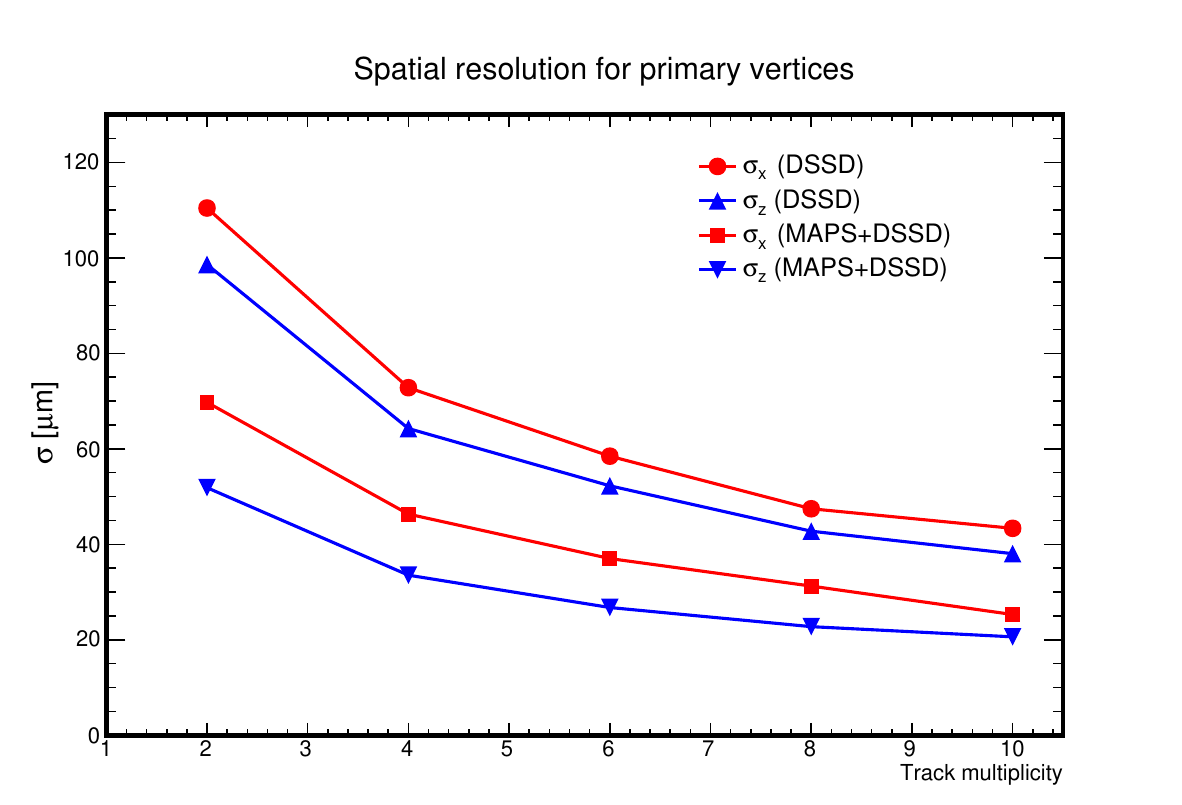} \\ (a)}
  \end{minipage}
  \hfill
  \begin{minipage}{0.49\linewidth}
    \center{\includegraphics[width=1\linewidth]{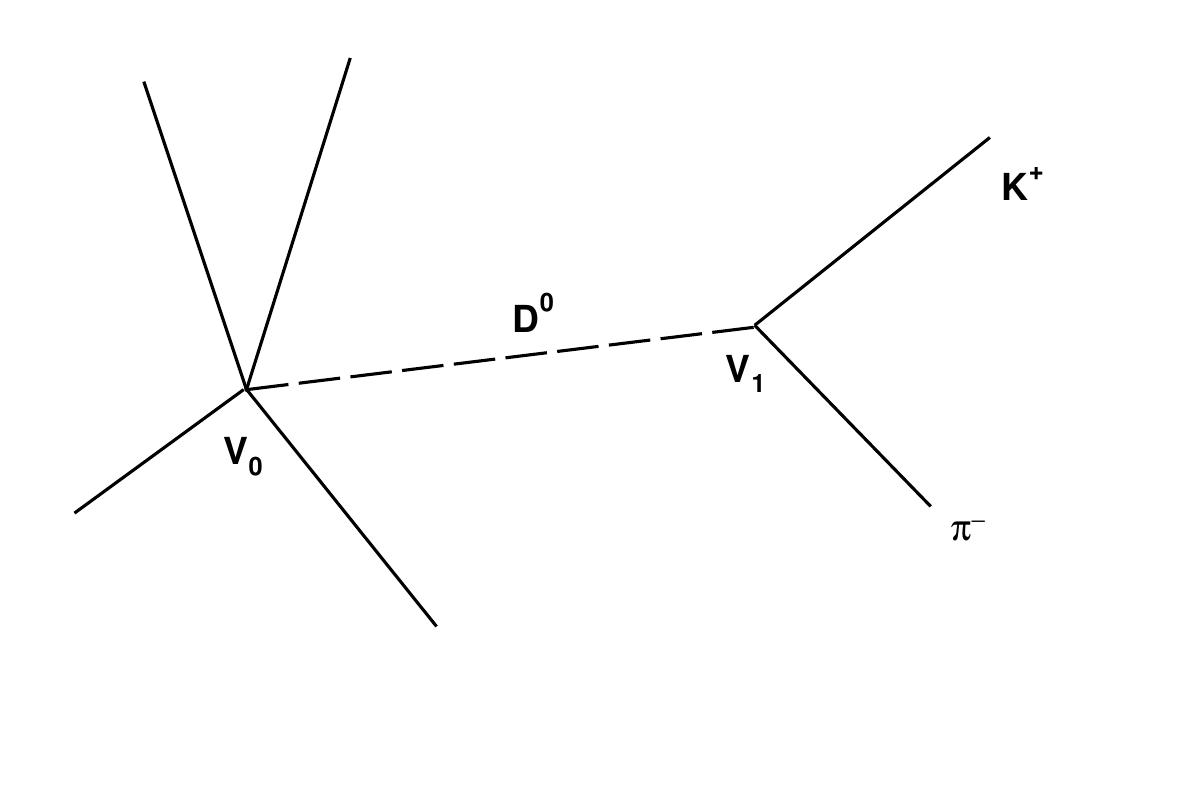} \\ (b)}
  \end{minipage}
  \caption{(a) Accuracy of the primary vertex position reconstruction as a function of the number of outgoing tracks for two configurations of the vertex detector. (b)Sketch of $D^0$-meson production and decay.}
  \label{fig:vertexing1}  
\end{figure}

The only subsystem that defines reconstruction of primary vertices is the silicon vertex detector. Its impact on the accuracy of the vertex reconstruction depends on the baseline (the radial distance between layers), the amount of passed material producing the multiple scattering effects, and the spatial resolution of the detector. The latter is a rather complex function of the number of fired strips (or pixels). We estimate the effective spatial resolution of the DSSD layer as $\sigma_{\phi}=11$ $\mu$m, $\sigma_{z}=23$ $\mu$m, while the resolution of the MAPS layer is $\sigma_{\phi,z}=4$ $\mu$m. $\sigma_{\phi}$ here denotes the resolution in the direction perpendicular to the beam line. The effective values are about two times smaller than the corresponding pitch divided by $\sqrt{12}$.  The amount of the material  corresponds to 300 $\mu$m and 50 $\mu$m of silicon per one layer of the DSSD and MAPS, respectively.  Figure \ref{fig:vertexing1}(a) shows the accuracy of the primary vertex position reconstruction as a function of the number of outgoing tracks for two configurations of the vertex detector: (i) 5 layers of the DSSD, (ii) 3 layers of the MAPS and 2 layers of the DSSD. In both cases the accuracy becomes better with an increasing number of outgoing tracks, as expected. The DSSD+MAPS configuration demonstrates a 1.5 times better precision.

\begin{figure}[!h]
    \begin{minipage}{0.49\linewidth}
    \center{\includegraphics[width=1\linewidth]{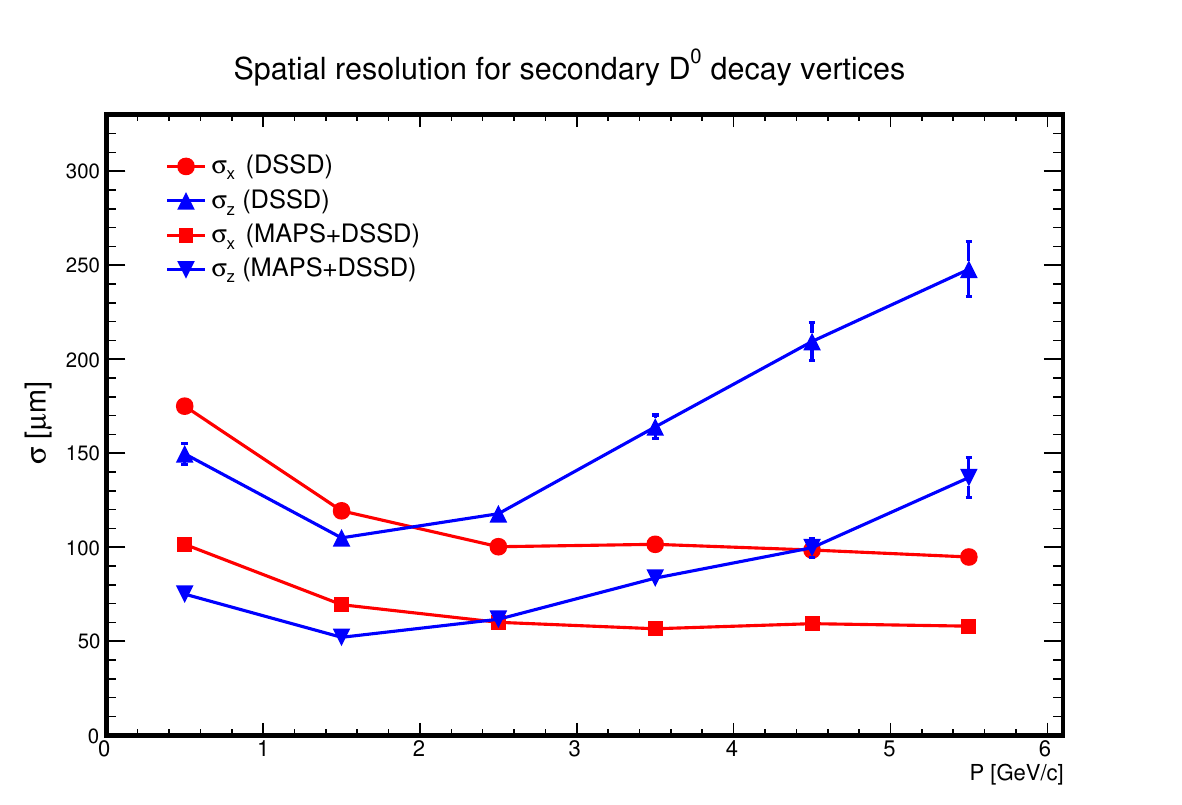} \\ (a)}
  \end{minipage}
  \hfill
  \begin{minipage}{0.49\linewidth}
    \center{\includegraphics[width=1\linewidth]{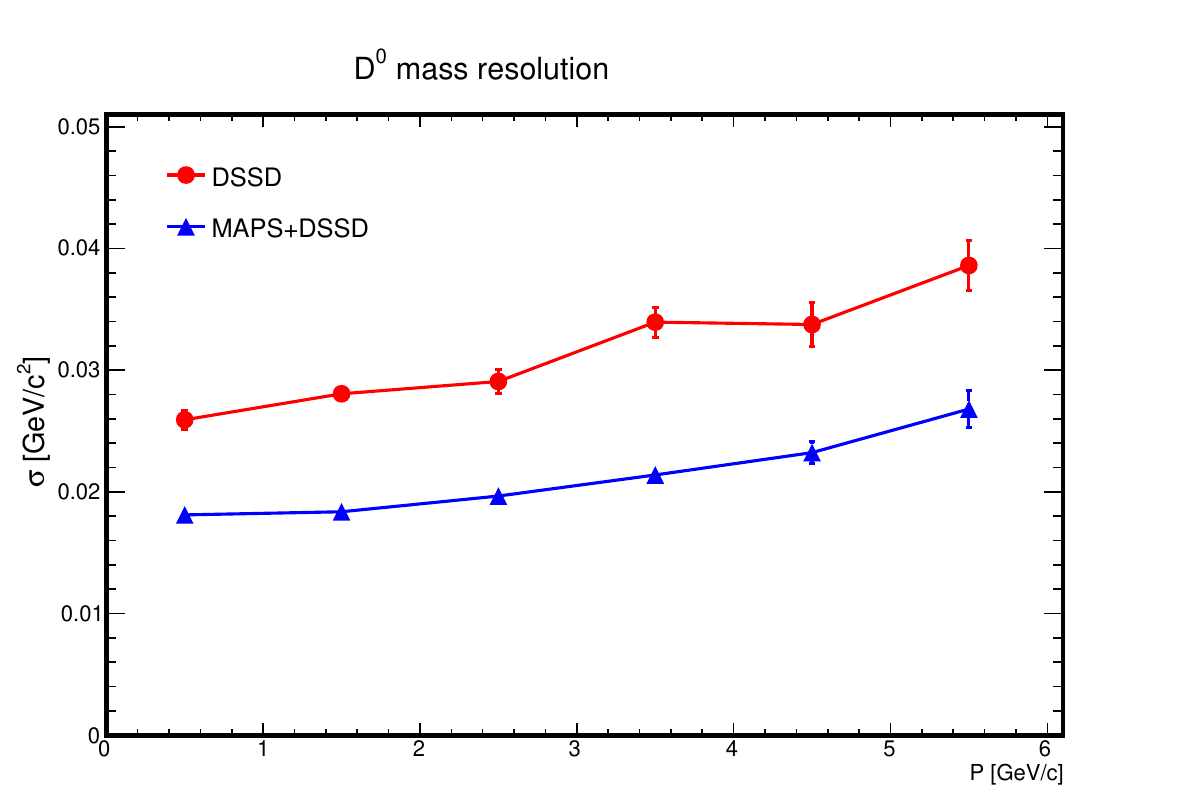} \\ (b)}
  \end{minipage}
  \caption{(a) Accuracy of the $D^0$-decay vertex reconstruction as a function of the  $D^0$  momentum. (b) $D^0$ peak width as a function of the $D^0$ momentum.}
  \label{fig:vertexing2}  
\end{figure}

\begin{figure}[!h]
    \begin{minipage}{0.45\linewidth}
    \center{\includegraphics[width=1\linewidth]{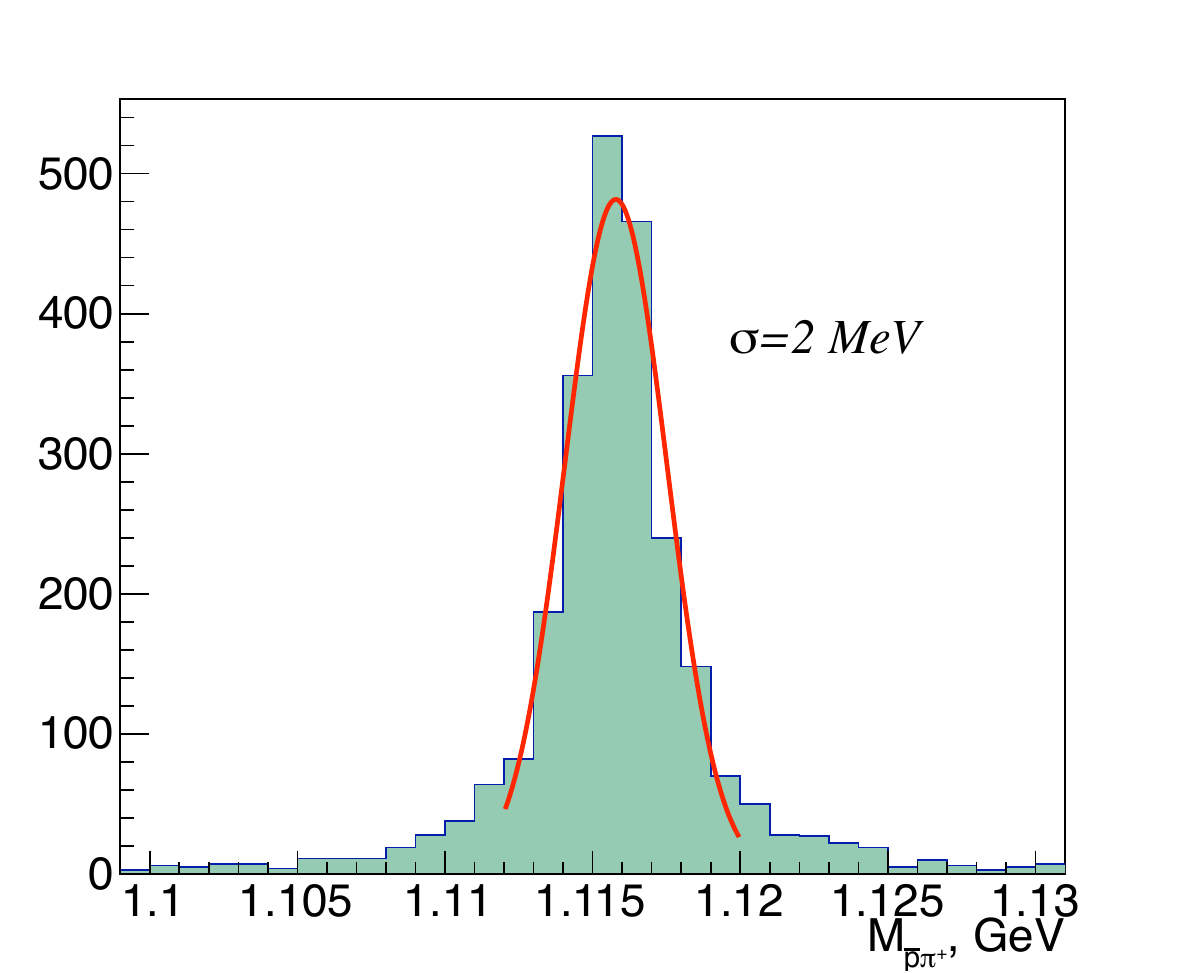} \\ (a)}
  \end{minipage}
  \hfill
  \begin{minipage}{0.54\linewidth}
    \center{\includegraphics[width=1\linewidth]{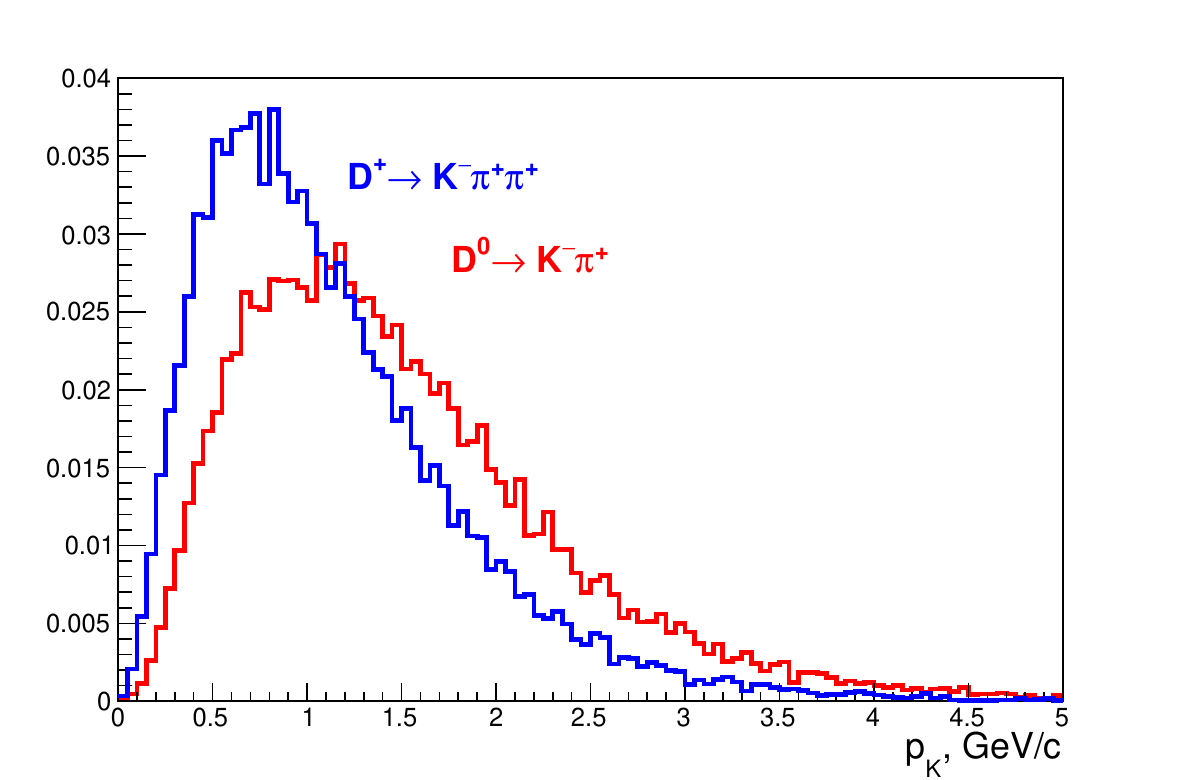} \\ (b)}
  \end{minipage}
  \caption{(a) $\Lambda$ peak in the $p\pi$ mass spectrum. (b) Kaon spectra from the decays $D^0\to K^-\pi^+ $ (red) and $D^+\to K^-\pi^+\pi^+$ (blue) produced at $\sqrt{s}=27$ GeV/$c$.}
  \label{fig:vertexing3}  
\end{figure}

The silicon vertex detector is also fully responsible  for reconstruction of the decay vertices of short-lived ($c\tau<1$ cm) particles. We use the $D^0\to K^+ \pi^-$ decay as an example (see the sketch in Fig. \ref{fig:vertexing1}~(b)), but all the conclusions are also valid qualitatively for the decays like $D^+ \to K^{-}\pi^+\pi^-$, $\Lambda_c^+ \to p \pi^+ K^-$ etc. The accuracy of the $D^0$-decay vertex reconstruction as a function of the  $D^0$ momentum is shown in the Fig. \ref{fig:vertexing2}~(a).
The Gaussian width of the $D^0$-meson peak in the $K^+\pi^-$ mass spectrum determined by the tracking accuracy (mainly by the momentum resolution) is  27.2 and 25.0 MeV for the DSSD and DSSD+MAPS configurations, respectively. The constrained fit of the $D^0$-decay, where the angle between the reconstructed $D^0$ momentum and the line connecting  the primary and  secondary vertices is forced to be zero and the found vertex is included in the track fitting, reduces the width to 21.4 and 18.0 MeV. That improves, respectively, the signal-to-background ratio by the factor of 1.3 and 2.4. 
 The $D^0$-peak width obtained from the constrained fit as a function of $D^0$ momentum is shown in Fig \ref{fig:vertexing2}~(b).
 The impact of the secondary vertex reconstruction procedure on our expectations for the asymmetries measurement is discussed in Sec. \ref{sec:acc}.

The decays of relatively long-lived unstable particles like $\Lambda^0$, $K^0$, $\Sigma^-$ etc. occur mainly within the straw tracker. The $\Lambda^0\to p^+\pi^-$ peak is presented in Fig. \ref{fig:vertexing3}(a) as an example.

\subsection{Particle identification \label{sec:PID}}

The effective $\pi$/K/p separation is required first of all for such physics tasks as the open charm production (identification of kaons form $D^{\pm,0}$ decays, see Fig. \ref {fig:vertexing3}~(b)); measurement of the SSA for charged kaons, and the study of the antiprotons yield. The proposed SPD setup has three instruments for identification of charged hadrons. The energy deposit $dE/dx$ in the straw tubes of the ST can be used for particle identification at lowest momenta. The TOF system could extend the separation range up to 1.5 GeV/$c$ and $\sim$ 3 GeV/$c$  for $\pi/K$ and $K/p$, respectively. The separation for the higher range of hadron momenta could be provided only by the aerogel detector.

\subsubsection{Identification using $dE/dx$}
The energy deposit $dE/dx$ in the straw tubes is plotted in Fig \ref{fig:t0rec}~(a) for the particles emitted at $\theta=90^{\circ}$ in respect to the beam axis as a function of their momenta. The truncated mean approach, where 20\% of the measurement results from the individual straw tubes is discarded, was applied. The straw tracker is able to provide $\pi/K$ and $K/p$ separation up to 0.7 GeV/$c$ and 1.0 GeV/$c$, respectively. With respect to the TOF method (see below), the efficiency of the $dE/dx$ method does not degrade at the low polar angles, since the end-cup part of the ST has enough layers for the precision measurement of the energy deposit.

\subsubsection{TOF performance}
Particle identification with a TOF detector is based on comparison between the time of flight of the particle  from the primary vertex to the TOF detector and the expected time under a given mass hypothesis. For particle identification, the presence of only one plane of the TOF detector requires precise knowledge of the event collision time $t_0$.
 It can be estimated by the TOF detector on an event-by-event basis, using the $\chi^2$
minimization procedure for the events with two and more reconstructed tracks. Having $N$ tracks in the event, which are matched to the corresponding hits on the TOF plane, it is possible to define certain combinations of masses $\vec{m_i}$, assuming the p, K, or p mass for each track independently. The index $i$ indicates one of the possible combinations ($m_1$, $m_2$, . . . , $m_{N~tracks}$) among the 3N-track ones \cite{Adam:2016ilk}.

To each track the following weight is attributed
\begin{equation}
W_i = \frac{1}{\sigma^2_{TOF}+\sigma^2_{t_{exp.~i}}}.
\end{equation}
Here $\sigma_{TOF}$ and $\sigma_{t_{exp.~i}}$ are the time resolution of the TOF detector and the uncertainty of the  expected time of flight under a given mass hypothesis $t_{exp.~i}$, respectively. The latter is defined by the uncertainty of the momentum and track length measurements.

The following $\chi^2$ function has to be minimized

\begin{equation}
\chi^2(\vec{m_i}) = \sum_{N} W_i (( t_{TOF} - t_0(\vec{m_i}))-t_{exp.~i})^2.
\label{t0_chi2}
\end{equation}
Here
\begin{equation}
t_0(\vec{m_i}) = \frac{ \sum_{N} (t_{TOF} - t_{exp.~i})}{\sum_{N} W_i}.
\label{t0_t0}
\end{equation}
The mass vector $\vec{m_i}$ that minimizes $\chi^2$ in Eq. \ref{t0_chi2} can be used in Eq. \ref{t0_t0} for determination of the event collision time $t_0$. For unbiased particle mass determination, each track has to be subsequently excluded from the $t_0$ calculation procedure. 

\begin{figure}[!h]
    \begin{minipage}{0.54\linewidth}
    \center{\includegraphics[width=1\linewidth]{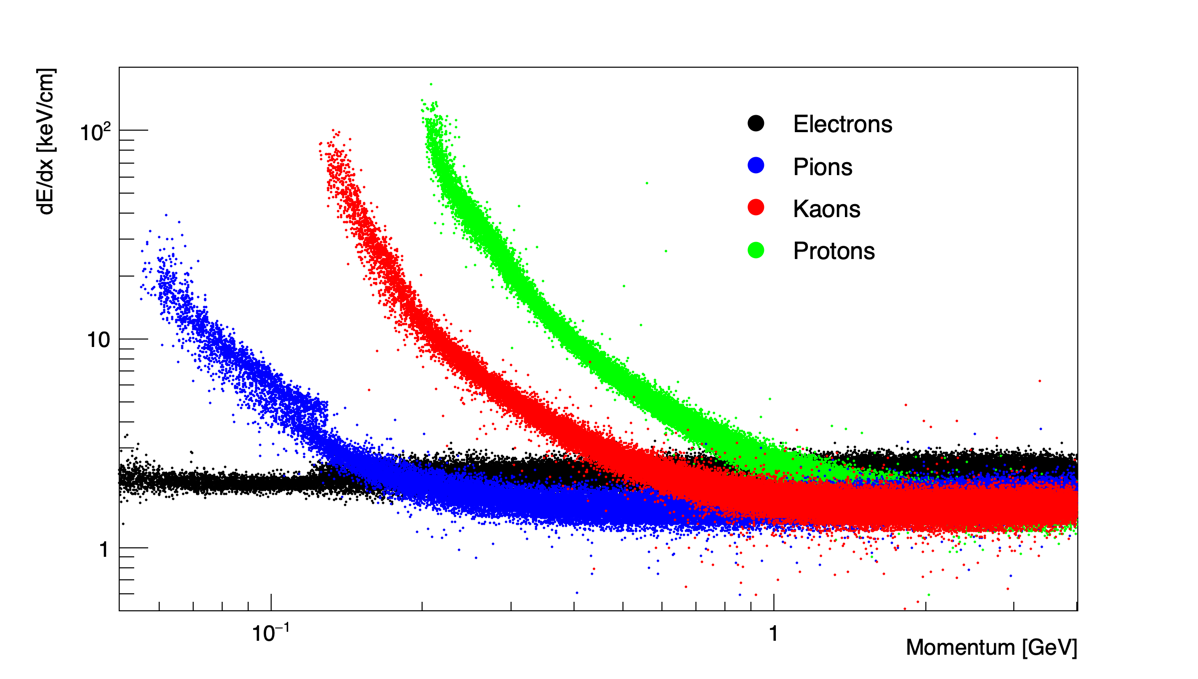} \\ (a)}
  \end{minipage}
  \hfill
  \begin{minipage}{0.45\linewidth}
   \center{\includegraphics[width=1\linewidth]{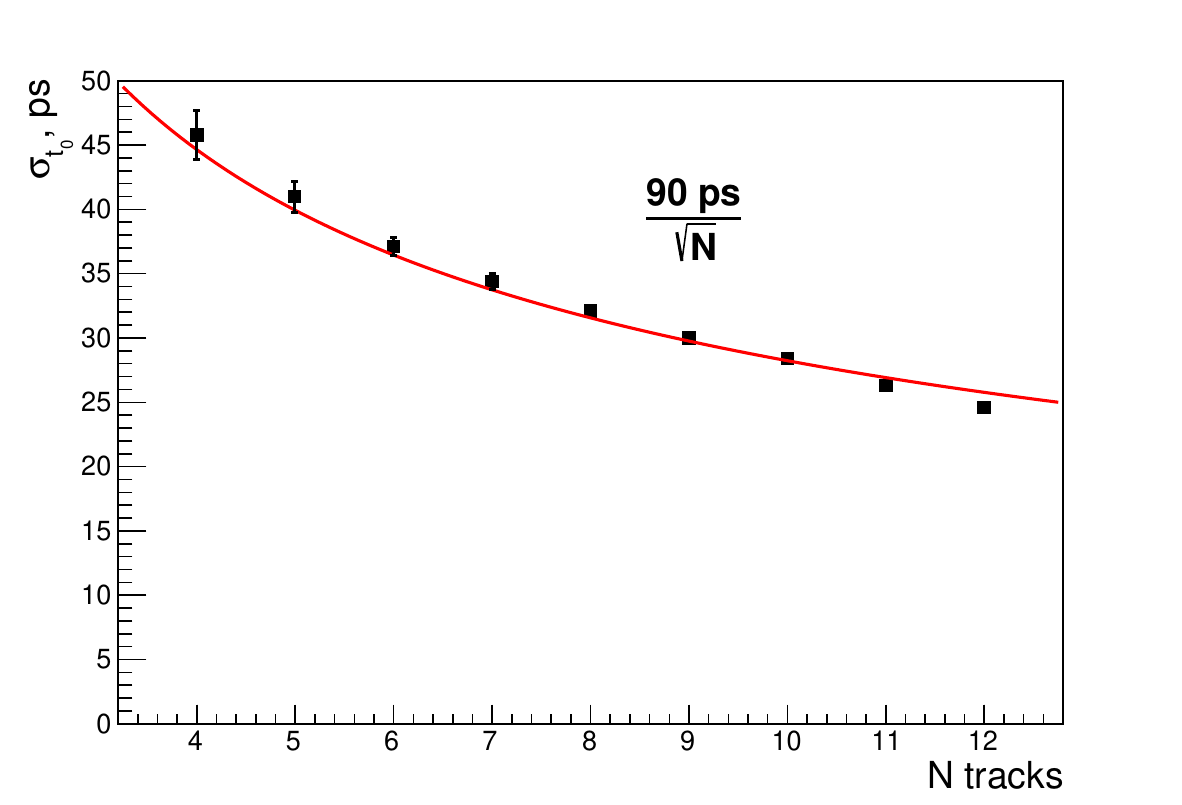} \\ (b)}
  \end{minipage}
  \caption{(a) Energy deposit $dE/dx$ at the ST for particles emitted at $\theta=90^{\circ}$ in respect to the beam axis. 
  (b) Accuracy of the $t_{0}$ reconstruction as a function of the number of tracks in the primary vertex.}
  \label{fig:t0rec}  
\end{figure}

\begin{figure}[!h]
    \center{\includegraphics[width=1\linewidth]{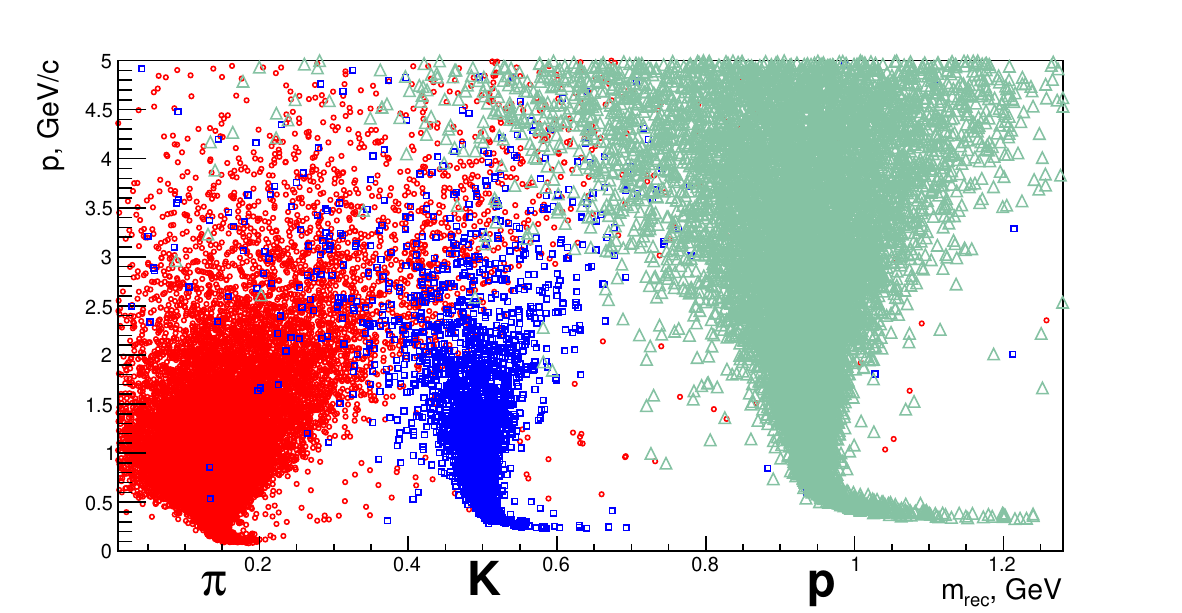}}
  \caption{ Reconstructed mass vs. the particle momentum for pions, kaons, and protons.}
  \label{fig:TOF_m}  
\end{figure}

\begin{figure}[!h]
    \begin{minipage}{0.49\linewidth}
    \center{\includegraphics[width=1\linewidth]{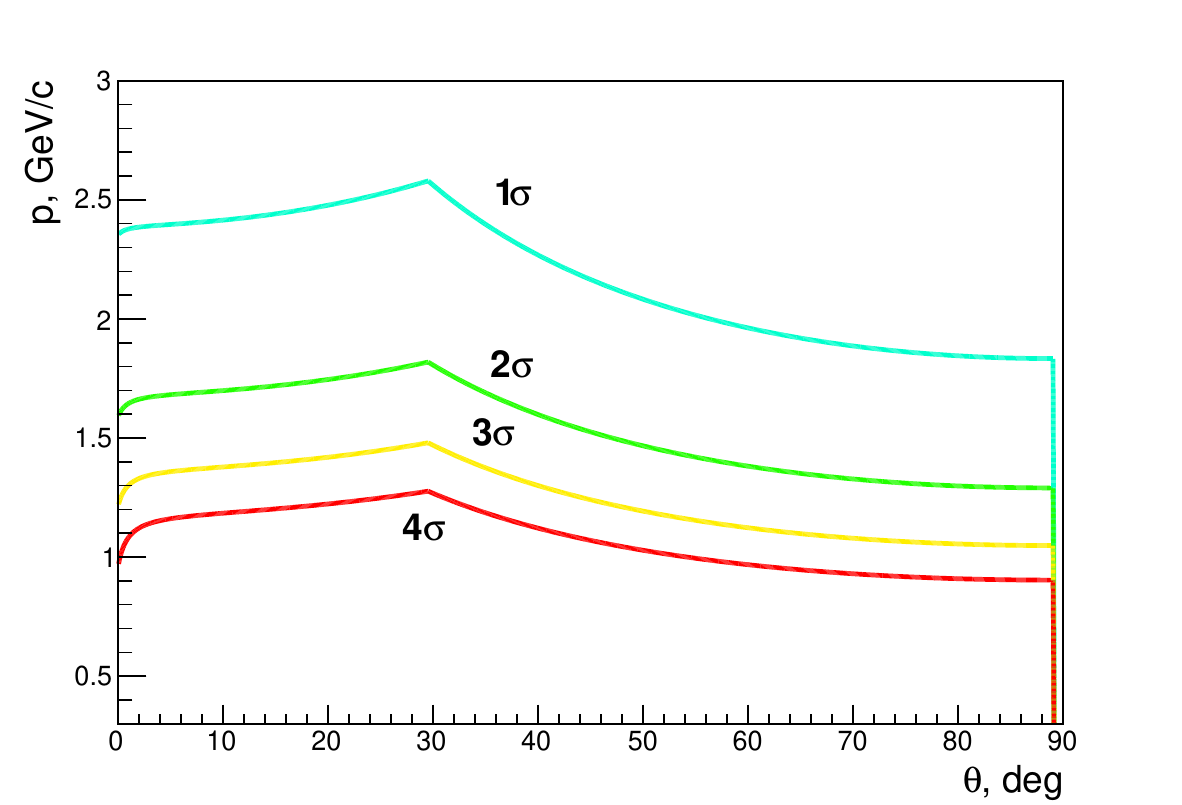} \\ (a)}
  \end{minipage}
  \hfill
  \begin{minipage}{0.49\linewidth}
    \center{\includegraphics[width=1\linewidth]{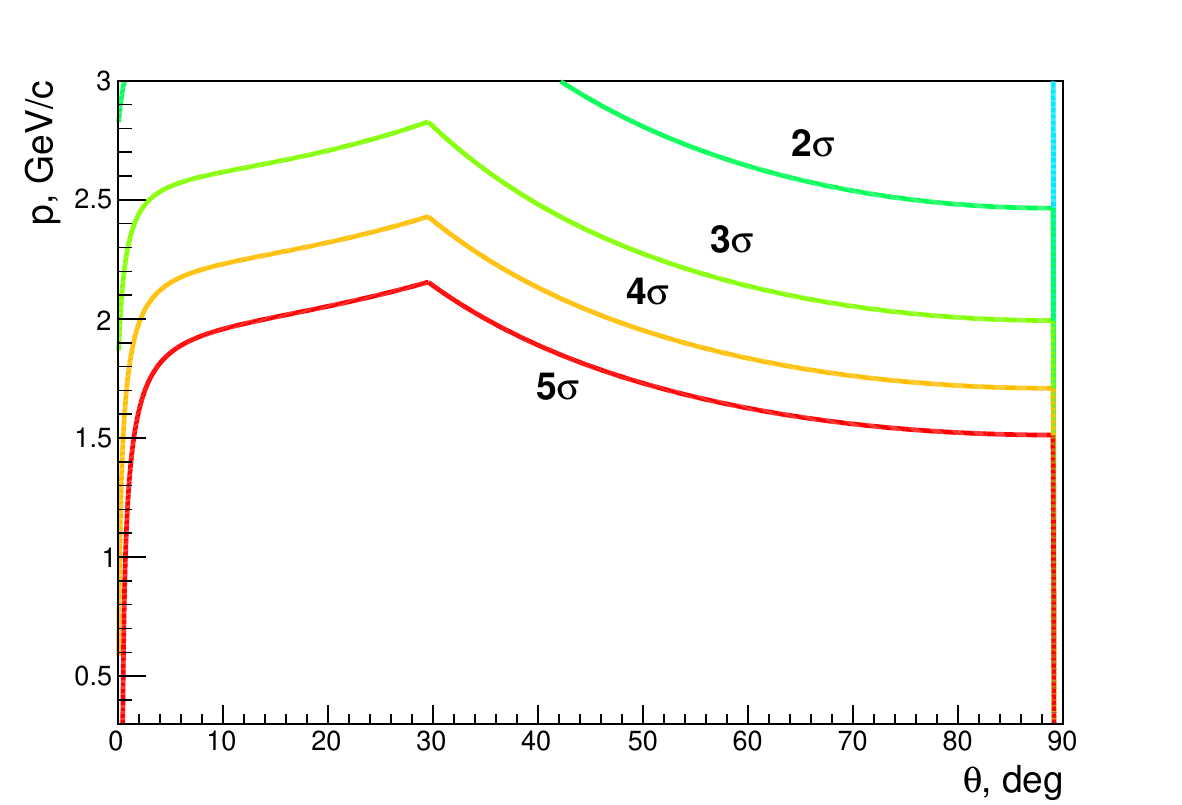} \\ (b)}
  \end{minipage}
  \caption{$\pi/K$ (a) and $K/p$ (b) separation power of the TOF system as a function of the particle momenta and the emission angle.}
  \label{fig:tof_sep}  
\end{figure}

Figure \ref{fig:t0rec}~(b) illustrates the accuracy of  $t_0$ reconstruction as a function of the number of tracks for $\sigma_{TOF}=70$ ps. One can see that $\sigma_{t_0}$ is proportional to $1/\sqrt{N}$ and for the track multiplicity 10 (typical for hard interaction events) is about 30 ps. Pion, kaon, and proton separation with the TOF detector is shown in Fig. \ref{fig:TOF_m}. The $\pi/K$ and $K/p$ separation power as a function of the particle momenta and the emission angle $\theta$ in the primary vertex is presented in Fig. \ref{fig:tof_sep}~(a) and (b), respectively,  for the time of flight ($t_{TOF}-t_0$) resolution equal to 80 ps. It is mostly defined by the time measurements, while the accuracy of the momentum reconstruction becomes sizable only for $\theta<10^{\circ}$.


\subsection{Calorimetry}
The electromagnetic calorimeter is one of the main detectors for the SPD gluon program. Its functions are: (i) to measure  the energy and the position of the hard prompt photons, and the photons from the radiative decays of $\pi^0$- and $\eta$-mesons; (ii) to reconstruct the soft photons ($\sim$ 0.5 GeV) from the decays $\chi_{c1,2}\to J/\psi \gamma$; (iii) to provide identification of the electrons and positrons by comparing the energy deposit in the ECal and their momentum measured in the tracking system.  The end-cup part of the ECal participates also in the online polarimetry with the inclusive $\pi^0$ production at high $x_F$ (see Sec. \ref{po0polar}).

\begin{figure}[!h]
    \begin{minipage}{0.49\linewidth}
    \center{\includegraphics[width=1\linewidth]{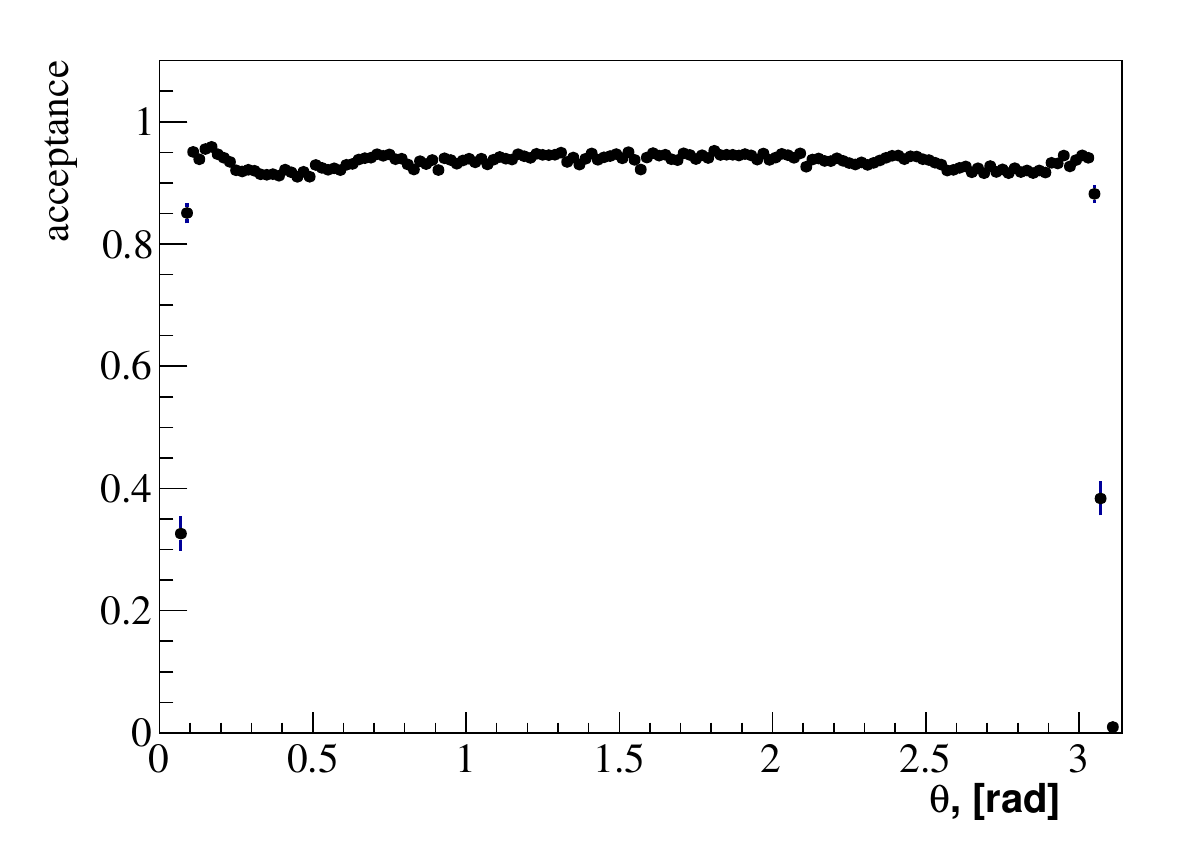} \\ (a)}
  \end{minipage}
  \hfill
  \begin{minipage}{0.49\linewidth}
    \center{\includegraphics[width=1\linewidth]{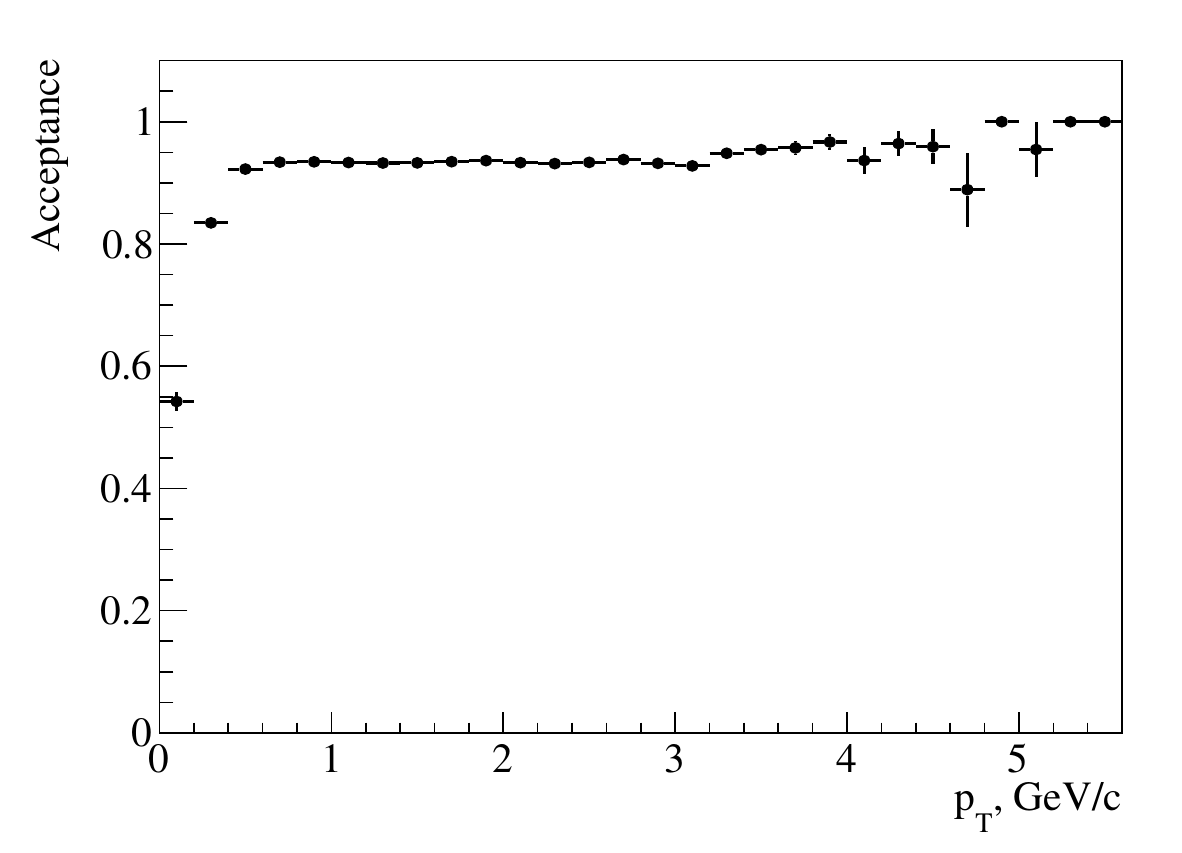} \\ (b)}
  \end{minipage}
  \caption{Efficiency of photon detection as a function of  (a) polar angle $\theta$ and (b) transverse momentum $p_T$.}
  \label{fig:calo1}  
\end{figure}

\begin{figure}[!h]
    \begin{minipage}{0.49\linewidth}
    \center{\includegraphics[width=1\linewidth]{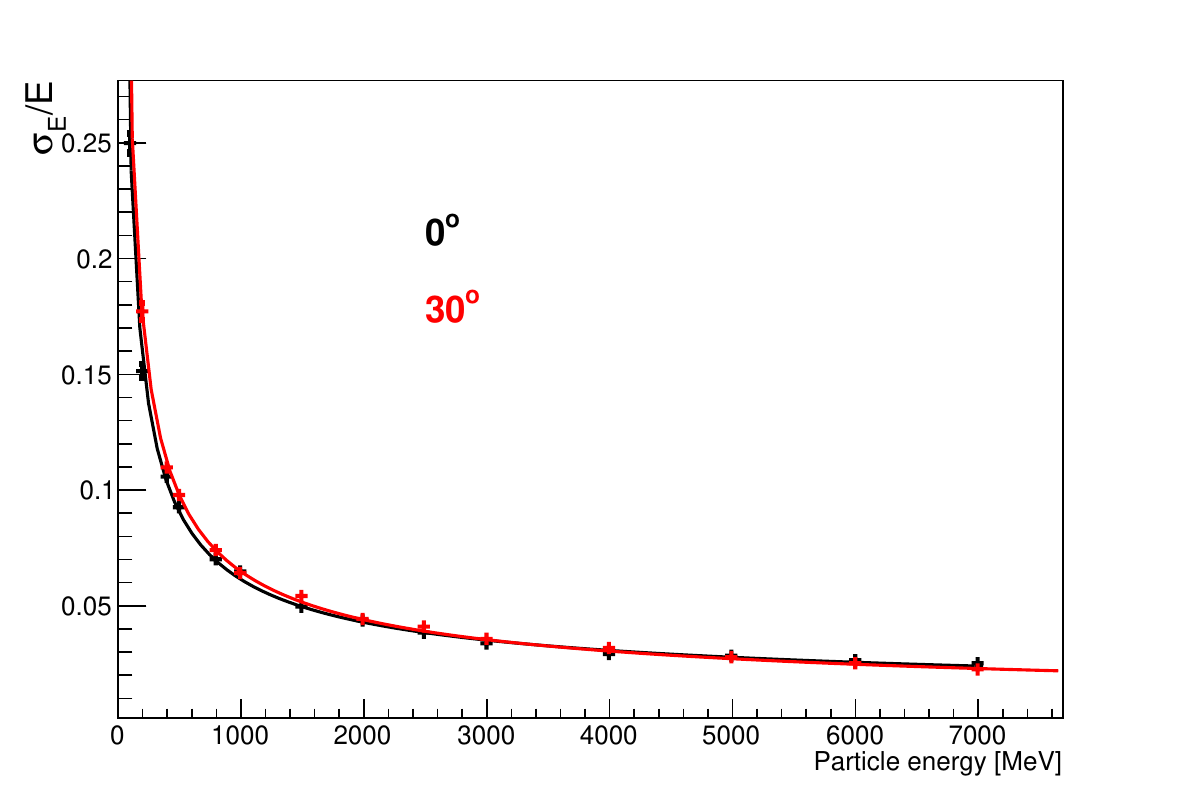} \\ (a)}
  \end{minipage}
  \hfill
  \begin{minipage}{0.56\linewidth}
    \center{\includegraphics[width=1\linewidth]{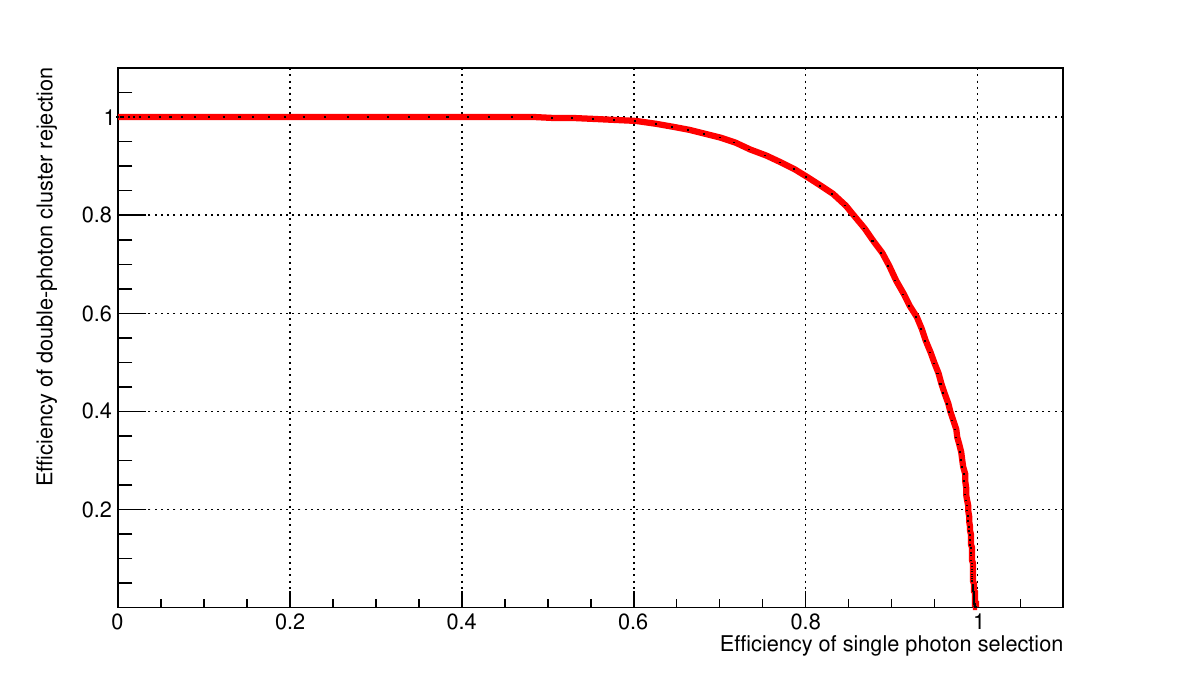} \\ (b)} 
  \end{minipage}
  \caption{(a) Energy resolution of the ECal for the normal incidence of photons and for the angle of 30$^{\circ}$. (b) Purity of the double-photon clusters rejection vs. the efficiency of the single-photon reconstruction for 6 GeV photons and two 3 GeV photons separated by a distance of 4 cm, based on the cluster shape analysis.}
  \label{fig:calo2}  
\end{figure}

The transparency of the SPD setup allows us to detect photons produced in the interaction point in a wide kinematic range. The efficiency of photon detection as a function of the production angle $\theta$ with respect to the beam direction and as a function of the transverse momentum $p_T$ is shown in Fig. \ref{fig:calo1}(a) and (b), respectively. The expected energy resolution of the ECal obtained from the Geant4-based Monte Carlo simulation for the normal incidence of photons and for the angle of 30$^{\circ}$ with respect to the normal line  is shown in Fig. \ref{fig:calo2}(a). Such effects as the individual cell energy threshold at the level of 50 MeV, the light absorption in the optic fibers, and the fluctuation of the number of photons are taken into account. The fitted curve has the shape:
\begin{equation}
\sigma_{E}/E = A\oplus\frac{B}{\sqrt{E/GeV}}\oplus\frac{C}{E/GeV},
\end{equation}
where the parameters $A$, $B$, and $C$ are 0.9\%, 5.9\%, 1.7\% and 0.0\%, 6.0\%, 2.2\%, respectively, for 0$^{\circ}$ and 30$^{\circ}$ of the incidence angle. The superconducting coils of the magnetic system (about 1.2 $X_0$ of the material) placed in front of the calorimeter practically do not reduce its acceptance for the hard photons and do not produce any sizable impart on the energy resolution. For instance,  the average resolution for the 1 GeV photons passing through the coil changes from 6.1\% to 6.3\%.

As long as the internal longitudinal and transverse size of the ECal is small,  there is a probability for photons from the high-energy pions decay ($E_{\pi^0}\gtrsim 6$ GeV) to produce a single cluster and be misidentified as a single high-energy photon. That is especially important for the prompt-photon part of the physics program. However,  it is possible to identify such clusters with a certain precision performing the cluster shape analysis. The cluster shape can be characterized using such variables as dispersion or the second-order moment (in one or two dimensions), the fourth-order moment, the ratio of the major and the minor semiaxes of the cluster ellipse, etc. Machine learning classification techniques are planned to be applied (the multilayer perceptron, the k-nearest neighbors, etc.), using these variables as an input to distinguish between single- and double-photon clusters.
Figure \ref{fig:calo2}(b) illustrates the purity of the double-photon rejection vs. the efficiency of the single-photon reconstruction for 6 GeV photons and two 3 GeV photons separated by a distance of 4 cm (exactly the ECal cell size), based on the cluster shape analysis.

\begin{figure}[!h]
    \begin{minipage}{0.49\linewidth}
    \center{\includegraphics[width=1\linewidth]{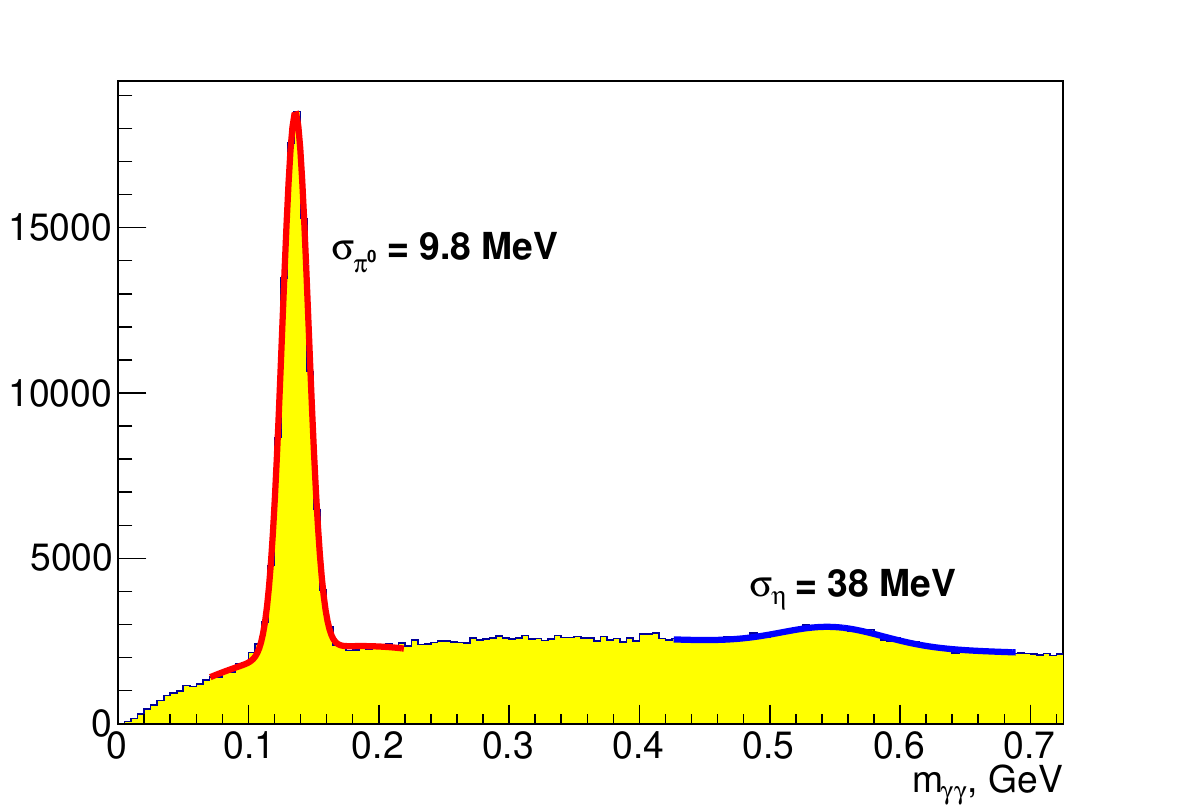} \\ (a)}
  \end{minipage}
  \hfill
  \begin{minipage}{0.49\linewidth}
    \center{\includegraphics[width=1\linewidth]{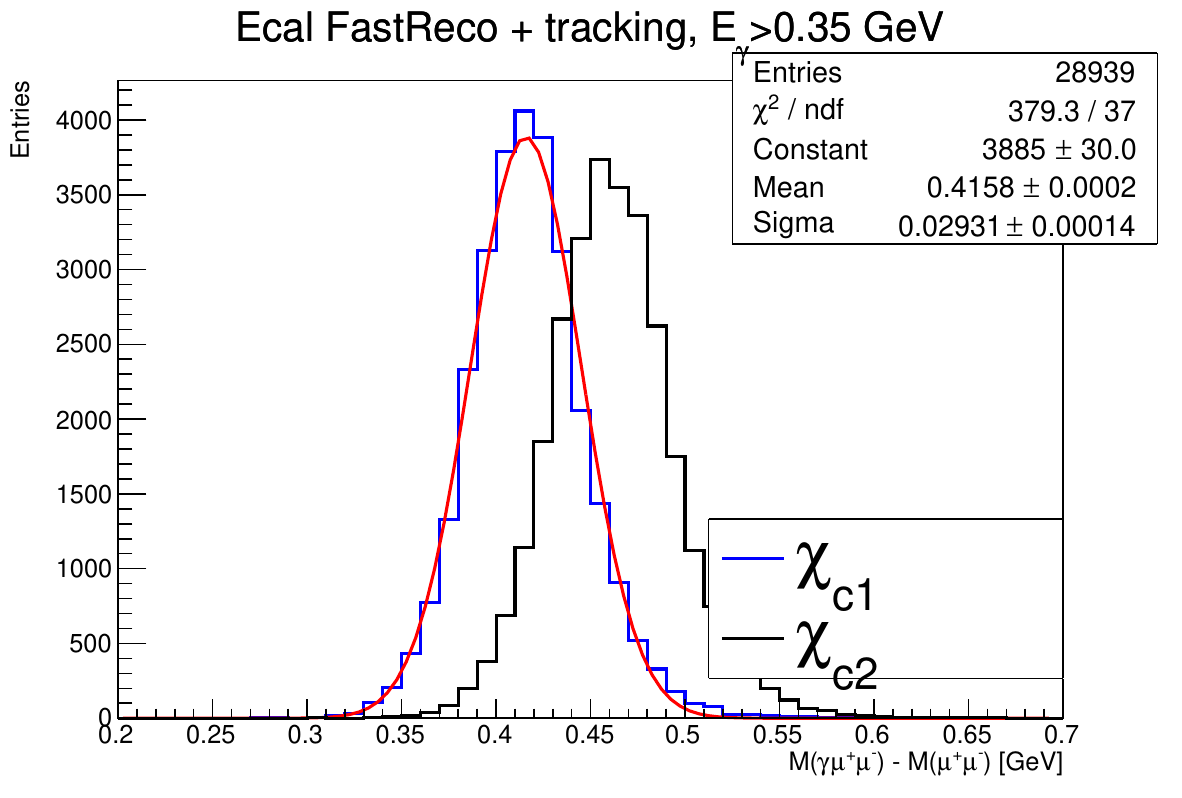} \\ (b)}
  \end{minipage}
  \caption{(a) $\pi^0$-peak in the $\gamma\gamma$ mass spectrum. (b) Mass resolution for $\chi_{c1,2}$ reconstructed via their decay into $J/\psi \gamma$ final state.}
  \label{fig:calo5}  
\end{figure}

The impact of the ECal energy resolution on the reconstruction of such states as $\pi^0$, $\eta$ is shown in Fig. \ref{fig:calo5}(a). The relative width of the  $\pi^0$  and $\eta$ peaks is 7.3\% and 6.9\%, respectively, for $E_{\gamma}>0.5$ GeV. The reconstruction of the charmonium states $\chi_{c,1,2}$ via their radiative decays is presented in Fig. \ref{fig:calo5}(b). The $\chi_{c1}$ and $\chi_{c2}$ peaks cannot be fully resolved ($\Delta{M}/\sigma_{M}\approx 1.5$), but, nevertheless, the relative contribution of these states could be estimated based on the detailed peak shape analysis.

\section{Accuracies of the asymmetries measurement \label{sec:acc}}
The results presented in this section illustrate general possibilities of the SPD setup to deal with spin asymmetries and should be treated as preliminary. The most of them require further optimization of the selection criteria and could be significantly improved. 

The single transverse ($A_N$) and the double longitudinal and transverse ($A_{LL}$ and $A_{TT}$) spin asymmetries are the main observables to be accessed at SPD. The asymmetry $A_N$ is denoted as
\begin{equation}
A_{N}=\frac{\sigma^{\uparrow}-\sigma^{\downarrow}}{\sigma^{\uparrow}+\sigma^{\downarrow}},
\end{equation}
where $\sigma^{\uparrow}$ and $\sigma^{\downarrow}$ denote the inclusive production cross-sections with the opposite transverse polarization of one of the colliding particles. In practice, taking into account the $2\pi$ coverage of the SPD setup in azimuthal angle $\phi$, the $A_N$ can be extracted from the azimuthal modulation amplitude of the differential cross-section $d\sigma/d\phi$
\begin{equation}
d\sigma/d\phi \propto 1+P A_N \cos(\phi-\phi_0),
\label{eq:An_phi}
\end{equation}
where $P$ and $\phi_0$ are the beam polarization and its direction. All estimations for the accuracy of the SSA measurement are performed under assumption that only one of the two beams is transversely polarized. In the case of  two polarized beams the statistical uncertainty could be reduced by the factor of $\sqrt{2}$.

The double longitudinal spin asymmetry can be expressed via the number of events for the same ($N^{++}$) and the opposite ($N^{+-}$)  spin orientations of colliding protons:
\begin{equation}
A_{LL}=\frac{\sigma^{++}-\sigma^{+-}}{\sigma^{++}+\sigma^{+-}} =  \frac{1}{P_1 P_2} \times \frac{N^{++}-RN^{+-}}{N^{++}+RN^{+-}}.
\label{eq: ALL_1}
\end{equation}	
$\sigma^{++}$ and $\sigma^{+-}$ denote the cross-sections with the same and the opposite proton helicity combinations, respectively.  $P_1$ and $P_2$ are the absolute values of proton beams polarizations and $R={L_{++}/L_{+-}}$ is the ratio of the integrated luminosities for the samples with the same and the opposite spin orientations. Assuming the same amount of data collected with both spin orientations Eq. \ref{eq: ALL_1} can be rewritten as:
\begin{equation}
A_{LL}= \frac{1}{P_1 P_2} \times \frac{N^{++}-N^{+-}}{N^{++}+N^{+-}}.
\label{eq: ALL_2}
\end{equation}	
The aforesaid is also valid for the $A_{TT}$ asymmetry.

\subsection{Charmonia production}

According to the modern theoretical approaches, the charmonia production
at the SPD energies ($10\text{ GeV} \le \sqrt{s} \le 27$~GeV) is dominated by the gluon-gluon fusion process. The inclusive $J/\psi$ production has a large cross-section (200 $\div$ 250 nb at the maximum energy) and clear experimental
signature in the dimuon decay mode, and thus is a powerful probe of the internal
structure of proton and deuteron.  At the same time, the distinct $J/\psi$ signal allows us
to reconstruct excited charmonia states in the decays $\chi_{c1,2} \to
\gamma J/\psi$ and $\psi(2S)\to \pi^+\pi^-J/\psi$. There is also
a possibility to reconstruct $J/\psi$ from the $e^+e^-$ final state, but it looks
less promising due a much larger background, a larger observed $J/\psi$ width, and a 
more complicated shape of the peak, which will significantly affect both
statistical and systematic uncertainties. The study of the $\eta_c$
production properties in the $p\bar p$ and $\Lambda \bar \Lambda$ decay modes may also
be feasible.

Muons are identified in the RS. The system is expected to separate showers
from strongly interacting pions and muon tracks (using standard or machine
learning techniques). The main background
is muons from pion decays and pions that passed a large distance in the RS.
The pion decays result in a small kink of the charged track (about 2$^\circ$), and
the decay muon retains from 60\% to almost 100\% of the initial pion energy.
There is a possibility that a fraction of decay muons can be suppressed by the
search for a kink in the tracker or by considering the correlation between the particle
momentum and the amount of material it crossed. However, the results in this section are based
on a simplified model (gives a lower performance boundary). A particle is
identified as a muon based on the
amount of material it passes in the active part of the RS, this amount is
given as a number of proton nuclear lengths ($n_\lambda$). Two possibilities
are considered: a particle from the initial interaction and a muon from a
pion decay (the pion must be from the initial interaction). In the latter case, if the
pion decays in the RS, the amount of material is added for the pion and muon.

It is clear that higher running energies are preferable for the physics with charmonia
due to larger production
cross-section, a stronger boost for pions and more energetic muons. All estimates
in this section assume a $pp$ collision energy of 27~GeV, $10^7$~s time of
data taking (one year) with the maximum luminosity and a polarization $P$ of 0.7.
At these conditions one expects about 12 million $J/\psi \to \mu^+\mu^-$
decays in the SPD detector.

\begin{figure}[htb]
	\begin{minipage}[ht]{0.49\linewidth}
		\center{\includegraphics[width=\linewidth]{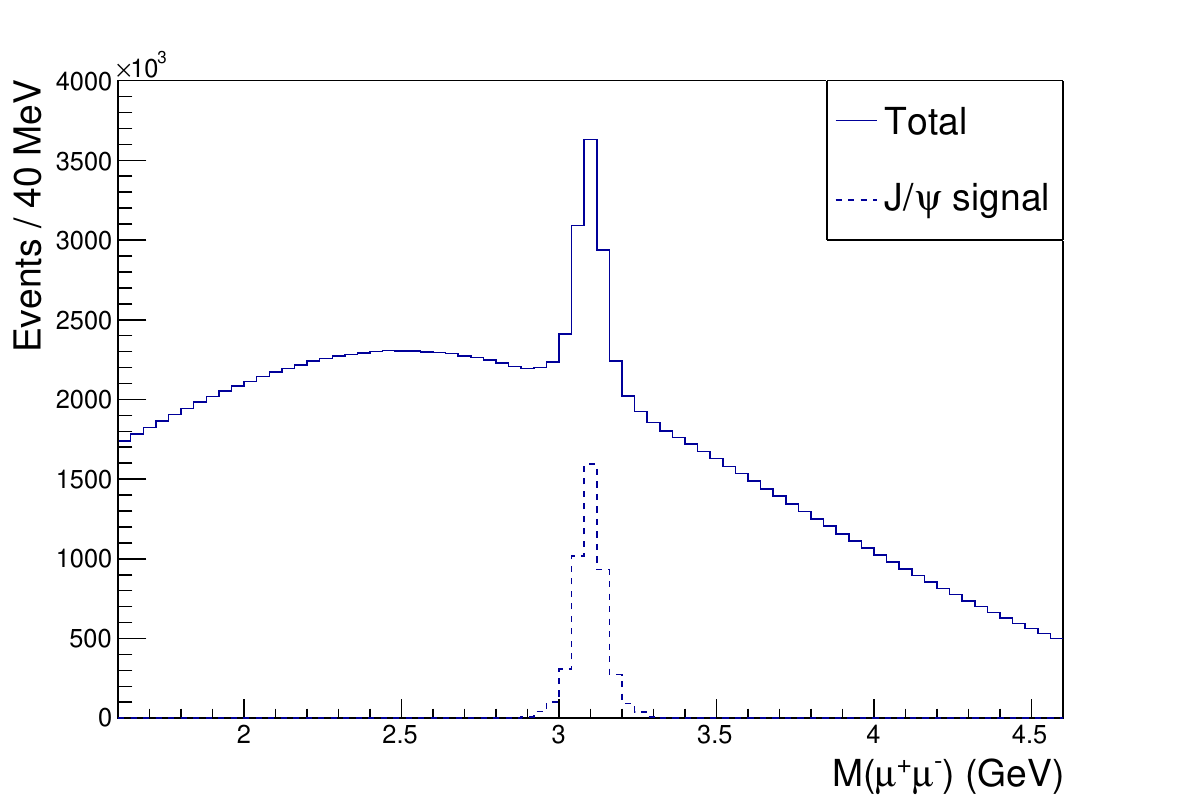}\\ (a)}
	\end{minipage}
	\hfill
	\begin{minipage}[ht]{0.49\linewidth}
		\center{\includegraphics[width=\linewidth]{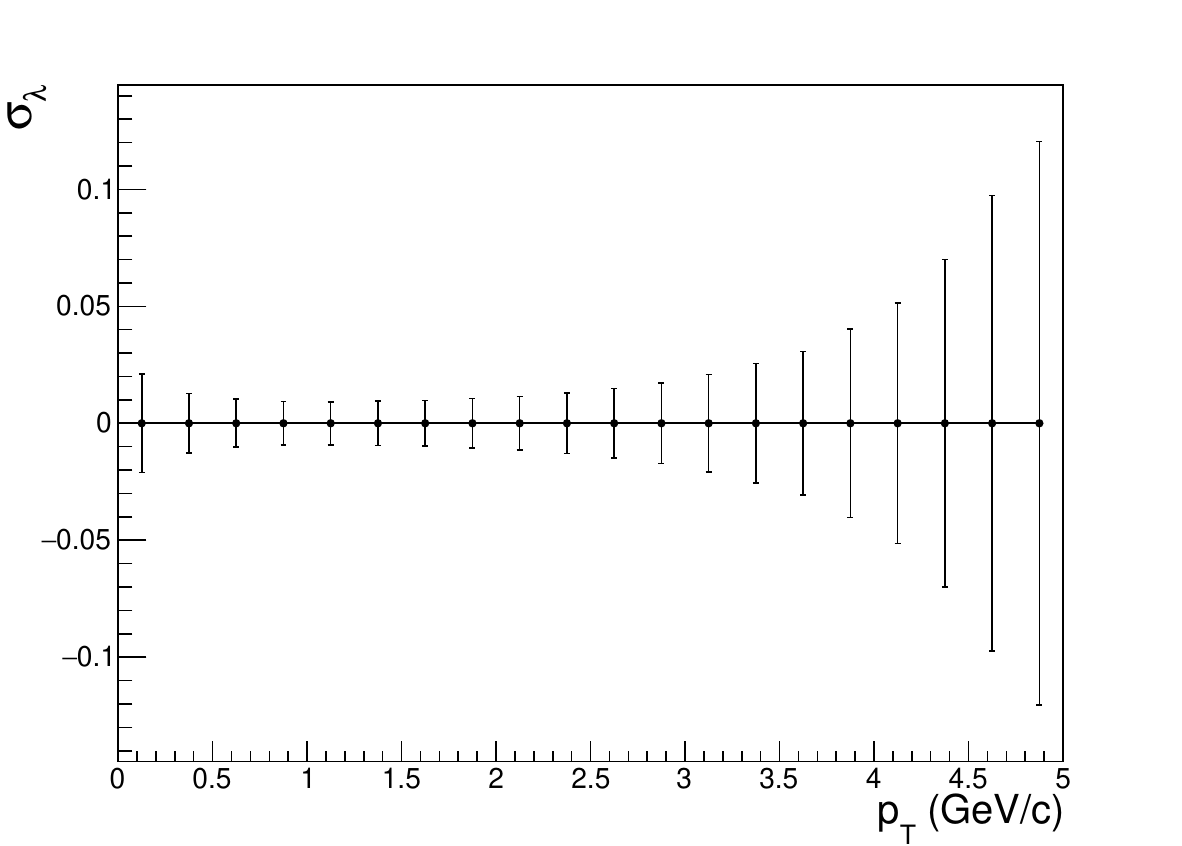}\\ (b)}
	\end{minipage}
		\caption{\label{fig:mmumu-stat}  (a)  Dimuon candidate spectrum and the $J/\psi$ peak after one year of data taking. (b) Expected statistical precision for polarization $\lambda_{\theta}$ as a function of the $J/\psi$ transverse momentum.
		}
\end{figure}

The $J/\psi$ events are simulated using Pythia8 and their number is normalized
to a production cross-section of 200~nb. For the background minimum bias events generated
with Pythia6 and Pythia8 are considered (giving almost the same predictions
around the $J/\psi$ peak). Approximately half of the background events are produced
in the hard interaction, but a sizable fraction also comes from the diffraction
processes. It appears that significant amount of background events can be
suppressed by the requirement on the polar angle of a muon candidate. The $\mu^+\mu^-$
invariant mass spectrum for the muon candidates with $n_\lambda>3$ and satisfying
$|\cos\theta|<0.9$ after one year of data taking is shown in Fig~\ref{fig:mmumu-stat}~(a). 
Approximately half of the background events under the $J/\psi$ peak are two 
decay muons, the other half mostly consists of one decay muon and one pion 
that passed a large distance in the RS, and the rest are two pions misidentified 
as $\mu^+$ and $\mu^-$.
The selection efficiency can be estimated
to be around 35 $\div$ 45\%, depending on the cut on $\theta$, resulting in
4 $\div$ 5 million selected events. The barrel part of the RS is essential for reconstruction
of approximately 90\% of $J/\psi$ events: for more than 50\% of $J/\psi$ events one lepton is 
reconstructed in the barrel and the other one in the end-caps, more than 35\% of events are
reconstructed solely in the barrel part of the detector.
The statistical errors for observables
can be estimated using a linear LSM fit~\cite{Zyla:2020zbs}. As an example, the estimated
statistical precision for the $J/\psi$ polarization $\lambda$  as a function of the transverse momentum $p_T$ is shown in Fig.~\ref{fig:mmumu-stat}~(b).
The estimation was done under the assumption  $\lambda \ll 1$ based on the angular modulation of the differential cross-section
\begin{equation}
d\sigma/d\cos\theta_{\mu} \propto 1 + \lambda_{\theta}\cos^2\theta_{\mu},
\end{equation}
where $\theta_{\mu}$ is the angle between the muon momentum and the $J/\psi$ momentum in the helicity frame.  

\begin{figure}[htb]
	\begin{minipage}[ht]{0.49\linewidth}
		\center{\includegraphics[width=.8\linewidth]{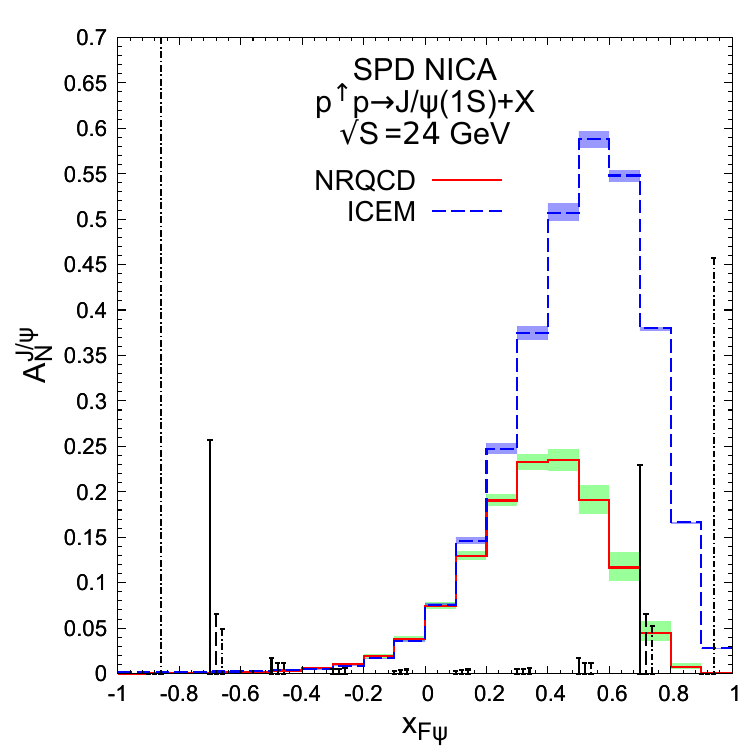}\\ (a)}
	\end{minipage}
	\hfill
	\begin{minipage}[ht]{0.49\linewidth}
		\center{\includegraphics[width=.8\linewidth]{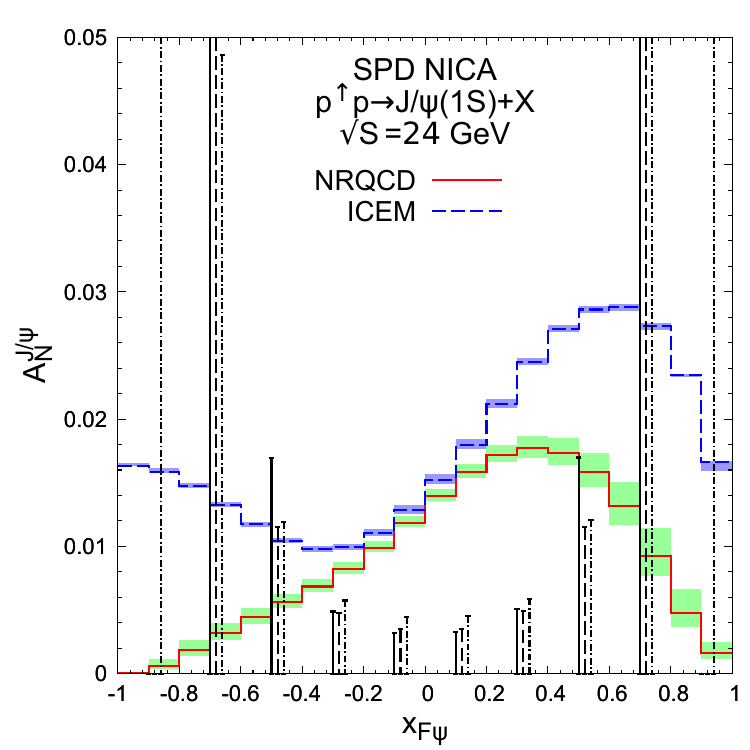}\\ (b)}
	\end{minipage}
	\label{fig:psi2Schic}
		\caption{\label{fig:AN} Projection of the estimated statistical
		uncertainties for $A_N$ compared to the GPM predictions from Ref.~\cite{Karpishkov:2020brv}
		for SIDIS1 (a) and D'Alesio PDF parameterizations (b).}
\end{figure}

The transverse single-spin asymmetry $A_N$ in $J/\psi$ production
probes the Sivers function.
 At $\sqrt s = 200$~GeV it was measured by the PHENIX
Collaboration and found consistent with zero~\cite{Adare:2010bd,Aidala:2018gmp}.
To estimate our statistical precision, 8 bins in
$\phi$ are considered (see Eq. \ref{eq:An_phi}). The same linear fit is used to, firstly, estimate the error
in the bins based on the expected $J/\psi$ number and, secondly, to extract $A_N$.
The projected statistical uncertainties for $A_N$ as a function of $x_F$
are compared to the GPM model predictions from Ref.~\cite{Karpishkov:2020brv} in
Fig.~\ref{fig:AN} (preliminary CGI-GPM calculations indicate lower asymmetries).
Compared to the PHENIX
measurement, we expect a much better precision and a much wider kinematic range
in $x_F$. Our rapidity range is approximately $|y|<2$.

\begin{figure}[htb]
	\begin{minipage}[ht]{0.49\linewidth}
		\center{\includegraphics[width=\linewidth]{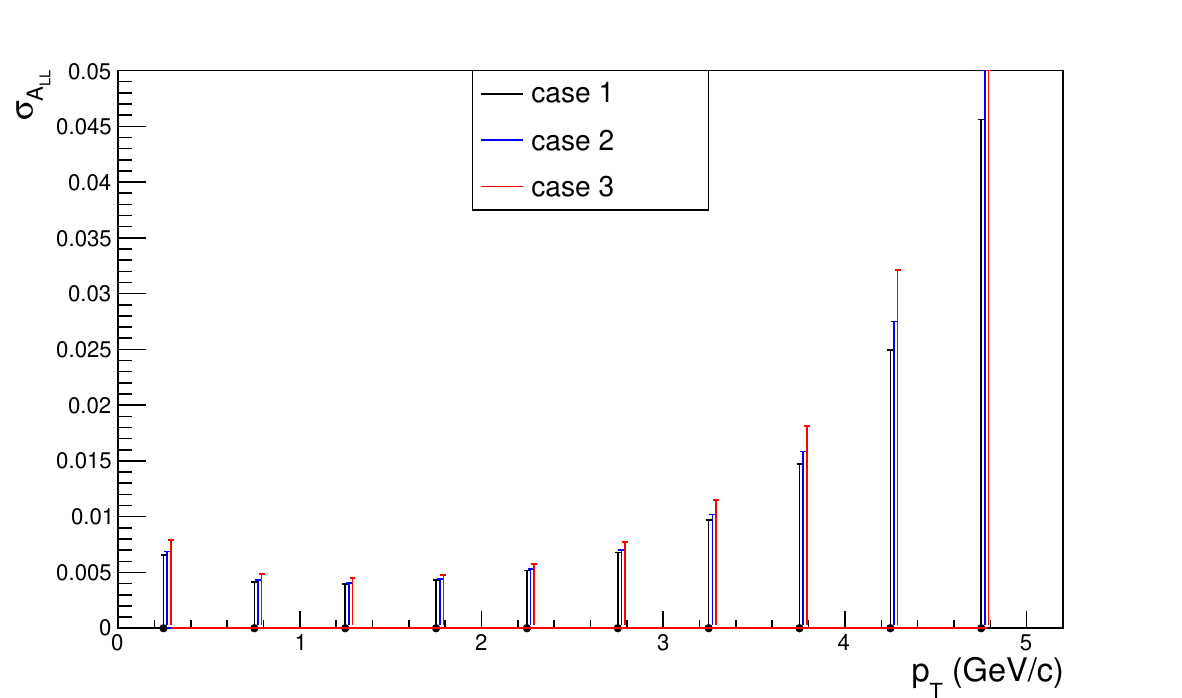}\\ (a)}
	\end{minipage}
	\hfill
	\begin{minipage}[ht]{0.49\linewidth}
		\center{\includegraphics[width=\linewidth]{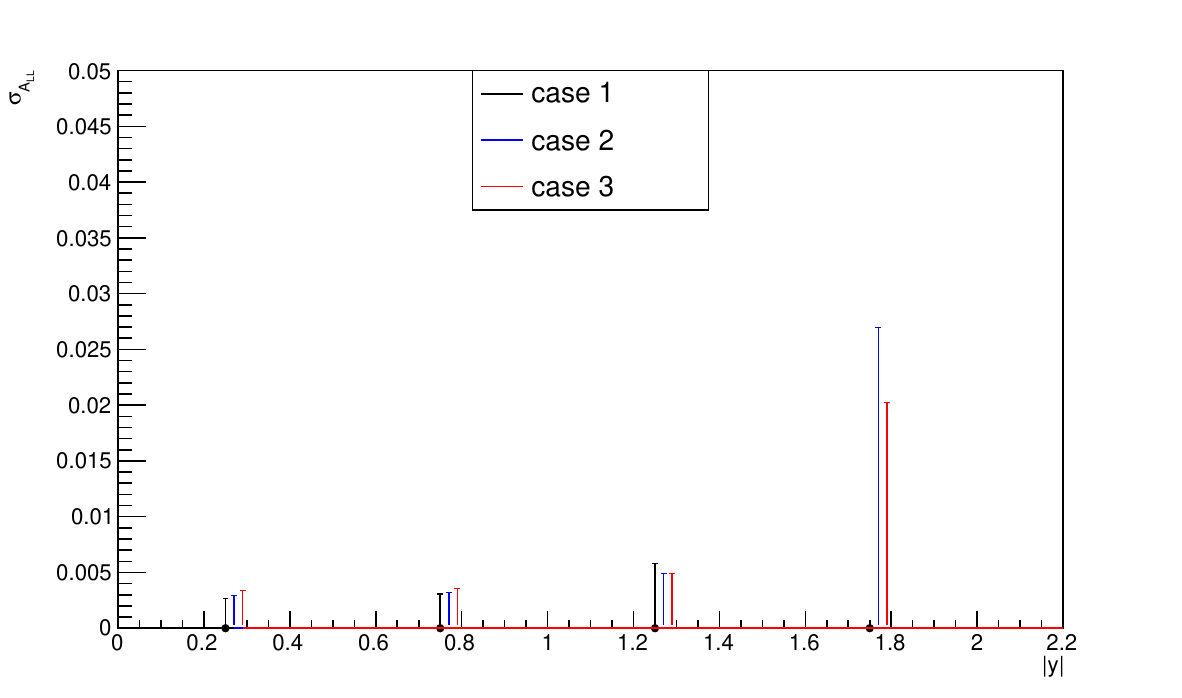}\\ (b)}
	\end{minipage}
		\caption{\label{fig:ALL} Estimated statistical precision of $A_{LL}$
		as a function of $p_T$ (a) and rapidity (b). Three different cuts on the polar angle of 
muon candidates are considered.}
\end{figure}

The statistical error of the longitudinal double spin asymmetry $A_{LL}$ sensitive to the
polarized gluon distribution was estimated basing on Eq. \ref{eq: ALL_1} and \ref{eq: ALL_2}. In these formulas, we neglect the uncertainties
of the measurement of the relative integrated luminosities and the beam polarizations.
The projection of the statistical uncertainties as functions of $p_T$
and $|y|$ are shown in Fig~\ref{fig:ALL}. Compared to the previous results obtained by
the PHENIX Collaboration at $\sqrt{s}=510$ GeV~\cite{Adare:2016cqe}, we have a much better
precision and probe a wider kinematic range.

The study of the associated  $J/\psi$ production  will be strongly restricted
by the insufficient expected statistics. The double $J/\psi$ production cross-section
was measured by the NA3 Collaboration~\cite{Badier:1985ri}
and was found to be $27\pm10$~pb
in the proton-nucleus interaction at $\sqrt{s}\approx 27$~GeV. Optimistically,
such a cross-section would result in 50$\div$100 reconstructed events, if both
$e^+e^-$ and $\mu^+\mu^-$ modes are used to reconstruct $J/\psi$. It may be
enough to determine low-$p_T$ cross-section dependence, but the study of
any angular modulation will not be possible. The study of $\gamma J/\psi$
production will
be challenging experimentally due to a lack of statistics
and a high expected background. Reasonable statistics might be 
expected for $J/\psi D$ production.

\begin{figure}[htb]
	\begin{minipage}[ht]{0.49\linewidth}
		\center{\includegraphics[width=\linewidth]{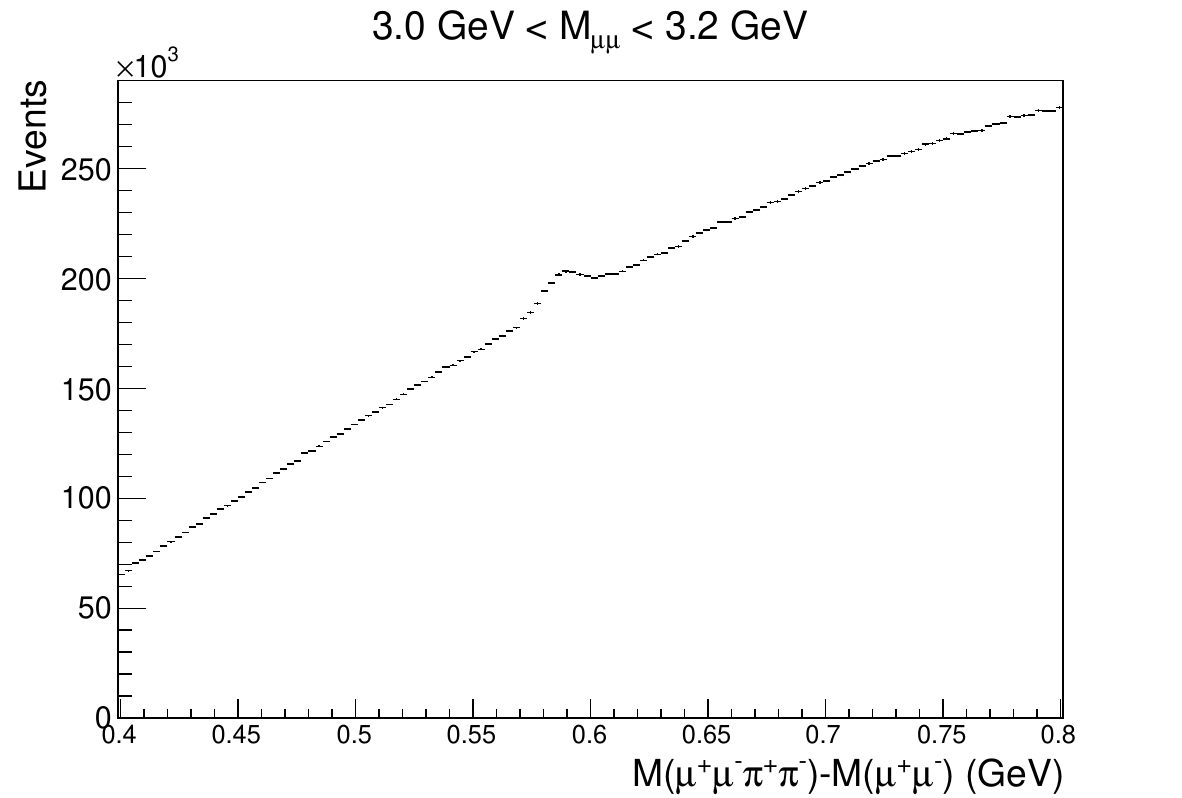}\\ (a)}
	\end{minipage}
	\hfill
	\begin{minipage}[ht]{0.49\linewidth}
		\center{\includegraphics[width=\linewidth]{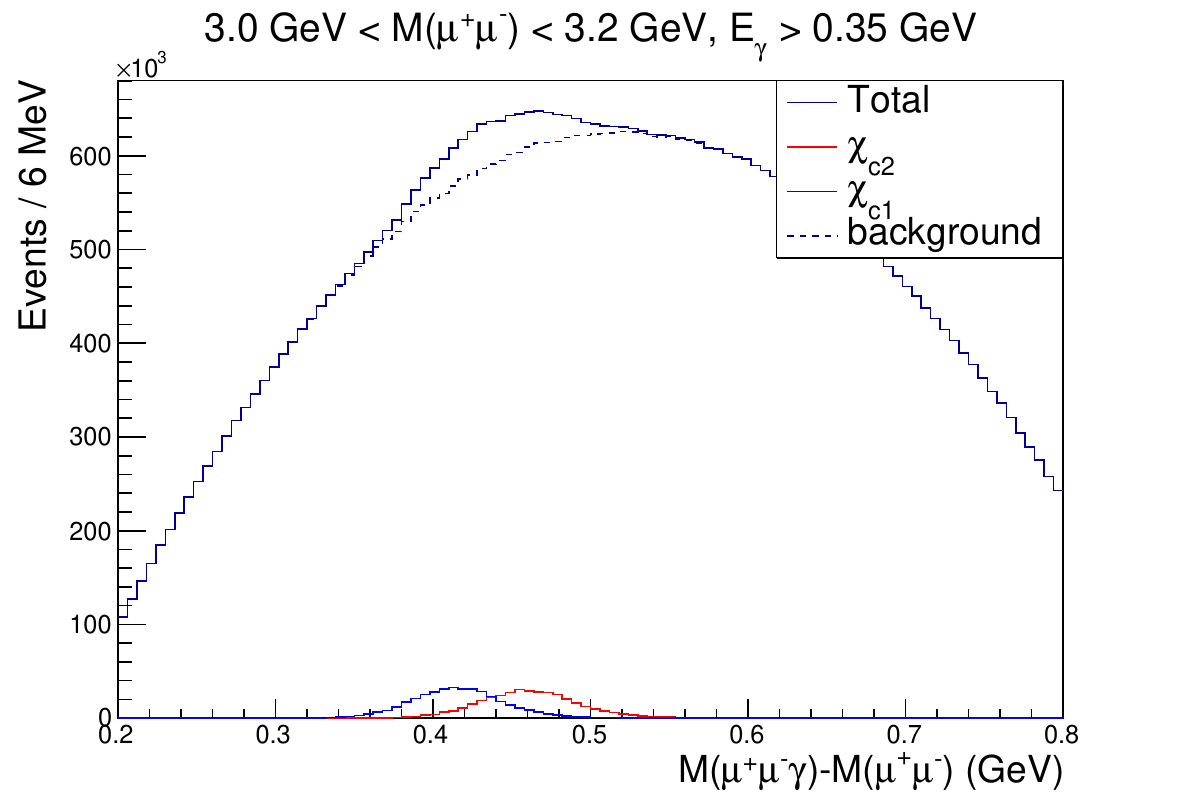}\\ (b)}
	\end{minipage}
	\caption{(a) $\psi(2S)$ signal in the $M_{\pi^+\pi^-\mu^+\mu^-} - M_{\mu^+\mu^-}$
		distribution.
		(b) Preliminary estimation of the background for the $\chi_{c1}$ and $\chi_{c2}$
		reconstruction. For this plot a feed-down fraction of 15\% is assumed for both states.}
	\label{fig:chic}
\end{figure}

The $\psi(2S)\to\mu^+\mu^-$ decay is suppressed, as compared to
$J/\psi\to\mu^+\mu^-$ by approximately a factor of 50 and its
reliable extraction may not be feasible. At the same time, the decay
$\psi(2S)\to\pi^+\pi^- J/\psi$ can be reliably identified as a narrow (about 10~MeV$/c^2$ wide)
peak in the $M_{\pi^+\pi^-\mu^+\mu^-} - M_{\mu^+\mu^-}$
distribution. This distribution is shown in Fig.~\ref{fig:chic}~(a).
The expected statistics is about $1\times10^5$ selected events.

The $\chi_{c1}$ and $\chi_{c2}$ states have a large partial width of 
decay to $J/\psi\gamma $ and can be reconstructed using it. The production
properties of these states at low energies are poorly known (e.g. see the review of the
experimental results in Ref.~\cite{Abt:2008ed}).
The identification of these decays at SPD relies on the ECal performance. The result of the MC
simulation for $M_{\gamma\mu^+\mu^-} - M_{\mu^+\mu^-}$ is shown in Fig.~\ref{fig:calo5}(b). 
It will not be possible to separate $\chi_{c1}$ from $\chi_{c2}$, but their relative
fractions should be well-measurable. For the expected statistics of approximately 0.5~million
reconstructed decays per year (for both states together) it should be possible to measure the
cross-section kinematic dependencies of these states. The major difficulty in the studying of
these states is the high expected background. Its very rough estimation is shown in
Fig.~\ref{fig:chic}~(b).

The $\eta_c$ production cross-section is highly uncertain. At $\sqrt{s}=24$~GeV it is
estimated that $\sigma_{\eta_c}\cdot B(\eta_c \to p\bar p)=0.6_{-0.4}^{+0.8}$~nb~\cite{Nefedov2020} or $6\times10^{5}$ events per year. The typical momenta of $p$ and $\bar p$
is 1.5~--~2~GeV/$c$, where these particles should be well-identified by the TOF system
(see Fig.~\ref{fig:TOF_m}). The feasibility of the differential cross-section measurements requires
detailed MC-simulations due to the high expected background. A very limited statistics, but a
clearer signal may be also expected in the $\Lambda \bar \Lambda$ decay mode.

\subsection{Prompt photon production}
\begin{figure}[!hb]
  \begin{minipage}[ht]{1\linewidth}
		    \center{\includegraphics[width=\linewidth]{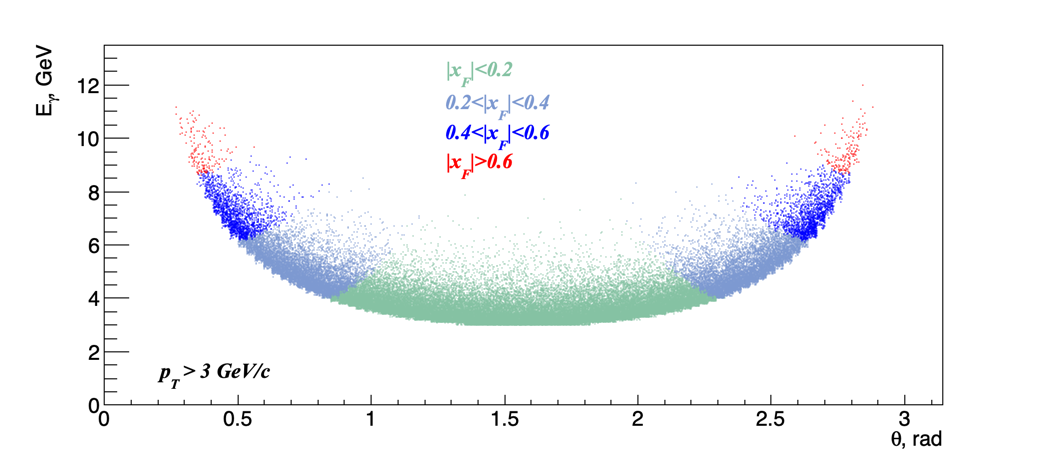}}
  \end{minipage}
	\caption{Expected uncertainty of the unpolarized cross-section $E d^3\sigma/dp^3$ measurement as a function of polar angle $\theta$. 
	}
	\label{PP_ACC_0}
\end{figure}

\begin{figure}[h!]
  \begin{minipage}[ht]{0.49\linewidth}
		    \center{\includegraphics[width=\linewidth]{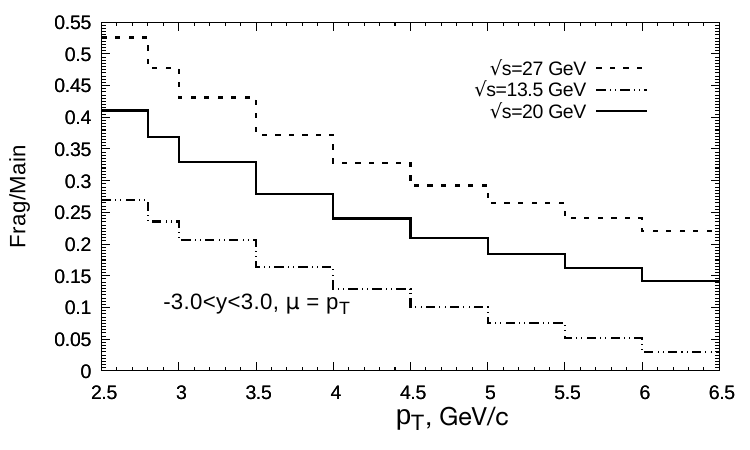}\\ (a)}
  \end{minipage}
  \hfill
  \begin{minipage}[ht]{0.49\linewidth}
		\center{\includegraphics[width=\linewidth]{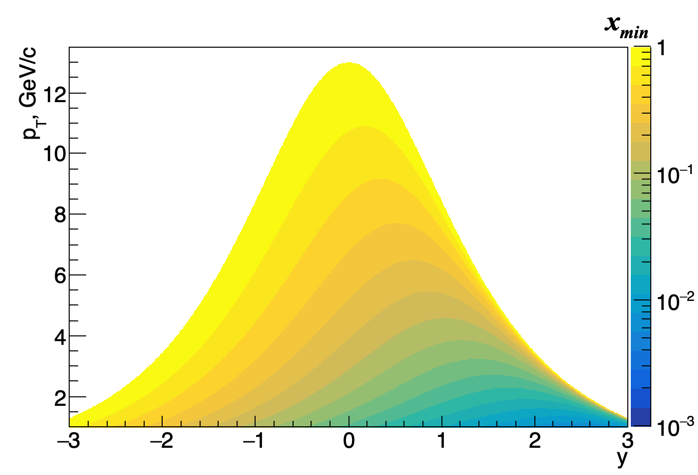}\\ (b)}
  \end{minipage}
	\caption{(a) Relative contribution of the fragmentation photons for the $p$-$p$ collision energies $\sqrt{s}=$13.5, 20, and 27 GeV.
	(b) Minimal value of gluon $x$ accessible via registration of prompt photon with rapidity $y$ and transverse momentum $p_T$ at $\sqrt{s}=27$ GeV. 
	 }
	\label{kinematic3}
\end{figure}

As it was already mentioned in Chapter \ref{chapter2}, two hard leading-order processes determine the production of prompt photons in the $p$-$p$ collisions in the leading order: gluon Compton scattering  $g q (\bar{q}) \to \gamma q (\bar{q})$ and quark-antiquark annihilation: $q\bar{q}\to g \gamma$. The contribution of the latter process to the total cross-section does not exceed 20\% in the discussed energy range. That is what makes the prompt photons a convenient probe for gluons inside the nucleon. The energy and angular distributions for prompt photons with $p_T>3$ GeV/$c$ are shown in Fig. \ref{PP_ACC_0}. There we do not take into consideration fragmentation photons, 
photons produced by fragmentation of final-state parton,
whose contribution is significant in the discussed range of  $p_T$, and which should also be treated as a part of the signal. Figure \ref{kinematic3}~(a) gives one an idea concerning the relative fraction of the fragmentation photons contribution with respect to the leading-order one. 
In the ultrarelativistic approximation the minimal value of the longitudinal momentum fraction of a struck parton $x_{min}$ accessible by detection of the prompt photon with the normalized transverse momentum $x_T=2p_T/\sqrt{s}$ and the rapidity $y$ can be expressed as \cite{Aurenche:1989gv}
\begin{equation}
x_{min}=\frac{x_T e^{-y}}{2-x_Te^y}.
\end{equation}
For fixed $x_T$, the minimal $x_{min}=x^2_T$ is reached at $y_0=-ln(x_T)$. The value $x_{min}$ as a function of rapidity $y$ and $p_T$ of photon for $\sqrt{s}=27$ GeV is shown in color in Fig.  \ref{kinematic3}(b). Here a low-$x$ parton goes against the $z$-axis. One can see that the possibility of accessing the low-$x$ region is limited by our capability to detect the prompt-photon signal at a low $p_T$ and the angular acceptance of the experimental apparatus. The latter is especially important for collider experiments like SPD, where large absolute values of $y$ correspond to the blind area  near  the beam pipe.

     The huge rate of decay photons makes the determination of the prompt photon production cross-section rather difficult. The main source of the decay photons is the two-body decay $\pi^0\to\gamma\gamma$. The second most important source is the decay $\eta\to\gamma\gamma$. In the kinematic range $p_T>3$ GeV/$c$ at $\sqrt{s}=27$ GeV there are about 0.18 photons from the $\eta$-decay per one photon from the $\pi^0$ decay. The relative contribution of all other decay photons ($\omega, \rho, \phi$ decays) does not exceed 0.03.
     
\begin{figure}[]
  \begin{minipage}[ht]{0.53\linewidth}
		    \center{\includegraphics[width=\linewidth]{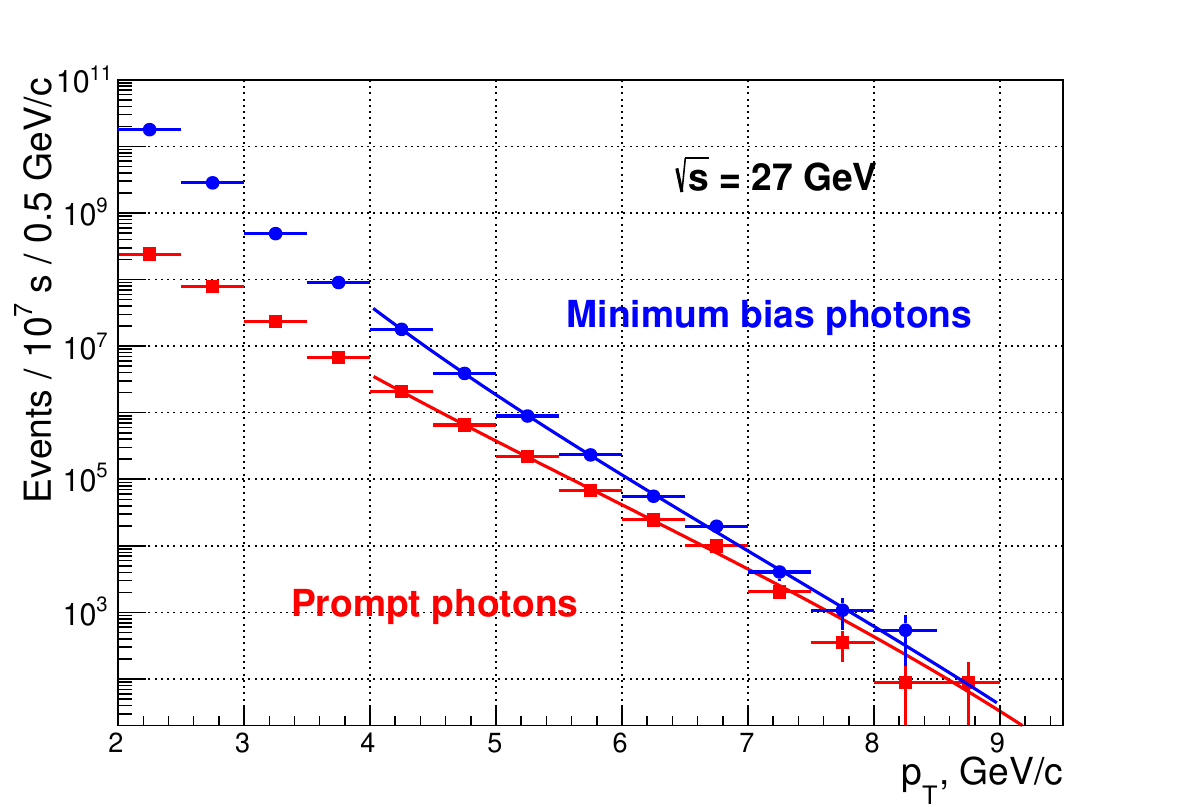}\\ (a)}
  \end{minipage}
  \hfill
  \begin{minipage}[ht]{0.46\linewidth}
		\center{\includegraphics[width=\linewidth]{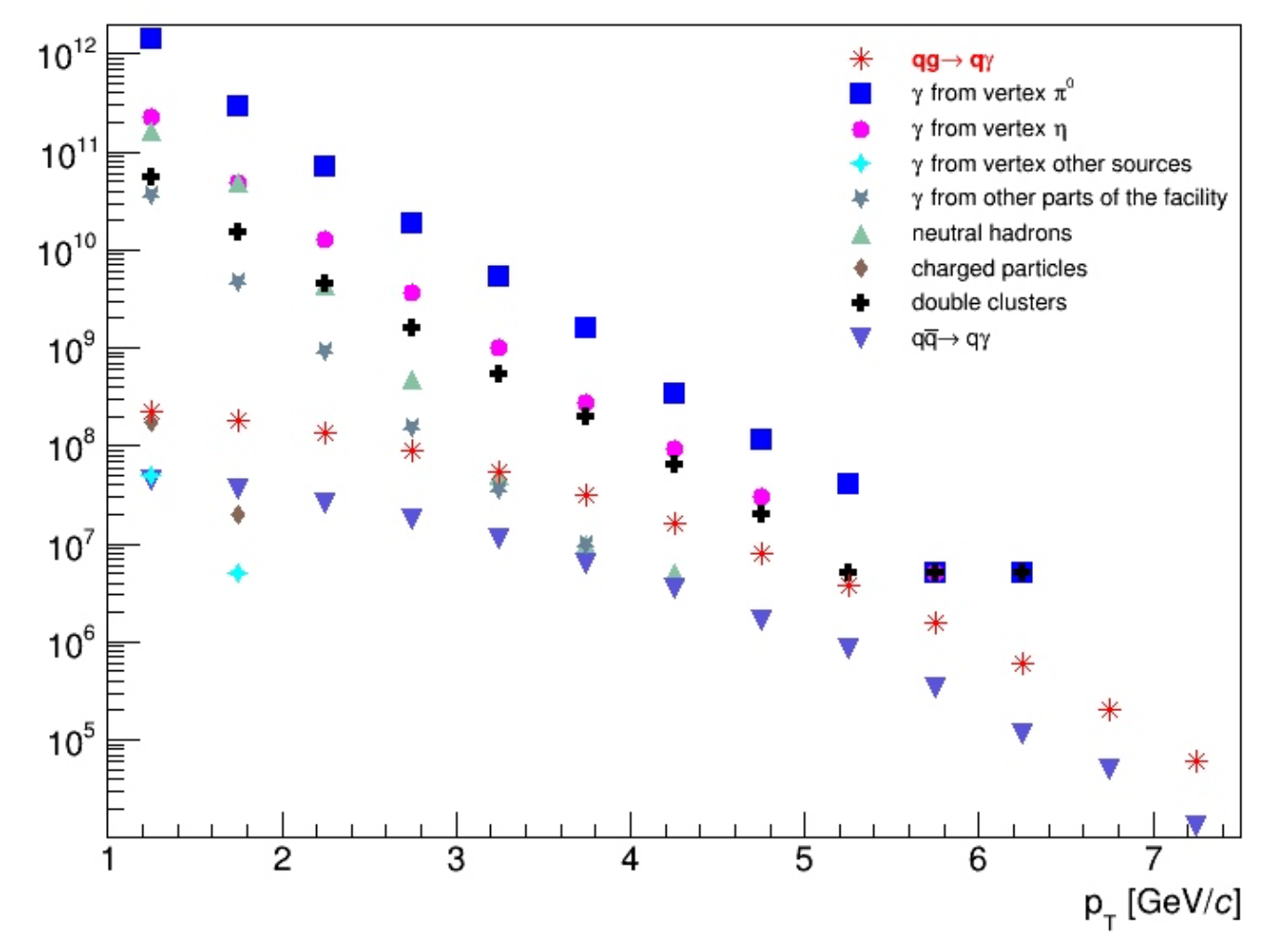}\\ (b)}
  \end{minipage}
	\caption{ 
	(a) $p_T$ spectra of the produced prompt (red) and the decay or minimum bias (blue) photons in $p$-$p$ collisions at $\sqrt{s} = 27$ GeV. Distributions are scaled to one year of data taking ($10^7$ s). (b) Contributions of different background components for the prompt photon production in $p$-$p$ collisions at $\sqrt{s}=27$ GeV (events per effective year of data taking) \cite{Rymbekova:2019rqd}.
	}
	\label{PP_ACC_2}
\end{figure}

      The $p_T$ spectra for the prompt and decay photons expected at SPD after one year of running at $\sqrt{s}=27$ GeV are presented in Fig. \ref{PP_ACC_2} (a). The result was obtained using the Pythia8 generator with the parameters tuned to reproduce the high-$p_T$ spectra of the  $\pi^0$ and prompt photons measured at similar energies by the WA70 ($\sqrt{s}=22.96$ GeV) \cite{Bonesini:1987ju,Bonesini:1987bv} and the UA6 ($\sqrt{s}=24.3$ GeV) \cite{Antille:1987kr} experiments, respectively. One can see that the $p_T$ spectrum of the decay photons goes down with the growth of the $p_T$ faster than for the prompt photons and their rates become comparable at $p_T\approx 7$ GeV/$c$. The fitted functions presented on the plot have the shape
\begin{equation}
N(p_T)=A(1-x_T)^n (p/p_0)^{-m}.
\end{equation}

Each cluster of the energy deposition in the ECal with the energy above the threshold $E_{0}=100$ MeV  that is not associated with any reconstructed tracks is treated as the  prompt photon candidate. The momentum of such a photon is reconstructed under the assumption of its production in the primary vertex. In order to reject photons from the $\pi^0\to\gamma\gamma$ decay the invariant mass of each two photons is calculated. If the difference between the reconstructed mass and the nominal mass of $\pi^0$ is smaller than 10 MeV, both photons are removed from the list of candidates. Nevertheless, this procedure removes just about 40\% of false candidates. The photons from the $\pi^0\to \gamma\gamma $  decay, whose partner  was not reconstructed due to conversion in the material, too low energy or the acceptance issue, remain on the list of candidates. The photons from the radiative decays of other particles are also in the list. The list of candidates also includes  photons associated with two or more overlapping clusters, first of all, the clusters from the decay of energetic $\pi^0$s. A significant part of such false candidates could be rejected by a sophisticated analysis of the cluster shape. The clusters produced by the charged particles whose tracks are lost, the clusters deposited by the photons originated from the elements of the setup and the clusters induced by the neutral hadrons are also taken into account as the background. The typical contributions of each source of the background mentioned above are presented as a function of $p_T$  in Fig.  \ref{PP_ACC_2}(b). 

As one can see, the photons from the unreconstructed decays of neutral pions are the main source of the background. The fraction of such unreconstructed decays can be estimated from the Monte Carlo simulation and is about 50\%. Based on the number $N_{\pi^0}$  of the reconstructed $\pi^0\to\gamma\gamma$ decays, the corresponding number of the remaining background photons $k\times N_{\pi^0}$ should be subtracted from the number of prompt-photon candidates $N_{\gamma}$ in order to get an estimation of a true number of prompt photons:
\begin{equation}
N_{prompt} = N_{\gamma} - k\times N_{\pi^0}.
\label{Nevents_eq}
\end{equation}

Here $k\approx 0.3$ is a coefficient calculated from the MC simulation that takes into account not only the inefficiency of the $\pi^0\to\gamma\gamma$ decay reconstruction, but also the overall contribution of all other background photons including the photons from the radiative decays of $\eta$, $\omega$, $\rho$,$\phi$, etc. The described subtraction procedure has to be performed for each bin of $p_T$ and $x_F$ ranges. One should keep in mind that the background of the decay photons is also spin-dependent, since there is an indication of non-zero asymmetries $A_{LL}$ and  $A_{N}$ in the inclusive $\pi^0$ and $\eta$ production \cite{Adams:1991rw, Adams:1991cs, Adare:2014hsq,Adare:2008aa,Adare:2008qb,Adare:2014qzo}.

The expected accuracy of the unpolarized cross-section $E d^3\sigma/dp^3$ measurement after one effective year ($10^7$ s) of data taking is shown in Fig.  \ref{PP_ACC_2}(b). 
At a low $p_T$ range the main contribution to the total uncertainty comes from the systematics of the $\pi^0$ background subtraction procedure while at  high $p_T$ the statistical uncertainty dominates. To estimate the systematics  $dk/k=1$\% is assumed as a realistic value. 

\begin{figure}[]
  \begin{minipage}[ht]{0.49\linewidth}
		    \center{\includegraphics[width=\linewidth]{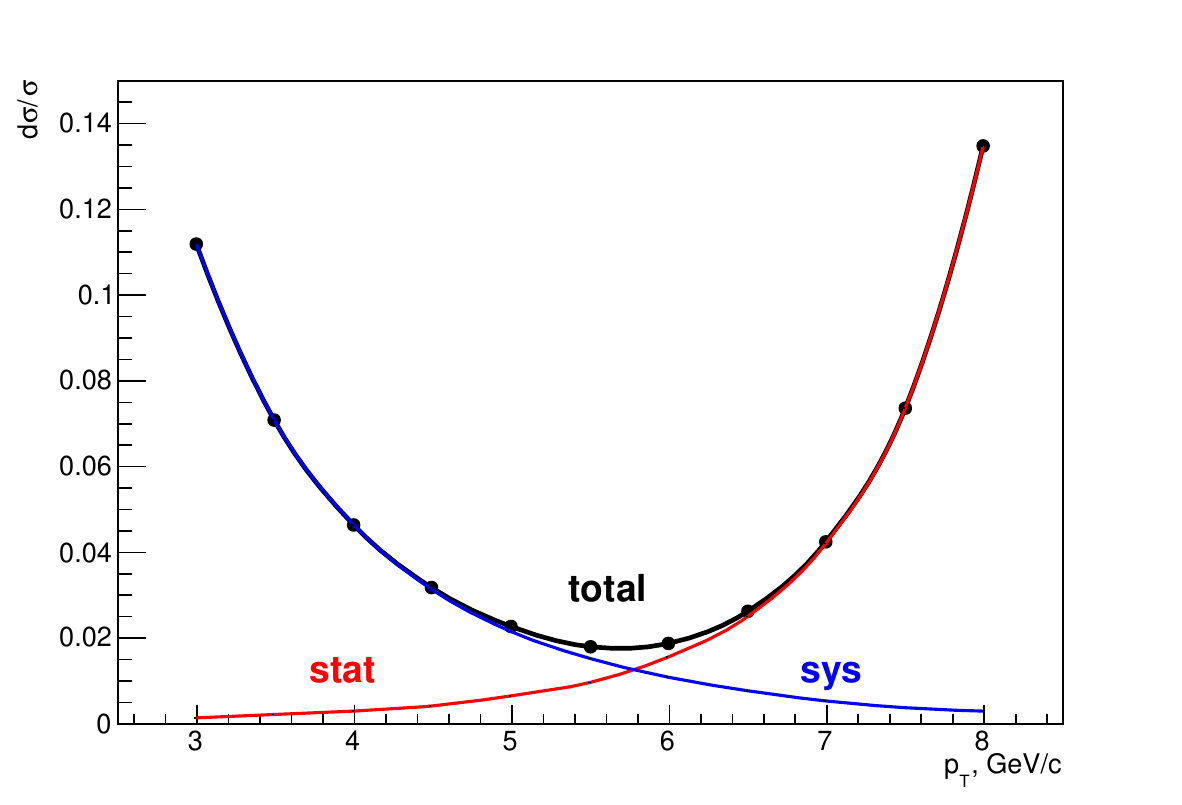}\\ (a)}
  \end{minipage}
  \hfill
  \begin{minipage}[ht]{0.49\linewidth}
	\center{\includegraphics[width=\linewidth]{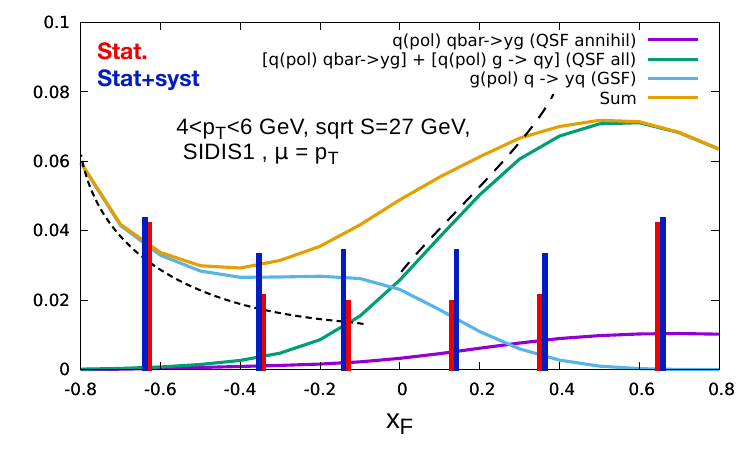}\\ (b)}
  \end{minipage}
	\caption{ 
	(a) Expected uncertainty of the unpolarized cross-section $E d^3\sigma/dp^3$ measurement as a function of $p_T$. 
	(b) Expected accuracy of the $A_N$ measurement for the prompt photons with $p_T> 4$ GeV/$c$ at $\sqrt{s}=27$ GeV as a function of $x_F$. The theoretical predictions are also shown. 
	}
	\label{PP_ACC_3}
\end{figure}

To estimate the $A_N$ asymmetry the function 
\begin{equation}
f(\phi) =  C+ P\times A_N cos~\phi
\end{equation}
is fitted to the expected  acceptance-corrected azimuthal distribution of the prompt-photon events. Here $\phi$ is the azimuthal angle of the photons produced  in the laboratory frame in respect to the direction of the proton beam polarization. The expected accuracy of the $A_{LL}$ and $A_N$ measurement as a function of $x_F$ is shown in Fig. \ref{PP_ACC_3}~(b). The statistical part is shown in red, while the total error that also includes the systematics related to the background subtraction ($dk/k=1\%$) is shown in blue. The error does not include the uncertainties related to luminosity and beam polarization measurement.

The systematic uncertainty related to the background subtraction could be partially reduced by the simultaneous study of the asymmetries in the production of prompt photons, $\pi^0$-, and $\eta$-mesons like it was demonstrated in the recently published analysis \cite{Acharya:2021fsd} . 


\subsection{Open charm production}

\begin{figure}[!h]
  \begin{minipage}[ht]{0.49\linewidth}
		    \center{\includegraphics[width=\linewidth]{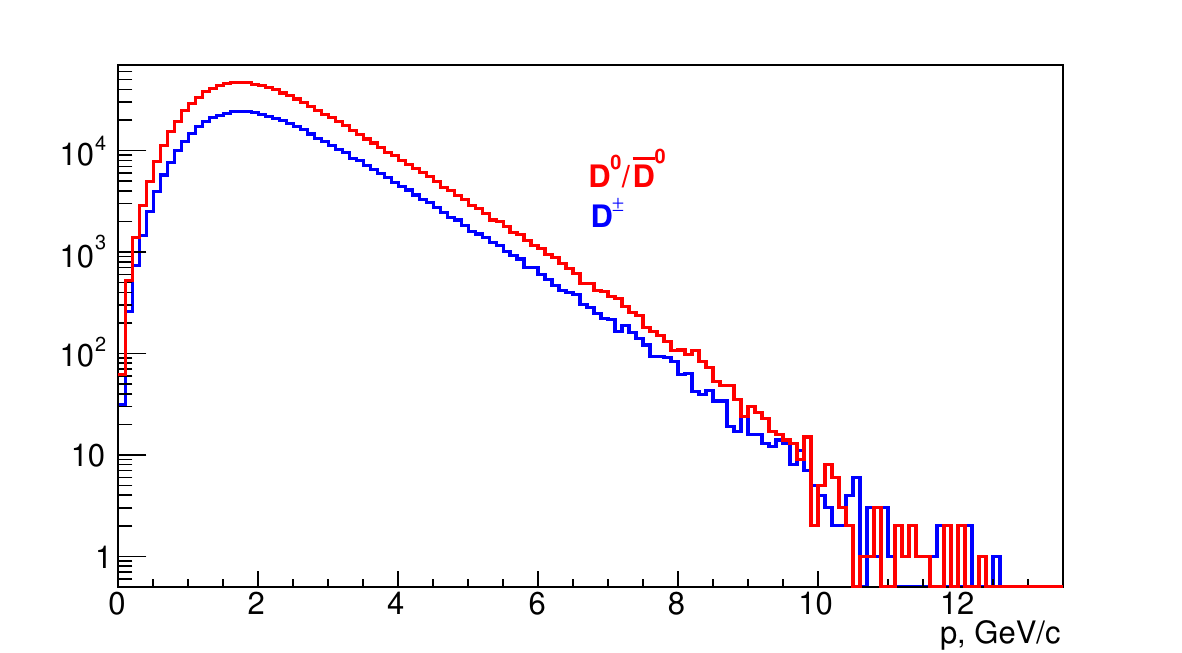}\\ (a)}
  \end{minipage}
  \hfill
  \begin{minipage}[ht]{0.49\linewidth}
	\center{\includegraphics[width=\linewidth]{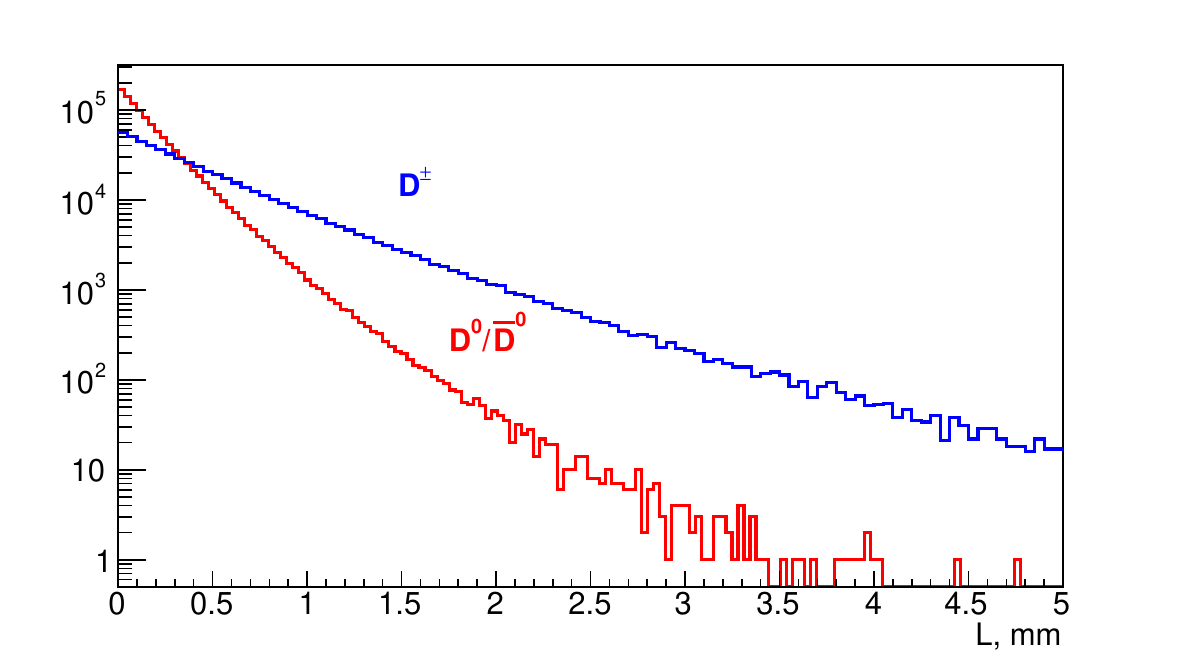}\\ (b)}
  \end{minipage}
	\caption{ (a) Momentum distributions for $D^{\pm}$ and $D^0 / \overline{D}^0$ produced in $p$-$p$ collisons at $\sqrt{s}=27$ GeV. (b) Spatial distance between the production and the decay vertices for $D$-mesons.
	}
	\label{fig:OC_1}
\end{figure}

In spite of the relatively large cross-section of the open charm production, most of the $D$-meson decays cannot be reconstructed easily. The ''golden'' decay channels are: $D^0\to K^-\pi^+$ and $D^+ \to K^-\pi+\pi^+$ (BF=3.95\% and  9.38\%, respectively). The momentum distributions for $D^{\pm}$ and $D^0 / \overline{D}^0$ produced in $p$-$p$ collisions at $\sqrt{s}=27$ GeV are shown in Fig. \ref{fig:OC_1}(a). The difference between the red and blue curves reflects the fact that the probability for the $c$-quark to hadronize into the neutral $D$-meson is 2 times higher than into the charged one.
Since the decay length $c\tau$ is 311.8 and 122.9 $\mu$m, respectively, which is larger  than the spatial resolution of the vertex reconstruction, the VD allowing one to reconstruct the secondary vertex of the $D$-meson decay is the key detector for the open charm physics at SPD. The distribution for the spatial distance between the primary (production) and secondary (decay) vertices for $D^{\pm/0}$ mesons is presented in Fig. \ref{fig:OC_1}(b).

We demonstrate the ability of the SPD setup to deal with the open-charm physics using the $D^0/\overline{D}^0$ signal. The following quantities  can be used as selection criteria in order to  suppress the combinatorial $\pi K$ background together with the kaon identification by the PID system:
\begin{itemize}
\item the quality of the secondary vertex reconstruction ($\chi^2$);
\item the distance between the primary and secondary vertices (spatial or in projections) normalized to the corresponding uncertainty;
\item the angle between the reconstructed momentum of the $\pi K$ pair and the line segment connecting the primary and secondary vertices;
\end{itemize}
The kinematics of the $\pi K$ pair (the angular and momentum distributions) could also be used for discrimination.

\begin{figure}[!h]
  \begin{minipage}[ht]{0.44\linewidth}
		    \center{\includegraphics[width=\linewidth]{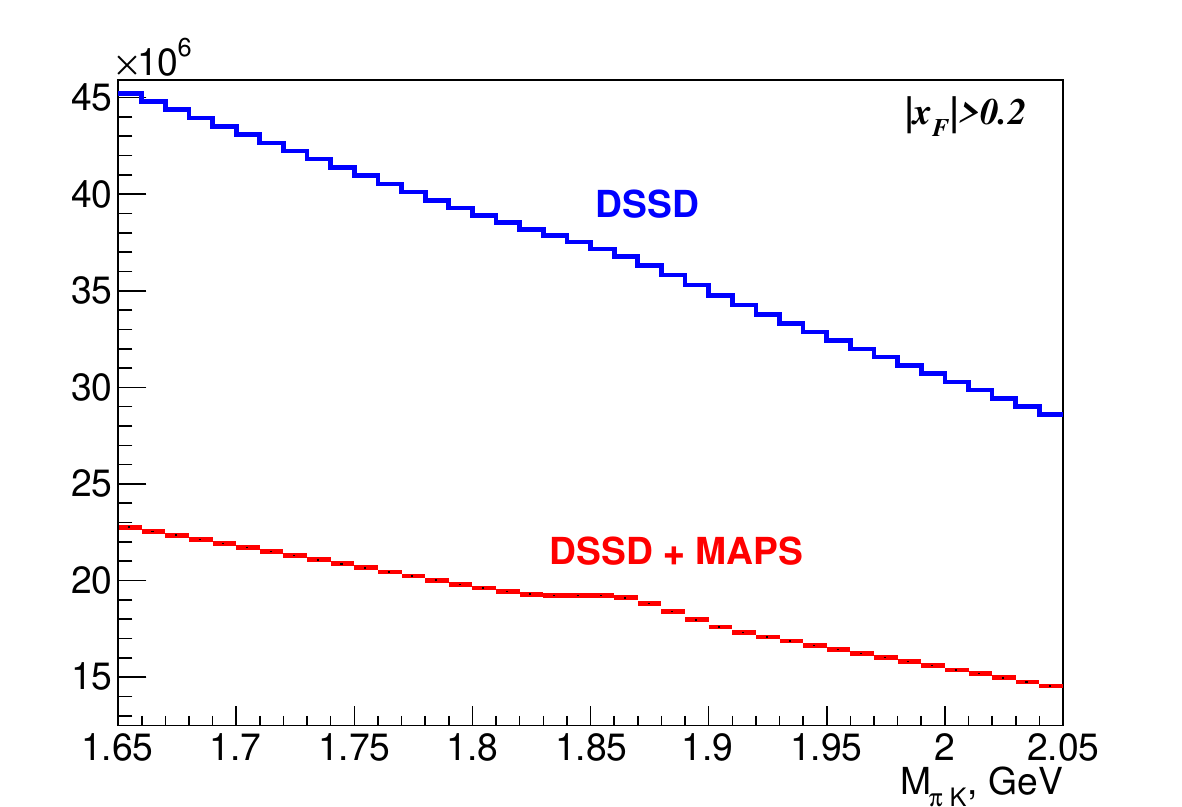}\\ (a)}
  \end{minipage}
  \hfill
  \begin{minipage}[ht]{0.54\linewidth}
	\center{\includegraphics[width=\linewidth]{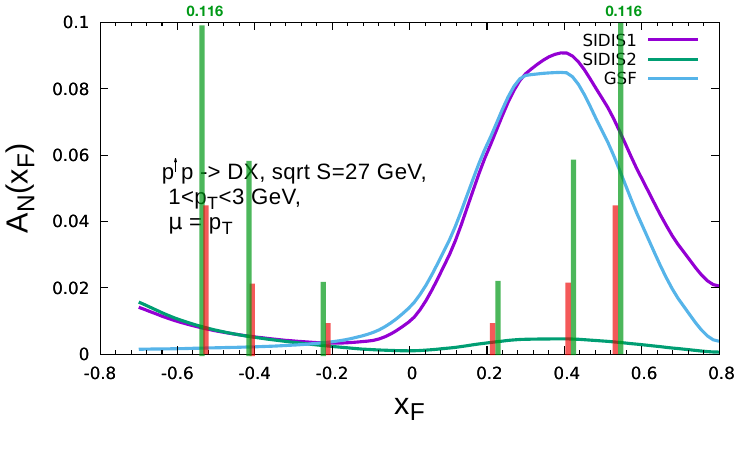}\\ (b)}
  \end{minipage}
	\caption{ (a) $K^{-}\pi^{+}$  invariant mass spectrum  after 1 year of data taking ($\sqrt{s}=27$ GeV, $|x_F|> 0.2$). (b) Corresponding statistical accuracy of the  $A_N$ measurement for the $D^0+\overline{D}^0$ mesons. The expected Sivers contribution to the SSA is also shown. 
	}
	\label{fig:OC_2}
\end{figure}

Figure \ref{fig:OC_2}~(a) presents the $K^{-}\pi^{+}$  invariant mass spectrum obtained as the result of such a selection for the $D^0$-signal in the kinematic range  $|x_F|> 0.2$ as an example for both variants of the VD after one year of data taking. About 96\% of the $D^0\to K^-\pi^+$ events  were lost, while the combinatorial background under the $D$-meson peak was suppressed by 3 orders of magnitude. The signal-to-background ratio for  $D^0$ is about 1.3\% for the DSSD configuration and about 3.9\% for the DSSD+MAPS one. Improving the signal-to-background ratio is the subject of further optimization of the selection criteria, as well as the reconstruction algorithms. The corresponding statistical accuracy of the SSA $A_N$ measurement is illustrated by Fig. \ref{fig:OC_2}~(b), where both signals, $D^0$ and $\overline{D}^0$, are merged. Similar or even better results (due to a larger $c\tau$ value) could be expected for the charged channel $D^{\pm}\to K^{\mp}\pi^{\pm}\pi^{\pm}$.

Another way to improve the signal-to-background ratio is tagging the $D$-mesons by their origin from the decay of a higher state $D^*\to D\pi$. The complexity of this approach lies in the need for detection of the soft  pion ($p_{\pi}\sim 0.1$ GeV/$c$).

One more possibility to reduce the background is the tagging a leptonic decay of the second $D$-meson in the event via reconstruction of the corresponding muon in the RS. The corresponding branching fractions ($\mu$ + anything) are $6.8\pm 0.6$\% and $17.6\pm3.2$\% for $D^0$ and $D^{\pm}$, respectively.
\chapter{Integration and services}

\begin{figure}[h!]
\begin{center}
\includegraphics[width=14cm]{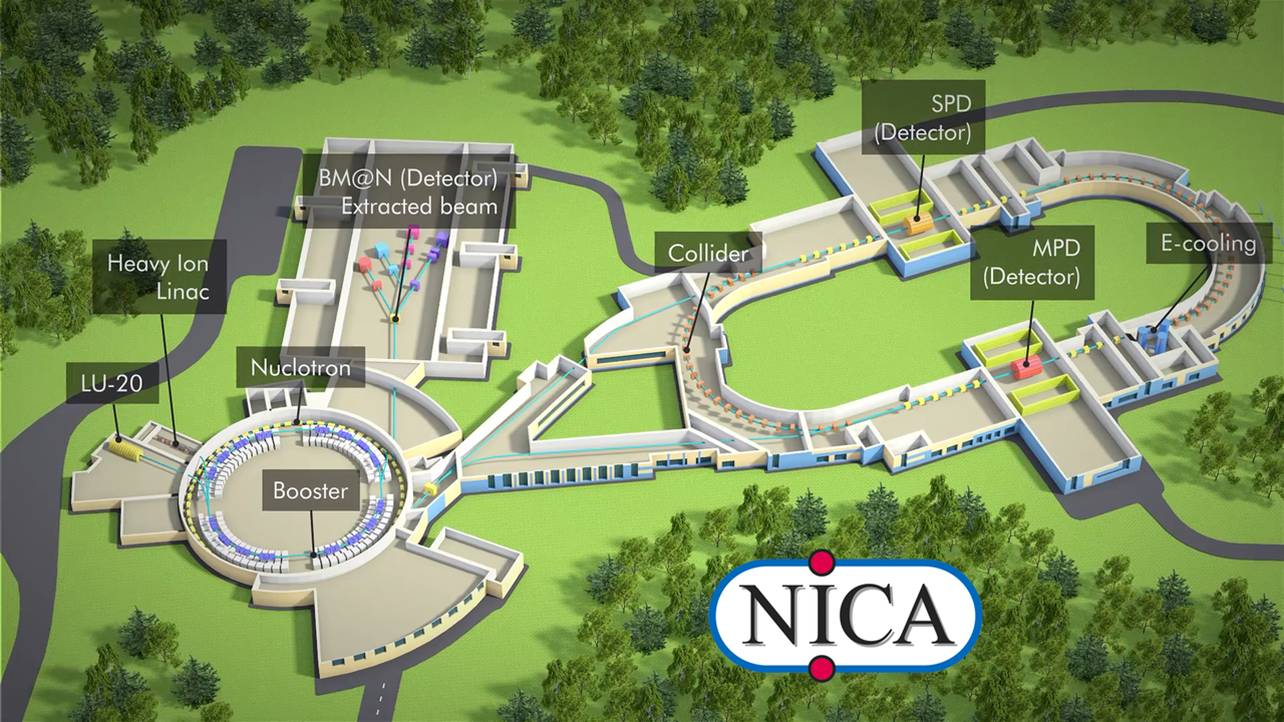}
\caption{Schematic layout of the NICA facility with SPD.}
\label{fig:nica}
\end{center}
\end{figure}

According NICA TDR \cite{NICA-TDR:2015} 
the SPD area is allocated in the southern point of beam collisions. 
The NICA complex with the SPD location  is shown in Figure \ref{fig:nica}.

The design of the SPD experimental hall is very similar to that of the MPD  \cite{MPDCDR2014}.


\section{Experimental building of SPD}

\begin{figure}[ht]
\begin{center}
\includegraphics[width=0.6\textwidth]{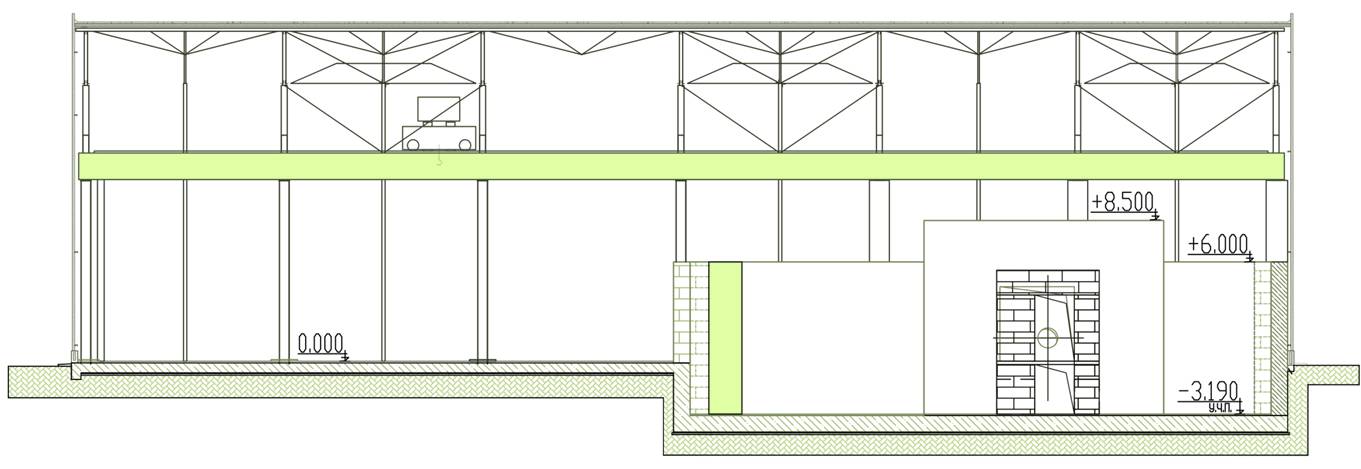}
\caption{Side view to the SPD building.}
\label{fig:spd-hall-1}
\vspace{0.5cm}
\includegraphics[width=0.6\textwidth]{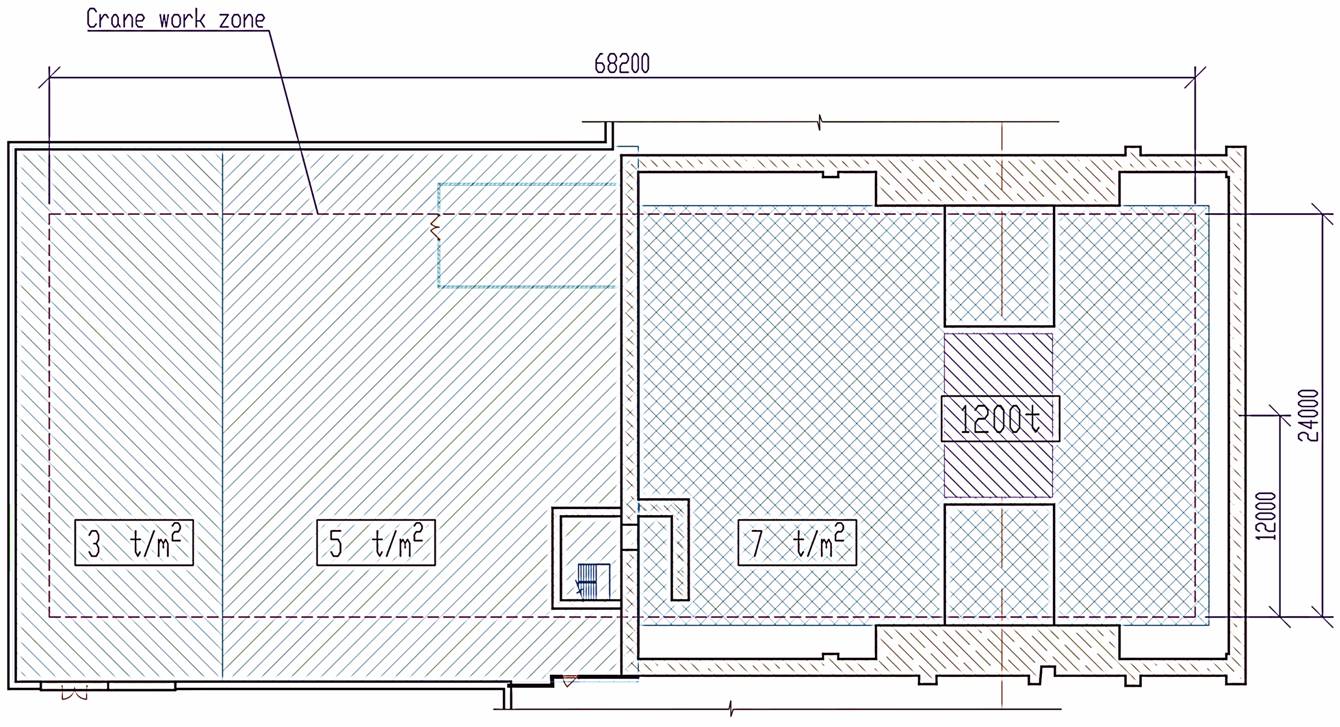}
\caption{The view from the top of the SPD building.}
\label{fig:spd-hall-2}
\end{center}
\end{figure}

The top and side views of the SPD building are shown in Figures \ref{fig:spd-hall-1}, \ref{fig:spd-hall-2}, respectively. 
The building is divided into  production and experimental sites with total area of about 2000~m$^2$. 
Since the production site has a gate, it will be used to unload materials.
The main gate for trucks is  \mbox{$4\,{\rm m} \times 4\,\rm{m}$}, 
while the dismantling part of wall for wider equipment is \mbox{$8\,{\rm m} \times \rm{8\,m}$}.
The production site will also be used for preparation, test and maintenance of the detector subsystems.
The floor of production site coincides with the ground level.

The experimental site is shown in the right side of the Figures \ref{fig:spd-hall-1}, \ref{fig:spd-hall-2},
below the level of the production site by 3.2\,m. 
It constructed in such a way in order to match the levels of the beam-line and the symmetry axis of the detector.
Two main rails are installed along the site  to provide transportation of the SPD detector from
the assembly to the operating (or beam) position.
There are additional rails in the detector assembly area, transverse to the main one. 
They have a maximum load capacity of 130~ton and will be used 
to insert inner subdetectors  using rail-guided support.
All rails are electrically connected to the metal reinforcement of the building which, in turn, 
is connected to numerous rod-electrodes deepened down to the earth’s surface by 3\,m 
below the building for grounding.

The concrete solid floor of the hall has sufficient load-bearing capacity to allow the assembly and 
operation of the SPD detector, including auxiliary structures.
It will be quite enough:

\begin{enumerate}
\item to withstand the weight of the fully assembled detector with all necessary services;
\item to preserve the integrity of the detector during its transportation on the rail-guided lodgement 
between the assembly and beam positions;
\item to provide a stable position of the detector with high accuracy during operating cycles.
\end{enumerate}

An overhead traveling crane with a maximum lifting capacity of 80 tons will be installed in the hall of the building.
The crane will be equipped with linear and H-shaped traverses with lifting capacity from few to 75~tons.
The crane service area covers both the production and experimental sites.

\section{Auxiliary equipment and services}

\begin{figure}[th]
    \begin{minipage}{0.49\linewidth}
    \center{\includegraphics[width=1\linewidth]{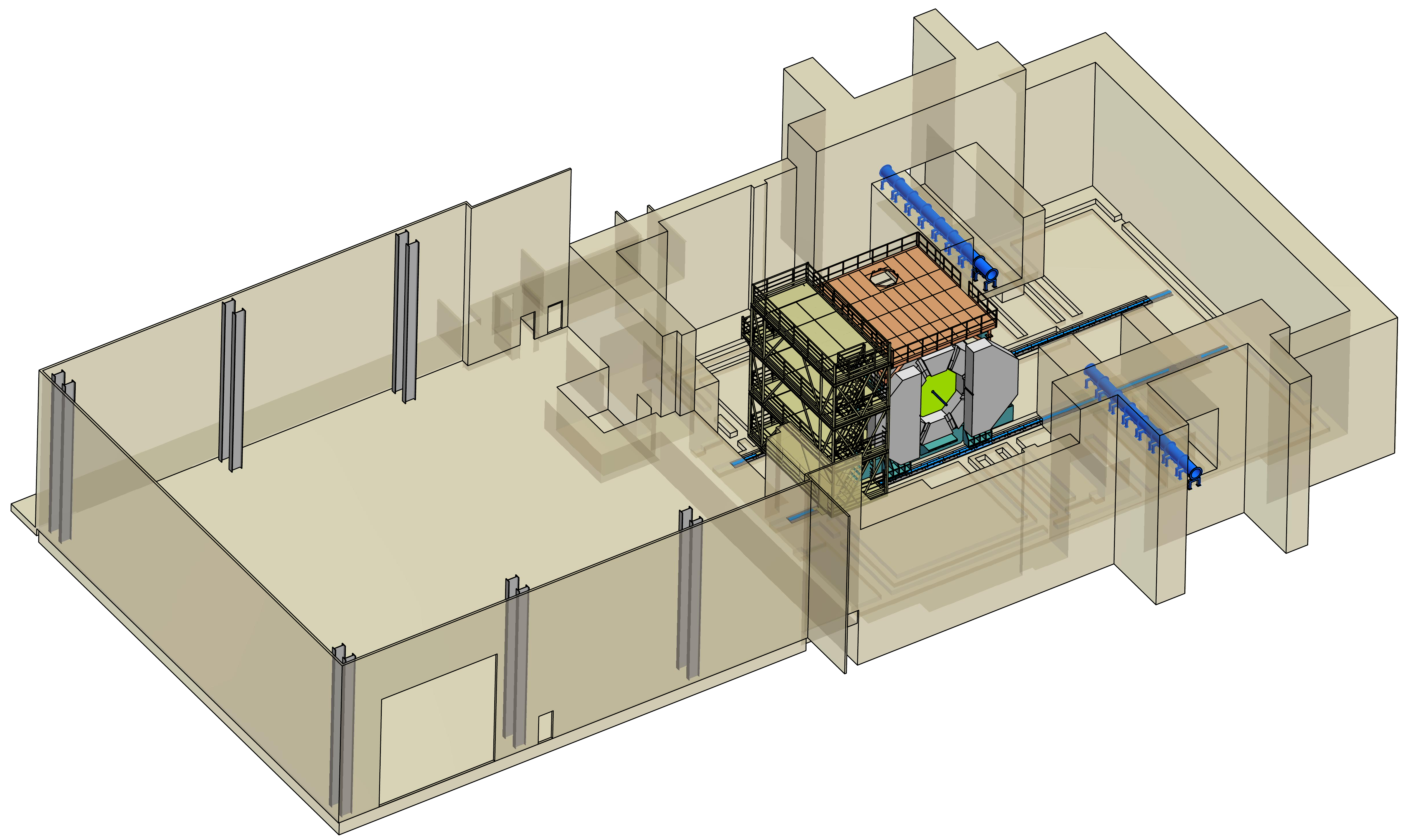} \\ (a)}
  \end{minipage}
  \hfill
  \begin{minipage}{0.49\linewidth}
    \center{\includegraphics[width=1\linewidth]{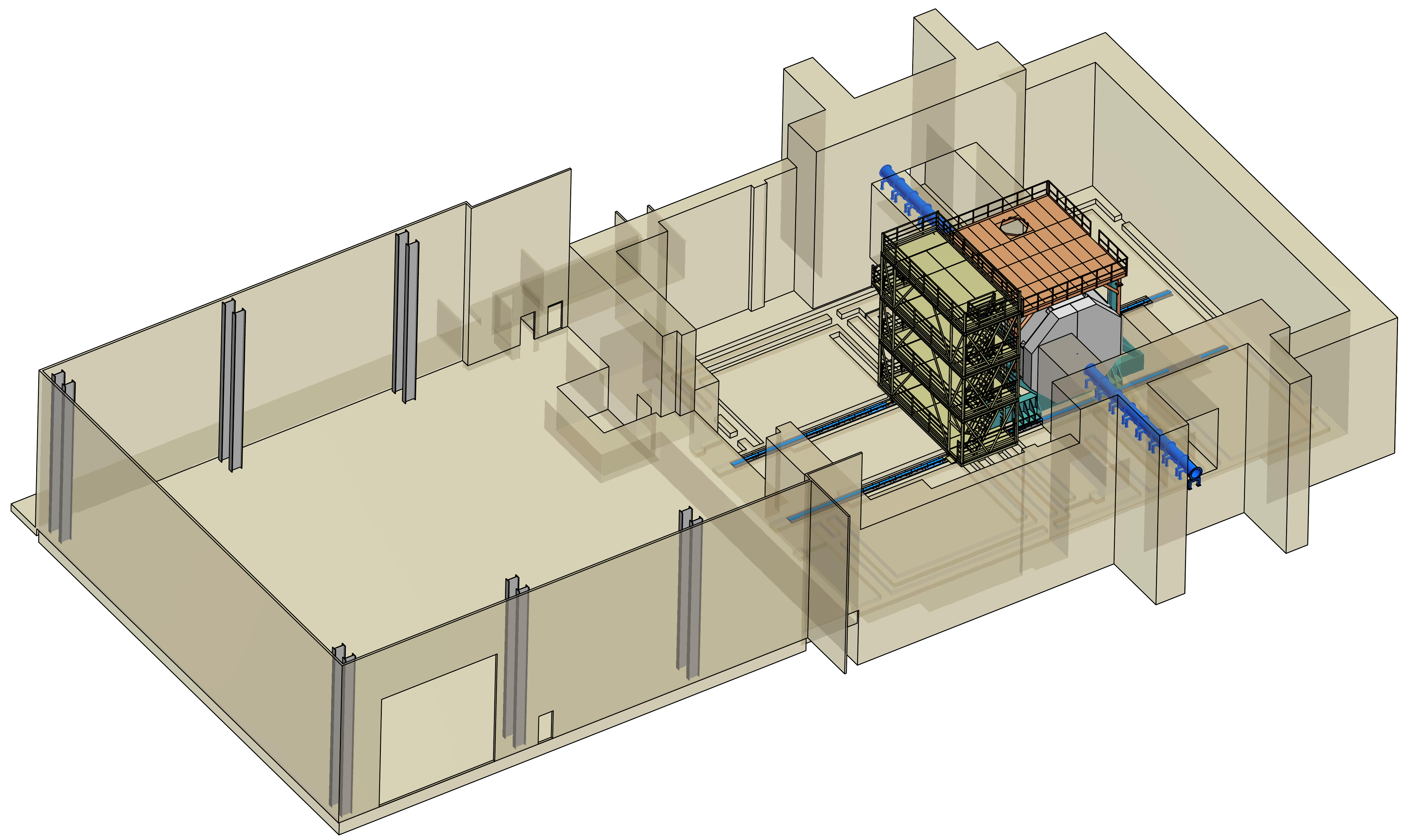} \\ (b)}
  \end{minipage} 
  \caption{The 3D view of the SPD building. The detector assembly position (a) and the beam position of the detector (b) are shown.
The beam line is highlighted  in blue.}
  \label{fig:spd-hall-3D}  
\end{figure}

Helium refrigerators have to be close to the detector to provide cryogenic fluids and gases for the magnet operation 
(see chapter 3.2). Presumably, they will be installed on a platform that will be located on top of the detector,
as shown in Figure~\ref{fig:spd-hall-3D}.
Crates with electronics for the data acquisition systems and power supplies of the detector subsystems
should also be placed in the vicinity of the detector. 
They will be located on a special platform next to the detector. The platform will be attached to the detector,
thus moving along the same rails as the detector itself.
Bottles with with Ar and CO$_2$ for suppling the RS and ST detectors, 
the corresponding gas recirculation compressors and monitoring systems, 
will be located on a similar platform from the opposite side of the detector (not shown in Figure~\ref{fig:spd-hall-3D}).
A dedicated radiation shielding can be installed between the detector and the platforms 
in order to provide access to the racks with the detector electronics and the gas system during data taking.

The total weight of the SPD detector is close to the maximum permissible limit of 1200~ton.
The weight of each detector subsystem is presented in Table~\ref{tab:spd-tech-req}.
The power consumption of subdetectors is currently under study. 
It is assumed that the maximum power supply capacity of the SPD hall should not exceed 1.2~MW.

\begin{table}[h!]
\caption{Technical requirements of the SPD detector subsystems (to be completed).}
\begin{center}
\begin{tabular}{|l|c|c|c|}
\hline
Subsystem  &  Weight, t & Power, kW  &  Special  \\
                    &                 &                     &  requirements \\
\hline
\hline
VD              &     $<0.1$ &    1$\div$1.5          &   water cooling                    \\
ST               &    0.1        &    2$\div$4             &  Ar+CO$_2$                 \\
ECAL          &    58         &    63              &                        \\
RS              &  800         &   31          &  Ar+CO$_2$   \\
TOF (sci)           &  0.4          &  4             &  \\
BBC           &    0.1         &    12                   &                        \\
\hline         
Magnet      &   50          &           $<100$            & cryogenics      \\
Supporting &&&\\
structure    &  200         &        $ \div$             &                         \\
\hline
DAQ          &                 &   46                 &                          \\
\hline
Total           & $\sim$1110      &   $\sim 200$              &                         \\ 
\hline
\end{tabular}
\end{center}
\label{tab:spd-tech-req}
\end{table}%

\chapter{Beam test facilities}
Two dedicated beam test facilities are planned to be operated for the benefit of the SPD project. 
The first one using secondary beams from the Nuclotron is foreseen to be constructed in building 205 (LHEP). 
It will be used for testing and certification of detector elements, electronics, data acquisition and slow-control systems under conditions close to those anticipated at NICA.
Some elements can be studied using the SPD straight section of NICA before the Spin Physics Detector construction at the early phase of the collider running. 
The beam test facilities will be also used for  training young specialists for the SPD project.

\section{Test zone with extracted Nuclotron beams }
\begin{figure}[!h]
  \begin{minipage}[ht]{1\linewidth}
    \center{\includegraphics[width=1.0\linewidth]{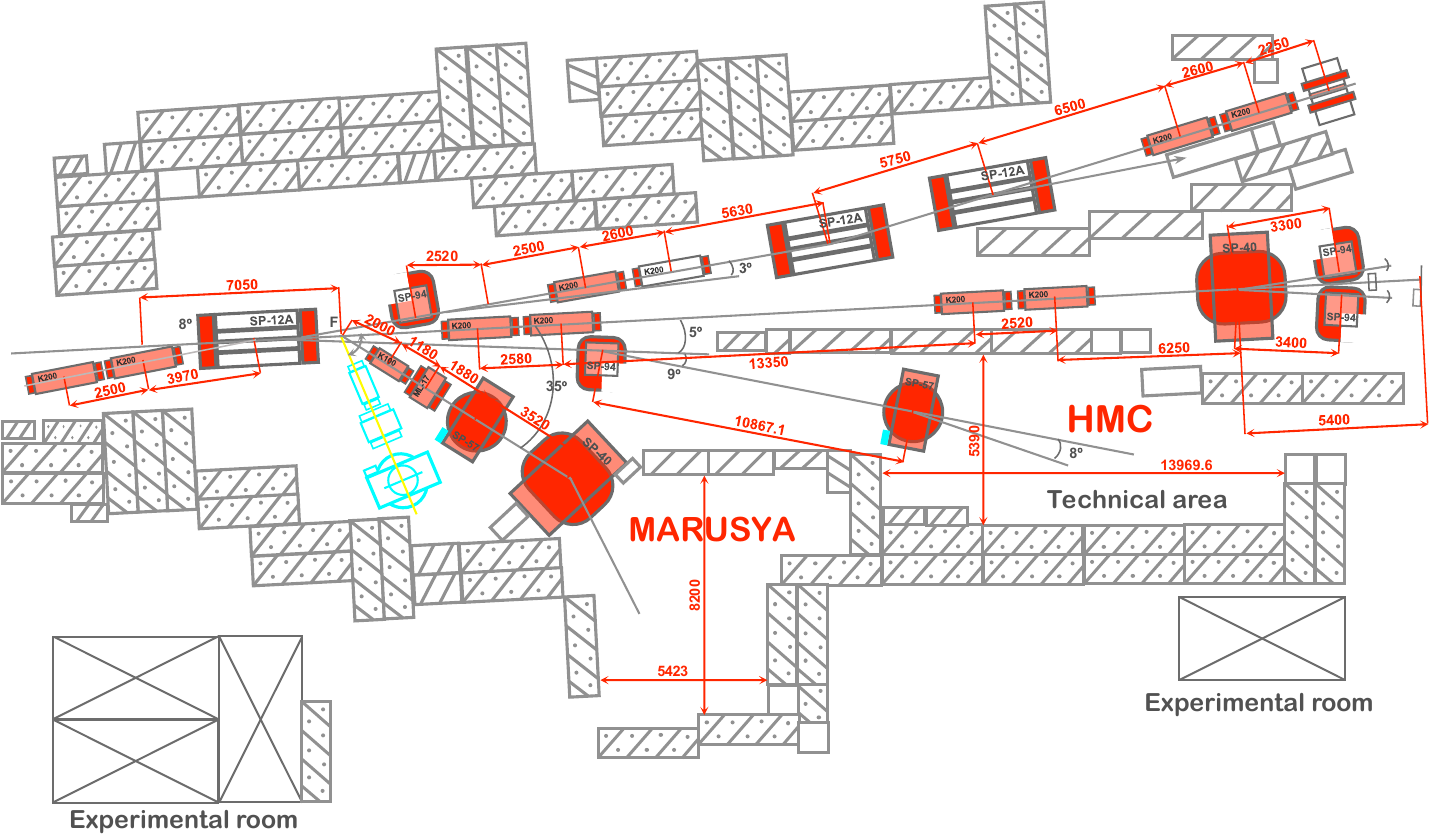}}
  \end{minipage}
  \caption{Layout of the main elements of the SPD test zone. All dimensions are in millimeters.}
  \label{fig:zone_1}  
\end{figure}

Two specialized channels for secondary particles (electrons, muons, pions, kaons, protons, neutrons, light nuclei) will be organized: the Low Momentum Channel (LMC) and the High Momentum Channel (HMC) in the region of focus F4 of extracted Nuclotron beams. The LMC is designated for secondary beams  with a momentum range from 100 MeV/$c$ to 2 GeV/$c$, while the momentum range of secondary particles at the HMC is from 1 GeV/$c$ to 10 GeV/$c$. After upgrading, the existing channel of the MARUSYA installation  \cite{Baldin2007,Baldin2006,Baldin2010,Baldin1998} will be used in the LMC construction.  It is advantageous that there exists positive experience in working with extracted polarized beams at MARUSYA  \cite{Baldin20}.
This would ensure physical measurements at the extracted beams using the existing experimental installation and infrastructure.
 The installation MARUSYA is well-suitable for the applied studies with secondary beams at the maximum possible intensity of the primary beam 
 extracted to building 205 up to $10^{11}$ protons per acceleration cycle. The development of the HMC requires two new magnetic 
 elements; therefore, it is considered as an independent installation to be put in operation at the second stage of upgrade in
  accordance with the existing regulations for the commissioning of experimental facilities. The layout of the main elements of the SPD test zone is shown in  Fig. \ref{fig:zone_1}. 
   
It is planned to use the SP12 magnet of the VP1 extraction channel situated directly in front of the F4 focus in order to turn the primary extracted beam toward the HMC. The calculations show that the primary beam can be turned to the required angle in a proton momentum range of 1$\div$7 GeV/$c$. For higher-energy particles, it is necessary to use a target in the F4 focus. In this operation mode, secondary beams are formed at the LMC and HMC simultaneously. Note that this operation mode is possible with parallel operation of the other installations at the VP1 extraction channel, in particular,  the BM\&N setup. 
For the primary 5 GeV/nucleon deuteron beam extracted to carbon targets with a thickness from 0.005 to 5 g/cm$^2$, the typical intensities for different species available at the LMC and HMC are shown in Tab. \ref{zone_tab}. There is also a possibility of forming a secondary quasi-monochromatic neutron beam via interaction of a deuteron beam in a target and deflecting out the charged component.
  
  \begin{table}[h]
\caption{Beam intensities feasible at the channel of the LMC and the HMC.}
\begin{center}
\begin{tabular}{|l|c|c|c|c|c|c|c|c|}
\hline
Channel &p, MeV/$c$ & d & p,n & $\pi^{\pm}$ & $K^+$ & $K^{-}$ & $\mu^{\pm}$ & $e^{\pm}$ \\
\hline
LMC &400 & $10^3$ & $10^5$ & $10^5$ & $10^3$ & $10^2$ & $10^3$ & $10^3$ \\
LMC &800 & $10^3$ & $10^4$ & $10^4$ & $10^3$ & $10^2$ & $10^3$ &  $10^3$\\
LMC &1500 & $10^2$ & $10^4$ & $10^4$ & $10^3$ & $10^2$ & $10^2$ &  $10^2$ \\
\hline
HMC &2000 & $10^4$ & $10^5$ & $10^4$ & $10^3$ & $10^2$ & $10^2$ &  $10^2$ \\
HMC &7000 & $10^4$ & $10^6$ & $10^3$ & $10^3$ & $10^2$ & $10^2$ &  $10^2$ \\
\hline
\end{tabular}
\end{center}
\label{zone_tab}
\end{table}%

Each channel-spectrometer provides spatial registration, identification, and tagging of each particle hitting the detector, provided 
 the electronic registration system of the installation matches the tested detector or the data acquisition system element. A prototype time-of-flight system based on  scintillation hodoscopes demonstrated reliable identification of protons, pions, kaons in a momentum range of 600$\div$1200 MeV/$c$ for the LMC. TOF scintillation hodoscopes providing a momentum resolution of 0.5\% and a time resolution at a level of 100 ps are capable of online detection and identification of secondary pions, kaons, and protons at the HMC. In order to extend the testing capabilities of the SPD test zone, it will be equipped with new coordinate detectors, a Cherenkov counter, and a  BGO-based electromagnetic calorimeter.

\section{Tests at the SPD straight section of the collider}

\begin{figure}[h]
  \begin{minipage}[ht]{1\linewidth}
		    \center{\includegraphics[width=\linewidth]{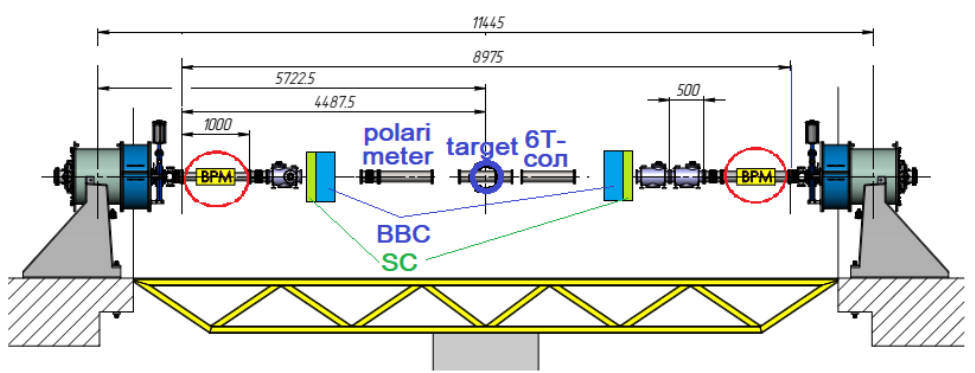}}
  \end{minipage}
	\caption{ The SPD straight section equipped with  the diagnostic and control units. All dimensions are in millimeters.
	}
	\label{fig:strategy4}
\end{figure}

We also suppose, that the beam test experiments and preparation for getting a luminosity of $10^{32}$ cm$^{-2}$ s$^{-1}$ at $\sqrt{s}=$ 27 GeV, including the proton polarization control will be demonstrated by the SPD commissioning. For that reason we propose to install of the diagnostic and control equipment at the SPD straight section (see Fig. \ref{fig:strategy4}). 
\chapter{Running strategy}
\section{Accelerator}

\begin{figure}[h]
  \begin{minipage}[ht]{1\linewidth}
		    \center{\includegraphics[width=0.7\linewidth]{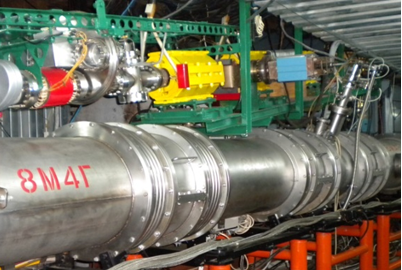}}
  \end{minipage}
	\caption{ View of the Nuclotron ring.
	}
	\label{fig:strategy1}
\end{figure}

We consider the strategy of the SPD operation as  chain of successive experimental works with polarized proton and deuteron beams aimed at the obtaining of the ultimate polarized proton beam parameters at the collider  and  the use of the unique existing  polarized deuteron beam for physics experiments from the early beginning of  the collider commissioning.
 Polarized deuterons  were first accelerated at the old LHEP proton accelerator  Synchrophasotron in 1986 and much later at the new superconducting synchrotron -- Nuclotron  in 2002 (see Fig. \ref{fig:strategy1}).

Polarized protons were first obtained in 2017. The first test was performed after the analysis of proton spin resonances in 2018. The first dangerous proton spin resonance in Nuclotron corresponds to the beam momentum of about 3.5 GeV/c, whereas in the deuteron case the spin resonance will occur at the particle kinetic energy of 5.6 GeV/nucleon. This limit is practically equal to the maximum achievable energy corresponding to the magnetic rigidity of the Nuclotron dipoles.

\begin{figure}[h]
  \begin{minipage}[ht]{1\linewidth}
		    \center{\includegraphics[width=\linewidth]{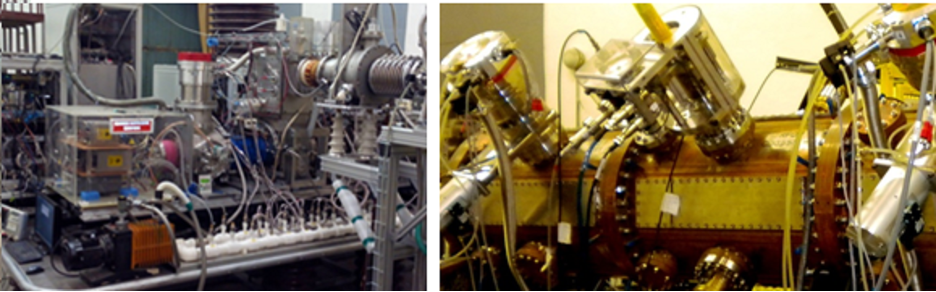}}
  \end{minipage}
	\caption{ View of the SPI (left)  and the existing RFQ (right).
	}
	\label{fig:strategy2}
\end{figure}
The existing polarized proton and deuteron ion source SPI provides up to  a 3-mA pulse current over $t\approx 100$ $\mu$s. Thus, up to $1.5\times10^{11}$ particles can be injected in the Nuclotron during the injection time (8 $\mu$s). The spin modes (pz, pzz): $(0,0)$, $(0,-2)$,      $(2/3, 0)$, and  $(-1/3,+1)$ were adjusted. The polarization degree of 80 \% was achieved.

The existing pre-accelerator of RFQ-type gives the limit on the achievable proton energy  in the next element of the injector chain - linac LU-20. We can obtain only 5 MeV at its output instead of 20 MeV that we have had in the past years. The new proton and light ion linac “LILAc” is now being manufactured \cite{Holtermann:2018hdg}. The LILAc output energy will be 12 MeV. Its commissioning is scheduled for  2025--2026. The photos of the SPI and the  existing RFQ are presented in Fig. \ref{fig:strategy2}.

The following tasks for the period of 2021--2025 are reasonable and necessary to start the SPD operation at the ultimate beam parameters:
\begin{itemize}
\item continuation of operation and further improvement  of the polarized ion source SPI, waiting beam time at the Nuclotron – 2021--2022;
\item	upgrade of the polarimeters: linak output; coasting beam; extracted beam; new polarimeter for a proton energy of above  6 GeV – 2020--2023;
\item	manufacturing of the 6T SC-solenoid model for the SPD test bench -- 2021--2022; 
\item	design and manufacturing of the equipment for the SPD test bench at the collider -- 2020--2023;
\item	LILAc manufacturing and tests – 2020--2025;
\item	analysis of the $^3$He (2+) polarized ion source based on the SPI upgrade.
\end{itemize}

\section{Spin Physics Detector}
The physics program proposed in Chapter \ref{chapter2} covers at least 5 years of the SPD running. A tentative running plan is shown in Tab. \ref{tab:summary}.

\begin{table}[htp]
\caption{Tentative running plan for the Spin Physics Detector.}
\begin{center}
\begin{tabular}{|l|c|c|}
\hline
Physics goal & Required time& Experimental conditions  \bigstrut \\
\hline
\hline
  \multicolumn{3}{|c|}{First stage}\\
\hline
Spin effects in $p$-$p$ scattering  & 0.3 year    & $p_{L,T}$-$p_{L,T}$, $\sqrt{s}<$7.5 GeV \bigstrut \\
dibaryon resonanses & & \\ 
\hline
Spin effects in $p$-$d$ scattering,      & 0.3 year    & $d_{tensor}$-$p$, $\sqrt{s}<$7.5 GeV \bigstrut \\
non-nucleonic structure of deuteron,  & & \\
$\bar{p}$ yield & & \\
\hline
Spin effects in $d$-$d$ scattering  & 0.3 year    & $d_{tensor}$-$d_{tensor}$, $\sqrt{s}<$7.5 GeV \bigstrut \\
hypernuclei & & \\ 
\hline
Hyperon polarization, SRC,  ...      & together with MPD    & ions  up to Ca \bigstrut \\
multiquarks & & \\
\hline
\hline
  \multicolumn{3}{|c|}{Second stage}\\
\hline
Gluon TMDs,      & 1 year    & $p_T$-$p_T$, $\sqrt{s}=$27 GeV \bigstrut \\
SSA for light hadrons & & \\
\hline
TMD-factorization test, SSA, & 1 year & $p_T$-$p_T$, 7 GeV$<\sqrt{s}<$27 GeV\bigstrut \\
charm production near threshold, & &  (scan) \\
onset of deconfinment, $\bar{p}$ yield & & \\
\hline
Gluon helicity, & 1 year & $p_L$-$p_L$, $\sqrt{s}=$27 GeV \bigstrut \\
... & & \\
\hline
Gluon transversity, & 1 year & $d_{tensor}$-$d_{tensor}$,  $\sqrt{s_{NN}}=$ 13.5 GeV \bigstrut \\
non-nucleonic structure of deuteron,    &            & or/and $d_{tensor}$-$p_T$,  $\sqrt{s_{NN}}=$ 19 GeV \bigstrut \\
''Tensor porlarized'' PDFs    &   &  \\
\hline
\end{tabular}
\end{center}
\label{tab:summary}
\end{table}%

\chapter{Cost estimate}

 The estimated cost of the Spin Physics Detector at current prices is 94.9 M\$. Taking into account the general uncertainty of the project timescale and  that at the moment the SPD project  is in the early stages of implementation, this value should be treated as a very preliminary estimation. This value does not include the construction of the SPD Test zone and possible R\&D expanses. Any expanses related with development and construction of an infrastructure for polarized beams at NICA are also out of this estimation.
 The detailed contribution to the total cost is presented in Tab. \ref{tab:cost}. 
 
\begin{table}[htp]
\caption{Preliminary cost estimate of the SPD setup.}
\begin{center}
\begin{tabular}{|l|l|c|c|}
\hline
\hline
& Subsystem       &  Option          &  Cost, M\$ \\     
\hline
\hline
SPD setup  &Vertex detector: & & \\
                   &~~~~  -- DSSD              &VD1 & ~9.5+6.5 (FE) \\
                   &~~~~  -- DSSD+MAPS  & VD2 & ~9.5+7.0 (FE) \\
\hline
&Straw tracker    &    & 2.4 \\
\hline
&PID system:      &   &  \\
&~~~~ -- RPC-based TOF     &PID1  & 5 \\
&~~~~ -- Scintillator-based TOF& PID2 & 4 \\
&~~~~ -- Aerogel PID system & PID3 & 5 \\
\hline
&Electromagnetic   &  &  21.1 \\
&calorimeter           & & \\
\hline
&Range system  &    &14.2 \\
\hline
&ZDC                  & & 2 \\
\hline 
&BBC                  & &0.4 \\
\hline
&Magnetic system & &10 \\
\hline
&Beam pipe           &  & 2 \\
\hline
\hline
General infrastructure & & & 5 \\
\hline
Slow control system  &  &  & 0.8\\
\hline
Data acquisition system &  & &2.6 \\
\hline
Computing & & &10 \\
\hline
\hline
TOTAL COST & VD2+PID2+PID3& & 96.0 \\
\hline
\hline

\end{tabular}
\end{center}
\label{tab:cost}
\end{table}%

\chapter{Participating institutions and author list}



\begin{center}
{\large \bf National Science Laboratory, Armenia}\\
Akopov N., Movsisyan A.
\end{center}

\begin{center}
{\large \bf Institute of Applied Physics 
of the National Academy of Sciences of Belarus, Minsk, Belarus }\\
Shulyakovsky R.
\end{center}

\begin{center}
{\large \bf Research Institute for Nuclear Problems of Belarusian State University, Minsk, Belarus}\\
Korjik M.V., Lobko A.S., Makarenko V.V., Solin A.A., Solin A.V.
\end{center}

\begin{center}
{\large \bf Universidad Andrés Bello, Santiago, Chile}\\
Kuleshov S., Zamora J.
\end{center}

\begin{center}
{\large \bf China Institute of Atomic Energy, Beijing, China}\\
Chen Lei, Jia Shi-Hai, Li Pei-Yu, Li Xiao-Mei, Song Jin-Xin, Sun Hao, Sun Peng-Fei, Zhang Yun-Yu, Zhang Jun-Wei, Zhao Ming-Rui, Zhi Yu
\end{center}

\begin{center}
{\large \bf Higher Institute of Technologies and Applied Sciences (InSTEC), Havana University, Havana, Cuba }\\
Guzman F., Gars\'{i}a Trapaga C.E. 
\end{center}

\begin{center}
{\large \bf Charles University, Prague, Czech Republic }\\
Finger M., Finger M. (jr.), Prochazka I., Slunecka M., Sluneckova V.
\end{center}

\begin{center}
{\large \bf Czech Technical University in Prague, Czech Republic }\\
Havranek M., Jary V., Lednicky D., Marcisovsky M., Neue G., Popule J., Tomasek L., Virius M., \fbox{Vrba V.}
\end{center}

\begin{center}
{\large \bf University of Turin and INFN Section, Turin, Italy}\\
Alexeev M., Amoroso A., Chiosso M., Denisov O.Yu., Kotzinian A., Maggiora A., Panzieri D., Parsamyan B., Tosello F.
\end{center}

\begin{center}
{\large \bf Warsaw University of Technology, Warsaw, Poland}\\
Buchowicz A., Dygnarowicz K., Galiński G., Klekotko A., Kurjata R., Marzec J., Pastuszak G., Rychter A., Zaremba K., Ziembicki M.
\end{center}

\begin{center}
{\large \bf Joint Institute for Nuclear Research, JINR, Dubna, Russia}\\



{\bf Laboratory of High-Energy Physics}\\

Akhunzyanov R.R., Alexakhin V.Yu., Anosov V.A.,  Azorskiy N.I., Baldin A.A., Baldina E.G., Barabanov M.Yu., 
 Beloborodov A.N., Belyaev A. V., Bleko V.V., Bogoslovsky D.N., Boguslavsky I.N.,  Burtsev V.E.,
 Dunin V.B., Enik T.L., Filatov Yu.N.\footnote{also Moscow Institute of Physics and Technology, Dolgoprudny, Russia}, Gavrishchuk O.P., Galoyan A.S., Glonti L., Golubykh S.M., Grafov N.O., Gribovsky A.S., Gromov S.A., Gromov V.A., Gurchin Yu.V., Gusakov Yu.V., Ivanov A.V., Ivanov N.Ya., Isupov A.Yu., Kasianova E.A., Kekelidze G.D., Khabarov S.V., Kharusov P.R., Khrenov A.N.,  Kokoulina E.S., Kopylov Yu.A.,  Korovkin P.S., Korzenev A. Yu., Kostukov  E.V., \fbox{Kovalenko A.D.}, Kozhin M.A., Kramarenko V.A., Kruglov V.N., Ladygin E.A., Ladygin V.P., Lednick\'{y} R., Livanov A.N., Lysan V.M., Makankin A.M., Martovitsky E.V., Meshcheriakov G.V.,  Moshkovsky I.V., Nagorniy S.N., Nikiforov D.N. Nikitin V.A., Parzhitsky S.S., Pavlov V.V.,  Perepelkin E.E.,  Peshekhonov D.V., Popov V.V., Reznikov S.G., Rogachevsky O.V., Safonov A.B., Salamatin K.M., Savenkov A.A.,   Sheremeteva A.I., Shimanskii S.S.,
  Starikova S.Yu., Streletskaya E.A., 
 Tarasov O.G., Terekhin A.A., Teryaev O.V., Tishevsky A.V., Topilin N.D., Topko B.L., Troyan. Yu.A., Usenko E.A., Vasilieva E.V., Volkov I.S.,Volkov P.V., Yudin I.P., Zamyatin N.I., Zemlyanichkina E.V., Zhukov I.A., Zinin A.V., Zubarev E.V.

{\bf Laboratory of Nuclear Problems} \\

Abazov V.M., Afanasyev L.G., Alexeev G.D., Balandina V.V., Belova A.P., Bobkov A.V., Boltushkin E.V., Brazhnikov E.V., Datta A., Denisenko I.I.,  Duginov V.N., 
Frolov V.N., Golovanov G.A., Gridin A.O., Gritsay K. I., Guskov A.V., Kirichkov N.V., Komarov V.I., Kulikov A.V., Kurbatov V.S.,  Kutuzov S.A., Maltsev A., Mitrofanov E.O.,  Nefedov Yu.A., Pavlova A.A., Piskun A.A., Prokhorov I.K., Rezvaya E.P.,  Romanov V.M., Rudenko A.I., Rumyantsev M.A., Rybakov N.A., Rymbekova A.,  Samartsev A.G., Shaikovsky V.N., Shtejer K., Sinitsa A.A., Skachkov N.B., Skachkova A.N., 
 Tereschenko V.V., Tkachenko A.V., Tokmenin V.V., Trunov N.O., Uzikov Yu.N., Verkheev A.Yu., Vertogradov L.S., Vertogradova Yu.L., Vesenkov V.A., Zhemchugov A.S., Zhuravlev N.I.

{\bf Laboratory of Theoretical Physics} \\
Anikin I.V., Arbuzov A.B., \fbox{Efremov A.V.}, Goloskokov S.V., Klopot Ya., Strusik-Kotlozh D., Volchansky N.I. 

{\bf Laboratory of Information Technologies} \\
Goncharov P.V., Oleynik D.A., Ososkov G.A., Petrosyan A.Sh., Podgainy D.V., Pelevanyuk I.S., Polyakova R.V., Uzhinsky V.V., Zuev M.I.

\end{center}

\begin{center}
{\large \bf St. Petersburg Nuclear Physics Institute, Gatchina, Russia}\\
Barsov S.G., Fedin O.L., Kim V.T., Kuznetsova E.V., Maleev V.P., Shavrin A.A., Zelenov A.V.
\end{center}

\begin{center}

{\large \bf  Lebedev Physical Institute of the Russian Academy of Sciences, Moscow, Russia}\\

Andreev V.F., Baskov V.A., Dalkarov O.D., Demikhov E.I., Gerasimov S.G., L'vov A.I., Negodaev M.A., Nechaeva P.Yu., Polyansky V.V.,  Suchkov S.I., Terkulov A.R.,  Topchiev N.P.

\end{center}

\begin{center}
{\large \bf  Skobeltsin Institute of Nuclear Physics of the Moscow State University,
Moscow, Russia}\\
Aleshko A.M., Belov I.N., Berezhnoy A.V., Boos E.E., Bunichev V.E., Chepurnov A.S., Gribkov D.Y., Merkin M.M., Nikolaev A.S. 

\end{center}

\begin{center}
{\large \bf  Institute for Theoretical and Experimental Physics, Moscow, Russia}\\
Akindinov A.V.,
Alekseev I.G., 
Golubev A.A.,
Kirin D.Yu.,
Luschevskaya E.,
Malkevich D.B.,
Morozov B.,
Plotnikov V.V.,
Polozov P.,
Rusinov V.,
Stavinskiy A.V.,
Sultanov R.I., 
Svirida D.N.,
Tarkovskyi E.I.,
Zhigareva N.M.
\end{center}

\begin{center}
{\large \bf  Institute for High-Energy Physics, Protvino, Russia}\\
Vorobiev A.
\end{center}

\begin{center}
{\large \bf Samara National Research University, Samara, Russia}\\
Karpishkov A.V., Nefedov M.A., Saleev V.A., Shipilova  A.V.
\end{center}

\begin{center}
{\large \bf St. Petersburg Polytechnic University, St. Petersburg, Russia}\\
Berdnikov A.Ya., Berdnikov Ya.A., Borisov V.S., Larionova D.M., Mitrankov Yu.M., Mitrankova M.M.
\end{center}

\begin{center}
{\large \bf St. Petersburg State University, St. Petersburg, Russia }\\
Feofilov G.A., Kovalenko V.N., Valiev F.F., Vechernin V.V., Zherebchevsky V.Yo.
\end{center}

\begin{center}
{\large \bf Tomsk State University, Tomsk, Russia }\\
 Chumakov A., Lyubovitskij V., Trifonov A.,  Zhevlakov A. 
\end{center}

\begin{center}
{\large \bf Belgorod State University, Belgorod, Russia }\\
Kaplii A.,
Kishchin I.,
Kluev A.,
Kubankin A., 
Kubankin Yu.
\end{center}

\begin{center}
{\large \bf University of Belgrade, Institute of Physics Belgrade, Belgrade, Serbia}\\
Jokovic D., Maletic D., Savic M.
\end{center}             
                
\begin{center}
{\large \bf V.N. Karazin Kharkiv National University,  Kharkiv, Ukraine}\\
Kovtun V.E., Lyashchenko V.N., Malykhina T.V., Reva S.N., Sotnikov V.V.
\end{center}              

\begin{center}
{\large \bf Institute for Scintillation Materials National Academy of Sciences of Ukraine (Kharkiv, Ukraine)}\\
Boyarintsev A.Yu., Eliseev D.A., Grinyov B.V., Zhmurin P.N.
\end{center}

\begin{center}
{\large \bf Individual contributors:}\\
Abramov V. (IHEP, Russia), El-Kholy R. (Cairo University), Haidenbauer J. (IAS and IKP, Forschungszentrum J{\"u}lich), 
Kondratenko A.M. (STL ''Zaryad'', Novosibirsk, Russia and Moscow Institute of Physics and Technology, Dolgoprudny, Russia),
Kondratenko M.A. (STL ''Zaryad'', Novosibirsk, Russia and Moscow Institute of Physics and Technology, Dolgoprudny, Russia),
Konorov I. (Technical University of Munich), Richard J.-M. (Universit\'e de Lyon), Semak A.A. (Institute of High Energy Physics, Protvino, Russia) Strikman M. (Penn State University),    Temerbayev  A. ( L.N. Gumilyov Eurasian National University, Nur-Sultan), Tomasi-Gustafsson E. (DPhN, IRFU, CEA, Universit\'e Paris-Saclay),  
Tsenov R. (University of Sofia), Wang Q. (South China Normal University, Guangzhou), Zhao Q. (Institute of High Energy Physics, CAS, Beijing)
\end{center}

\begin{center}
We are also grateful to  Kukhtin V. V., Nagaytsev A.P.,  Savin I.A., \fbox{Shevchenko O. Yu.}, and \fbox{Sisakyan A.N.} for their invaluable contribution to the SPD project at earlier stages of its implementation.
\end{center}

\begin{center}
{\large \bf
Acknowledgements.\\}
This work is supported by the RFBR research projects No. 18-02-40097 and 18-02-40061.
\end{center}








\chapter{Conclusion}
The proposed physics program and the concept of the SPD facility presented in the document, are open for exciting and challenging ideas from theorists and experimentalists worldwide.  The Conceptual Design Report of the Spin Physics Detector was presented and discussed at the 54th meeting of the JINR Program Advisory Committee for Particle Physics on January 18, 2021. The following recommendations were obtained: 
{\it 
      The PAC thanks the SPD (proto-)collaboration for the preparation of the comprehensive CDR and recommends the NICA management to appoint an appropriate Detector Advisory Committee for a thorough review of the CDR and its subsequent evolution into an SPD TDR (Technical Design Report). The PAC encourages the team to pursue every effort to form an international collaboration, find adequate resources and attract students and young scientists.
}

The international Detector Advisory Committee (SPD DAC) was formed in April 2021. 
The Committee is chaired by A. Bressan (University of Trieste, Italy) and includes P. Di Nezza (National Institute for Nuclear Physics, Frascati, Italy) and a member of the PAC P. Hristov (CERN, Switzerland). The report was presented by A. Bressan on behalf of the SPD DAC at the 56th meeting of the JINR Program Advisory Committee for Particle Physics on January 24, 2022. As an outcome of the report, the PAC issued the following recommendations:

{\it 
The PAC notes with interest the evaluation report presented by A. Bressan on behalf of the SPD DAC. The Committee held several meetings with the SPD team and asked several questions concerning the SPD detector concept. In addition to that, issues related to the NICA complex infrastructure for polarized beams and possible interactions between the SPD and MPD experiments were discussed in depth. The answers to the DAC's questions were satisfactory and the presentations during the joint meetings were well received by the committee. The DAC particularly appreciated the improvements in the SPD conceptual design with respect to the original CDR, namely changing the magnet location to be outside the ECAL, the possible use of a full MAPS inner tracker, and the clarifications on the straw detectors and the ZDC. On the basis of all that, and following the recommendation of the DAC, \textbf{the PAC approves the SPD CDR and asks the SPD team to move forward to the TDR preparation.} The PAC appreciates the important role of the DAC in the SPD project evaluation and requests periodic DAC reports.

}

  \chapter{List of abbreviations}
  ADC -- Analogue-to Digital Converter\\
  ASIC -- Application-Specific Integrated Circuit\\
  BBC -- Beam-Beam Counter\\
  BF -- Branching Fraction\\
  BPM -- Beam Position Monitor \\
  CCR -- Constituent Counting Rules\\
  CEM -- Color Evaporation Model\\
  CM -- Center-of-Mass\\
  CPM -- Collinear Parton Model\\
  CPQ -- Chromomagnetic Quark Polarization\\
  CR -- Cosmic Rays\\
  CT -- Color Transparency\\
  DAC --  Digital-to-Analogue Converter\\
  DAQ -- Data AcQuisition\\
  DIS -- Deep Inelastic Scattering\\
  DS(-mode) -- Distinct Spin mode\\
  DSSD -- Double-Sided Silicon Detector\\
  DY (process) -- Drell-Yan process \\
  ECal -- Electromagnetic Calorimeter\\
  ExEP -- Extended Equivalence Principle\\
  FBBC (monitor) -- Fast Beam-Beam Collisions (monitor)\\
  FEE -- Front-End Electronics\\
  FF -- Fragmentation Function\\
  GF -- Gluon Fusion\\
  GPD -- Generalized Parton Distributions\\
  GPM -- Generalized Parton Model\\
  GSF -- Gluon Sivers Function\\
  GTMD (PDF) -- Generalized  Transverse Momentum Dependent  (PDF)\\
  HMC -- High Momentum Channel\\
  ICEM -- Improved Color Evaporation Model\\
   IP -- Interaction Point\\
   ISM -- InterStellar Medium\\
  LDME -- Long-Distance Matrix Element\\
  LMC -- Low Momentum Channel\\
  MAPS -- Monolithic Active Pixel Sensor\\
  MCP -- MicroCannel Plate\\
  MDT -- Mini Drift Tubes \\
  ML -- Machine Learning \\
  MIP -- Minimum Ionizing Particle \\
  MPD -- MultiPurpose Detector\\
  MPPC - MultiPixel Photon Counter\\
  MRPC - Multigap Resistive Plate Chamber\\
  MS -- Magnetic System\\
  NICA -- Nuclotron-based Ion Collider fAcility\\
  NPE -- Number of PhotoElectrons\\
  NPQCD -- Non-Perturbative QCD \\
  NRQCD -- Non-Relativistic QCD \\ 
  PDF -- Parton Distribution Function\\
  PGF -- Photon-Gluon Fusion\\
  PID -- Particle IDentification\\
  pQCD -- perturbative QCD\\
  PRA -- Parton Reggeization Approach\\
  QGP -- Quark Gluon Plasma\\
  QSF -- Quark Sivers Function\\
  RMS -- Root Mean Square\\
  RS -- Range System\\
  SF -- Spin Flipping \\
  SIDIS -- Semi-Inclusive Deep-Inelastic Scattering\\
  SiPM -- Silicon PhotoMultiplier\\
  SN -- Spin Navigator (solenoid)\\
  SPD -- Spin Physics Detector\\
  SRC -- Short Range Correlations\\
  ST(-mode) -- Spin Transparency (mode)\\ 
  SSA -- Single Spin Asymmetry\\
  ST -- Straw Tracker\\
  STSA -- Single Transverse Spin Asymmetries\\
  TMD (PDF) -- Transverse Momentum Dependent  (PDF)\\
  TMDShF -- TMD Shape Function\\
  TOF (system) -- Time-Of-Flight (system)\\
  VD -- Vertex Detector\\
  WLS - WaveLength Shifter\\
  ZDC -- Zero Degree Calorimeter\\


\bibliographystyle{./sty/hunsrt} 


\bibliography{./bib/case_DY,./bib/case_PPa.bib,./bib/jpsi.bib,./bib/DAQ,./bib/case_GPD,./bib/polarimetry,./bib/ECAL,./bib/general,./bib/summary,./bib/htp,./bib/central,./bib/csw,./bib/t0,./bib/other.bib,./bib/quark.bib,./bib/pid.bib,./bib/vectors,./bib/physics1.bib,./bib/straw,./bib/ZDC,./bib/MC,./bib/reham,./bib/slow,./bib/charmoniaMC,./bib/RS,./bib/tof}


\end{document}